\documentclass[epj, final]{svjour}
\usepackage{graphics}
\usepackage{lineno}
\input{ds.def}
\begin{document}
\title{DarkSide-20k: A 20 Tonne Two-Phase LAr TPC for Direct Dark Matter Detection at LNGS}
\authorrunning{The DarkSide Collaboration}
\titlerunning{DarkSide-20k}
\institute{Pacific Northwest National Laboratory, Richland, WA 99352, USA \and Fondazione Bruno Kessler, Povo 38123, Italy \and Trento Institute for Fundamental Physics and Applications, Povo 38123, Italy \and Department of Physics, University of Houston, Houston, TX 77204, USA \and Instituto de F\'isica, Universidade de S\~ao Paulo, S\~ao Paulo 05508-090, Brazil \and Centro Fermi Museo Storico della Fisica e Centro Studi e Ricerche ``Enrico Fermi'', Roma 00184, Italy \and INFN Bologna, Bologna 40126, Italy \and Physics Department, Universit\`a degli Studi di Bologna, Bologna 40126, Italy \and Physics Department, Augustana University, Sioux Falls, SD 57197, USA \and Civil and Environmental Engineering Department, Politecnico di Milano, Milano 20133, Italy \and INFN Milano, Milano 20133, Italy \and CIEMAT, Centro de Investigaciones Energ\'eticas, Medioambientales y Tecnol\'ogicas, Madrid 28040, Spain \and INFN Pisa, Pisa 56127, Italy \and Physics Department, Universit\`a degli Studi di Pisa, Pisa 56127, Italy \and INFN Milano Bicocca, Milano 20126, Italy \and INFN Sezione di Roma, Roma 00185, Italy \and Budker Institute of Nuclear Physics, Novosibirsk 630090, Russia \and Novosibirsk State University, Novosibirsk 630090, Russia \and INFN Laboratori Nazionali del Gran Sasso, Assergi (AQ) 67100, Italy \and INFN Cagliari, Cagliari 09042, Italy \and Gran Sasso Science Institute, L'Aquila 67100, Italy \and Physics Department, Universit\`a degli Studi di Genova, Genova 16146, Italy \and INFN Genova, Genova 16146, Italy \and INFN Roma Tre, Roma 00146, Italy \and Mathematics and Physics Department, Universit\`a degli Studi Roma Tre, Roma 00146, Italy \and Physics Department, Universit\`a degli Studi di Cagliari, Cagliari 09042, Italy \and Institute for Particle Physics, ETH Z\"urich, Z\"urich 8093, Switzerland \and Chemistry and Pharmacy Department, Universit\`a degli Studi di Sassari, Sassari 07100, Italy \and INFN Laboratori Nazionali del Sud, Catania 95123, Italy \and Physics Department, Universit\`a degli Studi ``Federico II'' di Napoli, Napoli 80126, Italy \and INFN Napoli, Napoli 80126, Italy \and Virginia Tech, Blacksburg, VA 24061, USA \and Skobeltsyn Institute of Nuclear Physics, Lomonosov Moscow State University, Moscow 119991, Russia \and Physics Department, Universit\`a degli Studi di Torino, Torino 10125, Italy \and INFN Torino, Torino 10125, Italy \and Department of Mechanical, Chemical, and Materials Engineering, Universit\`a degli Studi, Cagliari 09042, Italy \and LPNHE, Universit\'e Pierre et Marie Curie, CNRS/IN2P3, Sorbonne Universit\'es, Paris 75252, France \and Chemistry, Materials and Chemical Engineering Department ``G.~Natta", Politecnico di Milano, Milano 20133, Italy \and INAF Capodimonte Astronomical Observatory, Napoli 80131, Italy \and Interuniversity Consortium for Science and Technology of Materials, Firenze 50121, Italy \and Saint Petersburg Nuclear Physics Institute, Gatchina 188350, Russia \and Physics Department, Princeton University, Princeton, NJ 08544, USA \and Physics Department, Sapienza Universit\`a di Roma, Roma 00185, Italy \and TRIUMF, 4004 Wesbrook Mall, Vancouver, British Columbia V6T2A3, Canada \and National Institute for R\&D of Isotopic and Molecular Technologies, Cluj-Napoca, 400293, Romania \and Joint Institute for Nuclear Research, Dubna 141980, Russia \and APC, Universit\'e Paris Diderot, CNRS/IN2P3, CEA/Irfu, Obs de Paris, USPC, Paris 75205, France \and Department of Chemistry, University of Crete, P.O. Box 2208, 71003 Heraklion, Crete, Greece \and Physics Department, Temple University, Philadelphia, PA 19122, USA \and National Research Centre Kurchatov Institute, Moscow 123182, Russia \and Institute of High Energy Physics, Beijing 100049, China \and School of Natural Sciences, Black Hills State University, Spearfish, SD 57799, USA \and Civil and Environmental Engineering Department, Universit\`a degli Studi di Enna ``Kore'', Enna 94100, Italy \and Department of Physics and Engineering, Fort Lewis College, Durango, CO 81301, USA \and Department of Physics, University of California, Davis, CA 95616, USA \and Fermi National Accelerator Laboratory, Batavia, IL 60510, USA \and Radiation Physics Laboratory, Belgorod National Research University, Belgorod 308007, Russia \and Electronics, Information, and Bioengineering Department, Politecnico di Milano, Milano 20133, Italy \and Energy Department, Politecnico di Milano, Milano 20133, Italy \and Physics Institute, Universidade Estadual de Campinas, Campinas 13083, Brazil \and National Research Nuclear University MEPhI, Moscow 115409, Russia \and Physics Department, Universit\`a degli Studi di Trento, Povo 38123, Italy \and Department of Physics and Astronomy, University of Hawai'i, Honolulu, HI 96822, USA \and Department of Physics, Royal Holloway, University of London, Surrey TW20 0EX, UK \and Amherst Center for Fundamental Interactions and Physics Department, University of Massachusetts, Amherst, MA 01003, USA \and Chemistry, Biology and Biotechnology Department, Universit\`a degli Studi di Perugia, Perugia 06123, Italy \and INFN Perugia, Perugia 06123, Italy \and M. Smoluchowski Institute of Physics, Jagiellonian University, 30-348 Krakow , Poland \and Physics Department, Universit\`a degli Studi di Milano, Milano 20133, Italy \and Chemical, Materials, and Industrial Production Engineering Department, Universit\`a degli Studi ``Federico II'' di Napoli, Napoli 80126, Italy \and Physics and Astronomy Department, University of California, Los Angeles, CA 90095, USA \and Department of Physics, Carleton University, Ottawa, ON K1S 5B6, Canada \and Department of Physics, University of Alberta, Edmonton, AB T6G 2R3, Canada \and Department of Physics and Astronomy, Laurentian University, Sudbury, ON P3E 2C6, Canada \and SNOLAB, Lively, ON P3Y 1N2, Canada \and Department of Physics, Engineering Physics and Astronomy, Queen's University, Kingston, ON K7L 3N6, Canada \and Physics and Astronomy, University of Sussex, Brighton BN1 9QH, UK \and Physik Department, Technische Universit\"at M\"unchen, Munich 80333, Germany \and Instituto de F\'isica, Universidad Nacional Aut\'onoma de M\'exico (UNAM), M\'exico 01000, Mexico}

\author{C.~E.~Aalseth\inst{1} \and F.~Acerbi\inst{2, 3} \and P.~Agnes\inst{4} \and I.~F.~M.~Albuquerque\inst{5} \and T.~Alexander\inst{1} \and A.~Alici\inst{6, 7,8} \and A.~K.~Alton\inst{9} \and P.~Antonioli\inst{7} \and S.~Arcelli\inst{7, 8} \and R.~Ardito\inst{10, 11}
% ...
}
\date{Submitted: \today}
\abstract{
Building on the successful experience in operating the \DSf\ detector, the DarkSide Collaboration is going to construct \DSk, a direct \WIMP\ search detector using a two-phase Liquid Argon Time Projection Chamber (\LArTPC) with an active (fiducial) mass of \DSkActiveMass\ (\DSkFiducialMass).  The \DSk\ \LArTPC\ will be deployed within a shield/veto with a spherical Liquid Scintillator Veto (\LSV) inside a cylindrical Water Cherenkov Veto (\WCV).   Operation of \DSf\ demonstrated a major reduction in the dominant \ce{^39Ar} background when using argon extracted from an  underground source, before applying pulse shape analysis.  Data from \DSf, in combination with MC simulation and analytical modeling, shows that a rejection factor for discrimination between electron and nuclear recoils of \DSkUArPSDRejectionFromGFourDS\ is achievable.  This, along with the use of the veto system, is the key to unlocking the path to large \LArTPC\ detector masses, while maintaining an ``instrumental background-free'' experiment, an experiment in which less than \BackgroundFreeRequirement\ (other than $\nu$-induced nuclear recoils) is expected to occur within the WIMP search region during the planned exposure.  \DSk\ will have ultra-low backgrounds than can be measured {\it in situ}.  This will give sensitivity to \WIMP-nucleon cross sections of \DSkSensitivityOneGeVUnit\ (\DSkSensitivityTenGeVUnit) for \WIMPs\ of \WIMPMassOneTev\ (\WIMPMassTenTev) mass, to be achieved during a \DSkRunTimePlanned\ run producing an exposure of \DSkExposure\ free from any instrumental background.  \DSk\ could then extend its operation to \DSkExtendedRunTimePlanned, increasing the exposure to \DSkExtendedExposure, reaching a sensitivity of \DSkExtendedSensitivityOneGeVUnit\ (\DSkExtendedSensitivityTenGeVUnit) for \WIMPs\ of \WIMPMassOneTev\ (\WIMPMassTenTev) mass.
}
\maketitle
\clearpage
\tableofcontents
\clearpage
\listoffigures
\clearpage
\listoftables
\clearpage
\newpage
\section{Introduction}
\label{sec:Introduction}

The existence of dark matter in the Universe, proposed by Jan~Oort~\cite{Oort:1932tb} and Fritz~Zwicky~\cite{Zwicky:1933ub,Zwicky:1937ec} in the 1930s, is commonly accepted as the explanation of many astrophysical and cosmological phenomena, ranging from internal motions of galaxies~\cite{Faber:1979em} to the large scale inhomogeneities in the cosmic microwave background radiation~\cite{Spergel:2003ci} and the dynamics of colliding galaxy clusters~\cite{Clowe:2006hr}.  It is accepted today in the scientific community that roughly \OmegaNonBaryonicPercent\ of the matter in the universe is in some non-baryonic form that neither emits nor absorbs electromagnetic radiation.  Alternatives, such as theories involving modifications to Einstein's theory of gravity, have not been able to explain the observations across all scales.  A favored hypothesis that explains these observations is that dark matter is made of weakly interacting massive particles (\WIMPs), heavy enough to have been in non-relativistic motion when they decoupled from the hot particle plasma in the early stages of the expansion of the Universe.  However, no such particles exist in the Standard Model~\cite{Steigman:1984ac,Bertone:2005bi} and none have  been directly observed at particle accelerators or elsewhere.  Hence the nature of the dark matter remains unknown.

While the Standard Model of particle physics does not include a viable dark matter candidate, several models of physics beyond the Standard Model ({\it e.g.}, supersymmetry~\cite{Ramond:1971ju,Golfand:1971iw,Volkov:1972jx,Wess:1974do,Fayet:1975et} with R-parity conservation) do propose the existence of \WIMPs.  Thus, besides accounting for the dark matter in the universe, the discovery of \WIMPs\ would give observational evidence for new physics beyond the Standard Model, paving the way for a new ``Particle Physics Paradigm''.

There are three broad classes of \WIMP\ searches: direct detection in shielded underground detectors; indirect detection with satellites, balloons, and ground-based telescopes looking for signals of dark matter annihilation; and detection at particle colliders where dark matter particles may be directly produced in high-energy collisions.

At the LHC, the ATLAS and CMS experiments search for dark matter particles pair produced in \PP\ collisions. A hard jet emitted as initial state radiation provides the trigger, and a large visible momentum imbalance betrays the unseen stable particle.  The range of dark matter particle masses accessible at the LHC is limited by the total center of mass energy, which at \SI{13}{\TeV\per\square\c} results in decreasing sensitivity for WIMP masses above a few hundred~\SI{}{\GeV\per\square\c}~\cite{Aaboud:2016cx,TheCMSCollaboration:2016tx}.  As the luminosity in the LHC Run II rapidly accumulates, it is expected that the sensitivity of ATLAS and CMS will remain competitive with existing or planned direct dark matter search experiments for very low ($<$~\SI{10}{\GeV\per\square\c}) \WIMP\ masses and provide powerful constraints for many classes of dark matter interactions for high \WIMP\ masses. The WIMP mass coverage and interaction sensitivity of LHC searches are often complementary to those of direct dark matter searches.

\begin{figure*}[t!]
\centering
\includegraphics[width=\textwidth]{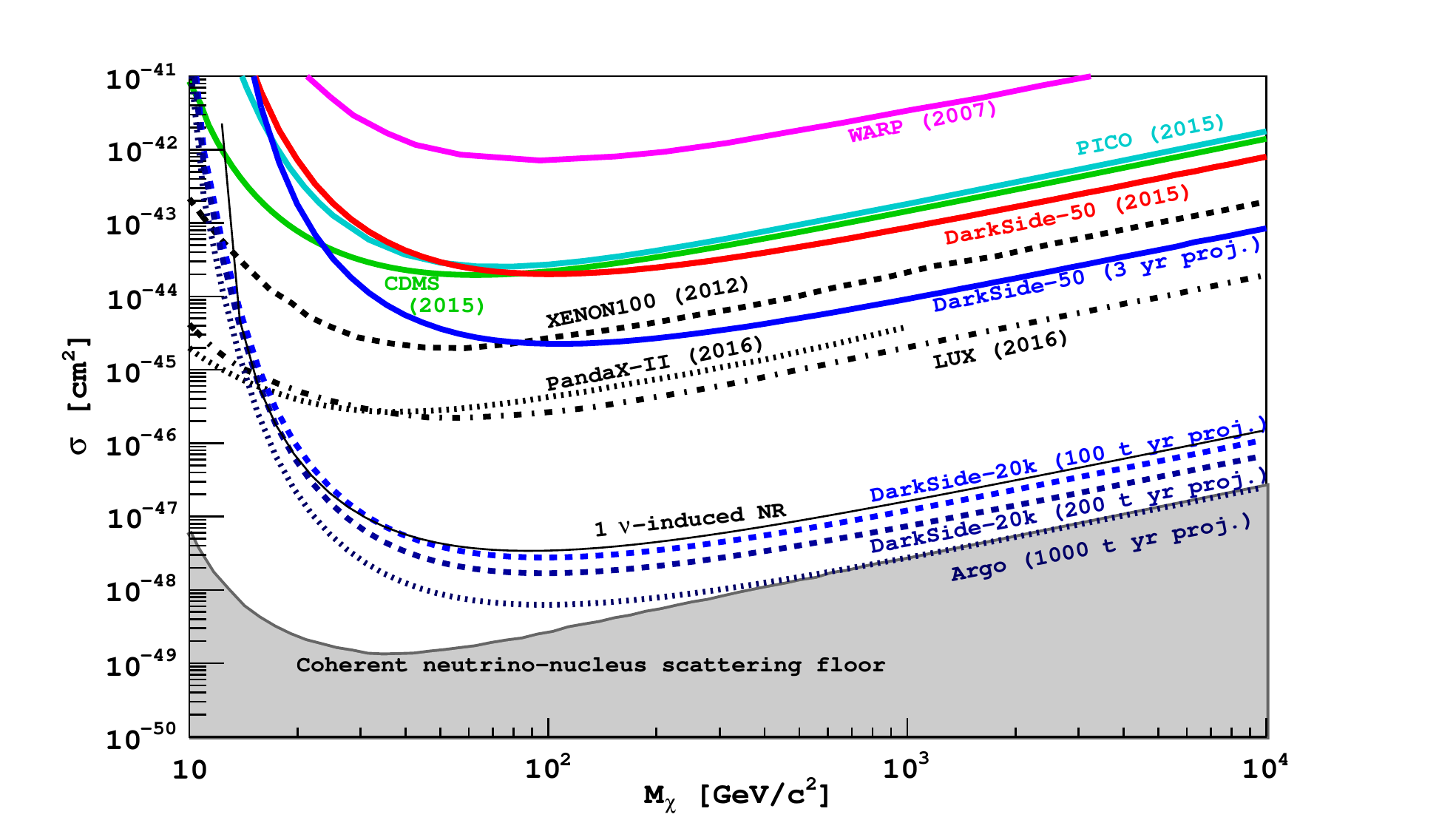}
\caption[\DSk\ WIMP cross section vs. mass sensitivity]{Current results of direct dark matter search experiments (inspired by the corresponding figure in Ref.~\cite{Cushman:2013ug} and adapted to include the most recent results from references cited elsewhere in this section).  The mean exclusion sensitivities for the full exposure of \DSf, for \DSk, and for \Argo\ (accounting for the $\nu$-induced background) are shown.  For comparison, the mean exclusion sensitivity for a generic argon-based experiment with a \SI{30}{\keVr} threshold, \SI{100}{\percent} acceptance for nuclear recoils, and expectation of one coherent neutrino-nucleus scatter during the lifetime of the experiment is also shown.  The grey shaded region is bounded from above by the ``coherent neutrino-nucleus scattering floor", the ultimate experimental reach for a xenon based experiment with arbitrary exposure, limited by the uncertainty on the $\nu$-induced background, introduced in Ref.~\cite{Billard:2014cx}.}
\label{fig:ArgoDSkDSf-ProjectedExclusion}
\end{figure*}

Indirect searches for dark matter annihilation products are prompted by models in which the dark matter particles can interact with one another, converting their rest mass into ordinary particles.  Searches for final states including neutrinos, antiprotons, positrons and \grs\ are performed from a variety of astrophysical sources expected to harbor gravitationally confined populations of dark matter particles.  Notably, data from the PAMELA satellite-borne experiment show a positron excess in the cosmic rays with dark matter annihilation suggested as a possible cause~\cite{Adriani:2009ci}.  The Alpha Magnetic Spectrometer (AMS) measurements of the primary cosmic-ray electron and positron fluxes each require more than a single power-law spectrum to be correctly described~\cite{Aguilar:2014cj}, and while this constrains certain dark matter annihilation channels, it still leaves viable some dark matter decay channels~\cite{Feng:2014ji}.  The Fermi Large Area Telescope (LAT) is an imaging high-energy \gr\ telescope ranging from \SI{20}{\MeV} to more than \SI{300}{GeV} and has shown an emission from the Galactic center that cannot be fully described by background models~\cite{Atwood:2009ha,Ackermann:2017gx}.  This emission has been suggested to be the result of dark matter annihilation at the Galactic center (see Ref.~\cite{Ajello:2015wz} and references cited therein).  The Cherenkov Telescope Array (CTA), which passed the first phase of its Critical Design Review in 2015 and is to be constructed during 2018-2023 would be sensitive to high energy photons from the inner Milky Way and dwarf galaxies due to thermally annihilating dark matter particles up to \SI{10}{\TeV\per\square\c} in mass~\cite{Carr:2015vv}. Annihilation of low-mass \WIMPs\ (down to \SI{5}{\GeV\per\square\c}) in the Sun and the galactic center would be detectable with the PINGU extension to ICECUBE~\cite{Chen:2014jc}, if it is funded.  Indirect astrophysical searches can reach high sensitivity for annihilating dark matter particles.  They have the advantage of associating any signals with cosmic accumulations of dark matter, but most of which are subject to the kind of model uncertainties inherent to astrophysical observations.

A great variety of direct dark matter searches, utilizing different technologies, have been running in recent years and are planned for the future.  These include cryogenic bolometers with ionization or scintillation detection (CDMS~\cite{Akerib:2010dr,Ahmed:2011kd,Agnese:2013hy,Agnese:2015cv}, SuperCDMS~\cite{Agnese:2014cq}, EDELWEISS~\cite{Armengaud:2012bn}, CRESST~\cite{Angloher:2012kl}), sodium/cesium iodide scintillation detectors (DAMA/LIBRA~\cite{Bernabei:2008ik,Bernabei:2010gs,Bernabei:2017kc}, KIMS~\cite{Kim:2012jh}), bubble chambers (PICASSO~\cite{Archambault:2009bx}, COUPP~\cite{Behnke:2011ip,Behnke:2014eq}, PICO~\cite{Amole:2015eu,Amole:2015tf}), point contact germanium detectors (CoGeNT~\cite{Aalseth:2008ij,Aalseth:2011cg,Aalseth:2013bg,Aalseth:2014wc}, MALBEK~\cite{Giovanetti:2015ke}), detectors using liquid xenon (ZEPLIN~\cite{Alner:2007je}, XENON-100~\cite{Aprile:2012kx,Aprile:2015jo,Aprile:2016hn}, LUX~\cite{Akerib:2014jv,Akerib:2016kr,Manalaysay:2016vd,Akerib:2017kg}, XMASS~\cite{Abe:2013fs}, PandaX-I~\cite{Xiao:2014gu}, PandaX-II~\cite{Ji:2016ua}, XENON1T~\cite{Aprile:2016cr,Aprile:2017tt}, LZ~\cite{Nelson:2014wy,Kudryavtsev:2015hy}, XENONnT~\cite{Aprile:2015wv}), detectors using liquid argon (ArDM~\cite{Marchionni:2011kg,Badertscher:2013vr,Calvo:2015vu}, MiniCLEAN~\cite{Hime:2011tt}, DEAP-3600~\cite{Boulay:2012er}, WArP~\cite{Benetti:2007fg,Benetti:2008kd}, and, described in this proposal, \DSf~\cite{Agnes:2015gu,Agnes:2016fz}).  These detectors all share the common goal of achieving a sufficiently low threshold energy to detect the collisions of \WIMPs\ with target nuclei, as well as  low enough background to identify these extremely rare events as  from non-standard sources.  Recent results from these experiments are shown in Fig.~\ref{fig:ArgoDSkDSf-ProjectedExclusion}, along with projected sensitivities from the \DS\ program.

Among all dark matter detectors to date, the XENON-1T collaboration has obtained the most sensitive limits on spin-independent interactions of \WIMPs\ with nuclei, using a xenon target to place a limit on the \WIMP-nucleon interaction cross section of \SI{7.7E-47}{\square\cm} for a WIMP mass of \SI{35}{\GeV\per\square\c}~\cite{Aprile:2017tt}.  Evidence for a low-mass dark matter signal has been claimed by the DAMA/LIBRA~\cite{Bernabei:2010gs} and CoGeNT~\cite{Aalseth:2008ij,Aalseth:2011cg,Aalseth:2013bg,Aalseth:2014wc} collaborations, and may be consistent with observations of the CDMS collaboration using Si detectors~\cite{Agnese:2013hy}.  However, these claims are in tension with a number of other strong results, including those of LUX~\cite{Manalaysay:2016vd,Akerib:2017kg}, XENON-100~\cite{Aprile:2012kx,Aprile:2015jo,Aprile:2016hn}, the low-energy analysis of SuperCDMS Ge detectors data~\cite{Agnese:2014cq}, and the result from MALBEK~\cite{Giovanetti:2015ke}.  The positive claims are  also in strong contradiction with the results from XENON-1T, which directly rule out the interpretation of COGENT and CDMS/Si~\cite{Agnese:2013hy} results in terms of low-mass \WIMPs.  With these results in hand, the region of cross section around \SI{E-39}{\square\cm} expected for $Z$-mediated scattering~\cite{Kuflik:2010kd} seems to be fully explored now - and excluded.

The motivation for direct \WIMP\ searches remains extremely strong, especially for high (above a few hundred~\SI{}{\GeV\per\square\c}) masses that will be out of the LHC reach, and the region of low cross sections (\SIrange{E-45}{E-47}{\square\cm}) corresponding to H-mediated scattering~\cite{Kuflik:2010kd}.  To discover the nature of dark matter particles, it is important for direct and indirect detection to reach new levels of sensitivity, improving current sensitivity levels by a few orders of magnitude.  The complementarity between direct and indirect search strategies will play an important part in the discovery of dark matter.  For the direct-detection searches, the ability to build experiments able to operate in a background-free mode will be crucial for a possible discovery of dark matter.

In this spirit, the four world-leading argon dark matter projects (ArDM at LSC, \DSf\ at LNGS, DEAP-3600 at SNOLab and MiniCLEAN at SNOLab) agreed on joining forces to carry out a unified program for \LAr\ dark matter direct detection in the framework of the \DSk\ project~\cite{Boulay:2017tn}.  In December~2015, the \DS\ Collaboration submitted a proposal to Istituto Nazionale di Fisica Nucleare (\INFN) and to the National Science Foundation (\NSF) for the funding of the \DSk\ experiment.  The \DSk\ experiment aims at a significant improvement - a factor of at least \DSkSensitivityImprovementOneGeV\ with respect to the recent LUX result~\cite{Akerib:2017kg} - in the sensitivity for the direct detection of \WIMPs, reaching \DSkSensitivityOneGeVUnit\ for \WIMPs\ of \WIMPMassOneTev\ mass.  It is proposed to achieve this goal with a liquid argon time projection chamber (\LArTPC) experiment with an active (fiducial) mass of \DSkActiveMass\ (\DSkFiducialMass), for a total exposure of \DSkExposure\ to be accumulated in a run of \DSkRunTimePlanned.  Thanks to its exceptionally low instrumental background, \DSk\ could extend its operation to \DSkExtendedRunTimePlanned, increasing the exposure to \DSkExtendedExposure\ and reaching a sensitivity of \DSkExtendedSensitivityOneGeVUnit.  In April~2017 the \DSk\ experiment was officially approved by INFN, with expectation that formal agreements with international partners will soon be in place.

.
\section{The Case for \DSk: An Instrumental Background-Free Dark Matter Search}
\label{sec:BackgroundFree}

\begin{figure*}[h!]
\centering
\includegraphics[width=\columnwidth]{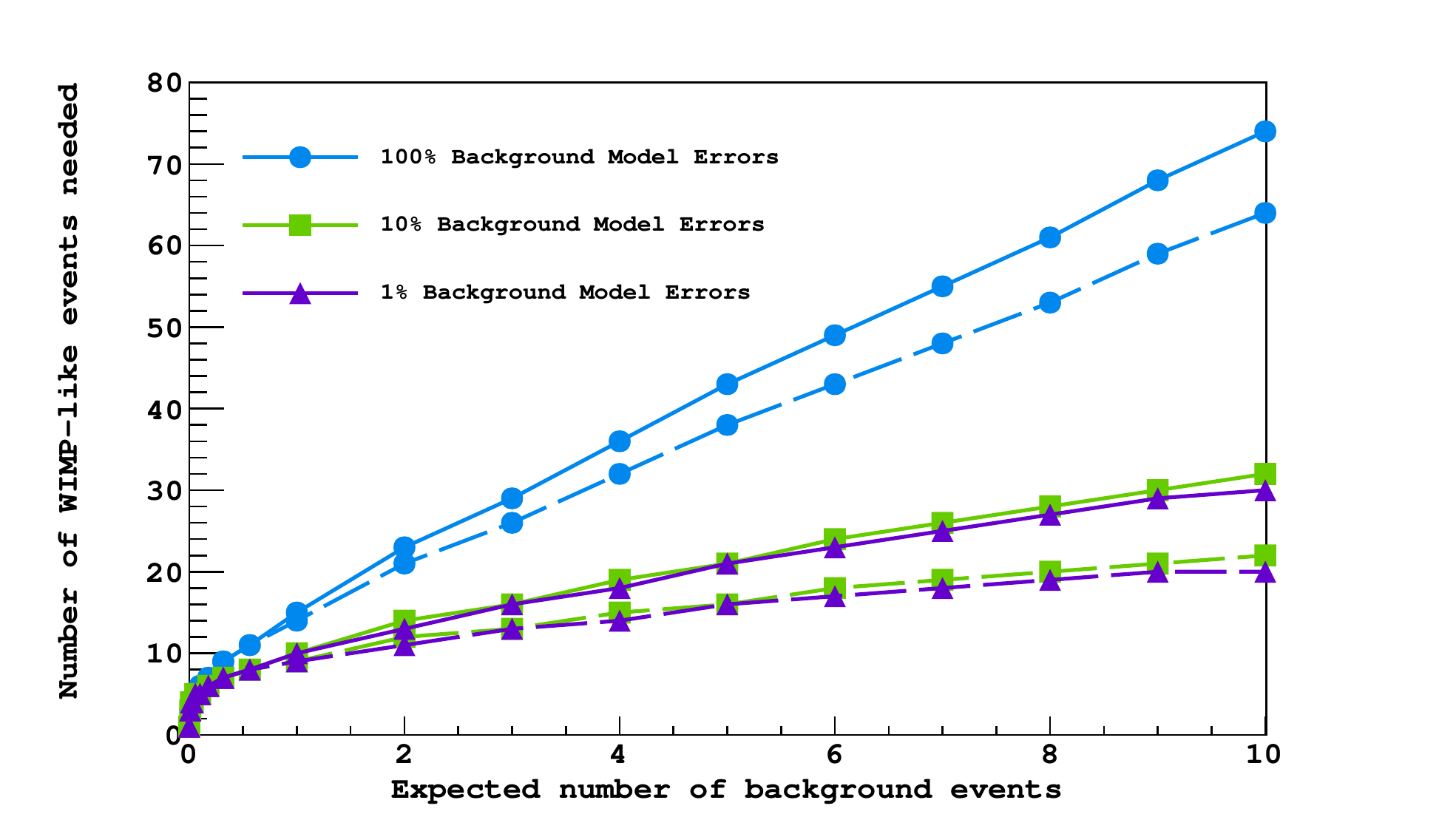}
\caption[Number of dark matter-like events needed to claim a WIMP observation (linear-linear scale).]{Number of dark matter-like events needed to claim a WIMP observation at the \SI{5}{\sgm} level, based on the predicted background rate of the experiment, in a linear-linear scale.  Solid lines show the number of dark matter-like events needed, including backgrounds, while dashed lines show the number of dark matter events after subtraction of the expected background.  Blue, green, and purple curves were made assuming uncertainty on the background model of \SI{100}{\percent}, \SI{10}{\percent}, and \SI{1}{\percent}, respectively.}
\label{fig:Discovery-SigmaLin}
\end{figure*}

The goal of \DSk\ is to discover dark matter.  With dark matter interactions very rare, and the onset of $\nu$-induced nuclear recoils for exposures of \DSkExposure\ and beyond, it is of essence to contain the number of instrumental background interactions to \BackgroundFreeRequirement, so that a positive claim can be made with as few events as possible.  A discovery for dark matter could come at any exposure level, and even a low but non-zero instrumental background can hinder the task of discovering dark matter.

Before going into the details of how backgrounds affect sensitivity, it is first important to draw a firm distinction: it is statistically much easier to exclude a region of parameter space than it is to make a detection.  Typical exclusion relies on a standard of \SI{90}{\percent} likelihood, while a positive detection requires a \SI{5}{\sgm} standard: this is equivalent to saying that, once a reliable background model for the experiment has been built, the observed signal has a probability of \num{3E-7} of being an artifact of statistical fluctuations in the background.  A signal \SI{3}{\sgm} above the null hypothesis has a probability of \num{E-3}, and is generally referred to as ``evidence'', while a \SI{1}{\sigma} deviation, with a probability of \num{0.16}, is ``uninteresting''.

Fig.~\ref{fig:Discovery-SigmaLin} shows the number of \WIMP-like events an experiment would need to observe in order to reject the null hypothesis to \SI{5}{\sgm} as a function of the number of expected background events, when the background model predicts the background with \SI{100}{\percent}, \SI{10}{\percent}, and \SI{1}{\percent} uncertainty.  As is evident from the figures, there is fairly little difference between \SI{10}{\percent} and \SI{1}{\percent} uncertainty. This results from the uncertainty in the model being small enough, so that the Poisson fluctuations dominate the statistics.  When the model predicts \num{0} or \num{0.1} background events, just \num{5} \WIMP-like events can be statistically significant enough to claim a detection at the \SI{5}{\sgm} level.  However, if an experiment expects to see even one background event, the number of \WIMP-like events needed for a discovery goes up to \numrange{10}{15}, depending on the uncertainty in the background model.  This means that a dark matter experiment expecting one background event will need up to \num{3} times more \WIMP\ interactions to claim a detection, compared to an experiment which plans to run background-free.  The requirement becomes increasingly more restrictive if the expected number of background events is greater than \num{10}.  More details of this calculation can be found in~\cite{Westerdale:2016ub}.

Among the variety of detector technologies, noble liquid TPCs, which detect both the scintillation light and the ionization electrons produced by recoiling nuclei, have significant advantages for direct dark matter searches.  For liquid argon (\LAr), the powerful discrimination against background using the time-development of the primary scintillation signal (pulse-shape discrimination or \PSD), the precise determination of event positions in all three dimensions given by the \TPC, and the effectiveness of chemical and cryogenic purification of the argon have all been demonstrated (see Refs.~\cite{Alexander:2013jn,Agnes:2015gu} and references cited therein).  The specific goal of searching for and discovering high-mass dark matter particles allows the search to be concentrated in the range of recoil energies from \DSkROIEnergyRange\ (\WIMP\ search region).  In this region, argon detectors outperform other technologies with their outstanding \bg\ background rejection, as first pointed out by Boulay and Hime~\cite{Boulay:2006hu} and verified by WArP~\cite{Benetti:2008kd} and \DSf~\cite{Agnes:2015gu}.

The \DS\ Collaboration has published background-free \WIMP\ search results (\BackgroundFreeRequirement\ expected and no observed events in the search region) from an exposure of \DSfAArExposure\ with atmospheric argon (\AAr)~\cite{Agnes:2015gu} and from a separate exposure of \DSfUArExposure\ with underground argon (\UAr)~\cite{Agnes:2016fz}.  The combined result of the \AAr\ and \UAr\ data analysis, as discussed below, leads to the expectation that a result free from instrumental background can also be obtained from a much larger exposure with a multi-tonne detector.

In addition to choosing a detector technology and target, the exact configuration of the target is an important choice.  The light quenching in \LAr\ introduced by the presence of a drift field was first discovered and then precisely measured by the SCENE experiment~\cite{Cao:2015ks}.  The precise measurements obtained with SCENE leads to the choice of a drift field of 200 V/cm for, which minimally penalizes the light yield of nuclear recoils, containing its loss within 10\%, and resulting in a barely noticeable loss of performance.  Ultimately the reconstruction of energy of the events is based on the consideration of both the scintillation signal, S1, and the ionization signal, S2, and so a very good resolution can be obtained.  

The choice of using single- or dual-phase geometry is driven by the need to obtain the best performance in \WIMP\ dark matter searches, and not a need for optimizing event reconstruction in terms of \SOne\ and \FNine, the fraction of \SOne\ light detected in the first \SI{90}{\nano\second} of the pulse.  The dual-phase \LAr\ detector \DSf\ has demonstrated an impressive performance in background rejection by producing two independent background-free results, with \AAr~\cite{Agnes:2015gu} and \UAr~\cite{Agnes:2016fz}.  The dual-phase xenon detectors have also achieved an impressive performance and lead the race for \WIMP\ dark matter sensitivity.  No single-phase detector has yet to produce a \WIMP\ dark matter result comparable.  The main difficulty encountered by the first generation of single phase detectors can be traced to the lack of an \STwo\ signal, without which the precise fiducialization that is typical of dual-phase detectors is no longer possible.  In absence of this fiducialization, surface background and background from Cherenkov events originating in the PMTs fused silica and in the Teflon reflector dominate.  Experience with the \DSf\ detector is indeed very similar, where as the two campaigns of data mentioned above would be far from background-free if the \STwo\ signals were not present, for the very reason stated above.  In fact, when artificially turning off the \STwo\ signal in the \DSf\ data analysis stage, the background-free condition is lost, and the surface and Cherenkov backgrounds contribute events to the final data sample, within the \WIMP\ search region.

\begin{figure*}[t!]
\centering
\includegraphics[width=\textwidth]{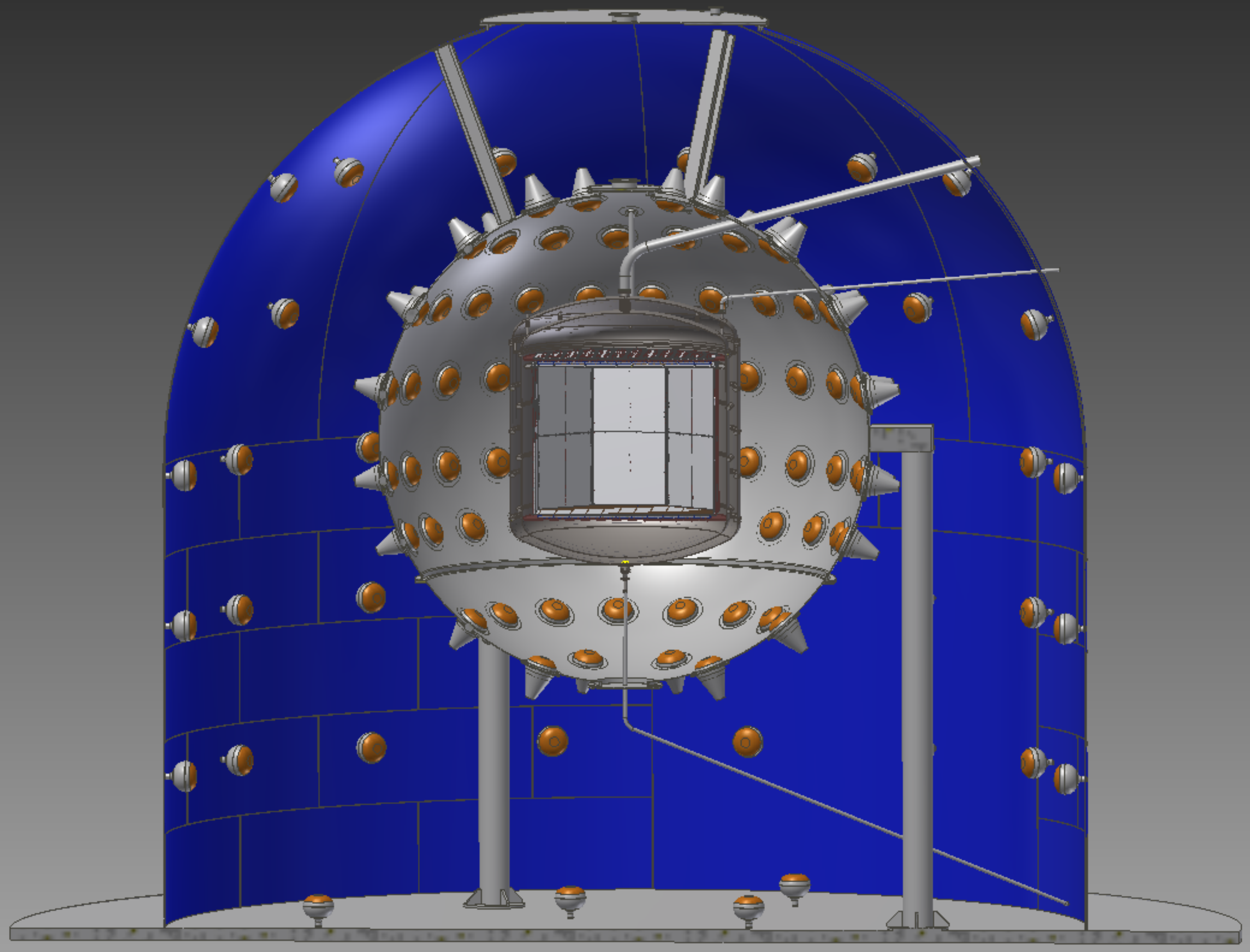}
\caption[Cross sectional view of \DSk\ through its center plane.]{Cross sectional view of the \DSk\ experiment through its center plane, showing the water tank and the \WCV\ detector, the stainless steel sphere and \LSV\ detector, and the cryostat and \LArTPC.}
\label{fig:Facilities-DSk3DCrossSection}
\end{figure*}

On this basis, an enlarged \DS\ Collaboration proposes the construction of \DSk, a direct \WIMP\ search using a \LArTPC\ with an active (fiducial) mass of \DSkActiveMass\ (\DSkFiducialMass) of \UAr.  As shown in Fig.~\ref{fig:Facilities-DSk3DCrossSection}, the \DSk\ \TPC\ will be deployed underground at \LNGS.  \DSk\ will be a detector with ultra-low background levels and the ability to measure its backgrounds {\it in situ}.  It is designed to achieve an exposure of \DSkExposure, accumulated during a run of approximately \DSkRunTimePlanned, free of instrumental background, giving a sensitivity to \WIMP-nucleon interaction cross sections of \DSkSensitivityOneGeVUnit\ (\DSkSensitivityTenGeVUnit) for \WIMPs\ of \WIMPMassOneTev\ (\WIMPMassTenTev) mass.  This is a factor of~\DSkSensitivityImprovementOneGeV\ improvement over the corresponding currently published limits, and covers a large fraction of the mass-cross section parameter space currently preferred by supersymmetric theories~\cite{Bagnaschi:2015hd}.  The projected sensitivity of \DSk\ is compared with other current and planned projects in Table~\ref{tab:Introduction-Sensitivity}.

\begin{table*}
\rowcolors{3}{gray!35}{}
\small
\centering
\caption[Sensitivity comparison of current and future direct detection dark matter experiments.]{Comparison of sensitivity for current dark matter experiments leading the search for high mass \WIMPs\ and of future approved and proposed experiments.  Also included are the calculated sensitivities for hypothetical xenon and argon-based experiments with an expectation of one coherent neutrino-nucleus scatter during their operation and with thresholds of \SI{10}{\keVr} and \SI{30}{\keVr} respectively.}
\begin{tabular}{lcrccc}
Experiment		&Target	&\multicolumn{1}{c}{Exposure/[\si{\tonne\yr}]}
										&$\sigma$/[\si{\square\cm}] @\WIMPMassOneTev\
													&$\sigma$/[\si{\square\cm}] @\WIMPMassTenTev\
																	&Reference\\
\hline
\DSf				&Ar		&0.011		&\num{9E-44}		&\num{8E-43}		&\cite{Agnes:2016fz}\\
XENON-100		&Xe		&0.021		&\num{2E-44}		&\num{2E-43}		&\cite{Aprile:2012kx}\\
LUX				&Xe		&0.092		&\num{1E-45}		&\num{1E-44}		&\cite{Akerib:2017kg}\\
ArDM			&Ar		&1.5			&\num{8E-45}		&\num{7E-44}		&\cite{Calvo:2015vu}\\
DEAP-3600		&Ar		&3.0			&\num{5E-46}		&\num{5E-45}		&\cite{Boulay:2012er}\\
XENON1T		&Xe		&2			&\num{1E-46}		&\num{1E-45}		&\cite{Aprile:2016cr}\\
LZ				&Xe		&15			&\num{3E-47}		&\num{3E-46}		&\cite{Dobson:2016wi}\\
XENONnT		&Xe		&20			&\num{2E-47}		&\num{2E-46}		&\cite{Aprile:2015wv}\\
\minitab{l}{1\,$\nu$-induced}{nuclear recoil}
				&\minitab{c}{Xe}{Ar}
						&\minitab{c}{34}{52}
									&\minitab{c}{\num{5.9E-48}}{\num{1.6E-47}}
													&\minitab{c}{\num{5.8E-47}}{\num{1.5E-46}}
																	&this document\\
\DSk\			&\UAr\	&100			&\num{\DSkSensitivityOneGeVBare}
													&\num{\DSkSensitivityTenGeVBare}
																	&this document\\
\DSk\			&\DAr\	&200			&\num{7.4E-48}	&\num{6.9E-47}	&this document\\
\Argo\			&\DAr\	&1000		&\num{\ArgoSensitivityOneGeVBare}
													&\num{\ArgoSensitivityTenGeVBare}
																	&this document\\
\end{tabular}
\label{tab:Introduction-Sensitivity}
\end{table*}

In the longer term, the aim of the \DS\ collaboration is to develop a path towards a dark matter detector to be built with a \ArgoTotalMass\ (\ArgoFiducialMass) active (fiducial) mass of depleted argon (\DAr), \UAr\ with an \ce{^39Ar} content further depleted by processing the \UAr\ through a cryogenic distillation column.  For now, this ultimate experiment is called \Argo.  A successful \DSk\ experiment would represent a fundamental milestone toward the realization of \Argo.  \Argo\ is conceived to accumulate an exposure of \ArgoExposure, free of background other than that induced by coherent scattering of neutrinos, and thus be sensitive to dark matter cross sections at the ``neutrino floor''~\cite{Billard:2014cx}.  

As well as providing the most sensitive \WIMP\ search, \Argo\ would carry out an ambitious program of precision measurements on low energy solar neutrinos (\ce{^7Be}, \PEP, and \CNO\ neutrinos) through neutrino-electron elastic scattering~\cite{Franco:2016ex}.  \LAr\ has a scintillation light yield ten times larger than that of the organic liquid scintillator target used for solar neutrino spectroscopy in Borexino~\cite{Bellini:2014ke}, resulting in much better energy resolution.  Currently, Borexino provides the most sensitive measurement of the low energy solar neutrino spectra, but is still unable to measure the \CNO\ neutrinos.  The higher energy resolution and tighter control of surface backgrounds could make a precision measurement of the \CNO\ neutrino flux possible in \Argo~\cite{Franco:2016ex}. The use of the two-phase argon technology allows for a sharp definition of the fiducial volume, significantly reducing the systematic error that dominates the measurement of \ce{^7Be} neutrinos in Borexino.  The anticipated precision of the rates are \ArgoBeSevenSolarNeutrinosAccuracy\ for \ce{^7Be}, \ArgoPEPSolarNeutrinosAccuracy\ for \PEP, and \ArgoCNOSolarNeutrinosAccuracy\ for \CNO\ neutrinos~\cite{Franco:2016ex}.                                                             

The low recoil energies and cross sections targeted with \DSk\ represent an enormous experimental challenge, especially in the face of daunting backgrounds from electron recoil interactions and from neutrons that mimic the nuclear recoil signature of \WIMPs.  To meet its challenge, \DSk\ will exploit the auxiliary facilities, including radon-free clean rooms, already built at LNGS as part of the \DS\ program.  The cryostat will be placed inside a \DSkLSVDiameter\ diameter stainless steel sphere filled with boron-loaded liquid scintillator, serving as an active neutron veto (\LSV), which in turn will sit inside a newly constructed \DSkWCVDiameter\ diameter \DSkWCVTotalHeight\ tall stainless steel tank, filled with ultrapure water, functioning as an active muon veto (\WCV).  The \LArTPC\ will be instrumented with Silicon PhotoMultipliers (\SiPMs) as photosensors.  To provide the \UAr\ target, the \DS\ Collaboration is already establishing, through the \Urania\ and \Aria\ projects, the entire chain of extraction, purification, depletion, transport, and storage of low-radioactivity argon at the multi-tonne level.  In short, \DSk\ will perform the most sensitive search for dark matter yet proposed and will also provide a convincing foundation for a \SI{100}{\tonne}-scale detector.

In order to deliver a result free from instrumental background (\BackgroundFreeRequirement\ in the \DSkExposure\ exposure), \DSk\ must be built with the goal of completely eliminating or minimizing backgrounds from the following sources:

\vspace{6mm}
\noindent{\bf Nuclear Recoils}

\begin{itemize}
\item[\it Neutrino-induced coherent nuclear scattering:\/]

\hfill \break\ With the nuclear recoil energy thresholds needed to achieve the excellent electron recoil rejection in \LAr\ from pulse shape discrimination (PSD), only atmospheric neutrinos and the diffuse supernova neutrino background are energetic enough to produce nuclear recoils in the \WIMP\ region of interest.  Individual nuclear recoils from coherent scattering of neutrinos from the argon nuclei in the target are indistinguishable from \WIMP-induced nuclear recoils.  Calculations checked against those in Ref.~\cite{Billard:2014cx} predict \DSkNuInducedBackgroundUnit\ from this source in the full \DSk\ exposure.  Though these events are irreducible background to the \WIMP\ search, they are also an as-yet-unobserved physics signal. 

\item[\it Neutron scattering:\/]

\hfill \break\ Since individual elastic neutron scatters are essentially indistinguishable from elastic \WIMP\ scatters, background from all sources of neutrons must be reduced well below the \BackgroundFreeRequirement\ level.  The main system that allows this is a highly efficient neutron veto system, similar to that used in \DSf, which drew from the design of Borexino solar neutrino detector.  Direct measurements of the cosmogenic backgrounds in Borexino and comparison with Monte Carlo simulations~\cite{Bellini:2013kr,Empl:2014ih} lead to the expectation that cosmogenic neutron backgrounds can be kept under control by using veto signals from the \LSV\ and \WCV, with no \WIMP-like events expected for exposures much larger than that expected for \DSk.

Background due to radiogenic neutrons from ($\alpha$,n) and fission decays in the construction materials of the \LArTPC\ must also be suppressed to below \BackgroundFreeRequirement\ in the \DSkExposure\ exposure.  The rate of neutrons entering the detector will be reduced by using radiopure silicon-based photosensors and careful screening and selection of construction materials.  It is currently foreseen that the biggest contributor to this background will be the cryostat and the reflective panels of the the \LArTPC, therefore, building the cryostat from either stainless steel or titanium will be investigated.  Any residual neutron rate will be measured and efficiently rejected using the \LSV.  The sub-centimeter position resolution of the \LArTPC~\cite{Agnes:2015gu} also helps reducing background from radiogenic neutrons by fiducialization and rejecting events with multiple-sited energy deposition.  With careful selection and screening of all construction materials and detector components, it is foreseen that the background due to radiogenic neutrons from ($\alpha$,n) and fission decays can be kept below the \BackgroundFreeRequirement\ level (see Sec.~\ref{sec:Physics-Background}).

\end{itemize}

\noindent{\bf Electron Recoils}

For electron recoils, the successful strategy developed in \DSf\ was to reduce the raw rates by stringent materials screening and selection, then to suppress the remaining events by analysis cuts (fiducial volume, multiple-site energy deposition, etc.), and finally to apply \PSD\ to identify residual electron recoil background events.  The energy region of interest (ROI) for the \WIMP\ search is \DSkROIEnergyRange, roughly equivalent to \DSkROIElectronEquivalentEnergyRange.  Given the effectiveness of self-shielding and fiducial cuts against external backgrounds for large detectors, the most important sources of electron recoil background for \DSk\ are those uniformly distributed throughout the fiducial volume and those which come from radioactive decays within the detector materials.  These include decays of radioactive nuclides in the noble liquid target material itself, and electron scatters induced by solar neutrinos.  Based on results from the \DSf\ experiment and Monte Carlo simulations, it has been demonstrated that these backgrounds can be held to below the required level (\BackgroundFreeRequirement) in a \DSkExposure\ exposure, if operated with a low-radioactivity argon target.  The crucial sources of electron recoil background are:

\begin{itemize}
\item[\it Solar Neutrino-induced Electron Scatters:\/]

\hfill \break\ Electron scatters from \PP\ neutrinos within the energy ROI occur at a rate of \DSkROIPPRate, giving \DSkROIPPUnit\ in the \DSkExposure\ exposure.  The \bg\ rejection power demonstrated with \DSf\ in its \AAr\ run~\cite{Agnes:2015gu} was better than one part in \DSfAArROIEventsNumber, more than sufficient to reject this \PP\ neutrino background for the entire planned exposure. This would also hold true for the \ArgoExposure\ exposure planned for \Argo.

\item[\it \ce{^238U}, \ce{^232Th}, and Daughters:\/]

\hfill \break\ The radioactive noble element radon may enter and dissolve in the \LAr\ after being produced in the  \ce{^238U} and \ce{^232Th} decay chains.  The most important background contribution comes from $\beta$ decays of the radon daughters whose spectra fall partly within the energy ROI.  The \ce{^222Rn} specific activity in \DSf\  was measured to be below \DSfRnTwoTwoTwoSpecificActivityLimit.  The \ce{^222Rn} concentration in \DSk\ is expected to be much lower than the upper limit obtained for \DSf\ due to the reduction in the surface-to-volume ratio.  However, even if the \ce{^222Rn} contamination were at the \DSf\ upper limit,  only \DSkROIBiTwoOneFourRate\ from \ce{^214Pb} $\beta$ decays would be expected in the energy ROI, giving \DSkROIBiTwoOneFourUnit\ in the total \DSkExposure\ exposure.  Once again the \bg\ rejection power demonstrated with \DSf\ in its \AAr\ run~\cite{Agnes:2015gu}, better than one part in \DSfAArROIEventsNumber, is more than sufficient to exclude this source of background from the \WIMP\ search region at the level required for an instrumental background free result.

\item[\it \ce{^39Ar}:\/]

\hfill \break\DSf\ accumulated an exposure of \DSfAArExposure\ with \AAr\ (\ce{^39Ar} specific activity: \DSfAArArThreeNineActivity), followed by an exposure of \DSfUArExposure\ with \UAr\ (\ce{^39Ar} specific activity: \DSfUArArThreeNineActivity).  Both campaigns provided background-free results, and with \DSfAArROIEventsNumberleft\ events left in the energy ROI, but located in the electron recoil band, provided a limit on the suppression factor for electron recoil events $>$\DSfAArROIEventsNumberleft, provided by PSD alone.  Argon extracted from the atmosphere is not an option for \DSk\ with its maximum drift time of \DSkDriftTime, due to a constant pile-up of events that would exist, but using \UAr\ with a much reduced \ce{^39Ar} component solves this constant pile-up problem.  To better understand what ultimate electron recoil rejection factor can be achieved, the \Geant-based MC simulation package tuned using \DSf\ data was used to simulate large exposures of \ce{^39Ar} events inside the geometry of the \DSk\ detector.  The simulated events included the readout characteristics, including \SiPM\ PDE, noise rate and cross-talk, and were then fit using an analytic model of the PSD parameter to be used in \DSk.  A suppression for \ce{^39Ar} events using only PSD is estimated to be \DSkUArPSDRejectionFromGFourDS, more than sufficient to discriminate against the expected \DSkUArArThreeNineROIBare\ remaining \ce{^39Ar} events coming from the \UAr.  This is further supported by the recent results from the DEAP-1 experiment, which show that a reduction fraction for electron recoils on the order of $10^{10}$ can be achieved in a \LAr\ detector with a scintillation light yield of at least \SI{8}{\pe/\keVee} \cite{Amaudruz:2016dq}.

Argon extracted from the atmosphere with a specific \ce{^39Ar} activity of \DSfAArArThreeNineActivity\ is not an option for \DSk\ with its maximum drift time of \DSkDriftTime\ due to constant pile-up of events. This solves the constant pile-up problem, but still gives \DSkUArArThreeNineROIBare\ beta decays in the design exposure in the \WIMP\ energy region of interest.  A MC simulation of \ce{^39Ar} events, analytically modeling the PSD parameters to be used, showed that the achievable discrimination against (any) electron recoils is estimated to be \DSkUArPSDRejectionFromGFourDS\ (Sec.~\ref{sec:Physics-Background}), validating the background-free goal entirely from the PSD mechanism. However, further depletion of \UAr\ via cryogenic distillation is planned for the \DS\ program, in sight of \Argo. The scaled-up extraction of \UAr\ and the production of \DAr\ are described in Sec.~\ref{sec:Argon}.

\item[\it \ce{^85Kr}:\/]

\hfill \break\ Recent data from \DSf\ show that the \UAr\ contains \ce{^85Kr} at a specific activity of \DSfUArKrEightFiveActivity, a rate comparable to the \ce{^39Ar} activity just discussed.  The \ce{^85Kr} in \UAr\ comes either from atmospheric leaks during the collection and purification, or from deep underground natural fission processes.   No attempt was made to remove \ce{Kr} from the \UAr\ for the  \DSf\ target, simply because the presence of \ce{^85Kr} was not expected at the time of purification.  For the \LAr\ target, it is expected that the \Urania\ \UAr\ extraction plant will be able to reduce the \ce{^85Kr} to a level resulting in a specific activity much less than the residual \ce{^39Ar}.  However, if for some reason the \Urania\ plant is unable to reduce the \ce{^85Kr} to the desired level, \SeruciOne\ at \Aria\ will be capable of making the necessary chemical purification of the \UAr.  \ce{Kr} in \ce{Xe} has been reduced by a factor $\sim$1000 per pass by cryogenic distillation~\cite{Wang:2014bk}, which should be even better for \ce{Ar}.  Calculations show that the \Aria\ cryogenic distillation column can reduce \ce{^85Kr} and other chemical impurities by a similar factor of more than \AriaChemicalPerPass\ per pass, at a rate of \AriaChemicalRate, making this source of contamination negligible in \DSk.

\end{itemize}

\noindent{\bf Surface Recoils and Cherenkov Backgrounds}

Surface recoils occur when a radioisotope decays on the interior surface of the detector. During the decay, it may eject an alpha particle toward the wall and recoil into the active argon volume, producing a nuclear recoil signal.  In \DSf, a surface recoil rate of \SI{0.8}{\milli\becquerel} was observed (across \SI{~0.75}{\square\meter} of surface area) and used to do a study of the alpha light yield in \LAr~\cite{Agnes:2017cl}. For the \WIMP\ search all the surface events were rejected through the use of fiducial cuts, specifically a drift time cut. No radial fiducial cut was needed, due to a reduced charge collection efficiency along the side wall preventing the events from passing the required \STwo\ cuts, and to a much lower rate of surface events on the side wall relative to the top and bottom.  In the case that the surface backgrounds do not exactly scale when moving to \DSk, a radial fiducial volume cut, in addition to the drift time cut, is already envisioned and is taken into account when making the total exposure calculation.  Finally, current research by \DS\ collaborators with the TPB wavelength shifter shows a long-lived signal associated with surface nuclear recoils that is not present in bulk nuclear recoils. The long data acquisition windows will be capable of capturing these signals and discriminating them.  With the effectiveness of the drift-time cut seen in \DSf, implementation of a radial fiducial volume cut, and the presence of a long-lived signal component, surface backgrounds will contribute a negligible amount to the background budget.

Cherenkov background occurs when an electron recoil event is contaminated with Cherenkov light coming from a \bg\ interaction in the surrounding detector materials (acrylic or PTFE), either resulting from multiple Compton scattering of a single particle or a separate particle interacting in time-coincidence with the first.  The Cherenkov light can overlap with the \SOne\ signal from the \LAr\ interaction producing an \FNine\ value that a nuclear recoil would take.  While the distribution of \SOne\ light is a major factor in rejecting these types of events, the additional information given by the position reconstruction using the \STwo\ signal allows for a more stringent cut to be placed, ensuring that no Cherenkov background event will survive the cut.

\section{Project Overview}
\label{sec:ProjectOverview}

Many fundamental design parameters for \DSk\ are based on the successful experience of the \DS\ collaboration in constructing, commissioning, and operating the \DSf\ detector in a background-free mode. The \LArTPC\ will also be deployed at \LNGS, in the underground Hall\,C, in the center of a newly constructed active veto system. Fig.~\ref{fig:Facilities-DSk3DCrossSection} shows the projected experimental arrangement, similar to an expanded version of the \DSf\ setup. This section gives a general overview of the project and introduces the features of the experiment. The \LArTPC, the production of its \UAr\ target, its \SiPM\ photosensors, and its cryostat and cryogenics system are described in Secs.~\ref{sec:Argon}-\ref{sec:LArTPC}.

\subsection{The \LArTPC}
\label{sec:ProjectOverview-LArTPC}

Energy deposits in the \LAr\ target result in a characteristic production of excited and ionized argon atoms, according to the underlying process of a recoiling electron or nucleus. Excited argon atoms, which can also be produced by recombining ionization charge, lead to an efficient formation of argon excimers decaying via the emission of scintillation light containing two components with different time constants of emission. Both components combined together to yield an instant light signal \SOne. Due to the deep UV nature (around \ArWaveLength) of this scintillation light, which is absorbed by most materials, a thin layer of wavelength shifter must cover all exposed surfaces. Ionization electrons escaping recombination are drifted by an applied electric field to the top of the \LAr, where a stronger applied field extracts the electrons into the gas pocket above the liquid. Here the strong field accelerates the electrons, enough for them to excite (but not ionize) the argon gas, producing a secondary scintillation signal \STwo, proportional to the ionization charge.  Photosensors placed behind wavelength shifter coated windows at the top and bottom of the \TPC, read out both scintillation signals in each event. \SOne\ is used for energy determination, as well as for PSD, the latter is derived from the ratio of the prompt and delayed light fractions. \STwo\ is used for energy and 3D position measurement of the event, the vertical coordinate from the drift time between \SOne\ and \STwo, and the horizontal coordinates from the light pattern in the top photosensors.

The octagonal \LArTPC\ will have a height of \DSkActiveHeight\ and a distance between parallel walls of \DSkActiveDiameter.  It will be instrumented with a new kind of photosensors, arrays of \SiPMs, arranged in assemblies called photodetector modules (\DSkPdms). Each \DSkPdm\ has an area comparable to that of a \SI{3}{\inch} photomultiplier tube (\PMT ), with the \LArTPC\ containing \DSkTilesNumber\ \DSkPdms\ in total.  Substantial effort was put in the developments of this technology, since \SiPMs\ promise a higher effective quantum efficiency, higher reliability at \LAr\ temperature, and a much higher radiopurity than \PMTs. All of these properties are crucial for \DSk: PSD, in \LAr\ the most important mechanism for background discrimination, depends critically on the light yield, while the large material budget of \PMTs\ is often the limiting factor for neutron- and gamma-induced backgrounds. The \LArTPC\ will be equipped with arrays of \SiPMs, totaling \DSkTilesArea\ in area.  The development of the \SiPMs, readout electronics, and high-radiopurity packaging is described in Sec.~\ref{sec:PhotoElectronics}.  

In comparison to \DSf, where a full digitization and recording of the waveform of each \PMT\ could be achieved, a custom scheme for the sampling of the two-orders-of-magnitude higher channel count of the \SiPM\ \DSkPdms\ has to be developed. Design parameters for the data acquisition (DAQ) system are driven by rates, occupancies, as well as leading edge timing and limited charge information from each channel, to compromise as little as possible on energy resolution and pulse-shape discrimination. The DAQ system is described in Sec.~\ref{sec:DAQ}.

All components of the detector, above all the inner components, like the \LArTPC, the cryostat, the \SiPM\ arrays, and cables, must be made from materials of highest radiopurity to keep backgrounds as small as possible. Sec.~\ref{sec:MatAssay} describes plans for the assay of materials and the determination of their suitability.  Facilities for cleaning the materials and keeping them clean, and, specifically, avoiding contamination from  radon and radon daughters, are described in Sec.~\ref{sec:Facilities}.

\subsection{The \DSk\ Prototype: \DSp}
\label{sec:Overview-Prototype}

In order to demonstrate and test technological developments on a relevant scale, it is planned to build and operate a \DSpAproximateMass\ prototype. The project is called \DSp\ and is described in Sec.~\ref{sec:Proto}. \DSp\ will be at an intermediate scale between \DSf\ and \DSk, able to test a few number of full size components intended for use in \DSk.  The cryostat therefore will be produced in two versions, firstly in stainless steel, to allow for a fast start of activities, while a second version will be realized from low-activity titanium, to demonstrate the feasibility of this innovative construction method for the larger \DSk\ cryostat, described in Sec.~\ref{sec:Cryogenics}. The size of \DSp\ was chosen to be able to validate mechanical and \DSkPdm\ components of the \DSk\ \LArTPC, including their readout system. The full size cryogenic system is also planned to be tested in this way, especially the argon condenser, the gas pump, the fast drain recovery system and the heat recovery system. Functionality and stability control, as well as safety issues during power failures, purification flow rates and general controls are meant to be explored. The prototype project also serves to validate the readiness of the various production lines in the different institutions of the \DS\ collaboration.

\subsection{The Active Vetoes: Water Cherenkov Veto \WCV\ and Liquid Scintillator Veto \LSV}
\label{sec:ProjectOverview-Detectors-Vetoes}

The veto system (Sec.~\ref{sec:Vetoes}) is crucial for the suppression of cosmogenic and radiogenic backgrounds originating from internal detector components, as well as from the surrounding environment. At the \LNGS\ underground site, a depth of \LNGSDepthMWE, the rate of cosmic rays is reduced to \LNGSMuonResidualFlux. The outer veto \WCV\ provides general tagging of cosmic rays and shielding from radioactivity in Hall\,C and the surrounding rock. The inner \LSV\ system targets events induced by internally- and externally-generated neutrons with high efficiency and is also effective to detect gamma-induced events in the \LArTPC.  Neutrons are detected primarily via capture on \ce{^10B} present in the scintillator cocktail, but also via their thermalization signal and/or detection of the gamma rays from neutron capture in the \TPC\ or the cryostat. The interaction rate in the \LSV\ is a strong criterion for material selection in \DSk\ to allow for the lowest possible veto threshold.  Although the veto threshold will dictate what the ultimate efficiency will be for the detection of neutrons in the final detector configuration, the requirement for the neutron detection efficiency is set as \SI{>99}{\percent}, a target that was achieved in \DSf.

\subsection{Calibrations}
\label{sec:Overview-Calibrations}

For the \DSk\ detector to reach its physics goals a comprehensive plan of calibration for the \LArTPC, \LSV, and \WCV\ detectors is necessary.  The calibration equipment and techniques are described in detail in Sec.~\ref{sec:Calibrations}.

In the \SCENE\ experiment\,\cite{Alexander:2013ke,Cao:2015ks} a monochromatic, low energy, pulsed neutron beam at the Notre Dame Institute for Structure and Nuclear Astrophysics was used to study the scintillation light yield from recoiling nuclei in a small \LArTPC, but only up to \SI{40}{\keVr}. The \DS\ collaboration is expanding this work with a new parallel project, the Recoil Directionality (\ReD ) experiment, described in Sec.~\ref{sec:ReD}. The \ReD\ detector is a small \LArTPC\ (similar in size to \SCENE ) equipped with the same \SiPM\ tiles planned for use in \DSk. \ReD\ is designed to select and measure neutron single elastic scatters on argon nuclei, by means of a large acceptance neutron spectrometer, composed of an array of liquid scintillator detectors.  Kinematic requirements of the neutron interactions allow for the precise detection of nuclear recoils in the \LAr, since the neutron energy, scattering angle, and drift field direction can all be precisely measured with the use of external neutron detectors.

\section{Procurement and Purification of the Underground Argon Target}
\label{sec:Argon}

\subsection{Overview and Goals} 
\label{sec:Argon-Overview}
The collaboration has developed a broad strategy to increase the production of \UAr\ to procure the target required for \DSk.  The \Urania\ project will extract and purify the \UAr\ from the \ce{CO2} wells at the Kinder Morgan Doe Canyon Facility located in Cortez, CO at a production rate of 100 kg/day.  It will be necessary to make a final chemical purification of the \UAr\ before deployment into the \LArTPC\ (driven by the filtration capacity of the getter purification unit), bringing the chemical impurity levels to those shown in Table~\ref{tab:getter}.  Additionally, it would be beneficial to further deplete the \UAr\ of \ce{^39Ar}, giving extended sensitivity to \DSk\ and a level of \ce{^39Ar} that is acceptable to be used in an experiment such as \Argo.  The \Aria\ project will serve to chemically purify the \UAr\ to better than the levels shown in Table~\ref{tab:getter} using a cryogenic distillation column called \SeruciOne.  \Aria\ could also potentially further deplete the \UAr\ of \ce{^39Ar} by a second, and larger cryogenic distillation column called \SeruciTwo. The ultimate goal of the \Aria\ project is to process about \AriaSeruciTwoRate\ of argon through \SeruciTwo\ to achieve an additional depletion factor between \AriaDepletionPerPass\ and \AriaDepletionPerTwoPass\ (in addition to the reduction of \ce{^39Ar} already seen in the \UAr).  However, the first objective of the \Aria\ project is to chemically purify the \UAr\ using \SeruciOne.  Therefore, the procurement of the \UAr\ for \DSk\ is broken into two main operations, extraction of the \UAr\ by \Urania\ and then chemical purification by \Aria\ using \SeruciOne.

\begin{table}[t!]
\rowcolors{3}{gray!35}{}
\centering
\caption{Urania/Aria: Inlet purity required by the getter of \DSk.}
\label{tab:getter}
\begin{tabular}{cc}
Element		&Inlet Purity Requirements (ppm)	\\
\hline
\ce{CH_4}		&\SI{<0.25}{}\\
\ce{CO}		&\SI{<0.1}{}\\
\ce{CO_2}		&\SI{<0.1}{}\\
\ce{H_2}		&\SI{<1}{}\\
\ce{H_2O}		&\SI{<1}{}\\
\ce{N_2}		&\SI{<1}{}\\
\ce{O_2}		&\SI{<1}{}\\
\end{tabular}
\end{table}

\subsection{The Urania Project}
\label{sec:Argon-Urania}

The Urania project will extract at least \UraniaTotalProduction\ of low-radioactivity \UAr, providing the required \DSkTotalMass\ of \UAr\ to fill \DSk.   The \Urania\ project will also lay the groundwork for \UAr\ procurement for future, larger argon-based detectors such as \Argo.  The goal of the \Urania\ project is to build a plant capable of extracting and purifying \UAr\ at a rate of \UraniaUArRate, from the same source of \UAr\ that was used for the \DSf\ detector.  

The opportunity to build \Urania\ has grown from the strong relationships between the \DS\ Collaboration and the Kinder Morgan and Air Products corporations. Based on gas analysis of the Cortez stream provided to Kinder Morgan by the \DS\ Collaboration during the extraction of the \DSf\ \UAr\ target, a major industrial partnership between Air Products and Kinder Morgan was established to extract helium from the \ce{CO2} at Kinder Morgan's Doe Canyon facility.  The Air Products helium plant began operation in \UraniaHeStartDate\ and presently supplies \UraniaHeNationalReserveFractionEquivalentRate\ of the production rate at the National Helium Reserve. 

The \DS\ Collaboration reached an agreement with Kinder Morgan to feed the \Urania\ plant with a small fraction (\UraniaGasFeedFraction) of the gas stream returned to Kinder Morgan by Air Products after helium extraction.  This gas stream holds two significant advantages over the gas stream used to extract the \UAr\ for \DSf: it is completely dehydrated,  and it contains only trace amounts of helium. These features greatly simplify the process for \UAr\ extraction by the \Urania\ plant.

Argon from active  \ce{CO2} wells in southwestern Colorado have been found to contain very low levels of the radioactive isotope  \ce{^39Ar}, with the concentration shown to be a factor of \DSfUArArThreeNineDepletion\ below that of argon derived from the atmosphere \cite{Agnes:2016fz}. In an effort lasting more than \SI{5}{\years}, \DSf\ collaborators at Princeton and Fermi National Accelerator Laboratory (Fermilab) extracted and purified \DSfUArMassDelivered\ of \UAr, slightly more than the  \DSfTotalMass\ needed for the target material in the \DSf\ detector. 

The \Urania\ feed gas stream is \UraniaCOTwoFeedFraction\ \ce{CO2}, plus a few percent of \ce{N2}, one percent \ce{CH4}, and \UraniaArFeedFraction\ of \UAr.  The processing scheme of the \UAr\ extraction plant is optimized for this feed composition in order to achieve an \UAr\ purity of better than \SI{99.9}{\percent}.  A modular plant consisting of skid-mounted units deployable on concrete platforms is being designed to carry out the processing.

\begin{figure*}
\centering
\includegraphics[width=\textwidth]{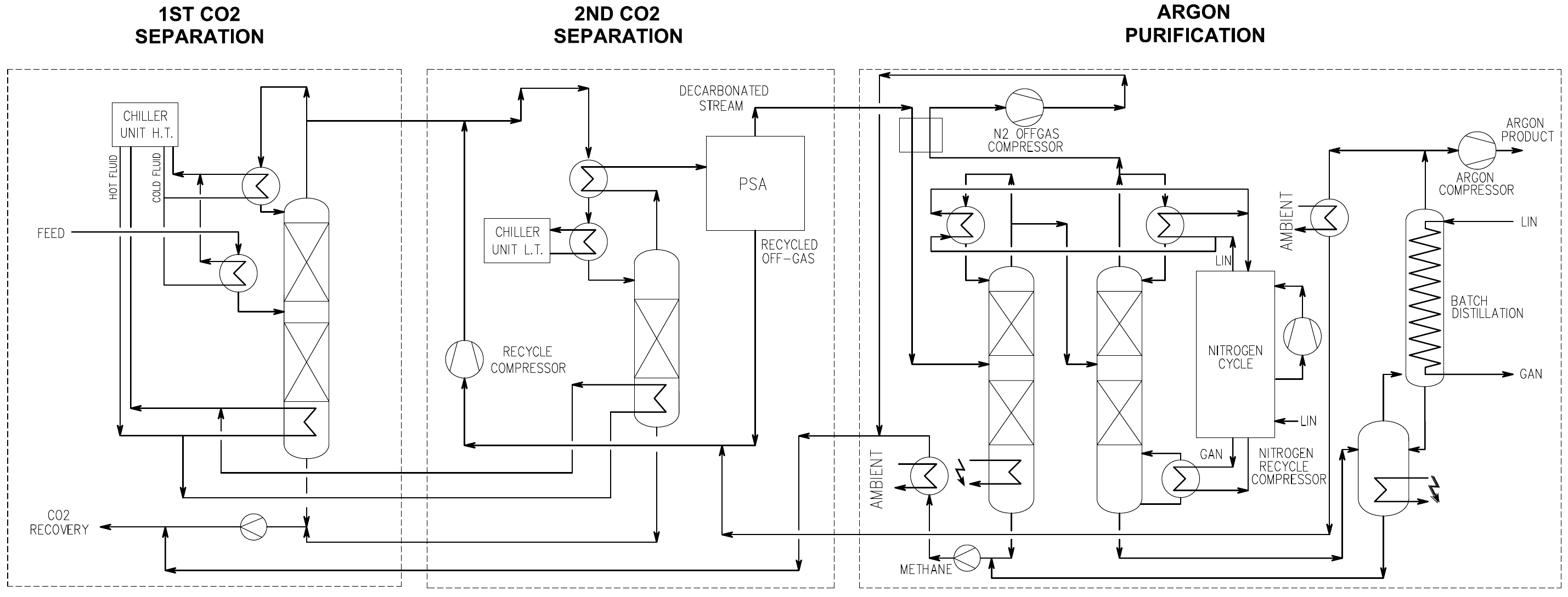}
\caption{Process flow diagram (PFD) for the \Urania\ \UAr\ extraction plant.}
\label{fig:Urania-PID}
\end{figure*}

The \UAr\ extraction plant will consist of three gas-processing units, as shown in Fig.~\ref{fig:Urania-PID}, followed by a cryogenic distillation unit.  The gas-processing units are two \ce{CO2} liquefier/strippers followed by a pressure swing adsorption unit (\PSA).   The first liquefier accepts gas at \UraniaGasFeedPressure, with a flow rate of \UraniaGasFeedFlow\ and a temperature of \SI{5}{\celsius}.  At these conditions the \ce{CO2} partially condenses and the stream is separated into 2-phases (gas/liquid) as it goes to the first stripper.  In the column a controlled quantity of heat is given by a hot fluid working between the chiller condenser and the column reboilers.  The light products are vaporized and recovered from the top of the column in gas phase.  The heavy products (mainly \ce{CO2}) are collected from the bottom, compressed to \UraniaGasReturnPressure\ and returned to Kinder Morgan as a liquid. The light products coming from the column head are cooled down in the second step to approximately \SI{-50}{\celsius} and sent to the second stripper.  The first column produces \UraniaFirstUnitOutputFlow\ of product flow, a factor of \UraniaFirstUnitFlowReduction\ reduction in the amount of gas to be processed by the more complex downstream units.  

The second liquefaction and stripping unit further reduces the \ce{CO2} content, in a similar process as the first stripping unit. The separated \ce{CO2} is joined with that from the first unit and returned to Kinder Morgan.  The product gas from the second stripper is re-heated in a heat exchanger and delivered to the \PSA\ unit, which separates the light fractions, including the argon, from the remaining \ce{CO2}.   The \PSA\ is composed of four adsorption beds to allow continuous operation with short time adsorption cycles.  The desorption of \ce{CO2} is made decreasing the pressure on the bed.  To optimize the performances, the operation of the adsorbers are combined by coupling the purge and pressure swing phases.  At the outlet of the \PSA\ adsorption tanks, one buffer tank is provided in order to dampen process fluctuations and allow for continuous operations of the final distillation process.  The \PSA\ off-gas is delivered to a recycle compressor and sent back to the second \ce{CO2} stripper inlet for reprocessing.  

\begin{figure*}
\centering
\includegraphics[width=0.45\textwidth]{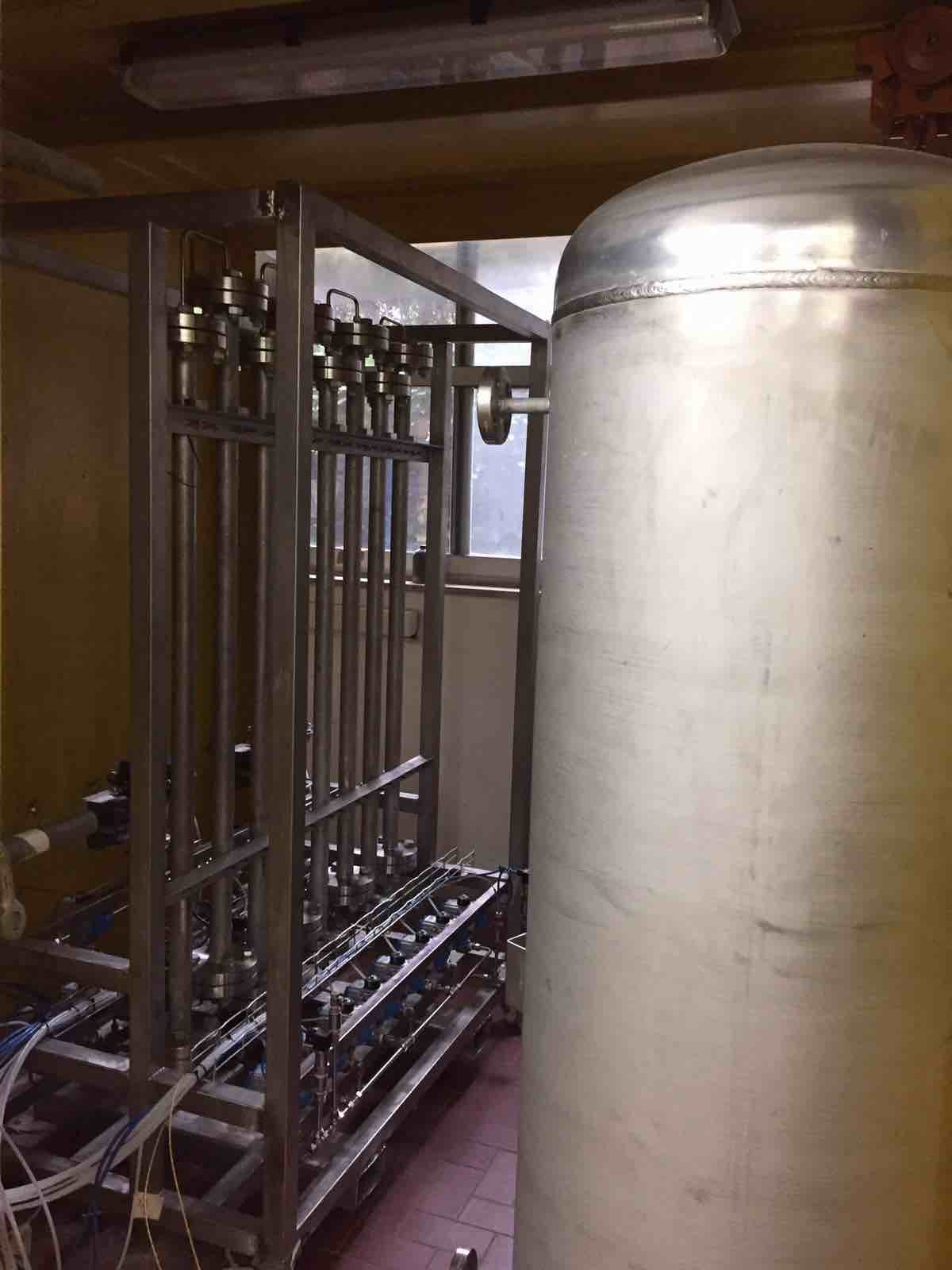}
\caption{Test plant used for characterizing and selecting sorbents for the \UAr\ extraction plant PSA unit.}
\label{fig:Urania-PSATestPlant}
\end{figure*}

The \PSA\ is the most critical unit of the entire process since the dynamic adsorption conditions are the most difficult to simulate and predict.  Optimization of the sorbent and other operational parameters is being done via a small scale lab setup in which breakthrough tests are being performed for a variety of gas species.  A picture of the test setup being used is shown in Fig.~\ref{fig:Urania-PSATestPlant}.  Sorbent screening relies on measurements of the breakthrough curves of the different gas species for the candidate sorbents.  A selection of sorbents which could work for the \PSA\ unit have already been defined, while the final selection of the exact sorbent to be used in the \UAr\ extraction plant will be determined by the test results.

The final unit of the \UAr\ extraction plant consists of three cryogenic distillation columns.  The \ce{CO2}-free product coming from the \PSA\ plant is pre-cooled and sent to the first column, which works at a lower pressure (\SI{\sim9}{\barg}), for the removal of \ce{CH4}.  The second column is used to remove the remaining light fractions from the resulting \ce{N2}-rich stream, and the third to perform the final purification of the \UAr\ using a batch distillation process.  In addition to removing the \ce{CH4} and \ce{N2} at this point, any \ce{^85Kr} present in the stream will also be removed by three cryogenic distillation columns.  The \ce{CH4}-rich and \ce{N2}-rich distillation wastes are returned to Kinder Morgan along with the \ce{CO2}.  The final product, \UraniaArFinalPurity\ pure \UAr, will be taken in liquid form from the top of the last column and a small portion collected into a tank to check the quality of the argon.  The majority of the liquid \UAr\ will be sent to the appropriate cryogenic vessels for shipment to Sardinia, where it will undergo final chemical purification by the \SeruciOne\ column.

\subsection{The Aria Project}
\label{sec:Argon-Aria}

The aim of the \Aria\ project is to perform chemical purification of the \UAr\ extracted by \Urania.  \Aria\ will also be the test bench to develop active depletion of \ce{^39Ar} from the \UAr\ to possibly provide \DAr\ targets for \LAr\ detectors.  \Aria\ consists of two \AriaSeruciHeight\ tall distillation columns of different processing diameters, \SeruciOne\ and \SeruciTwo, capable of separating isotopes by means of cryogenic distillation, a process that exploits the tiny difference in volatility due to the difference in isotopic mass~\cite{Lindemann:1919bq,Urey:1932gl,deBoer:1948br,deBoer:1948fc,deBoer:1939cs,Bigeleisen:1961cm}.

Design of the plant started in April~2015 with seed funding from the US NSF through \grant{PHY}{1314507}.  \Aria\ is to be installed in underground  vertical shafts of diameter \AriaMonteSinniDiameter\ and depth \AriaSeruciHeight, located at the Seruci mine campus of CarboSulcis, a mining company owned by the Regione Autonoma della Sardegna (\RAS).  In February~2015 a proposal was submitted to \INFN\ and \RAS, and the funding for the \SeruciOne\ column was approved on July 24, 2015.  Construction of \SeruciOne\ started in September 2015 at Polaris in Italy.    

Measurements of the relative volatility of argon isotopes~\cite{Boato:1962hg,Boato:1961hb,Boato:1959bn} and their theoretical interpretation~\cite{Casanova:1964gm,Casanova:1960dj,Fieschi:1961cd} marked the birth of the Italian school of condensed matter in Genoa and Milan.  The study of the relative volatility of argon isotopes was recently revisited~\cite{CanongiaLopes:2003ju,Calado:2000iq}, and shows a promising path for the separation of \ce{^39Ar} from \ce{^40Ar}.  

Algorithms developed by to calculate the relative volatility of argon isotopes, based on the extensive and detailed models available in the literature, predict that the volatility of \ce{^39Ar} relative to \ce{^40Ar} is \AriaArVolatiityRatio, and that it stays constant within theoretical uncertainties in the range of temperatures practical for the distillation of argon (\AriaArDistillationTemperatureRange).  The small volatility difference can be used to achieve active isotopic separation by using a cryogenic distillation system with thousands of equilibrium stages.

Design of the \Aria\ plant was optimized on the basis of high-precision numerical methods for estimating the isotopic separation of \ce{^39Ar} from \ce{^40Ar}.  \DS\ Collaborators developed two independent numerical codes, one based on the McCabe-Thiele method~\cite{McCabe:1925be}, and a second based on the Fenske-Underwood-Gilliland (FUG) method and its derivative, the Wynn-Underwood-Gilliland (WUG) method~\cite{Underwood:1949dw,Gilliland:1940ja,Fenske:1932do}.  Calculations for the isotopic separation power of \ce{^39Ar} from \ce{^40Ar} and of the processing rate were performed with the custom codes as well as with software routines supported by commercial chemical engineering CAD programs, such as Aspen~\cite{AspenTechnologyInc:2015ux}.

\begin{figure*}[!t]
\centering
\includegraphics[width=\textwidth]{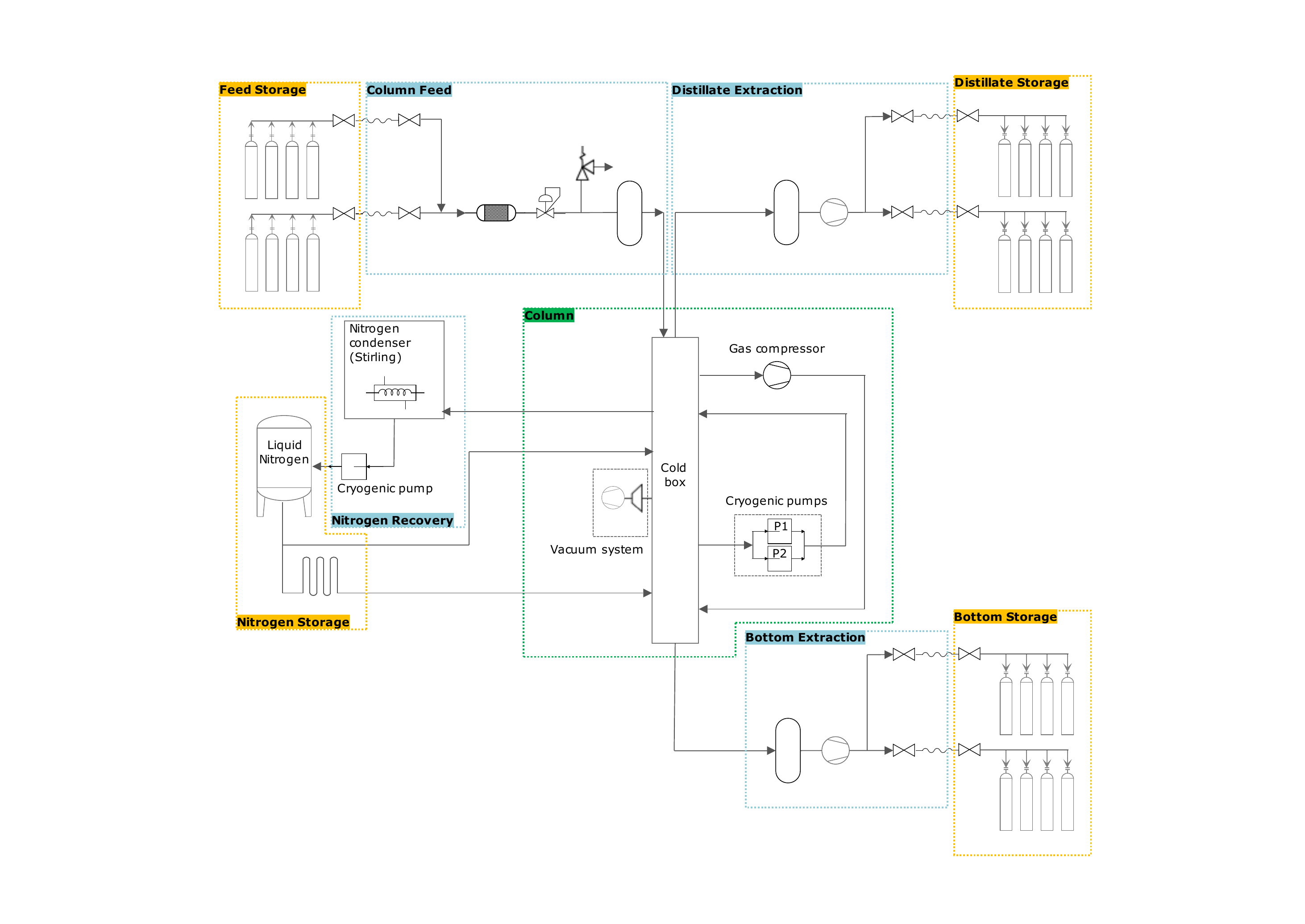}
\caption{Block diagram of the \UAr\ isotopic separation process.}
\label{fig:Aria-Block-Diagram}
\end{figure*}

Fig.~\ref{fig:Aria-Block-Diagram} illustrates the core of the process for isotopically separating \ce{^39Ar} from \ce{^40Ar} in the \UAr. The process consists mainly of two loops: the process loop, where the argon is distilled and the \ce{^39Ar} is separated from the \ce{^40Ar} and the refrigeration loop where nitrogen gas and liquid is used to evaporate and to condense the argon. Most of the heat is recovered, thanks to the compressor that pumps the nitrogen gas evaporated in the condenser to the reboiler and to the pumps that move the liquid nitrogen produced in the reboiler to the condenser, making the system as efficient as possible.  In Fig.~\ref{fig:Aria-Block-Diagram} all the sub-parts of the plant are represented:
\begin{compactitem}
\item Feed station, to filter and regulate the feed to the column;
\item Compressor station, to bottle the distillate at the bottom;
\item Vacuum system, to keep a good vacuum in the cold-box, in order to minimize the heat losses;
\item \ce{LN2} storage;
\item Nitrogen condenser system, consisting of 4 Stirling cryo-refrigerators needed to re-condense the nitrogen, used in a close loop.
\end{compactitem}

The two \Aria\ columns, \SeruciOne\ and \SeruciTwo, will consist of \AriaCentralModulesNumber\ modules of \AriaCentralModulesHeight\ height, plus a top module (condenser) and a bottom module (reboiler).  \SeruciOne\ is being constructed as a working prototype for the main column, \SeruciTwo, and will serve to prove the isotopic separation power of the cryogenic distillation method before the construction of the large and more expensive \SeruciTwo.  \SeruciOne\ will have the same separation capabilities as \SeruciTwo, but a factor of \num{15} less in the overall production rate of \DAr.  All modules will be pre-assembled at the factory and ready for deployment in the shaft.  In November 2015, Carbosulcis started the refurbishment of the \SeruciOne\ mine shaft, in order to make it suitable to host both the \SeruciOne\ and \SeruciTwo\ columns.


Calculations indicate that \SeruciOne\ will be able to process \UAr\ at a rate of \AriaChemicalRate, while removing all chemical impurities (including traces of \ce{N2}, \ce{O2}, and \ce{Kr}) with separation power better than \AriaChemicalPerPass\ per pass.  Additionally, \SeruciOne\ will be used to test the isotopic separation of the argon, in order to further reduce the \ce{^39Ar} content in the \UAr.  The same models which have been used to calculate the chemical purification rate, have also been used to show that \SeruciOne\ would be able to isotopically separate the \UAr\ at a rate of \AriaSeruciOneRate, while obtaining an \ce{^39Ar} depletion factor of \AriaDepletionPerPass\ per pass. Scaling the model to the size required for \SeruciTwo\ predicts a processing rate of \AriaSeruciTwoRate, while maintaining the same depletion factor. 

\begin{figure*}[!t]
\centering
\includegraphics[width=0.45\textwidth]{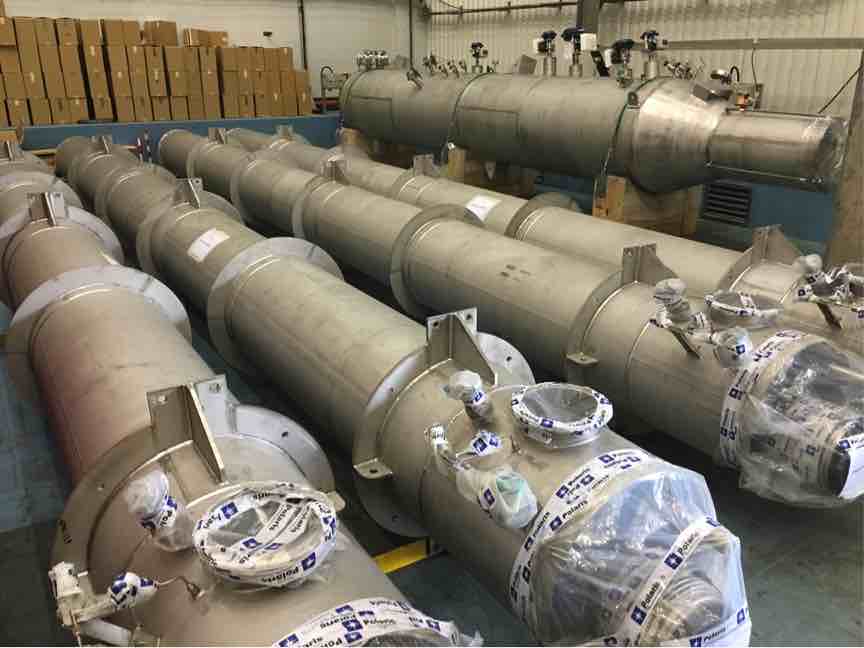}
\caption{Top and first four central modules of the \SeruciOne\ column of \Aria.}
\label{fig:Aria-Modules}
\end{figure*}

Work on \Aria\ is well underway, both at the places where the modules are being constructed and tested and at the installation site.  All modules for \SeruciOne\ have already passed through the phase of pre-construction, with some of them have already been constructed and tested.  After construction of the modules, two complementary leak checks are performed.  During the first, the process column and all the service pipes are individually checked for leaks at room temperature.  After the first check, the pipes are wrapped with superinsulation and everything assembled into the cold box.  The second check is a full module check, with an additional check done on the bottom reboiler module at \SI{77}{\kelvin}.  So far, the bottom reboiler, top condenser and one central module have passed both checks at room temperature and the bottom module is undergoing the final cold temperature check.  After the cold test is finished, the bottom, top and first central modules will be sent to Carbosulcis for installation in a first test configuration consisting of just these three initial modules (top, bottom and one central).  Four additional modules are now also ready for the second leak check.  A picture of the top and first 4 central modules of \SeruciOne\ are shown in Fig.~\ref{fig:Aria-Modules}, as they await the second leak test at CERN.  

In parallel with the construction and testing of the models, preparation at the installation site has been underway for more than two years.  Carbosulcis has already obtained all authorizations for the installation of the initial test of the top, bottom and one central module of \SeruciOne, and the building to house the test has already been prepared and equipped with the required utilities.  Refurbishment of the building housing the winze for the Seruci shaft has been completed, along with the refurbishment of the winze itself.  The refurbishment of the shaft and the surface buildings to support installation and operations has also started.

\subsection{Measuring \ce{^39Ar} Composition in \UAr\ and \DAr}
\label{sec:Argon-DART}

To qualify the natural \ce{^39Ar} depletion in gas extracted from the \ce{CO2} wells at Cortez, Colorado, two measurements of the argon in the liquid phase were performed in the past by \DSf\ collaborators: one reported in~\cite{Xu:2015do}, based on a device measuring \SI{0.5}{\kilo\gram} of \UAr\ that was able to set an upper limit for the residual radioactivity of \SI{<0.65E-2}{\becquerel\per\kilo\gram}, followed by the direct measurement in  \DSf~\cite{Agnes:2016fz}, which showed a residual \ce{^39Ar} specific activity of \DSfUArArThreeNineActivity.

It is not known and not proven that the reduction factor will be the same for all samples of gas extracted from underground wells, as local rock conditions can significantly alter the composition. Therefore, all future batches of \UAr\ will have to be tested individually. Moreover, the effectiveness of \Aria\ to further suppress the \ce{^39Ar} content must be experimentally demonstrated.  Preliminary studies are ongoing to explore the possibility of exploiting the ArDM detector, whose collaborators recently joined \DSk, at the Laboratorio Sotteraneo de Canfranc, Spain. It could be possible to insert a small O(kg) chamber filled with the \UAr\ (or eventually \DAr) to be measured in the middle of the ton scale ArDM vessel filled with \AAr.  This would allow for the use of ArDM as an active veto for the gamma background, allowing for the sensitivity that is required for the measurement to be reached.  It is also conceived that, in a second stage, a similar detector could be built and operated at the Seruci mine site.

\subsection{\UAr\ Delivery Plan and Management}
\label{sec:Argon-Delivery}

The design for \DSk\ relies on procurement of the \UAr\ using the \Urania\ argon extraction plant.  In this scenario, the background rejection demonstrated with \DSf, and extended to \DSk\ using MC simulations and analytical techniques, is already sufficient to ensure successful reduction of \ce{^39Ar} below the \BackgroundFreeRequirement\ goal for background.  \SeruciOne, and eventually \SeruciTwo, could then be used to further deplete the \UAr\ of \ce{^39Ar} to extend its sensitivity and begin procurement of the required target for a larger experiment such as \Argo.

\DSk\ will need \DSkTotalMass\ of \UAr. The \UAr\ extracted by the \Urania\ plant in Colorado will be shipped to Sardinia for further chemical purification with the \SeruciOne\ column. Several simulations have been performed in order to evaluate the production rate from \SeruciOne, the purification performance and the argon recovery in the cryogenic distillation plant. Assuming a reduction factor of chemical impurities of \AriaChemicalPerPass\ and an argon recovery factor of about \SI{85}{\percent} each pass, the total amount of \UAr\ needed to produce \DSkTotalMass\ that could be used for \DSk\ is thus about \AriaUArToGetDAr. 

\begin{figure*}[!t]
\centering
\includegraphics[width=\textwidth]{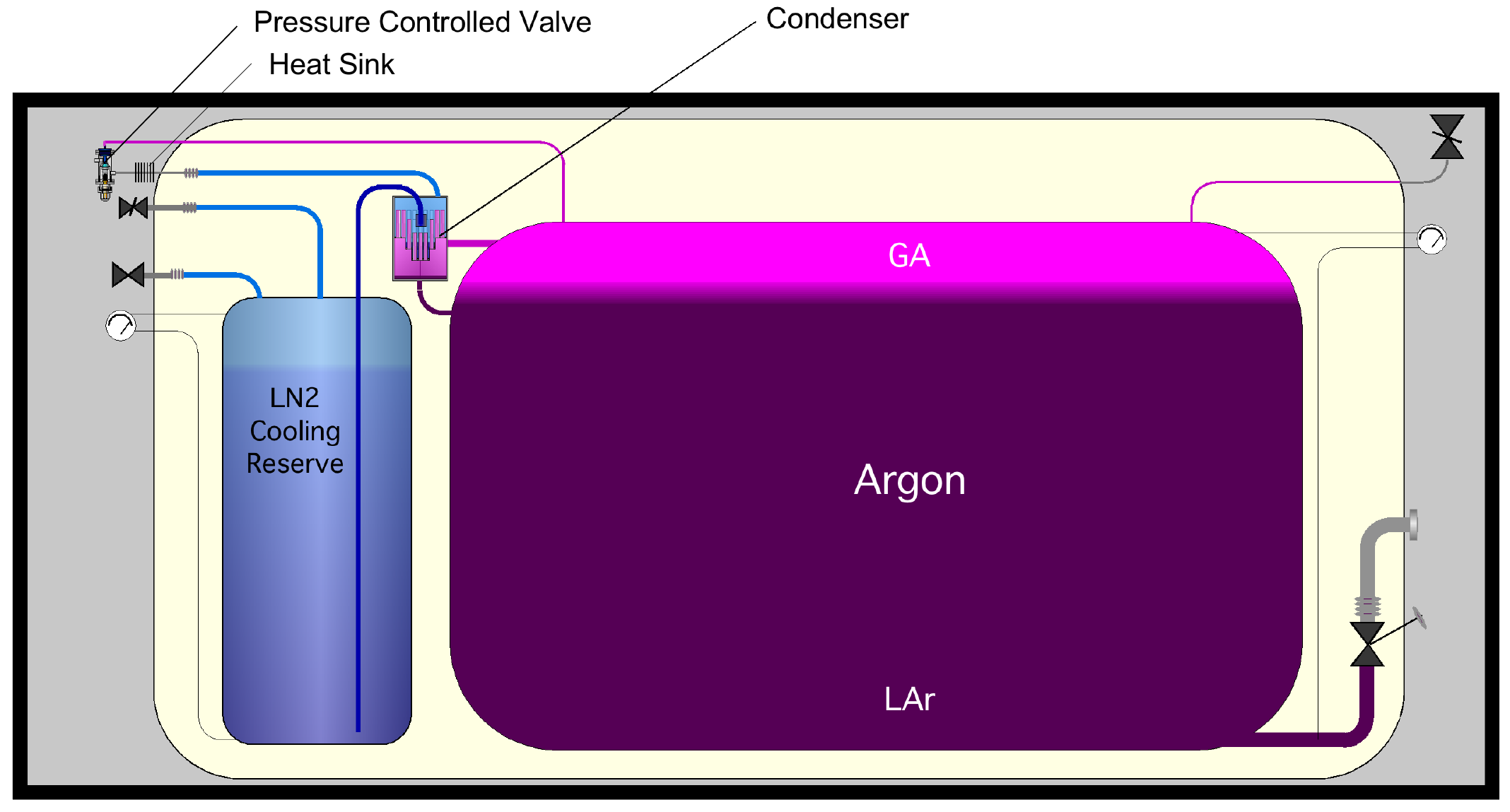}
\caption{Schematic view of the cryogenic shipping vessel design.}
\label{fig:Urania-UArShipping}
\end{figure*}

The shipment from Colorado to Sardinia will be done by boat in order to minimize the cosmic activation of the argon. For the shipment of the \UAr\ from Colorado, several options have been analyzed, and the collaboration has decided that the baseline method will be to ship the \UAr\ in liquid phase using custom built cryogenic vessels.  This is a more efficient and cost effective method, compared to shipping the \UAr\ in gas phase. The custom built cryogenic vessels will have a double wall structure, there will be two inner volumes with the larger one containing the \UAr\ and the smaller containing \LIN.  The outer volume would be at ultra-high vacuum to thermally insulate the two inner volumes from the atmosphere at ambient temperature.  During the transport, a \LIN\ fed condenser would slowly re-liquify the \UAr\ as it evaporated away, ensuring that none of the \UAr\ would be lost during the trip.  A schematic view of the cryogenic vessel design is shown in Fig.~\ref{fig:Urania-UArShipping}.  A minimum of five cryogenic vessels is foreseen:
\begin{compactitem}
\item One for \Urania\ production;
\item One for \Aria\ feed;
\item One for \Aria\ production;
\item Two traveling between Colorado, Sardinia and LNGS.
\end{compactitem}

The extracted \UAr\ will be shipped to Sardinia in batches of \UraniaUArShipmentMass\ (roughly every \SI{100}{\days}) in liquid phase, also eliminating the need to liquify the \UAr\ to be reprocessed by the \Aria\ column. After chemical purification in \Aria, the \UAr\ will then be shipped to Gran Sasso, also in batches of \UraniaUArShipmentMass, and stored in the argon recovery system.  In order to minimize the exposure to cosmic rays, the argon will be stored underground at Seruci and/or Gran Sasso.

\section{Cryogenics System and Cryostat}
\label{sec:Cryogenics}

\subsection{System Overview}
\label{sec:Cryogenics-Overview}

The design of the cryogenics system is largely based on the success of the \DSf\ system, with added improvement and scaling to handle much larger argon mass.  The cryogenics system must be able to efficiently cool and purify \DSkCryogenicsCapableLArMass\ of \LAr\ to a purity level similar or better than in \DSf, while also maintaining similar detector pressure stability. The robust safety features of the \DSf\ cryogenics system will be implemented fully in the \DSk\ system, in particular the feature that keeps the system immune to total power failure (including failure of the UPS system), see Sec.~\ref{sec:Cryogenics-Safety}.  

\begin{figure*}[h!]
\centering
\includegraphics[width=\columnwidth]{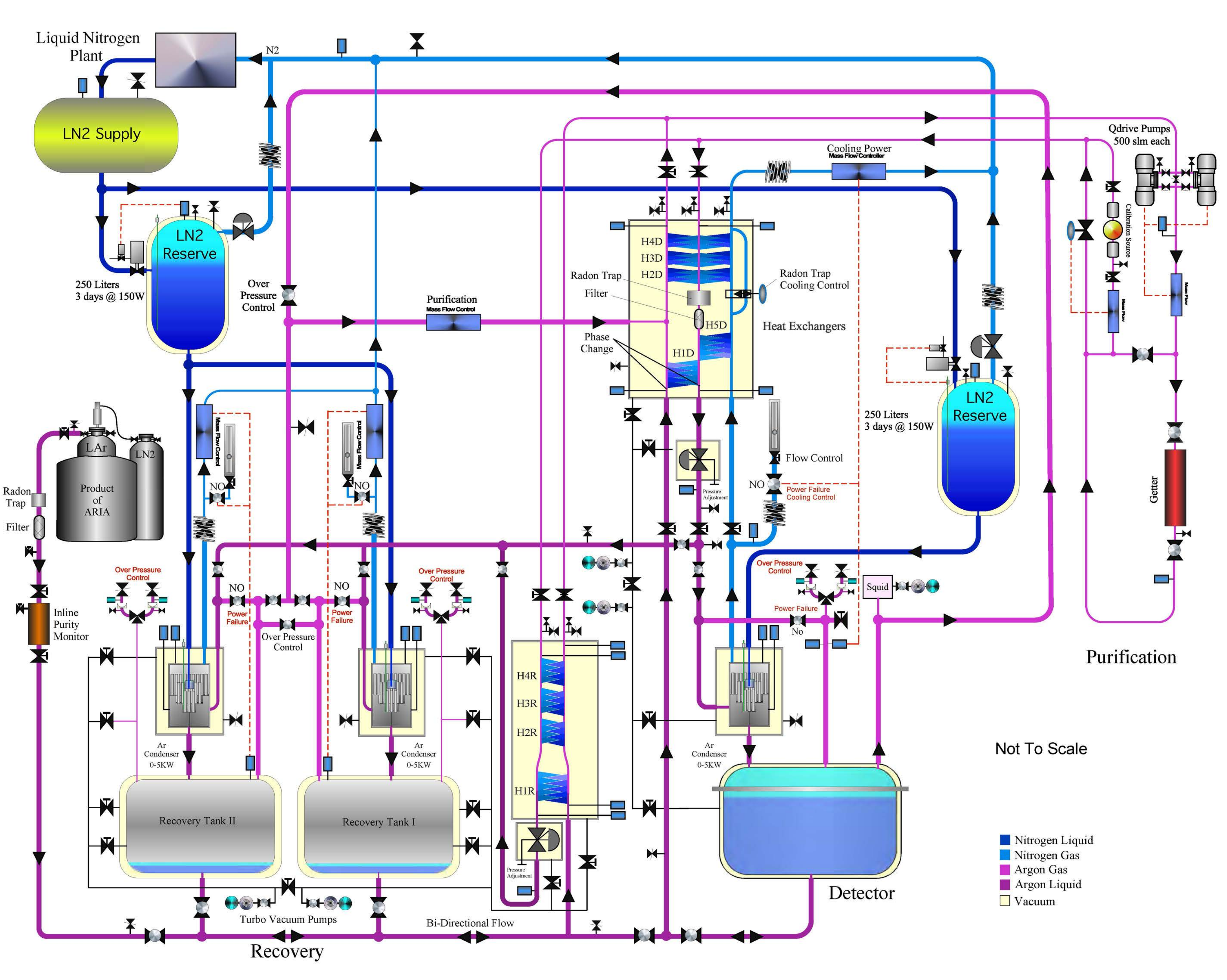}
\caption[Schematic of the \DSk\ cryogenics and gas handling system.]{Schematic of the \DSk\ cryogenics and gas handling system.  The violet and blue lines indicate the argon and nitrogen loops, respectively, while the lighter and darker colors indicate the liquid and gas phases, respectively.  Thicker lines are vacuum insulated transfer lines for cold media and thinner lines are single-wall lines for room temperature media.}
\label{fig:Cryogenics-DSkSketch}
\end{figure*}

Since the amount of contamination which enters the \LAr\ is proportional to the surface area of materials inside the cryostat, the purification capacity of the system that is initially needed to bring the \LAr\ to the level required for physics running was estimated based on an approximate surface area scaling between the \DSf\ and \DSk\ detectors.  The new system should have a total gas recirculation flow equivalent to \DSkCryogenicsGasFlowTotal\ to initially reach the desired purity of \LAr\ during the course of about one month.  A key feature added to the design concept for \DSk\ is the introduction of high efficiency heat exchangers to recover cooling power during fast purification circulation.  The cryogenics system will include a new custom-designed argon condenser, again based on the successful \DSf\ system.  The cooling power of the argon condenser is variable from \DSkCryogenicsCondenserCoolingPowerRange.  This wide cooling power range is designed to cover fast initial argon filling (up to \DSkCryogenicsCondenserCoolingPowerFilling), as well as normal operations (at about \DSkCryogenicsCondenserCoolingPowerNormal\ due to the need to remove the power consumption of the cold electronics and cryostat heat leak) and \TPC-off mode ($\sim$\DSkCryogenicsCondenserCoolingPowerTPCOff).  With efficient thermal management, the total external cooling power needed to maintain the system in \TPC-off mode is about \DSkCryogenicsCondenserCoolingPowerTPCOff.  A complete overview of the system is shown in Fig.~\ref{fig:Cryogenics-DSkSketch}.

The argon gas (\GAr) purification relies on the proven commercial SAES getter system, as in \DSf, but with a much larger getter to handle the needed flow.  A SAES heated-\ce{Zr} getter with a flow-rate capacity $>$\DSkCryogenicsGasFlowTotal\ will be able to reduce impurities to the required level.  The getter will be integrated into the cryogenics system so that it can be used to purify \GAr\ simultaneously from the detector or the necessary \LAr\ recovery system (see Sec.~\ref{sec:Cryogenics-Recovery}).  The cryogenics system will be located far from the central detector (outside the \LSV\ and the \WCV).  Only the small transfer lines \SIrange{10}{15}{\meter} long will provide the necessary link for the \LAr\ delivery from the recovery system.

The main system parameters of the cryogenics system are given in Table~\ref{tab:Cryo-parameters}.  The parameters that are listed in bold are considered to be system requirements, most often driven from experience gained during operations of the \DSf\ cryogenics system.  

\begin{table*}
\rowcolors{3}{gray!35}{}
\centering
\caption[\DSk\ cryogenics system parameters and requirements.]{\DSk\ cryogenics system parameters and requirements.  Requirements are those parameters listed in bold.}
\begin{tabular}{lc}
{\bf Parameter}															&{\bf Value}\\ 
\hline
\LAr\ mass during normal operations											&\DSkTotalMass\ \\
{\bf Maximum \LAr\ mass that can be purified}									&\DSkCryogenicsCapableLArMass \\
{\bf Commissioning time}													&$\le$\SI{60}{\days} \\
\LAr\ boiling threshold at \SI{3}{\meter} depth									&\SI{60}{\milli\watt\per\centi\meter\squared} \\
{\bf Condenser cooling power during normal operations}							&\DSkCryogenicsCondenserCoolingPowerNormal\ \\
Maximum heat load in \LAr\ before bubbling at cryostat bottom						&\DSkCryostatBottomLArNoBubblingHeatDissipation\ \\
{\bf Total max. cooling power of the condenser}									&\DSkCryogenicsCondenserCoolingPowerMax\ \\
Minimum condenser cooling power to hold \LAr\ inventory							&\DSkCryogenicsCondenserCoolingPowerTPCOff\ \\
{\bf Minimum heat recovery efficiency of heat exchanger}							&$>$\DSkCryogenicsRecoveryExchangerEfficiency\ \\
																	&\SI{<0.1}{\ppb} \ce{O2} \\
\multirow{-2}{*}{\bf \LAr\ purity required for stable \STwo\ generation}\cellcolor{gray!35}	&\cellcolor{gray!35}($>$\DSkCryogenicsDSfElectronLifetime\ equiv.) \\
\cellcolor{white}{\bf Max. total radioactivity of the cryostat}							&\cellcolor{white} \DSkCryostatSSThUActivityRequirement\ \ce{238U} and \ce{232Th} \\
\cellcolor{gray!35}														&\cellcolor{gray!35} \num{15.6}\DSkCryogenicsPressureStability\ 	\\
\multirow{-2}{*}{\bf Pressure inside cryostat during normal operations}				&(\num{1.075}$\pm$\SI{0.007}{\bar}) \\
Pressure stability achieved in \DSf\											&\SI{0.023}{\psi} (RMS) \\
{\bf Vent pressure of spring-loaded pressure relief valves}							&\DSkCryogenicsReliefValveSetting\ \\
{\bf Max. pressure of cryostat safety rupture burst disks}							&\DSkCryogenicsRuptureDiskLimit\ \\
{\bf Max. \LAr\ head height before \DSkCryogenicsHOne}							&\DSkCryogenicsLArHeadHeight\ \\
{\bf Max. flow of gas from liquid and gas withdrawal loops}							&\DSkCryogenicsGasFlowTotal\ \\
Max. flow rate through gas getter purifier										&$>$\DSkCryogenicsGasFlowTotal\ \\
{\bf Flow rate of cool \GAr\ from detector ullage}								&\DSkCryogenicsGasFlowGWL\ \\
																	&\DSkCryogenicsLiquidFlow\ \\
\multirow{-2}{*}{\bf Flow rate of \LAr\ from cryostat bottom}	\cellcolor{gray!35}			&\cellcolor{gray!35}(\DSkCryogenicsGasFlowLWL\ \GAr\ flow equiv.) \\
\cellcolor{white}Flow rate of \GAr\ through single Q-drive pump						&\cellcolor{white}\DSkCryogenicsQdriveSpeed\ \\
\cellcolor{gray!35}														&\cellcolor{gray!35}\num{2} pumps in parallel \\
\multirow{-2}{*}{Total number of Q-drive pumps for max. flow}						&(\num{+1} extra as a spare) \\
\LAr\ pressure at input of \DSkCryogenicsHOne\								&\DSkCryogenicsHeatExchangerArHeadPressure\ \\
Pressure of \GAr\ at output of Q-drive pumps									&\DSkCryogenicsAfterQDriveArPressure\ \\
Pressure of \GAr\ after gas getter purifier										&\DSkCryogenicsAfterGetterArPressure\ \\
Pressure of \LAr/\GAr\ after heat exchangers									&\DSkCryogenicsAfterRegulatorArPressure\ \\
Total mass of \LIN\ storage in cooling system									&\DSkCryogenicsLINStorageMass\ \\
Efficiency of radon purification by activated charcoal trap							&\SI{<2}{\micro\becquerel\per\kilo\gram} after trap \\
{\bf Max. pressure of insulating vacuum volume and lines}							&\DSkCryogenicsInsulationVacuum\ \\
{\bf Max. He leak rate at all welds and joints}									&\DSkCryogenicsHeliumLeakCheckRequirement\ \\
\end{tabular}
\label{tab:Cryo-parameters}
\end{table*}

\subsection{System Description}
\label{sec:Cryogenics-Description}

The cryogenics system will consist of the following subsystems:
\begin{inparaenum}
\item Gas handling system;
\item \LAr\ handling system including the recovery units, cold liquid transfer lines and other thermal insulation components;
\item Cryostat;
\item Liquid nitrogen (\LIN) reserve system as the cold source;
\item Radon trap.
\end{inparaenum}

As previously mentioned, the cryogenics system will be designed to maintain the stable working pressure of argon and continuously purify the argon to maintain the designed purity level.  The required  argon pressure stability in the cryostat is \DSkCryogenicsPressureStability.  This requirement is derived from the required stability in the \STwo\ signal and was successfully achieved with \DSf\ under a variety of heat loads.  In fact, the \DSf\ system has maintained a pressure stability within \SI{0.023}{\psi} for a period of more than two years.  The required \LAr\ purity that is needed to achieve a $\geq$\DSkCryogenicsDSfElectronLifetime\ electron lifetime is less than \SI{<0.1}{\ppb} \ce{O2}.  An electron drift lifetime \SI{>11}{\milli\second} was achieved and measured in \DSf, resulting in a purity so good that no significant loss of electrons was seen across the \SI{400}{\micro\second} drift.

The \DSf\ gas purification system philosophy was based on the efficient removal of cold gas from the ullage and purification of this gas using a getter system. Clean \LAr\ is then returned to the detector after purification and condensing of the \GAr.  Outgassing from \TPC\ materials is minimal once the surfaces are cooled to \LAr\ temperature.  Hence, after an initial cleanup, the steady-state source of argon contamination is mostly the warm areas of the detector.  In \DSf, the cryostat design kept both the \LAr\ and the \GAr\ in the ullage cold, and this will be duplicated in \DSk\ since it was proven to be extremely effective.  However, due to the large total argon mass of \DSk, if initial purity is low, purification using only the ullage gas withdraw may take too long to achieve the required purity.  Therefore, the plan is to maintain the \DSf\ philosophy, but in addition add a fast liquid withdraw point from the bottom of the cryostat.  Once the system purity reaches the design value, the \LAr\ withdraw can be either reduced or stopped, while the gas would continue to recirculate and continuously purify the argon coming from the detector ullage.
 
The \GAr\ recirculation from the \DSk\ cryostat ullage will be realized as was done for \DSf.  The gas flow from the ullage will have a maximum rate of \DSkCryogenicsGasFlowGWL, almost entirely boiled off by the \DSkCryogenicsCondenserCoolingPowerNormal\ heat load resulting from the combination of the cryostat heat leak and the power dissipated by the cold electronics inside the \LArTPC.  This gas flow is pumped out along the paths of the detector cables in order to flush out all possible outgassing from the cables.  The gas is then pushed through the getter for purification followed by the condenser where it is re-liquified before returning to the detector.  The cabling routes and gas circulation route, as well as the gas route tubing size, are carefully designed so that at the minimum circulation speed all outgassing of any warm part will still be flushed out.  The minimum flow speed will be larger than the back diffusion speed.

In order to reach the minimum electron lifetime goal (\DSkCryogenicsDSfElectronLifetime) in a reasonable time after the detector has been filled (about one month), the total combined purification speed is designed to be \DSkCryogenicsGasFlowTotal.  The cryogenics system will use a set of heat exchangers to form a high efficiency heat recovery system to boil off \LAr\ being drawn out of the bottom of the cryostat using heat from the clean warm \GAr, at the same time cooling the incoming \GAr\ as it heads towards the condenser.  So, in addition to the \GAr\ being drawn from the top of the detector as is done in \DSf, at the same time \LAr\ will also be pumped from the bottom of the cryostat, boiled off using clean warm \GAr\ flowing toward the cryostat, and then joined with the \GAr\ coming from the top of the cryostat to continue through the purification system, eventually all returning to the cryostat as \LAr.  

Recent detailed studies and lab tests show that an efficient approach to pumping the argon could be to realize a high-flow-rate QDrive gas pump, as shown in Fig.~\ref{fig:Cryogenics-DSkSketch}.  The QDrive pump pulls the \LAr\ from the bottom of the cryostat up to the bottom of the heat exchangers from the detector, while at the same time creating higher pressure on the return side of the heat exchangers.  The gas circulation system is driven by two QDrive pumps (variable speed up to \DSkCryogenicsQdriveSpeed\ each, see Fig.~\ref{fig:Cryogenics-DSkSketch}).  The QDrive pumps, custom-built in cooperation with QDrive, Inc., are magnetically driven resonance pumps.  A smaller version of such pump is being used in \DSf's \DSkCryogenicsDSfCirculationSpeed\ gas system.  The QDrive pump is sealed within a stainless steel body and hence is intrinsically safe against loss of the valuable \UAr\ in case of an operational failure.  The QDrive pump uses a set of long-life reed valves and a spring loaded piston moving in resonance mode, making it intrinsically wear free and maintenance free.  Two such pumps will be operated in parallel to achieve the required \DSkCryogenicsGasFlowTotal\ flow, with only one needed at a lower flow rate.  A third pump will be built as a spare.  With the QDrive pumps, \GAr\ streaming from the ullage and liquid from the detector bottom are pumped and joined together in the heat exchanger system between \DSkCryogenicsHTwo\ and \DSkCryogenicsHOne\ in Fig.~\ref{fig:Cryogenics-DSkSketch}.  A mass flow controller maintains the balance of the two flow streams.

The maximum height that the pump should push the \LAr\ up to is determined to be \DSkCryogenicsLArHeadHeight, to avoid a vapor pressure drop at the outgoing side of \DSkCryogenicsHOne\ to near the triple point of argon.  This means that \DSkCryogenicsHOne\ must be no more than \DSkCryogenicsLArHeadHeight\ above the \LAr\ pump.  A dedicated test is planned to confirm the argon head height.  This important parameter will be used to determine the location (vertical elevation) of \DSkCryogenicsHOne\ and hence the rest of the cryogenics system.

The differential pressure between incoming warm \GAr\ and the outgoing cold argon provide the ideal heat exchange condition in the heat exchanger stages \DSkCryogenicsHTwo, \DSkCryogenicsHThree, and \DSkCryogenicsHFour, to achieve better than {\DSkCryogenicsRecoveryExchangerEfficiency} heat recovery efficiency.  The chilled incoming cold \GAr\ is then liquefied by the outgoing \LAr\ in \DSkCryogenicsHOne.  This approach recovers most of the \LAr\ enthalpy and can be operated in a very stable condition thanks to the naturally self-balancing heat loads applied on the two counter-current streams.  The total \LAr\ withdrawal is only about \DSkCryogenicsLiquidFlow\ in order to provide the required \DSkCryogenicsGasFlowLWL\ gas flow needed for an initial fast and efficient cleanup of the \LAr.

An important feature of the cryogenics system design is its flexibility.  Once the desired purity is reached during the commissioning phase, it will be possible to turn off or to reduce the total cooling power to the minimum amount required for the operation of the detector with its anticipated \DSkCryogenicsGasFlowGWL\ \GAr\ flow.  The enthalpy contained in the ullage gas is also recovered in the high-efficiency heat exchangers \DSkCryogenicsHTwo, \DSkCryogenicsHThree, and \DSkCryogenicsHFour, resulting in a requirement of \DSkCryogenicsCondenserCoolingPowerNormal\ for the argon condenser to balance the cryostat heat leak and the power dissipated by the cold electronics inside the \LArTPC.

As previously mentioned, cold \GAr\ is pumped from the top ullage of the cryostat and into the heat exchanger just before H2D.  Room temperature \GAr\ exiting the heat exchanger will then be pumped through the Zr-based hot getter purifier and then returned to the heat exchanger.  From there, the returning warm argon will then be cooled in the tri-media heat exchanger by the outgoing cold \GAr\ and the outgoing \ce{N2} boiled off from the condenser.  This gas reaches near \LAr\ temperature before entering the radon trap.  Further, the cold \GAr\ will be mostly liquefied in H1D by the outgoing \LAr.  The argon condenser then liquefies any residual \GAr, and the fully liquefied argon will be returned to the \TPC\ using a long vacuum-insulated transfer line that will pass through the \WCV\ and \LSV.  

The purification of argon will happen only in the gas phase.  No liquid purification is being considered: the experience of WARP, ICARUS, and MicroBOONE has shown that resins and adsorbers in use for liquid phase purifications source enormous amounts of radon into the system.  The experience with noble liquid TPCs such as XENON, \DS, and LUX shows that hot zirconium-based getters from SAES are the best solution on the market to remove the major sources of chemical contamination (\ce{O2}, \ce{CO2}, \ce{H2O}, \ce{N2}).  These getters typically require an inlet purity at the ppm level and guarantee an outlet purity at the ppb level.  Since they work in a closed loop on a definite batch of noble liquid, they eventually can reach purities in the ppt level.  For the flow rate necessary for the recirculation system, a commercial unit from SAES satisfying the required needs has already been identified .  The active purification element is a large, replaceable zirconium cartridge.  There are no consequences to the cryogenics system due to the need of operating the purifying cartridge at high temperature since the getter is engineered by SAES to take care of the required thermalization and to return the gas at room temperature.  There is no single point outside the getterÕs own enclosure which is operated at high temperature.  In the course of the experimentation with the \DSf\ system, significant experience with these units was gained, including the construction and commissioning of a custom-built set of high capacity cartridges for the purification of the \UAr\ as it was injected into the \DSf\ cryostat.  The experience with \DSf\ was very positive and successful, and the \UAr\ immediately obtained a purity corresponding to an equivalent electron lifetime $>$\DSkCryogenicsDSfElectronLifetime, at the completion of fill.

The radon trap is a volume filled with activated synthetic charcoal which the clean argon from the getter must pass through, prior to returning to the detector.  Radon originating from any warm surfaces of the cables or the gas handling system will be efficiently removed by the radon trap.  The location of the radon trap, right above the \DSkCryogenicsHFive\ heat exchanger, allows the cold \GAr\ to be the cooling source of the trap.  Once in thermal equilibrium with the temperature achieved at the point between \DSkCryogenicsHTwo\ and \DSkCryogenicsHFive\ (near the \LAr\ temperature independent of the actual argon flow rate) the activated charcoal radon trap is effectively \DSkCryogenicsRadonTrapEfficiency\ efficient, meaning no detectable radon makes its way passed the radon trap~\cite{Simgen:2009dv}.  The key to the radon trap design is its operation in a range of temperature just above the argon liquefaction temperature.  With a \SI{2}{\micro\becquerel\per\kilo\gram} contamination (easily achieved in \DSf) the physics goals of \DSk\ are achievable.  

A feature identical to the one used in the \DSf\ gas system is the \ce{^{83m}Kr} source insertion system, which is integrated directly into the gas panel.  An adjustable percentage of the flow of the recirculating \GAr\ can be sent through a \ce{^{83m}Kr} source to provide a temporary distributed calibration signal in the \TPC.  Again as in \DSf, the \ce{^{83m}Kr} will pass through the radon trap.  To make sure the krypton will pass through the trap before decaying, the radon trap will be designed to handle the large flow by increasing its cross section but not the argon path length,  compared to that of \DSf.  \LAr\ should be avoided in the radon trap, so a bypass is added to adjust the amount of \ce{N2} flow into the heat exchangers \DSkCryogenicsHTwo, \DSkCryogenicsHThree, and \DSkCryogenicsHFour.

The cold source of the entire system is provided by \LIN.  The \LIN\ cooling loop is a closed system that consists of a \DSkCryogenicsLINStorageMass\ \LIN\ source supply tank outside Hall~C, a local reserve tank capable of storing \DSkCryogenicsLINBackupTime\ worth of cooling power in emergency mode, a \LIN/argon condenser, and a set of heat exchangers to recycle the spent \ce{N2} cold gas to pre-cool the incoming argon.  The spent \ce{N2} gas is then returned to the \LIN\ recycling system and the re-liquefied \LIN\ returned to the large source tank, see Sec.~\ref{sec:Facilities-LNGSUndergroundLab}.

\subsection{Pressure Balance in the Heat Exchangers}
\label{sec:Cryogenics-PressureBalance}

The pressure at a few critical locations must be carefully balanced in order to achieve heat exchanger efficiency better than \DSkCryogenicsRecoveryExchangerEfficiency.  The detector cryostat working pressure is set at \DSkCryogenicsDetectorOperationPressure\ (\DSkCryogenicsDetectorOperationImperialPressure) so that \LAr\ can be pumped to reach and enter H1D of the multi stage heat exchanger.  The \LAr\ head pressure is reduced to \DSkCryogenicsHeatExchangerArHeadPressure\ (see Fig.~\ref{fig:Cryogenics-DSkSketch}) at that point.  The argon then passes through the three tri-media heat exchanger stages labelled as \DSkCryogenicsHTwo, \DSkCryogenicsHThree, and \DSkCryogenicsHFour\ as it is brought up to room temperature.  The output gas of the \DSkCryogenicsHFour\ is at room temperature as it then enters the input of QDrive pumps.  The QDrive pumps increase the pressure to \DSkCryogenicsAfterQDriveArPressure\ as the gas is pumped to the getter purifier.  After passing through the getter purifier at room temperature, the return gas goes back into the \DSkCryogenicsHFour\ exchanger at \DSkCryogenicsAfterGetterArPressure\ pressure (reduced by the flow resistance of the getter purifier) where it begins to be cooled down.  By the time the argon exits the cold output of the \DSkCryogenicsHOne\ heat exchanger it has been partially re-liquefied.  The liquid/gas mixture then passes through a cryogenic regulator where the pressure is reduced to \DSkCryogenicsAfterRegulatorArPressure.  The mixture then travels to the condenser where it is completely condensed into liquid and then sent to the detector.  The cold \GAr\ coming from the detector or from the recovery system that is to be purified is injected into the gas stream between \DSkCryogenicsHOne\ and \DSkCryogenicsHTwo\ of the multi stage heat exchanger.  The design of the four stage heat exchanger is better than \DSkCryogenicsRecoveryExchangerEfficiency\ efficient.

\subsection{Condensers}
\label{sec:Cryogenics-Condensers}

\begin{figure}[h!]
\centering
\includegraphics[width=0.4\columnwidth]{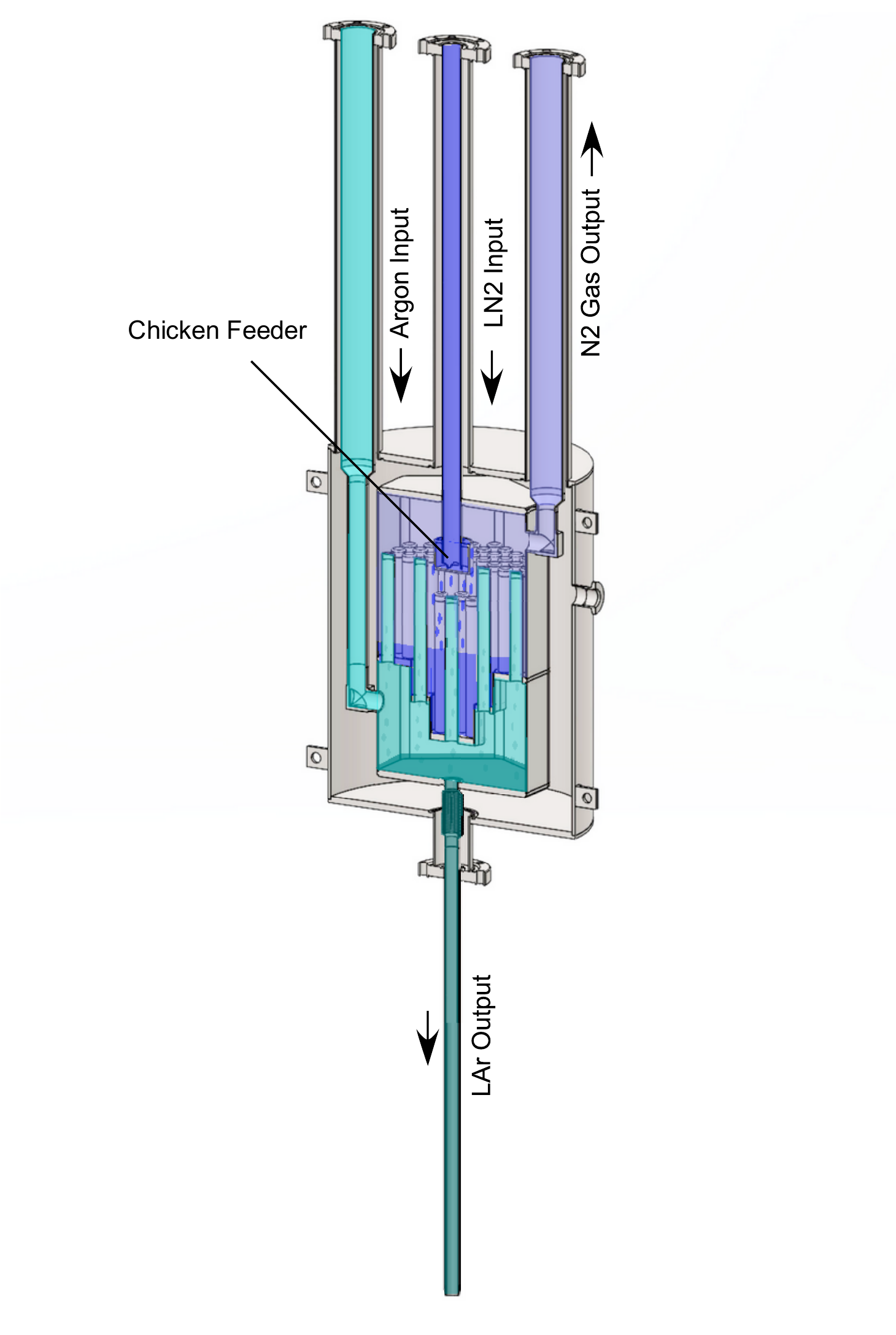}
\caption{3D rendering of the cryogenics system \LAr\ condenser.}
\label{fig:Cryogenics-Condenser}
\end{figure}

The \LAr\ condenser is the key part of the cryogenics system.  A safe, reliable and performance flexible condenser is critical to the success of the experiment.  The condenser, with \SI{100}{\percent} stainless steel construction, is a much improved version of the condenser used for the \DSf\ cryogenics system~\cite{Agnes:2015gu}.  The condenser design is shown in Fig.~\ref{fig:Cryogenics-Condenser}.  There will be three condensers in the cryogenics system, one for the detector and one each for the two recovery systems.  One of these condensers has already been completely fabricated at UCLA, and is currently being prepared for initial testing. 

The main heat exchange surfaces are composed of the walls of many short stainless steel tubes.  There are a total of \DSkCryogenicsCondenserTubesNumber\ \DSkCryogenicsCondenserTubesDiameter\ diameter tubes of two different lengths, \DSkCryogenicsCondenserTubesLenghtLong\ and \DSkCryogenicsCondenserTubesLenghtShort, with their tops blinded.  The bottoms of these tubes are welded onto a three-step holder.  Above the holder and outside the tubes (\ce{N2} volume) is filled with \LIN\ at a level that depends on the cooling power requirement, while the volume below the holder and inside the tubes (argon volume), isolated from the \LIN\ section, is filled with incoming \GAr\ to be liquefied.  The heat exchange between the \GAr\ and \LIN\ is done through the \DSkCryogenicsCondenserTubesWallThickness\ thick walls of the \DSkCryogenicsCondenserTubesDiameter\ diameter tubes  The size, spacing and number of tubes was derived from experience, detailed multi-physics simulations and laboratory tests.  The maximum heat exchange would occur when the \ce{N2} volume is filled with \LIN\ and has been designed to be approximately \DSkCryogenicsCondenserCoolingPowerMax.

\LIN\ is fed into the condenser through a half-inch line that ends in a feature called the ``chicken feeder'' and controls the flow of \LIN\ into the volume automatically by the pressure balance.  The chicken feeder does not allow gas to flow back into the liquid supply line which provides flow stability.  A mass flow controller is used on the \LIN\ vent line to control the outgoing flow of the boiled-off \ce{N2}, which is proportional to the cooling power of the condenser.  By controlling the \ce{N2} flow the cooling power is precisely controlled, as successfully done in \DSf\ where a detector gas pressure stability of \DSkCryogenicsDSfArPressureStability\ (RMS) was achieved.  The more \LIN\ that fills the condenser, the more surface area of the tubes that are in contact with the \LIN, which increases the cooling capacity.  The \LIN\ level in the condenser volume is automatically balanced once the outgoing flow (hence cooling power) is set.  The \LIN\ feed line pressure to the condenser is controlled by the back pressure regulator (located after the \LIN\ supply vessel) which keeps the pressure at \DSkCryogenicsLINSupplyPressure.

The \GAr\ is fed to the lower part of the condenser from the side (see Fig.~\ref{fig:Cryogenics-Condenser}) and condenses as it comes in contact with the tube array.  The liquid then drops down to the bottom of the condenser and flows into a vacuum-insulated transfer line that transports the liquid to the detector.  Once \LAr\ drops down from the bottom of the half inch tubes, the cooling to this element of \LAr\ stops, preventing further cooling.  The choice of the length of the half inch tube, and hence the time in contact with the cold stainless steel surface, is critical to avoid over-cooling by the much colder \LIN.  As \LAr\ drops away, it is quickly replaced with incoming \GAr, which continuously sends heat through the wall to the \LIN\ side.  On the \LIN\ side, the spaces between the tubes are carefully determined such that the boil off gas is constrained and pushes away \LIN, so the surface of the tube is not covered by \LIN\ \SI{100}{\percent} of the time.  This design, combining the above considerations, eliminates the need for temperature balancing using high pressure \LIN\ to match the argon temperature. The cooling power of the condenser is variable from zero up to the full power by simply controlling the boil-off \ce{N2} gas flow rate which is proportional to the latent heat of the \LIN\ and hence the cooling power.  This flow rate is controlled using the information from the detector pressure feedback.

\subsection{Low Background Cryostat}
\label{sec:Cryogenics-Cryostat}

\begin{figure}[h!]
\centering
\includegraphics[width=0.4\columnwidth]{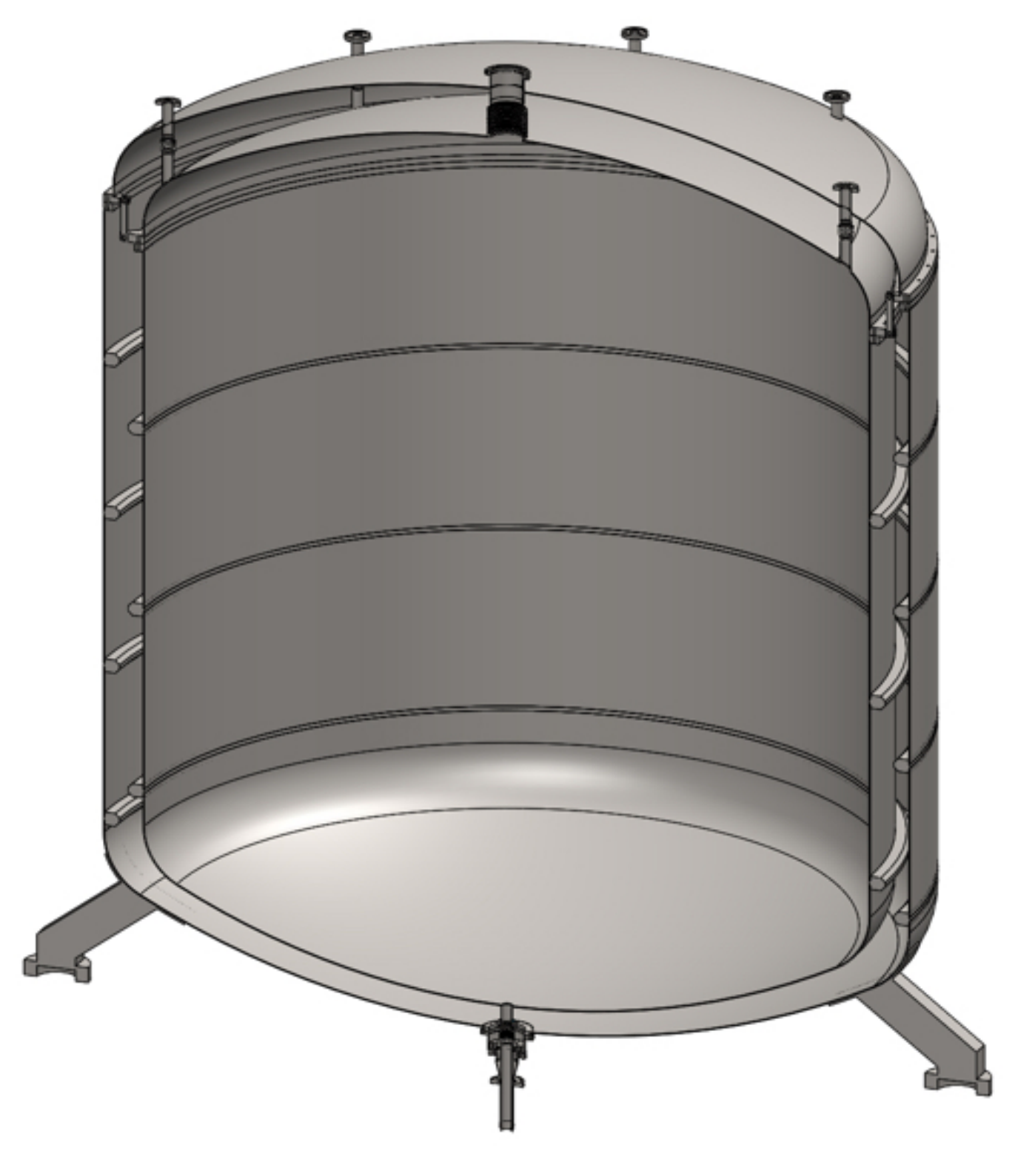}
\caption{3D rendering of the \DSk\ cryostat.}
\label{fig:Cryogenics-Cryostat}
\end{figure}

The conceptual design of the \DSk\ cryostat, as shown in Fig.~\ref{fig:Cryogenics-Cryostat}, follows closely that of the successful \DSf\ cryostat: a $4\pi$ vacuum-insulated vessel made of three separate parts,  the top assembly, the inner cryostat vessel, and the vacuum insulation vessel.  The top assembly is formed by an outer dome and an inner dome mechanically linked with solid low-thermal-conduction rods.  The outer dome of the top assembly has a flange that matches the insulation vessel flange and uses a metal o-ring as the vacuum seal.  The inner dome has a flange that matches the inner cryostat vessel and uses a V-groove indium wire seal.

Due to the large size of the two vessels, their lateral walls are reinforced by a number of strengthening ribs.  Thermo-mechanical finite element analysis is ongoing to finalize the strengthening elements design.  The inner vessel strengthening ribs are placed outside of itself, while for the insulation vessel they are placed inside (both inside the insulating vacuum volume).  With the cryostat and the insulation vessel ribs designed to be offset vertically when the vessels are in their final positions, the multi-layered super-insulation placed in between blocks all lines of sight for thermal radiation.

The cryostat port designation and locations will also follow the design used in \DSf.  The ports for various services penetrating the two top domes use bellows links to eliminate mechanical stress due to thermal loads.  The services include all signal readouts for light sensors, instrumentation control cables, \GAr\ purification recirculation ports, and the clean \LAr\ return port.  Both the gas-out and liquid-in ports are double-wall bayonets, to avoid thermal loss.  Unlike in \DSf, at the bottom of the cryostat there is a \LAr\ fast-drain port, which is a specially designed bayonet assembly connecting to both the inner vessel and the outer insulation vessel.

\TPC\ mounting anchor points are integrated into the flange of the inner dome so that the \TPC\ can be mounted easily before the vessels are closed.  The detector leveling will be done similarly to that of \DSf, by adjusting the level of the cryostat.  The leveling system will be integrated into the anchor points on the cryostat support structure.

The \DSf\ cryostat was made of stainless steel selected to have low levels of trace radioactivity (\DSkCryostatSSThUActivityRequirement\ of \ce{^238U} and \ce{^232Th}), and the baseline design of the \DSk\ cryostat assumes the same material, which gives input for the activity of the cryostat in the MC simulations.

It is known from commercial use that titanium alloys have properties sufficient for cryogenic use and so there is no need to develop a special group of alloys for cryogenic applications, if the conditions in Table~\ref{tab:Cryogenics-Titanium} are taken into account and the gas composition impurities are kept below (wt. \si{\percent}); \ce{O2}:\num{0.10}; \ce{N2}:\num{0.03}; \ce{H2}:undetectable; \ce{C}:\num{0.08}~\cite{Moieseyev:2006tq}.  

\begin{table*}
\rowcolors{3}{gray!35}{}
\centering
\caption{Properties of various titanium alloys at three different temperatures.}
\begin{tabular}{clcccccc}
Temperature		&Properties							&VT-1-0	&OT-4	&VT-5-1	&VT-6-S	&VT-14	&VT-16\\
\hline
\cellcolor{white}		&$\sigma_{\text{B}}$ [\SI{}{\mega\pascal}]		&470		&830		&820		&860		&980		&870\\
\cellcolor{white}		&$\sigma_{0.2}$ [\SI{}{\mega\pascal}]		&400		&770		&800		&810		&890		&780\\
\cellcolor{white}		&$\delta_{5}$ [\SI{}{\percent}]				&30		&24		&21		&17		&15		&20\\
\cellcolor{white}		&$\psi$ [\SI{}{\percent}]					&65		&50		&55		&55		&60		&60\\
\multirow{-5}{*}{\RoomTemperature}		
				&KCV [\SI{}{\mega\joule\per\meter\squared}]	&2.0		&0.9		&1.0		&1.4		&1.3		&1.4\\
\hline
\cellcolor{white}		&$\sigma_{\text{B}}$ [\SI{}{\mega\pascal}]		&920		&1430	&1320	&1310	&1440	&1380\\
\cellcolor{white}		&$\sigma_{0.2}$ [\SI{}{\mega\pascal}]		&700		&1400	&1310	&1270	&1380	&1310\\
\cellcolor{white}		&$\delta_{5}$ [\SI{}{\percent}]				&48		&13		&16		&16		&10		&15\\
\cellcolor{white}		&$\psi$ [\SI{}{\percent}]					&60		&19		&27		&48		&45		&40\\
\multirow{-5}{*}{\cellcolor{white}\LINNormalTemperature}
				&KCV [\SI{}{\mega\joule\per\meter\squared}]	&2.2		&0.5		&0.4		&0.6		&0.4		&0.6\\
\hline
\cellcolor{white}		&$\sigma_{\text{B}}$ [\SI{}{\mega\pascal}]		&1310	&1560	&1580	&--		&--		&--\\
\cellcolor{white}		&$\sigma_{0.2}$ [\SI{}{\mega\pascal}]		&920		&1410	&1400	&--		&--		&--\\
\cellcolor{white}		&$\delta_{5}$ [\SI{}{\percent}]				&24		&16		&15		&--		&--		&--\\
\cellcolor{white}		&$\psi$ [\SI{}{\percent}]					&17		&10		&9		&--		&--		&--\\
\multirow{-5}{*}{\SI{20}{\kelvin}}
				&KCV [\SI{}{\mega\joule\per\meter\squared}]	&1.3		&0.4		&0.4		&--		&--		&--\\
\end{tabular}
\label{tab:Cryogenics-Titanium}
\end{table*}

The LUX collaboration has pioneered the use of titanium in low-background applications~\cite{Akerib:2015is}, and recent results of experimental studies performed within the \DS\ Collaboration have shown great promise in establishing titanium as a low-background structural material: a titanium cryostat is a possible alternative to the baseline option of a stainless steel cryostat.  Titanium's higher strength-to-mass ratio means that for the same specific activity, titanium will give a lower gamma background contribution than stainless steel.  However, titanium has a higher ($\alpha$,$n$) neutron yield, resulting in about the same neutron production rate for structures made of titanium or stainless-steel with the same specific activity.  This sets a goal of using titanium with \DSkCryostatTitaniumThUActivityRequirement\ of total activity, and if unachievable, stainless steel with the lowest possible activity (and at least as good as was accomplished in \DSf) should be used (assuming the cryostat masses given in Table~\ref{tab:Materials-Activity}).

Initial investigations by members of the \DS\ Collaboration of radioactive contaminants in samples of rolled titanium and titanium-containing products were carried out during~2012 and 2013.  A new approach was proposed during 2014 and 2015 based on industrial processes implemented at the Solikamsk Magnesium Plant in Russia to manufacture titanium sponge.  Research results showed that the widely used Kroll process can be modified to produce titanium sponge with a predictable \ce{^238U} and \ce{^232Th} activity \DSkCryostatTitaniumThUActivityRequirement.

First tests suggest that the ultra-low radioactivity (ULR) titanium sponge can be converted into metal sheets with properties equal to VT-1-0 grade industrial construction material, while maintaining the original radioactive purity.  A plan to make industrial scale volumes of the ULR titanium from raw material includes the following steps:

\begin{compactenum}
\item Production of ULR titanium sponge;
\item Crushing sponge with further magnet separation;
\item Isostatic compacting of the sponge;
\item Vacuum arc-melting or electron-beam melting with special attention paid to recontamination, producing cylindrical and rectangular ingots (\DSkCryostatTitaniumIngotDiameter\ diameter and up to \DSkCryostatTitaniumIngotLength\ long);
\item Cutting long semi-industrial ingots to pieces \DSkCryostatTitaniumIngotCutLength\ long;
\item Rolling with annealing, leading to the production of sheets and massive titanium rods with optimal mechanical properties and low deformation for large flanges;
\item Laser welding.
\end{compactenum}

\begin{figure*}[h!]
\centering
\includegraphics[width=\textwidth]{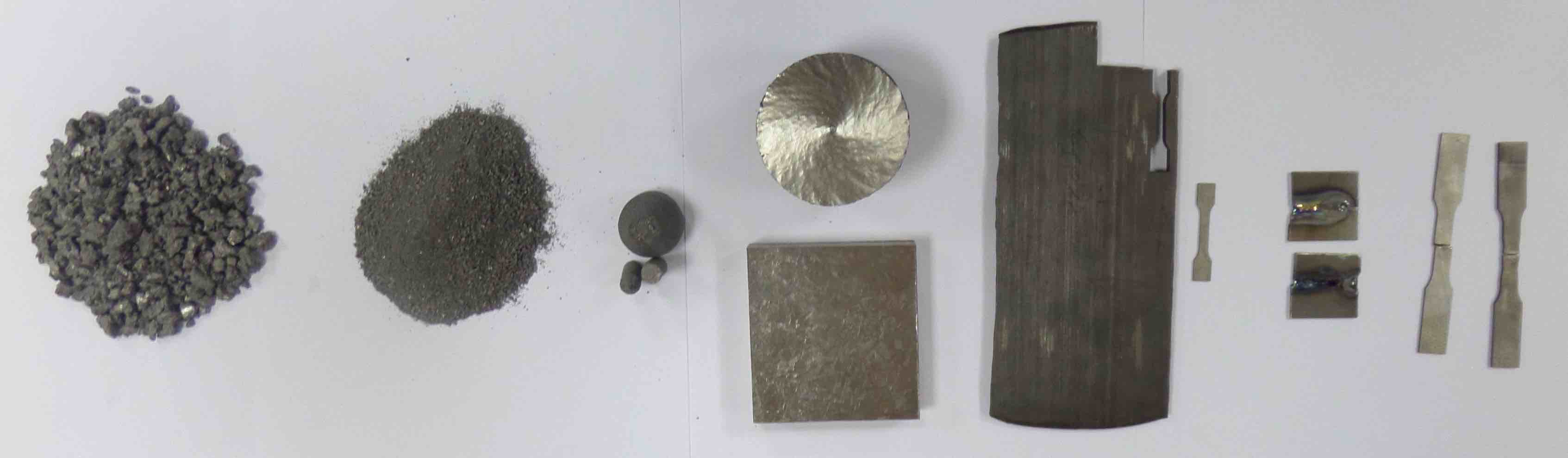}
\caption[Samples of ULR titanium at the different stages of production.]{Samples of ULR titanium at the different stages of production:
\begin{inparaenum}
\item ULR titanium sponge;
\item magnetically separated crushed sponge;
\item isostatic compacted sponge;
\item vacuum arc-welded (cylinder) and electron-beam melted (rectangular block) ingots; and
\item samples for mechanical tests.
\end{inparaenum}
}
\label{fig:Cryostat-TitaniumSamples}
\end{figure*}

Fig.~\ref{fig:Cryostat-TitaniumSamples} shows samples of the ULR titanium at different stages in this production process.

The procedure resulted in samples of ULR material with mechanical strength better than industrial titanium and close to the mechanical properties of stainless steel grade 12X18N10, usually used for low temperature applications.  In fact, one of the samples produced out performed titanium alloy VT-1-0, giving uranium and thorium upper limits of \DSkCryostatTitaniumSampleThreeULimit\ and \DSkCryostatTitaniumSampleThreeThLimit\ respectively.

Since it is certain that stainless steel complying with the background requirements can be obtained, this is taken as the baseline material to be used for the \DSk\ cryostat.  However, to reduce the mass of the cryostat, and to be able to provide a structural material with known, low, and controlled radioactive contaminants, the production of ULR titanium is actively being pursued by the \DS\ Collaboration.  After performing key mechanical tests at room and \LAr\ temperatures, radiopurity screening to assess the final radioactivity of the building material, and computer simulations of a titanium cryostat design for structural integrity and neutron production, a final decision of whether or not this ULR titanium can be used for the \DSk\ cryostat will be made.

\subsection{Argon Storage and Recovery System}
\label{sec:Cryogenics-Recovery}

\begin{figure}[h!]
\centering
\includegraphics[width=0.4\columnwidth]{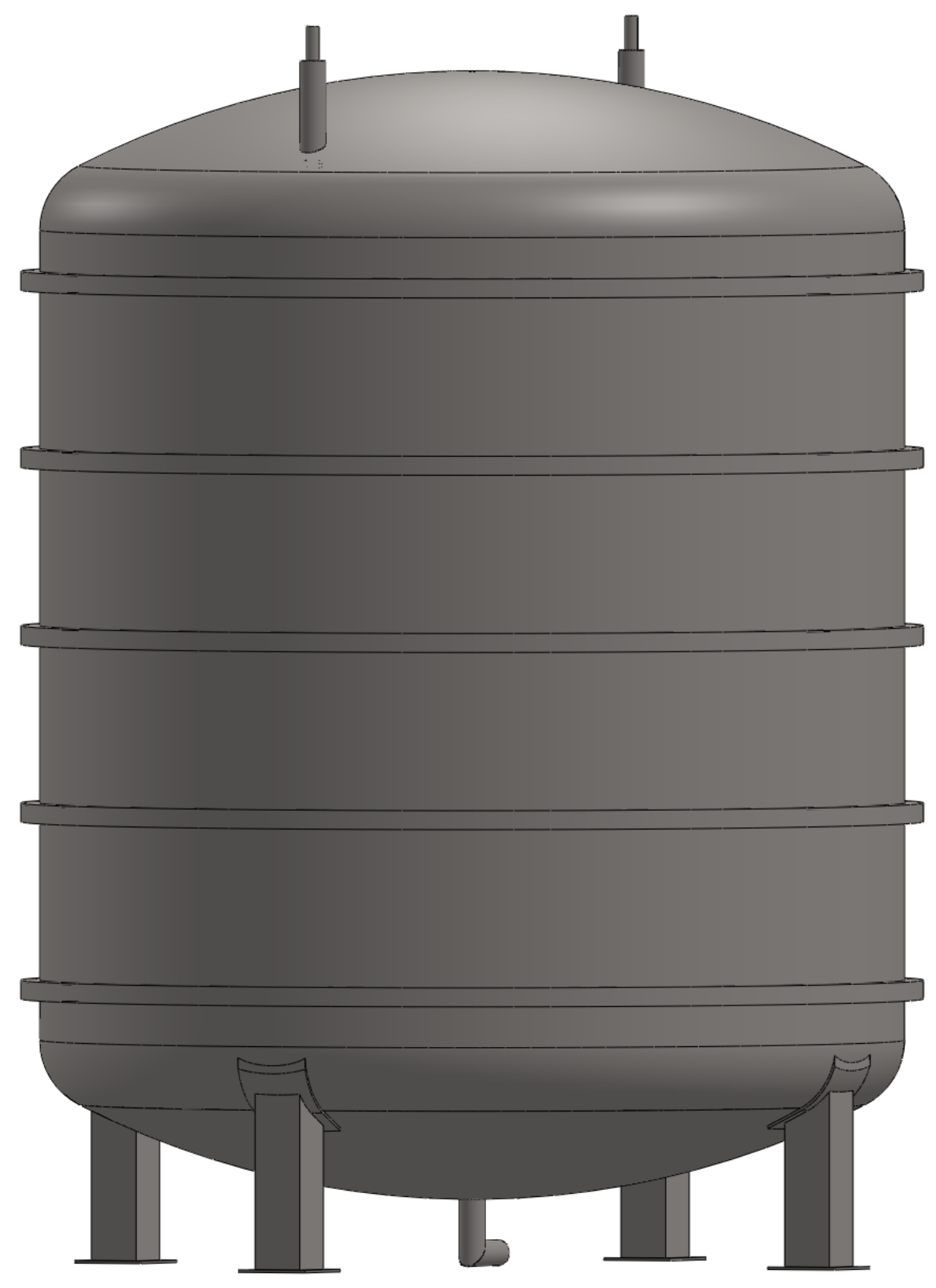}
\caption{3D rendering of the \DSk\ LAr recovery tank.}
\label{fig:Cryogenics-RecoveryTank}
\end{figure}

\DSk\ will be equipped with a zero boil-off recovery and storage system (\DSkCryogenicsRecoveryCryostatOne), capable of recovering and storing the total inventory of \LAr\ (see Fig.~\ref{fig:Cryogenics-RecoveryTank}).  Because of the value of the \LAr, a second recovery system labeled \DSkCryogenicsRecoveryCryostatTwo\ is in a constant cold standby condition ready to back up \DSkCryogenicsRecoveryCryostatOne\ and/or the detector.  Each recovery system is equipped with independent active cooling, capable of handling the full inventory of argon.  Recovery System I will be used when there is a need to empty the detector.  Each recovery system is capable of the rapid transfer of \LAr\ from the detector, simply driven by gravity, in case of unforeseen problems.  The system will also allow recovery of the \LAr\ target for possible scheduled procedures, such as upgrades of the inner detector or maintenance interventions on the cooling loop.  The recovery system will allow the cryogenics system to circulate argon and purify the \LAr\ while the full inventory is stored in the recovery cryostat.  Another set of heat exchangers are placed at \DSkCryogenicsLArHeadHeight\ above the recovery tanks, performing a similar function as the heat exchangers in the gas system described above.  The clean warm gas is then returned through the same heat exchanger to the recovery tank.  The designed overall gas flow rate in the recovery system is also \DSkCryogenicsGasFlowTotal, with the heat exchanging efficiency conservatively expected to be on the order of \DSkCryogenicsRecoveryExchangerEfficiency.  Therefore, the total cooling power required for operation is the power needed to condense \DSkCryogenicsRecoveryExchangerInefficiency\ of the overall gas flow rate, or \DSkCryogenicsRecoveryExchangerCoolingPower, and from a heat load expected to be similar to that of the detector cryogenics system (\DSkCryogenicsRecoveryCryostatPowerLeak).  An argon/\LIN\ condenser, identical to that of the detector will be installed on each of the recovery systems.

When the argon first arrives at \LNGS\ and is fed into the cryogenics system, it will undergo a quality check by passing through a purity monitor that will check the electron drift lifetime.  This test is an effective way of verifying the purity of the \LAr.  As the test is taking place, the gas and \LAr\ will be collected in \DSkCryogenicsRecoveryCryostatTwo\ isolating it from the main inventory of stored \LAr.  Once the content of a batch of argon has been tested and found to be acceptable, the argon will be transferred to \DSkCryogenicsRecoveryCryostatOne\ where it will be stored long-term.

While the liquid is being stored in \DSkCryogenicsRecoveryCryostatOne, it can be further purified by circulating the argon through the getter purifier system as discussed above.  Both liquid and gas in \DSkCryogenicsRecoveryCryostatOne\ or \DSkCryogenicsRecoveryCryostatTwo\ can be purified simultaneously.  In the event that argon is found not to have high purity, as found by the purity monitor test, it can be purified while the argon is isolated in \DSkCryogenicsRecoveryCryostatTwo.  Once the argon has reached an acceptable level of purity it can be transferred to \DSkCryogenicsRecoveryCryostatOne.

Each recovery system has an independent insulating vacuum system and a \LIN\ condenser for cooling.  If a problem arises with one of the recovery systems, the argon can be transferred to the other recovery system.  Each recovery system, as well as the detector, has as a simple standalone control system that can override the main control system in the event of an overpressure condition.  When an overpressure condition occurs, the assumption is made that there is a problem with the main control system and  local control takes over.  In this condition, the override control system only cools the volume liquid to a default temperature.  If the detector pressure does not drop, a transfer of gas will take place by venting the detector to \DSkCryogenicsRecoveryCryostatOne\ where it will be condensed.  This will take place below the pressure limit of the rupture disks.  At that point, the situation would be inspected and a decision would be made if the problem was serious of whether or not to transfer the liquid from the detector to the Recovery System.

\subsection{Additional System Considerations}
\label{sec:Cryogenics-AdditionalConsiderations}

\begin{asparaenum}
\item[\bf Power Dissipation:] It is inferred from Ref.~\cite{Petersen:1990vt} that convection in free \LAr\ is able to carry away up to \LArNoBubblingHeatDissipation\ before the onset of bubbling.  The maximum heat that can be withstood before the onset of boiling of the \LAr\ at the bottom of the cryostat is therefore about \DSkCryostatBottomLArNoBubblingHeatPerAreaDissipation, for a total maximum heat load of \DSkCryostatBottomLArNoBubblingHeatDissipation.  
\item[\bf Cleaning and Handling:] All of the cryogenic components are fabricated from stainless steel.  All surfaces in contact with \LAr\ or \GAr\ will be electropolished and cleaned with \DS\ standard cleaning procedures either done in house or externally by a selected trusted vendor(s).  All piping will be flushed with dry \ce{N2} gas to avoid dusts.  Before assembly, all stainless steel surfaces will be visually inspected for quality control.  The cleaning applies to all surfaces of the cryogenics system, with all materials being handled in a standard clean room environment.
\item[\bf Thermal Insulation:] All cryogenic volumes will be thermally insulated with super-insulation combined with vacuum better than \DSkCryogenicsInsulationVacuum.  Dry foreline pumps will be use throughout the system combined with turbomolecular pumps.  All cryogenic surfaces will be covered with multi-layer-insulation blanket which will consist of 20+ layers of radiation reflector film separated with perforated spacer sheets.  The radiation barrier is made of polyimide and/or polyester films that are vapor deposited with \SI{99.99}{\percent} aluminum on both sides.
\item[\bf Welding Specifications:] Where applicable, welding shall be performed in accordance with all applicable codes such as the ASME Boiler and Pressure Vessel Code Section IX and AWS D1.6.  After a tube has been welded, the weld shall be cleaned with a stainless steel brush to remove all weld oxidation.  Pickling and/or passivation are required for all welded surfaces and near welded areas.  All welding shall be done by using non-thoriated tungsten electrodes.  All welds shall be internally purged with \SI{100}{\percent} argon.  All welds will be inspected by a qualified weld inspector and documented.
\item[\bf Leak Detection Requirements:] All welds and joints will be leak tested with a helium leak detector with a sensitivity of \DSkCryogenicsHeliumLeakCheckRequirement.  No leak larger that \DSkCryogenicsHeliumLeakCheckRequirement\ is permitted.
\item[\bf Pressure Testing:] All fabricated components such as the charcoal trap vessel, the heat exchanger, the condenser vessel and the piping shall be pressure tested according to the requirements of ASME Section VIII.  Pneumatic testing with dry nitrogen gas is admissible.
\end{asparaenum}

\subsection{System Safety Considerations}
\label{sec:Cryogenics-Safety}

\subsubsection{Power Outages}
\label{sec:Cryogenics-Safety-PowerOutages}

It is critical to keep the entire \LAr\ inventory safe.  Extensive tests performed with the \DSf\ system provided the basis for the \DSk\ design.  A set of critical valves are arranged to be at the proper locations to serve as emergency pressure relief for the system.  When total loss of power including the UPS system occurs, these valves should be held open to provide continuous cooling and venting of the \LIN.  The local \LIN\ reserve will be guided through the condenser to provide the required cooling to maintain a safe system.  These features have been realized in the \DSf\ cryogenics system and tested extensively prior to the \LAr\ fill.  In the case of total electrical power loss, the \LIN\ loop will automatically become an open loop, while all active sources of heat inside the cryostat (mainly the cold electronics) will be shut off: the only substantial passive source of heat remaining is the \DSkCryogenicsCryostatPowerLeak\ cryostat heat leak.  As already proven with the \DSf\ cryogenics system, the \LIN\ reserve maintained at near full level, combined with the carefully designed pneumatic valve configuration, is immune to complete power failure, including the loss of the UPS system control power.  In this situation, the \DSkCryogenicsLINBufferVolume\ of \LIN\ stored in the \LIN\ tank will be enough to maintain the detector in a stable and unperturbed condition, for about \DSkCryogenicsLINBackupTime.  In addition, the large \LIN\ source tank is also linked to the local storage for extended cooling should the emergency situation continue for more than a few days.

\subsubsection{Pressure Relief}
\label{sec:Cryogenics-Safety-PressureRelief}

\begin{figure}[h!]
\centering
\includegraphics[width=0.4\columnwidth]{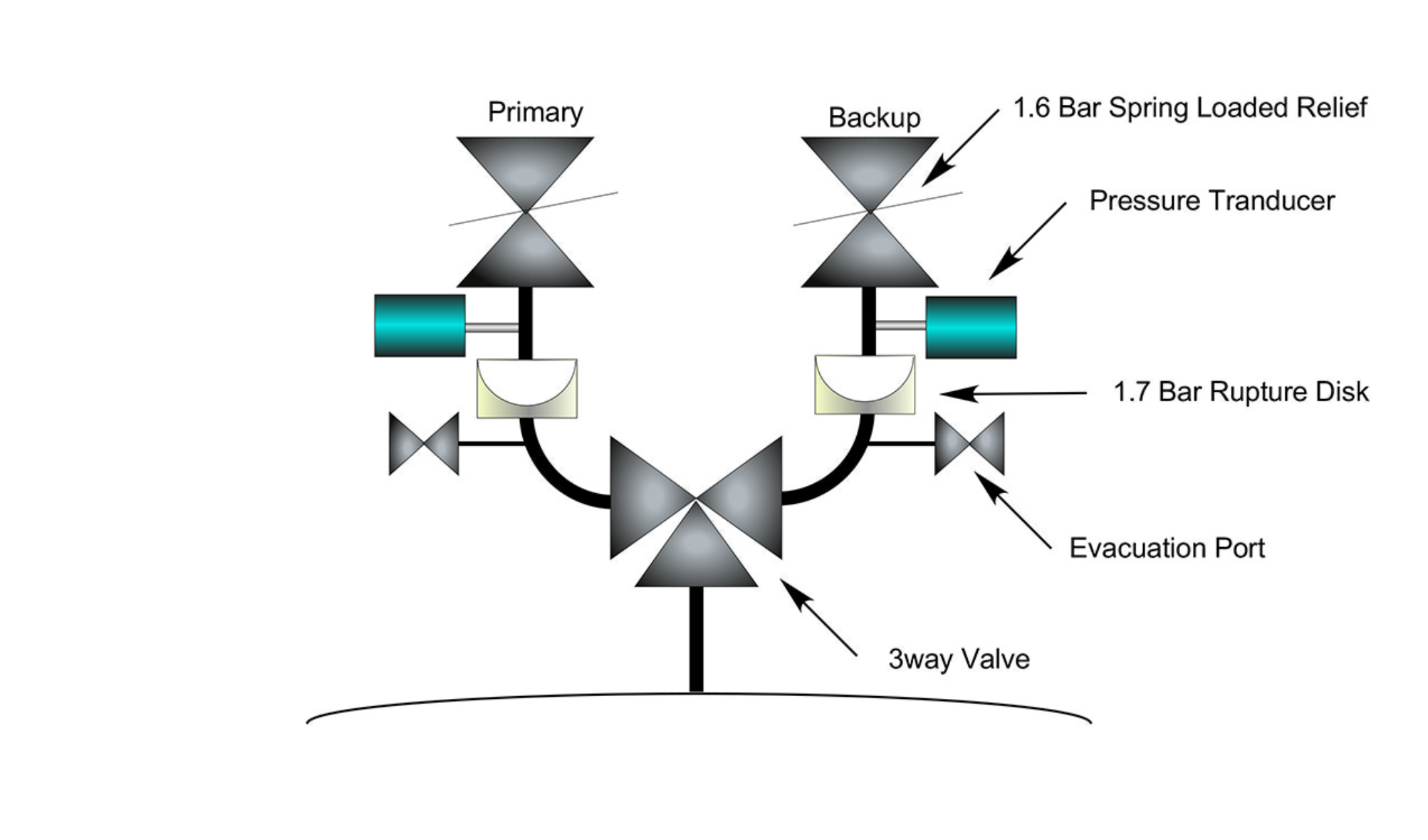}
\caption{Schematic of the \DSk\ cryogenics rupture disk relief assembly.}
\label{fig:Cryogenics-RuptureDisk}
\end{figure}

There are three volumes that have the ability to store the full inventory of \LAr\ (\DSkCryogenicsRecoveryCryostatOne, \DSkCryogenicsRecoveryCryostatTwo, and the detector cryostat).  Each is protected by a rupture disk relief assembly, as shown in Fig.~\ref{fig:Cryogenics-RuptureDisk}.  The rupture disks are a welded all-metal design that eliminates the possibility of permeation contamination into the system.  Each rupture disk has a pressure rupture limit of \DSkCryogenicsRuptureDiskLimit.  Each rupture disk is in series with a certified spring loaded o-ring sealed pressure relief valve located downstream of the rupture disk that is set for \DSkCryogenicsReliefValveSetting.  In the event that a rupture disk bursts, the spring loaded relief valve will only allow enough gas to vent to lower the pressure to a safe condition and will minimize the gas loss.  There is a pressure transducer located between these two reliefs that is continuously monitored by the slow control system for any change more than \DSkCryogenicsReliefValveTransducerMonitor.  If the pressure between the relief valves exceeds the set pressure limit, the rupture disk is considered to be leaking and will be replaced.  Every rupture disk has an online backup in the event that a rupture takes place or is found to have a leak.  In the event of a failure the spare will be brought on line by means of repositioning a 3-way hand valve.  This hand valve will be locked in the primary rupture disk position and will need to be unlocked to bring the backup rupture disk online.  The 3-way valve does not have an off position (at least one rupture disk is always linked to the system), it is only used to choose which rupture disk is in use.  There is a pump out port allowing the evacuation between the rupture disk and the 3-way valve to perform the clean replacement of a new rupture disk.

There are many cold spaces between control valves that can be potentially trap volumes if the associated valves are closed.  As the liquid or gas warms up the pressure within the confined space can potentially exceed the pressure ratings.  In these spaces small spring loaded reliefs are installed that have a pressure setting of \DSkCryogenicsSpringLoadedReliefPressure.  These valves have the potential of venting a significant amount of gas.  The output of all of these trapped volume relief valves will be connected to a common manifold that returns the gas to the recovery system where it will be re-condensed preventing and losses.

A full Risk Assessment Study of the entire cryogenics system will be carried out by an external company, and will be certified to abide by all environment and safety regulations in place at \LNGS.

\section{PhotoElectronics}
\label{sec:PhotoElectronics}

\subsection{Introduction}
\label{sec:PhotoElectronics-Introduction}

Silicon photomultipliers (\SiPMs) are one of the key enabling technologies for large-scale \LAr-based dark matter experiments.  \SiPMs\ may also play an important role in the next generation of \LAr-based neutrino detectors, such as DUNE~\cite{Acciarri:2016wz}, and liquid xenon based detectors for neutrinoless double beta decay, such as nEXO~\cite{Ostrovskiy:2015jl}.  \SiPMs\ have a number of performance advantages over traditional \PMTs, including higher photon detection efficiency (\PDE) and much better single-photon resolution, all while operating at much lower bias voltage.  \SiPMs\ can also be efficiently integrated into tiles that cover large areas and feature better radio purity than \PMTs.

The \DS\ Collaboration committed to building the next detectors of its dark matter research programs with \SiPM-based photosensors.  For \DSk, the photosensing unit will be a ``photodetector module'' (\DSkPdm), consisting of a large tile of \SiPMs\ covering an area of \DSkTileAreaStd\ operating as a single detector.  Besides the \tile, each module will also contain a cryogenic preamplifier, which will amplify and shape the signal in the immediate proximity of the sensor.  The output of the cryogenic amplifier will be passed on to a signal transmitter, also integrated into the \DSkPdm, and responsible for transmission of the signal through the cryostat penetration.  Finally, the \DSkPdm\ will also include the mechanical structure required to assemble all components and to efficiently dissipate heat in the \LAr\ target, minimizing the production of bubbles. An intelligent power distribution system is also foreseen, capable of disabling individual \DSkPdms\ in case of failure.

\begin{figure}[t!]
\centering
\includegraphics[height=3cm]{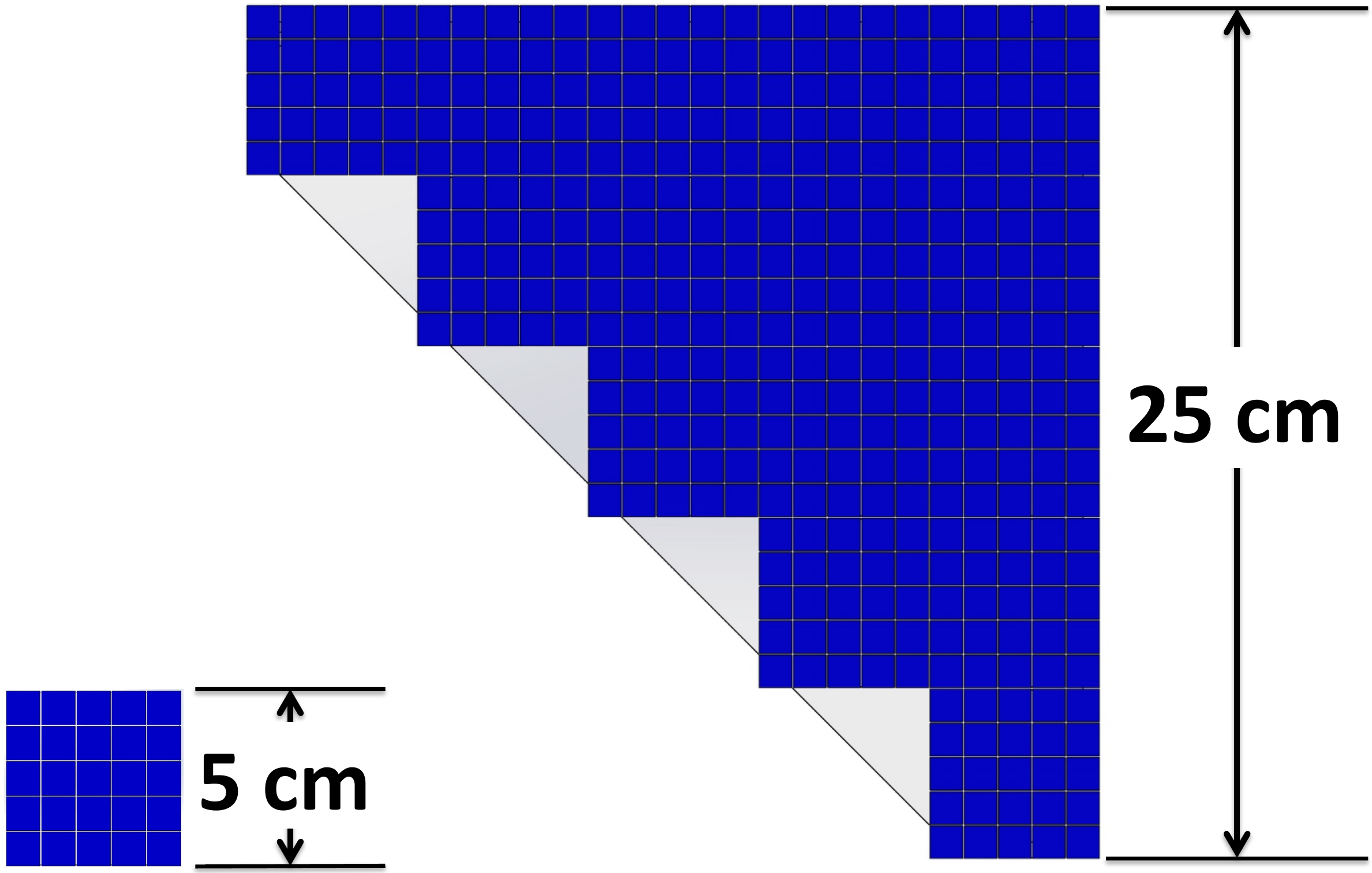}
\includegraphics[height=3cm]{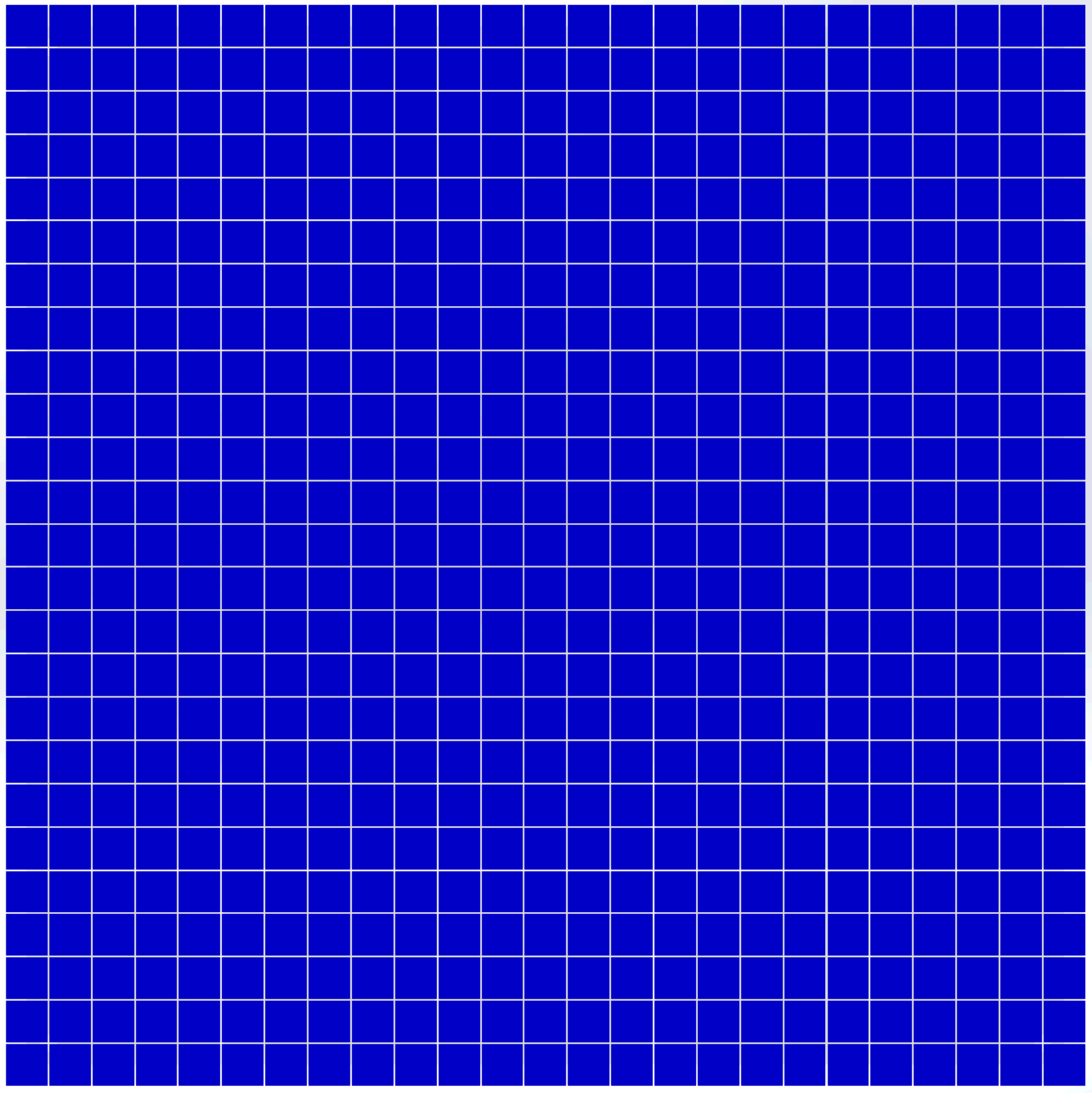}
\caption[\SiPM\ \tiles\ and photodetector modules.]{{\bf Left:} A single \SiPM\ \tile, the elemental photon readout channel in \DSk\ (tile shown is built from \DSkTileAreaStd\ \SiPMs).  {\bf Center:} A single \TRB\ (see text).  {\bf Right:} A single \SQB\ (see text).  \SQBs\ and \TRBs\ are assemblies of \DSkPdms\ and serve as building blocks of the \DSk\ photon readout planes.}
\label{fig:PhotoElectronics-Tile}
\end{figure}

The \DSkPdms\ are located above the anode and below the cathode and fully cover the top and bottom faces of the \LArTPC\ active volume, to detect both the \SOne\ and \STwo\ signals with high efficiency.  The top and the bottom photon readout assemblies consist of \DSkTilesHalfNumber\ \DSkPdms, each.  Multiple \DSkPdms\ are mounted to a single motherboard to form two larger basic mechanical units called the square board (\SQB) and the triangular board (\TRB), as shown in Fig.~\ref{fig:PhotoElectronics-Tile}.  The \SQB\ and \TRB\ have the same edge size of \DSkSQBTRBSize.  The \SQB\ and \TRB\ are then used to form the full readout planes (shown in Fig.~\ref{fig:LArTPC-AnodeAssembly}).  The total number of readout channels (top and bottom) is \DSkTilesNumber.

An important aspect to be considered when reading out large \SiPMs\ \tiles\ is the high output capacitance (compared to \PMTs), which is of the order of \DSkSiPMCapacitancePerArea.  This imposes a careful design of the read-out electronics and \SiPMs\ connection scheme on the \tile.  The signal to noise ratio (\SNR) is defined as the ratio of the mean value of the integrated single photoelectron (\SPE) response to the RMS of the integrated baseline.  The figure of merit of the entire system is the maximum \tile\ area that can be operated as a single \DSkPdm, connected to a single electronic channel, without deteriorating the signal to noise ratio.  The requirement for this figure of merit is given in the next section.

The requirements for the individual \DSkPdms\ are:

\begin{asparaenum}[\bfseries {PDM}-i]
\item\label{req:DSkDModPDE} \DSkPdm\ area covering a \DSkTileAreaStd\ surface.  This allows for the total number of channels to stay within \DSkTilesNumber;
\item Overall photon detection efficiency (\PDE) larger than \DSkPdmPDESpecification\ at \TPBWaveLength.  This is of primary importance for the light yield as the increased \PDE, along with the increase in effective photocathode coverage of the top and bottom of the TPC to \SI{>85}{\percent} (\SI{~80}{\percent} in \DSf), will provide a significant improvement over the \DSfPMT\ \PMTs\ used in \DSf\ and have a \QE\ of \DSfPMTQE. It is then required that the inactive gap between \SiPMs\ be smaller than \DSkSiPMInactiveGapSpecification\ and have a tile-level fill factor larger than \DSkTileFillFactorSpecification\ to meet the active area requirement;
\item Power dissipation density limited to \DSkTileDensityPower, corresponding to a total dissipated power of \DSkTilePower\ per \DSkPdm.  This allows for the use of up to four pre-amplifiers (40~mW each) and two differential transmitters (20~mW each) per \DSkPdm, while avoiding excessive heat load on the cryogenic system; 
\item Dynamic range higher than \DSkTileDynamicRangeSpecification\ detected simultaneously, without affecting the \SNR\ provided by the pre-amplifier.  Lower values would affect the reconstruction of \STwo;
\item Time resolution for the first hit on each \DSkPdm\ on the order of \DSkTileTimeResolutionSpecification.  Experience with \DSf\ and MC simulations have shown that this is crucial to maintain a very effective \PSD\ for the \LArTPC.
\item\label{req:DSkDModDCR} Overall noise rate below \DSkPdmDCRSpecification, this includes the dark count rate (\DCR) from the photo detectors and the fake pulses generated by electronic noise (plus possibly correlated noise injected in the system by external sources), both of which should be significantly below the mentioned rate.  Higher rates would impact both the trigger efficiency and the pulse shape discrimination power.
\end{asparaenum}

The last requirement sets very stringent limits on the \SNR\ levels, the discussion of which is a crucial part of this section.  Setting a charge threshold that gives a large single PE efficiency, while keeping the overall noise rate within the level indicated in {\bf PDM-vi}, requires an excellent SNR. Using a detailed simulation, constrained by real data, it was found that to obtain a \SI{95}{\percent} single PE efficiency, while keeping the electronic noise at a few percent of the \SiPM\ noise (see {\bf PDM-vi}), requires a SNR \num{>7}.

The goal is to build and operate \DSkTilesNumber\ \DSkPdms, each as a single analog element, whose output signal is digitized by the readout electronics.  As discussed in this section, work is being done to reach this objective in steps.  To date, \tiles\ with an area of \DSkTileAreaMin\ and \DSkTileAreaStd\ have been built and successfully operated.  The results described here are for the \DSkTileAreaMin\ tile, which meets the specifications.  Based on the preliminary results obtained with a \DSkTileAreaStd\ tile, the final goal of meeting all the \DSkPdm\ requirements could be achieved and have first been reported in~\cite{DIncecco:2017tj}.

\subsection{\SiPMs\ Specifications}
\label{sec:PhotoElectronics-SiPM}

Given the fact that the discrete electrical unit in will be the \DSkPdm, \SiPM\ specifications must ensure that each \DSkPdm\ will reach the stringent requirements (\ref{req:DSkDModPDE} to \ref{req:DSkDModDCR}) set forth by the \DS\ Collaboration in the previous section.  A list of preliminary, minimal \SiPM\ requirements for operation at \LArNormalTemperature\ is:

\begin{asparaenum}[\bfseries {SiPM}-i]
\item \PDE\ equal to or greater than \DSkSiPMPDESpecification\ at \TPBWaveLength, set by the amount required to offset the inactive gaps in the \DSkPdm\ \tile\ and the inefficiency of the hit detection, so that requirement {\bf \DSkPdm -\ref{req:DSkDModPDE}} is met;
\item Individual \SiPM\ size larger than \DSkSiPMAreaStd\ to ease the assembly of \SiPMs\ into tiles.  Ideal sizes would be \DSkSiPMAreaMaxcm;
\item \DCR\ lower than \DSkSiPMDCRSpecification, as per requirement {\bf \DSkPdm -\ref{req:DSkDModDCR}};
\item Total Correlated Noise Probability (\TCNP), crosstalk (CT) + afterpulsing (AP), lower than \DSkSiPMTCNPSpecification.  While the effects of total correlated noise on \SOne\ (energy and \PSD) can be minimized during the analysis, a high value for \TCNP\ can affect the dynamic range available for \STwo;
\item Effective output capacitance lower than \DSkSiPMCapacitancePerArea, in the frequency range of interest.  Higher values would negatively impact the \SNR.
\end{asparaenum}

\subsection{The Development of \SiPMs\ at \FBK}
\label{sec:PhotoElectronics-FBK}

Fondazione Bruno Kessler (\FBK) has developed two main \SiPM\ technologies with high density (suffix ``-\HD'') single photon avalanche diodes (\SPADs) that are sensitive to UV light:

\begin{asparaenum}
\item[\bf \NUVHdSf:] A family with peak efficiency in the near ultra-violet (\SI{410}{\nano\meter}) and standard electric field configuration~\cite{Piemonte:2016cj};
\item[\bf \NUVHdLf:] A family with peak efficiency in the near ultra-violet (\SI{410}{\nano\meter}) and low electric field configuration~\cite{Ferri:2016ky};
\end{asparaenum}

\begin{table*}[t!]
\rowcolors{3}{gray!35}{}
\normalsize
\centering
\caption[Characteristics of the two families of \FBK\ \SiPMs.]{Characteristics of the two families of \FBK\ \SiPMs\ (\NUVHdSf, and \NUVHdLf) at \RoomTemperature\ and at \LINNormalTemperature~\cite{Acerbi:2017gy}.  \OV: over-voltage (voltage above breakdown) required to match the \PDE\ requirement at \LINNormalTemperature; \DCR: dark count rate; \DiCT: direct cross talk; \DeCT: delayed cross talk; \AP: afterpulsing.  Samples were pre-selected based on their functionality at \RoomTemperature.  See Ref.~\cite{Piemonte:2012gl} for the test methodology adopted.}
\label{tab:FBK-SiPMs}
\begin{tabular}{lcc}
						&\NUVHdSf					&\NUVHdLf \\
\hline
\SPAD\ Size				&\DSkSiPMSPADSizeStd			&\DSkSiPMSPADSizeStd \\
\SiPM-level Fill Factor		&\NUVHdSfFillFactor			&\NUVHdLfFillFactor	 \\
\SiPM\ Size				&\DSkSiPMAreaMin				&\DSkSiPMAreaMin \\
\hline
\cellcolor{white}At \RoomTemperature \\
\hline
Breakdown Voltage			&\NUVHdSfRoomTempBreakdownVoltage
													&\NUVHdLfRoomTempBreakdownVoltage \\
\PDE\ $@$ \TPBWaveLength	&\NUVHdSfRoomTempPDE		&\NUVHdLfRoomTempPDE \\
Peak \PDE\ Wavelength		&\NUVHdSfRoomTempPeakPDEWavelength
													&\NUVHdLfRoomTempPeakPDEWavelength \\
DCR						&\NUVHdSfRoomTempDCR		&\NUVHdLfRoomTempDCR \\
DiCT						&\NUVHdSfRoomTempDiCT		&\NUVHdLfRoomTempDiCT \\
DeCT + AP				&\NUVHdSfRoomTempDeCTAP	&\NUVHdLfRoomTempDeCTAP \\
\hline
At \LINNormalTemperature \\	
\hline
Breakdown Voltage			&\NUVHdSfLINTempBreakdownVoltage
													&\NUVHdLfLINTempBreakdownVoltage \\
\OV\ $@$ Optimal \PDE		&\NUVHdSfLINTempOV			&\NUVHdLfLINTempOV \\
DCR						&\NUVHdSfLINTempDCR			&\NUVHdLfLINTempDCR	\\
DiCT						&\NUVHdSfLINTempDiCT			&\NUVHdLfLINTempDiCT \\
DeCT + AP				&\NUVHdSfLINTempDeCTAP		&\NUVHdLfLINTempDeCTAP \\
\end{tabular}
\end{table*}

Table~\ref{tab:FBK-SiPMs} summarizes the features of the \SiPMs\ of the two families with the standard \SI{25}{\micro\meter} cell size, at \RoomTemperature\ and at \LINNormalTemperature.  Other cell sizes have been produced at FBK with NUV-HD technology, ranging between \SI{20}{\micro\meter} and \SI{40}{\micro\meter}. They may be considered and tested in order to optimize detector performance.  When bringing the \SiPMs\ from \RoomTemperature\ to \LINNormalTemperature, the most dramatic changes in the \SiPMs\ characteristics occur in the breakdown voltage, in the \DCR, and the after=pulsing (\AP) rate.  The quenching resistor $R_q$ is made of polysilicon, with a resistance that depends on temperature.  By tuning the doping parameters of polysilicon, resistors $R_q$ whose resistance changes by a factor from \DSkSiPMPolysiliconResistanceTempVariationRange, passing from \RoomTemperature\ to \LINNormalTemperature, have been successfully manufactured.  Typically, $R_q$ increases with decreasing temperature, causing an increase in the \SPAD\ recharge time and, correspondingly, in the duration of the pulse.  For all \SiPMs, the choice of the value of $R_q$ is fundamental as this influences the recharge time of the \SPADs.  \SiPMs\ with a recharge time at \LINNormalTemperature\ in the range up to \LRqTimeScale\ are identified with the standard name \NUVHdLf, and \SiPMs\ with a recharge time at \LINNormalTemperature\ in the range of \HRqTimeScale\ with the suffix ``-\HRq'', hence \NUVHdLfHRq.  

\begin{figure}[!t]
\centering
\includegraphics[width=\columnwidth]{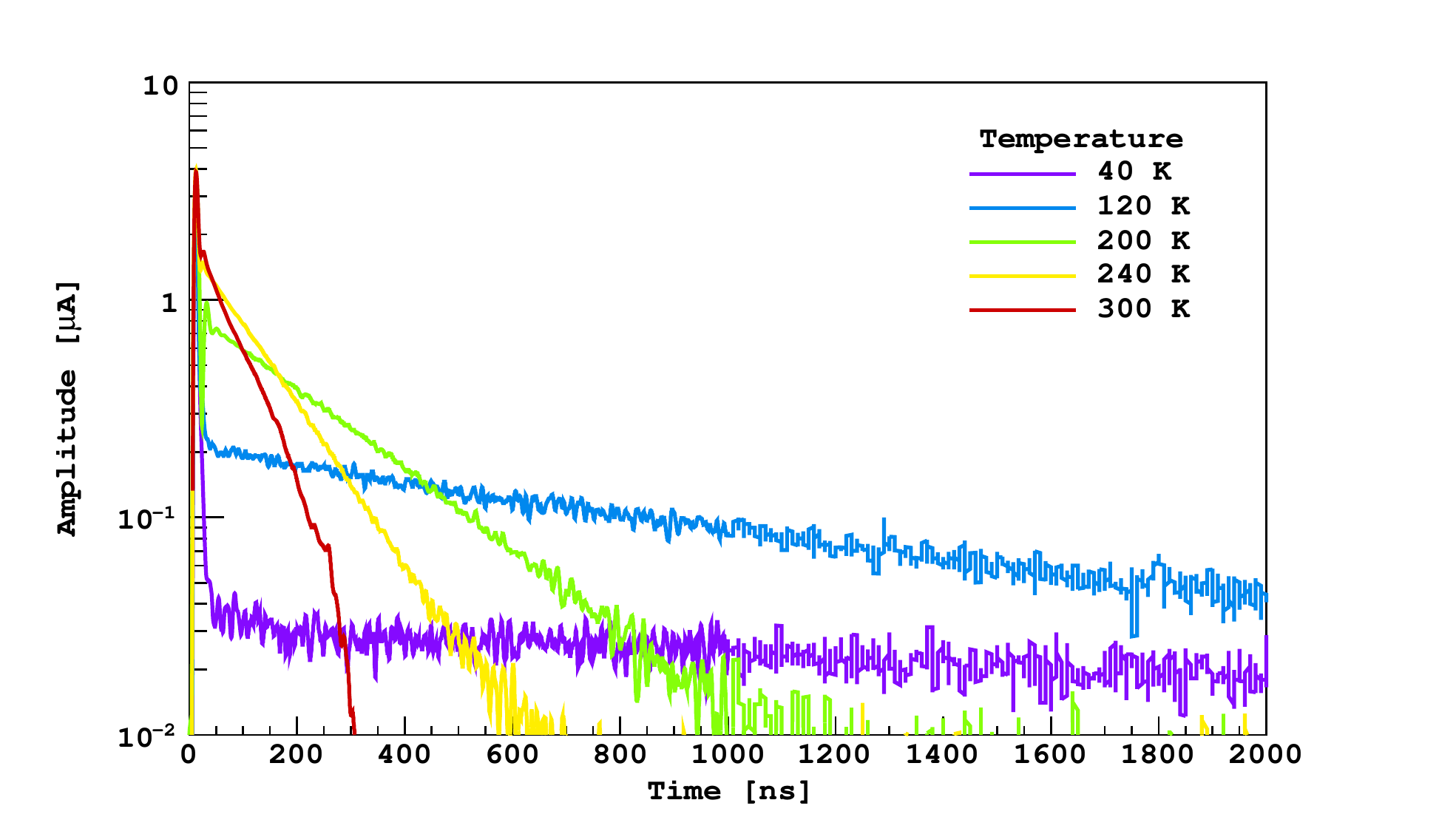}
\caption[Average of \NUVHdSfHRq\ \SCR\ in the range \LNGSCryoSetupTemperatureRange\ at constant over-voltage.]{Average of \NUVHdSfHRq\ \SCR\ in the range \LNGSCryoSetupTemperatureRange\ at constant \OV.  The presence of two components at all temperatures is evident, as well as the increase of the \SPAD\ recharge time with decreasing temperatures.  A similar behavior is observed with \NUVHdSf\ \SiPMs.}
\label{fig:NUVHdLf-SCRvsT}
\end{figure}

A long recharge time, as provided by the \HRq\ technology, offers significant advantages: strong suppression of the AP, higher \PDE\ (thanks to the higher applicable over-voltage (\OV)), and better reproducibility of the devices (thanks to the suppression of \AP).  On the other hand, the main drawback is that the signal charge is delivered during a longer time and, correspondingly, the tail of the single \SPAD\ response (\SCR) shows a decrease in its maximum amplitude.  Fig.~\ref{fig:NUVHdLf-SCRvsT} shows the change in the \SCR\ in the range \LNGSCryoSetupTemperatureRange, for a particular set of \NUVHdLfHRq\ \SiPMs\ with $R_q$ starting at \NUVHdLfRoomTempRq\ at \RoomTemperature\ and increasing up to \NUVHdLfLINTempRq\ as the temperature was lowered to \LINNormalTemperature.  There is a fast component that is almost constant with temperature (which accounts for about 10\% of the total charge), while the pulse tail becomes longer and longer with decreasing temperature (as does the recharge time).  As a result, the \SPAD\ recovery (recharge) time becomes quite long, \NUVHdLfHRqLINTempRechargeTime.  

\begin{figure}[!t]
\centering
\includegraphics[width=\columnwidth]{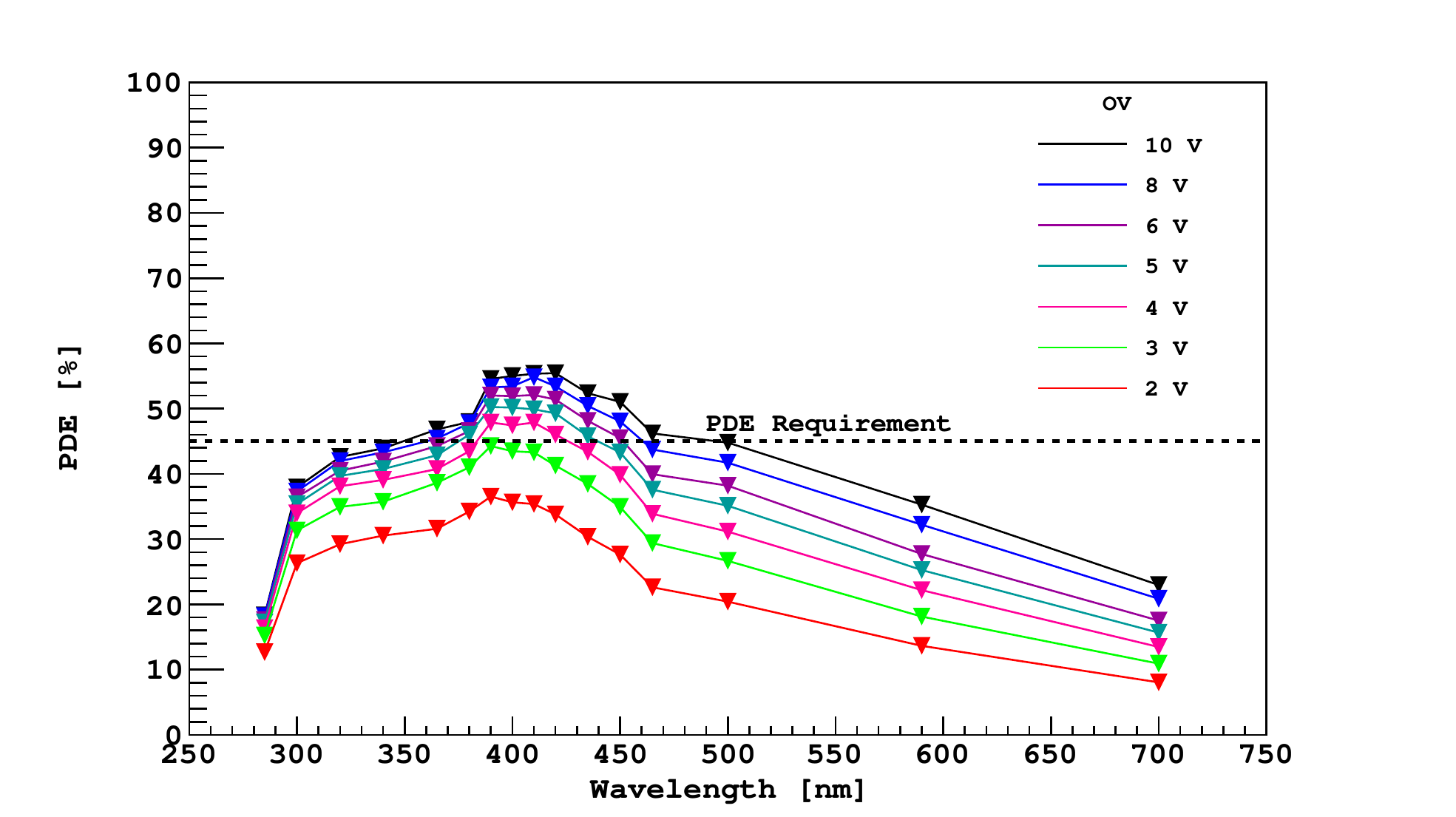}
\caption[\PDE\ versus wavelength for a \NUVHdSf\ \SiPM\ at \RoomTemperature.]{\PDE\ versus wavelength for a \NUVHdSf\ \SiPM\ at \RoomTemperature.  The definition of \PDE\ accounts for both quantum efficiency \QE\ and \SiPM-level fill factor.}
\label{fig:NUVHdLf-PDEVsWavelength}
\end{figure}

\begin{figure}[!t]
\centering
\includegraphics[width=\columnwidth]{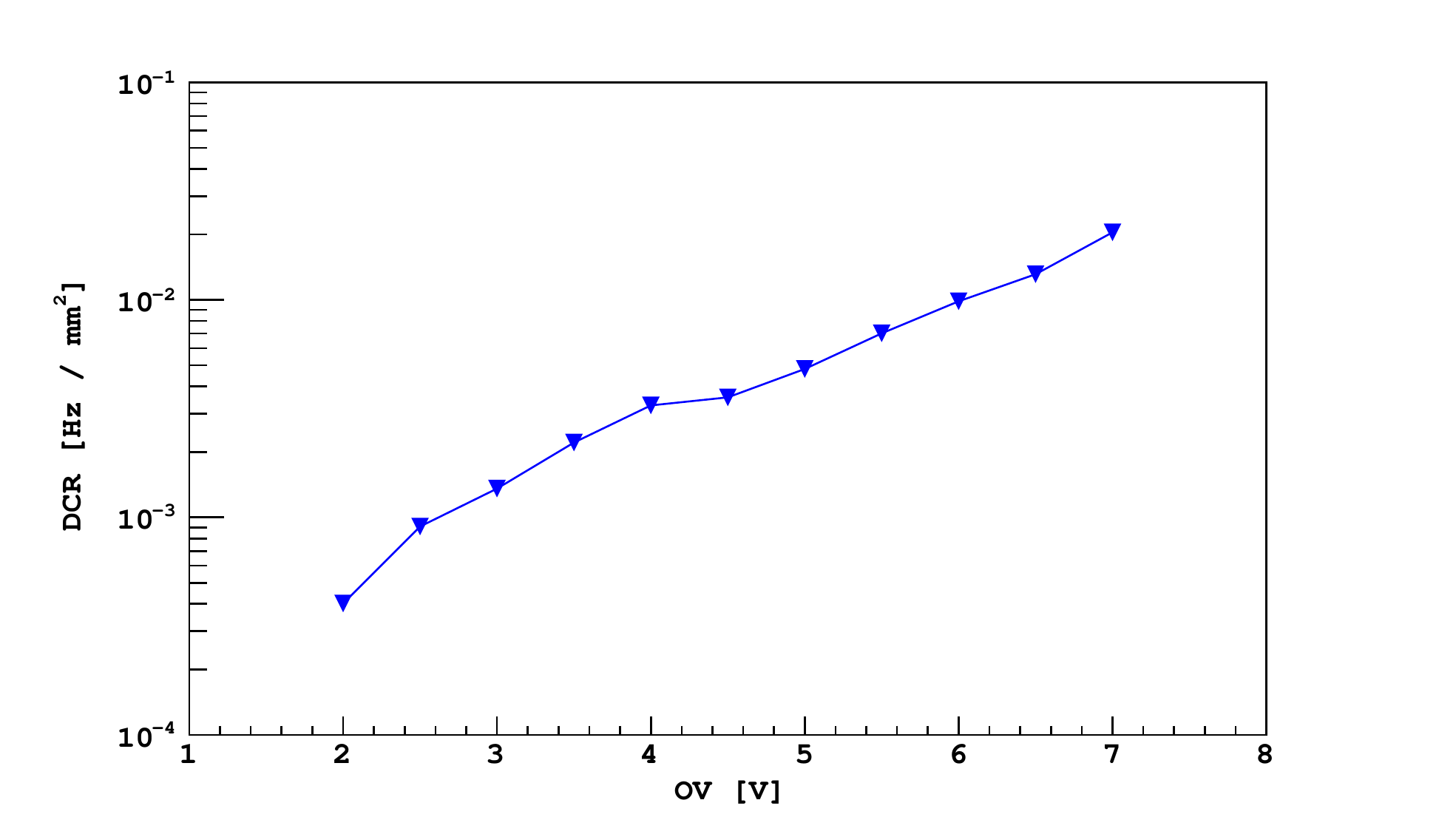}
\caption[\NUVHdLf\ \SiPMs\ \DCR\ vs. over-voltage at \LINNormalTemperature.]{\NUVHdLf\ \SiPMs\ \DCR\ vs. \OV\ at \LINNormalTemperature.  See text for details and interpretation on the dominating mechanism for \DCR.}
\label{fig:NUVHdLf-DCRvsOV}
\end{figure}

The \NUVHdLfHRq\ \SiPM\ is the \FBK\ technology with the best performance which meets all of the requirements.  In particular, at \LINNormalTemperature, operation at \NUVHdLfLINTempOV\ overvoltage (\OV) or greater allows to meet both the \PDE\ (with the hypothesis that the \PDE\ at \TPBWaveLength does not change with temperature) and the \DCR\ requirements.  The \PDE\ of the \NUVHdLfHRq\ is quite high and exceeds the design requirements by a significant amount, as shown in Fig.~\ref{fig:NUVHdLf-PDEVsWavelength}.  The \DCR\ is more than one order of magnitude lower than the requirement.  The study of the \NUVHdLf\ \DCR\ as a function of \OV\ at \LINNormalTemperature, seen in Fig.~\ref{fig:NUVHdLf-DCRvsOV}, shows an exponential growth of \DCR\ vs. \OV, which is typical of tunneling, the dominant carrier generation mechanism at this temperature.  The \DCR\ remains below \SI{1E-2}{\hertz\per\square\mm} at \SI{6}{\volt} \OV.

\begin{figure*}[!t]
\centering
\includegraphics[width=\textwidth]{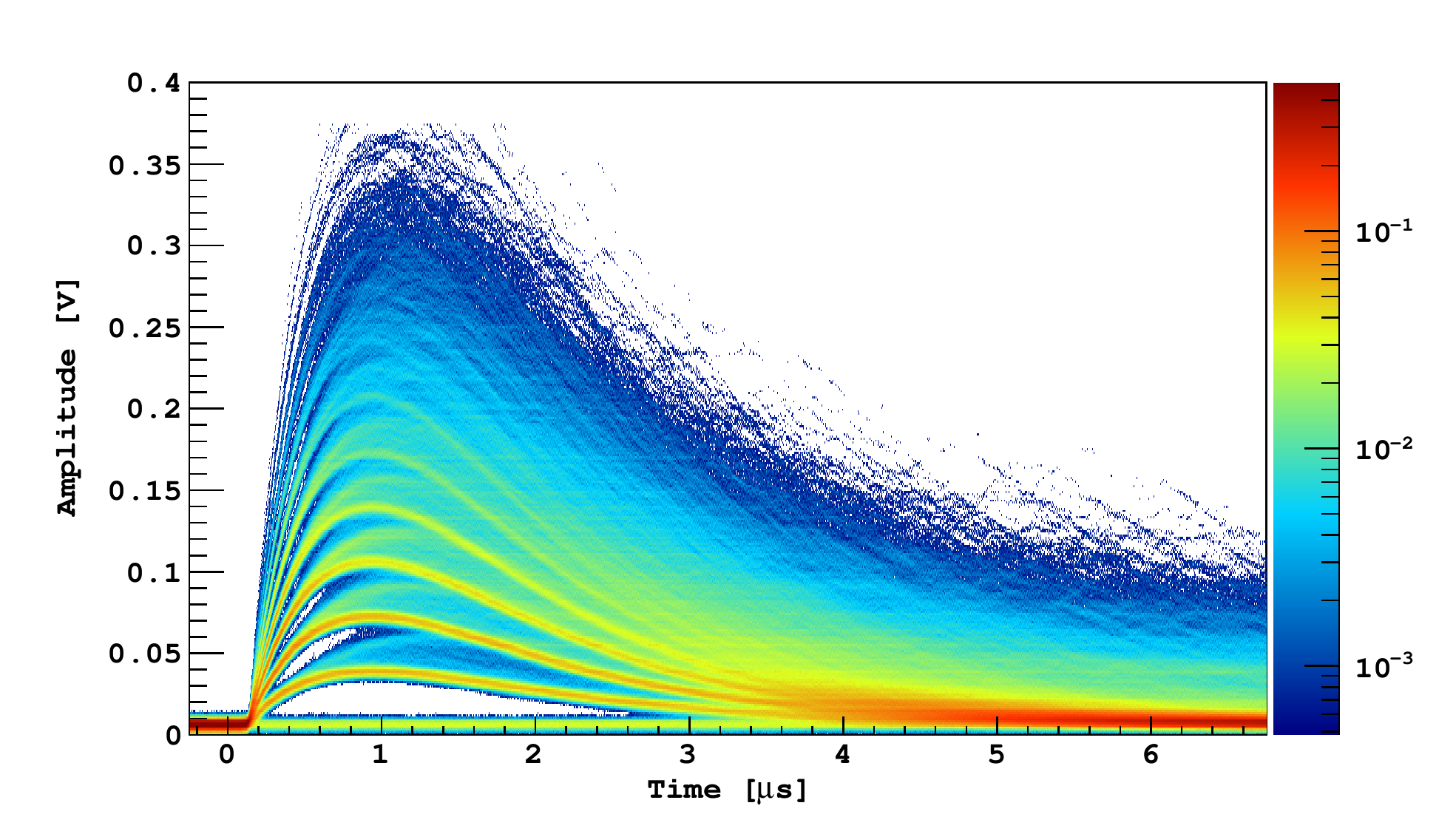}
\caption[Persistence plot for a \DSkSiPMAreaMax\ \NUVHdLfHRq\ \SiPM, operated at \SI{4}{\volt} \OV\ at \LINNormalTemperature.]{Persistence plot for a \DSkSiPMAreaMax\ \NUVHdLfHRq\ \SiPM, operated at \SI{4}{\volt} \OV\ at \LINNormalTemperature.  See text for details.}
\label{fig:NUVHdLfHRq-Persistence}
\end{figure*}

To confirm the results of the first \NUVHdLf\ prototypes, and to perform a specific evaluation, a dedicated \NUVHdLf\ wafer lot was produced at \FBK, manufacturing two \SiPM\ sizes, \DSkSiPMAreaStd\ and \DSkSiPMAreaMax, with different polysilicon quenching resistors, in order to tune the value at \LINNormalTemperature\ and check the difference in performance.  The first functional tests on \DSkSiPMAreaMax\ with high $R_q$ confirm the excellent performance of this technology.  Fig.~\ref{fig:NUVHdLfHRq-Persistence} shows a persistence plot measured with a \DSkSiPMAreaMax\ \NUVHdLfHRq\ \SiPM, operated at \SI{4}{\volt} \OV.  The measurement was performed at \FBK, by immersing the \SiPM\ in \LIN\ and using a non-cryogenic amplifier placed in close proximity and containing a \SI{160}{\milli\watt} power consumption.  To improve \SNR, a quasi-optimal filter is used, composed of a single, real pole low-pass with a time constant of \SI{400}{\nano\second} (the \SiPM\ microcell recharge time constant was \SI{300}{\nano\second}).  The resulting \SNRAmp\ was \num{14}.  Fig.~\ref{fig:NUVHdLfHRq-Persistence} also shows very good microcell gain uniformity over the \DSkSiPMAreaMax\ area, which contains approximately \num{160E3} individual \SPADs.

\begin{figure}[!t]
\centering
\includegraphics[width=\columnwidth]{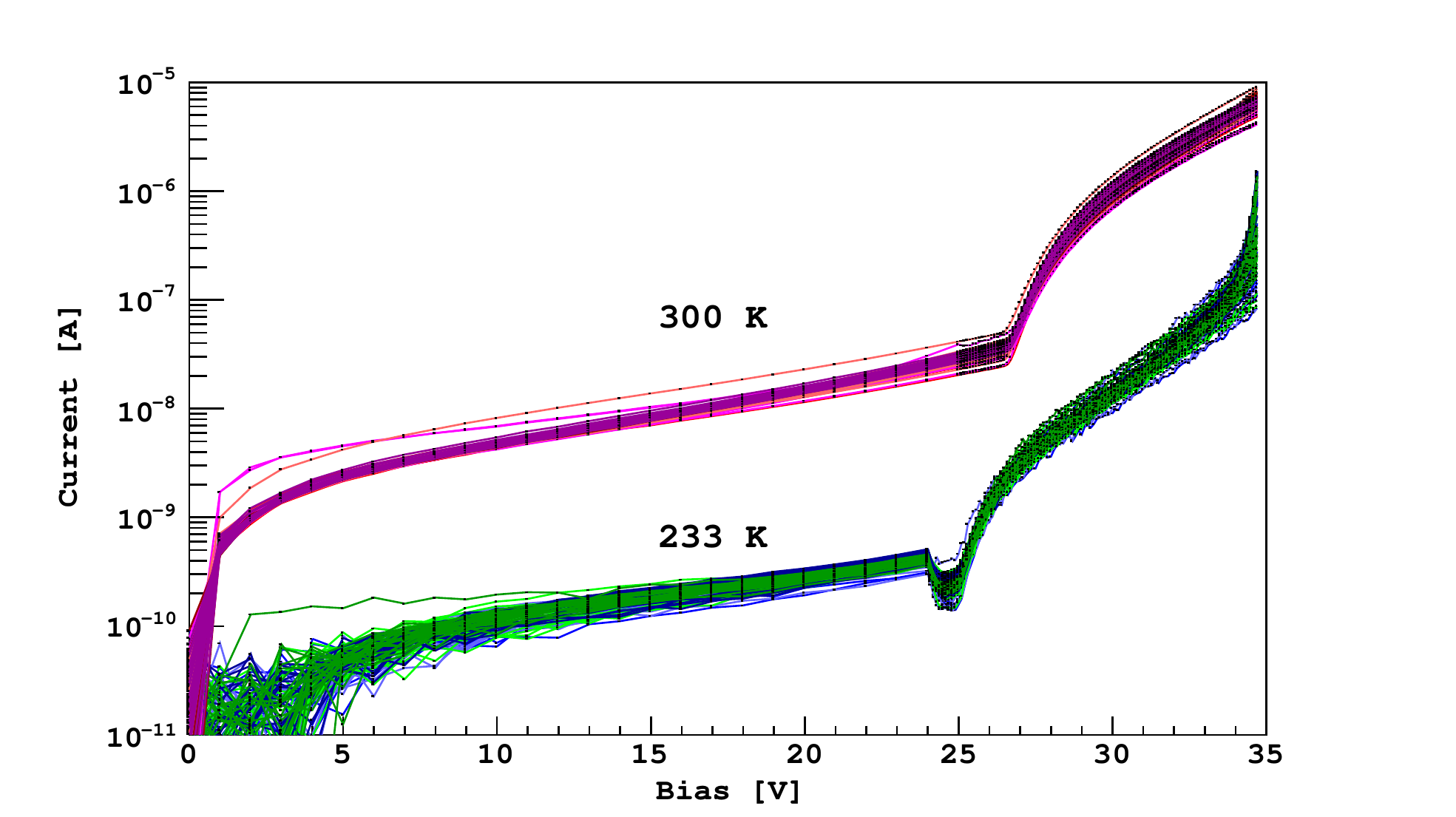}
\caption[$IV$ curves for \NUVHdSfHRq\ \SiPMs.]{$IV$ curves for \NUVHdSfHRq\ \SiPMs\ at \RoomTemperature\ and \FBKColdSetupTemperatureColdLimit.  See text for details.}
\label{fig:NUVHdLfHRq-IVvsT}
\end{figure}

In parallel with the basic functional characterization, statistical tests on yield and reliability have started at the wafer level.  At \FBK, full wafer testing in the range from \FBKColdSetupTemperatureRange\ was performed, thanks to an automatic prober that can cool the chuck holding the wafer.  As an example, Fig.~\ref{fig:NUVHdLfHRq-IVvsT} shows the reverse $IV$ curves of functional \SiPMs\ (\NUVHdSfHRq, \DSkSiPMAreaStd\ \SiPM\ size, \DSkSiPMSPADSizeStd\ \SPAD\ size) at \RoomTemperature\ and \FBKColdSetupTemperatureColdLimit.  The breakdown voltage and current scales as expected.

The large majority of devices followed this trend.  However, an important finding was that a few \SiPMs\ that are within specifications at \RoomTemperature\ show a slower decrease of the current with temperature.  Thus, the $IV$ curves at \FBKColdSetupTemperatureColdLimit\ are higher than the typical curves shown in Fig.~\ref{fig:NUVHdLfHRq-IVvsT}.  Although the difference may be moderate, this is an indication that, in a few \SiPM\ samples, the decrease of \DCR\ with temperature does not follow the exponential law as the majority of the population.  Thus, at \LINNormalTemperature, the \DCR\ could turn out to be higher and, potentially, outside of specifications.  This finding has very important implications for \DSk, as it points directly to the need of studying this potential yield reduction on much larger statistics and, if confirmed, to the need of managing the appropriate selection of \SiPMs\ that are fully conforming to specifications.  Ultimately, identification of defective \SiPMs\ cannot be performed at or near \RoomTemperature.  The solution identified is the use of a cryogenic wafer automatic prober machine, for statistical studies on large samples and, ultimately, for mass testing of all the \SiPMs\ to be used.  The cryogenic wafer probe will be part of the Nuova Officina Assergi (\NOA) packaging facility that will be built in the above-ground site of LNGS.

Also performed at \FBK\ were dedicated tests to check the \SiPM\ performance reliability upon a sudden large reduction of temperature.  Full silicon wafers were repeatedly immersed in liquid nitrogen (\LIN), following this procedure: immersion in \LIN\ for \SI{10}{\min}; immersion in \LIN\ \SI{2}{\hour}; and immersion in \LIN\ for \num{10} cycles lasting \SI{2}{\min} each.  The DC electrical properties of all the \SiPMs\ were cross-checked before and after the cooling/warming procedures.  No significant changes were observed in either the reverse or forward $IV$ measurements, for any family of \SiPMs.

\subsection{Precision Characterization of \FBK\ \SiPMs}
\label{sec:PhotoElectronics-Characterization}

\subsubsection{The \SiPM\ Cryogenic Test Setup}
\label{sec:PhotoElectronics-Characterization-Setup}

\begin{figure}[!t]
\centering
\includegraphics[width=0.45\columnwidth]{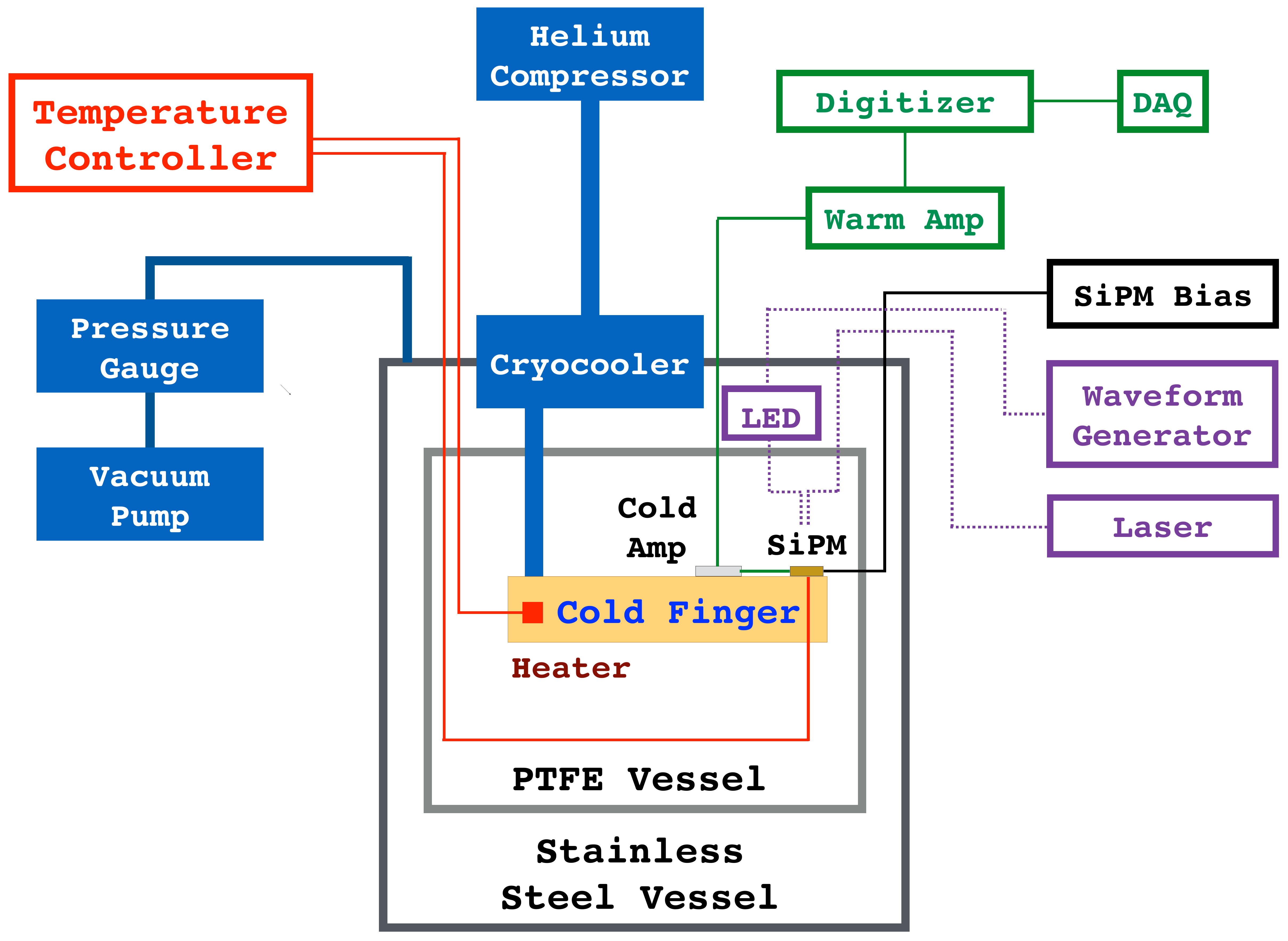}
\caption[Sketch of the experimental setup in use for the \SiPMs\ test program.]{Sketch of the experimental setup in use for the \SiPMs\ test program~\cite{Acerbi:2017gy}.}
\label{fig:LNGSSetup-Sketch}
\end{figure}

A special cryogenic setup for precision characterization of the \FBK\ \SiPMs\ was built and commissioned at LNGS.  The cryogenic setup is contained inside a stainless steel vacuum chamber \LNGSCryoSetupCryostatHeight\ in length and closed by two \LNGSCryoSetupCryostatFlangeModel\ flanges.  The cryostat is equipped with sufficient feedthroughs for the readout of many \SiPM\ signals via coaxial SMA connections, as well as the readout of the environmental parameters and supply of the laser light via optical fibers. An overall sketch of the system is shown in Fig.~\ref{fig:LNGSSetup-Sketch}, with full details presented in~\cite{Acerbi:2017gy}.  A \LNGSCryoSetupCryocoolerModel\ pulse tube cryocooler, capable of delivering \LNGSCryoSetupCryocoolerLINPower\ of cooling capacity at \LINNormalTemperature, is mounted on the top \LNGSCryoSetupCryostatFlangeModel\ flange.  

\begin{figure}[!t]
\centering
\includegraphics[width=0.45\columnwidth]{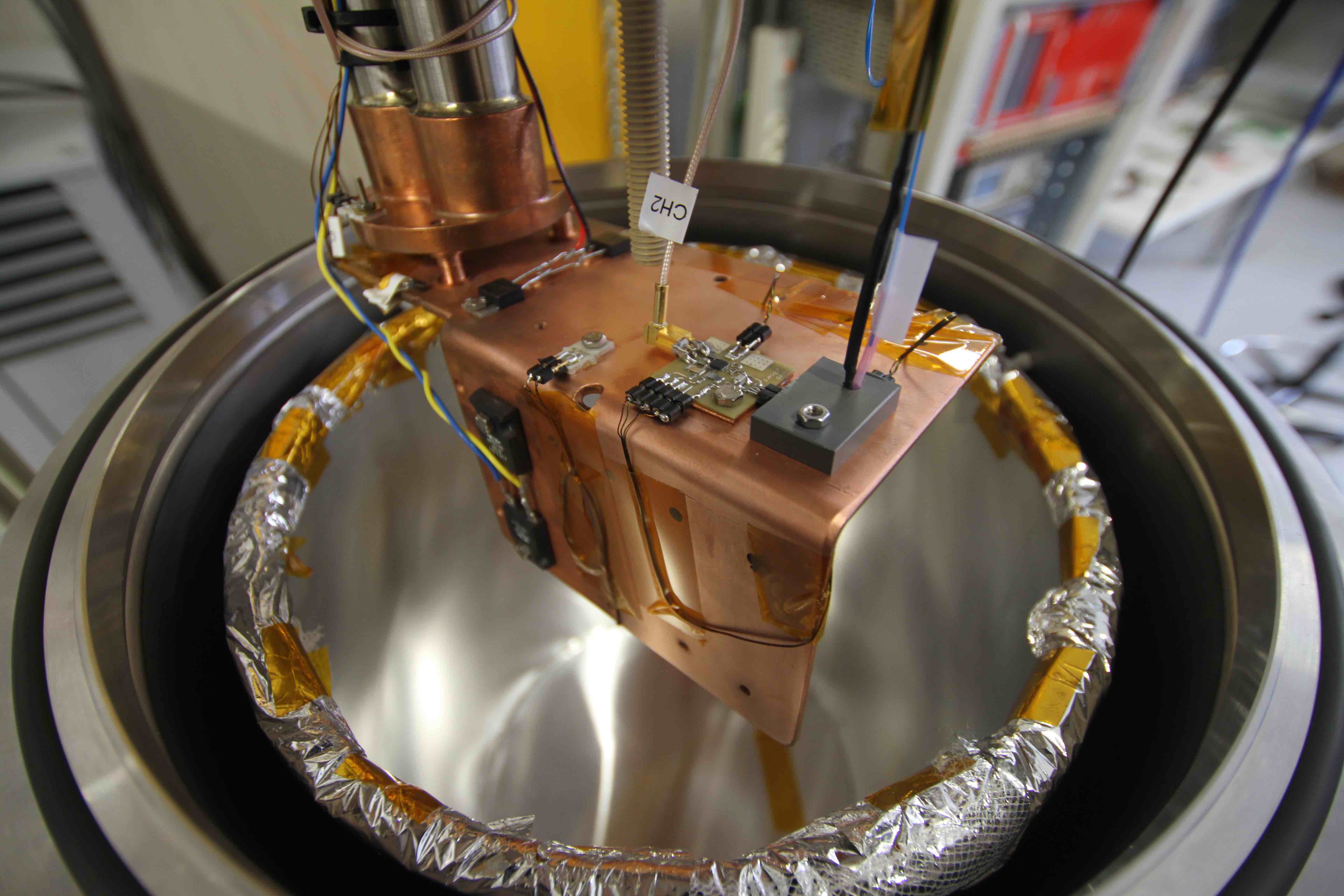}
\caption[Cold finger in use for the \SiPMs\ test program.]{Cold finger in use for the \SiPMs\ test program.  Visible is the \PTFE\ tube covered by the super-insulator, the black box containing the \SiPM\ under test, two unjacketed optical fibers (connected to a LED and to a laser source), cryogenic pre-amplifier, the cryocooler head, and the set of high power film resistors in use to control temperature~\cite{Acerbi:2017gy}.}
\label{fig:LNGSSetup-ColdFinger}
\end{figure}

The cold head of the cryocooler is equipped with a cold finger, which holds the \SiPM\ assembly under test, as shown in Fig.~\ref{fig:LNGSSetup-ColdFinger}.  This arrangement allows for a fast thermal cycle: the cold finger can be thermalized from room temperature to \LNGSCryoSetupTemperatureColdLimit \space in about \LNGSCryoSetupCooldownTime.  The use of a platinum RTD connected to a \LNGSCryoSetupTemperatureControllerModule\ temperature controller, which drives the thermal load required for temperature regulation directly to a set of high power metal film resistors mounted on the cold finger, allows the system to reach a temperature stability of about \SI{0.1}{\K} and an accuracy of about \LNGSCryoSetupTemperatureAccuracy\ in the range of interest \LNGSCryoSetupTemperatureRange.  To minimize unwanted thermal gradients, the \SiPM\ holders are made from \LNGSCryoSetupIMSPCBThickness-thick aluminum printed circuit board (PCB) using insulated metallic substrate (IMS) technology, to which the \SiPM\ sample is bonded with the silver-loaded conductive epoxy \LNGSCryoSetupConductiveEpoxyModel.  The holder is firmly connected to the cold finger by a screw and a thin layer of cryogenic grease ensuring a good thermal contact between the copper and the aluminum board.  A cryogenic pre-amplifier specifically designed for this application is connected to the \SiPM\ and is in thermal equilibrium with the cold finger.

The electronics chain used to readout the \SiPMs\ in the \LNGS\ cryogenic test setup is composed of:
\begin{compactitem}
\item A \SiPM\ bias source, feeding the \SiPM\ from outside the cryostat and allowing for direct current-voltage measurements in forward and reverse polarity;
\item A cryogenic pre-amplifier, directly connected to the device under test, and its bias source, built as a transimpedance amplifier (\TIA) capable of stable cryogenic operations;
\item A warm, low noise, single stage non-inverting amplifier outside the cryostat, based on a high speed, low noise operational amplifier.  This receives the signals from the pre-amplifier that is inside the cryostat;
\item A \LNGSCryoSetupDigitizerModel\ \LNGSCryoSetupDigitizerSamplingRate\ \LNGSCryoSetupDigitizerResolution\ digitizer, configured for interleaved acquisition and operated in auto-trigger mode;
\item A custom C++ program handles the configuration and readout of the digitizer, and allows for a time resolution of \LNGSCryoSetupTimeIndicatorFineResolution, within a given acquisition frame, and an adjustable trigger.  Data are saved in a custom file format, with each frame in the file including the digitized waveform and a header, which contains accurate timing information;
\end{compactitem}
Full details of the front-end electronics, the data acquisition (DAQ) system and the software used to analyze the data for the cryogenic characterization of the \SiPMs, are given in~\cite{Acerbi:2017gy}.

\subsubsection{Identifying \SiPM\ Pulses and Correlated Noise}
\label{sec:PhotoElectronics-Characterization-Analysis}

\begin{figure}[!t]
\centering
\includegraphics[width=\columnwidth]{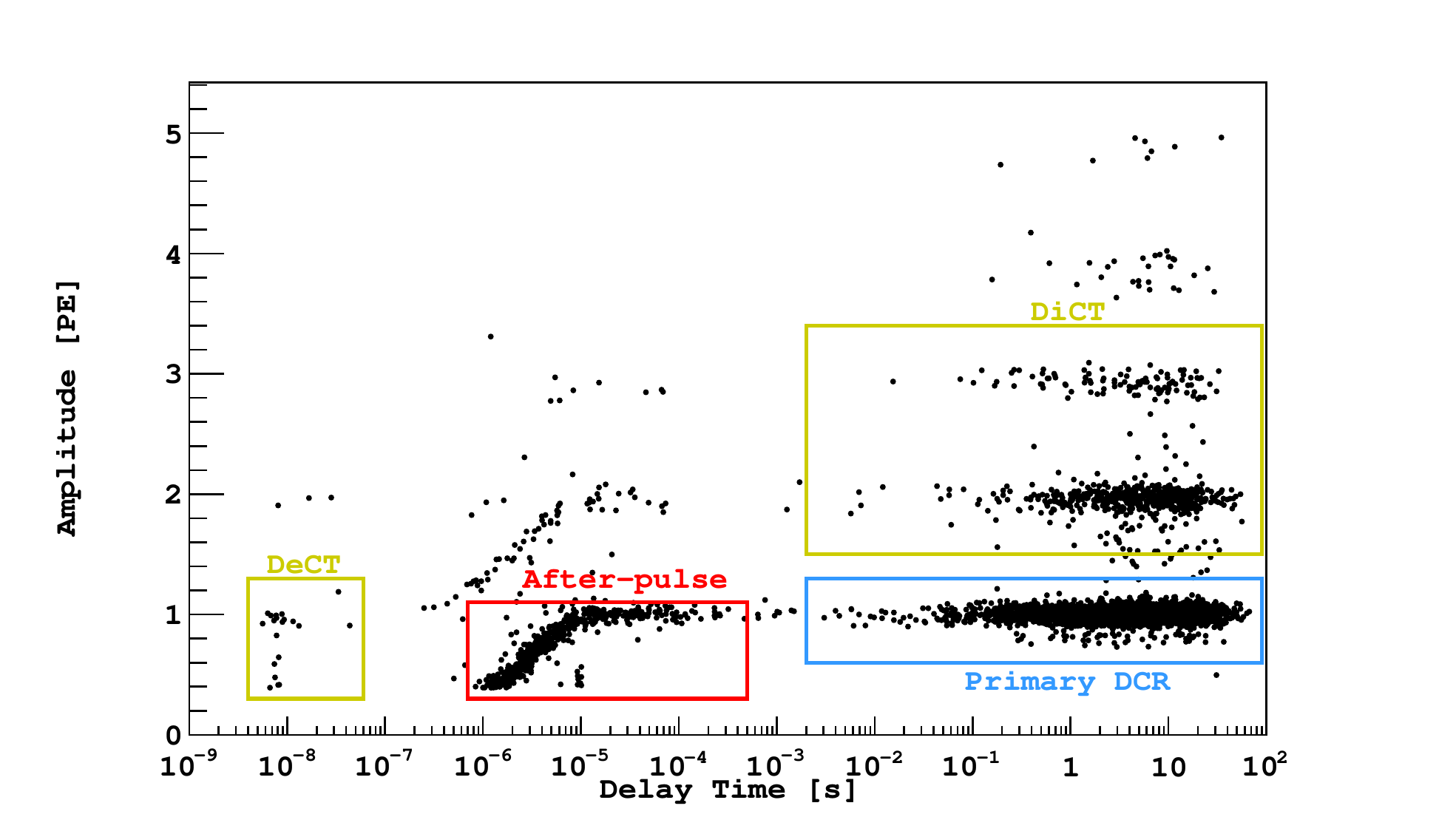}
\caption[\SiPM\ delay time versus amplitude at \LINNormalTemperature.]{Distribution of delay time versus amplitude for data taken at \LINNormalTemperature\ and in absence of light.  Few components of the noise response of the \SiPM\ can be clearly identified: primary dark count (part of the Dark Count Rate, \DCR), Direct CrossTalk (\DiCT), Delayed CrossTalk (\DeCT), and AfterPulsing (\AP)~\cite{Acerbi:2017gy}.}
\label{fig:NUVHdLf-TCNP}
\end{figure}

The data analysis software was developed at FBK following the procedure  described in~\cite{Piemonte:2012gl}.  It reads the data stored in the data frames and performs a detailed analysis of \SiPM\ response, with special emphasis on the time correlation of pulses.  The analysis capabilities are strongly enhanced by the addition of the time tagging of frames, described in~\cite{Acerbi:2017gy}, which is mandated by the significant difference in \SiPM\ dark pulse rates as a function of temperature that spans several orders of magnitude.  The time tagging of frames allows for the calculation of the time difference between consecutive pulses at the nanosecond scale using pulses within the same frame, for up to thousands of seconds, using different frames, with the only practical limit set by the time available for measurement.  This permits a seamless determination of the temperature-dependent \SiPM\ pulse rate, a detailed study of their correlation, and the extraction of the secondary noise probabilities.  The analysis is done the same way for data collected at all temperatures: each event is tagged with an ordered pair of values, {\it i.e.} the time distance from the previous pulse and the amplitude.  A typical scatter plot with all ordered pairs recorded for data taken at \LINNormalTemperature\ and in the absence of light is shown in Fig.~\ref{fig:NUVHdLf-TCNP}.  The plot allows for the identification of different families of pulses, composing the noise response of the \SiPM.

\begin{asparaenum}
\item[\bf \DCR:] The main group of events is determined by primary, Poisson-distributed, dark counts that make up the Dark Count Rate, \DCR.  The amplitude is centered at around \SI{1}{\pe}, and the distribution in time is an exponential with decay time corresponding to the inverse of the \DCR.
\item[\bf \DiCT:] Direct CrossTalk (\DiCT) events occur in a very short time interval given by the travel time needed by the crosstalk photon to reach a neighboring cell and trigger an independent avalanche: the time required is in the picosecond time scale, nearly impossible to be resolved with the readout electronics at hand.  As a result, single \SI{}{\pe} \DiCT\ pulses are superimposed with the \DCR\ pulses, since they have a similar time distribution, but \DiCT\ can be further characterized by pulses with greater amplitudes, corresponding to the detection of \num{2} or \SI{3}{\pe}, or more.
\item[\bf \DeCT:] The least populated of the noise groups, with characteristic delay time of few to tens of nanoseconds, are due by Delayed CrossTalk (\DeCT).  These events are caused by crosstalk photons absorbed in the non-depleted region of a neighboring cell.  The carriers diffuse for a short time before reaching the high-field region, finally triggering an avalanche.  The resulting pulse has the single cell amplitude but is delayed with respect to the previous pulse by the characteristic diffusion time, typically on the order of a few to tens of nanoseconds.  Such delayed pulses can further trigger \DiCT\ pulses, leading to events with the same time distribution but larger amplitudes.
\item[\bf \AP:] Finally, the group of events with intermediate delay times and amplitude of \SI{1}{\pe} or less, are identified as AfterPulsing (\AP).  AP occurs when, during an avalanche, an electron is trapped by some impurity in the silicon lattice and is then released after a characteristic time, generating a second avalanche.  Since the AP event and its primary avalanche occur in the same cell, the time distribution is determined by both the trap time constant and the recharge time constant of the microcell.  When the time interval is lower than the full micro-cell recharge, the resulting pulse has a reduced amplitude.  As with the other types of events, they can trigger \DiCT\ pulses, explaining the higher amplitude noise groups with same time distribution.
\end{asparaenum}

At the end of the analysis stage the following parameters are saved: primary \DCR, correlated noise probabilities (\DiCT, \DeCT, \AP), and single cell signal features (amplitude, recovery time, charge delivered in a fixed time gate).  Also stored is the waveform resulting from the application of the Differential Leading Edge Discriminator (\DLED) algorithm, as defined in~\cite{Gola:2012bj}, which is useful for the evaluation of the breakdown voltage ($V_{BD}$), as the resulting peak amplitude is linear with the applied \OV.  Consideration of the \DLED\ amplitude as a function of \OV\ typically allows a more precise determination of $V_{BD}$ than possible to derive from the $IV$ curve.  This method allows to determine precisely the \OV\ as a function of temperature.  In order to compare the \SiPM\ \DCR\ as a function of temperature, it is important that each point be taken at the same \OV.  Therefore, any plot of \DCR\ vs. \OV\ is re-corrected in the analysis stage, by performing a linear interpolation over the data to extract the correct \DCR\ at the exact value of \OV\ as determined with this procedure.  The same consideration applies for all the other \SiPM\ features.

\subsubsection{Characterization of \NUVHd\ \SiPMs}
\label{sec:PhotoElectronics-Characterization-NUVHd}

\begin{figure}[!t]
\includegraphics[width=\columnwidth]{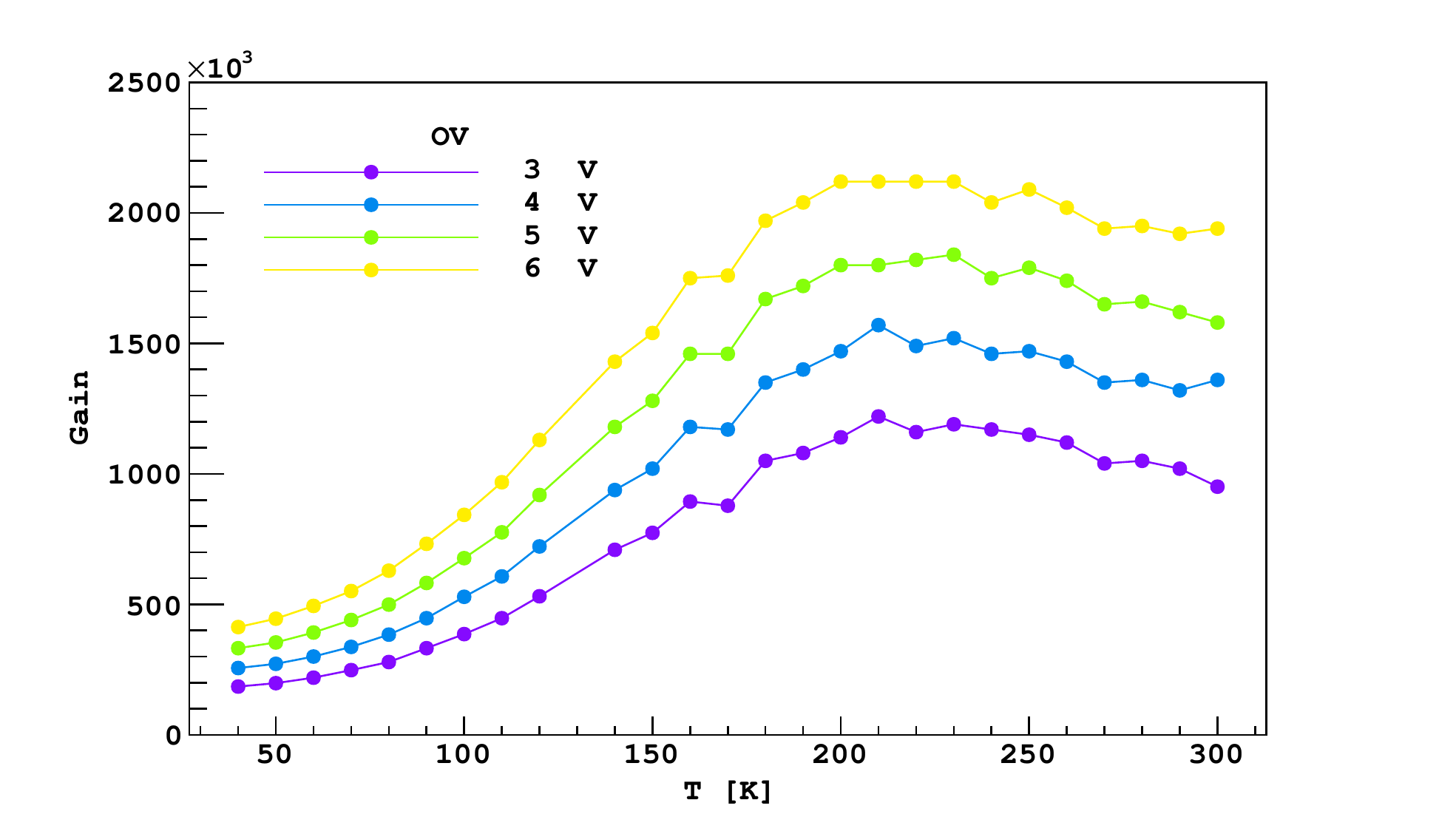}
\caption[Gain of \SiPMs\ versus over-voltage and temperature.]{Gain of the \NUVHdLf\ \SiPMs\ as a function of \OV\ and temperature~\cite{Acerbi:2017gy}.}
\label{fig:NUVHd-Gain}
\end{figure}

\begin{figure}[!t]
\includegraphics[width=\columnwidth]{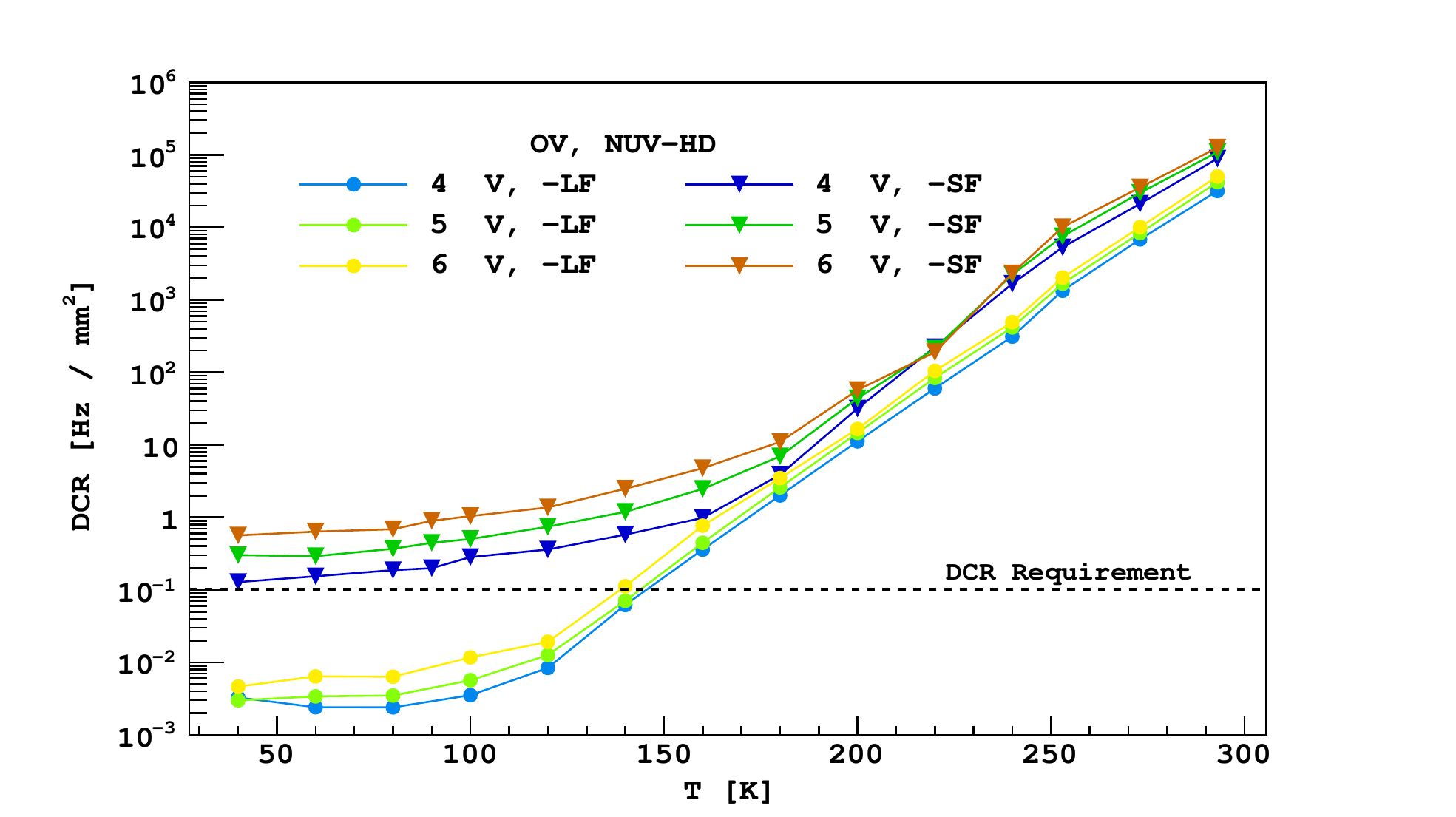}
\caption[\DCR\ of \SiPMs\ as a function of over-voltage and temperature.]{\DCR\ for \NUVHdSf\ (triangular markers) and \NUVHdLf\ (circular markers) devices as function of \OV\ and temperature~\cite{Acerbi:2017gy}.}
\label{fig:NUVHd-DCR}
\end{figure}

The Single Cell Response (\SCR) is defined as the signal generated at the output of the amplifier when a single microcell of the \SiPM\ fires. The impedance $R_q$ of the polysilicon resistor depends strongly on temperature, increasing with decreasing temperature, therefore affecting various signal features.  The typical \SiPM\ pulse is made up of two different components: a fast component, whose time constant $\tau_f$ is of the order of some nanoseconds, and a slow component, whose time constant $\tau_s$ can extend to several microseconds, this time constant corresponding to the \SPAD\ recharge time.  The electric model proposed in~\cite{Corsi:2007is} is well suited to describe the \SiPM\ behaviour: the slow time constant is expected to be proportional to the quenching resistance as in $\tau_s = R_q \cdot (C_{\rm SPAD} + C_q)$, where $C_q$ is the parasitic capacitance of the polysilicon resistor.  With decreasing temperature, the increase in $R_q$ translates into higher values of $\tau_s$.  This effect can be easily appreciated in Fig.~\ref{fig:NUVHdLf-SCRvsT}.

Using the \LNGS\ system for cryogenic characterization of \SiPMs\ and the analysis software developed for this purpose, the recharge (recovery) time for the individual \SPAD\ as a function of \OV\ and temperature could be determined.  The peak amplitude of the \SCR\ is mainly determined by the fast component of the response, the amplitude of which increases linearly with \OV\ and only very slowly with temperature (see Fig.~\ref{fig:NUVHdLf-SCRvsT}).  For each pulse in the system, the gain of the \SiPM\ is measured by comparing the integrated region of interest (ROI) around the pulse to an ROI of similar size just before the pulse (pedestal), and by measuring the distance between the peaks of the resulting distributions, it is possible to calculate the number of charge carriers generated in the \SPAD.  See Fig.~\ref{fig:NUVHd-Gain} for the measurement of the gain versus temperature and \OV.  The variation with temperature is due to the fact that, with decreasing temperature and longer pulses, an increasingly larger fraction of the total pulse sits out of the \LNGSCryoSetupDigitizerIntegrationGate\ integration gate.  Near room temperature, when the signal duration becomes shorter than the integration time, the gain is expected to saturate to a plateau: the residual dependence on temperature observed is due to the variation with temperature of the depletion width of each \SPAD\ for a given \OV. This happens because the breakdown voltage changes with temperature as well. Therefore, a variation of the \SPAD\ junction capacitance is also observed and, thus of its gain.

\begin{figure*}[!t]
\includegraphics[width=\columnwidth]{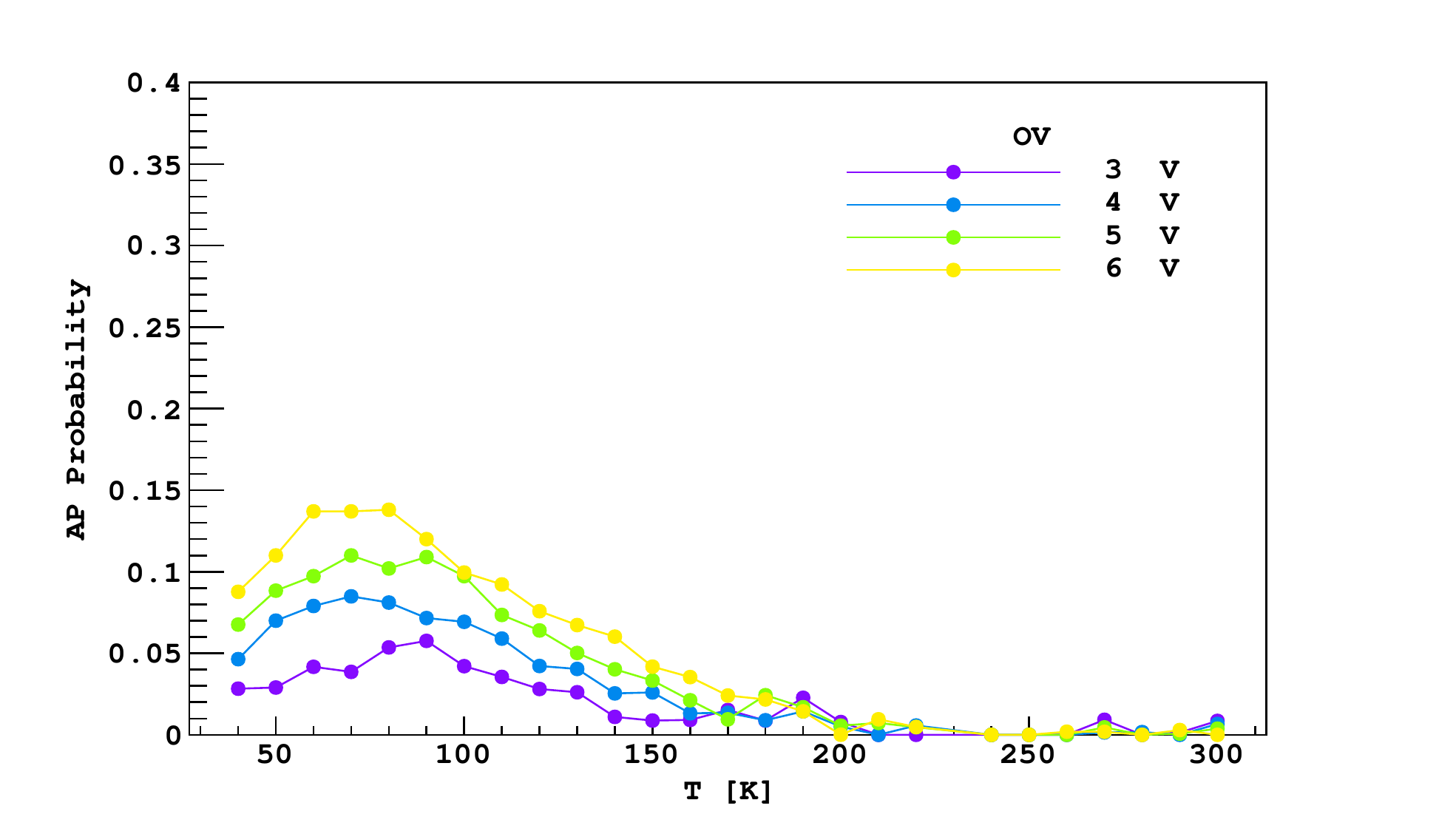}
\caption[\SiPM\ \AP\ probability versus temperature for different values of over-voltage.]{\AP\ probability as function of temperature for different \OV\ for \NUVHdLf\ \SiPMs~\cite{Acerbi:2017gy}.}
\label{fig:NUVHd-AP}
\end{figure*}

\begin{figure}[!t]
\includegraphics[width=\columnwidth]{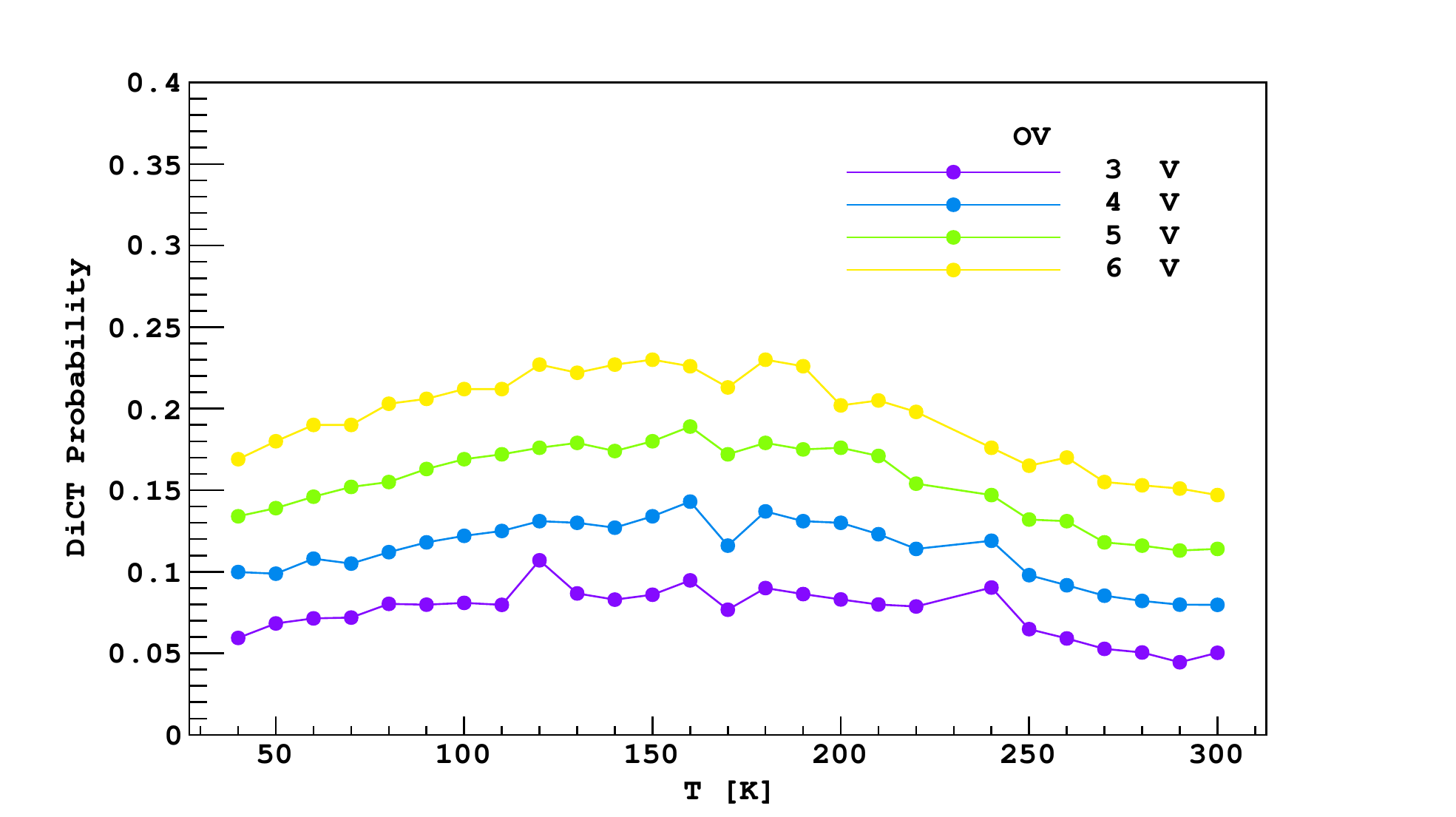}
\caption[\SiPM\ \DiCT\ probability versus temperature for different values of over-voltage.]{\DiCT\ probability as function of temperature for different \OV\ for \NUVHdLf\ \SiPMs.  \DiCT\ values exhibit a weak dependence on temperature.~\cite{Acerbi:2017gy}}
\label{fig:NUVHd-DiCT}
\end{figure}

\begin{figure}[!t]
\includegraphics[width=\columnwidth]{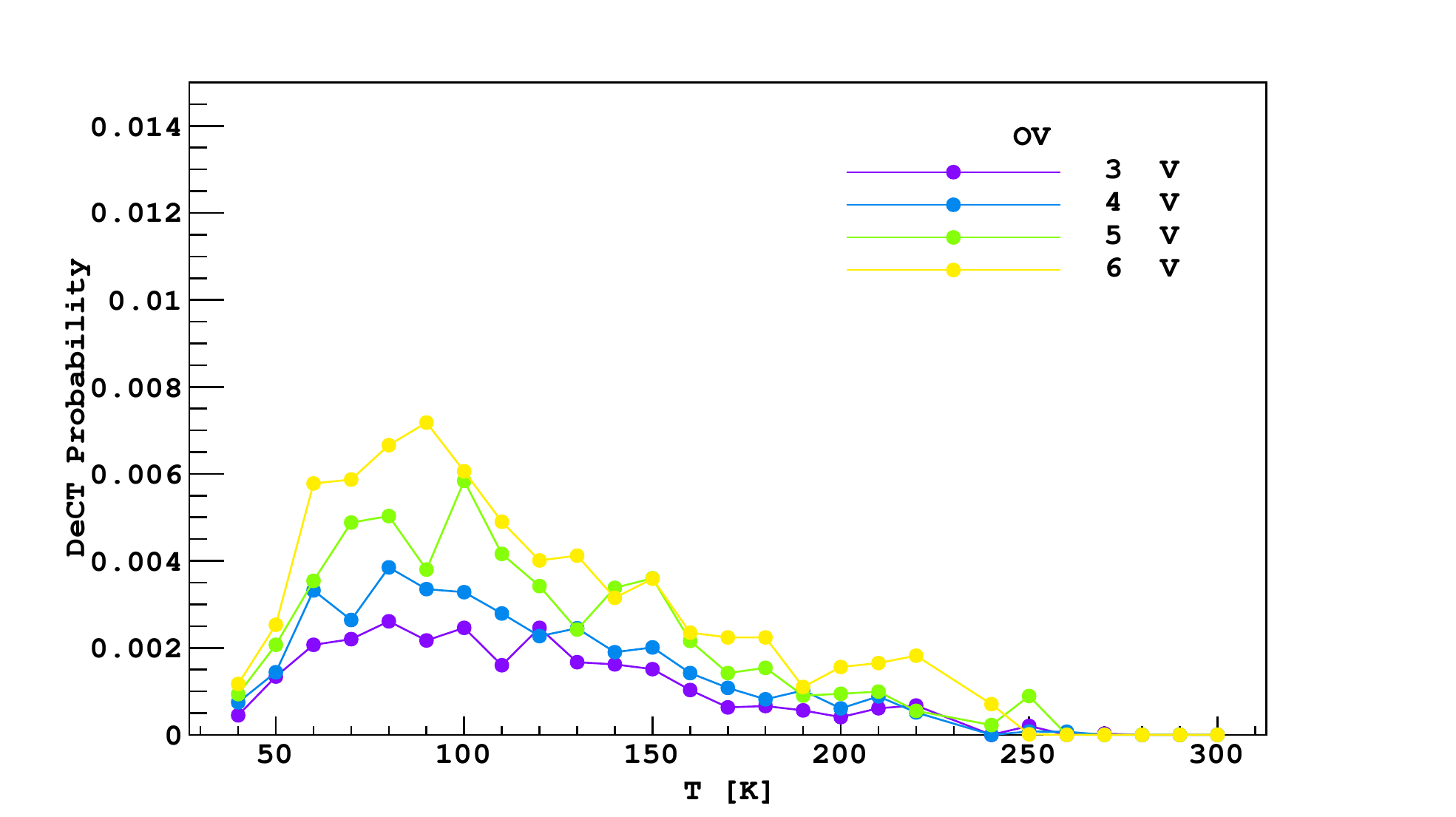}
\caption[\SiPM\ \DeCT\ probability versus temperature for different values of over-voltage.]{\DeCT\ probability as function of temperature for different \OV\ for \NUVHdLf\ \SiPMs, found to be negligible.~\cite{Acerbi:2017gy}}
\label{fig:NUVHd-DeCT}
\end{figure}

The \NUVHdLf\ \SiPMs\ are especially optimized to achieve good performances in terms of \DCR\ reduction at low temperature.  They exhibit a \DCR\ that is one to two orders of magnitude lower than the \NUVHdSf\ at the same value of \OV\ and temperature, see Fig.~\ref{fig:NUVHd-DCR}.  The \NUVHdLf\ \SiPMs\ show lower correlated noise with respect to the \NUVHdSf\ \SiPMs\ at the same \OV, see Fig.~\ref{fig:NUVHd-AP}. This is explained not only by the overall lower gain and avalanche triggering probability in the \NUVHdLf\ devices for the same \OV, but also by a suppression of some field effects contributing to this noise component. The \AP\ reaches a maximum in the temperature range from \SIrange{60}{80}{\kelvin} and then decrease to zero in the high temperature region.  For \AP, this peculiar dependence can be explained by the interplay of two different phenomena having opposite effects on the noise probability: at low temperatures the trapping time constants increase, enhancing the probability of a carrier to be released when the cell is at least partially recharged and is able to trigger a new avalanche, thus effectively producing an \AP\ event. On the other hand, the quenching resistance grows exponentially, partially suppressing the avalanche triggering probability during the recharge.  The \DiCT\ exhibits only a weak dependence on the temperature, while it is linear in \OV, see Fig.~\ref{fig:NUVHd-DiCT}. The \DeCT\ has an extremely small probability, see Fig.~\ref{fig:NUVHd-DeCT}.

\subsection{The Cryogenic Transimpedence Amplifier}
\label{sec:PhotoElectronics-TIA}

The unique challenge in readout presented by \SiPMs\ is mainly due to their capacitance: at \DSkSiPMCapacitancePerArea, a single \SiPM\ with \DSkSiPMAreaStd\ surface passes the \si{\nano\farad} scale for capacitance.  Yet, the experiment foresees a photosensing area of \DSkTilesArea.

The readout of a \LAr\ calorimeter faced a similar challenge, with capacitance for the individual cells often passing the \si{\nano\farad} scale~\cite{Willis:1974do}.  Early developments on \LAr\ calorimeter readout~\cite{Radeka:1974ca,Radeka:1988ku,Chase:1993dj,Chase:1997fk} demonstrated that the use of a transimpedance amplifier (\TIA) is preferable to a charge integration amplifier, whose noise and rise time are very strongly affected by the large detector capacitance.

The \DS\ Collaboration has developed and optimized a \TIA\ with performance at \LArNormalTemperature\ in mind~\cite{DIncecco:2017td}.  The specific goal was to maximize the amplification factor while preserving a stable signal bandwidth and \SNR.

\begin{figure}[!t]
\centering
\includegraphics[width=0.45\columnwidth]{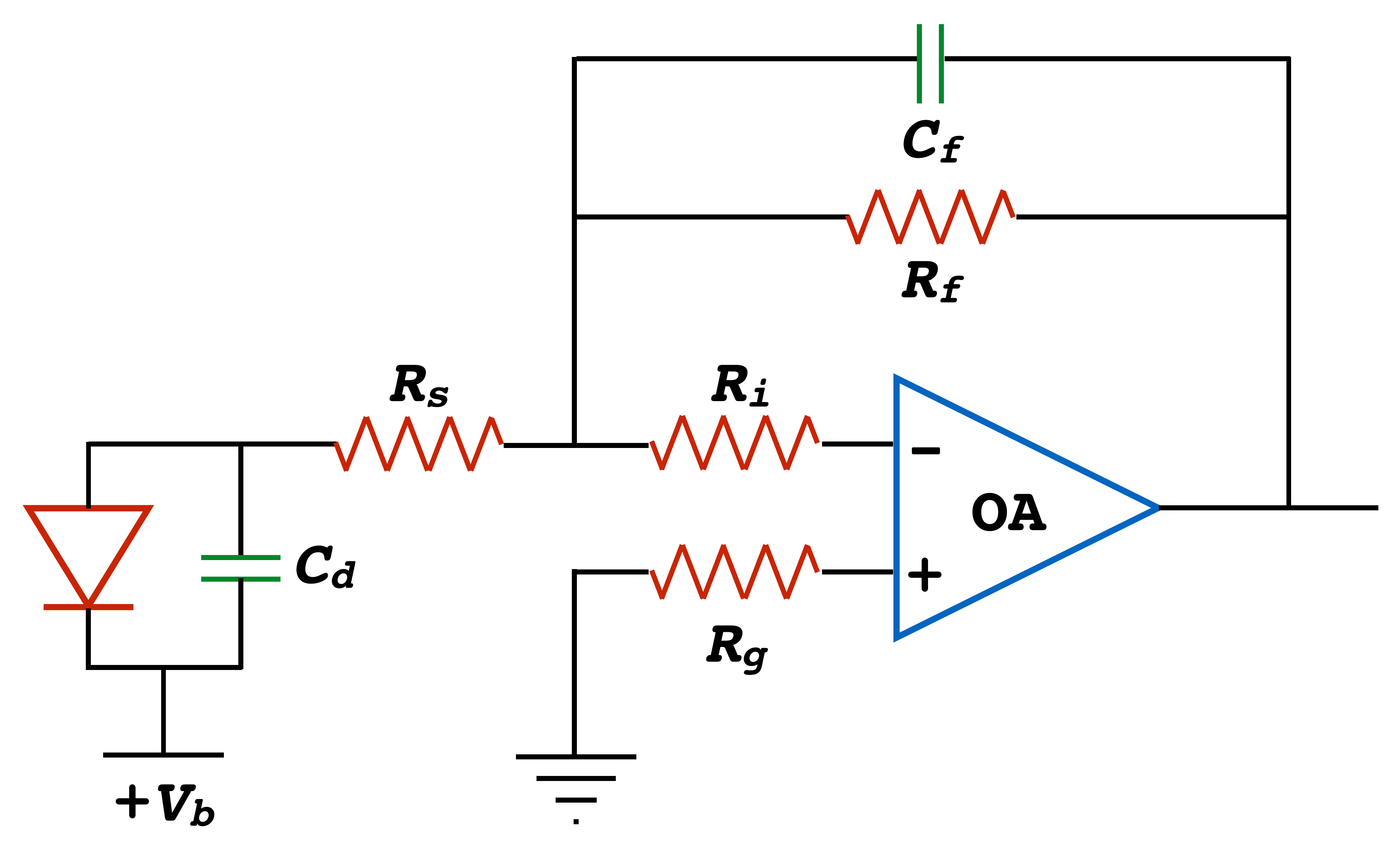}
\caption[Schematic view of a \SiPM\ plus \TIA\ model.]{Schematic view of a working model of a single \SiPM\ and associated \TIA\ acting as its pre-amplifier.}
\label{fig:TIA-Sketch}
\end{figure}

\begin{figure*}[!t]
\centering
\includegraphics[width=\columnwidth]{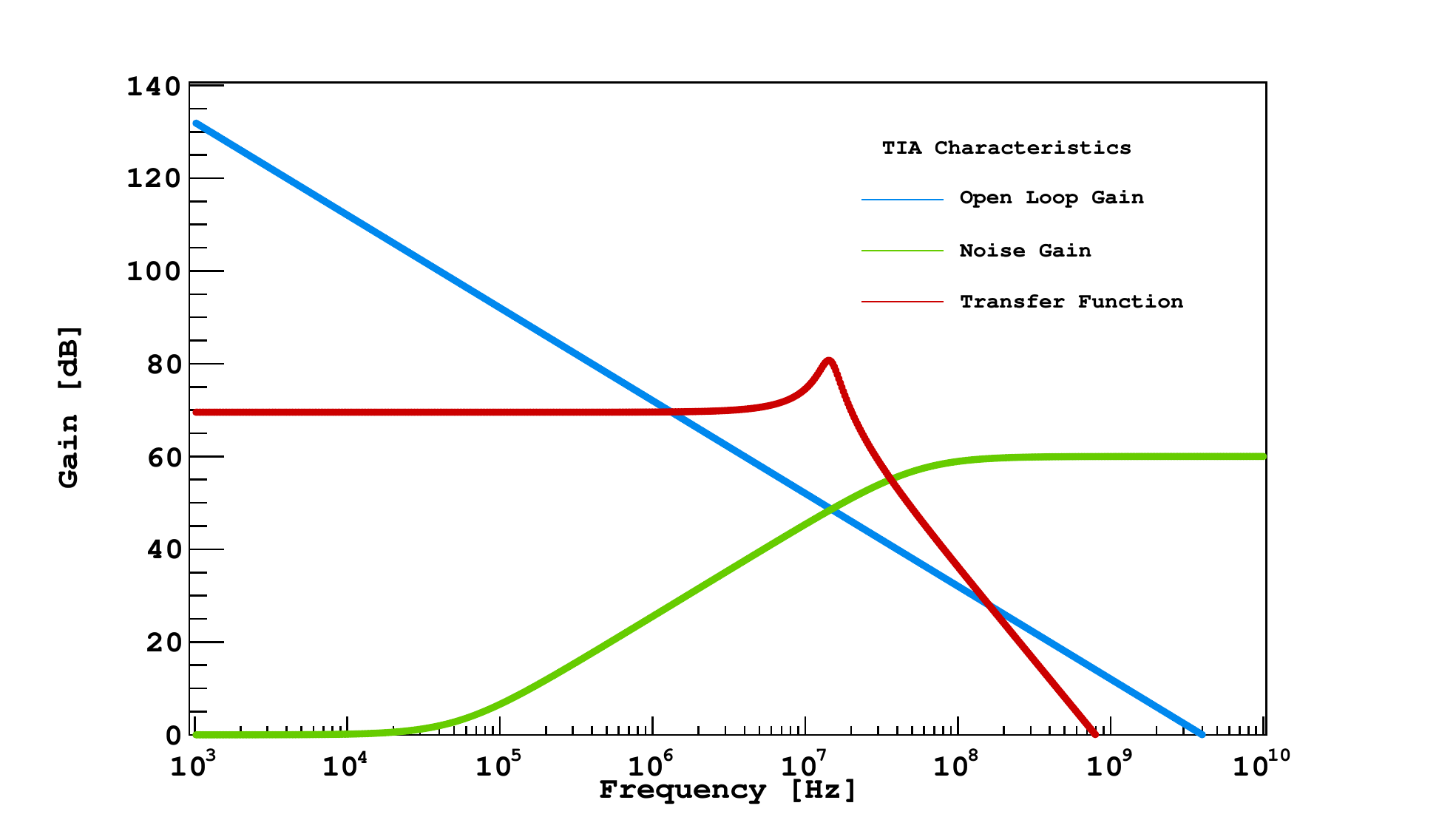}
\includegraphics[width=\columnwidth]{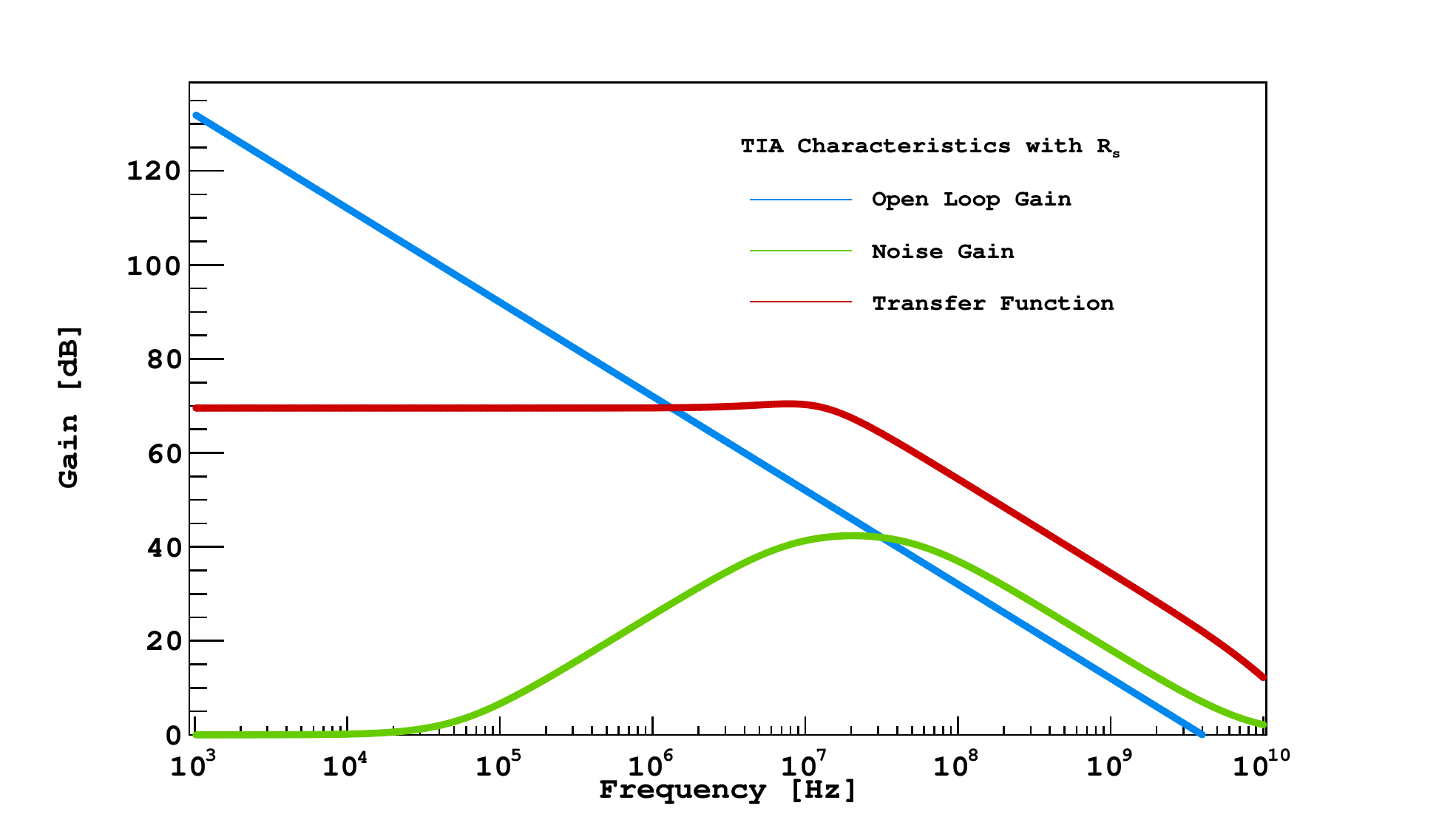}
\caption[\TIA\ characteristics.]{{\bf Top:} open loop gain (\AVol), noise gain (\NoG) and transfer Function (\Tz) of the \TIA\ configured with $C_d = \SI{1}{\nano\farad}$, $R_f = \SI{3}{\kilo\ohm}$, $C_f = \SI{1}{\pico\farad}$, and $R_s = \SI{0}{\ohm}$.  {\bf Bottom:} same distributions with $R_s = \SI{20}{\ohm}$.
}
\label{fig:TIA-CharacteristicsNoRs}
\end{figure*}

A schematic view of the TIA is shown in Fig.~\ref{fig:TIA-Sketch}.  The actual value of $R_f$ depends on the size and the characteristics of the \SiPMs\ (detector capacitance $C_d$ and signal bandwidth), with typical values from \DSkTIAFeedbackResistanceRange, while $C_f$ is only due to the parasitic capacitance of the circuit and can be estimated as about \DSkTIAFeedbackCapacitanceScale.  The implementation of the \TIA\ includes a few components required to stabilize the circuit at cryogenic temperature (given the very high Gain-Bandwidth Product (\GBP)).  The cryogenic \TIA\ consists of a high speed very low noise operational amplifier capable of operating in the temperature range \LNGSCryoSetupTemperatureRange.  Fig.~\ref{fig:TIA-CharacteristicsNoRs} shows the \AVol, \NoG, and \Tz\ for the \TIA\ with a specific configuration of the circuit components.  While full details of the TIA will be given in a later publication, it is stated here that the performance of the \TIA\ has met the required specifications of the \DSk\ experiment.

\subsection{Signal-to-Noise Ratio measurement on FBK \SiPMs}
The following measurements were performed at \LNGS\ in a dewar filled with liquid nitrogen, where the \SiPMs\ were illuminated by a \LNGSCryoSetupLaserModel\ \si{\pico\second} pulsed diode laser, which also served as the trigger sync. The signal, after further amplification at room temperature, is then measured and directly histogrammed by a \LNGSCryoSetupOscilloscopeModel\ oscilloscope.  The histogram of the signals includes the peaks from single and multiple photoelectrons and the baseline noise corresponding to events where no photons were recorded.  In order to better estimate this parameter, as many peaks as possible are fit within the spectrum; the standard deviation of the $n^{th}$ photoelectron peak is given by $\sigma_{n} = \sqrt{\sigma_b^2 + n \SigmaSPEt}$, where $\sigma_b$ is the baseline contribution and \SigmaSPE\ is the intrinsic \SiPM\ resolution in response to the single photoelectron signal.

\SNR\ is defined as the ratio of the average \SPE\ signal to the noise of the baseline: this definition is flexible and several quantities of use can be identified:

\begin{asparaenum}
\item[\bf \SNRAmp:] The \SPE\ amplitude is evaluated as the difference between the maximum signal amplitude and the median of the baseline (estimated in the pre-trigger window), while the noise is measured as the standard deviation of the baseline in the pre-trigger.  For this measurement technique, the first peak corresponds to the events where no photon was detected by the \SiPM\ under test: its shape is not gaussian since it is filled to the maximum value of the baseline in a fixed gate (\DSkSiPMBaselineAmplWindow) and its median is correlated to the crest factor of the noise of the baseline.  Therefore the first peak is not used in the analysis.
\item[\bf \SNRCharge:] The \SPE\ charge is evaluated as the difference between the integral of the signal in a fixed window (typically \DSkSiPMSNRChargeWindow) and the integral of the baseline in a gate of the same duration.
\item[\bf \SNRFast:] Defined as \SNRCharge\ but for a short integration window, corresponding to the fast discharge peak of high $R_q$ \SiPMs.
\item[\bf \SNRFilter:] Defined as \SNRAmp\ on a filtered signal.  An optimal filter technique is used with an $RC$ filter that has $\tau$ matched to the SiPM recharge time.
\end{asparaenum}

Table~\ref{tab:Photoelectronics-SNR} provides a summary of the performances of the devices tested throughout the \DS\ program.

\begin{table*}
\centering
\rowcolors{3}{gray!35}{}
\caption[Summary of read-out performances for \SiPMs\ and \tiles.]{Summary of read-out performances for \SiPMs\ and \tiles.  The absolute values of the peaks can not be compared since different amplifications were used in the different measurement campaigns, so what matters is only the quoted \SNR.}
\label{tab:Photoelectronics-SNR}
\begin{tabular}{llrrrc}
Device, Configuration, Size, Gain 				&\SNR\ Type	&\multicolumn{1}{c}{\SigmaBaseline }	&\multicolumn{1}{c}{$\SigmaSPE$} 	&\multicolumn{1}{c}{\SPE}		&\SNR\ Value\\
\hline
\cellcolor{white}{\NUVHdSfHRq, Single \SiPM}		&\SNRAmp	&\SI{2.5}{\milli\volt} 					&\SI{2.3}{\milli\volt} 				&\SI{29.9}{\milli\volt} 			&\num{12}\\
\cellcolor{white}{\DSkSiPMAreaStd, $7\times10^5$}	&\SNRCharge	&\SI{21.4}{\pico\volt\second} 			&\SI{8.0}{\pico\volt\second} 		&\SI{171.9}{\pico\volt\second} 	&\num{8} \\
\hline
\cellcolor{white}{\NUVHdLfHRq, Single \SiPM}		&\SNRAmp	&\SI{4.4}{\milli\volt} 					&\SI{1.8}{\milli\volt} 				&\SI{50.7}{\milli\volt} 			&\num{12}\\
\cellcolor{white}	{\DSkSiPMAreaMax, $1.4\times10^6$}	&\SNRFast	&\SI{104}{\pico\volt\second} 		&\SI{38}{\pico\volt\second} 		&\SI{1109}{\pico\volt\second} 	&\num{11} \\
\cellcolor{white}								&\SNRCharge	&\SI{325}{\pico\volt\second} 			&\SI{211}{\pico\volt\second} 		&\SI{11578}{\pico\volt\second} 	&\num{36} \\
\hline
\cellcolor{white}	{\NUVHdLfHRq, 4p4s \tile}		&\SNRAmp	&\SI{2.8}{\milli\volt} 					&\SI{3.0}{\milli\volt} 				&\SI{17}{\milli\volt} 			&\num{6}\\
\cellcolor{white}{\NUVHdLfHRqStdFourPFourSArea, $1.6\times10^6$}	&\SNRFast	&\SI{25}{\pico\volt\second} &\SI{25}{\pico\volt\second} 	&\SI{117}{\pico\volt\second} 	&\num{5} \\
\cellcolor{white}								&\SNRCharge	&\SI{121}{\pico\volt\second} 			&\SI{174}{\pico\volt\second} 		&\SI{1250}{\pico\volt\second} 	&\num{11} \\
\cellcolor{white}		   						&\SNRFilter	&\SI{1.6}{\milli\volt} 					&\SI{2.4}{\milli\volt} 				&\SI{33}{\milli\volt} 			&\num{20}\\
\end{tabular}
\end{table*}

\subsubsection{Single \SiPM}
\label{sec:PhotoElectronics-TIA-SingleSiPMSNR}

\begin{figure*}[!htpb]
\centering
\includegraphics[width=\columnwidth]{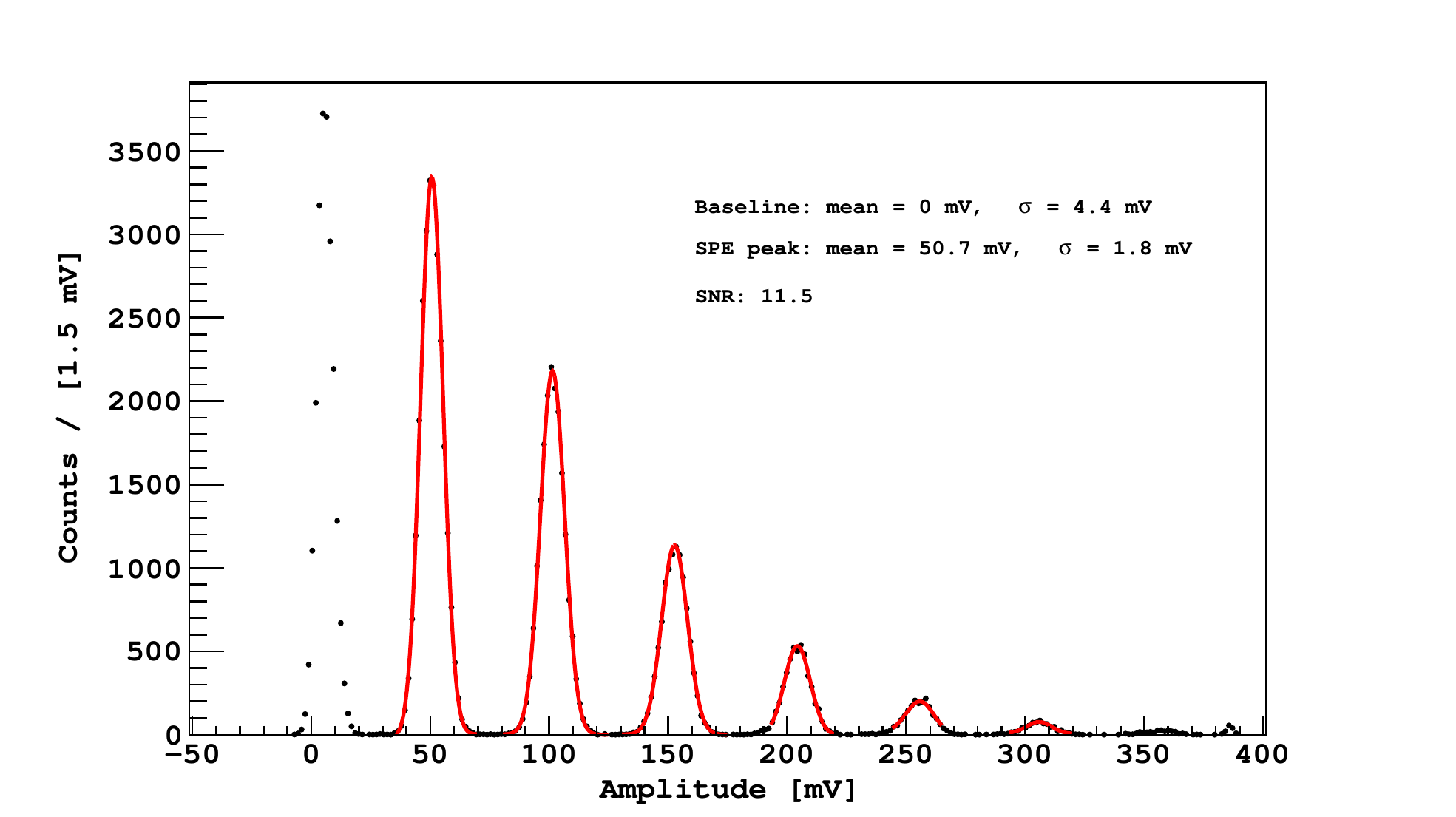}
\includegraphics[width=\columnwidth]{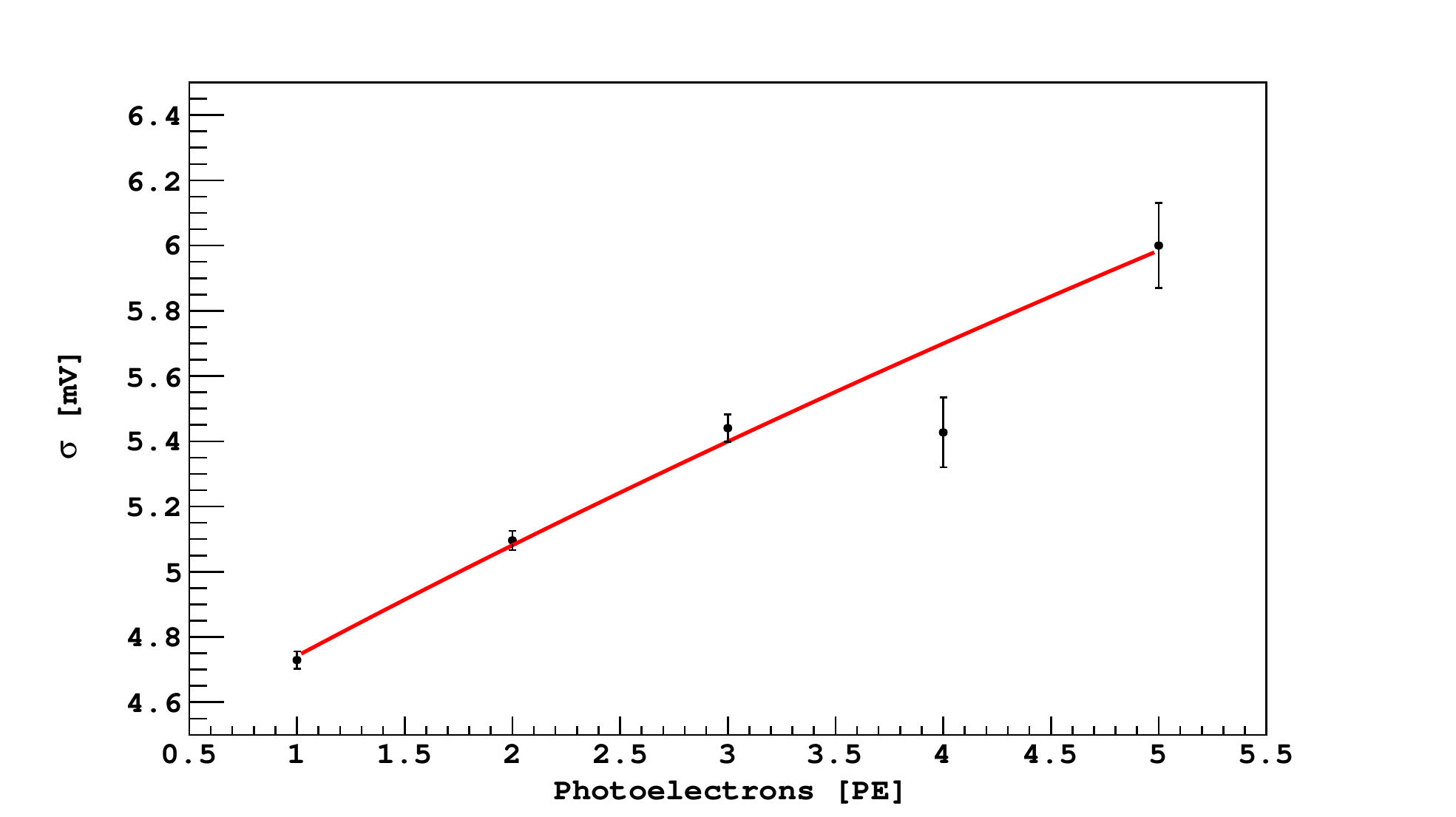}
\caption[Photoelectron peaks amplitude spectrum for \NUVHdLfHRq\ \SiPMs\ of \DSkSiPMAreaMax\ area.]{Amplitude spectrum {\bf (top)} and distribution of the standard deviations {\bf (bottom)} of the photoelectron peaks for \NUVHdLfHRq\ \SiPMs\ of \DSkSiPMAreaMax\ area.}
\label{fig:NUVHdLfHRqMax-Ampl}
\end{figure*}

\begin{figure*}[!htpb]
\centering
\includegraphics[width=\columnwidth]{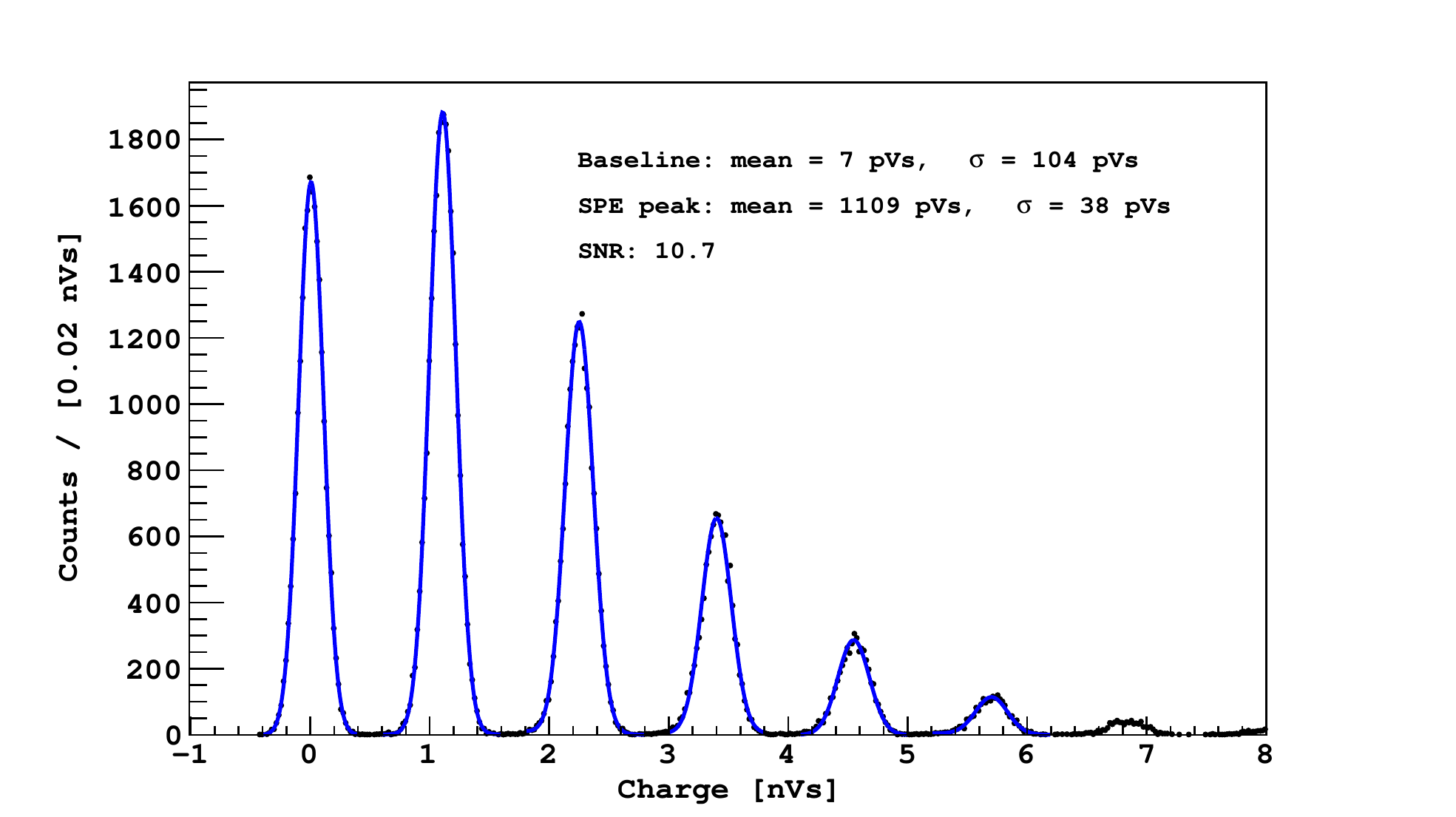}
\includegraphics[width=\columnwidth]{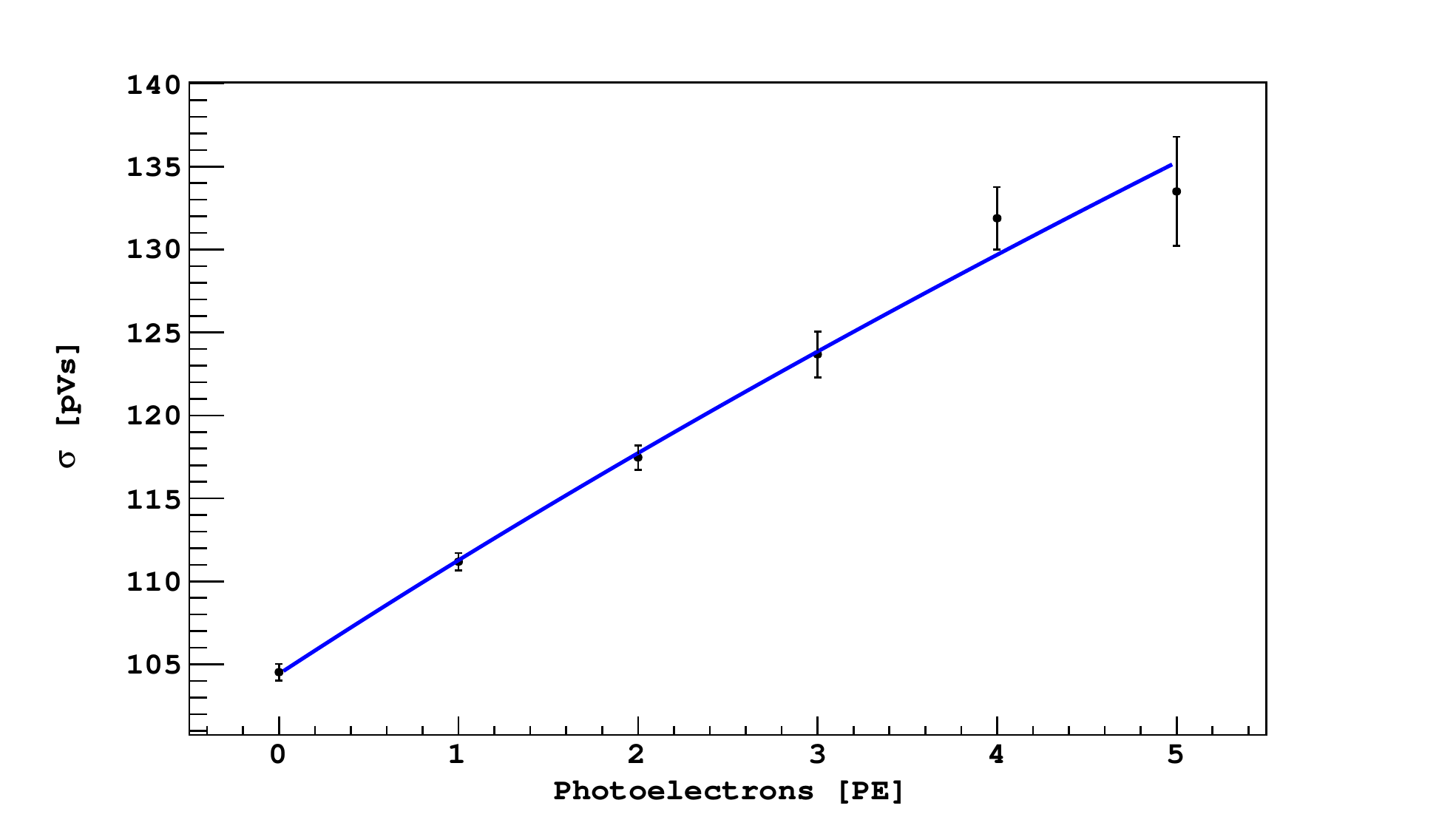}
\caption[Photoelectron peaks fast charge spectrum for \NUVHdLfHRq\ \SiPMs\ of \DSkSiPMAreaMax\ area.]{Fast charge spectrum {\bf (top)} and distribution of the standard deviations {\bf (bottom)} of the photoelectron peaks for \NUVHdLfHRq\ \SiPMs\ of \DSkSiPMAreaMax\ area.}
\label{fig:NUVHdLfHRqMax-30nsCharge}
\end{figure*}

\begin{figure*}[!htpb]
\centering
\includegraphics[width=\columnwidth]{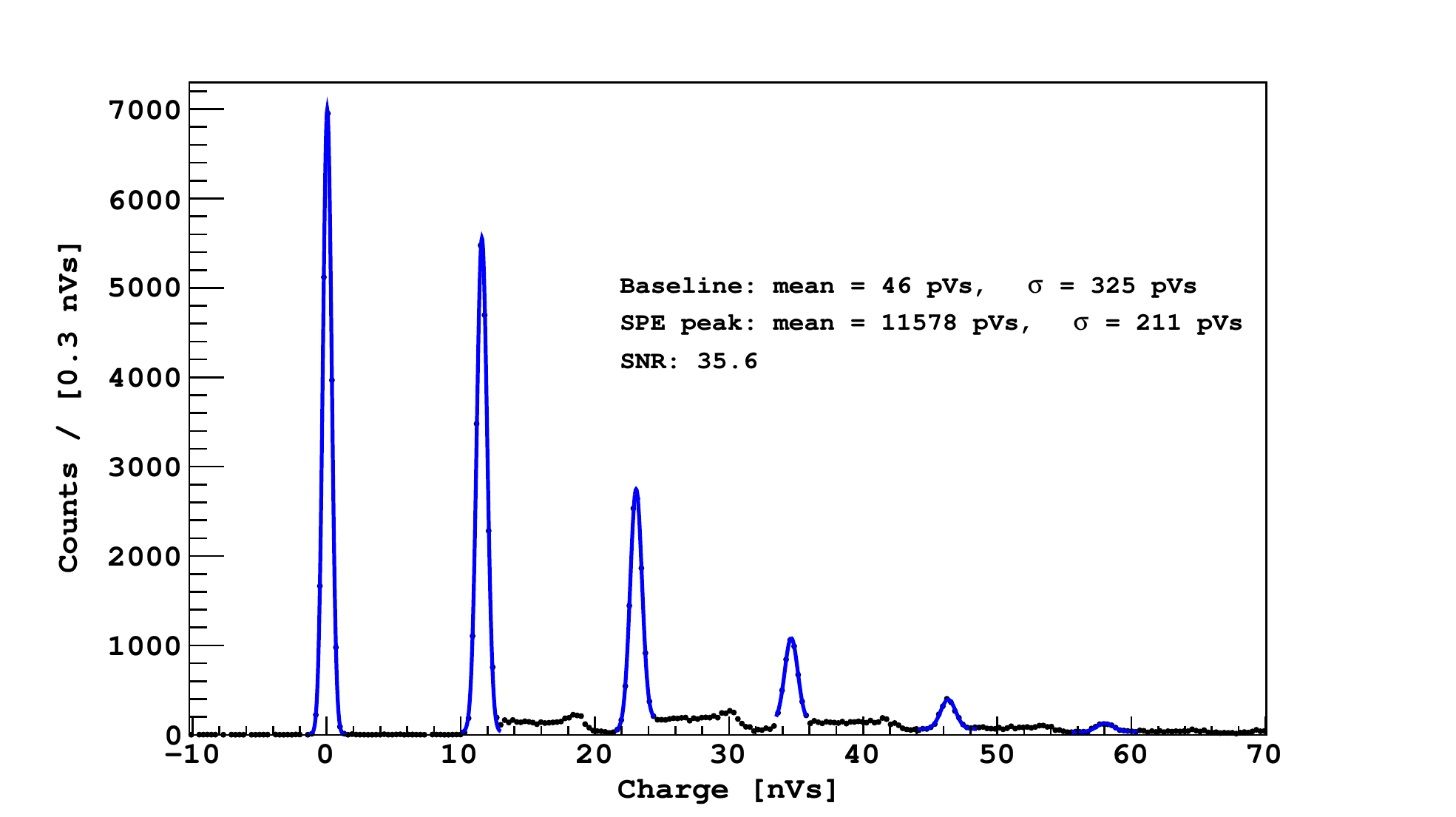}
\includegraphics[width=\columnwidth]{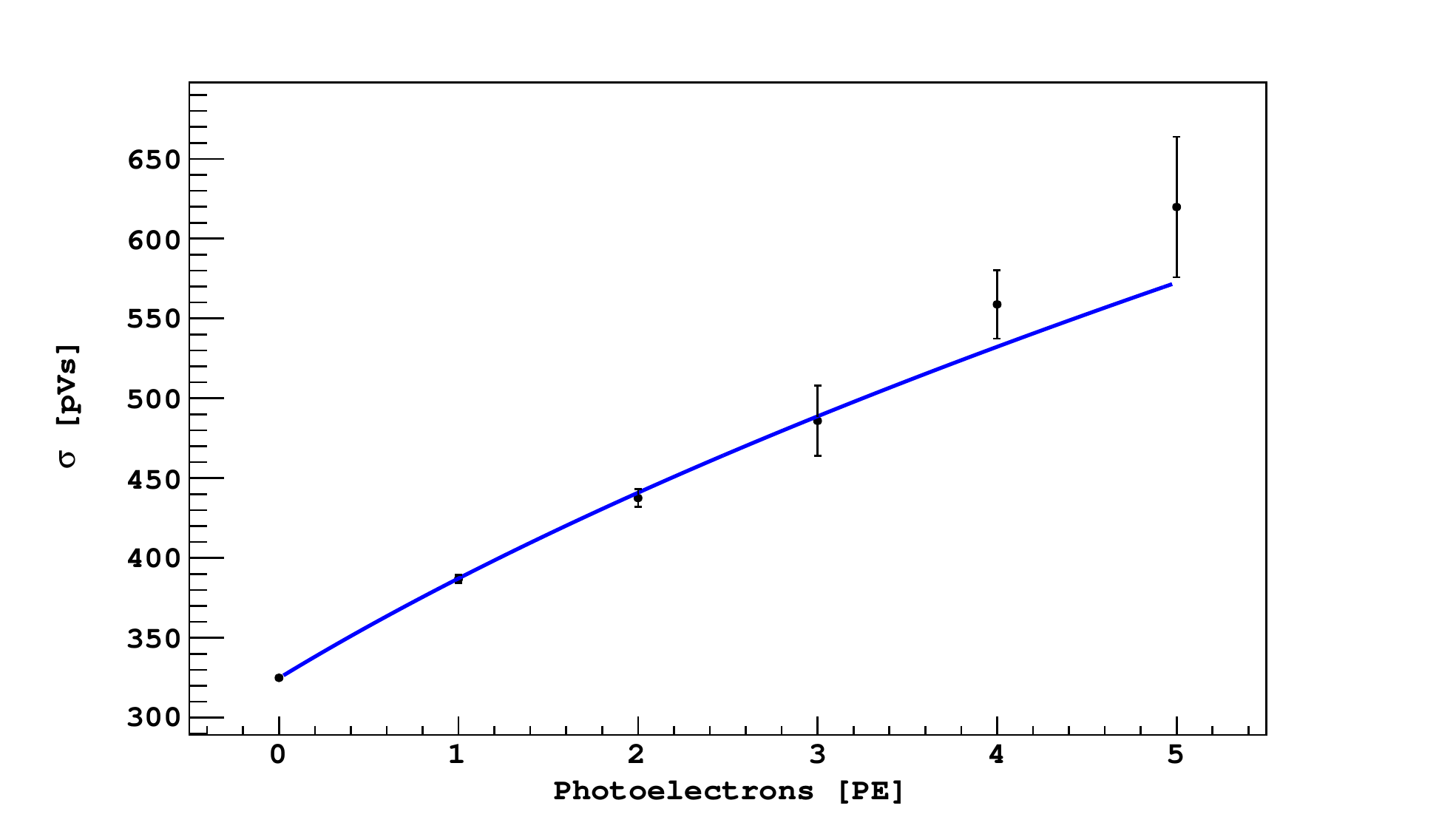}
\caption[Photoelectron peaks charge spectrum for \NUVHdLfHRq\ \SiPMs\ of \DSkSiPMAreaMax\ area.]{Charge spectrum {\bf (top)} and distribution of the standard deviations {\bf (bottom)} of the photoelectron peaks for \NUVHdLfHRq\ \SiPMs\ of \DSkSiPMAreaMax\ area.}
\label{fig:NUVHdLfHRqMax-1500nsCharge}
\end{figure*}

The first step to the measurement of the \SNR\ expected in \DSk\ consisted of testing single \SiPM\ devices connected to the \TIA, before performing the measurement with the \SiPM\ tiles.  \NUVHdLfHRq\ \SiPMs\ of area \DSkSiPMAreaMax\ were studied, results of the measurements are shown in Fig.~\ref{fig:NUVHdLfHRqMax-Ampl}, Fig.~\ref{fig:NUVHdLfHRqMax-30nsCharge}, and Fig.~\ref{fig:NUVHdLfHRqMax-1500nsCharge}.  The measurements of \DSkSiPMAreaMax\ \NUVHdLfHRq\ \SiPMs\ was performed with a \TIA\ specifically tuned to obtain the maximum \SNR\ with the device under test, allowing to reach a very high level of \SNR.  Two general outcomes of the measurement were:
\begin{asparaenum}
\item The fit to the standard deviations of the distribution of the photoelectron peaks is well described by the model introduced in the previous section;
\item The intrinsic dispersion of the \SPE\ peak ($\SigmaSPEt / \SPE$) is different between amplitude and charge but is quite compatible between the families: for charge it is approximately \FBKSiPMChargeDispersion\ while for amplitude is approximately \FBKSiPMAmplDispersion.
\end{asparaenum}



\subsubsection{Optimal \SiPMs\ Connection Scheme}
\label{sec:PhotoElectronics-TIA-SiPMConnection}

In the most trivial readout scheme, where all the \SiPMs\ are readout in parallel, $C_d$ is simply the sum of all \SiPMs\ capacities, resulting in a degraded \SNR.  An even more important effect of the very large input capacitance is the slowing down of the amplifier response.  The bandwidth of the amplifier is proportional to \fRes, which is defined as:

\begin{equation}
\fRes = \sqrt{ \frac{GBP}{2 \pi R_f (C_d + C_f)} }.
\end{equation}

$GBP$ is the gain-bandwidth product, $C_d$ is the detector capacitance, $C_f$ is the parasitic capacitance and $R_f$ is the recharge resistance.  It is fundamental to maintain an overall bandwidth in excess of \DSkTIABandwidthForTimeResolution\ to be able to reconstruct the timing of a photon with a resolution of \DSkTIATimeResolution\ (even in condition of a low \SNR).  Moreover, to appropriately treat the fast peak produced by high $R_q$, a \SiPM\ minimum bandwidth of \DSkTIABandwidthForFastPeak\ is required.  Lowering $R_f$ is not possible, since the peaking of the noise gain would increase, much to the detriment of the \SNR.  The only option is to lower $C_d$.

\begin{figure}[!t]
\centering
\includegraphics[width=\columnwidth]{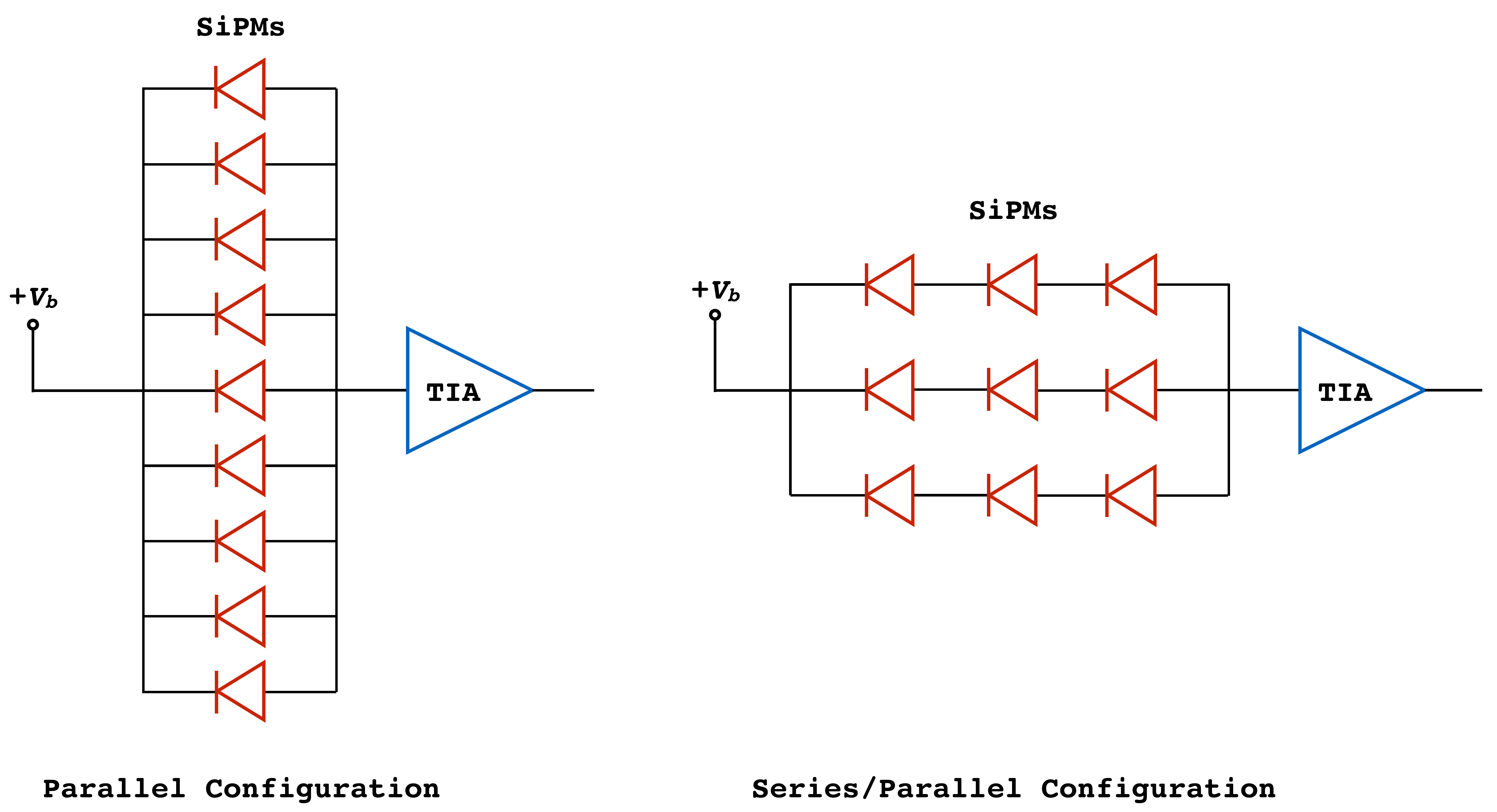}
\caption[Summed \SiPM\ readout scheme.]{Classic readout scheme with all \SiPMs\ summed in parallel (left) and hybrid scheme with \SiPMs\ summed in series-parallel (right).}
\label{fig:TIA-SiPMConnectionScheme}
\end{figure}

A possible improvement comes from the consideration of a hybrid readout scheme, with parallel-series combination of \SiPMs, as shown in Fig.~\ref{fig:TIA-SiPMConnectionScheme}, and first introduced for the MEG experiment~\cite{Ootani:2013cn,Cattaneo:2016dq}.  In this configuration, the output signal of an individual \SiPM\ is reduced by a factor equal to the number of \SiPMs\ put in series, but this disadvantage is offset by the attenuation of noise gain due to the reduction in the input capacitance (thus in this case ending in the same SNR obtainable with a single device).  Most importantly, the improvement of the bandwidth with respect to the parallel readout scheme is achieved, with nine \SiPMs\ easily readout connecting three sets of three parallel connected \SiPMs\ in series (know as the 3p3s configuration) with the same bandwidth of a single \SiPM\ in input.

\subsubsection{\NUVHd\ \tile}
\label{sec:PhotoElectronics-TIA-NUVHdLfHRqStd4p4sSNR}

\begin{figure*}[!t]
\includegraphics[width=\columnwidth]{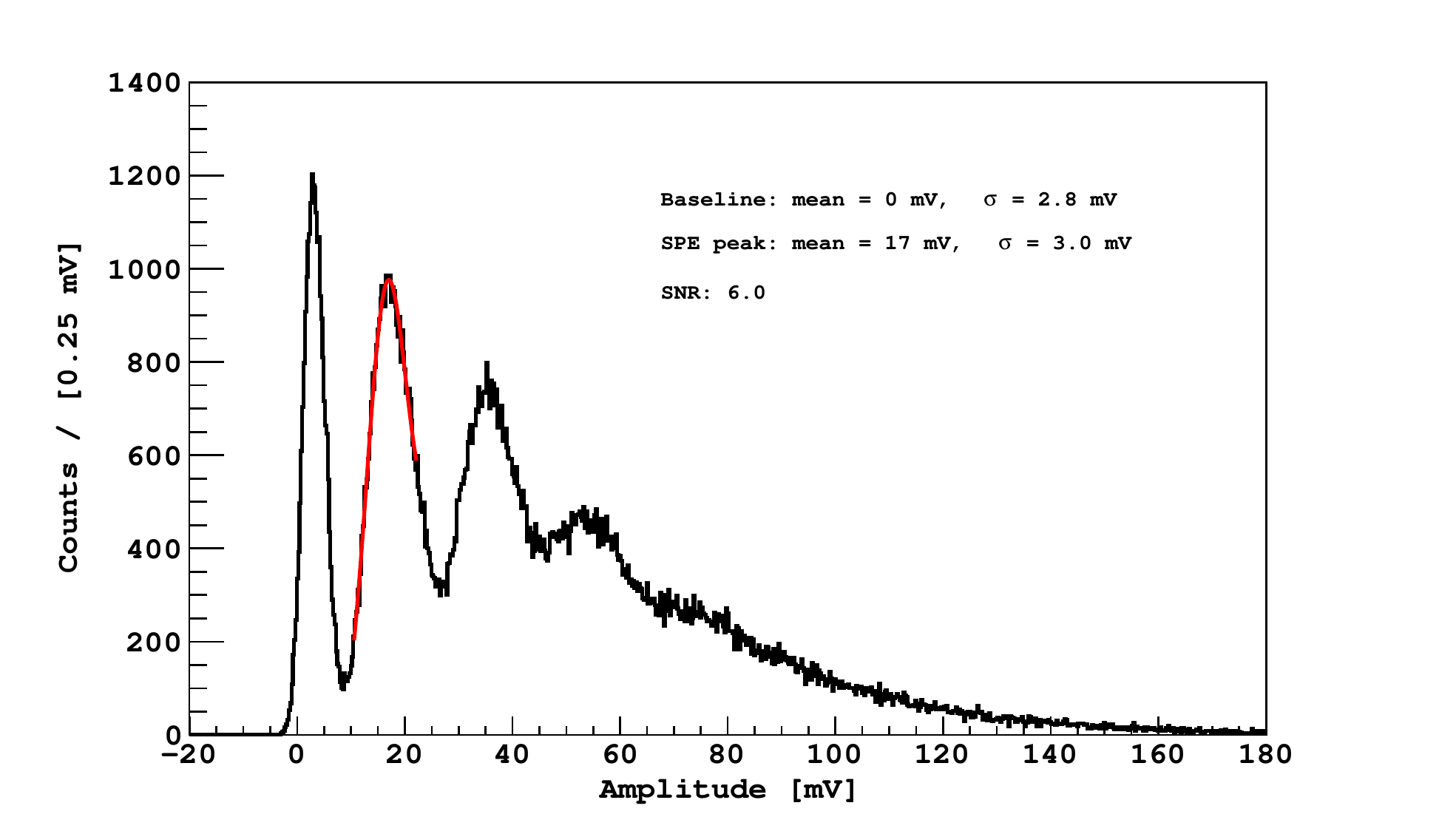}
\includegraphics[width=\columnwidth]{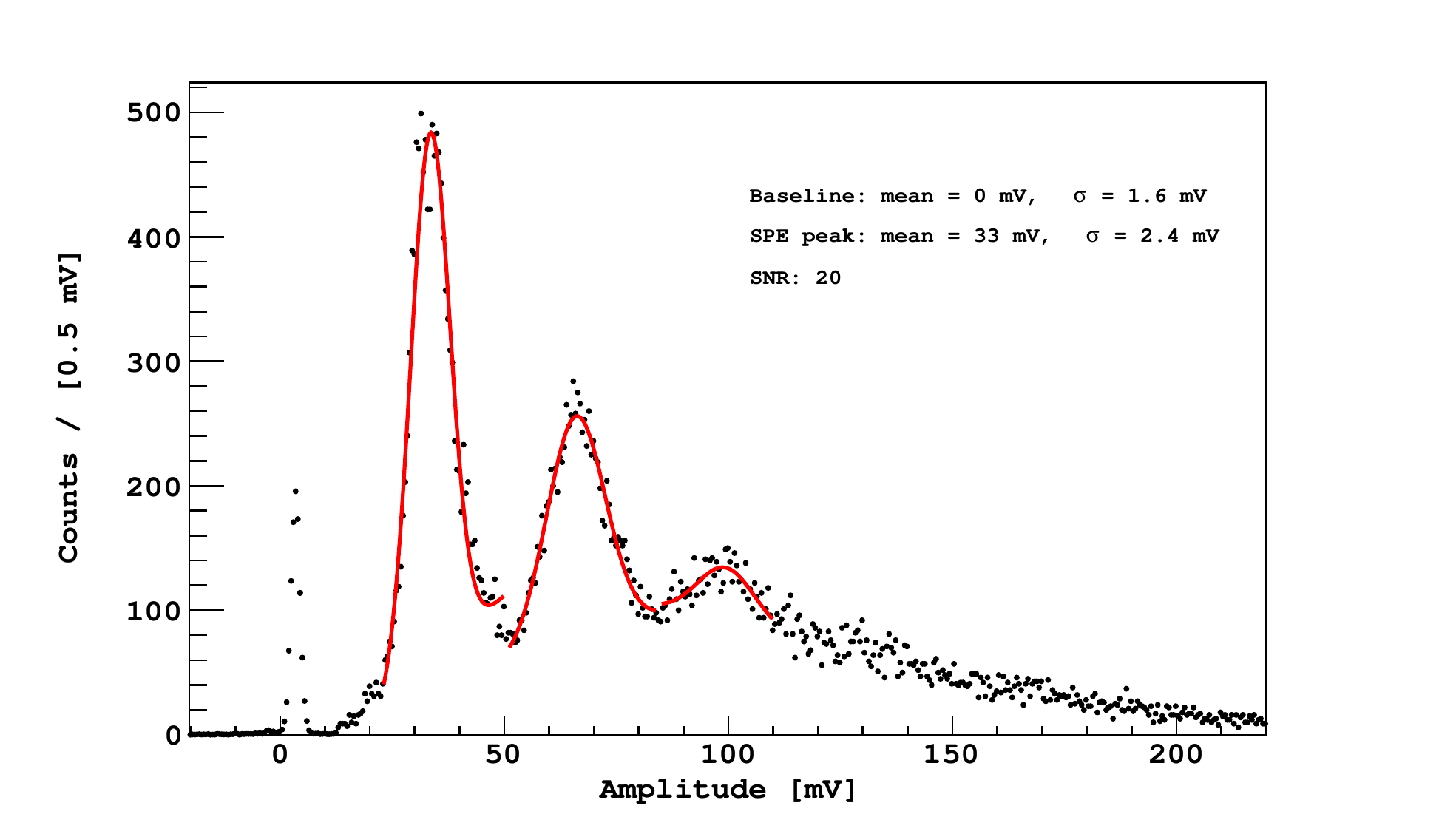}
\caption[Amplitude and filtered spectra for a tile \NUVHdLfHRq\ \SiPMs.]{Amplitude spectrum {\bf (top)} and (almost) optimal filtered spectrum {\bf (bottom)} for a tile of \NUVHdLfHRq\ \SiPMs, each of area \DSkSiPMAreaStd, arranged in a 4p4s configuration to cover a total area of \NUVHdLfHRqStdFourPFourSArea.}
\label{fig:NUVHdLfHRqStd4p4s}
\end{figure*}

For the next step, the measurement of \SNR\ for a tile of \NUVHdLfHRq\ \SiPMs\ was performed, each of area \DSkSiPMAreaStd, arranged in a 4p4s configuration (similar to the 3p3s configuration, but now with four sets of four parallel connected \SiPMs\ connected in series) and covering a total area of \NUVHdLfHRqStdFourPFourSArea.  For this tile, all the \SNR\ types were measured, see Fig.~\ref{fig:NUVHdLfHRqStd4p4s} and Table~\ref{tab:Photoelectronics-SNR}: in all cases the spectrum of the single photo-electrons remain clearly visible.  However, the noise in the measurement of the amplitude is excessive, leading to problems in the realization of larger tiles.  This is discussed in the next section.

\subsection{Noise-driven Limits on Tile Surface Scaling}
\label{sec:PhotoElectronics-TileScaling}

\subsubsection{Hit Identification Theory}
\label{sec:PhotoElectronics-TileScaling-HitIdentificationTheory}

As explained in the previous sections, electronic noise poses serious limitations to the design of an operating \SiPM\ tile.  The identification of the \SPE\ in the signal acquired from a \tile\ is of primary importance for the trigger algorithm, as well as for the downstream analysis.  A hit is typically identified when it crosses a threshold: different noise conditions can affect the number of noise hits crossing the threshold or affect the efficiency of the algorithm to identify a real photoelectron.

To estimate the rate of fake hits injected in the system due to baseline noise, it is necessary to introduce a noise model.  A white band-limited noise was assumed, with baseline RMS of \SigmaBaseline\ and an upper cutoff frequency ($f_{up}$), given by the pre-amplifier or by additional analog or digital filters.  In this condition, the expected rate of threshold-crossings of the threshold \VThreshold\ above baseline is~\cite{Rice:1944cm}:

\begin{equation}
R = \frac{f_{up}}{\sqrt{3}} \exp{\left( -\frac{1}{2} \left( \frac{\VThreshold}{\SigmaBaseline}\right)^2 \right)}.
\end{equation}

The inefficiency of the threshold is defined as the number of signals with peak value between \num{0} and the threshold of the gaussian best fitting the \SPE\ peak.  This inefficiency affects the overall light yield.

\begin{figure*}[!t]
\includegraphics[width=\columnwidth]{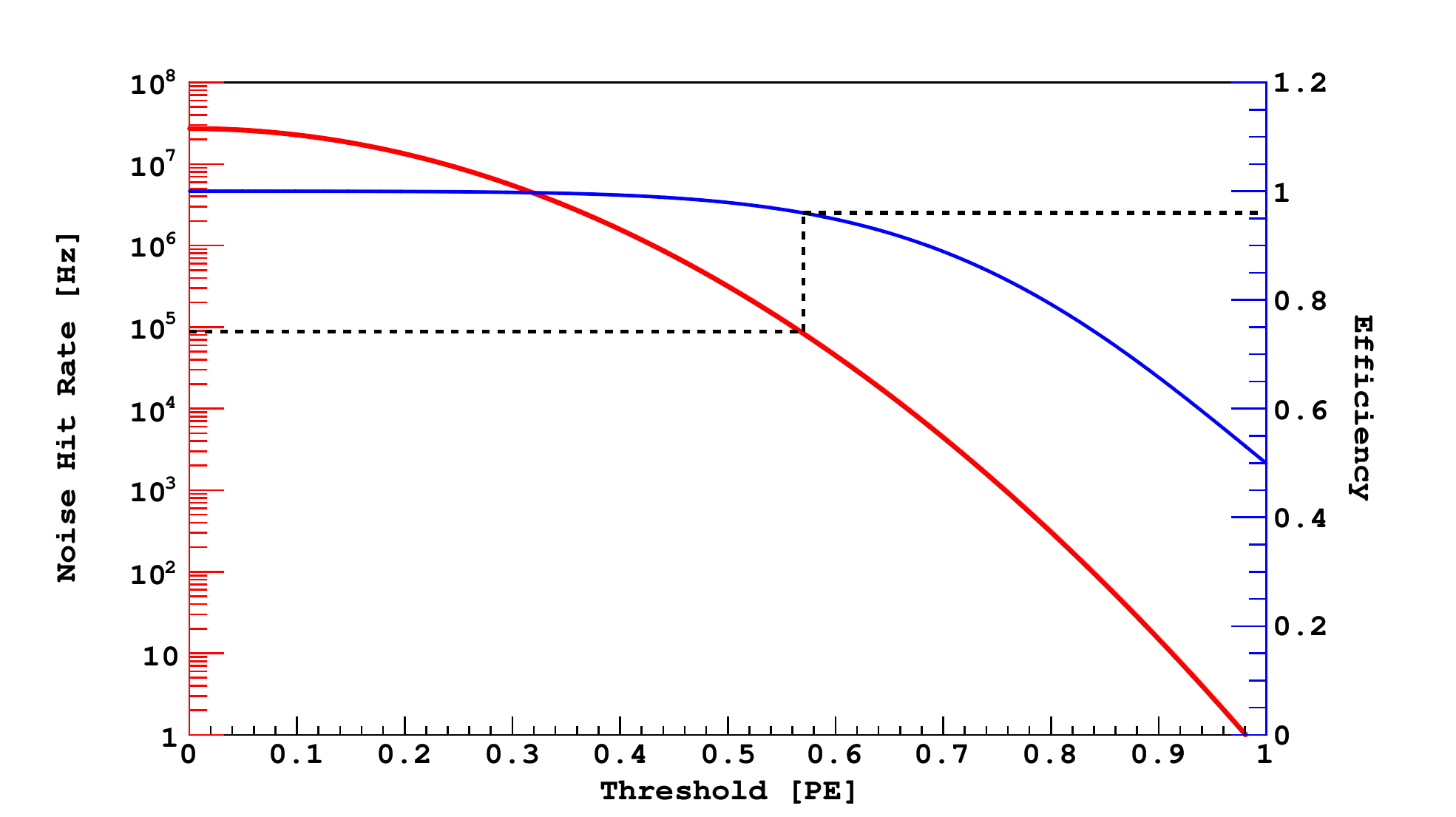}
\includegraphics[width=\columnwidth]{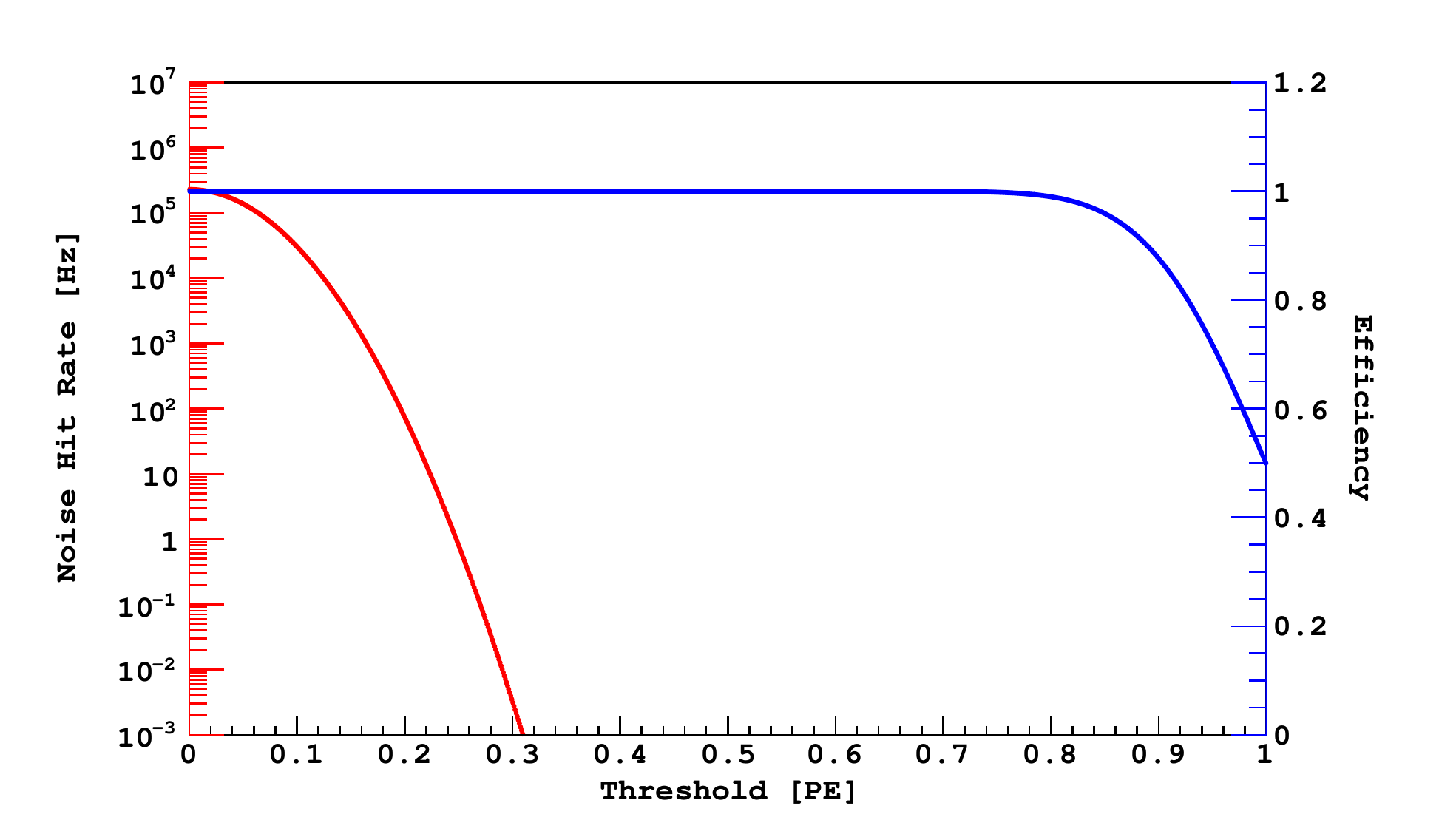}
\caption[\NUVHdLfHRq\ signal discrimination.]{{\bf Top:} rate of fake hits above the specified threshold due to electronic noise (in red) and threshold efficiency for real photoelectrons (in blue) for the signal from the TIA + \NUVHdLfHRq\ tile of total area of \NUVHdLfHRqStdFourPFourSArea, corresponding to the left panel of Fig.~\ref{fig:NUVHdLfHRqStd4p4s}.  The dashed line corresponds to a threshold efficiency of \SI{96}{\percent} with a fake hit noise of \SI{100}{kHz}. {\bf Bottom:} same plot with (almost) optimal filter, corresponding to the right panel of Fig.~\ref{fig:NUVHdLfHRqStd4p4s}.}
\label{fig:SNR-TileSignalEfficiency}
\end{figure*}

The noise hit and the threshold efficiency are shown as a function of the threshold level in Fig.~\ref{fig:SNR-TileSignalEfficiency} for the signal from the \NUVHdLf\ tile described earlier: the plot on the top panel corresponds to the bare signal (relative than to \SNRAmp\ measurement) while the plot on the bottom panel corresponds to the signal filtered by a low pass $RC$ filter with $\tau = \DSkMatchedRCFilterTau$.  As seen in the figure, one can achieve \SI{95}{\percent} efficiency while keeping a \SI{100}{\kilo\hertz} frequency tuning range, as well as a great improvement of performance with the addition of the filter.

\subsubsection{Hit Identification Experimentation}
\label{sec:PhotoElectronics-TileScaling-HitIdentificationExperimentation}

To verify the model, the darknoise of the \NUVHdLfHRq\ tile of total area of \NUVHdLfHRqStdFourPFourSArea\ was measured in liquid nitrogen.  The setup included the full \tile\ + \TIA, an external amplifier configured for optimal filter and a readout based on the \LNGSCryoSetupDigitizerModel.  The threshold was set to \SI{0.4}{\pe}.  The resulting rate was \SI{5.1(1)}{\ct\per\second}, corresponding to \SI{13}{\milli\hertz\per\square\mm} at \SI{6}{V} of \OV.  The measurement is compatible with the value measured for smaller device and proves that the fake hits can be significantly rejected (rate is much lower than the darknoise requirement of \DSkSiPMDCRSpecification).

\subsubsection{Surface Size for \NUVHdLfHRq\ \tiles}
\label{sec:PhotoElectronics-TileScaling-ExtrapolatedSurface}

The reason for the limited \SNRAmp\ of the \NUVHd\ tile is related to the intrinsic limitation of the TIA.  The total charge of the fast peak is about one tenth of the full gain of the SiPM, this means a charge of about \SI{2E5}{\electron}.  Considering the duration of the fast peak (about \NUVHdSfHRqPeakDuration\ for \NUVHdLfHRq\ devices with $R_s = \SI{60}{Ohm}$), the peak current results in around \SI{3}{\micro\ampere} with a bandwidth of about \SI{30}{\mega\hertz}.  Different values of $R_s$ can lower the bandwidth at a detriment to the peak current (the integral of the pulse remains constant).  Moreover, given the 4p4s configuration, the effective signal arriving to the pre-amplifier is reduced by a factor \num{4}, leading to a detectable peak current of \SI{0.75}{\micro\ampere}.

With a surface of \DSkSiPMAreaStd\ in the 4p4s configuration, and a $R_f$ of \SI{4}{\kilo\ohm}, the bandwidth amounts to \SI{30}{\mega\hertz} (in fully parallel configuration would instead be \SI{8}{\mega\hertz}).  The noise of the \TIA\ can be calculated from the formulas given in~\cite{Graeme:1996vs} or simulated.  In this configuration, the input equivalent noise of the \TIA\ is around \SI{100}{\nano\ampere}, corresponding to a $\SNRAmp = \SI{0.75}{\micro\ampere} / \SI{100}{\nano\ampere} = \num{7.5}$ (to be compared with the measured value of \num{6.0}, see Table~\ref{tab:Photoelectronics-SNR}).

A further optimization may be possible, for a gain in \SNRAmp\ of a factor of about \num{2}, but pushing the performance of the \TIA\ much further does not look feasible given the current technologies.  Since measurements have shown that \SNR\ should scale with the square root of surface area, a different \SiPM\ configuration must be used for \tiles\ built from the current \NUVHdLfHRq\ \SiPMs\ to extend the total area to \DSkTileAreaStd.  Preliminary test results show that this can achieve the required signal-to-noise ratio for a single \DSkPdm.  Improvement to the \NUVHdLf\ \SiPMs\ in order to achieve more stable operation at \LArNormalTemperature\ with lower $R_q$ is foreseen.

\subsubsection{Strategy for Tile Readout in \DSk}
\label{sec:PhotoElectronics-TileBaseline}

\begin{figure}[t!]
\centering
\includegraphics[width=\columnwidth]{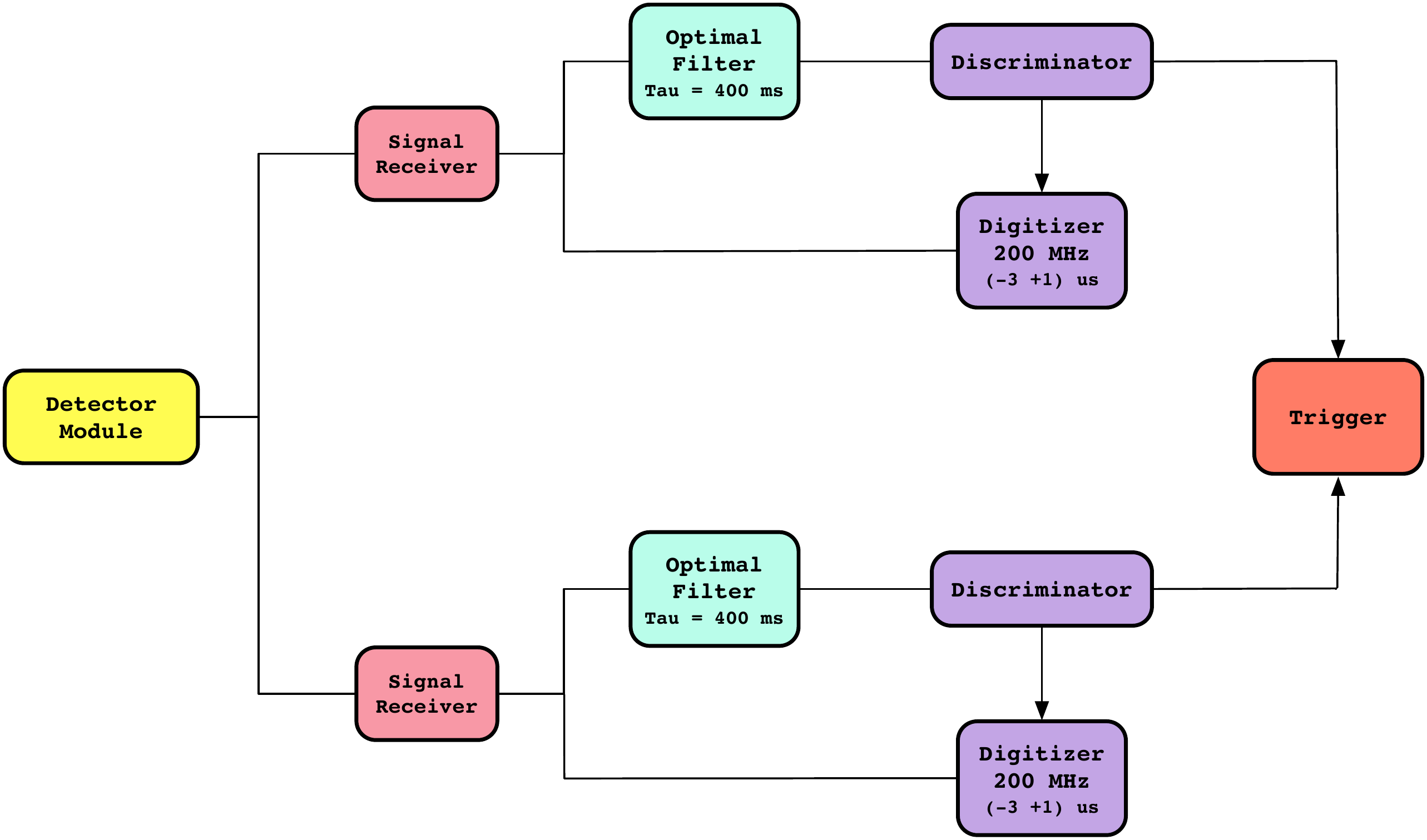}
\caption{Proposed strategy for signal trigger and discrimination.}
\label{fig:DAQ-DiscriminationTrigger}
\end{figure}

Considering the arguments discussed above about the hit rate at different \SNR, the readout requires a cautious design to maintain the effectiveness of the trigger and the downstream analysis.  Considering a photodetector module (\DSkPdm) with \num{2} analog channels each and a \DSkTileAreaStd\ \NUVHdLfHRq\ tile, a direct discrimination of the signal is not possible (as shown in the top panel of Fig.~\ref{fig:SNR-TileSignalEfficiency}, the rate would be excessive).  Therefore, discrimination will be made on the (optimal) filtered signal, while timing of the hit is done using the unfiltered signal.  As shown in the bottom panel of Fig.~\ref{fig:SNR-TileSignalEfficiency}, and as already experimentally proven, see discussion in Sec.~\ref{sec:PhotoElectronics-TileScaling-HitIdentificationExperimentation}, the hit discrimination of filtered signals guarantees a very strong rejection of uncorrelated electronic noise with a very high detection efficiency for real photoelectrons.  The readout design, as shown in Fig.~\ref{fig:DAQ-DiscriminationTrigger}, acquires the unfiltered signal in a window of \SI{4}{\micro\second} around the hit and produces no noise contribution from the baseline.  In fact, as shown in the top panel of Fig.~\ref{fig:SNR-TileSignalEfficiency}, a \SI{96}{\percent} detection efficiency corresponds to about \SI{100}{kHz} and would end up in a probability of \SI{5}{\percent} for one fake hit in the specified window.  Additionally, a more refined analysis like a likelihood based one can reduce this term further.  The identification of the hit in the unfiltered signal allows for the precise timing (better than \DSkTileTimeResolutionSpecification) of the photoelectron, as required for effective pulse shape discrimination.  This scheme has been implemented and in~\cite{DIncecco:2017td} the first prototype photodetector module meeting all of the \DSk\ requirements has now been demonstrated.
 
\subsection{\SiPM\ Packaging}
\label{sec:PhotoElectronics-Packaging}

Individual \SiPMs\ will be mounted on \DSkTileAreaStd\ patterned substrates to form detector \tiles.  A detector \tile\ provides bias distribution, a signal summing resistor network, and a low-mass connector for mating the tile to a front-end board (\FEB).  The substrate material will be selected for its compatibility with conventional lithography techniques, its mechanical robustness, and the availability of radio-pure wafers.  

\begin{figure}[t!]
\centering
\includegraphics[width=0.45\columnwidth]{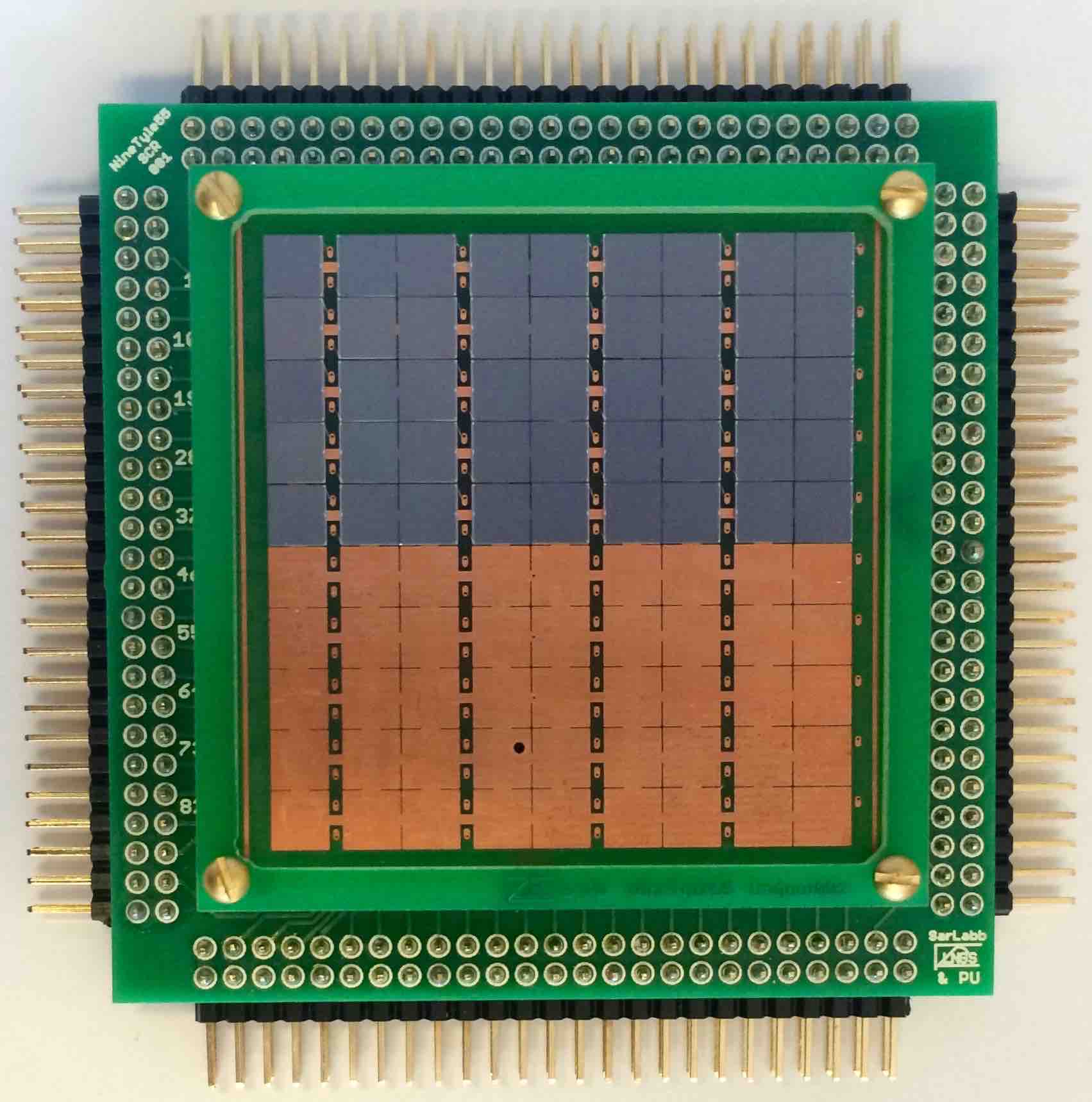}
\caption{Prototype \SiPM\ \tile\ on an FR-4 substrate.}
\label{fig:Packaging-Tile}
\end{figure}

A set of prototype \SiPM\ \tiles\ have been assembled on FR4 substrates, an example of which is shown in Fig.~\ref{fig:Packaging-Tile}.  These prototype \tiles\ route the anode and cathode of each \SiPM\ to a high density connector on the back-side of the board that mates to a carrier \PCB.  This flexible design allows the tile to be used to measure the cryogenic performance of individual \SiPMs, as well as to test different \SiPM\ summing schemes described in this section.  \SiPM\ dies are mounted on the prototype \tile\ using cryogenic silver loaded epoxy that has been tested for mechanical and electrical stability after thermal cycling.  The epoxy is screened onto the tile in a \DSkTileEpoxyThickness\ thick layer, creating a coplanar surface on which the \SiPM\ dies are placed with a manual die bonder.  An alternative to cryogenic epoxy is being developed in which a grid of \DSkTileGoldBallsDiameter\ gold balls are placed on the back of the \SiPM\ die, forming a bump-bond style surface-mount package.  Currently, the top-side \SiPM\ contact is made with gold ball wire-bonds.  Eventually, the production of \SiPM\ dies will move to an external foundry.  The next generation will use through silicon vias (\TSV) into the SiPM design, moving both contacts to the back-side of the die and eliminating the need for wire-bonding in lieu of an all bump-bonded interface.  The $IV$ curve of each SiPM on the tile is measured before and after the SiPM die is mounted, and a report is generated for each tile that details the properties of the tile, the tile assembly procedure, and individual \SiPM\ performance.  This information will eventually be included in a materials and parts tracking database currently under development.

The transition from conventional \PCB\ substrates to alternative substrates is currently underway.  The \tile\ will be double-sided, requiring through hole vias, and will have a network of thin film summing resistors deposited directly on the substrates.  Studies are also being done on the fabrication of thin film nickel-chromium summing resistors, focused on optimizing the resistor geometry and the nickel-chromium deposition parameters.

\subsection{The \DSk\ Photodetector Modules}
\label{sec:PhotoElectronics-PhotodetectorModule}

The photodetector module (\DSkPdm) will be the basic unit for photon detection in the \LArTPC.  The \DSkPdm\ is conceived to be equipped with a \tile\ array of \SiPMs\ covering a nominal surface area of \DSkTileAreaStd, along with a set of cryogenic \TIAs.  The \TIAs\ will receive and shape the signals from the \DSkTileAreaStd\ \tiles, then transmit them over dedicated transmission lines through the cryostat penetration to be received by the warm electronic modules residing outside the \LArTPC.

The top and bottom surfaces of the \LArTPC\ will be covered by \DSkPdms\ grouped in arrays of \DSkSQBDModArray\ and mounted on a common motherboard providing a convenient mechanical, electrical and thermal interface among the \LArTPC\ and the \DSkPdms.  The mechanical structure of the \DSkPdm\ and its assembly procedures must be as light and simple as possible, both to reduce the amount of dead material in the \LArTPC\ and to ease the production of the large number of \DSkPdms\ needed for the experiment.

The \DSkPdm\ will provide a reliable mechanical frame to connect the \tiles\ and the \FEBs. The structure will be able to withstand the thermal cycles from warm to cryogenic temperature envisaged for the testing, characterization and running phases of the experiment.  The \DSkPdm\ will also provide an adequate thermal path from the \TIA\ active components, whose thermal dissipation to the motherboard is foreseen to be below \DSkTilePower\ per \tile, in order to mitigate bubbling in the \LArTPC\ as much as possible.

\begin{figure*}
\centering
\includegraphics[width=\textwidth]{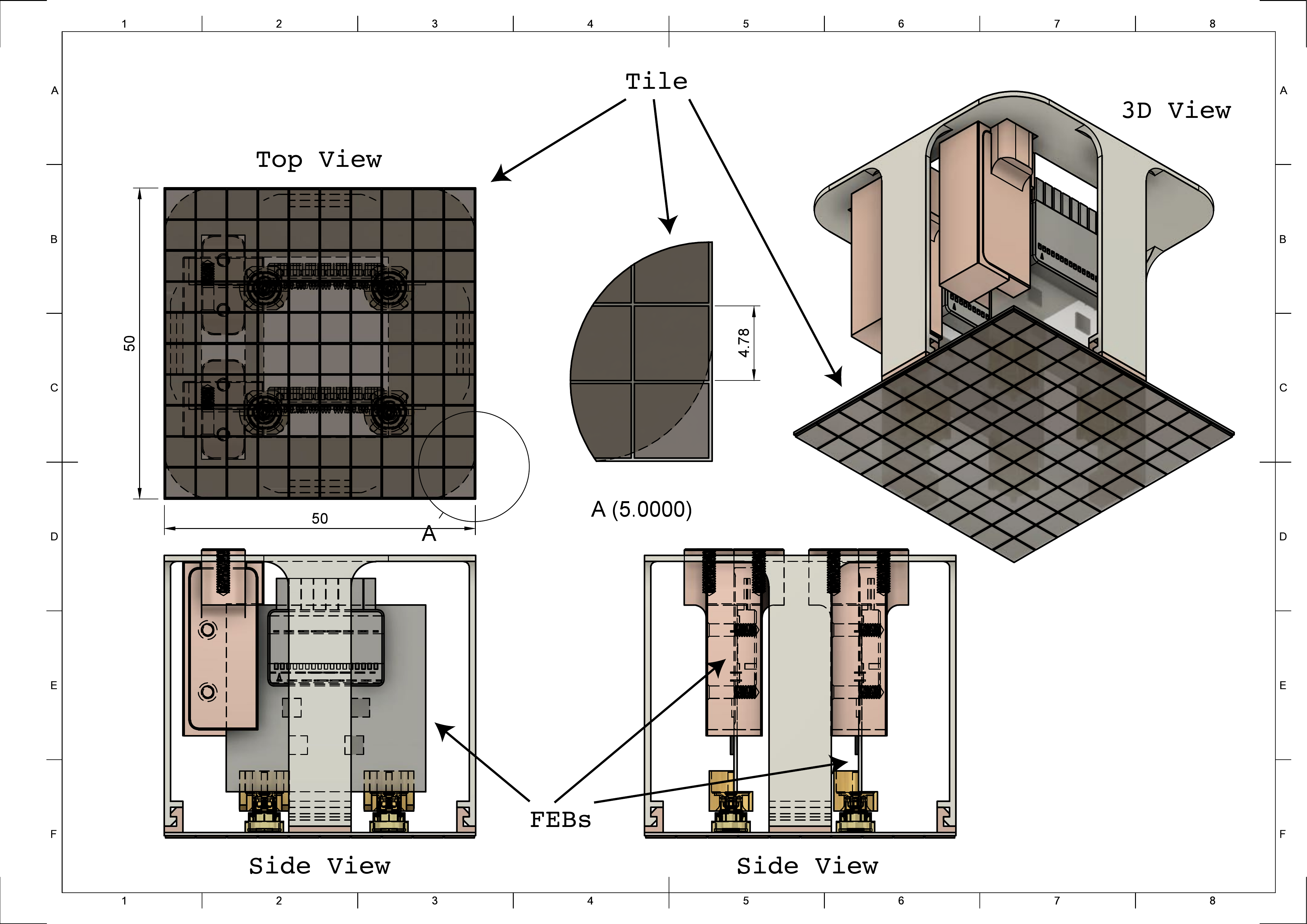}
\caption[Conceptual sketch of the \DSkPdm\ assembly.]{Conceptual sketch of the \DSkPdm\ assembly.  In this specific design, the \DSkTileAreaStd\ \tile\ is equipped with \num{2} \FEBs.}
\label{fig:PhotodetectorModule-Concept}
\end{figure*}

The current \DSkPdm\ assembly design is the one depicted in Fig.~\ref{fig:PhotodetectorModule-Concept}.  In the specific design presented in this figure, the \DSkTileAreaStd\ \tile\ is built from an array of \num[product-units=power]{5 x 5} \SiPMs.  The back side of the \tile\ hosts the electrical connectors to the \FEBs\ and four soldered profiles suitable for a snap-fit joint with the plastic cage devoted to latching the tile to the \FEBs.  The thermal contact is loose in order not to transfer heat from the \FEBs\ to the \SiPMs\ and to maintain their working temperature at \LArNormalTemperature.  The snap-fit profile material will be preferably OFHC copper because of its radiopurity and its good soldering properties.  The connectors will have to mate with the \FEB\ providing some axial float (roughly \SI{1}{\milli\meter}) to accommodate for the mechanical misalignment of the \TIA\ components and some radial float (roughly \SI{0.5}{\milli\meter}) to accommodate for the mismatch in the  thermal dilatation of different materials. 

Readily available connectors from reliable suppliers are not specified for operations in \LAr, hence extensive tests at cryogenic temperature are needed to identify a good working candidate.  Radiopurity qualification and detailed \GFDS\ studies are necessary to validate the impact of these components on the total background rate.

The \FEBs\ will be provided with mating connectors that attach to the \SiPM\ tile, to the power supplies (low voltages for the \FEBs\ and bias voltage for the \SiPM), and to the transmission lines for communicating with the DAQ electronic residing outside the \LArTPC.  In summary, each \FEB\ will contain:
\begin{inparadesc}
\item common ground;
\item two rails for the low voltage power supplies for the \TIA\ and transmitter;
\item a bias voltage for the \SiPM;
\item the output signal, a twisted pair cable or an optical fiber.
\end{inparadesc}

\subsubsection{\DSkPdm\ Thermal Model}
\label{sec:PhotoElectronics-PhotodetectorModule-ThermalModel}

Precise control of the temperature of the components of the \DSkPdms\ is of paramount importance.  The \SiPM\ performance strongly depends on the temperature, and hence a sizable effort is devoted to the thermal management design of the \DSkPdm.  As stated in Sec.~\ref{sec:Cryogenics-AdditionalConsiderations}, a heat load greater than \DSkCryostatBottomLArNoBubblingHeatDissipation\ will cause the onset of bubbling in \LAr\ contained in the bottom of the cryostat.  In this respect, the heat load from the \FEBs\ is not a big concern, provided that the \DSkPdm\ is able to evenly distribute the heat dissipated by its electronic components to the whole supporting motherboard.  A total power dissipation of about \DSkTIAPower\ for a first \FEB\ prototype assembled on FR4 was measured.  The worst conditions will be presumably the ones experienced by the top readout plane, where the pressure is limited due to the smaller \LAr\ depth, however, bubbling for the top array is not a problem, since any bubbles produced there will be on the top side of the anode window and hence will not enter the \TPC.

The heat is mainly dissipated by the operational amplifier whose surface area is only \DSkTIAOpAmpSurface.  Good thermal contact between the operational amplifier and the tile substrate is thus of paramount importance in order to prevent boiling.  To further improve the thermal exchange efficiency to the \LAr\ and reduce the risk of bubbling and thermal run-away, the \TIA\ substrate has a copper bridge to the motherboard that provides an additional thermal exchange surface area with the \LAr. 

\begin{figure}[t!]
\includegraphics[width=\columnwidth]{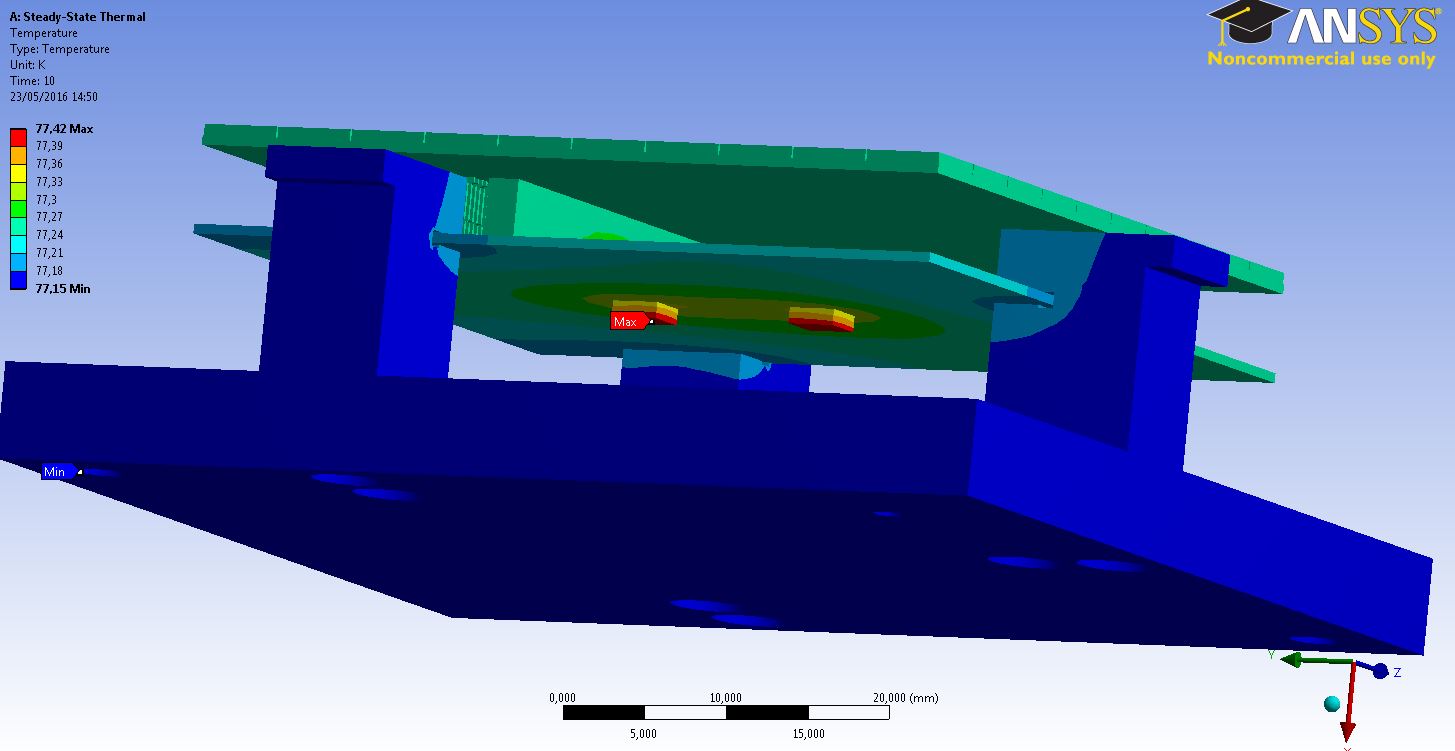}
\caption[ANSYS thermal steady state simulation of the \TIA-\SiPM\ mock-up assembly to be tested.]{ANSYS thermal steady state simulation of the \TIA-\SiPM\ mock-up assembly to be tested.  Boundary conditions are \SI{100}{\milli\watt} dissipated by each of the two active components (maximum temperature \SI{77.42}{\kelvin}) and a copper base.  In blue, the minimum temperature \SI{77.15}{\kelvin} is fixed at the normal \LIN\ boiling point, no heat transport across the surfaces of the tiles through convection or conduction in the surrounding medium is considered.  Tabulated data is used for sapphire and OFHC copper.  The device is mounted on a copper support, a picture of which is shown in Fig.~\ref{fig:PhotodetectorModule-MountingProtocol}.} 
\label{fig:PhotodetectorModule-ThermalSimulation}
\end{figure}

A detailed thermal simulation using the ANSYS FEM suite has begun to be implemented, in order to predict the device temperature reached under realistic running conditions and the resulting stresses due to combination of different materials. A preliminary result is reported in Fig.~\ref{fig:PhotodetectorModule-ThermalSimulation}, obtained in the worst case scenario in which thermal exchange by conduction and convection through the surrounding medium is neglected.  A working point of \LINNormalTemperature\ was chosen, in order to be able to easily replicate this result with a small laboratory setup.

In the ANSYS simulation, the heat transfer is only through the substrate/indium/copper joints and the boundary condition is set on the copper plate that is kept at \LIN\ temperature.  The electrical connections between the preamplifier and the \SiPM\ \tile\ were modeled as small copper prisms with an effective thermal conductance tunable by changing the contact surface. 

The results in Fig.~\ref{fig:PhotodetectorModule-ThermalSimulation} show a maximum $\Delta T \approx \SI{0.3}{\kelvin}$.  Considering the relation between temperature and vapor pressure of \ce{Ar}, and the hydrostatic pressure of the overlying \LAr, one finds a maximum allowable temperature difference of about \SI{2}{\kelvin}, which is well above the result obtained from the preliminary simulation.

\begin{figure}[t!]
\centering
\includegraphics[width=0.45\columnwidth]{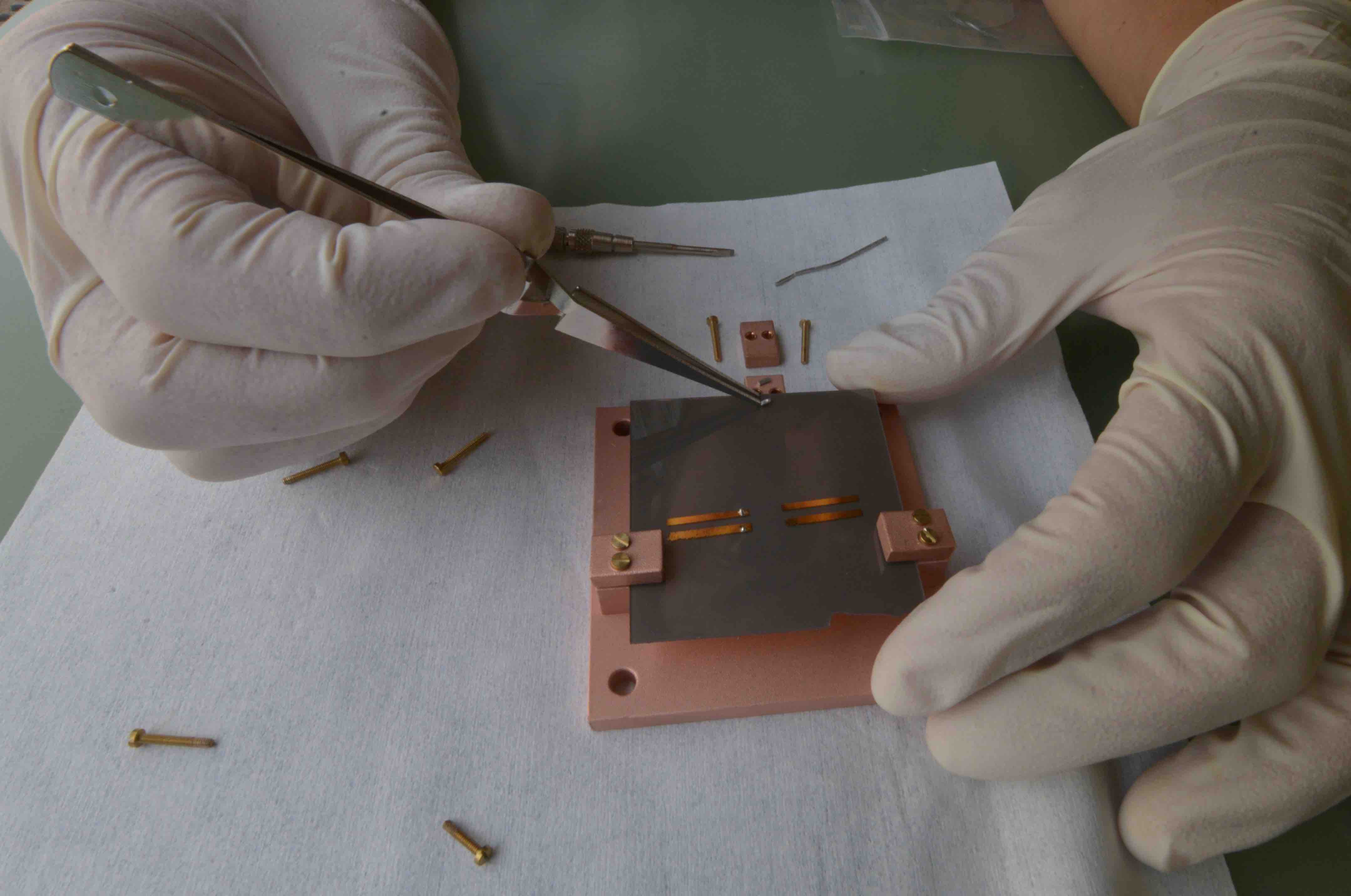}
\caption[Sapphire substrate being mounted on copper base to measure the temperature distribution.]{A sapphire substrate is being mounted on a copper base to measure the temperature distribution in an environment close to the final one.  Indium pellets are placed on top of the thermal joint and squeezed on top of the sapphire by the small copper vice.  The metallic pads are used to put a couple of PT100 sensors (one used as heater and one as thermometer) in close thermal contact with the sapphire itself.}
\label{fig:PhotodetectorModule-MountingProtocol}
\end{figure}

In order to validate the results from the simulation, a small setup was prepared in which a sapphire \tile\ of size \DSkTileAreaStd\ is equipped with resistors to simulate the thermal load and sensors to measure the temperature distribution (see Fig.~\ref{fig:PhotodetectorModule-MountingProtocol}).  PT100 RTDs were used as both heater and temperature sensors.  The heater PT100 is biased with a high current in order to dissipate \SI{100}{\milli\watt}.  In this way one could measure the self-heating of the heat load, which can be scaled to the active component.  Furthermore, the PT100 RTDs have the advantage of providing a highly linear $R(T)$ relationship that can be used to obtain a precision on the order of \SI{0.1}{\kelvin}.  In order to suppress the heat exchange through convection and conduction in the surrounding medium, a thermally insulating cover made of plastic foam was applied.  The sapphire tile is clamped in the OFHC structure made from a plate with three pillars whose base is kept in liquid nitrogen.  Each pillar has a \DSkPdmIndiumPelletVolume\ indium pellet squeezed between the copper and sapphire in order to have a thermal conductance as close as possible to the final one.  Preliminary results from the setup are in good agreement with the results of simulation.

\subsection{Motherboards}
\label{sec:PhotoElectronics-Motherboards}
The motherboards will be made of pure copper and will be fabricated in the format of square motherboards (\SQB s) and triangular motherboards (\TRB s).  The \SQB\ (\TRB) will host \num{25} (\num{15}) \DSkPdms, anchored by copper screws on one side and kapton strip printed circuits on the other side.  The kapton strips connect the steering module (see Sec.~\ref{sec:PhotoElectronics-PowerDistribution}) to the \DSkPdms, providing the TIA low voltages and the \SiPM\ bias voltage.  In addition, the analog differential signals from the \DSkPdms\ are routed into a single connector.

\subsection{Signal Transmission}
\label{sec:PhotoElectronics-SignalTransmission}

The signal transmission from the \DSkPdms\ to the warm electronics is of primary importance for the experiment. Given the high number of independent channels, the cables will introduce a large mass in the cryostat with the inherent problems of radio-purity and heat load.  Single ended technology over coaxial cables can provide high signal integrity and good immunity to electronic noise with a simple cryogenic transmission stage.  However, the use of traditional coaxial cables is impractical since coaxial braid may contain radioisotopes of lead and silver. To overcome these problems, two alternative solutions are being developed: a fully differential cryogenic analog transmitter and an optical analog cryogenic transmitter.

\subsubsection{Differential Transmitter}
\label{sec:PhotoElectronics-SignalTransmission-Differential}

A low power cryogenic analog differential transmitter and a matching room temperature receiver has been developed at \LNGS.  The specifications for operation at \LArNormalTemperature\ are:
\begin{asparaenum}[\bfseries {Transmitter}-i]
\item Gain of \DSkTransmitterGainVal\ with a flatness of \DSkTransmitterGainFlatness;
\item \FTDb\ of \DSkTransmitterBandwidth;
\item Input noise equivalent of \DSkTransmitterInputVoltageNoiseDensity, corresponding to \DSkTransmitterOutputNoiseDensityRMS\ of output noise RMS;
\item Output swing of \DSkTransmitterOutputSwing\ with \DSkTransmitterOutputCompression\ of compression.  This correspond to a dynamic range in excess of \DSkTransmitterDynamicRange, if the single photoelectron peak is placed at \DSkTransmitterSPEVal\ (\DSkTransmitterSNR\ times the transmitter noise).
\end{asparaenum}

A prototype transmitter was built and operated, at both \RoomTemperature\ and \LINNormalTemperature.  The power dissipation at \LINNormalTemperature\ is around \DSkTransmitterCryoPower.

Signal transmission is foreseen over twisted pair cables with a FEP (extrudable PTFE-variant) jacket, providing good radio-purity levels and low section copper conductors.  This solution provides strong immunity to ground loops and moderate immunity to noise pickup and cross-talk between channels.

\subsection{Cabling}
\label{sec:PhotoElectronics-Cabling}

The cabling system described in this section refers to signal and power cables necessary to operate the \SiPM\ sensors and the cold electronics.  These services transport:

\begin{compactitem}
\item Signal readouts from the cold electronics to the data acquisition system located outside the cryostat;
\item Control signals in and out of the cryostat;
\item Low voltage power from the remote power supplies to the sensor.
\end{compactitem}

The signal and power cable will penetrate the cryostat through a dedicated top flange, where conflat flange-mounted connectors will terminate the high-purity cryogenics cables and will connect them to standard warm cables on the outside of the detector.  The specifications discussed in this section refer to the cryogenic part of the cabling system only. From the cabling system standpoint, a large-area \tile\ is preferable to reduce the overall number of cables to be extracted from the cryostat as well as the heat transfer.

Based on the results of preliminary tests, specifications for the cryogenic cables are given as:

\begin{asparaenum}[\bfseries {Cable}-i]
\item Type: FEP shielded twisted pair;
\item Average length: \DSkCablingTwistedPairAvgLength;
\item Impedance: \DSkCablingTwistedPairImpedance;
\item Signal bandwidth (\FTDb): up to \DSkCablingTwistedPairFTDb;
\item Maximum attenuation at \DSkCablingTwistedPairFSDb: \SDb;
\item Crosstalk between two adjacent pairs: \DSkCablingTwistedPairCrossTalk;
\item Gauge: low section copper conductors, with gauge equal or less than \DSkCablingTwistedPairGauge\ to minimize heat transfer;
\item Shield: high radio-purity and low outgassing, FEP insulators acceptable;
\item Pair grouping: possible;
\item Minimum bending radius after grouping: not exceeding \DSkCablingTwistedPairMinBendingRadiusAfterGrouping.
\end{asparaenum}

Several signal cable manufacturers have been contacted and cable samples are being acquired so that cable material purity may be screened.  

\subsection{Power Distribution}
\label{sec:PhotoElectronics-PowerDistribution}

\begin{figure*}[t!]
\centering
\includegraphics[width=\textwidth]{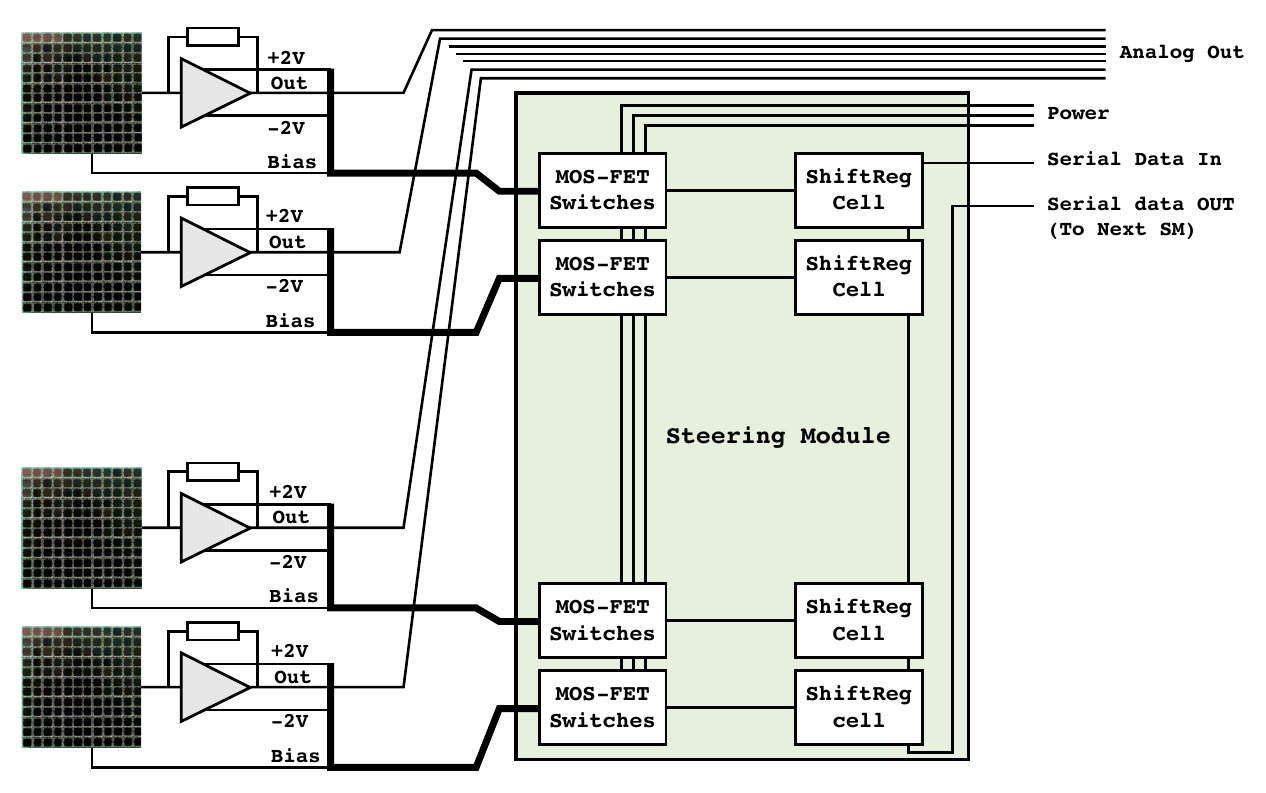}
\caption[Conceptual design of a \DSkSM\ powering a group of \DSkSMTilesNumber\ \DSkPdm.]{Conceptual design of a \DSkSM\ powering a group of \DSkSMTilesNumber\ \DSkPdm.  In \DSk, the \DSkTilesNumber\ \DSkPdms\ are managed by about \DSkSMsNumber\ \DSkSMs.}
\label{fig:SteeringModule-Concept}
\end{figure*}

The power required per \DSkPdm, up to the maximum value of \DSkTilePower\ established as a requirement, including the power requirement of the \TIA, the transmitter, and the \SiPMs, is too small to be distributed individually.  A more realistic solution is to build a steering module (\DSkSM) that acts as a power distribution network in proximity of the loads, {\it i.e.} the \DSkPdms.  So, the choice for low voltage power distribution cables presently considered for the experiment is based on the assumption that a single power channel will serve approximately \DSkSMTilesNumber\ \DSkPdms, connected to the same power supply and distributing low (\DSkSMOutputLowVoltage) and high voltage ($<$\DSkSMOutputHighVoltage) to all connected \DSkPdms.  An approximate number of \DSkSMsNumber\ units would be required for the entire detector.  This solution has the advantage of reducing the number of cables to be extracted from the cryostat and of optimizing the number of power supplies used in the experiment.  A conceptual design of this board, based on an output buffered shift register, was developed by ETHZ, see Fig.~\ref{fig:SteeringModule-Concept}.

The physical location of the \DSkSMs\ will be in the \LAr, in close vicinity to the \DSkPdms, requiring a cryogenic electronics design.  The signal output lines (copper/optical) of the \DSkPdms\ will be routed directly to feedthroughs in the periphery of the detector. \DSkPdms\ will be selected for similar breakdown voltage and associated to SM groups with a single bias voltage supply from the outside.  Besides grouping the power lines, this board should have the additional functionality of being able to turn off individual \DSkPdms\ in case of malfunctioning.  To disable a faulty \DSkPdm, the \DSkSMs\ can be programmed via a serial interface to disconnect individual modules from all supply voltages. 

The design shown in Fig.~\ref{fig:SteeringModule-Concept} consists of two blocks.  Memory cells from a shift register receive and store the state of operation of each \DSkPdm\ by a serial data bus.  The logic output levels (TTL) of the memory cells are then used to drive MOS-FETs  switches from individual \DSkPdm\ units.  The board has to operate reliably at cryogenic temperatures with a power consumption \DSkSMPower\ per board.  The possible integration of an FPGA in the design is under study.  If necessary, more complex tasks like fine tuning of the bias voltages or monitoring detector module performances (or others that are similar), could then be accomplished, but this would require an increase in power of the \DSkSM\ boards.

\subsection{The \SiPM\ and \DSkPdm\ Test Facility}
\label{sec:PhotoElectronics-DModTestFacility}



Testing of the large quantity of \SiPMs\ and \DSkPdms\ will be done in a clean room environment, in two test stations.  The first test station is dry, with the cryocooling provided by a \NaplesCryoDryStationCoolingPower\ pulse tube refrigerator.  It consists of a cylindrical cryostat, \NaplesCryoDryStationCryostatHeight\ high and \NaplesCryoDryStationCryostatDiameter\ in diameter, which can reach a vacuum level of \NaplesCryoDryStationCryostatVacuum\ before cryocooling.  The cold finger can be connected to up to \NaplesCryoDryStationPlatesHosted\ copper plates, of up to \NaplesCryoDryStationPlateDiameter\ in diameter, sufficient to accommodate \NaplesCryoDryStationSQBPerPlate\ \SQB\ motherboard each, for a total of up to \NaplesCryoDryStationSQBHosted\ \SQB\ hosted in the system at the same time.  The dry test station will be equipped with the necessary feedthroughs for the individual illumination, voltage supply and signal readout of the \DSkPdms.  The dry test station is already equipped with an external relay matrix system that enables, on special development \tiles, the readout of individual \SiPMs\ as well as their readout in a flexible series/parallel configuration.

\begin{figure}[t!]
\centering
\includegraphics[width=\columnwidth]{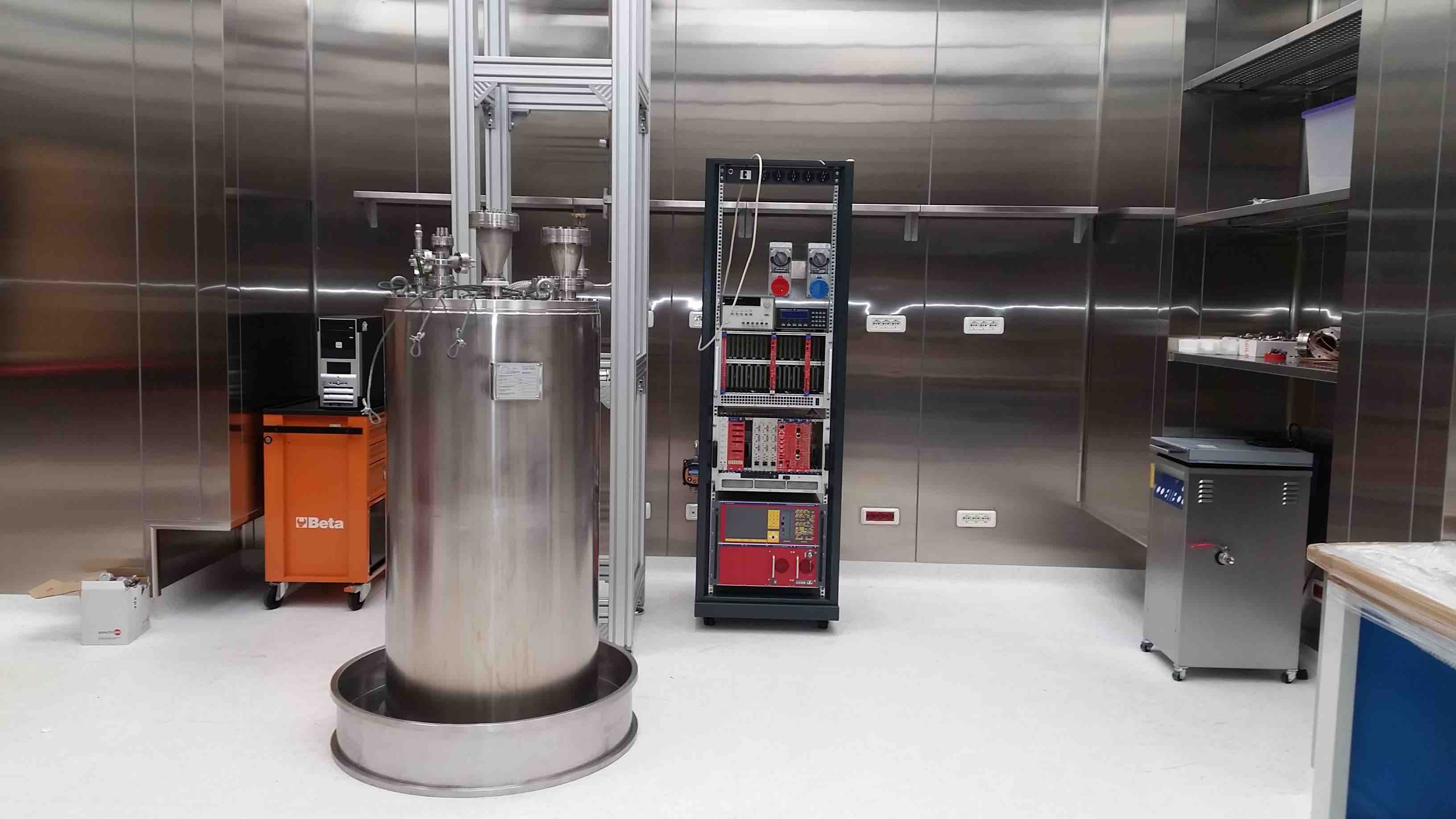}
\caption{Picture of the clean room and of the wet station for the mass test of \DSkPdms.}
\label{fig:DModTestFacility-WetStation}
\end{figure}

The second is a wet test station, fed with either \LIN\ or \LAr\ directly by either of the two storage tanks, see Fig.~\ref{fig:DModTestFacility-WetStation}.  It consists of a cylindrical cryostat, \NaplesCryoWetStationCryostatHeight\ high and \NaplesCryoWetStationCryostatDiameter\ in diameter, which can reach a vacuum level of \NaplesCryoWetStationCryostatVacuum\ before cryocooling.  The cryostat can host up to \NaplesCryoDryStationPlatesHosted\ holding plates, of up to \NaplesCryoWetStationPlateDiameter\ in diameter, sufficient to accommodate \NaplesCryoWetStationSQBPerPlate\ \SQB\ motherboards each, for a total of \NaplesCryoWetStationSQBHosted\ \SQB\ motherboards simultaneously.  This system will also be equipped with the necessary feedthroughs for the individual illumination, voltage supply and signal readout of the \DSkPdms.

Optical fibers facing the center of the \DSkPdms\ are connected to one of the outputs of two 1x25 fused couplers located in an external black-box.  Light pulses are generated by a Hamamatsu PLP10 light pulser, equipped with a \SI{408}{\nano\meter} laser head and pulse width of \SI{70}{\pico\second}.  The wavelength is chosen to be consistent with the wavelength-shifted light in the experiment.  The laser is piped to the the splitters through optical attenuator mounts. Discrete filters of different attenuation coefficients can be used.

The DAQ system provides each acquisition channel full signal digitization, in order to allow the detailed \DSkPdms\ characterization in terms of:
\begin{inparadesc} 
\item workfunction;
\item pulse shape (recharge time);
\item single photon response;
\item dark rate;
\item stability in time;
\item correlated pulses;
\item and, robustness to mechanical stress tests.
\end{inparadesc}

Optionally, \PDE\ measurements will be feasible in the wet test station for a limited fraction of about \NaplesCryoFractionPDEMeasurementInPdm\ of the \DSkPdm\ production run.

Since each of the two stations can accommodate up to \NaplesCryoDryStationSQBHosted\ \SQB\ motherboards at once, \NaplesCryoDryStationPdmHosted\ \DSkPdms\ can be tested simultaneous.  With one run per week, a test of all of the \DSkPdms\ required for \DSk\ could be performed in one year.

\subsection{Summary and Outlook}
\label{sec:PhotoElectronics-Conclusions}

The \NUVHdLfHRq\ already meets the stringent requirements of \DSk.  The final goal of developing tiles and \DSkPdms\ of area \DSkTileAreaStd\ was achieved in 2017~\cite{DIncecco:2017td}.  With these \SiPMs, the collaboration has achieved some of the first (perhaps the first) demonstration of large area, \DSkTileAreaMin\ and \DSkTileAreaStd, \SiPM\ assemblies operated as a single channel photodetector with very high \PDE\ and sensitivity to single scintillation photons.  

With the development of \SiPMs\ at \FBK 's own production facility, a milestone for the \DSk\ experiment has been reached.  The production of the large area, (\DSkTilesArea) required for \DSk\ necessitates the transfer of prototyping and production to an external foundry.  Details of the performance will depend on the outfit of choice, as the details of the implantation profiles and of the silicon contamination during its process will likely influence major characteristics of the \SiPMs, such as the \DCR.  In order to make further progress, the procedures required for the selection of an external foundry to accompany the entire development process of the \SiPMs\ must be followed.

In the next two years, it is planned to achieve further improvements on the \NUVHdLf\ technology to help manage the risks associated with the production of tiles and related electronics.  Depending on new results of the functional characterization of \SiPMs, and taking into account the constraints of the read-out electronics, new specific development runs may be needed.  For example, the first requirement may be an extension of the \OV, mainly to facilitate the signal read-out on very large area \SiPMs.  The most important action items as of today are:

\begin{asparaenum}[\bfseries {ACT}-i]
\item Transfer of the production to an external foundry able to accompany the entire development process of the \SiPMs;
\item Start of the process of mass characterization of the performance of \NUVHdLfHRq\ \SiPMs\ at \RoomTemperature\ and at \LArNormalTemperature\ with a cryogenic wafer probe in the \NOA;
\item Further reduction of the electric field for the \NUVHdLf\ technology;
\item Improvement of the production process of \NUVHdLfHRq\ \SiPMs\ in order to have stable behavior at cryogenic temperature with lower quenching resistors;
\item Increase of the charge of the fast component, maintaining a fixed total capacitance.
\end{asparaenum}

\section{Liquid Argon \TPC}
\label{sec:LArTPC}

\subsection{Introduction and Requirements}
\label{sec:LArTPC-Introduction}

The \DSk\ \LArTPC\ is the dark matter detector and the central element of the experiment, with all auxiliary detectors and systems specified and designed in support of the \LArTPC.  To achieve its stated physics goals, the \LArTPC\ must respond to a stringent set of requirements.  In particular, it must: 

\begin{asparaenum}[\bfseries {\LArTPC}-i]
\item Provide \DSkFiducialMass\ target mass after fiducial volume cuts;
\item Employ only radio-pure materials of well known performance in \LAr, such as OFHC copper (Cu),  Teflon (\PTFE), low radioactivity stainless steel, and titanium;
\item Minimize the use of hydrogen-containing materials to avoid compromising the effectiveness of the \LSV;
\item Provide \SOne\ light yield at least as high as the \DSkNullFieldLightYieldProjected\ at null field,  via employment of large-area, densely-packed custom arrays of photosensors;
\item Enable stable application of a cathode voltage of \DSkCathodePotential, required to generate the drift field;
\item Provide uniform drift, extraction, and electroluminesence fields and gas pocket thickness uniformity for high resolution of \STwo\ signals;
\item Allow tilt adjustment of the \TPC\ to allow leveling the anode plane;
\item Provide $x-y$ position resolution on the order of, or better than, \DSkXYResolution;
\item Provide tolerance for differential thermal contraction, amplified by any thermal gradients during cool-down;
\item Allow effective circulation of \LAr\ to maintain purity;
\item Allow rapid deployment of \ce{^{83m}Kr} throughout the active \TPC\ volume to facilitate calibration;
\item Provide access for optical fibers that will carry narrow, low-photon-occupancy light pulses for photosensor calibration.
\end{asparaenum}

\begin{figure}[t!]
\centering
\includegraphics[width=0.45\columnwidth]{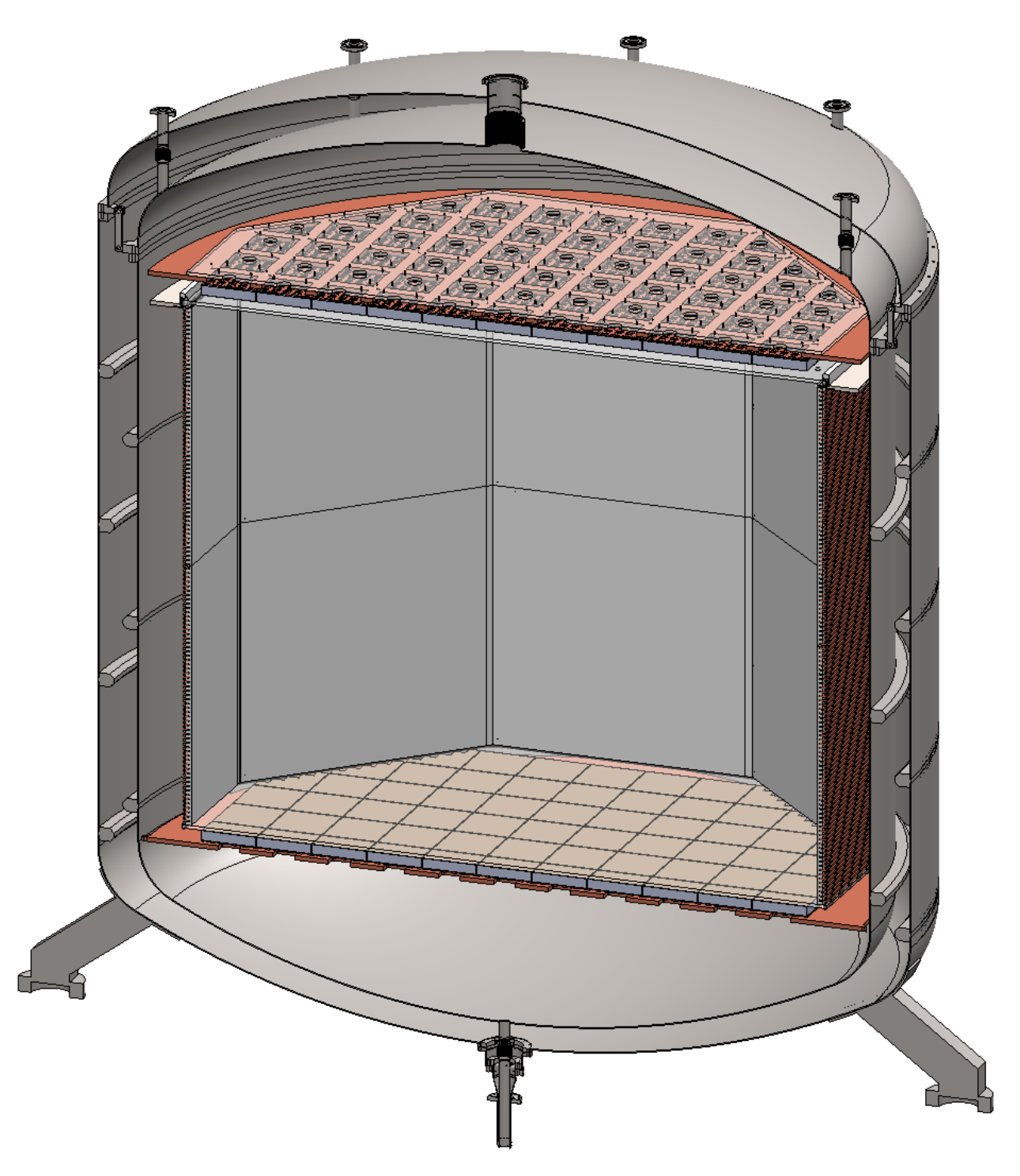}
\caption{3D rendering of the \DSk\ \LArTPC\ and cryostat.}
\label{fig:LArTPC-Detector}
\end{figure}

\subsection{Design Overview}
\label{sec:LArTPC-Overview}

The \LArTPC\ operates in a two-phase mode, utilizing both the liquid and gas phases of argon, as in \DSf.  Fig.~\ref{fig:LArTPC-Detector} shows a 3D representation of the \LArTPC\ contained inside the cryostat.  The cryostat holds $\sim$\DSkTotalMass\ of \LAr\ in total, with \DSkActiveMass\ inside the instrumented (active) \LArTPC.  Table~\ref{tab:LArTPC-Parameters} lists some of the key parameters of the  \LArTPC. 

The active \LAr\ volume is defined by a transparent conductive indium-tin-oxide (\ITO) cathode  at the bottom, a grid just below the liquid surface, separating the electron drift and extraction regions at the top, and a reflecting surface with an octagonal footprint on the sides.  The thin argon gas electroluminescence region above the liquid surface is bounded by a flat diving bell shaped ITO anode electrode at the top.  Interactions in the active \LAr\ volume generate prompt scintillation light that gives the \SOne\ signal.  A high voltage is applied between the cathode and grid to produce a vertical electric field to drift the ionization electrons created at the interaction site upward.  An independently-adjustable potential between the grid and anode creates the fields to extract the electrons into the gas and accelerate them to create secondary scintillation, called electroluminescence, which gives the \STwo\ signal.  Both the \SOne\ and \STwo\ signals are  detected by top and bottom arrays of \SiPM\ photosensors which view the active \LAr\ volume through large acrylic windows.  This design maximizes light yield and it is the same concept of \STwo\ generation as was tested in \DSf, namely a thin layer of gas generated below an anode  in the form of conductive and transparent  ITO film deposited on the transparent supporting window, allowing for liquid argon to occupy the space above the window. The cathode and anode plane designs require octagonal windows with a width of roughly \DSkActiveDiameter, and since there are no available ultra-high-purity fused silica plates of this size, acrylic is the material of choice.

The wavelength of the argon scintillation and electroluminescence signals is in the far UV region, at \ArWaveLength.  Therefore, it must be wavelength-shifted before being detected by light sensing devices.  To maximize the light yield, all inner surfaces delimiting the active volume are coated with the wavelength shifter tetraphenylbutadiene (\TPB), which absorbs the \ArWaveLength\ scintillation light and reemits it with high efficiency~\cite{Gehman:2011fc} in the blue region ($\sim$\TPBWaveLength).

The \SOne\ signal is distributed roughly equally over both arrays, while the \STwo\ signal, emitted in the gas pocket, is concentrated in the top array and within a few \SiPM\ tiles around the transverse position of the ionization drift, thus yielding a precise $x-y$ location.  The drift time (the time between the \SOne\ and \STwo\ signals) determines the $z$-location of the event in the \LArTPC.  Three dimensional precision imaging of an event can thus be achieved, with an expected resolution of \DSkXYResolution\ in $x-y$ and \DSkZResolution\ in $z$.

\begin{table}
\rowcolors{3}{gray!35}{}
\centering
\caption{\LArTPC\ detector characteristics.}
\begin{tabular}{lc}
{\bf \LArTPC\  Dimensions} 			&\\
\hline
Height							&\DSkActiveHeight\\
Effective Diameter					&\DSkActiveDiameter\\ 
Active \LAr\ Mass 					&\DSkActiveMass\\
\hline
\cellcolor{white}{\bf Nominal TPC Fields and Grid}
								&\cellcolor{white}\\
\hline
Drift Field 							&\DSkDriftField\\
Extraction Field  					&\DSkExtractionField\\
Luminescence Field 					&\DSkElectroLuminescenceField\\
Operating Cathode Voltage 			&\DSkCathodePotential\\
Operating Extraction Grid Voltage	  	&\DSkGridPotential\\
Operating Anode Voltage  				&\DSkAnodePotential\\
Electroluminescence Distance  			&\DSkGasPocketThickness\\
Grid Wire Spacing  					&\DSkLArOverGridPitch\\
Grid Optical Transparency  			&\DSkGridTrasparency\\
\hline
\cellcolor{white}{\bf \SiPM\ Tiles}		&\cellcolor{white}\\
\hline
Number of Tiles 					&\DSkTilesNumber\\
Size of Tiles 						&\DSkTileAreaStd\\ 
\end{tabular}
\label{tab:LArTPC-Parameters}
\end{table}

\subsection{Drift, Extraction and Electroluminescence System Design}
\label{sec:LArTPC-Fields}

The electric field configuration inside the \LArTPC\ consists of an active \LAr\ volume with an applied uniform drift field of \DSkDriftField, \DSkLArOverGridThickness\ of \LAr\ above the grid  with an extraction field of \DSkExtractionField, and a \DSkGasPocketThickness-thick argon gas pocket with an electroluminescence field of the order of \DSkElectroLuminescenceField.

\begin{figure}[t!]
\centering
\includegraphics[width=0.75\columnwidth]{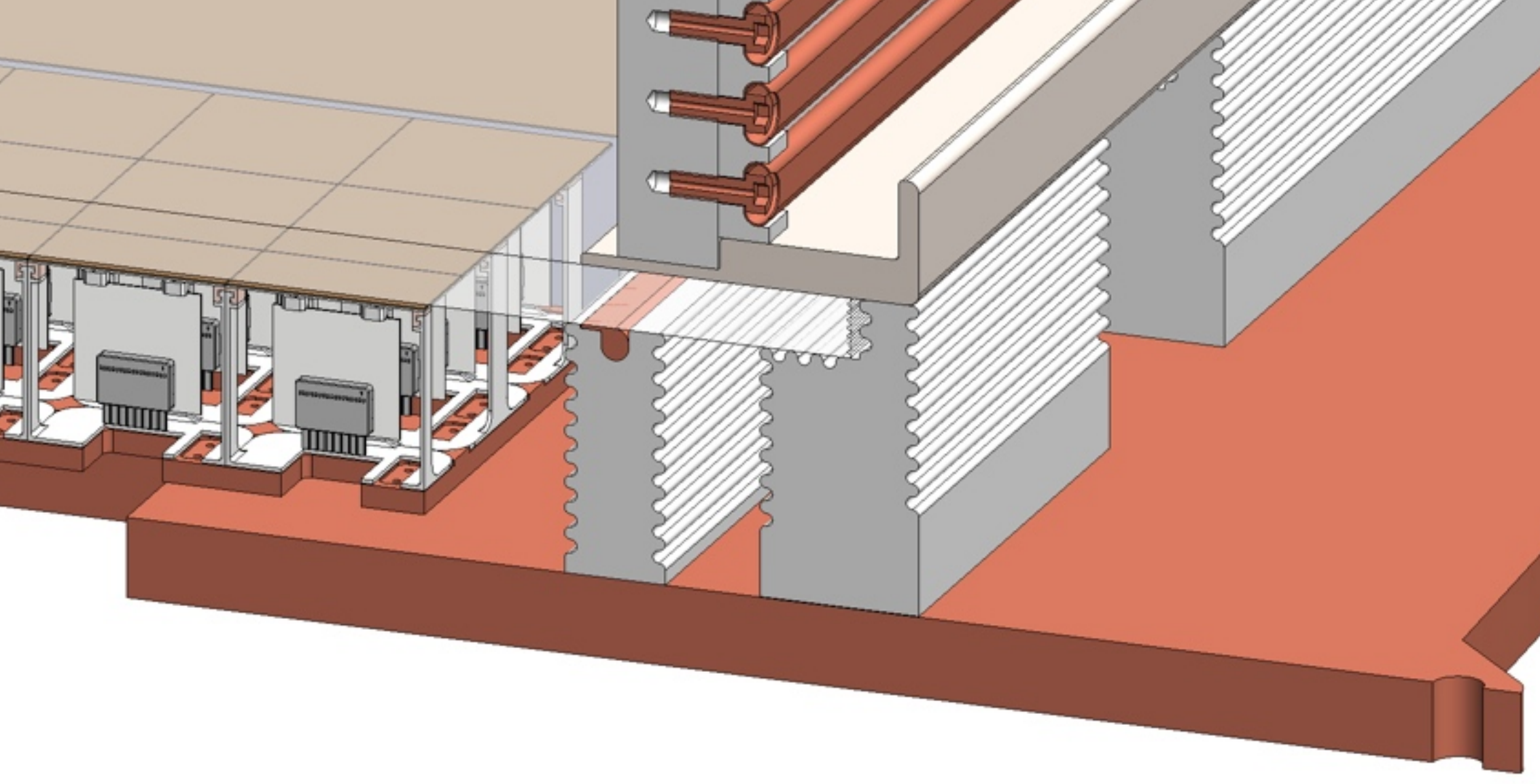}
\caption{Cathode region of the \LArTPC.}
\label{fig:LArTPC-CathodeRegion}
\end{figure}

The bottom boundary of the active volume, shown in Fig.~\ref{fig:LArTPC-CathodeRegion}, is a \DSkAcrylicWindowThickness\ thick acrylic window coated on both sides with a thin layer of \ITO.  The top \ITO\ layer, together with a contacting guard ring serve as the cathode.  The potential of the cathode plane is fixed at \DSkCathodePotential.  The bottom \SiPM\ array is placed \DSkBotSiPMToBotITOBotPlate\ below the bottom \ITO\ layer, which is held at \DSkAnodePotential.  The acrylic window can easily sustain the applied electric field, while the P-profile copper guard rings contact the bottom \ITO\ layers and serve to minimize the electric field at the \ITO\ edge, avoiding discharges in the \LAr.  This concept was tested to be electrically stable up to about \DSkWindowTestHHV\ with a full-scale mock-up of the \DSf\ cathode region, using a \DSkWindowTestFusedSilicaWindow\ thick fused silica window in an open dewar.  The solid contacting guard ring will be used to mechanically hold the \TPC\ above the cathode and make contact to the top \ITO\ layer on the acrylic with multi-spring-loaded pins.  The acrylic will be held by a sandwich-type structure, such that no load is applied onto the acrylic to avoid potential stress and breakage, see details in Fig.~\ref{fig:LArTPC-CathodeRegion}.  The structure also allows the acrylic to shrink and slide inward, yet still cover the full bottom \SiPM\ array.  The guard ring below the acrylic is also used to smooth the HV field lines and provide contact to the bottom grounded \ITO\ layer.  Note that Fig.~\ref{fig:LArTPC-CathodeRegion} shows the cathode structure details at room temperature.  At \LAr\ temperature, the edge of the acrylic, as well as the field cage, will shrink inwards.  The cathode plane, with its support structure, acts as a rigid frame such that it is strong enough to hold the \TPC.  The cathode is fixed on the bottom grounded frame by a number of PTFE pillars with walls grooved for cathode HV related concerns.  The edge of the acrylic and the flat faces of any insulator in this region will be grooved based on the same techniques applied in \DSf.

\begin{figure}[t!]
\centering
\includegraphics[width=0.45\columnwidth]{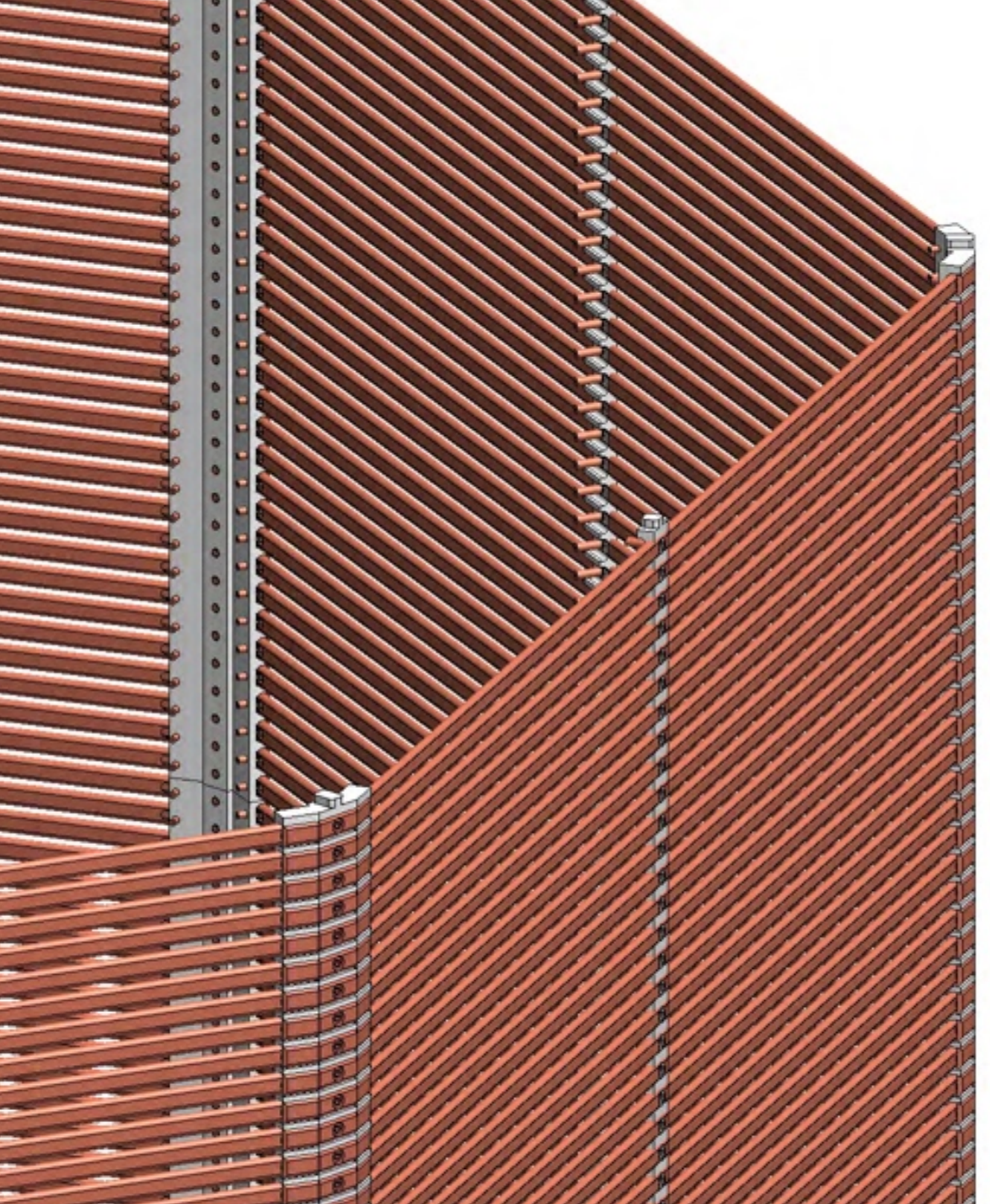}
\includegraphics[width=0.45\columnwidth]{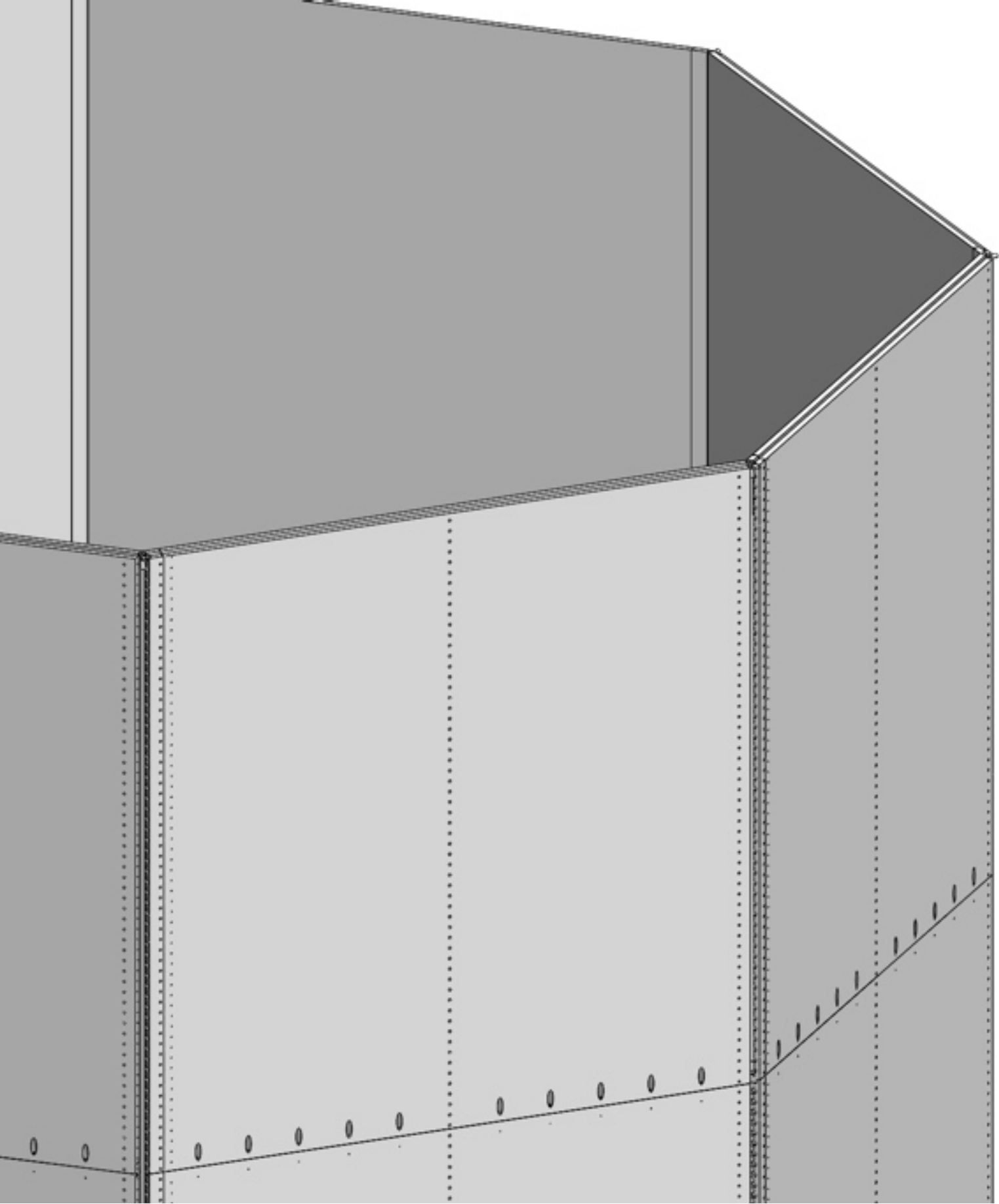}
\caption{Field cage and reflector panels of the \LArTPC.}
\label{fig:LArTPC-FieldCage}
\label{fig:LArTPC-PTFECage}
\end{figure}

When cold, the distance between the anode and the cathode is \DSkActiveHeight\ and the active effective diameter is \DSkActiveDiameter.  The total drift distance in liquid is \DSkDriftLength.  A uniform electric drift field in the \LArTPC\ is provided by a field cage consisting of a set of \DSkFieldCageRingsNumber\ octagonal field shaping rings, equally spaced along the $z$-axis and biased with a uniform voltage gradient.  Each field shaping ring is constructed from \DSkFieldCageRingCuTubesNumber\ overlapping \ce{Cu} tubes and held in place using \DSkFieldCageRingCuTubesNumber\ \ce{Cu} pins and \DSkFieldCageRingCuTubesNumber\ \ce{Cu} corner pieces set into the PTFE reflector cage (see Fig.~\ref{fig:LArTPC-FieldCage}).  The rings have an oval cross section with height of \DSkFieldCageRingsTubeHeight\ and are spaced at \DSkFieldCageRingsSpacing\ center-to-center.  They are separated from the active volume by the \DSkPTFEPanelsThickness\ thick \PTFE\ structure.  The rings are  hollow to help minimize the weight of the \ce{Cu} field cage, which is mounted onto the \PTFE\ reflector cage (see Fig.~\ref{fig:LArTPC-PTFECage}), which in turn is supported from the very bottom stainless steel flange that hangs from the cryostat.

The uniform voltage gradient is provided by a HV divider chain, consisting of 4 resistors in parallel (as a redundancy to allow stable operation if a resistor fails).  Each resistor (\DSkFieldCageMiddleResistorValue) will be mounted in the low electric field region between the inner surface of the \ce{Cu} corner piece and the \PTFE\ corner piece.

The side boundaries of the active volume are defined by an octagonal assembly of sixteen flat, highly-reflective \PTFE\ panels (each \DSkPTFEPanelsHeight\ high and \DSkPTFEPanelsWidth\ wide when cold).  Eight \PTFE\ panels form an octagon that is stacked onto another \PTFE\ octagon (see Fig.~\ref{fig:LArTPC-PTFECage}).  The \PTFE\ will shrink by $\sim$ \DSkPTFEShrinkInPercent\ linearly when cooled from room temperature to \LAr\ temperature and will be taken into account when the final materials selection has been made and exact coefficients of thermal expansion are measured.  The use of the Cu pins and corner pieces to hold the rings in place allows for the rings and \PTFE\ panels to shrink concentrically relative to each other, maintaining the uniform electric field at \LAr\ temperature.  The \ce{Cu} corner pieces and pins require high machining precision and will thus be fabricated using Wire EDM techniques.  The \ce{Cu} tubing will be rolled using a flat rolling mill with a custom profile.

\begin{figure}[t!]
\centering
\includegraphics[width=0.45\columnwidth]{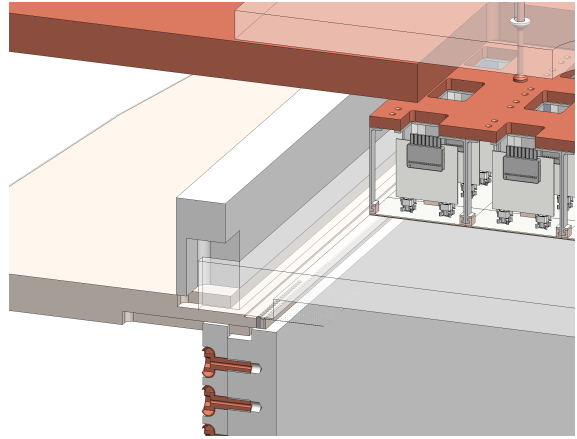}
\caption{Anode region of the \LArTPC.}
\label{fig:LArTPC-AnodeRegion}
\end{figure}

The top boundary of the active volume, shown in Fig.~\ref{fig:LArTPC-AnodeRegion}, is given by a \DSkAcrylicWindowThickness\ thick flat acrylic diving bell coated with a thin \ITO\ layer at the bottom.  The bottom \ITO\ layer serves as the anode for the \LArTPC.  The potential of the anode layer is fixed at \DSkAnodePotential.  An extraction fine wire grid sitting \DSkLArOverGridThickness\ below the liquid surface isolates the higher extraction and electroluminescence fields from the lower uniform drift field in the \LArTPC\ bulk volume.  The extraction wire grid potential is fixed at \DSkGridPotential\ to provide strong electric field in the gas region with nearly \SI{100}{\percent} extraction efficiency of electrons from the \LAr\ into the gas phase.  The liquid level is controlled by a diving bell system similar to that used in \DSf.  An array of PT-100 heaters will be used to uniformly boil off \LAr.  The resulting gas will be used to feed the diving bell to form the gas pocket, and done in such a way as to prevent gas bubbles from disturbing the \LAr\ surface. The height of the lip of the anode acrylic diving bell defines the \DSkGasPocketThickness-thick gas pocket.
 
\begin{figure*}[t!]
\centering
\includegraphics[width=\textwidth]{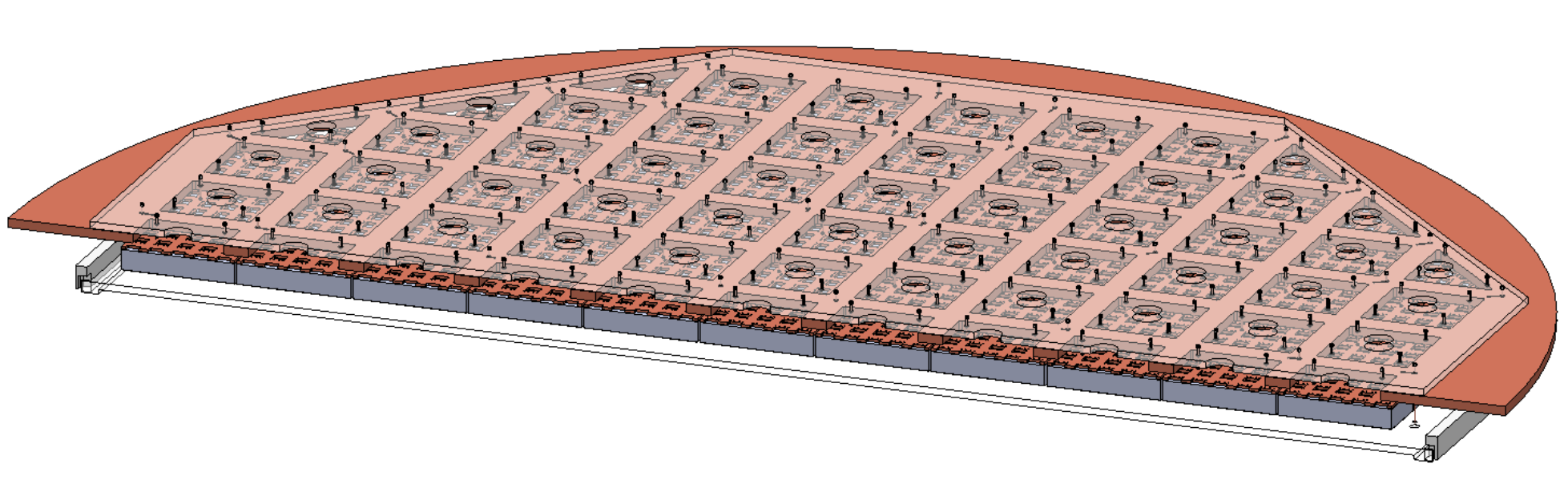}
\caption{Anode assembly of the \TPC.}
\label{fig:LArTPC-AnodeAssembly}
\end{figure*}

The extraction wire grid employs  \DSkExtractionGridThickness\  stainless steel wires placed with a pitch of \DSkLArOverGridPitch\ in a linear parallel grid.  Wires will be mounted onto a \DSkCathodeWireFrameDiameter\ diameter stainless steel frame.  Each end of each wire will be wrapped around a stainless steel post at the desired tension using a dedicated wire winding machine.  The stainless steel posts are mounted into precisely machined holes in the stainless steel frame.  The wires are guided using precision  Wire-EDM machined grooves.  The tension of each  wire must compensate for the deflection due to its weight and electrostatic attraction between the extraction grid and anode.  For a vertical deflection \DSkExtractionGridVerticalDeflection\ the tension on each wire is required to be \DSkCathodeWireTension.  The grid frame has an inner octagonal shape, an outer circular shape and  an L-shaped cross-section to enhance its rigidity and to keep its maximum radial deformation to \DSkCathodeWireFrameRadialDeflection.  The vertical deflection requirement of \DSkExtractionGridVerticalDeflection\ and a pitch of \DSkLArOverGridPitch\ for the extraction wire grid ensure the uniformity of the electroluminescence field near the gas-liquid interface.  Fig.~\ref{fig:LArTPC-AnodeRegion} shows the anode region with \SiPM\ motherboards and the support structure, which allows thermal shrinkage during cool down and expansion during warming up.  Fig.~\ref{fig:LArTPC-AnodeAssembly} shows the full anode assembly of the \TPC.  The density of acrylic is slightly less than that of \LAr, hence the anode window is subject to bowing upward when immersed in \LAr, due to buoyancy.  In addition, the displaced argon within the gas pocket translates directly into a force being exerted upward on the lower surface of the anode window, resulting in additional buoyancy.  To maintain uniformity of the \STwo\ response, the anode window is kept flat within \DSkAnodeVerticalDeflection\ using a set of \DSkAnodePillarsNumber\ \ce{Cu} pillars, each approximately \DSkAnodePillarsDiameter\ in diameter.

During cool down and warm up, the acrylic anode window and its supporting \ce{Cu} pillars might have relative shifts due to differential thermal contraction.  To prevent possible mechanical issues, the anode assembly is designed so that the \ce{Cu} pillars will be co-moving with the anode.  A similar-thickness  acrylic plate (coherent plate) is located on top of the copper mechanical support structure such that the coherent plate and the anode shrink at the same rate.  This coherent plate is constrained in the vertical direction on the copper plate near each pillar so that the coherent plate will not bow-up.  The \ce{Cu} pillars supporting the anode are fixed to the coherent plate.  The \ce{Cu} pillars must also be placed in between the densely packed SiPM motherboards.  At the \DSk\ scale, the \ce{Cu} pillar near the edge of the anode will shrink about \DSkAnodePillarsRadialThermalShift\ towards the center, hence the SiPM mother boards must also co-move with the rest.  This can be realized by mounting the SiPM \ce{Cu} motherboards also on the acrylic coherent plate.  Even within the dimensions of the motherboards, the four fixture bolts (fixed on the \ce{Cu} motherboard and incurring less shrinkage related to that of acrylic) will have relative motion with respect to the acrylic.  This design already implements a solution that allows for such relative motion.  The design of the cathode does not call for the supporting pillars and hence makes the handling of the thermal contraction is much easier to deal with.  The baseline design calls for the cathode window to be made slightly larger than the interior of the \LArTPC, which after cool down would then match the \LArTPC\ dimensions.  To make sure the window does not slide, relative to the TPC, at least one point would be held in place, thus controlling the direction of the thermal contraction.

The \ce{Cu} pillars are intentionally lengthened to provide a \DSkTopSiPMToBotITOTopPlate\ long liquid region above the anode acrylic window.  In \DSf, the top \PMT\ array was kept very close to the fused silica anode assembly.  This reduced sharing of the \STwo\ light among the \DSfPMTSize\ \PMTs, degrading the $x$-$y$ position resolution.  The length of \DSkTopSiPMToBotITOTopPlate\ for the offset of the top photosensor array away from luminescence region is driven by size of each SiPM tile which is about \DSkTileAreaStd.  A similar approach is used in liquid xenon dark matter  experiments such as  LUX, LZ and Xenon1T~\cite{Akerib:2014jv,Nelson:2014wy,Aprile:2016cr}, where position resolution of less than \SI{1}{\centi\meter} is achieved or expected using \SI{3}{\inch} \PMTs.

\begin{figure*}[t!]
\centering
\includegraphics[width=\textwidth]{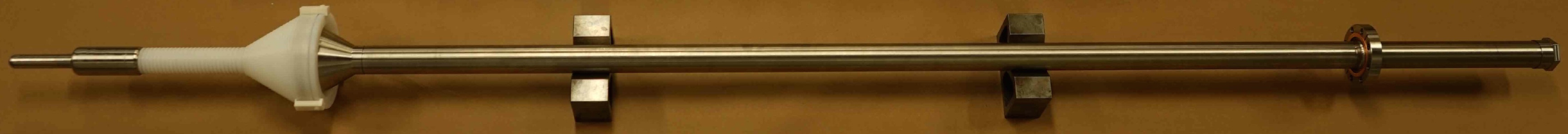}
\caption{Prototype HV feedthrough tested to up to \DSkHHVTest.}
\label{fig:LArTPC-HVFT}
\end{figure*}

Given the design drift field of \DSkDriftField\ over the drift length of \DSkDriftLength\ and the additional voltage required to generate the extraction and luminescence fields, the total voltage between anode and cathode will be \DSkCathodePotential.  The HHV will be delivered through the top flange of the cryostat by a custom-made high voltage feedthrough (\HVFT), similar in design to that used in \DSf.  The \HVFT\ design and fabrication follow a well established and reproducible procedure developed over many years at UCLA and is dependent on low-radioactivity materials: an ultra-high-molecular-weight polyethylene (\UHMWPE) tube serving as the insulator, stainless steel used for the outer grounding tube and inner conductor core, and an integral, conflat-flange-mounted custom connector terminating the cable connecting to the external power supply.  The outer grounding tube is extended deep into the \LAr\ region to reduce field in \GAr\ region and chance of \LAr\ boiling at its termination.  Below the ending of the outer grounding tube, the \UHMWPE\ tube and inner conductor core continue and are connected to the cathode guard ring.  The \UHMWPE\ tube will be sized to withstand \DSkHHVUHMWPEDesignBreakdown\ and minimize the electric field at its outer surface, which will be grooved to reduce charge transfer due to charge buildup.  Earlier versions of this \HVFT\ have been tested well beyond the required \DSkCathodePotential\ and operated successfully in \DSt, \DSf\ and the \DUNE~\SI{35}{\tonne} prototype, without any operational difficulty.  Fig.~\ref{fig:LArTPC-HVFT} shows a picture of a prototype feedthrough tested up to \DSkHHVTest.

All metal pieces, stainless steel and \ce{Cu}, that are exposed to electric fields will be designed without sharp edges and their surfaces will be additionally smoothed using various existing techniques such as optical quality mechanical polishing, electropolishing, chemical etching, and conditioning.  Detailed electrostatic and mechanical studies of the all \TPC\ subsystems will be performed.  Scaled-down versions of  the parts of these systems will be tested in the \LAr\ R\&D setups at various institutions within the collaboration and in an integrated test of all HV and electric field systems in the \LArTPC\ prototype (see Sec.~\ref{sec:Proto}).

\subsection{Light Readout}
\label{LArTPC-PhotoElectronics}

\begin{figure}[t!]
\centering
\includegraphics[width=0.45\columnwidth]{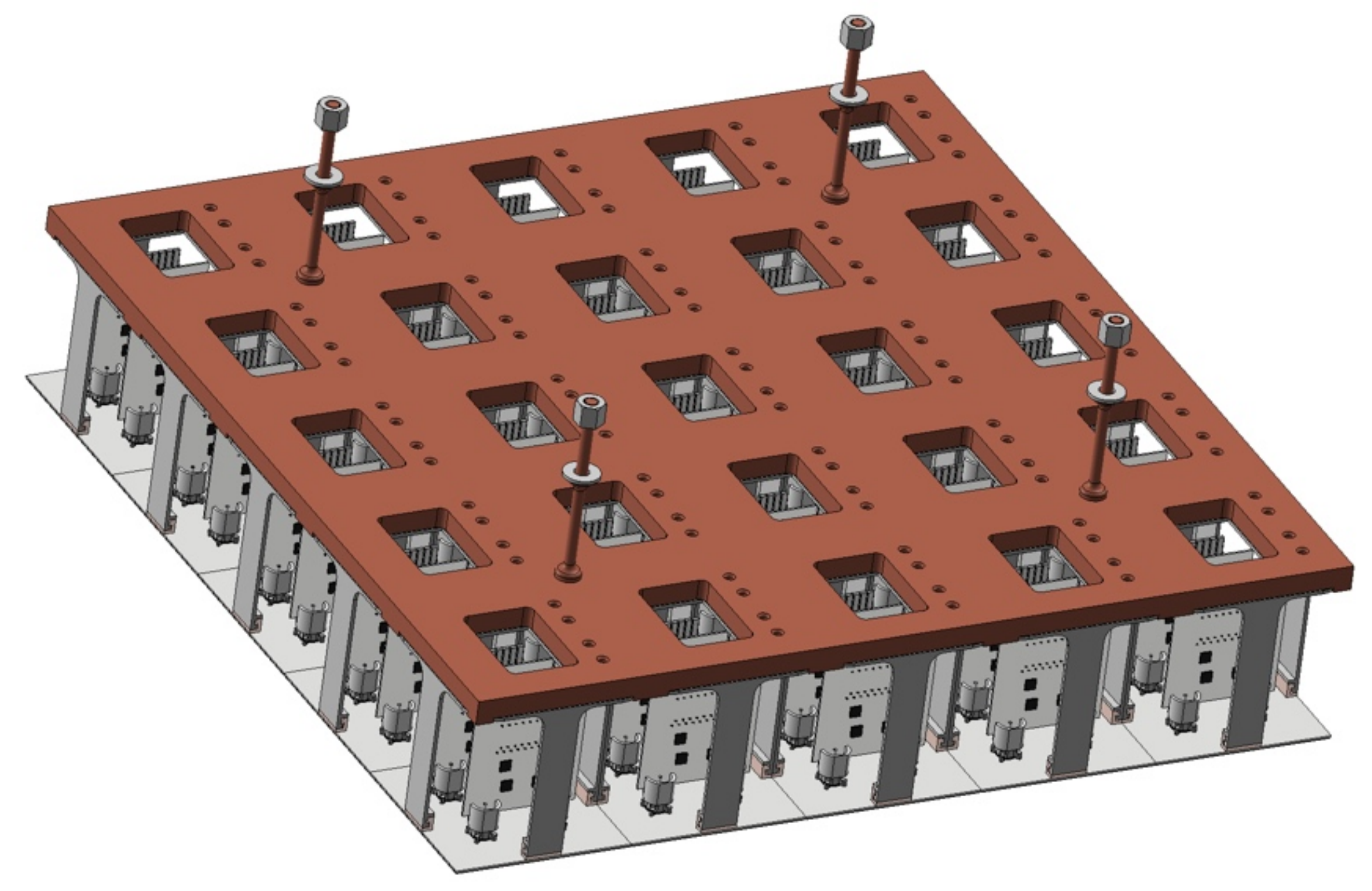}
\includegraphics[width=0.45\columnwidth]{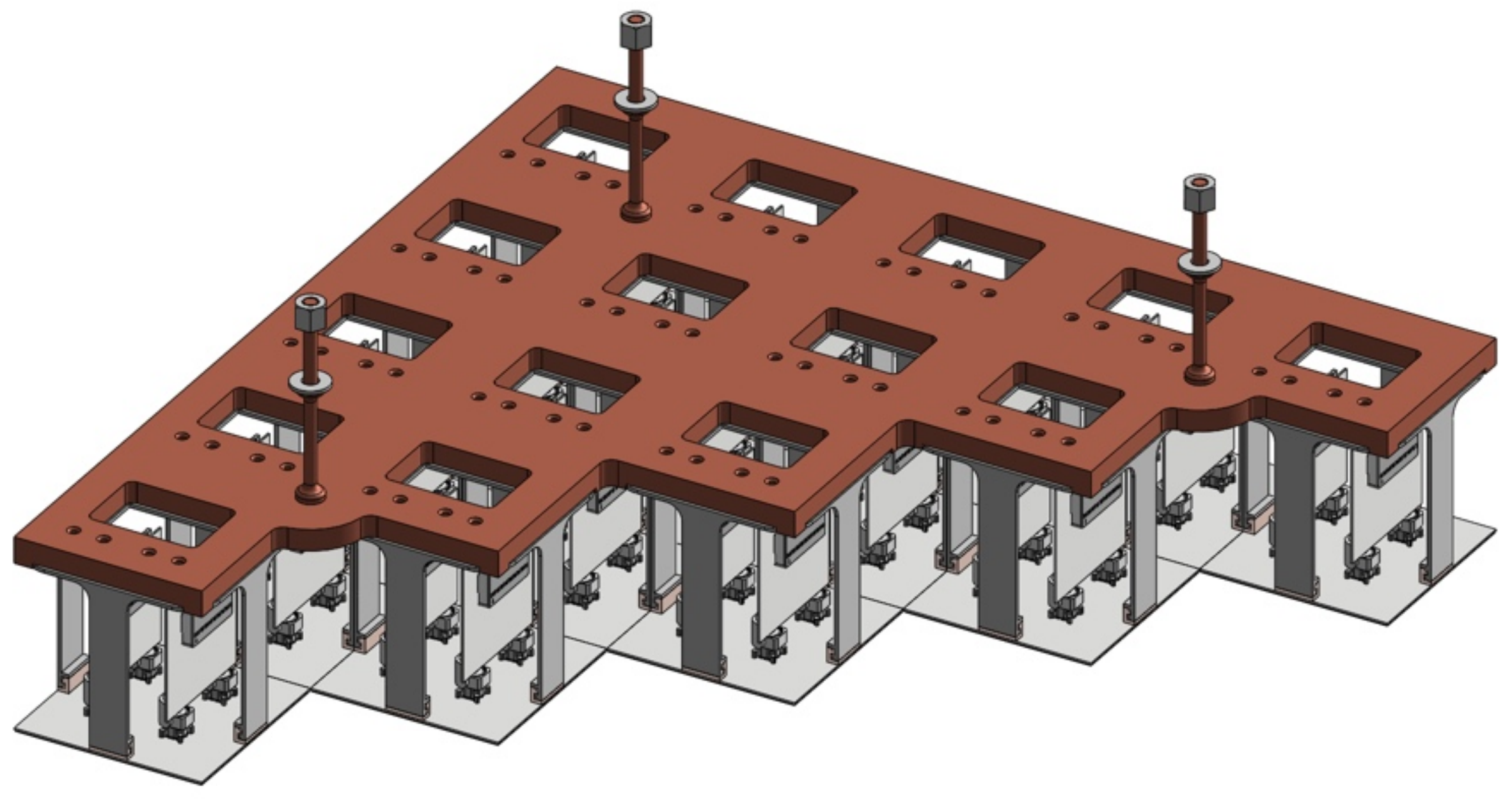}
\caption[Square and triangular \SiPM\ mother board assemblies.]{{\bf Left:}  A single \SQB\ (see text).  {\bf  Right:} A single \TRB\ (see text).  \SQBs\ and \TRBs\ are assemblies of tiles serving as building blocks of the photon readout planes.}
\label{fig:LArTPC-Tiles}
\end{figure}

Densely packed arrays of \SiPMs\  above the anode acrylic window and below the cathode acrylic window fully cover the top and bottom faces of the \LArTPC\ active volume, to detect both the \SOne\ and \STwo\ signals with high efficiency.  The top and the bottom photon readout assemblies consist of \DSkTilesHalfNumber\ photon detection channels each.  A single readout channel consists of a \DSkTileAreaStd\ \SiPM\ tile assembly containing an appropriate number of individual \SiPMs\ joined together to form a single light sensing unit.  The \SiPM\ tiles are used to form two larger basic mechanical units called the SQuare Board (\SQB) and the TRiangular Board (\TRB), as shown in Fig.~\ref{fig:LArTPC-Tiles}.  The \SQB\ and \TRB\ have the same edge size of  \DSkSQBTRBSize.  The \SQB\ and \TRB\ are then used to form the full readout octagonal planes.  The total number of photon readout channels (top and bottom) is \DSkTilesNumber.

\DSkTilesNumber\ channels require careful routing of all signal and power cables. The concept will be similar to that implemented in \DSf, where cables from the bottom of the \TPC\ are grouped and guided with vertical PTFE panels, with the cables lined between the PTFE panel and the cryostat wall to prevent HV issues near the cathode region. These cables then join the cables from the top readout array to form a single bunch which is fed-through out of the cryostat, the \LSV\ and finally the \WCV\ using a single tube. The path of the single tube from inside the cryostat to outside of the water tank is important, and must be made such that the out going gas of the gas purification loop passes through this tube and hence flushes out any impurities resulting from the outgassing of the cables. 

Based on the previous experience of \DSf\ in optimizing light collection, the \PTFE\ panels will be \DSkPTFEPanelsThickness\ thick and coated on the inside surface with a \DSkTPBThickness\ \TPB\ layer ~\cite{Agnes:2015gu}.  The \DSkAcrylicWindowThickness\ thick anode and cathode acrylic windows have better than \DSkAcrylicWindowTransparency\ optical transparency.  The thickness of \ITO\ films will be \DSkITOThickness\ as in \DSf, for an estimated optical transparency of \DSkITOTransparency~\cite{Alexander:2013jn}.  The inner surface of the acrylic plates will be coated with \TPB\ over the ITO.  The extraction grid optical transparency is estimated to be \DSkGridTrasparency.

Optical MC studies using the full \DSk\ geometry and the \GFDS\ simulation framework developed for \DSf\ have been performed to estimate the expected \SOne\ and \STwo\ light yield.  This is detailed in Sec.~\ref{sec:Physics}.

\subsection{Performance}
\label{sec:TPC-Performance}

\begin{figure}[h!]
\centering
\includegraphics[width=0.3\textwidth]{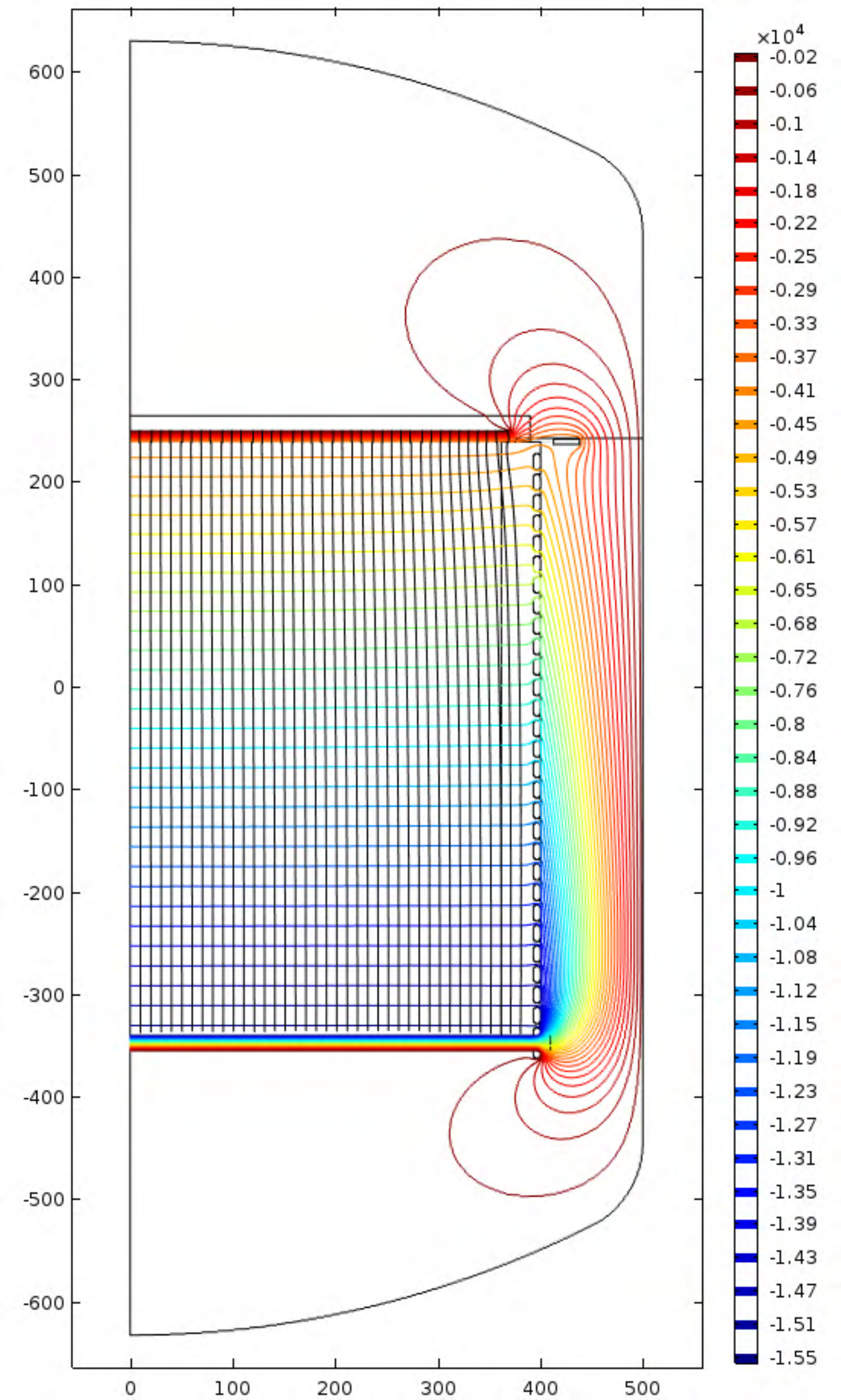}
\includegraphics[width=0.3\textwidth]{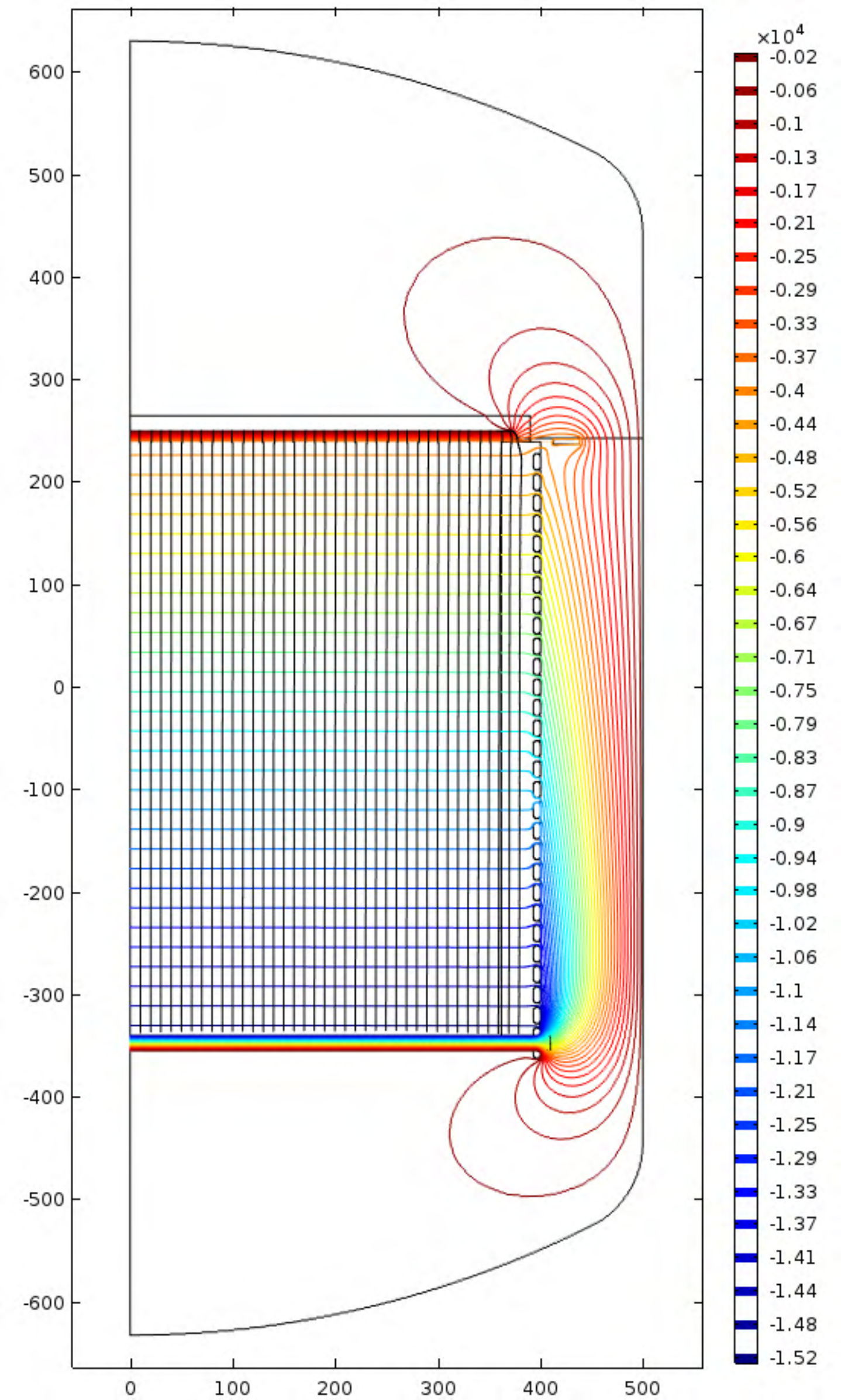}
\includegraphics[width=0.3\textwidth]{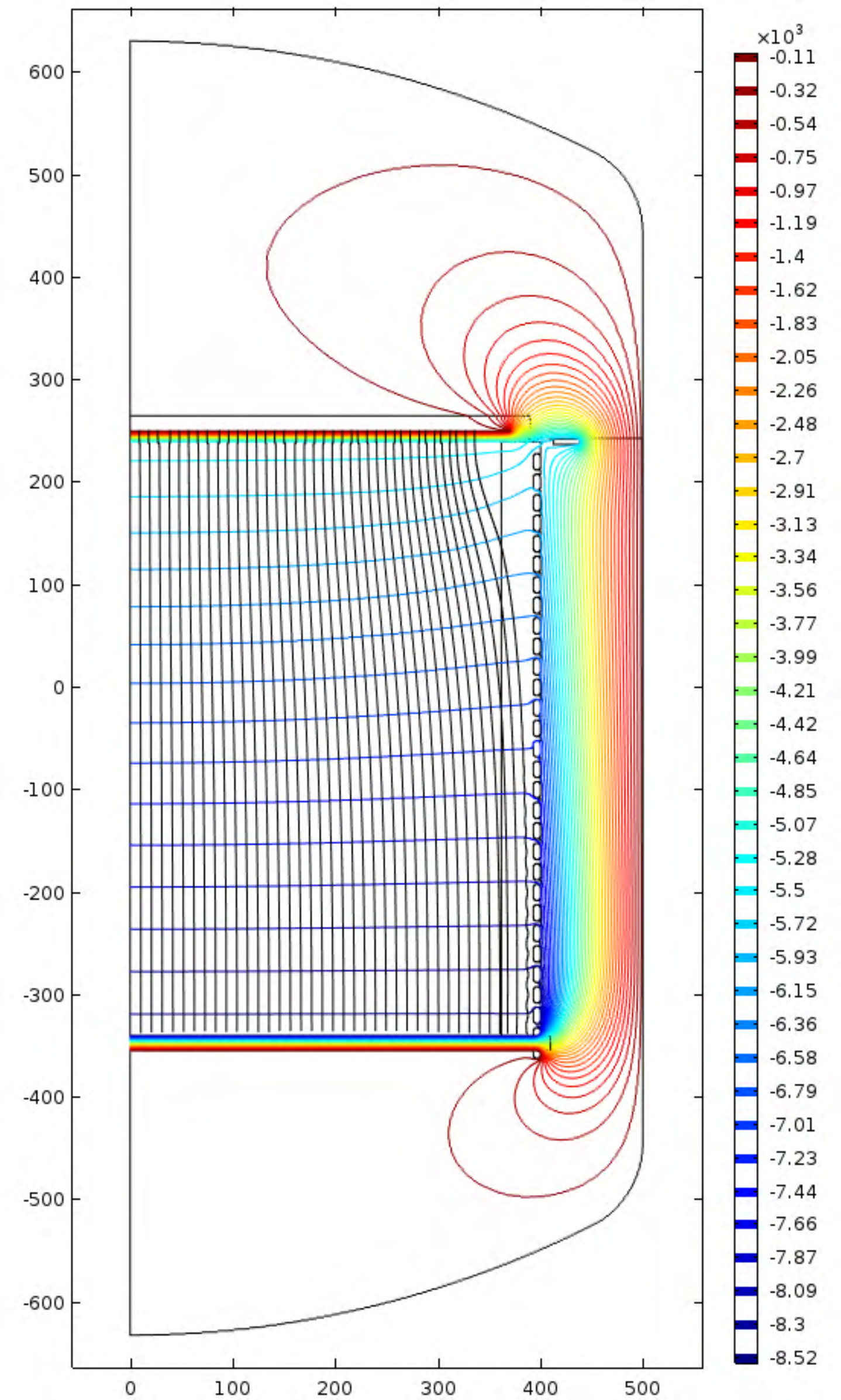}
\caption[Electrostatic simulations of the drift field with the Prototype \LArTPC\ geometry.]{Simulation of the electrostatic field for the \LArTPC\ prototype in a 2D geometry approximation with horizontal axis $r$ and vertical axis $z$, in units of millimeters.  Contours represent equipotential lines with the color bar on the right side showing the potential in units of volts, and streamlines represent the field lines inside the field cage.  {\bf Left:} Untuned $R_g$ = \DSkFieldCageMiddleResistorValue\ for desired field condition of \DSkDriftField\ for the drift field and \DSkElectroLuminescenceField\ for the electroluminescence field.  Field lines are distorted near the extraction grid; {\bf Middle:} Tuned $R_g$ = \DSkFieldCageTopResistorValue\ for \DSkDriftField\ drift field and \DSkElectroLuminescenceField\ electroluminescence field.  Field lines are much more uniformly distributed.  {\bf Right:} Tuned $R_g$ = \DSkFieldCageTopResistorValue\ for a quite undesired field condition of \SI{50}{\volt\per\centi\meter} for the drift field and \SI{6}{\kilo\volt\per\centi\meter} for the electroluminescence field to illustrate the dramatic effect of different field conditions on the drift field uniformity.
}
\label{fig:Field-Simulation-Baseline}
\end{figure}

\begin{figure}[h!]
\centering
\includegraphics[width=0.5\columnwidth]{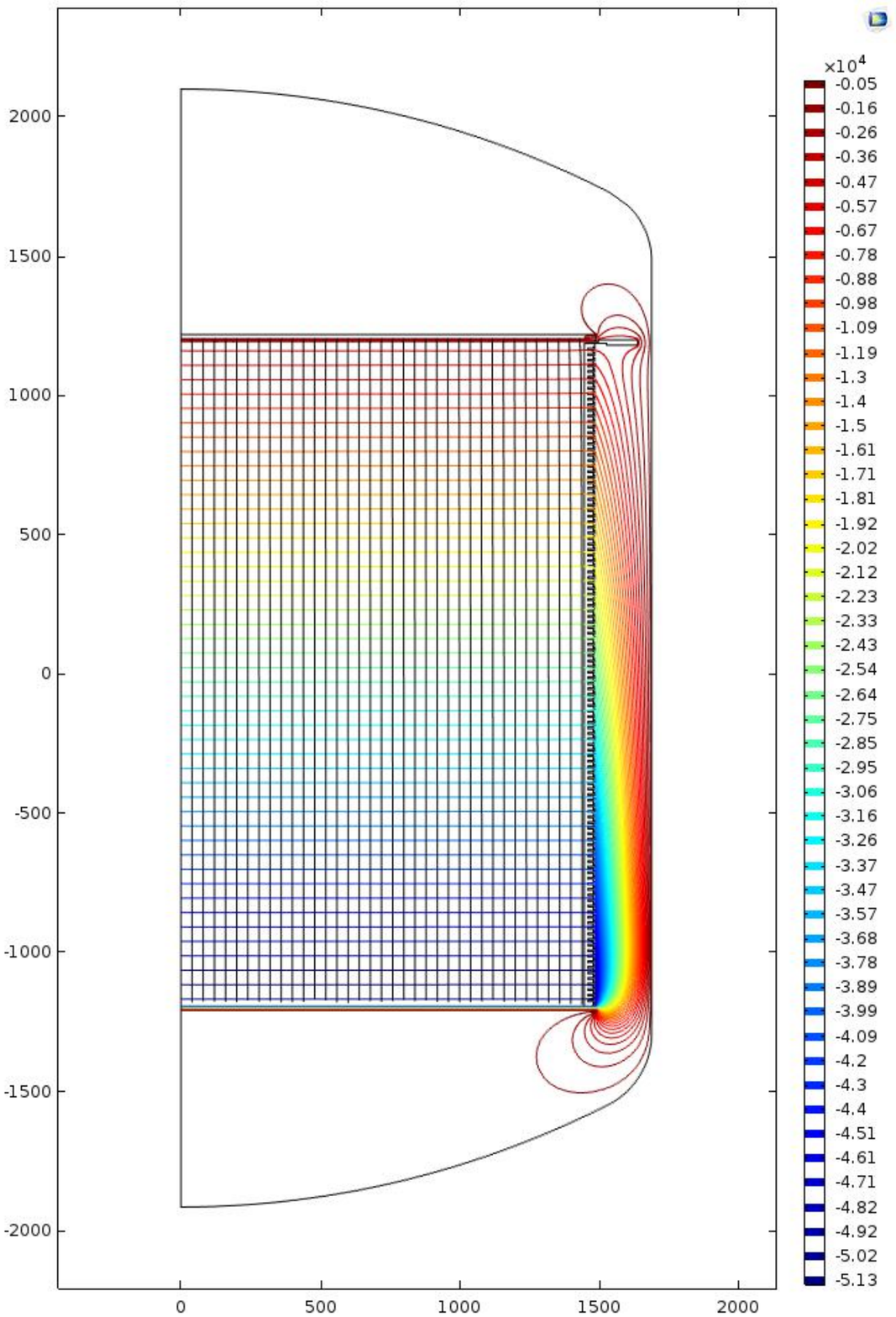}
\caption[Electrostatic simulations of the drift field with the DS-20k \LArTPC\ geometry.]{Simulation of the electrostatic field for the DS-20k \LArTPC\ in a 2D axial symmetric approximation with horizontal axis $r$ and vertical axis $z$, in units of millimeters.  Contours represent equipotential lines with the color bar on the right side showing the potential in units of volts, and streamlines represent the field lines inside the field cage.  Tuned $R_g$ = \DSkFieldCageTopResistorValue\ for \DSkDriftField\ drift field and \DSkElectroLuminescenceField\ electroluminescence field, resulting in field lines that are uniformly distributed.}
\label{fig:LArTPC-Field-Simulation}
\end{figure}

The electrostatic field of the \LArTPC\ is studied by finite element analysis (FEA) using the COMSOL Multiphysics software.  The simulation was first performed for the Prototype \TPC, introduced in Sec.~\ref{sec:Proto}, which is just a scaled-down version of \DSk\ with the same electrostatic features.  As shown in Fig.~\ref{fig:Field-Simulation-Baseline}, the geometry of the \LArTPC, including all relevant components, is simplified by the 2D approximation with coordinates of $r$ and $z$.  With the same operation configurations as \DSk, the drift field in the simulation is tuned to \DSkDriftField\ and extraction (electroluminescence) field is tuned to \DSkExtractionField\ (\DSkElectroLuminescenceField).  Since in the prototype the drift length is scaled down to \DSpDriftLength, the number of field shaping rings is reduced to \DSpFieldCageRingsNumber.  All shaping rings are evenly placed with vertical pitch of \DSkFieldCageRingsSpacing, and between every two adjacent field shaping rings a set of \num{4} resistors are connected in parallel to form \DSkFieldCageMiddleResistorValue. In order to eliminate effects of extraction field leakage, $\text{R}_{\text{g}}$, the first 4 parallel resistors in the divider chain, are tuned to have a total resistance of \DSkFieldCageTopResistorValue.

Comparing the left and middle plots in Fig.~\ref{fig:Field-Simulation-Baseline}, it is clear that the uniformity of the drift field is enhanced by this tuning.  As varying either drift or extraction (electroluminescence) field strength will affect the uniformity of drift field as well, simulations are also carried out to study how large the effect is.  It was found that with $R_g$  set at the tuned value, the distortion of drift field lines is acceptable with the drift field strength set at \num{100}, \num{300}, and \SI{400}{\volt\per\centi\meter} while the electroluminescence field remained fixed at \DSkElectroLuminescenceField, and the electroluminescence field set at \num{3}, \num{5} and \SI{6}{\kilo\volt\per\centi\meter} while the drift field strength remained fixed at \DSkDriftField.  Only with the drift field strength set at \SI{50}{\volt\per\centi\meter} while electroluminescence field remained at \DSkElectroLuminescenceField, do the drift field lines start to be noticeably distorted.  A combination of \SI{50}{\volt\per\centi\meter} drift field and \SI{6}{\kilo\volt\per\centi\meter} electroluminescence field is also simulated to compare with the desired field condition, illustrating what a quite undesired  field condition looks like, as shown in the right panel of Fig.~\ref{fig:Field-Simulation-Baseline}.  

Based on this study, the resistance of each set of resistors in the chain is determined, and this configuration is set as the baseline design.  In addition, since the \TPC\ is designed as octagonal instead of circular, besides the simulations described above which are performed using a radius normal to the \PTFE\ panels, another simulation was run in which the radius of the cylinder is set at the corner-to-corner width of the octagon and compared.  The two simulations show a consistent electrostatic field distribution inside the \TPC\ for all the configurations described above, which shows that the 2D approximation is valid for this study.  To verify that the uniformity of the field lines is still sufficient when scaling the geometry back up to that required for the \DSk\ \TPC, but still using the tuned values of the resistors and fields, the same FEA analysis as above was carried out for the full scale geometry.  The results of the COMSOL simulation are shown in Fig.~\ref{fig:LArTPC-Field-Simulation}, with the same visual representation used in the previous figures.  As is seen in this figure, the drift, extraction and electroluminescence fields all remain uniform, throughout the entire active volume of the \DSk\ \LArTPC.

\begin{figure}[h!]
\centering
\includegraphics[width=0.45\columnwidth]{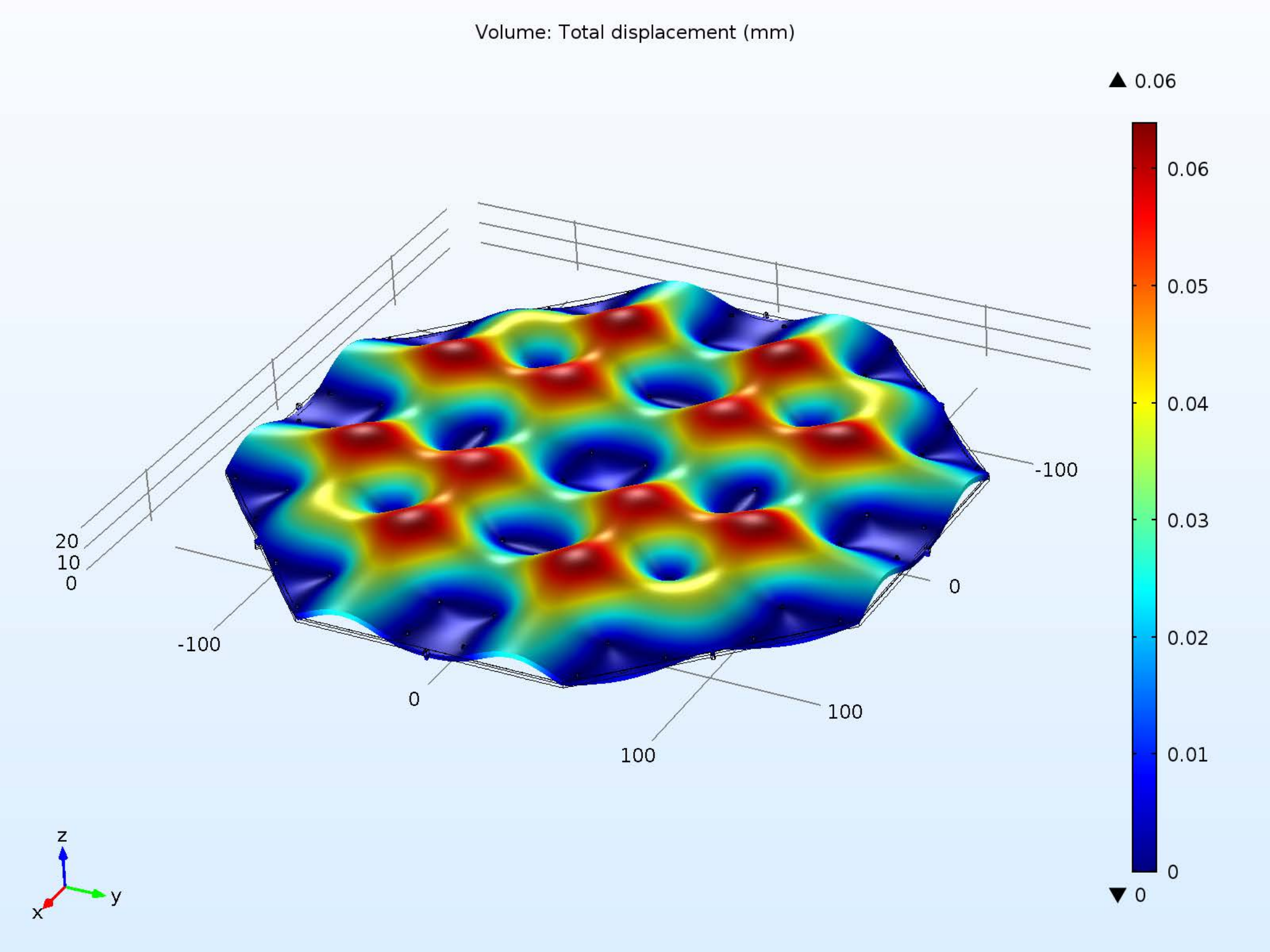}
\caption[Simulation of buoyancy forces on the anode window.]{Simulation of buoyancy forces on the anode window, showing the vertical displacement.}
\label{fig:Bowing-Simulation}
\end{figure}

The bowing of the anode acrylic window is studied via a FEA analysis  in order to optimize the thickness of the anode (as thin as possible for neutron thermalization) and minimize the deformation of the anode face (for \STwo\ uniformity).  Fig.~\ref{fig:Bowing-Simulation} shows the results of a \DSkAcrylicWindowThickness\ thick acrylic anode with a \DSkGasPocketThickness\ gas pocket.  Supporting \ce{Cu} pillars must be placed every \DSkAnodePillarsSpacing\ over the anode window to keep the gas pocket thickness variation to \DSkAnodeVerticalDeflection.  As a comparison, in an anode assembly with no \ce{Cu} pillars, the anode center will bow up more than \DSkAnodePillarsWindowBowingPrevented.

The physics performance of the \LArTPC\ will be described in Sec.~\ref{sec:Physics}.

\subsection{Assembly and Installation}
\label{LArTPC-AssemblyInstallation}

Within the design of the infrastructure, it is anticipated that the \WCV\ water tank itself will serve as a radon-free clean room for assembly and installation of both the \LSV\ and the \LArTPC\ detector.  The main flange of the \LSV\ will be located on the bottom, and so the \LArTPC\ will be assembled on the ground floor, then lifted up into the \LSV\ and placed in the desired position.  The fundamental assembly steps for the \LArTPC\ will be:

\begin{asparaenum}
\item Independently assemble the top \SiPM\ array, bottom \SiPM\ array and field cage;
\item From bottom to top, assemble the bottom \SiPM\ array, cathode acrylic plate, field cage, extraction grid, acrylic diving bell and top \SiPM\ array onto the bottom \ce{Cu} plate of the \LArTPC;
\item Hang the bottom \ce{Cu} plate of the \LArTPC\ from the top flange of cryostat, and level the \TPC\ relative to the cryostat;
\item Connect all the cables, insulated wires and optical fibers to the corresponding feedthroughs, check the conductivity and insulation;
\item Connect all the tubes and pipes inside the cryostat;
\item Close inner vessel and outer vessel of the cryostat successively, performing a helium leak check after each step;
\item Perform comprehensive signal check for all the cables and wires;
\item Lift up the cryostat and place it into the \LSV\ at the desired position, pre-level the cryostat and make all necessary connections.
\end{asparaenum}



\section{\DSp}
\label{sec:Proto}

\subsection{Introduction and Requirements}
\label{sec:Proto-Introduction}

The collaboration will benefit greatly from experience gained with \DSf, however, \DSk\ will be more than two orders of magnitude larger in size and will use new technologies.  Therefore the collaboration plans to build a prototype detector of intermediate size, \DSp, incorporating the new technologies for their full validation.  The choice of the \DSpAproximateMass\ mass scale allows a full validation of the technological choices for \DSk.

The construction, operation, and commissioning of \DSp\ will allow validation of the major innovative technical features of \DSk, including the mechanical and cryogenic design and the integration  of the custom photodetector modules and  the full read-out electronics and data acquisition chain.  It is expected that the developments in the different WGs preparing technology for \DSk\ will proceed at differing paces, due to different constraints and the nature  of their activities.  The plan for \DSp, in each of its phases, will be based on the status and progress of each sub-system.  

It is anticipated that the prototype will evolve as elements of the \DSk\ technologies are incorporated as they become available and usable, {\it i.e.} interface-able to the rest of the chain even with reduced functionality or containing still-developing technology.  \DSp\ is not intended to replace validation and tests made in laboratories, but rather to complement them.  The merit of such an approach is to identify sooner rather than later issues related to functionality and performance that might be revealed only during and after integration.  It will also help to develop a better view of the overall project time-line and of the resources needed, both human and financial, to verify the production readiness of the different institutions and to help the integration of the various teams working on the different \DS\ sub-systems.  For this reason, \DSp\ is scheduled during the first years of the project.

\begin{figure}[ht!]
\centering
\includegraphics[width=0.45\columnwidth]{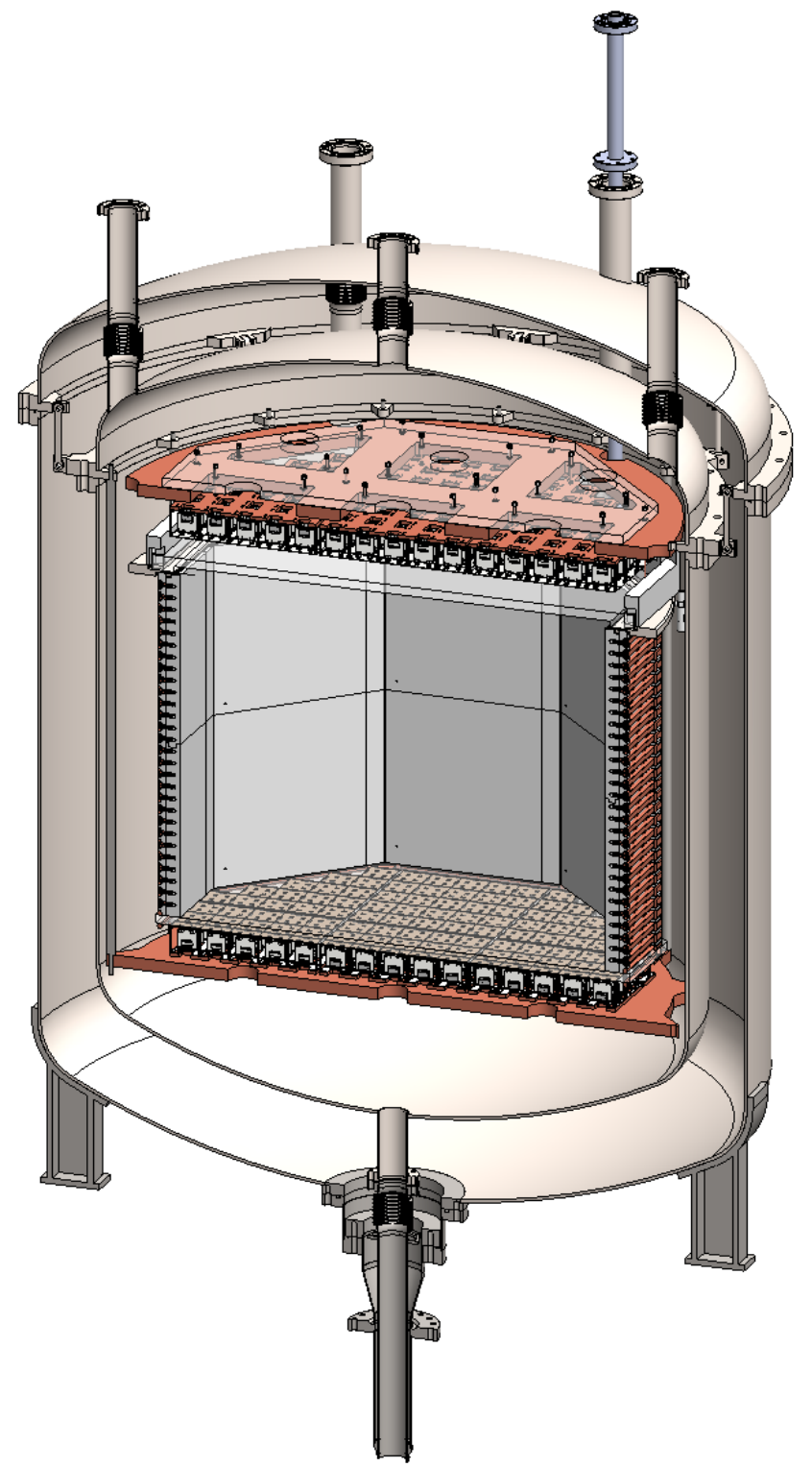}
\caption[3D drawing of the \DSp\ \LArTPC.]{3D drawing of the \DSp\ \LArTPC, shown inside the cryostat.}
\label{fig:Proto-LArTPCWithCryostat}
\end{figure}

The \DSp\ \LArTPC\ must respond to a stringent set of requirements.  In particular, it must:

\begin{asparaenum}[\bfseries {Proto}-i]
\item Be built on a scale sufficient to test the full capabilities of the \DSk\ cryogenics system;
\item Employ materials with the same performance in \LAr\ as those chosen for the \DSk;
\item Enable employment of large area  densely packed custom photosensors array providing high photocathode coverage and high \SOne\ light yield;
\item Enable employment of large scale coated acrylic windows for defining active \LAr\ volume and providing support for thin film transparent electrodes;
\item Enable employment of large area PTFE panels structure as light reflector, electric insulator and holding structure for \ce{Cu} field cage;
\item Enable employment of large-area fine pitch wire extraction grid and provide drift, electroluminescence field and gas pocket thickness uniformity for high resolution of \STwo\ signals;
\item Allow for an integrated test of the HV and electric field systems of \DSk;
\item Verify $x-y$ position resolution to be on the order of, or better than, \DSkXYResolution;
\item Verify coherent co-moving of all \LArTPC\ subsystems together during the detector cool down.
\end{asparaenum}

The full prototype program is broken into three separate phases, with the first set to start at the beginning of 2018 and the final phase being completed during the second half of 2019.  The three separate phases will consists of:

\begin{compactenum}
\item[\bf \DSpPhaseZero:] Test of cryogenic system concept at the test site; identification and preparation of full readout and DAQ of \DSkPdmsFirstBatchNumber\ pre-production \DSkPdms;
\item[\bf \DSpPhaseOne:] Design, construction and assembly at test site of cryostat and \LArTPC\ equipped with \DSkPdmsFirstBatchNumber\ pre-production \DSkPdms; assembly, commissioning, and operation of full read-out and DAQ for \DSkPdmsFirstBatchNumber\ \DSkPdms;
\item[\bf \DSpPhaseTwo:] Assembly and commissioning of full system, including \DSkPdmsSecondBatchNumber\ first production \DSkPdms; full readout and DAQ operational; evolution towards final configuration.
\end{compactenum}

\subsection{Prototype Design Overview}
\label{sec:Prototype-Overview}

The prototype \TPC\ and cryostat are a down-scaled version of the corresponding \DSk\ systems.  The choice of the size has been driven by the requirement of a minimal configuration \TPC\ with the same electrostatic features of the \DSk\ \TPC, large enough to validate the mechanical choices for the modular \SiPM\ assembly concept and allowing for a full functionality test of the \DSk\ gas and liquid recirculation system. Fig.~\ref{fig:Proto-LArTPCWithCryostat} shows a 3D representation of the \DSp\ \LArTPC\ and its cryostat.  

\begin{figure}[t!]
\centering
\includegraphics[width=0.75\columnwidth]{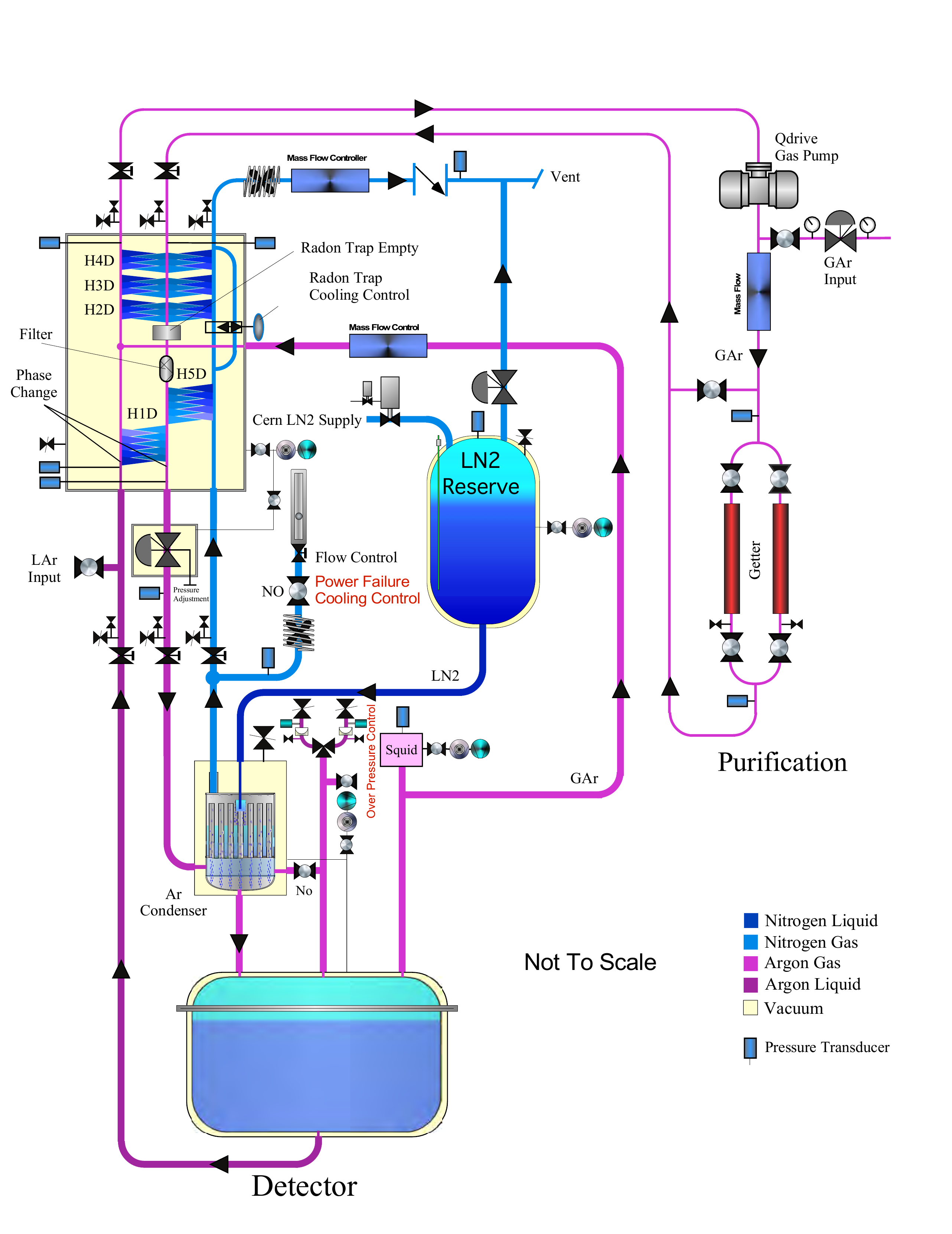}
\caption[Schematic view of the cryogenics system for the prototype detector.]{Schematic view of the design of the cryogenic LAr recirculation system for the prototype detector.}
\label{fig:Proto-Cryogenics}
\end{figure}

\DSp\ will use the final cryogenic system of \DSk, but without the recovery tanks.  All elements required for the \DSp\ cryogenics, shown in Fig.~\ref{fig:Proto-Cryogenics}, are identical to those of the cryogenics of \DSk, shown in Fig.~\ref{fig:Cryogenics-DSkSketch}.  The recirculation rate will be \DSpCryogenicsGasFlowTotal, with the system as shown in Fig.~\ref{fig:Proto-Cryogenics}.  Crucial elements and features that can be tested include:
\begin{compactenum}
\item New upgraded argon condenser, heat exchanger and the \DSkCryogenicsQdriveSpeed\ gas pump;
\item Fast drain recovery and heat recovery systems;
\item Functionality and stability of controls and the TPC level control loop;
\item System safety during power failure;
\item \DSpCryogenicsGasFlowTotal\ purification flow rate.
\end{compactenum}

The baseline option for a low-background structural material for the cryostat remains stainless steel: the plan is to build a stainless steel cryostat to test the cooling and recirculation system during the first phase.  But since the cryostat is one of the major contributors to the neutron background budget in \DSk, it is also planned to explore titanium as a possible alternative for the final vessel.  Recent studies performed within the \DS\ Collaboration have shown that ULR titanium could be used as a low-background structural material, with mechanical strength better than industrial titanium and close to the mechanical properties of stainless steel.  The plan illustrated in Sec.~\ref{sec:Cryogenics-Cryostat} is to develop ULR titanium from the raw material to the industrial scale.  This development will be validated by the production of a second \DSp\ cryostat in radiopure titanium, thus allowing for the accurate test of all the mechanical features as well as for the validation of the construction process, including laser- or electron-beam welding of titanium sheets and rods to build the inner vessel, the outer vessel, and their flanges.

\begin{figure*}[t]
\centering
\includegraphics[width=\textwidth]{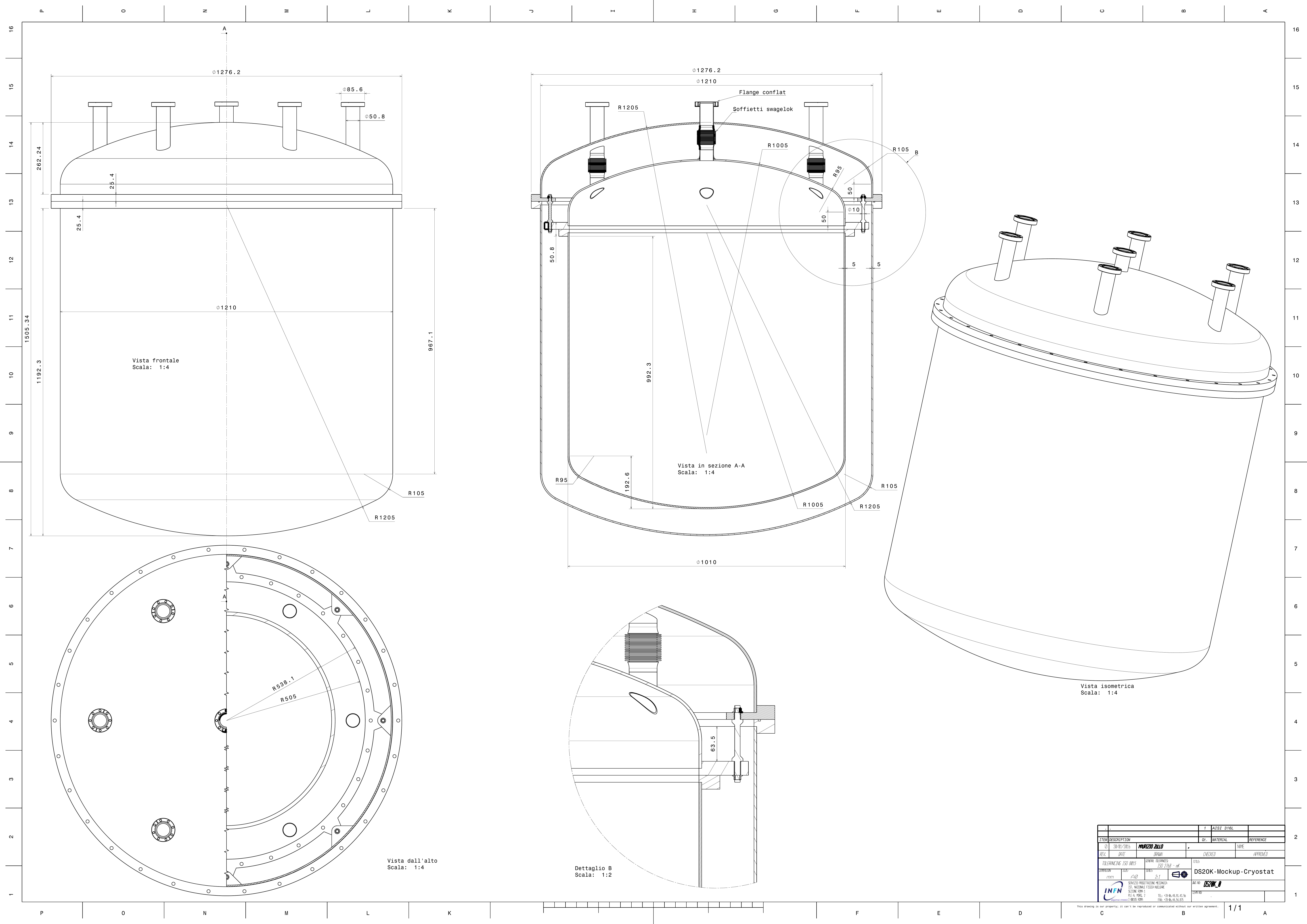}
\caption{2D drawing of the \DSp\ cryostat.}
\label{fig:Proto-Cryostat2D}
\end{figure*}

\begin{figure}[t]
\centering
\includegraphics[width=0.45\columnwidth]{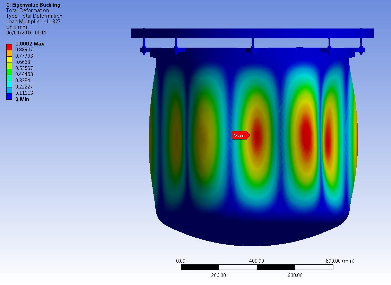}
\caption[FEA analysis of the \DSp\ cryostat.]{Results of the FEA mechanical stress simulation performed for the \DSp\ cryostat.}
\label{fig:Proto-CryostatFEA}
\end{figure}

The outer diameter of the inner and outer vessels of the cryostat for the prototype would be \DSpCryostatOuterVesselDiameter\ and \DSpCryostatInnerVesselDiameter, respectively.  With a choice of a \DSpCryostatWallThickness\ for the wall thickness of both vessels, results of the finite element analysis for the thermal and mechanical loads, including the weight of the \TPC\ has been performed in Rome, indicate that the same design is viable for both stainless steel and titanium as construction materials (for titanium, standard mechanical characteristics were used, corresponding to sample VT1-0 in Table~\ref{tab:Cryogenics-Titanium}).  A final validation of the design is necessary in order to match the mechanical properties of the actual ULR Titanium and of its welds, whose precise characterization is currently undergoing in Milano.  The stainless steel version of the cryostat would weight \DSpCryostatSSMass, while the titanium version is about \DSpCryostatTiMass.  Efforts to optimize the design for even lighter cryostast are ongoing.  Fig.~\ref{fig:Proto-Cryostat2D} shows a 2D drawing of the cryostat that will be used for the prototype detector, while Fig.~\ref{fig:Proto-CryostatFEA} shows the results of the finite element analysis calculation of mechanical stress performed for the titanium cryostat.

The \TPC\ for \DSk\ will be based on the \DSf\ \TPC\ design with several major modifications.  The major change is related to the proposed use of \SiPM\ tiles instead of \PMTs.  Another important change is the geometry of the active volume, changing from cylindrical to an octagonal prism.  In addition the anode, cathode and extraction grids are of new concept.  Finally, the larger \TPC\ also requires new mechanical designs.

\begin{figure}[t!]
\centering
\includegraphics[width=0.45\columnwidth]{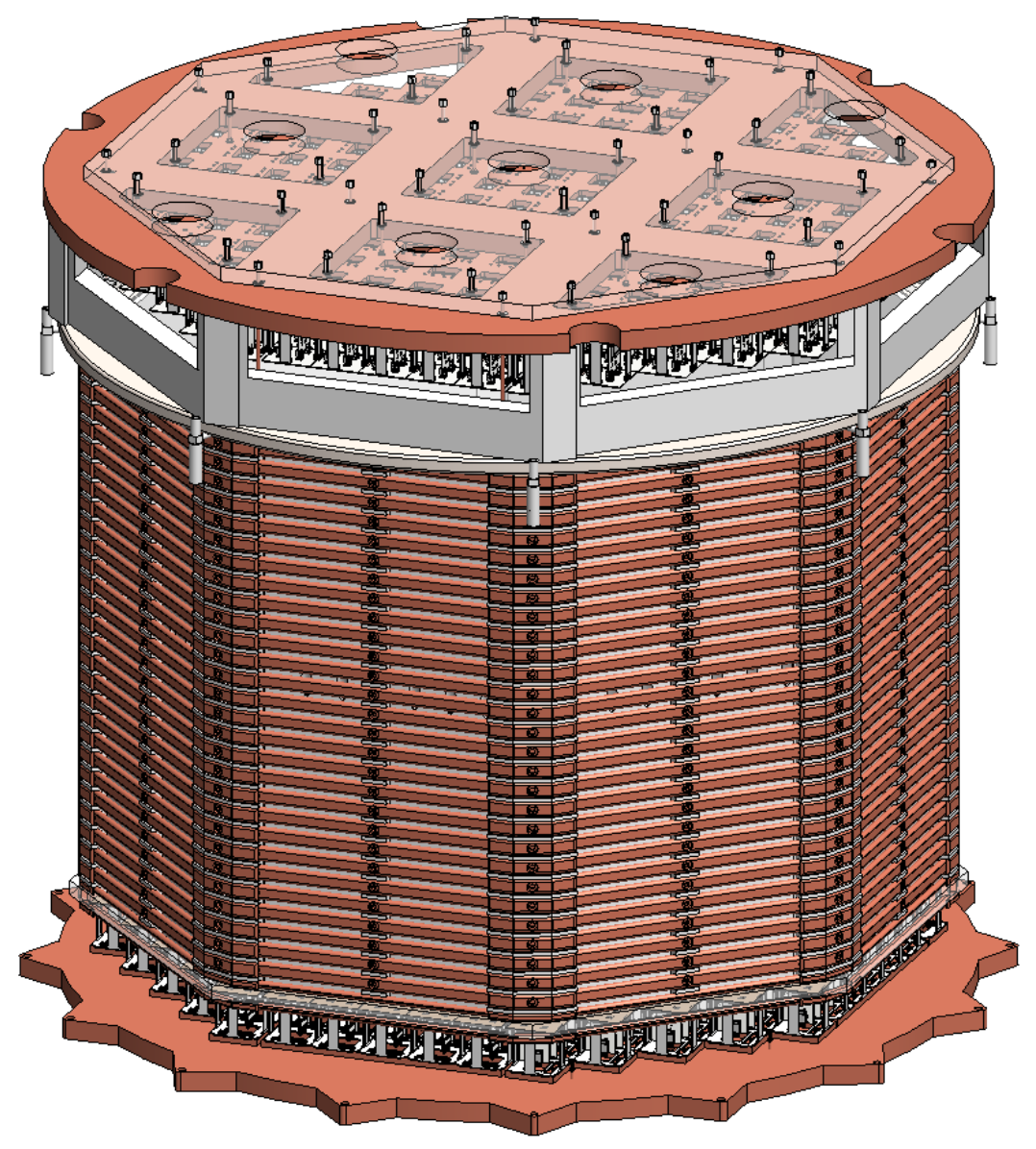}
\caption{3D drawing of the \LArTPC\ prototype detector.}
\label{fig:Prototype-LArTPCNoCryostat}
\end{figure}

Fig.~\ref{fig:Prototype-LArTPCNoCryostat} shows a 3D drawing of the prototype \LArTPC. The minimal configuration needed in order to validate the overall choices for mechanics is based on an active volume defined by an octagonal prism, with the base edge defined by the size of a \SQB\ size (\DSkSQBTRBSize).  The drift length is \DSpDriftLength. 

The top and bottom boundaries of the active volume are two acrylic planes of the same thickness, coated and mounted as in the \DSk\ \LArTPC, acting as anode and cathode.  The extraction grid is also scaled down in diameter from the final detector one, but with the same wire size and pitch.  As stated in Sec.~\ref{sec:LArTPC}, the desired extraction (electroluminescence) field is \DSkExtractionField\ (\DSkElectroLuminescenceField), and the prototype will have the same value for the potential at grid, since it has the same design as \DSk, with \DSkLArOverGridThickness\ \LAr\ extraction field region and \DSkGasPocketThickness\ GAr electroluminescence region.  With a drift field of \DSkDriftField\ over the drift length, the total voltage between anode and cathode will be \DSpCathodePotential.  

In order to achieve exactly the same electrostatic configuration as in the final detector, the number of field shaping rings in the \TPC\ is reduced to \DSpFieldCageRingsNumber.  All shaping rings are evenly placed with vertical pitch of \DSkFieldCageRingsSpacing, and between every two adjacent shaping rings \DSkFieldCageMiddleResistorComponentsNumber\ resistors of \DSkFieldCageMiddleResistorComponentsValue\ are connected in parallel to form a \DSkFieldCageMiddleResistorValue\ equivalent resistance.  The resistance between the topmost shaping ring and the extraction grid, $R_g$, is tuned to \DSkFieldCageTopResistorValue, formed also by \num{4} resistors (\SI{900}{\mega\ohm} each), in order to eliminate the effect of extraction field leakage.  A complete study of the electrostatic field in the prototype \TPC\ was reported and discussed in Sec.~\ref{sec:TPC-Performance}.  The key parameters of the prototype \TPC\ are summarized in Table~\ref{tab:Proto-Parameters}.

\begin{table}
\rowcolors{3}{gray!35}{}
\centering
\caption{\DSp\ \LArTPC\ detector characteristics.}
\begin{tabular}{lc}
{\bf \LArTPC\  Dimensions} 				&\\
\hline
Height									&\DSpTPCHeight\\
Effective Diameter						&\DSpTPCDiameter\\ 
Total \LAr\ Mass						&\DSpTotalMass\\
\hline
\cellcolor{white}Nominal TPC Fields and Grid
										&\cellcolor{white}\\
\hline
Drift Field 							&\DSkDriftField\\
Extraction Field  						&\DSkExtractionField\\
Luminescence Field 						&\DSkElectroLuminescenceField\\
Operating Cathode Voltage 				&\DSpCathodePotential\\
Operating Extraction Grid Voltage  		&\DSkGridPotential\\
Operating Anode Voltage  				&\DSkAnodePotential\\
Luminescence Distance  					&\DSkGasPocketThickness\\
Grid Wire Spacing  						&\DSkLArOverGridPitch\\
Grid Optical Transparency  				&\DSkGridTrasparency\\
\hline
\cellcolor{white}\SiPM\ Tiles			&\cellcolor{white}\\
\hline
Number of Tiles							&\DSpTilesNumber\\ 
Size of Tiles 							&\DSkTileAreaStd\\ 
\end{tabular}
\label{tab:Proto-Parameters}
\end{table}

\subsection{Prototype PhotoElectronics and Readout}
\label{sec:Proto-Readout}

\begin{figure}[t!]
\centering
\includegraphics[width=0.6\columnwidth]{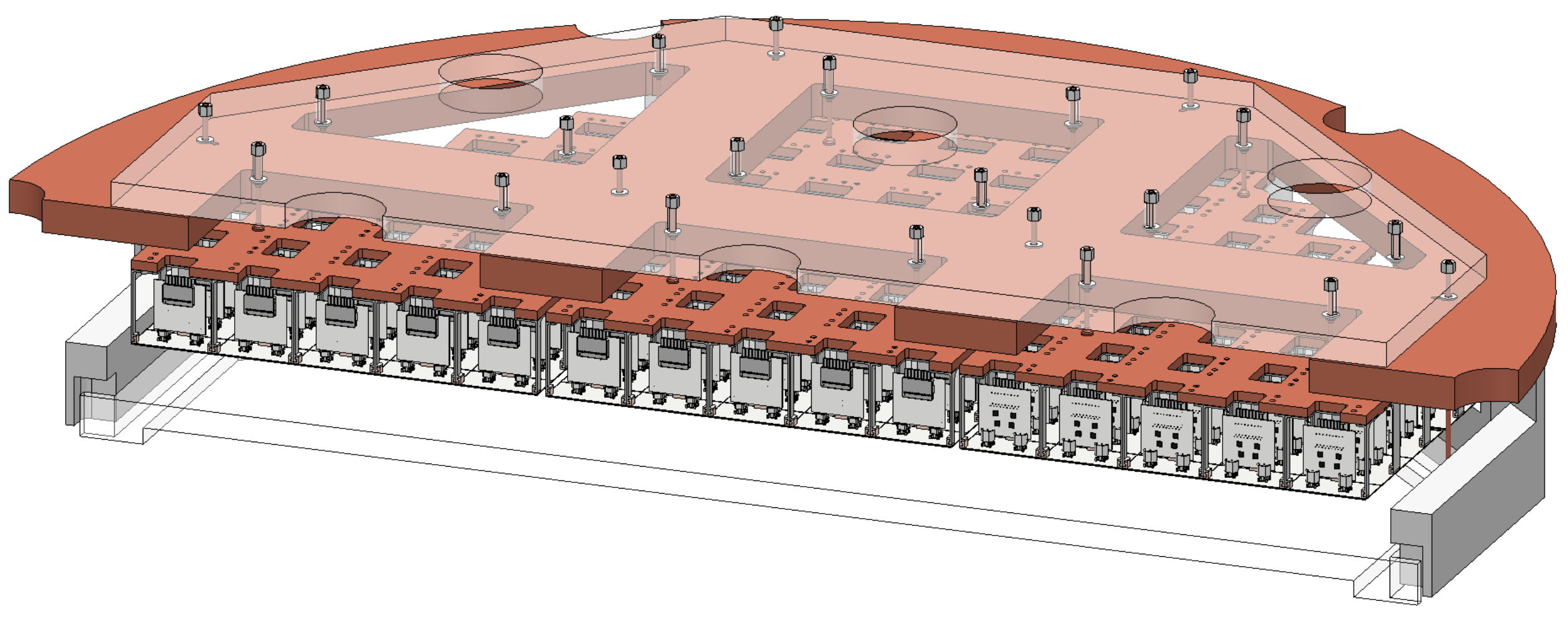}
\caption{Anode assembly of the \DSp\ \LArTPC.}
\label{fig:Proto-AnodeRegion}
\end{figure}

The top photon readout plane of the prototype is shown in Fig.~\ref{fig:Proto-AnodeRegion}.  It consists of \DSpTilesHalfNumber\ photodetector modules assembled into \DSpSQBsNumber\ \SQBs\ and \DSpTRBsNumber\ \TRBs.  The bottom plane is similarly equipped with \DSpTilesHalfNumber\ modules for a total of \DSpTilesNumber\ channels to be readout.

For the initial implementation in phase \DSpPhaseOne, \DSkPdmsFirstBatchNumber\ \tiles\ and \DSkPdms\ will be available, part of the \DSk\ pre-production.  The \tiles\ will be arranged on the cathode and anode to provide a uniform but incomplete coverage of the area.  The \tiles\ will be mounted on \DSkPdms, such as to provide the first stage of pre-amplification and of signal transmission through the cryostat penetration.  The readout of the first \DSkPdmsFirstBatchNumber\ \DSkPdms\ will be based on a prototype receiver board under development that includes a differential receiver, an optimal filter shaper, a signal discriminator, and an output to be fed to a digitizer.  It is planned to use \DSfADCSamplingRate\ digitizers and a majority trigger as done in \DSf.

Once the first prototypes of the final electronics and DAQ chain will be available, the \DSp\ detector will be equipped with them, in order to estimate their full performance for key parameters ({\it e.g.} zero suppression, software/advanced trigger algorithms, online DSP for timing) also in comparison with the response of the well known \DSf\ system. 

For the final implementation in phase \DSpPhaseTwo, \DSkPdmsSecondBatchNumber\ \tiles\ will have been produced as part of the first \DSk\ production, allowing for a full coverage of the readout planes.  Readout of the fully equipped \DSp\ will only be possible using the final electronics and DAQ chain.

\subsection{Infrastructure}
\label{sec:Proto-Infrastructure}

The infrastructure required to install and test the \DSp\ \LArTPC\ has minimal requirements:

\begin{compactenum}
\item A small, soft-wall and HEPA filter class-1000 clean room;
\item \ce{N2} \& \LIN\ to cool and liquefy the argon target;
\item Pressurized \ce{N2} or air for the actuation of the pneumatic valves of the \DSp\ the gas panel;
\item Ar gas to fill the 1-tonne prototype cryostat;
\item Electrical power;
\item Chilled water;
\item An overall footprint of \SI[product-units=power]{5 x 5}{\square\meter} with height of \SI{5}{\meter}.
\end{compactenum}

The assembly of the \DSp\ \LArTPC\ will proceed, as much as possible, following the sequence and approach developed and foreseen for the \DSk, such as to validate (or modify, where necessary) the \DSk\ assembly plan.  The plan for \DSp\ is to break the operations into three distinct steps, the first of which will test the cryogenics system along with the \DSp\ cryostat.  A second phase will include the \DSp\ \TPC\ inside the cryostat, instrumented with a limited number of \DSkPdms.  The third and final phase will have the \TPC\ fully instrumented with the total 370 \DSkPdm\ channels. 

\section{Veto Detectors: \LSV\ and \WCV}
\label{sec:Vetoes}

\subsection{Introduction}
\label{sec:Vetoes-Introduction}

Nuclear recoils induced by single neutron scatters are indistinguishable from \WIMP\ interactions.  Even the large size of the \LArTPC\ does not allow a fiducial volume completely shielded from neutron-induced backgrounds.  External passive shielding provides protection  against neutrons from outside the \TPC\ (cosmogenic or radiogenic from Hall C), but not from neutrons from the components of the \TPC\ itself.  Neutron induced recoils will be suppressed using a liquid-scintillator neutron veto~\cite{Wright:2011ig} (\LSV), a separate detector surrounding the \TPC\ in which the neutrons from both internal and external sources are detected with very high efficiency, and the corresponding recoil events in the \LArTPC\ are identified and rejected.  In addition to removing neutron backgrounds, the \LSV\ also provides in situ measurements of the neutron backgrounds, allowing for reliable predictions of the number of neutron-induced recoils expected in the data sample to be made.  It also has a substantial efficiency for detecting \grs\ from the \TPC\ and cryostat.

To lower the background rate in the \LSV\ in order to allow as low an energy threshold as possible for efficient neutron-detection, additional shielding is required surrounding the \LSV.  The \LSV\ will be surrounded by a large tank of ultra-pure water, instrumented as a Cherenkov detector to veto cosmic rays, the Water Cherenkov Veto (\WCV).  This layered veto concept has been used very successfully in \DSf~\cite{Agnes:2015gu,Agnes:2016fz,Agnes:2016fw}.

Neutrons that enter the \LArTPC\ come primarily from four sources:
\begin{compactenum}
\item Radioactivity in the environment outside the detector;
\item Cosmogenic interactions due to cosmic ray muons;
\item Spontaneous fission reactions in the detector materials;
\item ($\alpha$,n) reactions in the detector materials.
\end{compactenum}

Neutrons from the first two sources are suppressed by the shielding and signals generated by the veto detectors.  Fission reactions that produce neutrons often generate multiple neutrons and high energy gamma-rays, significantly increasing the probability of multiple coincident interactions in either the \LArTPC\ or in the neutron veto.  This leaves ($\alpha$,n) neutrons as the most challenging type of neutron to veto, and much of the design is targeted around vetoing this class of neutrons with high efficiency.

($\alpha$,n) neutrons are produced by the alpha decays of radioisotopes in the detector materials, in particular by the \ce{^238U}, \ce{^235U}, and \ce{^232Th} decay chains.  Cross sections for ($\alpha$,n) reactions result in a neutron yield usually between \NeutronsPerChainDecayUTh\ neutrons per equilibrium decay of the entire chain, with light elements giving the highest yields.  ($\alpha$,n) yield is strongly dependent on the $\alpha$ energy, making the lower sections of the \ce{^238U} and \ce{^232Th} decay chains particularly important.  This has serious implications for the \DSk\ materials assay campaign, discussed in Sec.~\ref{sec:MatAssay}.

It is planned to use a liquid scintillator doped with an isotope with a high cross section for thermal neutron capture.  In \DSf\ this isotope was \ce{^10B}, accomplished by a solution of 1,2,4-TriMethylBenzene (PseudoCumene or \PC) as solvent and 2,5-DiPhenylOxazole (\PPO) as fluor mixed with TriMethylBorate (\TMB)~\cite{Agnes:2015gu,Agnes:2016fz,Agnes:2016fw}.  The use of this mixture is the baseline solution thanks to the successful experience gained in operating \DSf.  However, the collaboration is aware of the safety concerns related to the handling of large amount of PC and TMB and for this reason alternative options are presently under investigation, as detailed in the next section.  

Neutrons can be detected by exploiting two signals, usually defined as prompt and delayed signals.  The prompt signal is produced during the thermalization of the neutrons.  Neutrons lose their energy by scattering off the nuclei of the scintillator, in particular hydrogen.  Protons that are scattered by neutrons produce scintillation light.  The scintillation light of the protons, although heavily  quenched, can be detected by the \PMTs.  The thermalization of the neutrons is a fast process, usually contained in a narrow time window of a few hundred \si{\nano\second} with respect to the scattering on argon, therefore a very low threshold can be used in a narrow prompt window.

The delayed signal is due to the neutron capture.  The thermalized neutron can capture on various isotopes in the scintillator.  \ce{^10B}, with a natural abundance of \BTenNaturalAbundance, is one of the isotopes with the highest thermal neutron capture cross section, \BTenNeutronCaptureCrossSection.  The time scale and the relative probability of capture in boron, hydrogen, or carbon nuclei depends on the amount on boron in the scintillator, with \ce{^10B} dominant at all but the smallest concentrations.  The capture on \ce{^10B} has two branches:

\begin{align}
\ce{^10B} + n \rightarrow~	&\ce{^7Li} + \alpha~(\BTenNeutronCaptureGroundDecayAlphaEnergy)~[\BTenNeutronCaptureGroundDecayBR] \nonumber\\
\ce{^10B} + n \rightarrow~	&\ce{^7Li^*} + \alpha~(\BTenNeutronCaptureExcitedDecayAlphaEnergy)~[\BTenNeutronCaptureExcitedDecayBR] \nonumber\\
							&\hookrightarrow \ce{^7Li^*} \rightarrow \ce{^7Li} + \gamma~(\BTenNeutronCaptureExcitedDecayGammaEnergy). \nonumber
\end{align}

The advantage of the \ce{^10B} final products is that the alpha and \ce{^7Li} have very short range.  Therefore if a neutron is captured on \ce{^10B} in the \LSV, the reaction always produces some scintillation light. An important feature of this signal is that it is independent of the energy of the neutron, meaning that a neutron that has too low an energy to produce a detectable prompt signal may still produce a detectable delayed signal. However, the light yield of alpha and \ce{^7Li} nuclei is highly suppressed due to ionization quenching, causing them to scintillate equivalent to a \LSVOldAlphaQuenchedEnergy\ electron. Detecting these reaction products therefore requires a high light collection efficiency. The time scale of the delayed signal depends on the TMB concentration.  With a \LSVOldTMBConcentration\ volume concentration, the neutron capture time is \LSVOldNeutronCaptureMeanLife, and \LSVOldBTenCaptureFraction\ of neutron captures happens on \ce{^10B} (the rest of the captures happens mostly on \ce{H}), while with a \LSVNewTMBConcentration\ concentration, the neutron capture time is \LSVNewNeutronCaptureMeanLife\ and \LSVNewBTenCaptureFraction\ of neutron captures happens on \ce{^10B}.

The delayed neutron tagging works in this way: a delayed veto window is opened in the \LSV\ after each event in the \TPC. If a scintillation signal above a certain threshold is detected in the \LSV\ during the delayed veto window, the \TPC\ event is discarded. The threshold must be low enough to reliably detect the signal due to the alpha plus \ce{^7Li} signal from the ground-state capture. The concentration of TMB drives the width of the delayed veto window needed to reach the required veto neutron efficiency.  The window width then determines the fraction of the \TPC\ events that will be accidentally discarded, due primarily to scintillation from radioactivity in the \LSV\ components.  The requirements on the radioactive contamination in the \LSV\ are derived requiring an acceptable level of \TPC\ event loss due to accidental background in the veto delayed window.

\subsection{Baseline Design}
\label{sec:Vetoes-BaselineDesign}

The geometry of the new \WCV\ water tank has been optimized to provide the required amount of shielding for the \LSV\ and \LArTPC, but still fit within the constraints of Hall~C.  The new stainless steel water tank will have a cylindrical shape base \DSkWCVDiameter\ in diameter and \DSkWCVBarrelHeight\ tall, with a dome-shaped ceiling having a radius of \DSkWCVDomeRadius.  The total height of the water tank is \DSkWCVTotalHeight.  The new \LSV\ will be \DSkLSVDiameter\ in diameter, lined with Lumirror reflective foil to increase light collection, and equipped with \PMTs\ as was done in \DSf.  However, the \LSV\ will use \DSkVetoPMTDiameter\ diameter \PMTs, as described in the next section.  The \LSV\ will be filled with the same boron-loaded liquid scintillator as was used in \DSf, but with the TMB concentration optimized for the new detector.  The water tank will also be instrumented with \DSkVetoPMTDiameter\ \PMTs\ (same type as the \LSV) and lined with Tyvek to provide ample light collection.  It will provide at least \DSkLSVWaterShield\ of active shielding to the \LSV\ in all directions.

To achieve at least the same light yield per unit of quenched energy inside the new \LSV, the surface area covered by \PMTs\ will be increased with respect to what was used in \DSf.  This will compensate for the increase in surface area and for the loss of light due to the moderate attenuation length (a few to ten meters) of the scintillator cocktail.  \DSkLSVPMTsNumber\ \DSkVetoPMTDiameter\ \PMTs\ will be used in the new \LSV, leading to $\sim$\DSkLSVPMTCoverage\ photocathode coverage, more than sufficient to see the $\alpha$ following a neutron capture on \ce{^10B} to the \ce{^7Li} ground state.  \DSkWCVPMTsNumber\ \DSkVetoPMTDiameter\ \PMTs\ are sufficient to instrument the \WCV.  In total, the veto system will require \DSkVetoPMTsNumber\ \DSkVetoPMTDiameter\ \PMTs.

As anticipated, there are safety concerns related to the handling of large amount of \TMB, due to its very low flash point, as well as \PC, due to its low flash point.  In addition, \TMB\ and \PC\ have substantial vapor pressures at room temperature, much higher than those of linear alkylbenzene (\LAB) and di-isopropylnaphthalene (\DIN).  In reaction to the hypothesis that LNGS safety regulations could disfavor the use of \TMB\ and \PC, the collaboration is actively investigating the possibility of using a scintillator with a safe choice for the solvent and for the dopant enhancing the neutron capture rate and the associated signal.  

\begin{table*}
\rowcolors{3}{gray!35}{}
\centering
\caption[Chemical properties of  solvents under consideration for the veto.]{Chemical properties of solvents under consideration for the veto. \PC\ is pseudocumene; \DIN\ is Di-isopropylnaphthalene; \LAB\ is Linear alkyl benzene; and \PXE\ is phenylxylylethane. \TMB\ is trimethylborate and is included in this table for the sake of comparison of the chemical properties.}
\begin{tabular}{lcccc|c} 
										&\PC				&\DIN			&\LAB		&\PXE   			&\TMB \\
\hline
Formula                 							&\ce{C_9H_12}		&\ce{C_16H_20}	&\minitab{c}{\ce{C_6H_5C_nH_{2n+1}}}{[\num{10}$<$\mbox{n}$<$\num{16}]}																				&\ce{C_16H_18} 	&\ce{B(OCH_3)_3} \\
\hline
Flash point [\si{\celsius}]						&\num{48}			&\num{>140}		&\num{130}	&\num{145}		&\num{-8}	\\
Vapor pressure at room temperature [\si{mbar}]		&\SI{1.6}{}			&\SI{0.77}{}		&\SI{<0.07}{}	&\SI{8E-4}{}		&\SI{4E3}{} \\
Density [\si{\gram\per\milli\liter}]					&\num{0.89}		&\num{0.96}		&\num{0.86}	&\num{0.99}		&\num{0.93} \\
Hydrogen atoms density [\SI{E22}{\per\cubic\cm}]	&\num{5.35}		&\num{5.45}		&\num{6.31}	&\num{4.34}		&\num{4.90} \\
Absorption maximum [\si{\nm}]					&\num{267}		&\num{279}		&\num{260}	&\num{269}		&\num{270} \\
\end{tabular}
\label{tab:solvent}
\end{table*}

Table~\ref{tab:solvent} shows the solvents under consideration and a few selected physical properties of interest for the selection.  \PC\ and \TMB\ have been included for the purpose of comparison.  The higher flash point and lower room temperature vapor pressure of \DIN, \LAB, and phenylxylylethane (\PXE) makes them particularly attractive as a choice for solvent; an additional benefit is that they are less chemical aggressive than PC.  \LAB\ is particularly interesting as it is gaining rapidly acceptance for use in many large volume detectors.  The loss of scintillation light yield when switching from a \PC-based scintillator to a \LAB-based scintillator is limited: the yield reported for \LAB+\PPO+bisMSB is \SI{70}{\percent} of that of \PC+\PPO~\cite{Buck:2016hk} and it can be compensated, where needed, by a modest increase in the number of \PMTs.  A recent development~\cite{Bentoumi:2012gn} identified the possibility of loading organic liquid solvents with an organoboron compound of fairly limited safety and environmental impact: ortho-carborane, (1,2-Dicarbadodecaborane(12), \ce{C_2B_10H_12}), a solid at room temperature with a melting point of \SI{260}{\celsius}, can be dissolved in \LAB\ and also in \DIN\ up to a concentration of \SI{8.5}{\percent} in weight~\cite{Chang:2015js}, thus providing a safer alternative to \TMB. 

Another nucleus  with large neutron capture cross section is \ce{^6Li}. Indeed, the  reaction with neutron produces charged particles with a very short range:
\begin{equation}
\ce{^6Li} + n \rightarrow t\,(\LiSixNeutronCaptureTritonEnergy) + \alpha\,(\LiSixNeutronCaptureAlphaEnergy)
\end{equation}
In this case, though, the higher energy of the $\alpha$ particle and the lower quenching of the triton result in an electron equivalent energy from \LiSixNeutronCaptureTritonAlphaQuenchedEnergy~\cite{Bass:2013jk}, nearly ten times as large as that for the lowest energy $\alpha$ particle from capture on \ce{^10B} cited above.  A number of authors~\cite{Bass:2013jk,Kim:2015kz,Kim:2015fn}, and among them the PROSPECT Collaboration~\cite{Ashenfelter:2015ku}, studied in detail the possibility of loading organic liquid solvents with aqueous solutions of \ce{^6LiCl}, and their collective work has established a sound and safe avenue for the development of \ce{^6Li}-loaded organic liquid scintillators.

In order to study the backgrounds induced by neutrons originating from ($\alpha$,n) reactions induced by the radioactive contaminants of the TPC are generated and tracked using  the Monte Carlo.  The fraction of these neutron events producing a signal in the \LArTPC\ and surviving a set of standard \TPC\ cuts cannot be distinguished by a true \WIMP\ energy deposit without using the information of the veto detectors.  The energy deposit in the \LSV\ is analyzed within a time window of about \num{7}~$\tau$'s around the TPC signal ($\tau$ is the thermal neutron capture time relative to the specific dopant of choice for the scintillator).  The fraction of events that are not tagged as neutrons as a function of the minimum threshold energy in the \LSV\ represents the residual background. 

The comparison between the four scintillator options has to consider what threshold can be applied vs deadtime. The short capture times of the two boron-based scintillators allow to lower the threshold down to a few photoelectrons (\SI{6}{\pe} in \DSf): even in presence of ~2.5 kBq of \ce{^14C}, the deadtime introduced by the veto will be \textless 8\%. For the other two scintillator options, the long capture times requires one to set the threshold above the \ce{^14C} Q-value (\SI{\sim156}{\keV}), where the neutron rejection efficiency is 3-4 times lower than in the boron case. 

\begin{figure}[t!]
\begin{center}
\includegraphics[width=0.45\columnwidth]{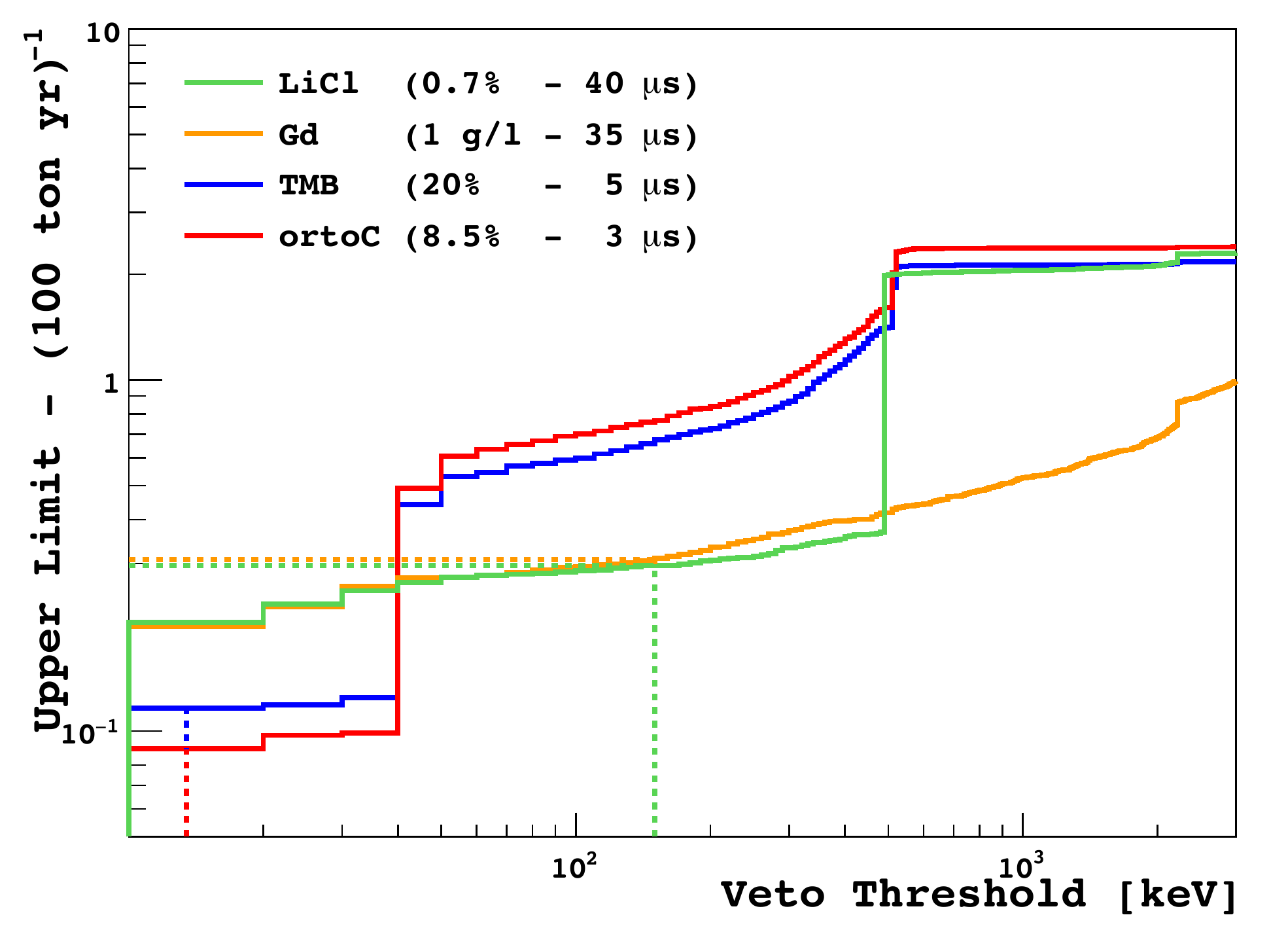}
\caption{Upper limit to residual neutron background after \LSV\  cuts.}{Upper limit to residual neutron background  after \LSV\ cuts for four different possible loading of the \LSV\ solvent with scintillator dopants, as indicated in the legenda, with the relative concentration and thermal neutron capture time. The dotted lines indicate realistic  thresholds values that can  be applied to limit the dead time, as explained in the text.}
\label{fig:LNGS-ScintillatorDopant}
\end{center}
\end{figure}

Fig.~\ref{fig:LNGS-ScintillatorDopant} shows the upper limit on the residual background in a \DSkExposure\ exposure, more about how this analysis was performed in Sec.~\ref{sec:Physics-Background}.  The curves were obtained with a fraction of dopant that has already been tested by other experiments.  

Indeed, the option of loading the scintillator with orto-carburane turns out to be particularly appealing. Further studies on the long term chemical and optical stability, chemical compatibility etc. are needed to fully validate the choice.  

\subsection{\PMTs}
\label{sec:Vetoes-PMTs}

\Geant-based MC simulation studies demonstrated that the choice of the \PMT\ diameter does not affect the physics performance of the \LSV\ and \WCV\ if the total photocathode coverage remains fixed.  The option to instrument the \LSV\ and \WCV\ with the same \DSkVetoPMTDiameter\ \MCPPMTs\ developed for the JUNO neutrino detector~\cite{An:2016el} has been decided as the baseline design.  The choice of \DSkVetoPMTDiameter\ \MCPPMTs\ allows to reduce the number of cables, connectors, and electronics channels, containing the cost and simplifying the design and construction of the experiment.  The expected performance of the \MCPPMTs\ is summarized in Table~\ref{tab:Veto-PMT}.  The \LSV\ will be equipped with \DSkLSVPMTsNumber\ \DSkVetoPMTDiameter\ \PMTs\ mounted on the stainless steel sphere;  the \WCV\ will be equipped with \DSkWCVPMTsNumber\ \DSkVetoPMTDiameter\ \MCPPMTs\ mounted to the inner surface of the water tank.

\begin{table}
\rowcolors{3}{gray!35}{}
\centering
\caption[Expected performance of the veto \PMTs.]{Expected performances of the \MCPPMTs\ in construction for JUNO and to be used in \DSk.}
\begin{tabular}{lc}
{\bf Item}							&{\bf Value}\\ 
\hline
Type								&\MCPPMT\\ 
Gain								&\num{E7}\\
QE at \TPBWaveLength\				&\SI{>25}{\percent}\\
Peak/Valley						&\num{>3}\\
TTS (top point)						&\SI{<15}{\nano\second}\\ 
Rise time							&\SI{2}{\nano\second}\\ 
Fall time							&\SI{12}{\nano\second}\\ 
Anode Dark Rate at \RoomTemperature	&\SI{<30}{\kilo\hertz}\\ 
After-Pulse Time					&\SI{4.3}{\micro\second}\\
After-Pulse Rate					&\SI{<3}{\percent}\\ 
\end{tabular}
\label{tab:Veto-PMT}
\end{table}

The \MCPPMTs\ of choice contain a double stage micro-channel plates (\MCP) instead of a dynode chain.  A voltage divider is still necessary to provide the proper voltage to the front and back side of the \MCP\ and to the anode which collects the multiplied electrons.  The socket with the voltage divider will be mounted on the back side of the \MCPPMT\ and a single cable will run from this socket to the front end board located outside the water tank.  In the front end board the \MCPPMT\ signal will be decoupled from the HV bias.  The cables from both the \LSV\ and \WCV\ \MCPPMTs\ will be immersed in deionized water for many years.  Based on the \BX\ and \DSf\ experience, cables and connectors designed for submarine operation that have demonstrated proven success and ensure reliable working conditions will be used.

With this choice of \PMTs, the signal from a single photoelectron (\SPE) is expected to be a pulse of \DSkVetoPMTSPEHeight\ in amplitude and \DSkVetoPMTSPEWidth\ FWHM.  Very preliminary simulations suggest that scintillation events with energy higher than a few hundred \si{keV} are no longer in the \SPE\ regime.  Simulations also suggest that most of the scintillation light is collected in \DSkSimulationGeantLSVCollectionTime, with tails extending to \DSkSimulationGeantLSVTailsTime.  The dark rate of the \LSV\ \MCPPMTs\ is expected to be \DSkVetoPMTDCR\ at \RoomTemperature.  Muons in the \LSV\ produce a huge scintillation signal that can last for several microseconds, blinding the \LSV\ \MCPPMTs\ for this time.  After-pulses typically occur in a time window of \DSkVetoPMTAfterPulseTime, with a probability which is \DSkVetoPMTAfterPulseProb\ per \SPE\ of signal.

The \MCPPMTs\ must be encapsulated using low radioactivity materials that have long-term chemical compatibility both with deionized water and with the organic liquid scintillator.  The sealing solution already used in the \BX\ water Cherenkov muon veto~\cite{Alimonti:2009dd}, which showed high performance over many years, will be adapted to these large \MCPPMTs\ .  To be specific, the baseline design uses this style of encapsulation for both the \LSV\ and \WCV\ \PMTs.

\begin{figure}[t!]
\begin{center}
\centering
\includegraphics[width =0.5\columnwidth]{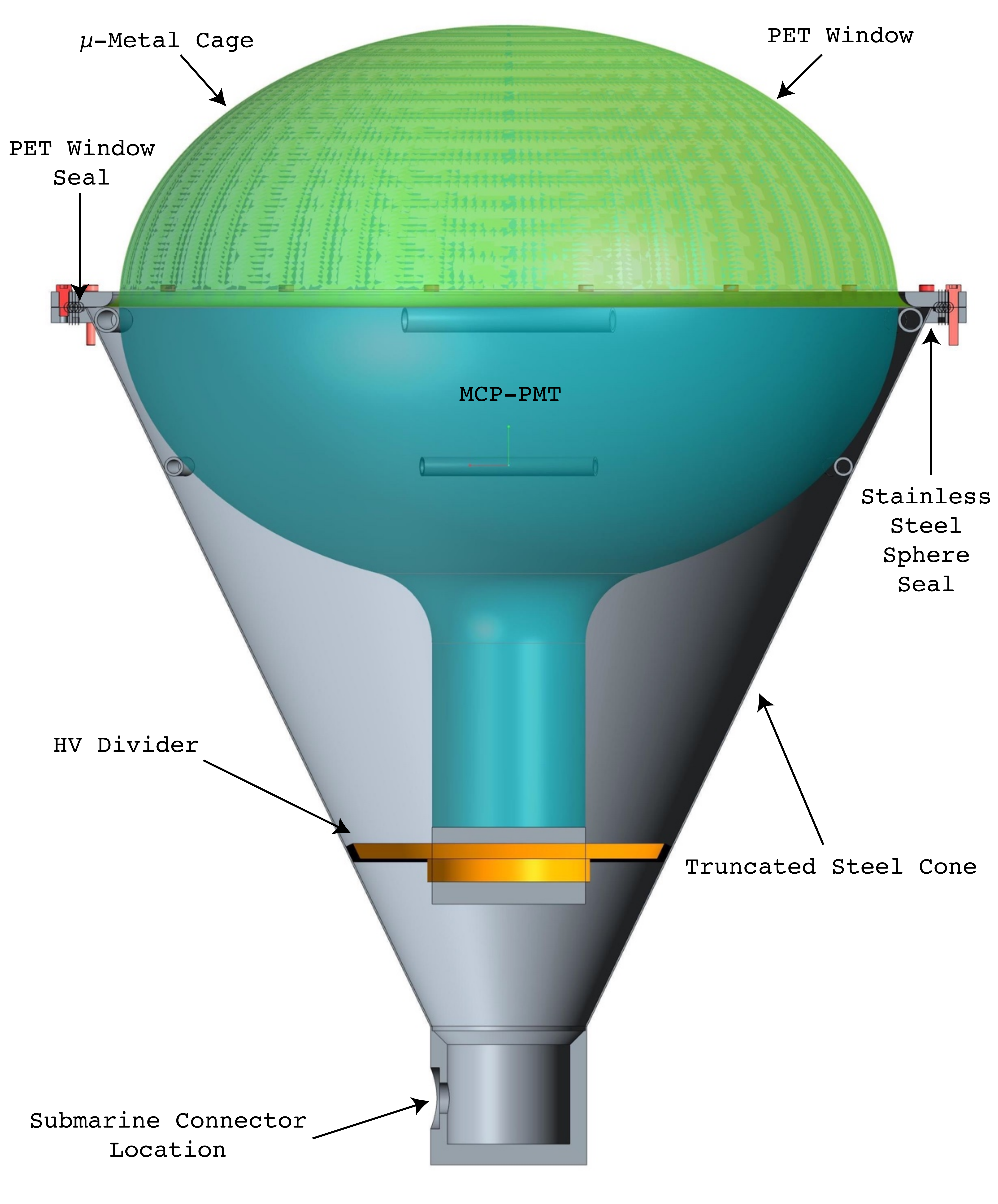}
\caption{Design principle of the encapsulation of a \MCPPMT.}
\label{fig:Vetoes-PMTEncapsulation}
\end{center}
\end{figure}

The \PMTs\ will be fully encapsulated within a leak-tight case made of two pieces:
\begin{inparaenum}
\item A truncated stainless steel cone, shown in Fig.~\ref{fig:Vetoes-PMTEncapsulation}, that surrounds the \PMT\ and the socket;
\item A thin, transparent, polyethylene terephthalate (PET) window, conforming to the shape of the photocathode and covering it.
\end{inparaenum}
The stainless steel truncated cone has a flange as shown in Fig.~\ref{fig:Vetoes-PMTEncapsulation} with a diameter slightly larger than the maximum \PMT\ diameter.  This flange has an initial sealing surface on the side facing the \MCPPMT\ photocathode, allowing the transparent window to make a leak-tight seal with the truncated cone.  The stainless steel truncated cone extends to host a second sealing surface, on the side opposite to the \MCPPMT\ photocathode, which anchors the encapsulated \MCPPMT\ to the stainless steel sphere and makes a leak-tight seal with the sphere itself.  For the mounting of the \MCPPMTs\ in the \WCV, the first sealing surface is still used to ensure a leak-tight seal between the stainless steel truncated and the thin transparent photocathode window, however, the second sealing surface is not used, and so its bolt holes are instead used to mount the encapsulated \MCPPMT\ to proper mechanical supports.

The stainless steel truncated cone is equipped, at the end opposite to the \MCPPMT\ photocathode, with a submarine connector, mating to the submarine connector of the cable.  As for the \PMTs\ of the \BX\ water Cherenkov muon veto~\cite{Alimonti:2009dd}, the volume delimited by the stainless steel truncated cone and by the thin transparent photocathode window will be filled by transparent mineral oil.  The voltage divider will be embedded within a proper resin, acting as another barrier between the water and the scintillator.  Extensive tests on prototypes and screening of materials to ensure both the chemical compatibility and the needed low radioactivity will be performed: a baseline solution which consists of a replica of the materials in use for the \PMTs\ of the \BX\ water Cherenkov muon veto~\cite{Alimonti:2009dd} already exists.

Following the lead of Ref.~\cite{Alimonti:2009dd}, optical fibers will be mounted such as to illuminate each \MCPPMT, allowing to perform precise, periodic calibrations and monitoring of the \MCPPMTs\ gain.  Fibers serving the \WCV\ \MCPPMTs\ will be spliced inside the water tank, and fibers serving the \MCPPMTs\ of the \LSV\ will be spliced inside the stainless steel sphere, also following the lead of Ref.~\cite{Alimonti:2009dd}, such as to minimize the number of fiber optic feedthroughs.  All fibers will be connected to a pulsed laser source with adjustable power, eventually tuned to provide an illumination level equivalent to a fraction of \si{\pe} per \MCPPMT\ per pulse.

\begin{table*}[t!]
\rowcolors{3}{gray!35}{}
\normalsize
\centering
\caption[Relative \PMT\ \PDE\ with and without magnetic field shielding.]{Relative \PDE\ for various applied magnetic fields, without and with $\mu$-metal cage, and with only half (hemisphere) of the $\mu$-metal cage.  See text for details.}
\begin{tabular}{cccc}
Field Strength and Orientation
					&No Cage			&Full Cage		&Half Cage Only\\
\hline
Null					&\num{1.00}		&\num{0.96(4)}		&\num{1.02(4)}\\
\SI{20}{\micro\tesla}, Dynodes Axis $\parallel$ $B$
					&\num{0.74(4)}		&\num{0.92(4)}		&\num{0.89(4)}\\
\SI{40}{\micro\tesla}, Dynodes Axis $\parallel$ $B$
					&\num{0.17(4)}		&\num{0.71(4)}		&\num{0.38(4)}\\
\TMF\ (\SI{55}{\micro\tesla}), Avg. of Random Orientations
					&\num{0.53(4)}		&\num{0.85(4)}		&\num{0.62(4)}\\
\end{tabular}
\label{tab:PMTMagneticFieldShielding}
\end{table*}

The terrestrial magnetic field (\TMF) causes the degradation of the characteristics of the large area \PMTs\ because of the deflection in the trajectories of the photoelectrons drifting from the photocathode to the first dynode.  Trajectories of secondary electrons in the dynode chain can also be affected, depending on the orientation of the \PMT\ relative to the \TMF.  The \TMF\ effects are increased with increased photoelectron path length inside the \PMT\ (longer path from the photocathode to the first dynode).  Photoelectrons are defocused causing losses in collection efficiency on the first dynode and degradation of gain at the first dynode, resulting in worse SPE resolution transit time spread.  The \TMF\ effects increase with the increasing size of the \PMTs~\cite{Aiello:2012hf}.  These effect can be compensated by shielding the \TMF\ in the entire detector, such as by covering the entire detector with large Helmoltz coils, or by shielding the \TMF\ locally, for example by the use of $\mu$-metal screens or cages.  For \PMTs\ with a classic dynode system, the effects can also be reduced by properly orienting the dynodes axis in a direction orthogonal to the \TMF.  The preferred option is the use of $\mu$-metal cages, in form of a hemispherical mesh of $\mu$-metal wires placed in front of the \MCPPMT\ photocathode.  These hemispherical cages were already successfully applied to compensate the \TMF\ for the smaller (\SI{8}{\inch}) \PMTs\ in the \DSf\ experiment, and are being developed for use in JUNO.

Initial tests were performed using a Hamamatsu \DSkVetoPMTDiameter\ \PMT\ with a classic dynode system, covered by a spherical cage composed by two hemispheres made by $\mu$-metal mesh.  Results are presented in the Table~\ref{tab:PMTMagneticFieldShielding}.  The relative \PDE\ is the main parameter to check, normalized to the first run performed in near-absence of magnetic fields (\TMF\ almost completely compensated).  Data at \SI{20}{\micro\tesla} and \SI{40}{\micro\tesla} were collected with the main dynode axis oriented in the direction parallel to the field, which maximizes the magnetic field disturbance.  Data for the \TMF\ represent the average of random orientations of the \PMT\ dynodes axis.  Blank runs with an ideal compensation of the \TMF\ were performed throughout the experiment, to provide a first order assessment of possible statistical errors.  As one can see from the table, the cage options provide a satisfactory path towards the goal of containing the loss in \PDE\ due to the \TMF\ within \DSkVetoPMTTMFPDELossRequirement.  

While the baseline solution for shielding the veto \PMTs\ from external magnetic fields is to used a $\mu$-metal cage, as was done for \DSf\ and is currently being developed for the \DSkVetoPMTDiameter\ \PMTs\ intended for JUNO, another option is also under investigation.  This second option consists of the use of $\mu$-metal foil, which would be rolled into a cone and wrapped around the bottom portion of the \MCPPMT.  This solution would have the advantage of eliminating the cage in front of the photocathode and the associated loss of light due to the blockage by the mesh.  Also, initial tests have shown that this solution can provide better shielding of the external fields without the need for annealing the material, as is required for the $\mu$-metal cages.  This option will also be studied in detail.  A final decision on the best solution will be made in advance of the time required for production and delivery of the components.  

\subsubsection{\PMTs\ Backup Solution}
\label{sec:Vetoes-PMTs-Backup}

A back-up solution would be necessary if the JUNO \PMTs\ were not able to be delivered on time or they did not meet the requirements.  The alternative option is that the same number of \DSkVetoPMTDiameter\ Hamamatsu \PMTs\ be used, which have been used in past experiments and will also be used inside the JUNO detector to provide improved timing resolution.  The performance of the \PMTs\ being developed for JUNO is set to match that of the Hamamatsu \PMTs\ in all but timing.  The encapsulation developed for the \DSkVetoPMTDiameter\ \MCPPMTs\ would still be compatible for use with the \DSkVetoPMTDiameter\ Hamamatsu \PMTs, modulo the necessary mechanical adaptations.

\subsection{\LSV\ and \WCV\ Front-End Electronics}
\label{sec:Vetoes-FrontEnd}

The veto detectors' front-end electronic system must be able to handle the \LSV\ and \WCV\ \PMTs\ signals, extracting charge and time information.  The baseline design follows that of the \DSf\ front-end system~\cite{Agnes:2016fw}, implementing different modularity and a better mechanical arrangement.  R\&D is currently underway to provide a prototype of the full system.

\begin{figure*}[t!]
\centering
\includegraphics[width=\textwidth]{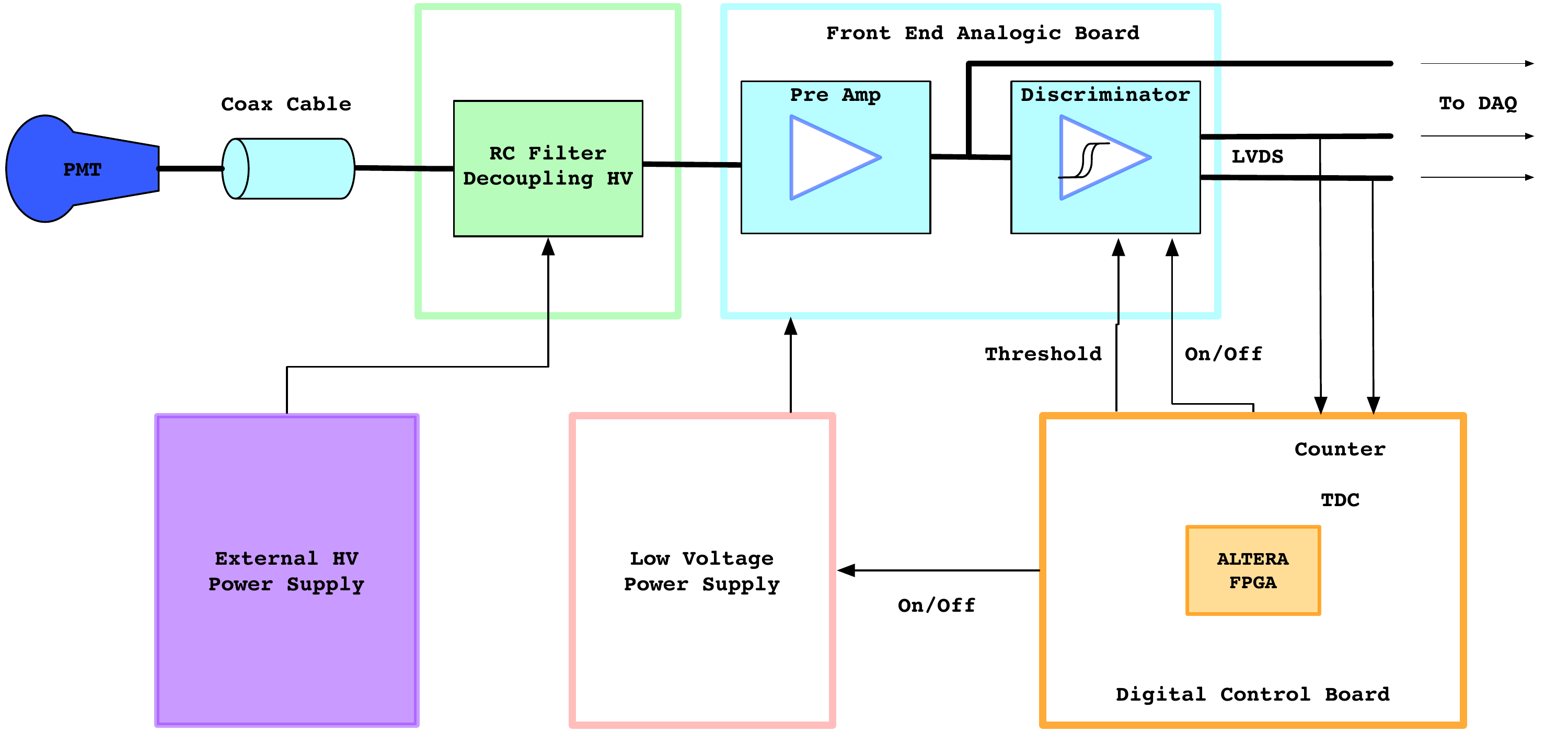}
\caption{Vetoes front end block diagram.}
\label{fig:VetoesFrontEndBlockDiagram}
\end{figure*}

Fig.~\ref{fig:VetoesFrontEndBlockDiagram} shows the block diagram of the front-end electronics.  Each \PMT\ will be connected to the electronic system with a single coaxial cable (RG213-like) used for both HV supply and analog signal collection (see green box in Fig.~\ref{fig:VetoesFrontEndBlockDiagram}).  The HV bias will be provided by an external power supply (gray box, CAEN-SY127-like) and will be decoupled from the signal with a simple $RC$ network.  The analog signal pulse will then be amplified and discriminated (see cyan box in Fig.~\ref{fig:VetoesFrontEndBlockDiagram}).

The specifications for the preamplifier are:

\begin{compactenum}
\item Bandwidth greater than \DSkVetoPreAmpBandwidth;
\item Gain of \DSkVetoPreAmpGain;
\item $Z_{\rm in} = \DSkVetoPreAmpInputImpedance$;
\item $\SNR > \DSkVetoPreAmpSNR$.
\end{compactenum}

The preamplifier output signal is output directly and also feeds a leading-edge discriminator with a programmable threshold.  Multiple discriminators feed a local veto digital control board (\VDCB) and the overall DAQ system, described in Sec.~\ref{sec:DAQ}.

A highly modular design is proposed, in which \DSkVetoVFEBChannels\ preamplifier-discriminator channels will be put on a single printed circuit board (\PCB) called the veto front end board (\VFEB).  The \VFEB\ \PCB\ dimensions will be \DSkVetoVFEBDimensions, a standard double Eurocard module.  On this card, the input \PMT\ signals come from the crate backplane connectors (right side in the picture); the analog output signals will be put on the board front panel to be easily connected to the external digitizers.  The discriminator digital signals will be sent to the backplane connectors to be handled by the \VDCB.  Moreover, pulses from four channels are summed together providing additional information that can be used for coarse information about the energy of an event. In total, two summed outputs will be available on the front panel.  Up to \DSkVetoPerCrateVFEBs\ \VFEBs\ will be housed in a standard \SI{19}{\inch} 6-U crate having a modular unit of \DSkVetoPerCrateChannels\ channels.

The HV decoupling circuits will be implemented in a separate board, handling \DSkVetoHVDecouplerChannels\ channels, placed in the rear part of the crate.  There will be \DSkVetoPerCrateDecouplers\ HV decoupling circuits per crate.  This will help a lot during system setup and cabling: the \PMTs' RG213 cables (heavy and rigid) will be connected once to the system and never touched again.  Analog signals will be distributed to the \VFEB\ via a dedicated backplane using tiny U.FL connector coax cables (state of the art in the mobile phone era).

The low voltage power supply and the \VDCB\ will also be housed in the crate.  The \VDCBs\ functions will include:

\begin{compactenum}
\item \VFEB\ parameter settings (On/Off, thresholds, etc.);
\item Slow control (measure of voltages, currents, temperatures, etc.);
\item Event counters (scalers);
\item Time-to-digital conversion (TDC) of discriminator signals;
\item Trigger logic.
\end{compactenum}

The \VDCB\ will be implemented using state-of-the-art field programmable gate array (FPGA) components: all logic functions are in a single chip, giving the flexibility to change the operational code at any time during the development process and also during the lifetime of the apparatus.  The firmware upgrade can also be done remotely, thus avoiding access of the board.  TDCs with \si{\nano\second} resolution, which is adequate for veto \PMT\ signals, can easily be implemented in the FPGA.  Current R\&D work is focusing on a commercial ALTERA Cyclone~IV device as the possible FPGA candidate for the \VDCB\ implementation.  The \VDCB\ will communicate to the control and DAQ systems using a high-speed serial standard interface like Fast Ethernet or Gigabit Ethernet, connected using CAT-6 copper cable or standard optical fiber.

In the current design of the veto detectors, the \LSV\ and \WCV\ together have \DSkVetoPMTsNumber\ \PMTs, so the electronic front-end system will have only \DSkVetoCratesNumber\ crates.  A small-scale, single-channel prototype has been developed to test the performance of the preamplifier-discriminator channels and also the HV decoupling.  All the functions are implemented using off-the-shelf components.

\section{Calibrations}
\label{sec:Calibrations}

\subsection{Requirements}
\label{sec:Calibrations-Requirements}

Calibration of the \DSk\ detectors are required to reach the physics goals of achieving a highly sensitive, multi-year, background-free \WIMP\ search. Calibrations range from low-level detector issues, such as the single-photoelectron response of individual photosensors, to high-level physics issues like the acceptance as a function of energy for nuclear recoils.  Calibrations required include:

\begin{compactitem}
\item Photosensor charge and time response;
\item \TPC\ and \LSV\ scintillation light yield;
\item Spatial uniformity of the \TPC\ detector response;
\item Pulse shape discrimination of nuclear and electron recoils;
\item Neutron detection efficiency in the \LSV;
\item Stability of the detectors' responses over time (\TPC, \LSV\ and \WCV).
\end{compactitem} 

\noindent
The combination of radioactive sources, neutron generators and light sources proposed here ensures a robust calibration plan.

\subsection{Overview}
\label{sec:Calibrations-Overview}

Laser calibrations, with laser pulses delivered through optical fibers to the \TPC, \LSV\ and \WCV\  (described in the \TPC\ and \LSV\ sections) are used to tune and characterize the \SiPM\ and \PMT\ charge and time response.  Calibrations with gamma sources calibrate the \TPC\ and \LSV\ energy scales and the \TPC\ position reconstruction and resolution for surface background rejection.  They also provide data for tuning the Monte Carlo detector response modeling and track detector stability with time.  Calibrations with neutron sources verify the detector's ability to detect \WIMP-induced nuclear recoils and distinguish them from electron recoils, along with measuring the effectiveness of the veto detectors to identify and reject neutron-induced nuclear recoils. Besides specific calibrations, neutron and gamma sources provide generally useful samples of signal-like and background-like events.  UV LED light calibration provides an alternative to sources in the \LSV\ and \WCV\ and vastly extends the possible range of energy calibrations (\CalLEDLowEnergy\ to \CalLEDHighEnergy) by varying the power output of the controllable LED and assuming that the number of emitted photons is proportional to the energy of the simulated physics event).

The \LArTPC\ light yield in the \DSkROIEnergyRange\ energy range of interest for \WIMP\ scatters  is another key detector response.  A distributed source of low energy electron recoils can be obtained by periodically introducing \ce{^{83m}Kr}  into the active argon volume~\cite{Lippincott:2010jb,Kastens:2009hz,Venos:2006gx,Venos:2005dw}.  This short-lived ($\tau$=\KrEightThreeMOneMeanLife) isotope gives a mono-energetic peak from its near-coincident \KrEightThreeMOneECEnergy\ and \KrEightThreeMTwoECEnergy\ conversion electrons, summing to \KrEightThreeQValue~\cite{Wu:2001gl}.  \ce{^{83m}Kr} is conveniently obtained by milking a \ce{^83Rb} ($\tau$=\RbEightThreeMeanLife) electron capture (EC) source connected to the gas system.  Injection of the \ce{^{83m}Kr} is done through the recirculation system's argon return line.  Such a source will be permanently installed in an accessible part of the external gas loop, as was done for \DSf, and can be replaced when necessary to compensate for its natural decay.  The \ce{^{83m}Kr} source will also allow the spatial dependence of the \SOne\ and \STwo\ scintillation responses to be studied in detail, since the \ce{^{83m}Kr} will be injected into the \LArTPC\ at multiple points to avoid laminar flow and give good spatial distribution of the source throughout the \LArTPC\ volume.

A variety of neutron and \gr\ sources will be deployed within the \LSV\ to calibrate the detectors.  \DSf\ used a vertical CALibration Insertion System (\calis) to insert and position calibration sources within the \LSV\ through ``organ pipes": vertical pipes that penetrate the \WCV, allowing access to the \LSV.  Based on the success of \calis\ in \DSf, \DSk\ will use a similar system, modified for the planned geometry.  Such system gives flexibility in the choice of calibration source and deployed position.  At the same time, such a system does involve direct contact with a fairly complex liquid scintillator mixed with TMB, making the calibration campaigns elaborate in terms of planning, execution and duration.

For this reason, the collaboration has also envisioned two closed guideline systems, one for the \LSV\ and another for the \TPC. The guideline systems consist of closed tubes that loop around the inner \LSV\ wall and outer cryostat wall for the \TPC\ calibration. The tubes used have a small diameter to avoid shadowing effects and can accommodate only miniature radioactive sources. On the other hand, they allow calibration of the \TPC\ and \LSV\ without any contact with the liquid scintillator, can be more efficient (hence calibrations could be done more frequently) and provide better localization of the deployed sources.

Calibrations with a neutron source are essential to determine and monitor the \LSV\ neutron detection efficiency and to measure the nuclear-recoil \SOne\ pulse shape in the \LArTPC\ used in the background discrimination.  A conventional \AmBe\ source has coincident high-energy gammas that allow event-by-event tagging of neutron events in the \LArTPC\ by detecting these gammas in the \LSV. However, coincidence high-energy gammas interfere with measurements of the prompt neutron detection efficiency in the \LSV.  The \DSf\ collaboration has produced and employed an alternate ($\alpha$,n) source based on \ce{^241Am} and \ce{^13C}, which  is free of these gammas.  Both types of sources will be used.

Accurate absolute neutron detection efficiency measurements require a tagged neutron source with no accompanying radiation.  For this purpose, a customized \DOneDOne\ tagged neutron generator based on the Thermoscientific~API-120~\cite{ThermoFisherScientificInc:2015vq,Chichester:2005kw}, which has already been procured for \DSf, will be used.  In addition, a small, compact, battery operated pyro-electric neutron generator and a tagged \ce{^252Cf} source are under development.

\subsection{Neutron Generators}
\label{sec:Calibrations-NeutronGenerators}

\subsubsection{Tagged neutron generator}
\label{sec:Calibrations-NeutronGenerators-Tagged}

Homeland Security applications of neutron imaging by the associated particle technique have led to commercial availability of the Thermoscientific API-120 \DOneDOne\ generator.  This reaction produces no penetrating radiation in coincidence with the neutrons.  The standard API-120 provides individually time- and direction-tagged, monoenergetic neutrons at very high intensity (\CalDDNeutronGeneratorCommercialFlux).  The tagging is accomplished by detecting the \ce{^3He} from the neutron production reaction as it strikes an internal scintillator screen, using an external multi-anode MCP-PMT closely coupled through a sapphire window.  A custom version was developed by Thermoelectron with neutron flux limited to \CalDDNeutronGeneratorFlux, in order to meet the local approval by LNGS radiation safety regulations without having to seek national approval.  The configuration was also modified to allow the device to fit through the \SI{6}{\inch} ID organ pipe into  the \DSf\ \LSV.  Neutron emission times are tagged within about \CalDDNeutronGeneratorTiming\ and the emission angles within \CalDDNeutronGeneratorAngleResolution. This will allow the \LSV\ neutron detection efficiency to be thoroughly mapped out in position and direction.

\subsubsection{Pyroelectric neutron generator}
\label{sec:Calibrations-NeutronGenerators-PyroElectric}

Pyroelectric crystals, such as \ce{LiNbO3} or \ce{LiTaO3}, produce an electric field up to \CalPyroGunEField, when exposed to a temperature gradient. \DS\ collaborators have confirmed experimentally that a setup with a \CalPyroGunTestCrystalSize\  crystal installed in a stainless steel chamber with a residual gas pressure of about \CalPyroGunPressure\ and grounded metal target, generates X-rays with energy up to \CalPyroGunXRayEnergy. The same setup could be used to generate \DDNeutronEnergy\ neutrons if the target is deuterated and the residual gas is \ce{^2D}. Thus, a pyroelectric neutron generators (PNG) seems to be a promising tool for calibration of a \LArTPC\ and other low background detectors, because of their On/Off mode of operation and the absence of radioactive materials. Such a controllable PNG will be periodically deployed in the detector for the calibration of the detector response to neutrons.

\begin{figure*}[!t]
\centering
\includegraphics[width=\textwidth]{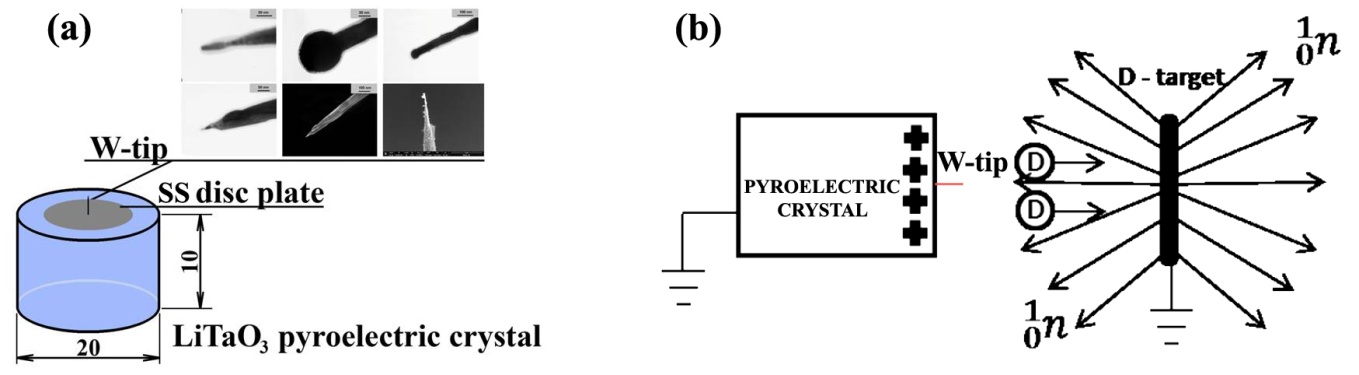}
\caption[Overview of the pyroelectric neutron generator.]{The PNG pyroelectric element (left)  and the mechanism of neutron generation (right).}
\label{fig:Cal-PNG}
\end{figure*}

\begin{figure}[!t]
\centering
\includegraphics[width=0.5\columnwidth]{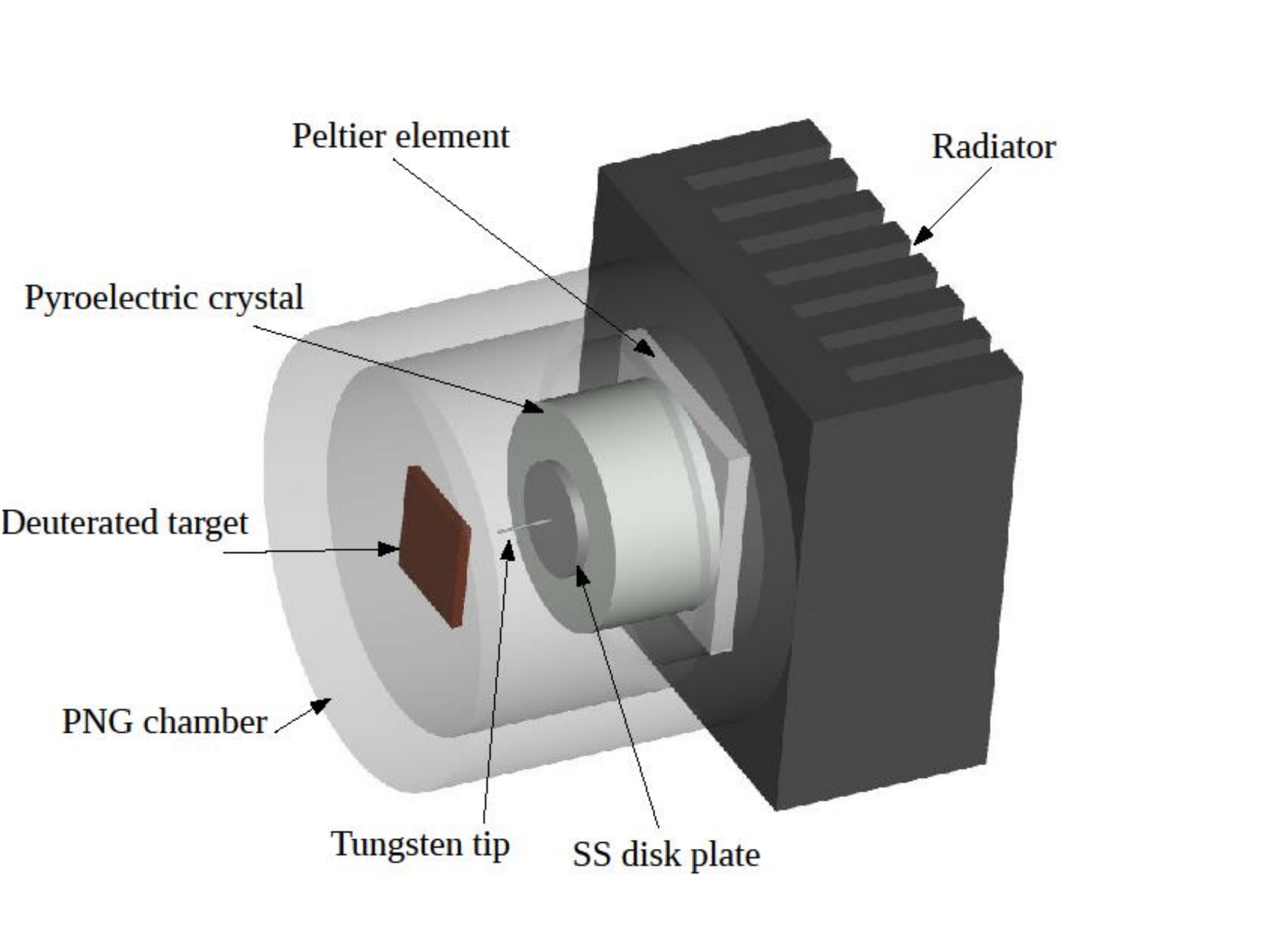}
\caption{Pyroelectric neutron generator layout.}
\label{fig:Cal-PNGSchematic}
\end{figure}

The assembly which was used to demonstrate the PNG-generated X-rays consists of a \ce{LiTaO3} crystal, and a stainless steel disk plate with tungsten W-tips welded to it, as shown in Fig.~\ref{fig:Cal-PNG} (left).  Fig.~\ref{fig:Cal-PNG} (right) illustrates the neutron generation process. A PNG prototype, shown in Fig.~\ref{fig:Cal-PNGSchematic}, is planned to be \CalPyroGunLength\ in diameter, \CalPyroGunLength\ long and have an expected neutron yield not less than \CalPyroGunFlux.

\subsection{Radioactive Sources}
\label{sec:Calibrations-RadioactiveSources}

\subsubsection{\AmC\ Neutron Source}
\label{sec:Calibrations-RadioactiveSources-AmC}

Neutron sources based on the \ce{^9Be}($\alpha$,n)\ce{^12C} reaction have a large probability of producing \ce{^12C} in the \AmBegrenergy\ excited state, due to the large Q value for this reaction.  The resulting prompt \AmBegrenergy\ \grs\ have been found in \DSf\ to be a convenient trigger for study of neutron captures in the \TPC\, which typically occur microseconds after the neutrons thermalize.  However, the necessary very high veto efficiency requires that the prompt signals from the thermalization process also be detected with high efficiency.  It turns out that a source based on \ce{^13C}($\alpha$,n)\ce{^16O} with $\alpha$'s of less than \CalAmCSourceAlphaEnergyLimit\ still has usable neutron yield, but gives essentially no excited \ce{^16O} final states and hence no prompt reaction \grs.  The choice of \ce{^241Am} as the $\alpha$ source is almost unique in that it also guarantees very few (\num{\sim E-4} per $\alpha$) nuclear \grs\ associated with the $\alpha$ emission.  \ce{^241Am} does emit copious \CalAmCSourceGammasEnergy\ gammas, but these are easily attenuated to the necessary level by \CalAmCSourcePbThickness\ of \ce{Pb}.  Such sources were independently developed by the Daya Bay experiment~\cite{Liu:2015ig}.  An  \AmC\ source using \CalAmCSourceDSfActivity\ of \ce{^241Am} was assembled and successfully deployed in \DSf, giving a few neutrons per second and permitting studies of the thermalization signals.  A new source with slight improvements is envisioned.

\subsubsection{\AmBe\ neutron source}
\label{sec:Calibrations-RadioactiveSources-AmBe}

While the \AmC\ source represents the best choice for calibration of the \LSV\ neutron detection efficiency, interactions in the source holder and materials between the source and the \TPC\ produced significant electron recoil background in the \DSf\ \TPC, problematic for characterization of the nuclear recoils of argon nuclei. The \AmBe\ source was shown to be more suitable for this calibration. The \AmBegrenergy\ \gr\ is promptly emitted in the \ce{^9Be}($\alpha$,n)\ce{^12C} nuclear reaction in about \SI{60}{\percent} of cases. By tagging the \AmBegrenergy\ \gr\ in the \LSV\ in a very tightly constrained time interval prior to the signal in the TPC, a very pure sample of nuclear recoils is obtained. The \AmBe\ calibration in \DSf\ provided the best available nuclear recoil calibration of the \DSf\ \TPC\ {\it in situ}, and was in excellent agreement with the nuclear recoil calibration from the stand-alone SCENE experiment~\cite{Cao:2015ks}.

\begin{figure}[!t]
\centering
\includegraphics[width=\columnwidth]{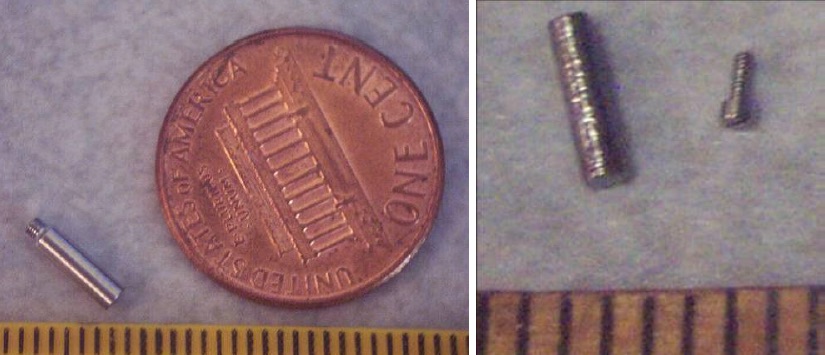}
\caption[Pictures of the miniature \AmBe\ neutron source.]{{\bf Left:} The outer capsule of the custom made \AmBe\ source from the Alabama group~\cite{Ostrovkskiy:2012wv}. {\bf Right:} The inner tungsten capsule of the custom made miniature \AmBe\ source.}
\label{fig:Cal-AmBeMiniatureUA}
\end{figure}

The main novelty planned for the \AmBe\ calibration is a custom production of a miniature (a few \SI{}{\milli\meter}) \AmBe\ source, since the typically available commercial sources cannot fit within the planned narrow \TPC\ and \LSV\ guide tube systems. The miniature \AmBe\ source will rely on the procedure developed by the neutrino group at the University of Alabama~\cite{Ostrovkskiy:2012wv}. The group successfully fabricated an \AmBe\ source with an outer capsule dimension of just \CalAmBeSourceDiameter\ in diameter and \CalAmBeSourceLength\ long, with an activity of \CalAmBeSourceNeutronYield~neutrons per second using \CalAmBeAmSourceAmActivity\ of \ce{^241Am}. Fig.~\ref{fig:Cal-AmBeMiniatureUA} shows the outer and inner source capsules. The inner capsule is made of tungsten to efficiently suppress the \CalAmCSourceGammasEnergy\ gamma emission from the \ce{^241Am}. A very pure \ce{Be} powder is mixed with \ce{^241Am} high purity powder in an alcohol solution in a miniature test tube with a special micropipette and then transferred into the tungsten capsule. After the alcohol evaporates, the procedure is repeated. In the end, the mixture of \AmBe\ is compressed with a wire for higher source activity. The wire is also used to seal the tungsten capsule. The tungsten capsule is placed inside the stainless steel capsule and welded shut.  The source is then ready for certification and use. The outside capsule has a threaded part to attach the source to the rest of the calibration system.

\subsubsection{Tagged \ce{^252Cf} source}
\label{sec:Calibrations-RadioactiveCf}

A tagged \ce{^252Cf} spontaneous fission source (under development by \DS\ collaborators in Russia) provides an interesting option for understanding the \LSV\ neutron detection efficiency, by accessing locations otherwise inaccessible to the API-120 \DOneDOne\ generator. Properties of \ce{^252Cf}  spontaneous fission (energy spectrum of neutrons and gammas, mean number of neutrons per fission \CalCfSourceNeutronsPerFission) are well known and such type of sources are widely used in neutron physics when studying materials with neutron diffraction and neutron spectroscopy. 

Tagging of the spontaneous \ce{^252Cf} fission is achieved by a  semiconductor surface-barrier  \ce{Si} detector. The low activity (\CalCfSourceActivity\ fissions/s) sample of \ce{^252Cf} is placed near the surface of the \ce{Si} detector. Fission products detected in the thin \ce{Si} detector are separated by energy from alpha-particles as well as from gammas and betas. The mean kinetic energy of the fission fragments is about \CalCfSourceFissionFragmentEnergy, while the maximum energy of the alphas is \CalCfSourceAlphaMaxEnergy. So it is easy to identify fission and produce an external trigger for the main detector. 

The total activity of the source (\CalCfSourceFissionActivity\ fissions/s,  \CalCfSourceNeutronFlux\ and \CalCfSourceAlphaActivity\ alphas/s) is below the level which requires special radioactive source certificate.

\subsubsection{External Miniature \gr\ and Positron Sources}
\label{sec:Calibrations-RadioactiveSources-GammaPositron}

Three external gamma sources were used in \DSf: \ce{^57Co}, \ce{^133Ba}, and \ce{^137Cs}. In addition, a \ce{^22Na} positron source was used. The combination of these sources and the \ce{^39Ar} available during the \AAr\ run gave an excellent calibration of the electron recoil \FNine\ band, energy scale and provided data for the tuning of the detector response in the Monte Carlo simulation. The gamma sources were chosen to span the energy range of interest in combination with the distributed \ce{^{83m}Kr} source: \KrEightThreeQValue\ for \ce{^{83m}Kr}, \CoFiveSevenQValue\ for \ce{^57Co}, \BaOneThreeThreeQValue\ for \ce{^133Ba}, and \CsOneThreeSevenQValue\ for \ce{^137Cs}.  A similar set of sources, but miniature in size, is planned to be used within the \TPC\ and \LSV\ guide tube systems. Source production procedures like the one used for the miniature \AmBe\ source production will be employed, except for replacing the inner tungsten capsule with one made out of stainless steel.

\subsubsection{Distributed \ce{^{83m}Kr} Source}
\label{sec:Calibrations-RadioactiveSources-Kr}

Due to the nature of \TPC\ operation, it is simply not possible to insert a solid source into the \TPC\ active volume.  A mechanical delivery close to the \TPC\ is possible but does not solve the uniformity issue due to the large detector size.  Uniformity of the response to gamma rays is not necessarily required, since the primary and sole necessary calibration comes from the SCENE data and in situ calibration with \ce{AmBe}, where the response of \LAr\ to nuclear recoils was very carefully characterized and calibrated against the response of \LAr\ to the monochromatic line resulting from the two EC in \ce{^{83m}Kr}.  A careful mapping with gamma rays is not one of the requirements of the experiment.
 
Monte Carlo studies of \DSk\ with \GFDS\ show that the interaction length for gamma sources spans from \SI{5}{\centi\meter} for \ce{57Co} to \SI{9}{\centi\meter} for \ce{137Cs}.  While their interaction length will allow probing just the edge of the \LArTPC, they can successfully be used (as in \DSf) for MC simulation tuning and crosschecks with the \ce{^{83m}Kr} calibration in the overlapping region.  Finally, the main physics result of \DSk\ requires careful calibration of the nuclear recoils response.  Calibration of the electron recoils is simply instrumental for the calibration of the nuclear recoils response, and the baseline solution relies on the use of \ce{^{83m}Kr}.

The monoenergetic decays of \ce{^{83m}Kr}, if distributed through the active volume of the \TPC, can give a key calibration point in the \WIMP\ recoil energy region. The 3D reconstruction of events in the TPC allows a full mapping of position-dependence of the light yield using this source as well.  This means that, while broad dissemination through the active volume is important, a uniform distribution is not required.  The \ce{^{83m}Kr} decays quickly ($\tau$=\KrEightThreeMOneMeanLife) to a stable nuclide and causes no long-term contamination or background to the \WIMP\ search.

The \ce{^{83m}Kr} source is based on the source used successfully in \DSf.  A tiny droplet of a solution of \ce{^83Rb} ($\tau$=\RbEightThreeMeanLife) is adsorbed into a piece of charcoal, which, after drying, is placed between two particulate filters in a branch of the argon recirculation system  which is normally isolated by valves and can separately be pumped to vacuum.  In \DSf, an initial \CalDSfKrSourceActivity\ of \ce{^83Rb}\ gave a \ce{^{83m}Kr} trigger rate of hundreds of \SI{}{\hertz}, even though the flow subsequently passed through a cooled radon trap.  The impact of the colder radon trap (see Sec.~\ref{sec:Cryogenics}) will need evaluation.

In \DSk, the challenge will be to get the \ce{^{83m}Kr}  broadly distributed in the \DSkActiveMass\ active mass before it decays.  To further this, \LAr\ will be returned to the TPC after repurification via numerous tubes whose endpoints will be distributed over the surface area of the \TPC\ PTFE reflectors.

\subsubsection{Distributed \ce{^220Rn} Source}
\label{sec:Calibrations-RadioactiveSources-Rn}

\ce{^220Rn} and its short-lived daughters produce several \grs, $\beta$'s, and $\alpha$'s of various energies.  This opens up a possibility to use \ce{^220Rn} as another distributed calibration source~\cite{Lang:2016jh}.  The source of \ce{^220Rn} may be prepared as an electroplated \ce{^228Th} on stainless steel or copper.  Enclosed in a small metal-sealed and vacuum-tight volume equipped with a VCR/CF port it may be hooked up to the gas circulation system in a similar way as the \ce{^{83m}Kr} source.  The advantages of the \ce{^220Rn} source are the following:

\begin{compactitem}
\item \ce{^220Rn} and its daughters are short-lived, the longest half-life in the chain is \BiTwoOneTwoHalfLife\ for \ce{^212Bi}, therefore the activity introduce into the detector will disappear after a few days;
\item The high energy alphas appearing in the chain (\BiTwoOneTwoAlphaTwoEnergy, \BiTwoOneTwoAlphaOneEnergy, \RnTwoTwoZeroAlphaEnergy, \PoTwoOneSixAlphaEnergy, and \PoTwoOneTwoAlphaEnergy);
\item The coincident \ce{^220Rn}-\ce{^216Po} and \ce{^212Bi}-\ce{^212Po} decays will allow to study the homogeneity of the distributed sources (affect also \ce{^{83m}Kr} and the performance of the distribution system in the large \LAr\ volume), and possible \LAr\ flow patterns as well as potential ``dead volumes'' that are not affected by the \LAr\ circulation;
\item Drift of the charged ions inside the \LAr\ volume (interesting also with respect to Po isotopes from the \ce{^222Rn} chain);
\item Low-energy $\beta$'s appearing in the chain may be used for calibrations of the low-energy response of the detector.
\end{compactitem}

To avoid removal of the \ce{^220Rn} source by the charcoal trap it should be by-passed during the calibration run.  Release of \ce{^220Rn} from the source should be done in a way that avoids contamination of the detector with residual \ce{^224Ra} (half-life \RaTwoTwoFourHalfLife) or \ce{^228Th} (half-life \ThTwoTwoEightHalfLife).  A similar risk exists for the \ce{^{83m}Kr} source, but has been thoroughly mitigated and never observed.

\subsection{Calibration Deployment Systems}
\label{sec:Calibrations-DeploymentSystems}

\subsubsection{\LSV\ and \TPC\ Guide Tube System} 
\label{sec:Calibrations-DeploymentSystem-GuideTube}

The \LSV\ and \TPC\ guide tube systems have been designed for simple, safe and efficient calibration of the \LSV\ and \TPC\ using miniature radioactive sources. They will allow deployment of the sources inside the \LSV\ without direct contact with the liquid scintillator, eliminating risks associated with it. 

The \TPC\ guide tube system will bring radioactive sources from the top of the \WCV, through the water, into the \LSV, through the liquid scintillator, and then to locations along the cryostat exterior surface. The sources will be delivered through an enclosed stainless steel tube coming vertically down from the top of the water tank, passing by the cryostat wall and around its bottom. To avoid any possibility of the source getting stuck in the detector, the enclosed tube will form a closed loop so that the sources can be pushed down around the cryostat and then back up on the other end in the unlikely event that a source gets detached from the guide wire.

\begin{figure}[!t]
\centering
\includegraphics[width=\columnwidth]{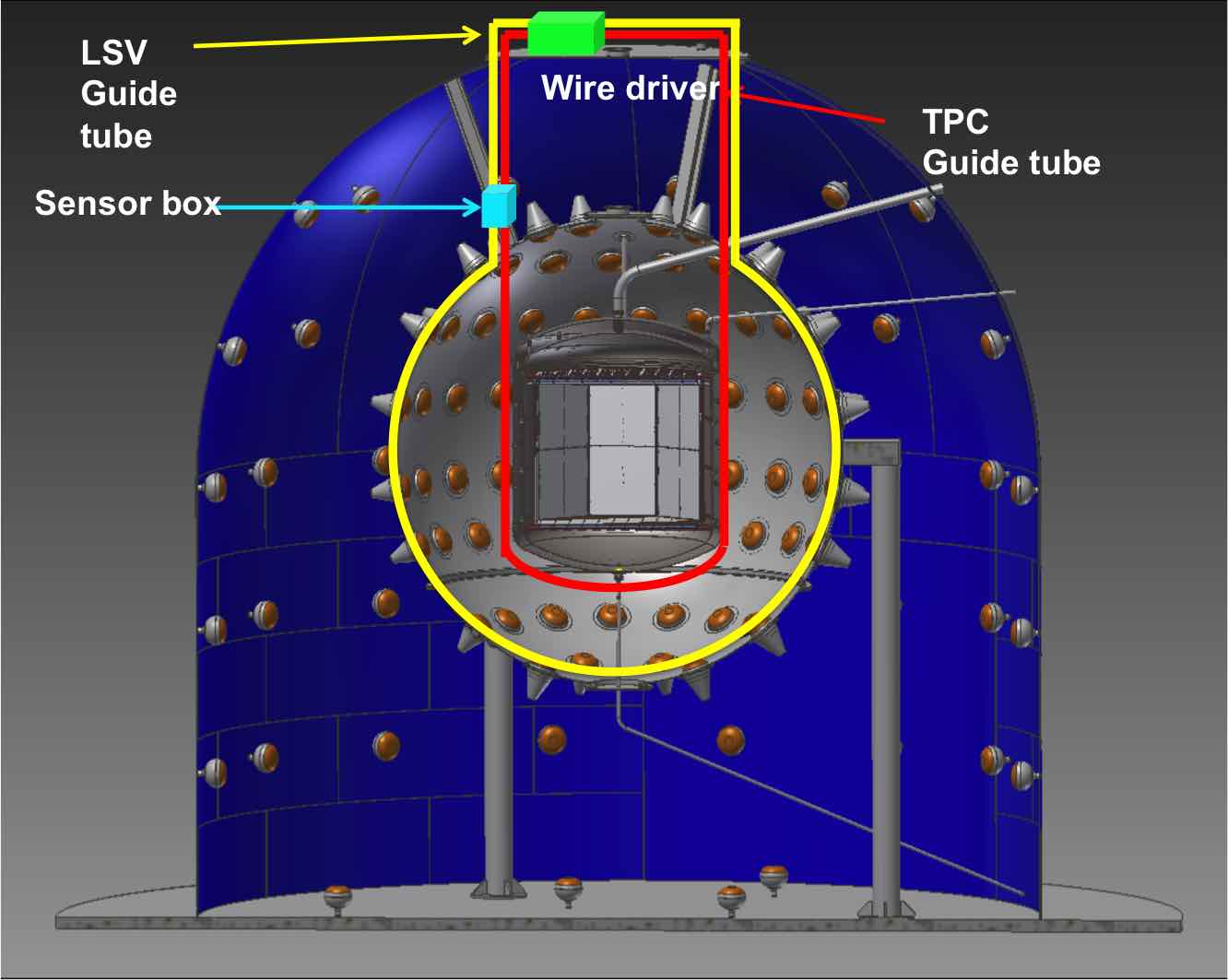}
\caption[\TPC\ and \LSV\ calibration source guide tube systems.]{\TPC\ and \LSV\ guide tube systems share a common path, enclosed in the pipe passing through the \WCV. Once the tubes enter the \LSV, the \LSV\ guide tube system goes along the wall of the stainless steel sphere in a circle, while the \TPC\ guide tube goes along the wall and the bottom of the cryostat.}
\label{fig:Cal-GuideTubeSystem}
\end{figure}

The \LSV\ guide tube system is conceptually the same, except that the guide tube inside the \LSV\ will be attached to the wall of the stainless steel sphere.  Fig.~\ref{fig:Cal-GuideTubeSystem} shows the conceptual path of the \TPC\ and \LSV\ guide tube systems.  The guide tubes will be \CalGuideTubeDiameter\ (\CalGuideTubeDiameterInches) in diameter to minimize shadowing, dead space and radioactive background coming from the stainless steel. The stainless steel tube segments will be connected with Swagelok connectors. In addition, relatively small tube diameter will help for tighter control of the source location, especially important in the case of the gamma sources. 

The miniature sources will be attached to the stainless steel guide wire that will be pushed with a wire driver powered by a stepper motor. A teflon sleeve, made out of the PTFE tubing, will envelope the guide wire and the source to significantly reduce friction while the source and its guide wire are being pushed around. The wire driver will be equipped with a set of encoders and two sensor boxes for accurate positioning of the source. A pair of ring type inductive proximity sensors placed inside the sensor box will be used to reset the encoders when sources pass through a sensor at exactly known locations. The location will preferably be close to the \LSV, since the proximity increases the accuracy of the source positioning. Finally, both systems will be purged with a low pressure nitrogen to eliminate any residual radioactivity that may have entered into the tubes.

Except for the sensor box, all other parts of the \LSV\ and \TPC\ guide tubes that require power, control and nitrogen flow will be at the top of the water tank. The sensor box should be located as close as possible to the calibration volume, preferable close to the top of the \LSV, but inside the \WCV. The sensor box will require a power cable and a communication cable that will be routed along the guide tubes on the outside. Since the sensor box will be integrated with the guide tube volumes inside the \WCV, its interior will be purged with nitrogen.

An elaborate testing program must be performed to ensure high quality and safety of the calibration when using the guide tube systems. All parts will be radio-assayed and the entire system will be tested off-site for robustness, precision and reliability. Dedicated survey is planned to record the exact guide tube path around the cryostat.
 
\subsubsection{Neutron Generators Deployment System}
\label{sec:Calibrations-DeploymentSystem-NeutronGenerators}

\begin{figure}[!t]
\centering
\includegraphics[width=\columnwidth]{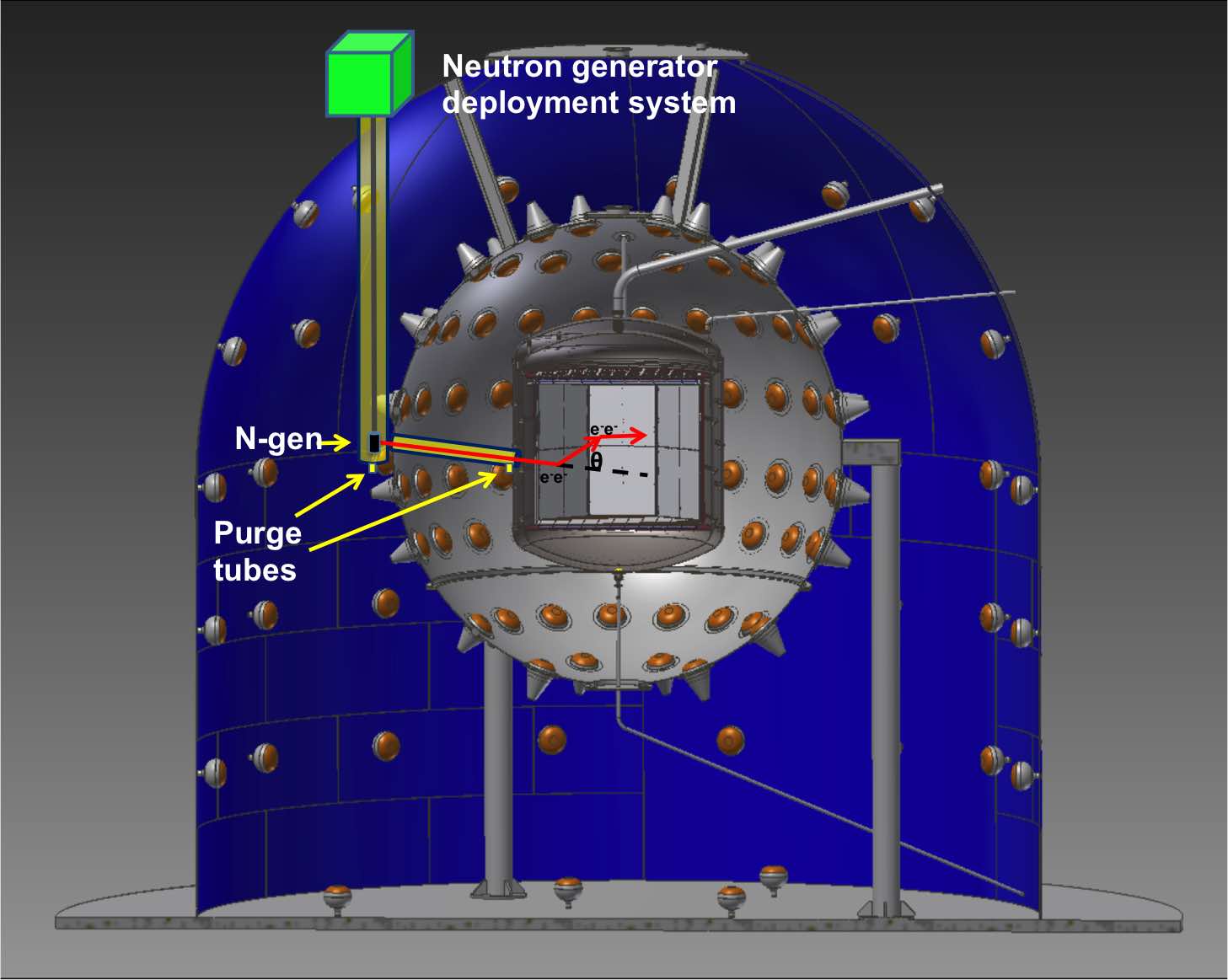}
\caption[Neutron generator calibration deployment system.]{Neutron generator deployment system consisting of a vertical clear, polycarbonate tube in the \WCV\ and nearly horizontal and a narrower, clear polycarbonate pipe inside the \LSV\ that provides a free path for the neutrons to enter the cryostat and the \TPC. The tubes have small diameter purge tubes at the bottom that allow entry and exit of water in the \WCV\ and liquid scintillator in the \LSV.}
\label{fig:Cal-NGdeployment}
\end{figure}

A well collimated, monoenergetic beam of neutrons is one of the most powerful tools for the calibration of argon nuclear recoils, with respect to the mean and width of their \FNine\ distribution in the \TPC. The conceptual design of the neutron generator deployment system is shown in Fig.~\ref{fig:Cal-NGdeployment}. This system allows deployment of the neutron generator without direct contact with the liquid scintillator, eliminating risk and safety concerns associated with handling TMB loaded liquid scintillator. 

The deployment mechanism on the top lowers the neutron generator through a vertical polycarbonate tube (\CalNeutronGunTubeDiameter\ in diameter and \CalNeutronGunTubeLength\ in length) to the detector equator level inside the \WCV.  There is a constant flow of nitrogen inside the tube, entering at the top and exiting at the bottom through a small diameter tube. During the calibration session, dry nitrogen gas will be used to push the water out of the tube through a small pig-tail tube at the bottom of the vertical tube.  During all non-calibration times, the dry nitrogen gas flow is turned off, allowing water to fill the transparent, vertical tube to preserve efficiency of the \WCV\ muon veto. 

In order to deliver neutrons from the neutron generator to the \TPC, a low collision path is needed for neutrons flying through the \LSV. For this purpose, a nearly horizontal (slightly angles to allow release of nitrogen gas) polycarbonate tube (\CalNeutronGunHorizontalTubeDiameter\ in diameter and \CalNeutronGunHorizontalTubeLength\ in length) will be  installed in the \LSV, extending from the stainless steel sphere wall at the equator level (the same level as the neutron generator when deployed in the \WCV) to the wall of the cryostat. During a calibration campaign, dry nitrogen gas will push liquid scintillator out from the tube through a small pig-tail tube at the bottom. Neutrons will easily pass through the nitrogen filled tube and reach the cryostat and the \TPC. At all other times, the nitrogen flow will be stopped, and liquid scintillator will completely fill the space inside the tube. The polycarbonate tube will be transparent and filled with liquid scintillator. In this way, the high efficiency of the \LSV\ will be preserved during normal data taking times due to the similarity of the indices of refraction between the liquid scintillator and polycarbonate material. An essential requirement for both tubes is full displacement of the liquids, which will be verified by a set of cameras observing bubbles coming out of the pig-tail tubes.  The horizontal polycarbonate tube will only be attached to the stainless steel sphere and not to the cryostat. The wall thickness will be chosen in such a way to ensure that the tube is nearly neutrally buoyant during calibration, and negatively buoyant when filled with liquid scintillator.

Thanks to the high neutron capture cross-section of the liquid scintillator, only neutrons passing through the tube will reach the \TPC, while the rest will be stopped/captured. In this way, excellent collimation of just \CalNeutronGunTubeColiimationAngle\ full opening angle is achieved. Even with this narrow opening angle, about \CalNeutronGunNeutronsToTPC\ neutrons per second from the API-120 will reach the \TPC, to provide millions of monoenergetic, directional, pure neutron events in a matter of days, in combination with the tagging information coming from the API-120's internal \ce{^3He} detector.

\subsubsection{Calibration Insertion System}
\label{sec:Calibrations-DeploymentSystem-Insertion}

CALibration Insertion System (\calis) is a flexible system that lowers different calibration sources/devices directly into the \LSV. It has very good reach inside the \LSV\ and can position sources/devices at various locations, both close to the cryostat and away from it. A very similar system already exists for \DSf\ and has been successfully used in many of the source calibration campaigns of the \DSf\ experiment. 

The system consists of the deployment mechanism on top of the \WCV\ that lowers the calibration sources and devices through an organ pipe that opens into the \LSV. The assembly that holds the source and that is lowered into the \LSV\ can rotate and bring the source off-axis from the organ pipe, giving greater flexibility in the source positioning. The system will be completely air tight, since the organ pipe is filled with liquid scintillator and so no exposure to air or humidity is allowed. The system will also be light tight, to avoid risk of damaging \LSV\ \PMTs\ during a calibration campaign. During normal data taking, the deployment assembly is situated inside the sealed housing on top of the \WCV, while the organ pipe is sealed with a gate valve. Two organ pipes are planned to provide good coverage of the \LSV\ volume. 

While \calis\ provides flexibility in the source/device positioning, its operation is more time consuming and requires direct contact with the liquid scintillator. Thus, calibration with \calis\ will be used for deployment to locations the other systems cannot reach, or for sources/devices that are too large for the guide tubes and need to be placed in locations other than the one achievable with the neutron generator deployment tube.

\subsection{The \LSV\ and \WCV\ Camera System}
\label{sec:Calibrations-CameraSystem}

Six wide-field-of-view cameras will be installed inside the \LSV, and two additional cameras will be installed inside the \WCV. Cameras will be placed inside liquid-tight housings and will have their own sets of LED lights for detector illumination while in use. The interior of the housings will be constantly purged with nitrogen.  Four out of six cameras in the \LSV\ will be placed around the equator at \SI{90}{\degree} locations, to provide visual coverage of the polycarbonate tube and sources deployed with \calis. The last two cameras in the \LSV\ will be placed close to the top and bottom of the stainless steel sphere. \DSf\ has three cameras inside its \LSV\ that have proven to be very useful during calibration campaigns with \calis\ and have operated flawlessly for several years now. While the new cameras are still to be identified (since better and cheaper models are available on the market), the general design of the camera housings will remain the same. One of the concerns with the cameras and housings is their contribution to the detector backgrounds, especially potassium which is often present inside glass. However, the larger size of the \LSV, combined with careful selection of components, will mitigate any risk associated with additional backgrounds from the cameras.

The two cameras inside the \WCV\ will be positioned close to the detector top to observe detector filling, and at the equator level to observe displacement of water from the neutron generator deployment tube during calibration. Cameras will be externally powered and controlled via software on a dedicated PC. Power cables, communication (ethernet) cables and nitrogen gas tubing will be delivered to the back of each camera, extending into the \WCV.

\subsection{Interfaces}
The various calibration systems directly interface with the \WCV, \LSV\ and \TPC.
\begin{compactitem}
\item The vertical polycarbonate tube will be located in the \WCV\ as a part of the neutron generator deployment system. The tube will have a bellows on the upper end that connects it to the top flange in order to eliminate high stress on the polycarbonate tube. The tube will also be fixed at the bottom to the stainless steel sphere for stability and exact position localization, as the location at the bottom needs to be well aligned with a horizontal tube in the \LSV;
\item The polycarbonate tube in the \LSV\ will be attached to the stainless steel sphere by special fixtures that will be able to withstand buoyancy changes between the states when the tube is filled with liquid scintillator and when the tube is purged with nitrogen. Nitrogen gas teflon tubing will be supplied from the \WCV\ top for this purpose and will use spare ports on the \WCV\ and \LSV;
\item \LSV\ and \TPC\ guide tubes will require two ports on the top of \WCV\ and two ports on the \LSV, assuming the ports can be shared by both systems.  In addition, the \LSV\ guide tube system will be attached to the stainless steel sphere wall in several places.  While the \TPC\ guide tube system will not be directly attached to the cryostat, welded brackets on the cryostat will be used to preserve its overall shape. Exact locations of the brackets will be determined and located after the cryostat fabrication;
\item \calis\ will require two organ pipes that will connect the top of the \WCV\ with the \LSV\ top, as was done for  \DSf. Organ pipes will have gate valves on the top to allow access to the \LSV\ during calibration;
\item Cameras in the \LSV\ and \TPC\ will be installed in the dedicated locations on the stainless steel sphere and \WCV\ wall. Nitrogen lines, power and ethernet cables will be attached to the camera housings and will be routed on the outside of the stainless steel sphere. Ports will be needed on the top of the \WCV\ to pass the cables and the nitrogen lines.
\end{compactitem}

\subsection{Performance}
\label{sec:Calibrations-Performance}

The expected performance of the calibration systems will be evaluated through a combination of simulations, bench and lab tests of the built systems prior to installation and commissioning.

Experience from \DSf\ showed that calibration with gamma sources is more sensitive to positioning precision than the neutron sources. Based on this experience, \LSV\ and \TPC\ guide tube systems should provide source position with a couple of cm accuracy. 

Activity of the gamma and neutron sources will be a compromise of the maximum triggering efficiency of the \DAQ\ and safety regulations at the \LNGS, which is especially stringent in the case of neutron sources. However, minimum activity required will be determined from the statistics needed to conduct high precision calibrations, without significantly affecting detector livetime for \WIMP\ search.

\subsection{Additional Calibration Resources}
\label{sec:Calibrations-AdditionalResources}

\subsubsection{Extending the \LSV\ and \WCV\ Energy Calibration Range with LED Light Sources}
\label{sec:Calibrations-AdditionalResources-LED}

Simulation of physics interactions in the liquid scintillator and water with a controllable double LED pulser~\cite{Chepurnov:2016du} can extend the energy range for calibration far beyond  \SI{\sim11}{\MeV} (which is the maximum achievable energy with more commonly used radioactive sources;). In addition, the use of radioactive sources requires special care in order to avoid damage to the ultra-pure medium of the detector. 

The concept of the calibration with LEDs implies an assumption that the total deposited energy is directly proportional to the number of emitted photons per bunch.  The LED is a preferred choice since it is significantly cheaper and provides a set of flexible parameters like amplitude, duration, rate of pulses and time delay between pulses, in a relatively easy manner. The calibration system design can accommodate LEDs of different wavelengths. There are at least two options: UV diodes for the \LSV\ and blue ones for the \WCV. The UV LEDs are required in order to increase isotropy of the radiation in the liquid scintillator and plausibly simulate point-like events. Anisotropy arises from the use of optical fibers for light injection. Diffusers provide additional tools for reduction or increase (in the case of the \WCV) in anisotropy, by placing it at the end of the optical fiber.

\begin{figure}[!t]
\centering
\includegraphics[width=0.5\columnwidth]{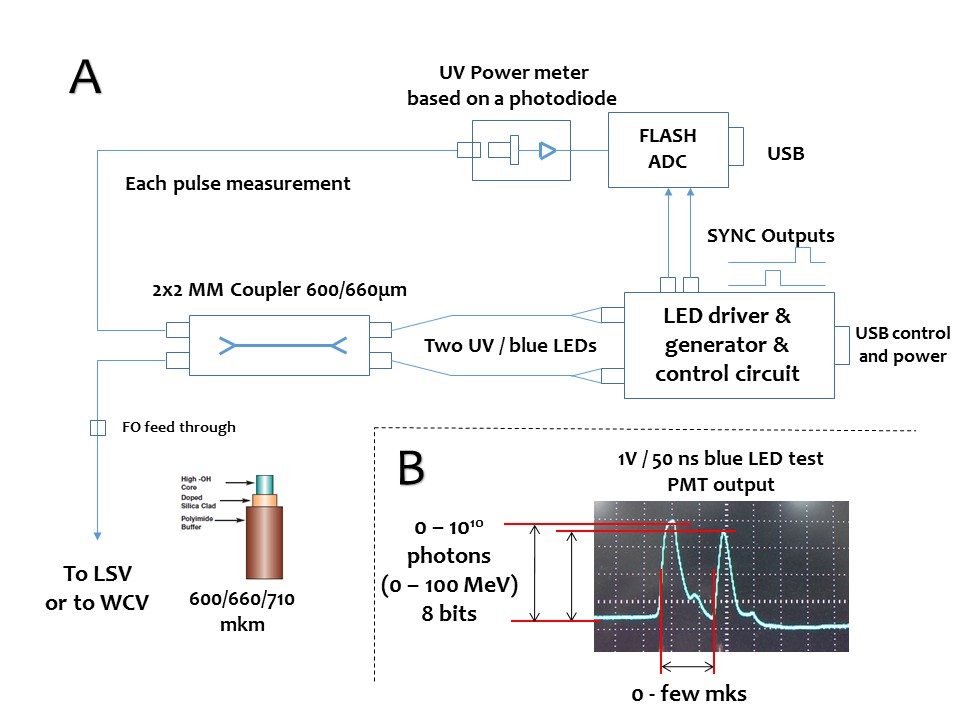}
\caption[Overview of the LED calibration system.]{A) The basic simplified scheme of the LED calibration system. B) An example of generated signals with ranges of parameters.}
\label{fig:Cal-LEDscheme}
\end{figure}

\begin{figure}[!t]
\centering
\includegraphics[width=0.5\columnwidth]{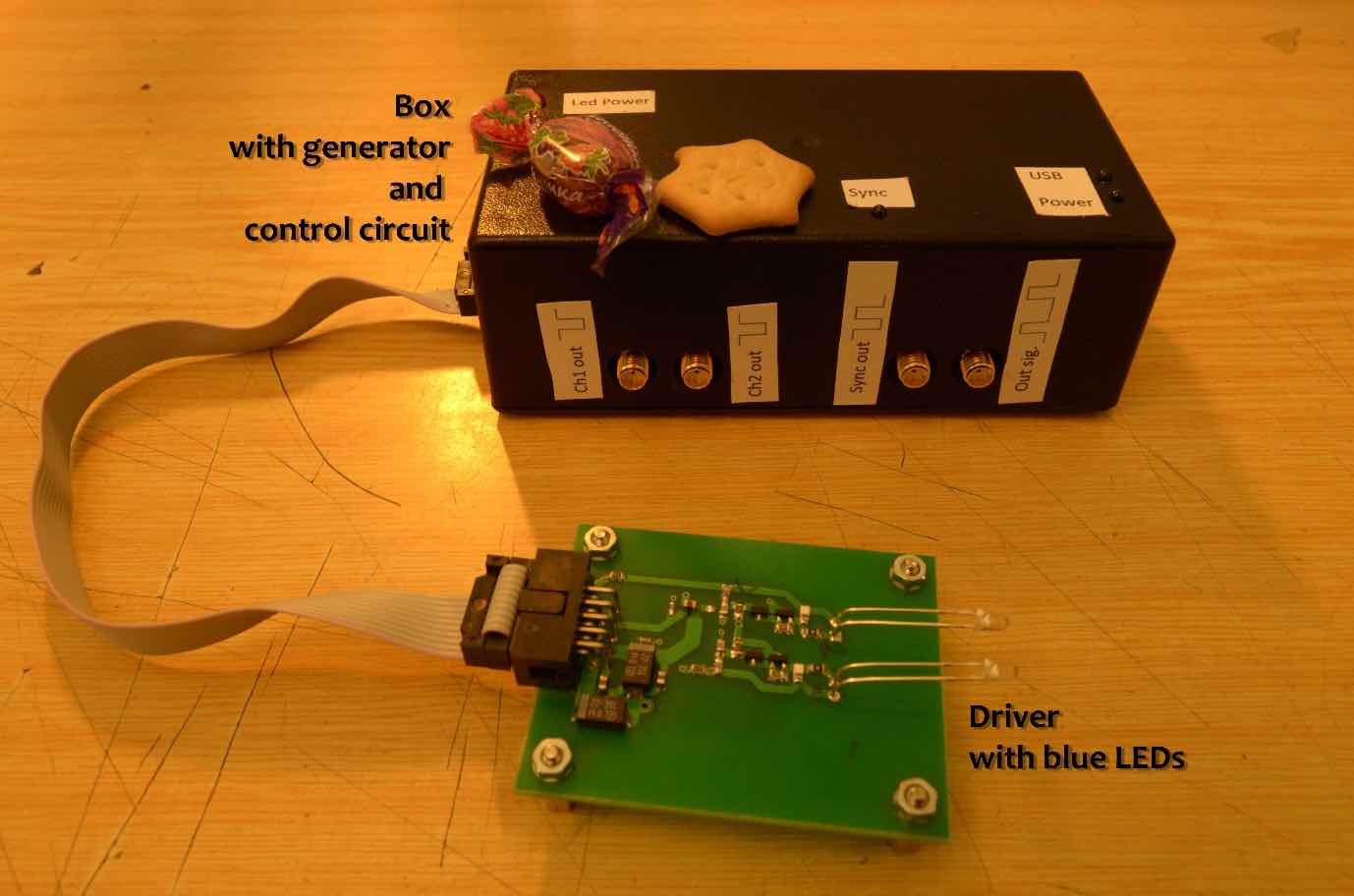}
\caption{Working prototype of the LED driver and a box with generator and control circuit.}
\label{fig:Cal-DriverPrototype}
\end{figure}

The original electrical scheme of a nanosecond LED pulser was proposed by J.S.Kapustinsky and his colleagues in 1985~\cite{Kapustinsky:1985jr} and it has gained in popularity since then~\cite{Belolaptikov:1997ds,Chepurnov:2016du}. The basic simplified sketch of the system is presented in Fig.~\ref{fig:Cal-LEDscheme}. There are four main features:
\begin{enumerate}
\item {\bf Adjustable input of the LED pulser:} The amplitude, rate and delayed time of the simulated signals are adjustable parameters. They are set remotely by the operator using the PC software. Fig.~\ref{fig:Cal-DriverPrototype} shows a working prototype of the driver and a box with generator and control circuit.
\item {\bf Double-LEDs scheme:} The LED driver has two diodes to provide two pulses very close in time. The delay time between the two pulses can be finely controlled. 
\item {\bf Controllable output of the LED pulser (line of control or power meter):} Allows measurement of the pulse energy. One uses the method of relative measurements. For calibration of the line of control a few test sets of the pulser signals are written and compared with the other calibration data collected during the cycle of measurements with radioactive sources. The energy scale up to \SI{10}{\MeV} is built and must be linear. Then the scale is extrapolated implementing the main assumption of the proposed procedure that the energy of the simulated signal is directly proportional to the number of emitted photons. 
\item {\bf All components except the end of the fiber with diffuser can be placed outside the cleanroom.}
\end{enumerate}

\begin{figure}[!t]
\centering
\includegraphics[width=0.5\columnwidth]{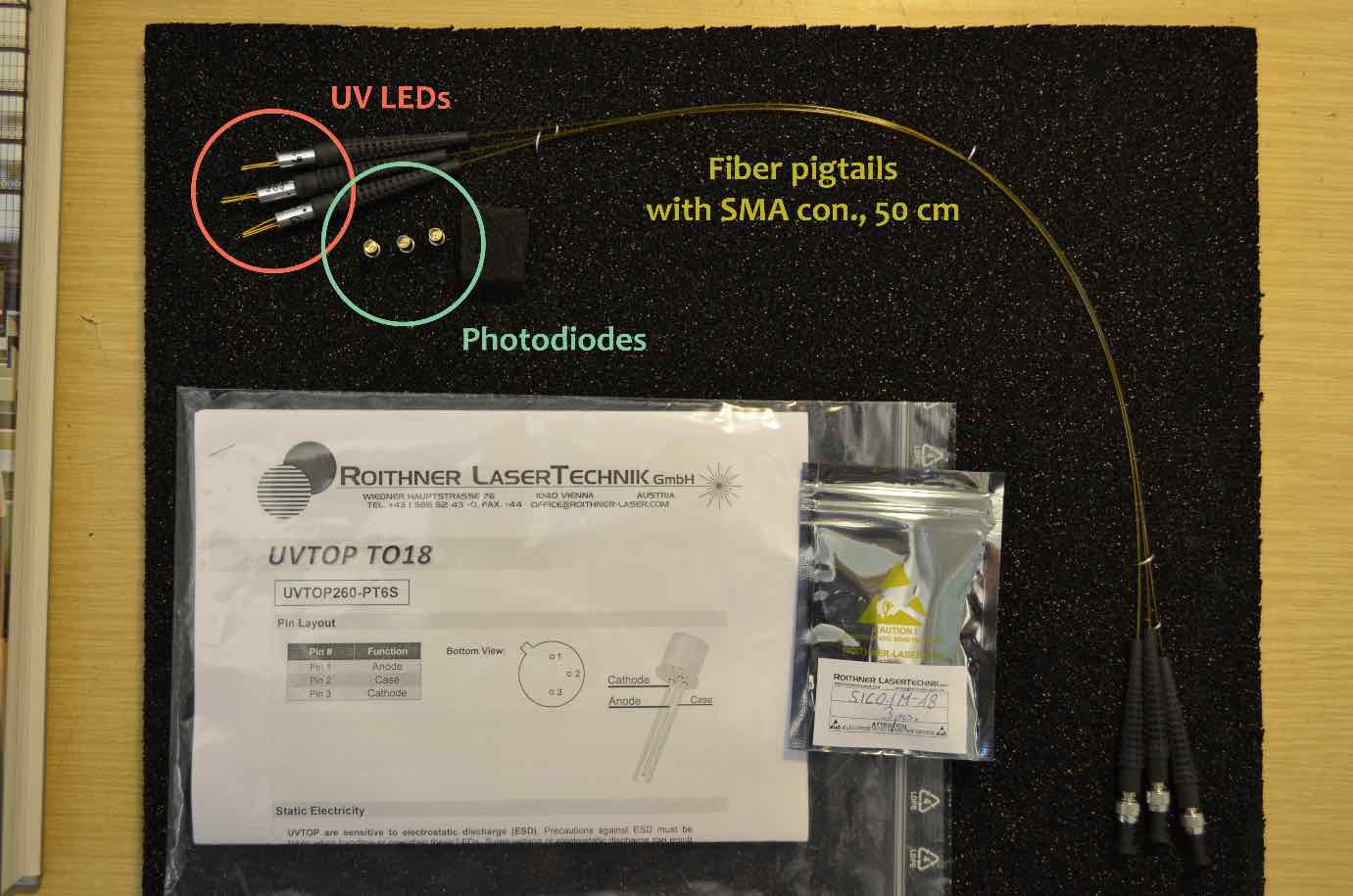}
\caption{UV LED and photodiode candidates for \DSk.}
\label{fig:Cal-LEDparts}
\end{figure}

The current design of the calibration system includes UV LEDs with wavelength \CalLEDWaveLength\ and a maximum pulse power of a few mW at \CalLEDCurrent\ (pulser mode). The diodes are shown in Fig.~\ref{fig:Cal-LEDparts}. It's worth noting that an off-axis calibration can be performed with the proposed calibration system. It could also be used for the development or tuning of spatial reconstruction algorithms. 

\subsection{Nuclear Recoil Calibration with R\lowercase{e}D}
\label{sec:ReD}

\subsubsection{Overview of R\lowercase{e}D}
Precise understanding of the scintillation and ionization signals produced by Ar nuclear recoils (NR) in \LAr\ will be necessary for the interpretation of data from \DSk.  Members of the \DS\ collaboration performed  the \SCENE\ experiment~\cite{Alexander:2013ke,Cao:2015ks} in~2013, providing information on the recoil energy and electric field dependence of these signals which was important in the analysis of \DSf~\cite{Agnes:2016fz}.  There is a plan to perform a dedicated set of experiments to obtain additional measurements needed to extend the range and improve the precision of the \SCENE\ results.

The plan is to expose a small \LArTPC\ to variable-energy neutron beams and to study single elastic neutron scatters from argon nuclei, as was done in \SCENE.  Improvements compared to \SCENE\ will include use of the Geiger-mode avalanche photodiode \TPC\ (\GAPTPC), a purpose-built small \LArTPC\ using \DSk\ photodetectors and having very high light yield for improved photoelectron statistics.  Neutron-generator and accelerator-produced beams will be used as needed to extend the recoil nucleus energy range from the \SCENERecoilsEnergyMax\ maximum of \SCENE\ up to about \ReDRecoilsEnergyMax.  Finally, the tantalizing possibility of a variation of the NR signals as a function of their directions relative to the applied drift (electric) field will be studied with a definitive high-precision and high statistics dataset.

Sensitivity to the direction of the nuclear recoils originating from \WIMP\ scatters would be a very highly desirable capability for a direct dark matter detection experiment.  It would provide a powerful signature (perhaps the only one) for identifying any observed signal with the galactic halo dark matter~\cite{Spergel:1988fg,Gondolo:2002fg}.  Detailed calculations have shown that in the standard isothermal halo model, just a few to several tens of events are required for a statistically significant directional signal~\cite{Morgan:2005hm}, depending on the angular resolution of the detection method.  Even structures beyond the isothermal halo, like galactic tidal streams, have been shown to give observable effects in detectors with very limited directional sensitivity~\cite{Freese:2005iy}   

Initial Recoil Directionality (\ReD) experiments will be carried out at Universit\`a degli Studi di Napoli ``Federico II'', using a $d-d$ neutron generator delivering monoenergetic neutrons of \DDNeutronEnergy.  Later the experiment will move to the INFN Laboratori Nazionali del Sud (\LNS) in Catania.  For \ReD, near-exclusive use of beam as well as dedicated target rooms  have been made available at Napoli and \LNS\ for long-term installation and running of the \ReD\ detector.  This mode of operation will enable lengthy runs with a stable proton/ion beam at the desired energy, see Sec.~\ref{sec:ReD-LNS}.  The inability to perform very long runs in a stable setup was another limitation of the \SCENE\ experiment, performed in user mode at the accelerator lab of Notre Dame University.

\begin{figure*}
\centering
\includegraphics[width=\textwidth]{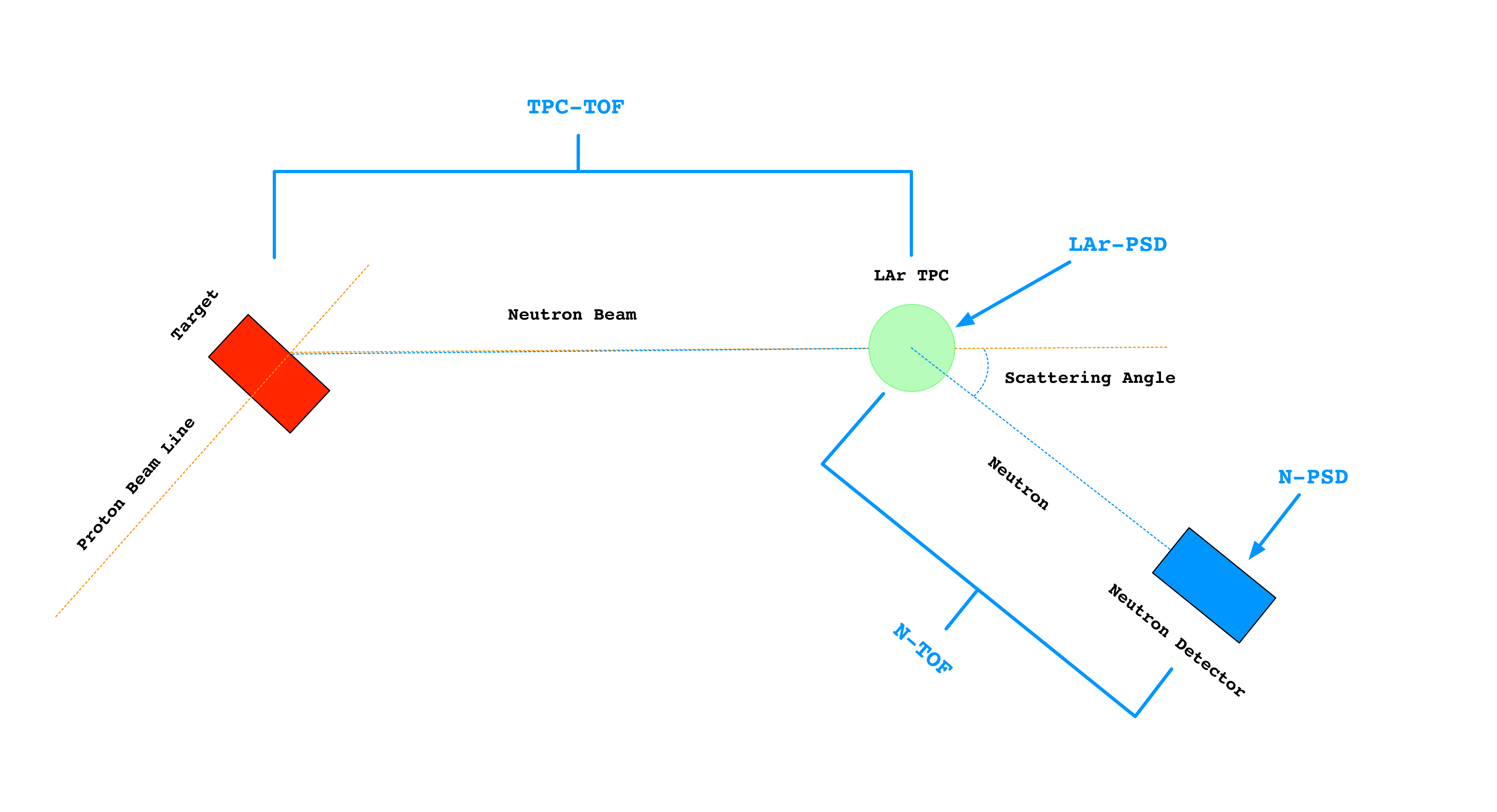} 
\caption{Schematic drawing of the \ReD\ experimental setup.}
\label{fig:ReD-Sketch}
\end{figure*}

Fig.~\ref{fig:ReD-Sketch} shows a schematic drawing of the experiment.  \LArTPC\ signals from neutron elastic scattering nuclear recoils of energy \ReDRecoilsEnergyRange\ in \LAr\ will be recorded in the \LArTPC, in coincidence with signals from the scattered neutron interacting in neutron detectors placed at various scattering angles relative to the beam.  The coincidence kinematics allow the energy and direction of the nuclear recoils occurring in the \LAr\ target to be precisely known event-by-event, given the beam energy and the scattered neutron angle.  As demonstrated in \SCENE~\cite{Alexander:2013ke,Cao:2015ks}, the coincidence technique combined with time-of-flight and the \PSD\ capabilities of both the \LArTPC\ and the neutron detectors  can give a very clean selection of single-scatter neutron elastic events.

\SCENE\ produced a study of the response of nuclear recoils in \LAr\ as a function of drift field~\cite{Alexander:2013ke,Cao:2015ks},  demonstrating that electron-ion recombination and primary scintillation yield for nuclear recoil tracks in \LAr\ depend strongly on the applied drift field.  \SCENE\ also characterized the energy dependence of ionization by nuclear recoils and \grs, and performed an initial search for a dependence of the nuclear recoil response upon the direction between the drift field and the initial momentum of the recoil.  This  gave tantalizing hints of a possible directional signal above \SCENEDirectionalityThreshold\ in \LAr, but no statistically significant result.  Factors limiting the precision of \SCENE\ included the use of  cryogenic \DSfPMT\ \PMTs, whose inherent instability at LAr temperature was exacerbated by the high-rate environment of the accelerator target room and the high dynamic range of the primary scintillation vs. ionization signals.  This limited the beam intensity that could be tolerated, and hence the attainable statistics, as well as introducing a number of corrections from the several monitoring systems.

The \ReD\ detector itself is a new type of \LArTPC, the \GAPTPC, whose innovative design takes advantage of the availability of \SiPMs\ to increase the photoelectron yield and greatly improve the stability of the photosensor response.  The \ReD\ \GAPTPC\ will be instrumented with tiles of the new-generation \SiPMs\ currently being developed for the \DS\ Collaboration, which are discussed in Sec.~\ref{sec:PhotoElectronics}.  The design of the \GAPTPC\ maximizes the light collection efficiency, with \SiPMs\ placed on the lateral surface of the detector as well as the top and bottom, see Sec.~\ref{sec:ReD-GAPTPC}.

The large-acceptance neutron spectrometer is designed to select and measure single-scatter neutron  elastic  events on argon (Sec.~\ref{sec:ReD-NeutronsSpectrometer}).  The spectrometer consists of an array of liquid scintillator neutron detectors on precision mounts which, combined with variable-angle neutron beams, allow the neutron-\TPC\ coincidences to select nuclear recoils at desired angles to the (perforce vertical) \GAPTPC\ drift field, including \num{0} and \SI{90}{\degree}.  A directional dependence would manifest itself as a difference in the scintillation and/or ionization response for nuclear recoils of the same energy but different track orientations, selected from data of a single run according to which neutron detector fired.

Directional effects on both scintillation light and ionization yield are expected in the recoil energy range relevant for dark matter searches, due to the phenomenon of columnar recombination in \LAr~\cite{Jaffe:1929gz,Jaffe:1913gs,Jaffe:1913cf,Swan:1965ha}.  Columnar recombination models suggest that the amount of ionization charge recombination (and therefore the  size of \STwo\ relative to \SOne) should, under some circumstances, vary with the angle between the field and the track direction.  When electrons are forced to drift through a column of ion-electron pairs produced by the ionizing track, the recombination probability is enhanced.  Larger effects are expected for tracks more energetic than the maximum measured with \SCENE.  Evidence for a subtle directional effect on $\alpha$-particle scintillation field-quenching  was seen long ago~\cite{Hitachi:1992gq}, but an effect useful for dark matter searches has not been found.

The \ReD\ experiment is also designed to provide a detailed characterization of the direction-averaged  nuclear recoils in \LAr\ in the presence of drift fields.  The optimized design of the \GAPTPC\ will minimize the probability of contamination from multiple neutron scatters inside the active volume.  The liquid scintillator neutron spectrometer will allow a large sample of nuclear recoil events to be collected in the energy region of interest for dark matter detection.  With a slight modification of the setup, electron recoils in \LAr\ will also be thoroughly studied, to help understand the response to electron recoil backgrounds.  Back-to-back \PositronAnnihilationGammaEnergy\ coincident \grs\ from a \ce{^22Na} positron source will provide a clean sample of events associated with a single Compton interaction in the argon detector, triggered on coincidences with a liquid scintillator behind the source.  A second liquid scintillator, shielded by a collimator, will be located at a known (variable) angle with respect to the line of travel of \grs, to select a precise electron recoil energy in \LAr.  By comparing the electron recoil energy with the nuclear recoil energy from the neutron beam measurements, one can map the ionization and scintillation efficiency as a function of the electric field and of the energy.

The success of the \ReD\ experiment could have a crucial impact on the broader program of the \DS\ Collaboration.  Aside from the added precision and range of detector response measurements, the discovery of sensitivity to the direction of recoils in \LAr\ would revolutionize the field of dark matter direct searches and would provide a formidable asset for \DSk, as well as future \LAr\ detectors.

\subsubsection{The \ReD\ \GAPTPC}
\label{sec:ReD-GAPTPC}

The ReD experimental setup is designed to minimize all neutron interactions other than single scatters in the liquid argon.  This is achieved in two ways: by reducing as much as possible the material between the neutron beam and the \LAr, and with an innovative structure for the \LAr\ detector which minimizes the presence of inactive volumes.

The conceptual design of the \GAPTPC\ detector has been described in Ref.~\cite{Rossi:2016jg}.  It is conceived to achieve a very high light of \GAPTPCLightYield, an exceptional single photon resolution of \GAPTPCSPEResolution, and spatial resolution of \GAPTPCSpatialResolution\ in each of the two directions in the $x-y$ plane, which will considerably improve the multi-hit events rejection.

A first step towards the \GAPTPC\ was taken at Naples by building a prototype \LArTPC\ with \SiPM\ readout.  In the present configuration, the detector design closely follows that used in \SCENE.  The active volume is contained within a \GAPProtoActiveDiameter\ diameter, \GAPProtoActiveHeight\ tall, \PTFE\ support structure lined with Lumirror enhanced specular reflector and capped by fused silica windows.  The \LAr\ is viewed through the windows by two \SiPM\ \DSkTileAreaStd\ tiles.  The windows are coated with \ITO, allowing for the application of electric field, and copper field rings mounted on the \PTFE\ structure maintain field uniformity.  All internal surfaces of the detector are evaporation-coated with \TPB.  A hexagonal stainless steel mesh is fixed at the top of the active \LAr\ volume and electrically grounded to provide the drift field (between the bottom window and the mesh) and the extraction and amplification fields (between the mesh and the top window).

This detector will be used for the preliminary tests of the neutron beam line.  Once installed into the experimental hall and coupled with a few neutron detectors of the neutron spectrometer, a preliminary test of the whole experimental facility will be made, with the initial objective of reproducing the results of the \SCENE\ experiment.  

\begin{figure}[t!]
\centering{} 
\includegraphics[width=0.5\columnwidth]{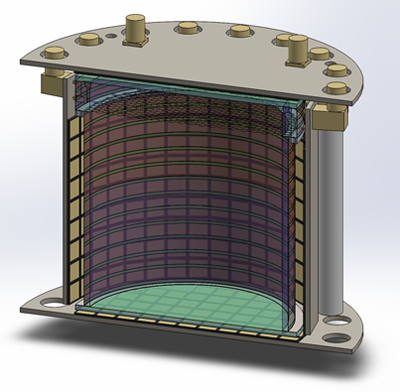}
\caption{A 3D rendering of the proposed \GAPTPC\ detector.}
\label{fig:ReD-TPC}
\end{figure}

The \GAPTPC\ (see Fig.~\ref{fig:ReD-TPC}) will have a cylindrical size of about \GAPTPCActiveDiameter\ diameter and height.  A fused silica vessel defines the active volume, with the top and bottom windows \ITO\ coated to allow for the application of the electric field.  Near the top, a hexagonal mesh grid terminates the drift field and allows the higher extraction field to be applied between it and the anode.  The drift field is kept uniform along the drift coordinate by means of field shaping rings, deposited by thin coating the walls of the fused silica vessel with \ITO.  The \GAPTPC\ will not make use of any reflectors. Different size optical readout detector modules will be installed all around the fused silica cylinder to achieve $4\pi$ coverage.  

A simple readout system based on CAEN~VT1730 14-bit 500 MS/s waveform digitizers is being setup for  
the \GAPTPC.  For the top array it is planned to individually read out each of the \SiPM\ dies in order to better study the \STwo\ signal distribution on the photosensors top plane, thereby enhancing the $x-y$ resolution.  The bottom plane, that does not take part in the $x-y$ reconstruction of the events, can be readout by a single channel or more, if required by signal to noise considerations.  The plan is to optimize the readout granularity of the lateral surface by Monte Carlo simulations.  The total number of signal output channels will be in the range 35-50.  Signal outputs will be fed to an external front-end amplifier and then to the waveform digitizer system.  The same readout will be used for the liquid scintillator channels of the neutron spectrometer, described below.  Digitized waveforms will be recorded based on a hardware trigger generated by a coincidence of channels above threshold.  In a later stage, a more sophisticated trigger will be deployed allowing more complex coincidence between the \LArTPC\ and the liquid scintillator counters.

\LArTPCs\ are usually housed within vacuum insulated (double walled) stainless steel cryostats, surrounded by a buffer region of inactive argon.  In this design, the main body of the TPC is a transparent fused silica vessel defining the active volume.  All photosensors and the ancillary equipment are placed outside the vessel so as to completely eliminate backgrounds due to the presence of a buffer of passive argon outside the \TPC.  The \GAPTPC\ will be housed in a single-walled stainless steel cryostat.  The cryostat will be equipped with thin windows for entry and exit of the neutron beam, if the simulation studies which are underway will suggest so.  An indium-sealed top flange will seal the cryostat and will mount a series of additional smaller flanges and feedthroughs needed for vacuum, \LAr\ fill, signal and HV cables, optical fibers for calibration laser signals, pressure, temperature and level monitors, safety valves, etc.  A Cryomech PT90 pulse-tube provides the cooling power through a heater block down to a condenser and cooling rods to the sensors.

Commercial argon gas will be initially passed through a commercial getter in order to remove impurities (mainly oxygen, nitrogen and water) from both the input gas and the detector components.  The system will be equipped with a gas recirculation loop, including the possibility of injection of a gaseous \ce{^{83m}Kr} source.

An additional innovative liquid recirculation system is available in Naples.  The system consists of a mechanical bellows pump coupled with a Trigon filter placed in the vacuum jacket of the cryostat.  Commercial pumps for cryogenic liquid make use of lubricants that can spoil the \LAr\ purity during the recirculation, and usually run only at a fixed speed.  The custom-designed Naples mechanical pump uses no lubricant, and furthermore cannot induce any electronic noise on the detector front end.  It also has the flexibility to adjust the recirculation speed as needed for the experiment, and finally it is much cheaper than commercial alternatives.

The \GAPTPC\ cryogenics will be equipped with an automated slow control system.  The \LAr\ filling and emptying procedure, the LAr and the gas recirculation and the exhaust safety  line will be automated by means of pneumatic valves controlled by the slow control system.  This allows parameters of the system to be changed while the neutron beam is on, without entering in the experimental hall.  It is also very beneficial in case of, {\it e.g.}, a power failure of the cryocooler, since it allows an automatic quick intervention on the cryogenic system to put the system in a safe and steady condition if needed.

The slow control system will be LabView based, allowing the necessary flexibility without excessive software effort.  The software architecture will be a state machine able to read, write, and display values using a graphical user interface.  The slow controls will interface with the DAQ system.  In particular, before each run there will be a complete check of the parameters and the status of the system will be logged in a file read by the DAQ.

\subsubsection{Neutron Beam at \LNS}
\label{sec:ReD-LNS}

A suitable neutron beam can be produced by the accelerator facilities of the INFN \LNS, using any one of a variety of production reactions.  Floor space is available to install the \ReD\ \LArTPC\  at about \SI{1}{\meter} from the interaction point, with all ancillary services and the neutron spectrometer.  The \RedLNSDistanceDedicatedLine\ beam line at the \LNS\ Tandem accelerator has been dedicated for this purpose by the \LNS\ Scientific Committee.  This will allow stable long-term operation of  the experiment and to accumulate statistics 
over many runs.

The \LNS\ Tandem accelerator has a maximum terminal voltage of \RedNaplesVoltage\ and good reliability.  It has been recently upgraded, replacing the original charging belt by a pelletron system, with excellent voltage stability verified both at low and high terminal voltages.  Many isotopes can be accelerated by the Tandem, with the exception of noble gases.  The maximum current ranges from \RedNaplesCurrent\ typically.

As the Tandem cannot be operated in pulsed mode, the energy of the produced neutron must be reconstructed event by event using kinematics, detecting the angle and the energy of the charged particle produced along with the neutron in the reaction.  The detection of the charged particle will provide the start for a time-of-flight measurement of the scattered neutron.  The coincidence between the charged particle detector and the \LArTPC\ will be recorded in the \DAQ, contributing an additional method for background reduction.  Among the variety of possible reactions for neutron production, the following set is considered:
\begin{inparaenum}
\item $p$(\ce{^7Li},\ce{^7Be})$n$;
\item $d$($d$,\ce{^3He})$n$;
\item \ce{^197Au}($d$,$np$)\ce{^197Au}.
\end{inparaenum}

Preliminary measurements will be performed using a simple neutron detector, in order to chose the most favorable reaction for the \ReD\ purpose and to study the room background, to minimize parasitic interactions which would spoil the global performance.  For this purpose, the \RedLNSDistanceDedicatedLine\ beam line must be completely refurbished, to optimize the layout and to accommodate a new beam dump and the \LArTPC.  Moreover, a new custom-made scattering chamber is being built to host the production targets and the Si detectors used for charged particle detection, featuring thinner walls and an exit window in the outgoing neutron direction, to minimize the neutron interactions.

\subsubsection{\lowercase{n}T\lowercase{o}F Spectrometer}
\label{sec:ReD-NeutronsSpectrometer}

To measure the scattered neutrons, the \ReD\ experiment will use an array of liquid scintillator neutron detectors that can measure the neutron scattering angles and the neutron time-of-flight (\TOF).  In this way, for  monochromatic neutrons single-scattered from \LAr, the kinematics is fully determined and the \ce{^40Ar} recoil energy and direction can be reconstructed, and background from direct neutrons coming from the production target or random coincidences of neutrons with signals in the \TPC\ can be suppressed.  The proposed  neutron spectrometer is based on \RedLNSNeutronSpectrometerDetectorsPMTsDiameter\ liquid scintillator counters from \RedLNSNeutronSpectrometerDetectorsManufacturer, using \RedLNSNeutronSpectrometerDetectorsScintillatorType\ liquid scintillator coupled to \RedLNSNeutronSpectrometerDetectorsPMTsType\ \PMTs.  These tubes are fast enough to provide accurate \TOF\ with a precision of around \RedLNSNeutronSpectrometerDetectorsPMTsTimeResolution.  These detectors can provide separation between neutron-induced and other signals, by using of pulse shape discrimination, from \RedLNSNeutronSpectrometerDetectorsPSD.  The plan is to use \RedLNSNeutronSpectrometerDetectorsNumber\ of such counters in the array, allowing one to record data for different recoil kinematics simultaneously.  The geometry is still to be finalized.  One scheme foresees the construction of a lightweight mechanical structure where counters can be placed at different angles on support guides held at constant radius from the \LArTPC\ center.  The support structure would allow simultaneous detection of  scattered neutrons corresponding to a given recoil energy but producing recoil argon atoms at different angles with respect to the electric field direction.

\begin{figure}[h!]
\centering
\includegraphics[width=0.5\columnwidth]{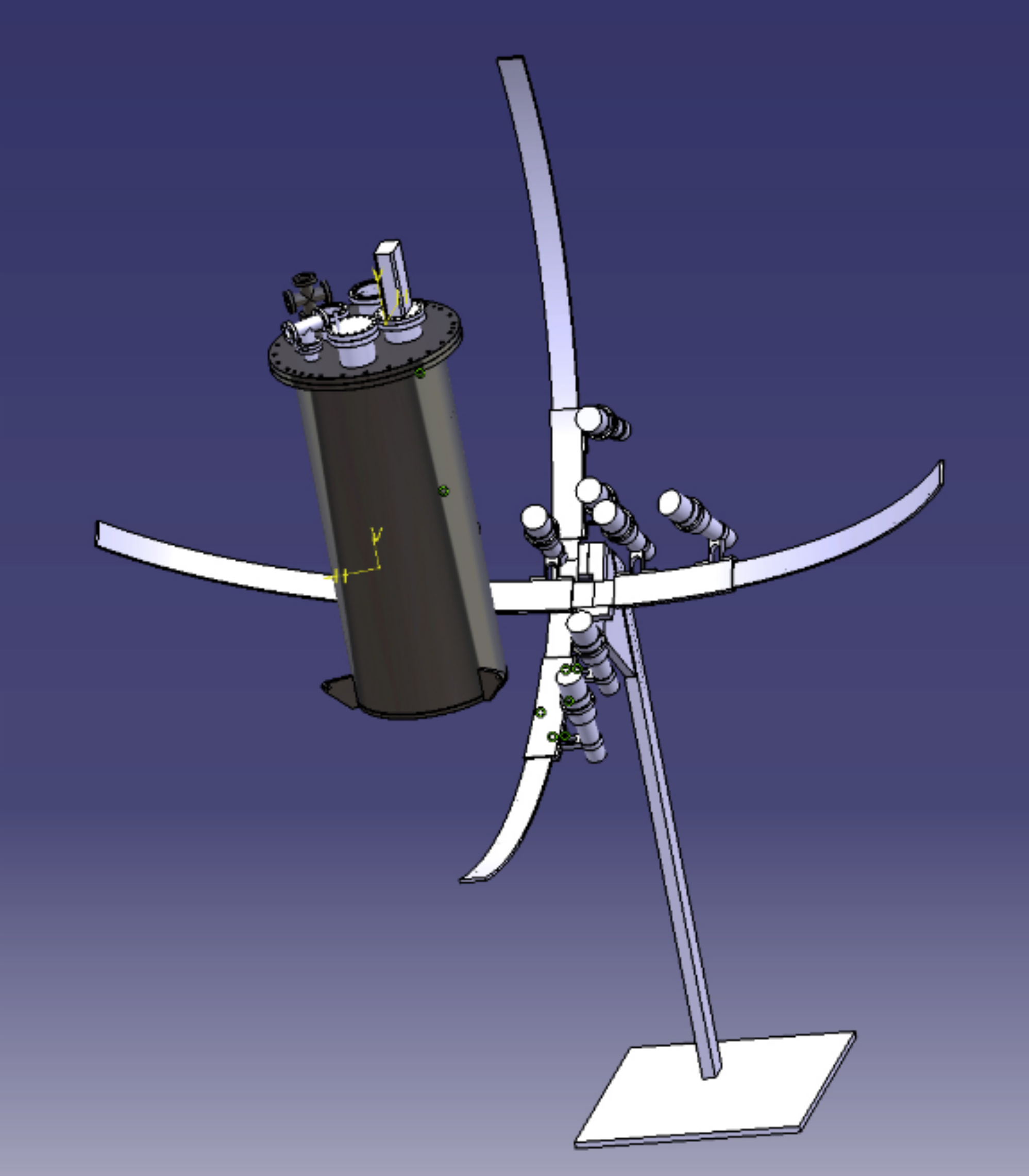}
\caption{Proposed ReD neutron spectrometer geometry and support structure.}
\label{fig:ReD-NeutronSpectrometerFront}
\end{figure}

\section{Trigger and \DAQ}
\label{sec:DAQ}

\subsection{Overview}
\label{sec:DAQ-Overview}

The present design for the \DSk\ electronics and its data acquisition system (\DAQ) accommodates both the large number of sensors and the long drift-time (expected maximum electron drift time is \DSkDriftTime) of the \LArTPC, and the readout of the veto detectors.  The design of the \DAQ\ has to be flexible enough to cope for the continuing progress in the \SiPMs\ performance and for the seamless integration of the veto detectors.

The trigger rate during dark matter search data taking has three major contributors: background events from detector materials, background events from \ce{^39Ar}, and random triggers.  To estimate the first term, the event rate measured in the \UAr\ running of \DSf\ is taken, excluding the contribution from the \PMTs\ and from the remaining \ce{^39Ar} in \DSf, and scaled by the ratio of surface areas of \DSk\ and \DSf, a factor of \DSkOverDSfAreaRatio, obtaining an expected rate of \DSkTriggerSurfaceRate; it is noted that these events will be concentrated at the surfaces of the active volume.  In the  assumption of a \ce{^39Ar} depletion of \DSkDArThreeNineDepletion\ from the use of \UAr, there will still be an additional rate of about \DSkTriggerDArThreeNineRate\ from \ce{^39Ar}, uniformly distributed throughout the active volume. In summary,  events with a correlated \SOne\ and \STwo\ signal at a rate of about \DSkTriggerTotalRate\ are expected in \DSk.  The average singles rate per channel is dominated on the other hand by the \DCR\ of \SiPMs.  With the required  \DSkSiPMDCRSpecification\ specification, this will imply a single rate for detector module of about  \DSkTileDCRSpecification.  It is assumed there is only a very small possibility of occasional light leakage between \DSkPdms\ which is neglected in the following, thus, for the \LArTPC\, a total of $\DSkTileDCRSpecification \times \DSkTilesNumber $=$ \DSkSingleTotRate$ singles rate is expected.

Data acquisition could be initiated by a coincidence of hits in the \TPC\ within a specified time window.  A coincidence of \DSkTriggerThresholdHits\ hits in \DSkTriggerThresholdWindow\ would result in a random trigger rate well below \DSkTriggerRandomRate.  Nuclear recoils at the trigger threshold, producing about \DSkWIMPMinPE\ in \DSkWIMPMinPETimeWindow, would result in the collection of \DSkWIMPMinTwoHundredns\ within the first \DSkTriggerThresholdWindow.  Thus, the trigger would be fully efficient for \WIMP-like signal of interest.  \SOne\ events for \WIMP\ candidates produce signals with no more than \SI{1}{\pe} per tile.  One should expect \DSkDarkCountHitsPerMaxDriftTime\ during the maximum \DSkDriftTime\ drift time.

A conceptual design for the \TPC\ \DAQ\ is described in Sec.~\ref{sec:DAQ-SignalDigitization}.  A flexible design could in principle lead to a common architecture for both the \TPC\ and vetoes, with the obvious simplification in the operations and commissioning of \DSk, however other options are also considered.  In any case the \DAQ\ will be unified at the level of the central event builder.

The event building and software trigger stage is described in Sec.~\ref{sec:DAQ-HLST} and is realized with modern commodity CPUs and connected through fast ethernet with front-end DAQ processors.  Given the low expected rate a trigger-less option is considered viable.  Digitized data would streamed to the event builder and software trigger for a detailed analysis and trigger selection.

Synchronization between the \TPC\ and veto \DAQ\ and between the different readout boards running the \TPC\ digitization is fundamental for the effectiveness of the design, and will be provided and maintained during data taking.  The same clock source of the \TPC\ \DAQ\ will be used and digital signals (like GPS time stamps or trigger IDs) will be generated to uniquely identify each event regardless of the trigger origin and the detector.  A pulsed signal to all channels will be used to check and correct the alignment of each channel among the three detectors.

The \DAQ\ system will be located in an electronics room that is to be placed close to the water tank, in an environment that allows personnel access, keeping to a minimum the cable length from the \LArTPC\ detector to the signal receivers.  The choice of this transmission design mitigates intrinsic radioactivity inherent in shielded coaxial cables, and has been demonstrated to keep noise at a manageable level.  The alternative optical transmission described in~\ref{sec:PhotoElectronics-SignalTransmission-Optical} would ease the issues related to the cable length and routing and would imply minor revision to the electronics  described in the following.

\subsection{Signal Digitization}
\label{sec:DAQ-SignalDigitization}

\begin{table*}
\rowcolors{3}{gray!35}{}
\centering
\caption[Estimated count rates and data rates for \TPC\ and veto readout.]{
Estimated count rates and data rates for \TPC\ and veto readout.   To estimate the typical rate from real signals in addition to the noise rate an \SOne\ signal of \DSkDAQSOneAvgPE\ and a true rate of \DSkTriggerTotalRate\ in the \LArTPC are assumed.  For the veto detectors, data will be collected with the main \LArTPC\ trigger at the same \DSkTriggerTotalRate\ rate.  The first line details the data rate when each channel is digitized at the \DSkDAQADCSamplingRate\ sampling rate for the full \DSkDAQVetoWindow\ window starting around the \LArTPC\ trigger.  The zero suppressed line takes into account the storage saving resulting from zero suppression, which would lead to store about \DSkDAQVetoHitSamples\ (\DSkDAQVetoHitSize) for each hit.  The veto hits are dominated by the \DSkVetoPMTDCR\ dark count rate of the veto \MCPPMTs, leading to about \DSkDAQVetoHitAvgNumber\ to be digitized in \DSkDAQVetoWindow.}
\begin{tabular}{ccccc}
Signal		&Noise Rate		&True Rate		&Data			&Rate\\				
\hline                         
\SOne
			&$\DSkTileDCRSpecification \times \DSkTilesNumber = \DSkSingleTotRate$
							&$\DSkTriggerTotalRate \times \SI{600}{\ch} = \SI{27}{\kilo\hertz}$
											&\SI{8}{\byte\per\ev\per\ch}
															&\SI{10.5}{\mega\byte\per\second}\\
\STwo
			&\textendash	&$\DSkTriggerTotalRate \times \SI{4000}{\ch} = \SI{180}{\kilo\hertz}$
											&\SI{300}{\byte\per\ev\per\ch}
																&\SI{54}{\mega\byte\per\second}\\
Veto
			&\textendash	&$\DSkTriggerTotalRate \times \SI{200}{\ch} = \SI{9}{\kilo\hertz}$
											&\SI{15}{\kilo\byte\per\ev\per\ch}
																&\SI{135}{\mega\byte\per\second}\\
Zero-suppressed Veto
			&\textendash	&$\DSkTriggerTotalRate \times \SI{300}{\hits} = \SI{14}{\kilo\hertz}$
											&\SI{30}{\byte\per\hit}
																&\SI{0.4}{\mega\byte\per\second}\\																
\end{tabular}
\label{tab:DAQ-Rate-new}
\end{table*}

\subsubsection{\LArTPC}

For what concerns the \LArTPC\ readout, one or two analog read-out channels for each photo detector module (\DSkPdm) are being considered, as explained in Sec.~\ref{sec:PhotoElectronics}: the analog signal will be extracted from the active volume of the detectors on twisted pairs or on optical fibers.  The baseline design of the DAQ for the \LArTPC\ foresees \DSkTilesNumber\ channels, or, as backup plan, twice that number.  Considering the geometry of the detector, the properties of the light emission of \SOne and the propagation of ligh in the active volume, the photons will be evenly distributed among the many \DSkPdms.  Consequently, in the region of interest for the \WIMP\ search, the typical occupancy of each channel will be significanly smaller than one.  In the single photon counting hypothesis the best read-out strategy is TDC based with a time resolution of O(\DSkTileTimeResolutionSpecification).  For the \STwo\ signal, where the total number of photons in the signal depends on the details of the \LArTPC\ design and may be a factor 20 higher, with approximately half the photons concentrated on a few \DSkPdms, the number of photoelectrons produced by one sensor may be as many as \DSkSTwoMaxPerChannel\ at a peak rate of \DSkSTwoRateMaxPerChannel.  The best strategy for the read-out of \STwo\ is the full digitization of the waveform, which is up to \DSkDAQSTwoDigitizationWindow\ long.

The proposed design intends to efficiently acquire both signals with a single data stream thanks to the use of highly optimized digital signal processing on chip (in the processors of the acquisition boards) or on the software trigger stage of the proposed \DAQ\ system~\ref{sec:DAQ-HLST}.  In the latter case the additional bandwidth needed is significant but would not be a bottle-neck given the present day performance of computing and networking hardware.

\begin{figure}[t]
\begin{center}
\includegraphics[width=\columnwidth]{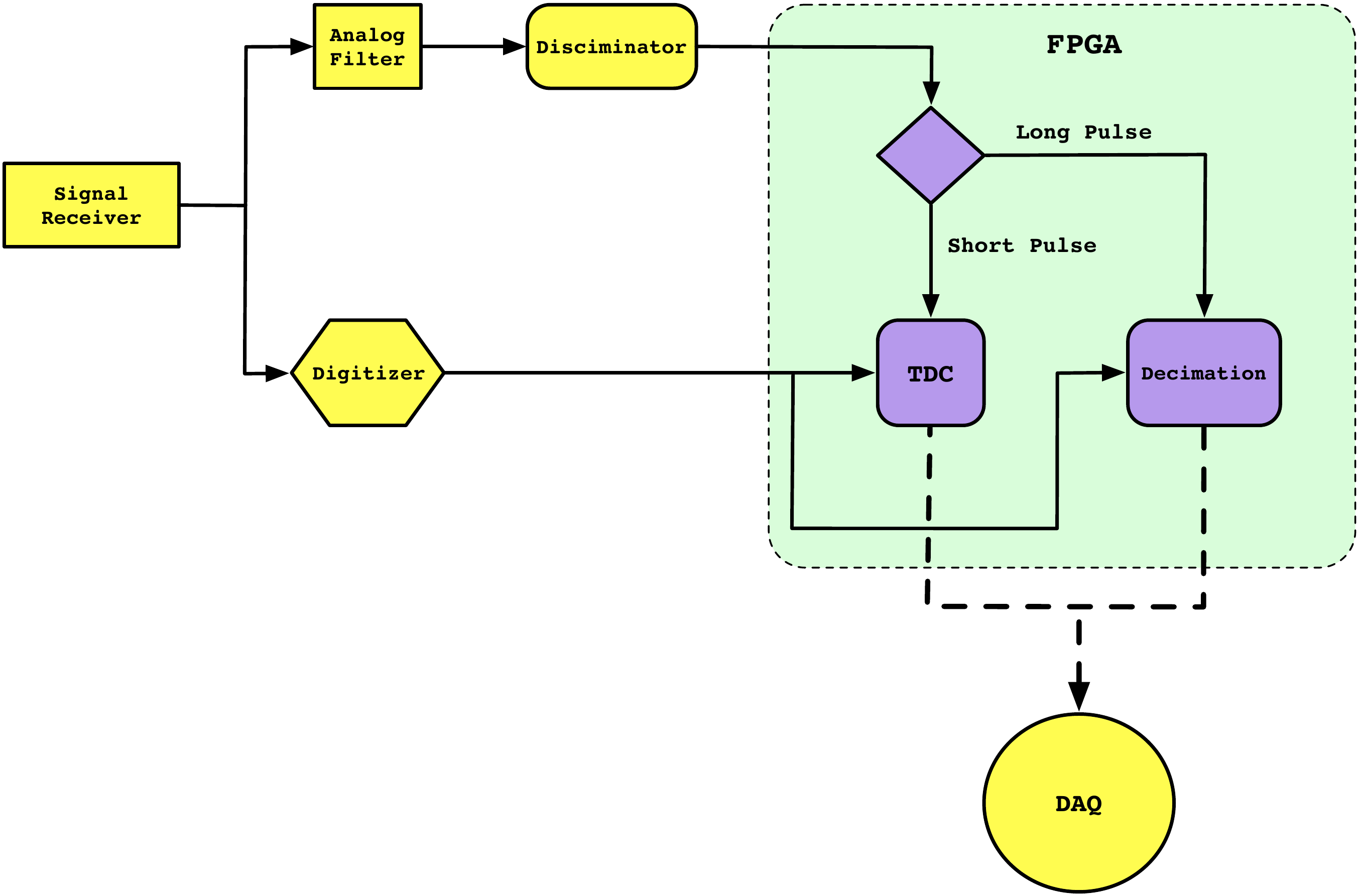} 
\caption[A schematic block diagram of the proposed \DSk\ DAQ readout channel.]{A schematic block diagram of the proposed \DSk\ DAQ readout channel. Multiple analog channel will be processed by a single FPGA.
\label{fig:DAQ-Flow}}
\end{center}
\end{figure}

The presently envisaged baseline scheme for the \TPC\ \DAQ\ electronics hardware foresees a signal receiver (differential or optical), a discriminator fed with a filtered signal, a high speed digitizer (\DSkDAQADCBits\ bit at \DSkDAQADCSamplingRate) connected to a field programmable gate array (FPGA) as shown in Fig~\ref{fig:DAQ-Flow}. The number of channels that will be served by a single FPGA, the required resources of the chip, and the number of channels per board are still subjects of further investigations, but we assume, based on similar modern projects, that a single board processing of the order of 30 analog channels is feasible with present-day technology.

As discussed in Sec.~\ref{sec:PhotoElectronics} the signal discrimination is performed after an analog filter that will minimize the rate of fake hits: depending on the actual \SNR\ provided by the pre-amplifier a strong filtering will be required (with a formation time up to hundreds of nano seconds).  The discriminator output is fed to an FPGA that will receive as well the data of a free running digitizer.  Depending on the length of the discriminated pulse, one can select between two different algorithms in the FPGA: for a short pulses (the case of \SOne) the TDC algorithm will search in a specific gate, defined by the properties of the analog filter, the time of the hit and its amplitude.  For long pulses (\STwo) the corresponding samples will be down-sampled with a CIC filter to a frequency compatible with the signal of \STwo\ (about \DSkDAQADCDownSampledRate).  To store \SOne\ hits \DSkDAQSOneSizeTotal\ are sufficient (\DSkDAQSOneSizeChannelID\ for channel identification, \DSkDAQSOneSizeTiming\ for timing, and \DSkDAQSOneSizeCharge\ for charge), while for storing the \STwo\ waveform about \DSkDAQSTwoSamples\ samples (equivalent to \DSkDAQSTwoSize) is required.  Various rate estimations are provided in Table~\ref{tab:DAQ-Rate-new}.

The DAQ is trigger-less and is implemented by timing with respect to a \DSKDAQClockCycle\ clock.  All module, crate, and event builder clocks are synchronized to a master clock tied to a global GPS system.  Thus each event is timed with respect to the start of a \DSKDAQClockCycle\ clock signal.  The DAQ records a maximum of \DSKDAQClockTicksMax\ clock ticks, equivalent to \DSKDAQClockTickSize\ of timing information.  Thus, depending on when the last \STwo\ event occurs, there is a maximum system deadtime of \DSkDAQADCMaxDeadTimePerSecond\ every \DSKDAQClockCycle, or \DSKDAQClockDeadTimeFraction.  The system can reset in a few \DSKDAQClockResetScale.  The clock and its fan-out will be similar to the clocking used in \DSf.  It is an ``in-house'' single width 6U~VME module which accepts TTL, LVTTL, or CMOS signals with selective \num{50} or \SI{100}{\ohm} termination.  It uses a Texas Instruments CDCE906 programmable 3-PLL clock synthesizer/multiplier/divider and provides the common \DSKDAQClockRate\ clock which is fanned-out to all crates.  It is synchronized by a GPS receiver and provides a data reset and data start every second.  The individual PLL multipliers in each FPGA in a module provide a synchronized \DSKDAQPLLClockRate\ clock.  It also removes any phase shifts and allows a measure event timing with respect to the data start time within \DSkTileTimeResolutionSpecification.  Note that this time resolution is more than adequate, and it is very easily synchronized across the system.

The University of Houston group is testing prototype boards based on the above design.  An FPGA board is available and is able to read \num{96} signals.  The board is similar in architecture to Fig.~\ref{fig:DAQ-Flow}, with the exception that the input stage employe a simple shaper rather than an optimal filter.  The state-of-the-art FPGA implements the timing measurement of the discriminated signal and perform zero suppression and other operation on the output of a free running digitizer at \SI{50}{\mega\byte\per\second}.  The board had been originally designed for the readout of the time and charge of hits in the Mu2e experiment tracker.

\subsubsection{Veto}

Several options are considered for the readout of the about \DSkVetoPMTsNumber\ PMTs of the veto detectors, as discussed also in Sec.~\ref{sec:Vetoes}.  In short, in order to work the prompt and delayed veto analysis, and to calibrate the energy response of the \LSV\ detector from few tens of keV to a few MeV, both charge and timing information of \PMT\ pulses are needed.  Preliminary simulations showed that \PMTs\ are no more in SPE regime for signals in the LSV above a few hundred MeV, therefore a charge readout is necessary.  A timing-only readout is adequate for the \WCV\ readout, where precise energy calibration is not fundamental.  With the aim of maximizing the electronics uniformity for the DAQ system the time information can be easily extracted using the same hardware architecture used for the \LArTPC\ readout (Sec.~\ref{sec:DAQ-SignalDigitization}).  The front-end board developed for the Veto system described in Sec.~\ref{sec:Vetoes-FrontEnd} can also perform the time measurement within the FPGA included in the design.  For what concern the energy measurement, due to the different signal time structure, between the \LArTPC\ and LSV,  it might be necessary to have dedicated systems for the LSV.  This option is still being explored.

\begin{figure*}[t!]
\centering
\includegraphics[width=0.45\columnwidth]{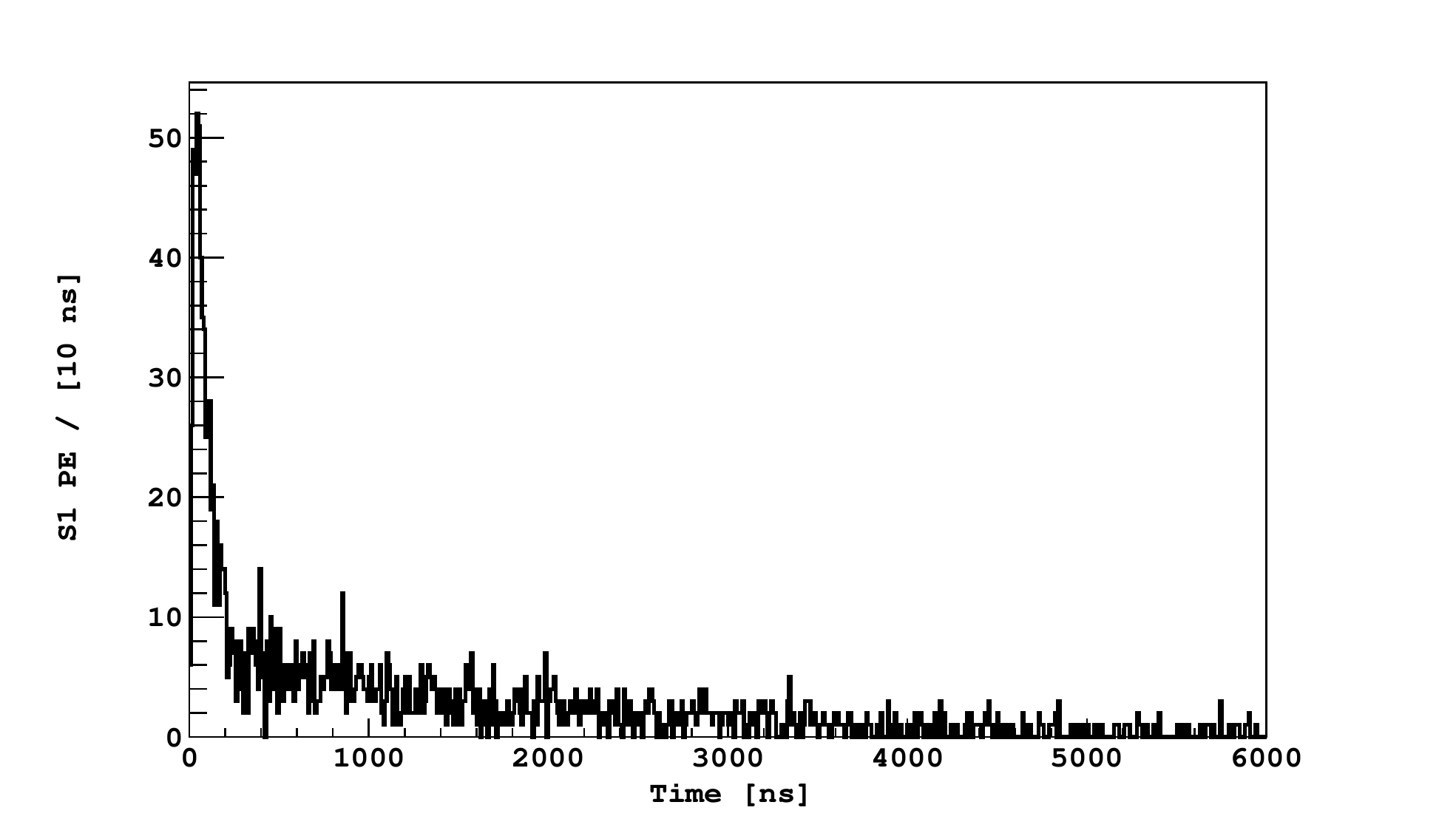}
\includegraphics[width=0.45\columnwidth]{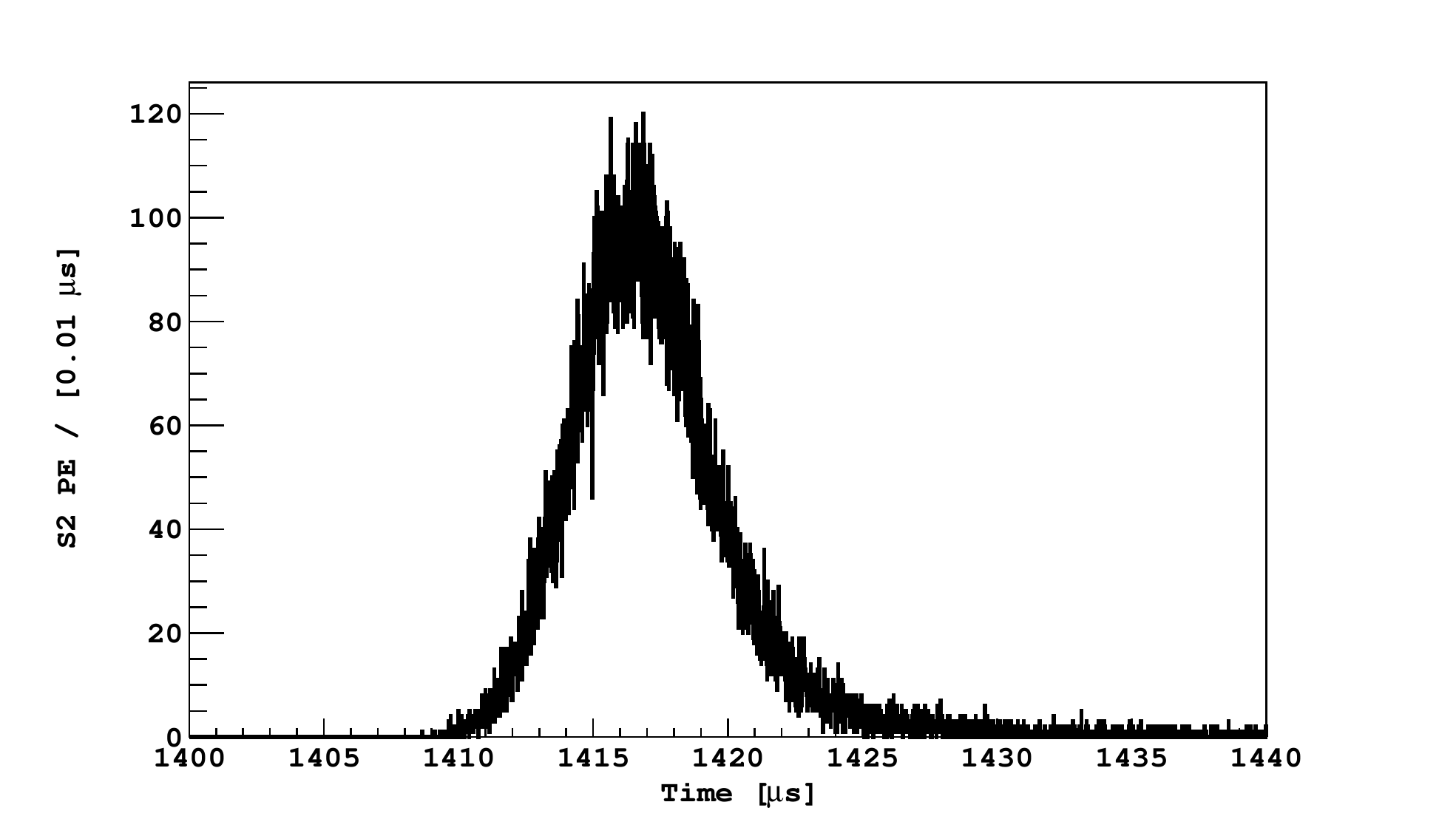}
\caption[Arrival time of \SOne\ and \STwo\ hits from a complete simulation of the detector, electronics, and DAQ.]{Arrival time of \SOne\ {\bf (Left)} and \STwo\ {\bf (Right)} in \DSk, resulting from a complete simulation of the detector, electronics, and DAQ.}
\label{fig:DAQ-ArrivalTime}
\end{figure*}

\begin{figure*}[t!]
\centering
\includegraphics[width=0.45\columnwidth]{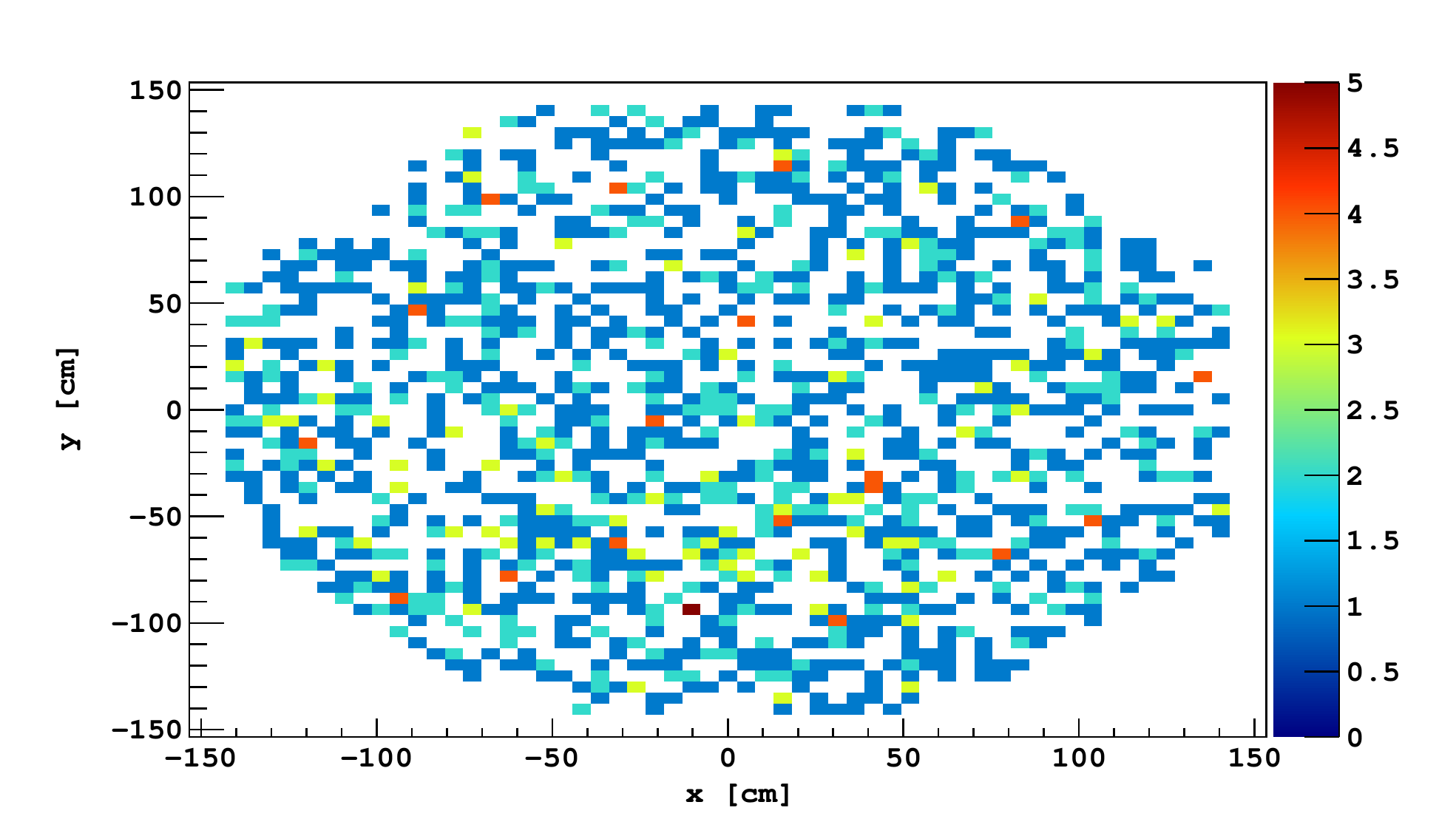}
\includegraphics[width=0.45\columnwidth]{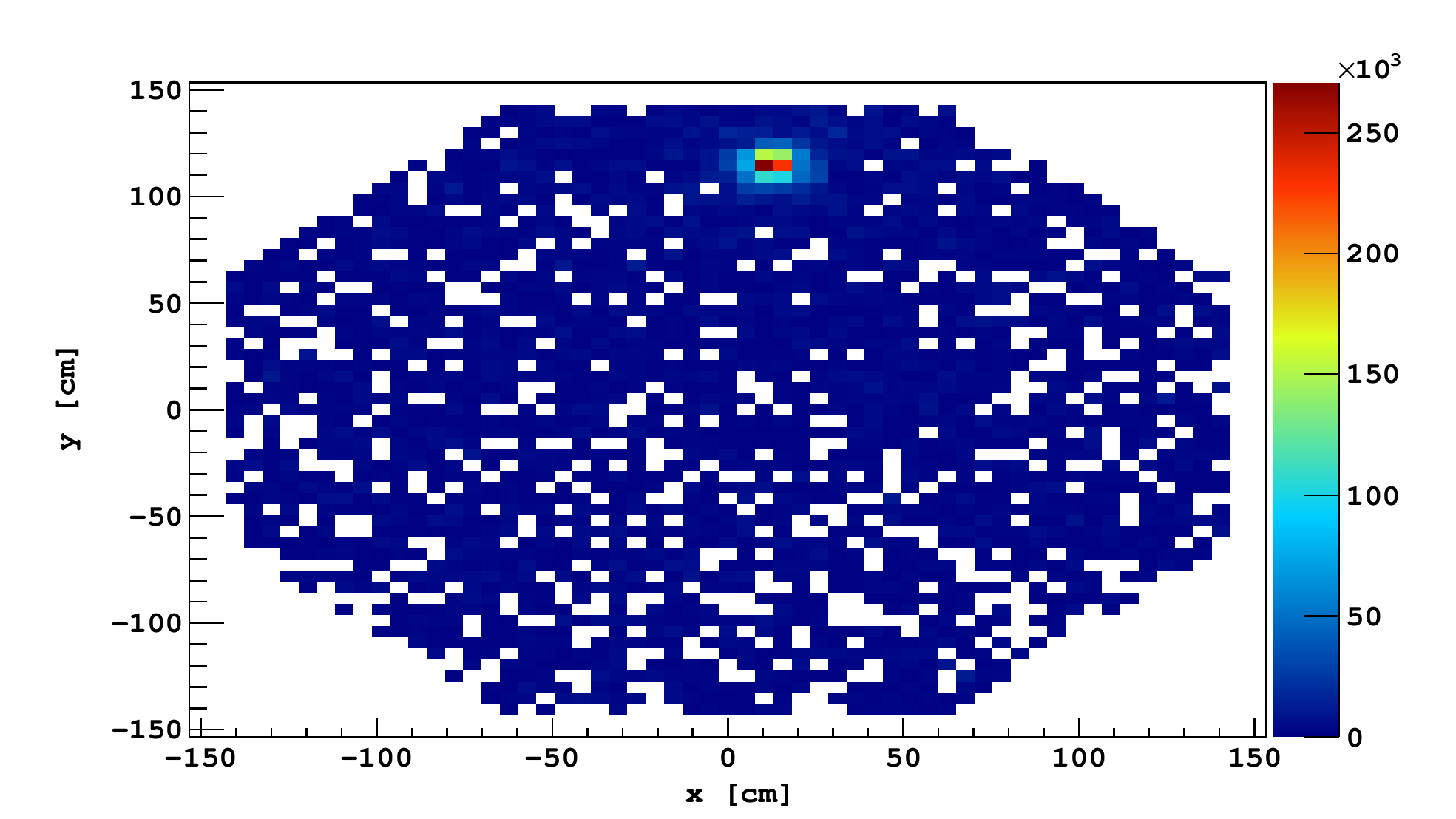}
\caption[Hit maps for for \SOne\ and \STwo\ events.]{Hit Maps for the bottom plane of a typical \SOne\ signal {\bf (Left)} and for the top plane of the \STwo\ signal {\bf (Right)}.}
\label{fig:DAQ-HitMap}
\end{figure*}

\subsection{Expected Performances from Simulation of the \LArTPC\ \DAQ}
\label{DAQ-Performance}

Simulation studies have been performed to understand the expected performance of the readout scheme discussed above.  Simulated samples of both electron recoils from \ce{^39Ar} decay and \ce{^40Ar} nuclear recoils have been produced using the \DSk\ geometry, including the detailed \DSkPdm\ layout on the top and bottom photosensor plane, including the \DSkPdmsGap\ wide gaps between \DSkPdms, resulting in a geometric efficiency of about \DSkPdmsGeometricEfficiency.  The photon detection efficiency for the \DSkPdm\ is set to \DSkPdmPDESpecification\ requirement, see Sec.~\ref{sec:PhotoElectronics-Introduction}.  In order to obtain a more accurate simulation of the system response \SiPMs\ noise is taken into account and injected in the data-flow: the dark count rate is set to \DSkTileDCRSpecification\ per readout channel uniformly distributed in time, while the correlated noises amount to \DSkDAQSimulationAP\ for \AP\ (after-pulse probability) and \DSkDAQSimulationDiCT\ for \DiCT\ (direct cross talk); \DeCT\ (delayed cross talk) has been neglected since it is negligible, see Sec.~\ref{sec:PhotoElectronics-Characterization}.  The readout simulation is performed on two parallel tracks: the first employs TDCs to produce ``hits'', the second one uses ADCs to sample the analogical waveforms in a temporal window, thus giving the charge.  The ADC simulation uses \DSkDAQADCBits\ bits digitization; the \SPE\ (single photo-electron) response is set to have on average an amplitude of $6$ bits, leaving the other $6$ free to have a dynamic range of $64$ PE.  Hits carry the arrival time information, performing well in the estimation of the \SOne\ prompt fraction of light, while ADC sampled signals are needed to measure the number of photons for \STwo\ signals due to the high occupancy of the readout channels in this case.  Events in the simulation are uniformly distributed in the \LAr\ volume, with the \ce{^39Ar} energy spectrum for electron recoils and a power law spectrum for \ce{^40Ar} nuclear recoils.  The time structure of \SOne\ and \STwo\ signals can be seen in Fig.~\ref{fig:DAQ-ArrivalTime}.  Typical hit maps are shown in Fig.~\ref{fig:DAQ-HitMap} for both \SOne\ and \STwo\ signals.  After a fiducial cut of \SI{5}{\centi\meter} from the top and bottom planes of the SiPMs in the TPC volume no saturation is observed for either \SOne\ or \STwo\ signals.

\begin{figure*}[t!]
\centering
\includegraphics[width=0.45\columnwidth]{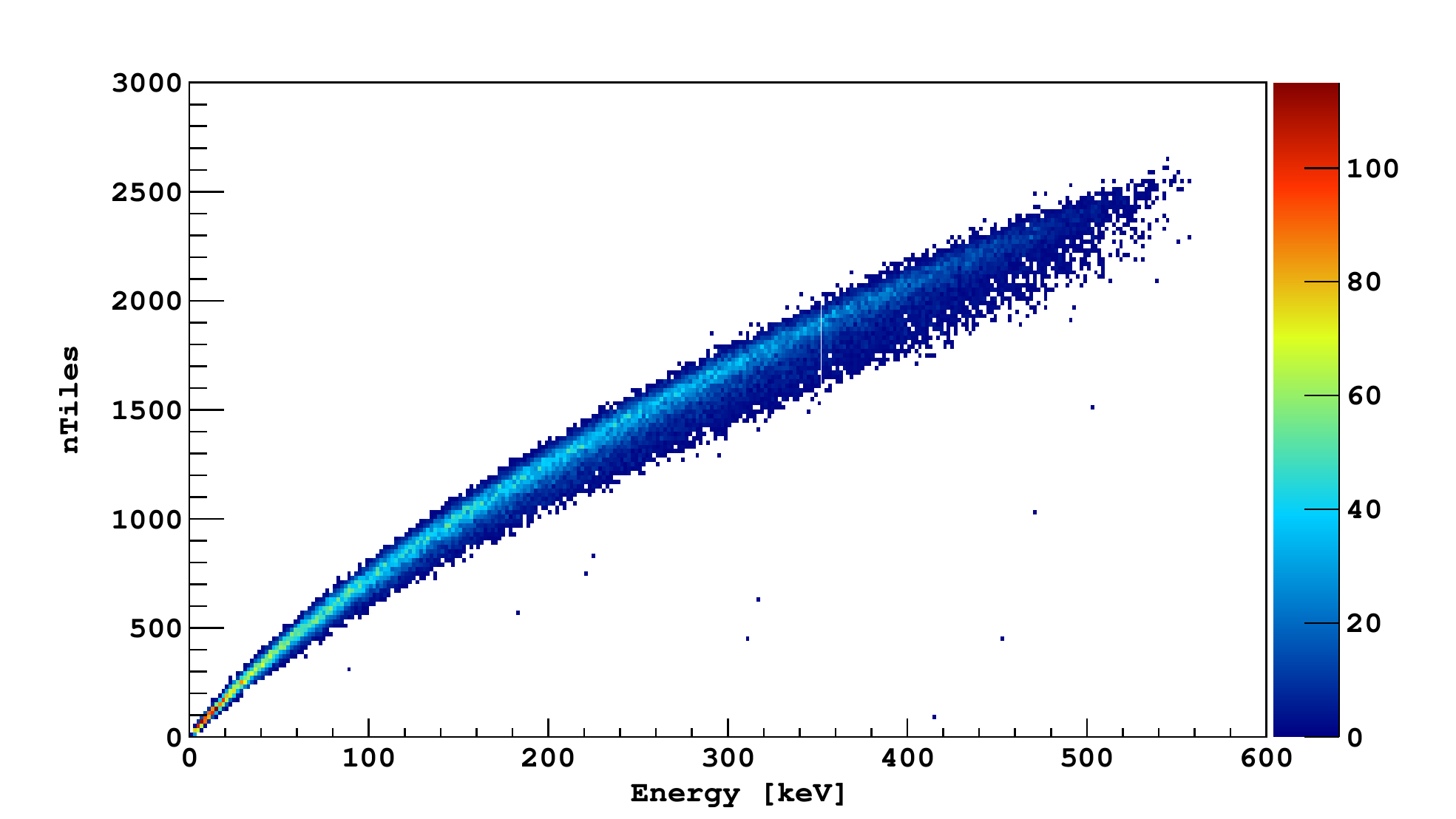}
\includegraphics[width=0.45\columnwidth]{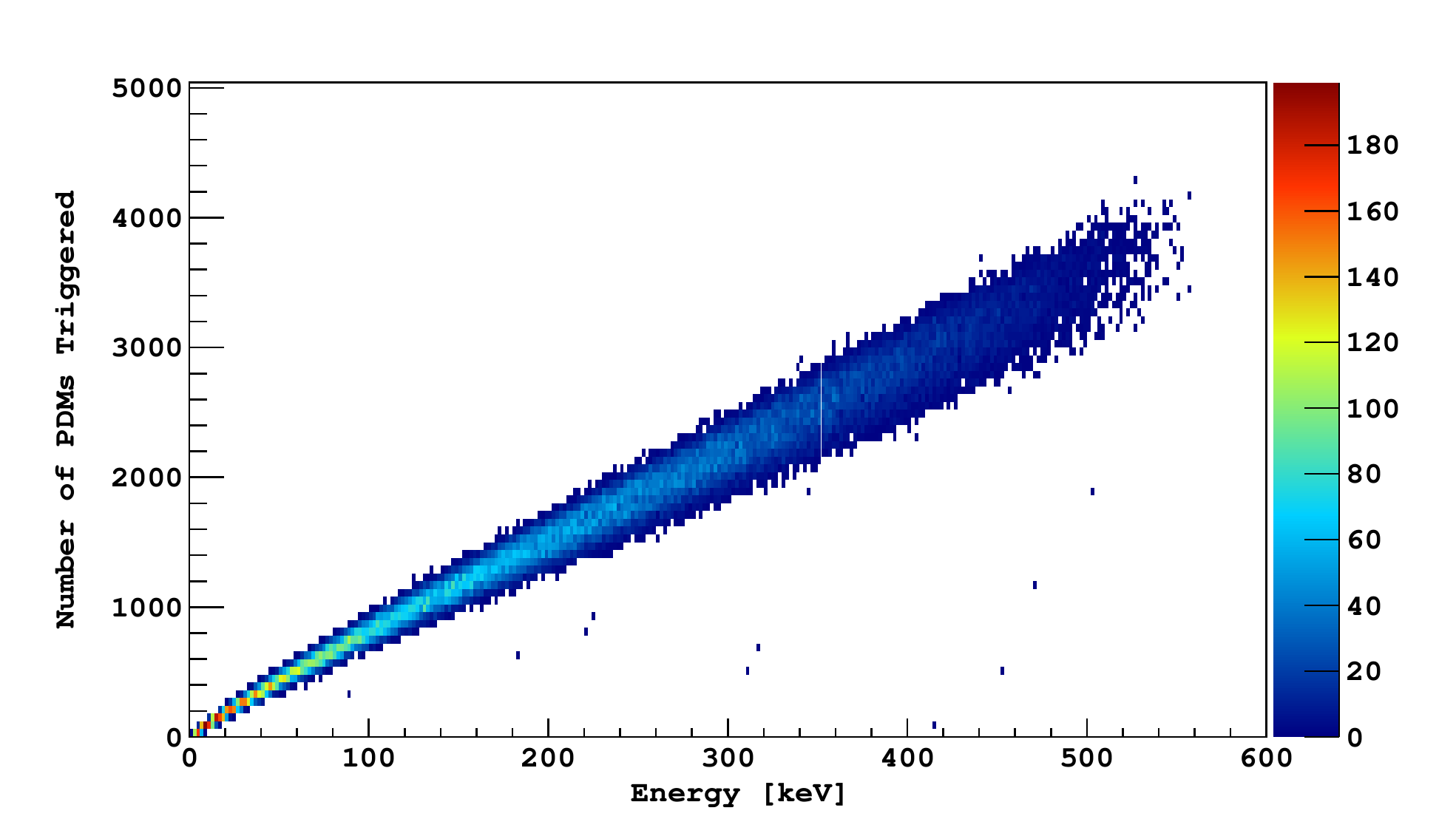}
\includegraphics[width=0.45\columnwidth]{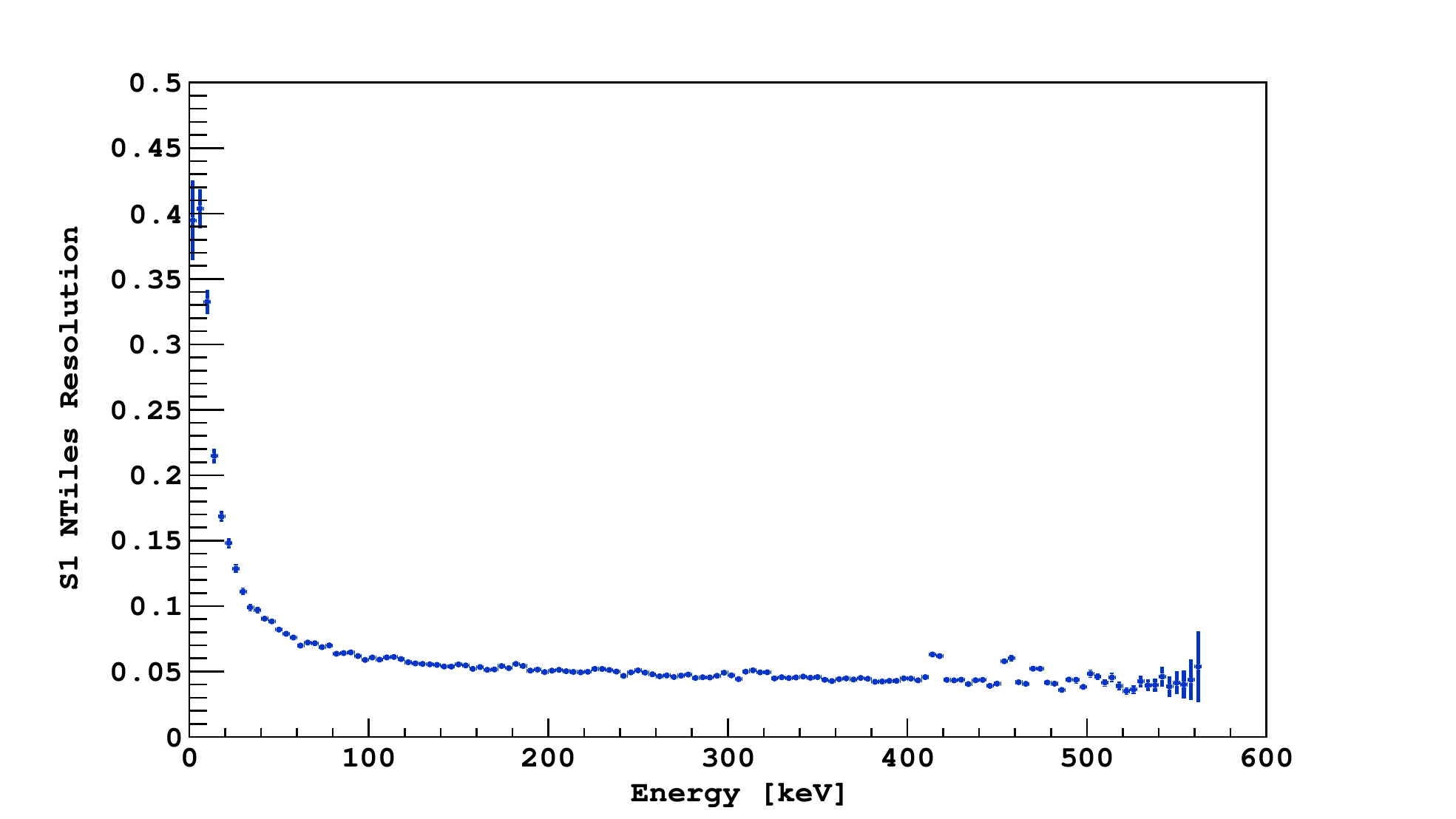}
\includegraphics[width=0.45\columnwidth]{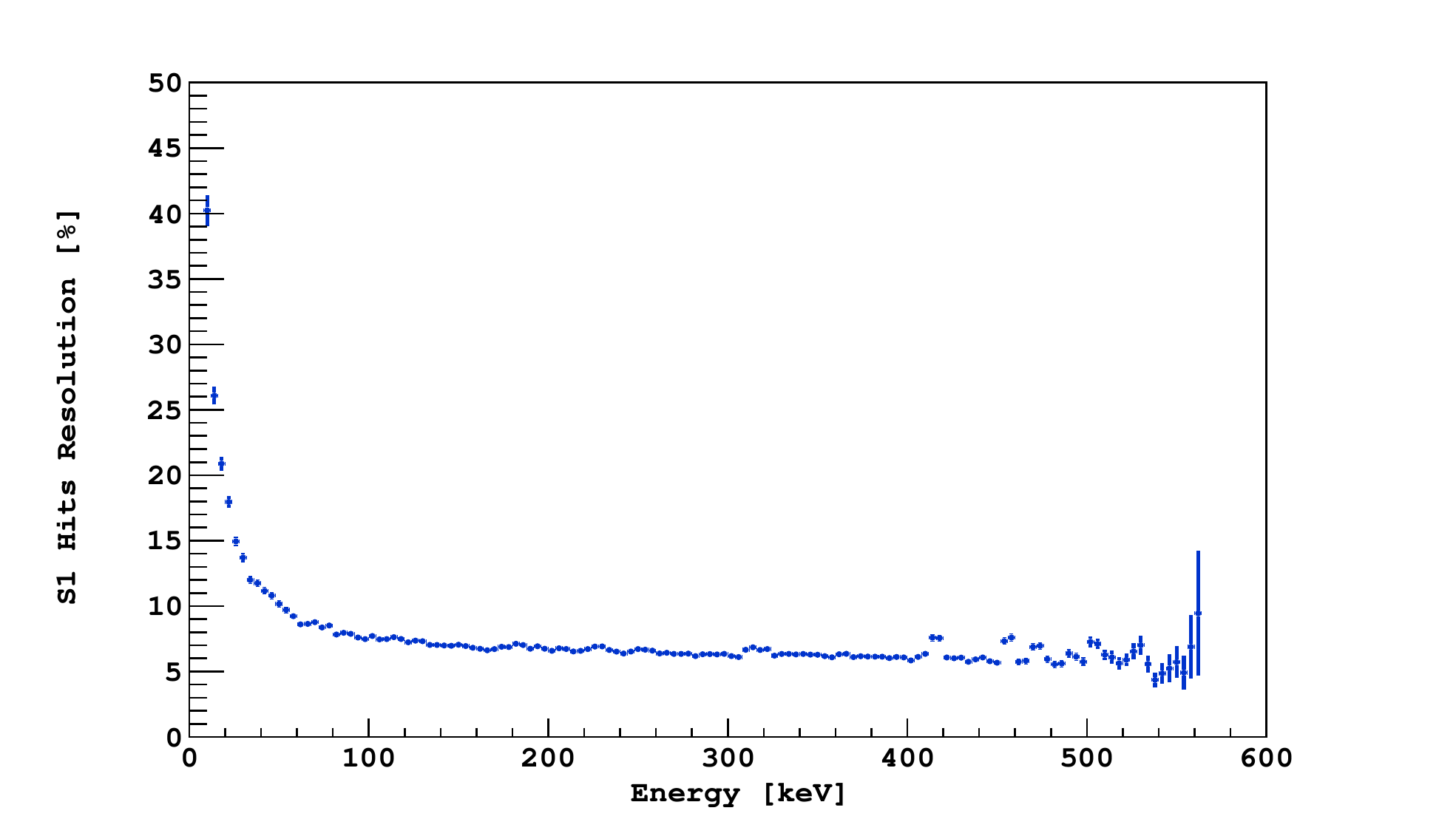}
\caption[Linearity and resolution of \SOne\ signals.]{Linearity and resolution of \SOne\ signals.  {\bf (Top Left):} Number of \DSkPdms\ above threshold as a function of event energy; {\bf (Top Right):} Number of hits  as a function of event energy;
{\bf (Bottom Left):} standard deviation divided by mean for \SOne\ obtained using \DSkPdms\ above threshold; {\bf (Bottom Right):} same but using the number of hits.}
\label{fig:DAQ-SOneLinearityResolution}
\end{figure*}

In the energy region relevant for dark matter searches, the great majority of detector modules report at most 1 PE, hence energy can be reconstructed by just counting the number of detector modules above threshold in an appropriate time window, here set to \SI{6}{\micro\second} after a trigger given by a coincidence of at least \DSkTriggerThresholdHits\ hits in \DSkTriggerThresholdWindow.  As shown in Fig~\ref{fig:DAQ-SOneLinearityResolution} the resulting distribution is quite narrow, thus resulting in a good resolution up to energies of few hundred of keV.  Taking into account also the hit multiplicity per \DSkPdm\ results in a good linearity for \SOne\ signal but in a slightly worse resolution.  This effect is due to the \SiPMs\ noise injected in the simulation, whose intrinsic nature spoils the hits counting keeping relatively unaffected the number of tiles above threshold.  The resolution of the S1 estimators is between $5\%$ and $10\%$, with higher values in the very low energy region as shown in Fig.~\ref{fig:DAQ-SOneLinearityResolution}.

\begin{figure*}[!ht]
\centering
\includegraphics[width=0.45\columnwidth]{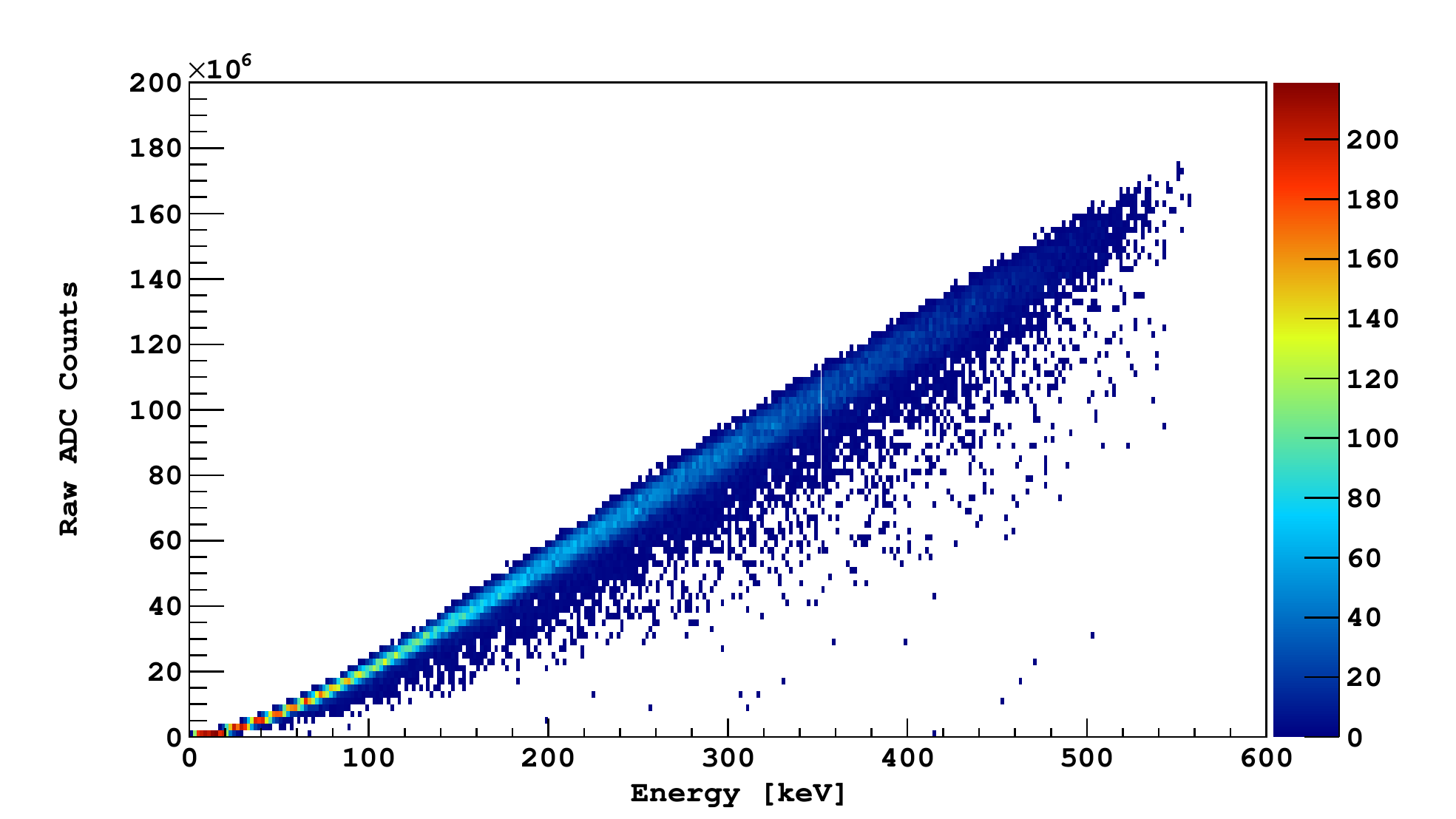}
\includegraphics[width=0.45\columnwidth]{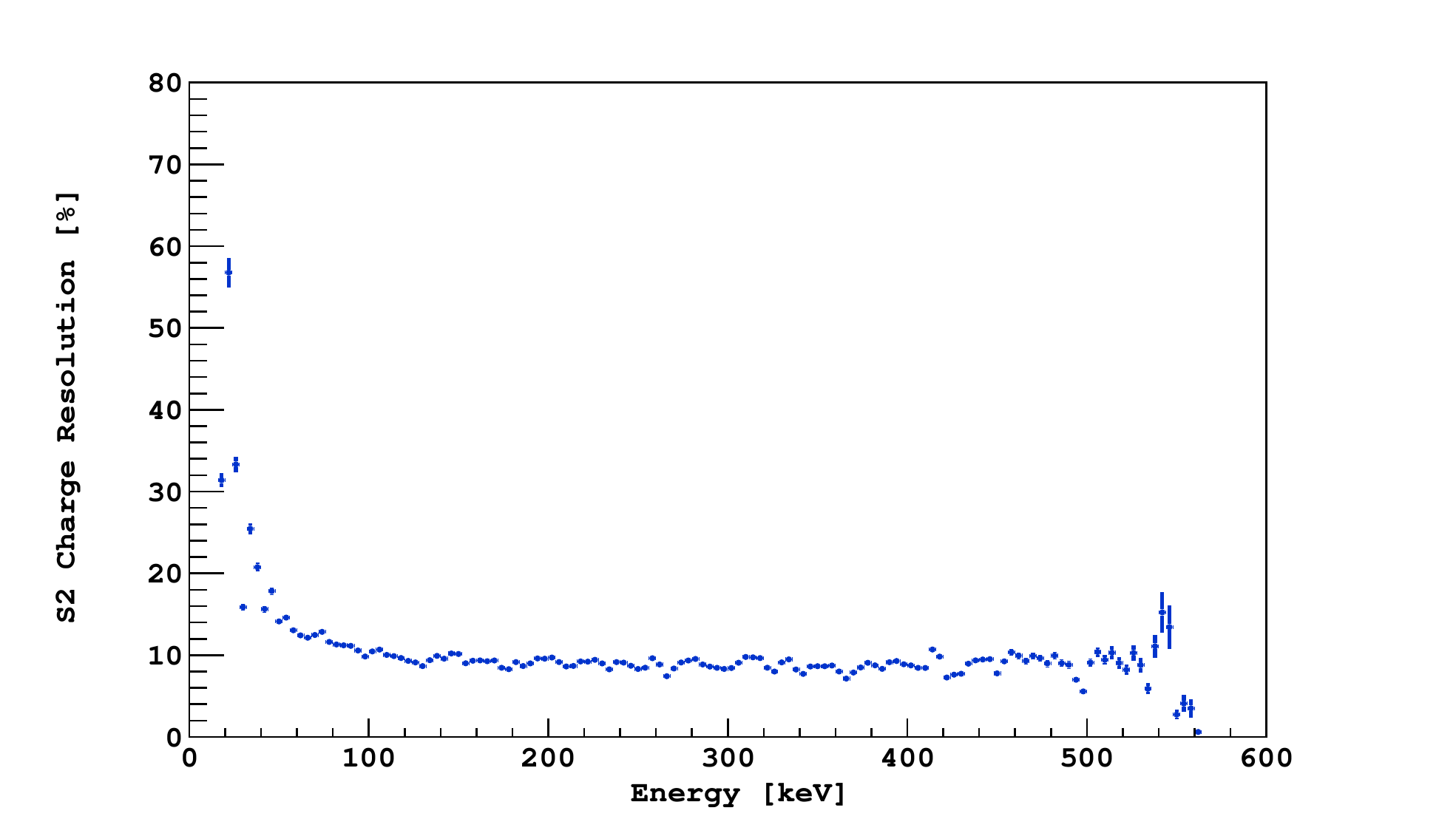}
\caption[Linearity and resolution of \STwo\ signals.]{Linearity {\bf (Left)} and resolution {\bf (Right)} as a function of event energy for \STwo\ signals.}
\label{fig:DAQ-STwoLinearityResolution}
\end{figure*}

For what concerns \STwo\ the signal sampling in a fixed time window may not integrate the whole charge of the pulses, whose duration can well exceed the microsecond.  Moreover the baseline noise further degrades the resolution of the estimation.  The measured \STwo\ pulse integral vs true event energy is shown in Fig.~\ref{fig:DAQ-STwoLinearityResolution}, left panel.  
The simulation of the \DAQ\ described above result in the resolution in \STwo\ of order $10\%$ in the relevant energy range as shown in the right panel of Fig.~\ref{fig:DAQ-STwoLinearityResolution}.

In conclusion, this preliminary study demonstrates that the readout and \DAQ\ envisaged allows the reconstruction of energy for both \SOne\ and \STwo\ signals, while the foreseen timing resolution on photon detection of O(\SI{10}{\nano\second}) would preserve the information needed for PSD.  A very preliminary algorithm for reconstruction of the XY position of hits using the digitized information has also shown to provide the required sub-centimeter resolution needed for efficient fiducialization in \DSk.

\subsection{Event Builder and Software Trigger}
\label{sec:DAQ-HLST}

Online event selection and event building in the trigger is planned to be performed in two steps, both of them implemented with two software-based triggers, Level-0 (L0T) and Full-Event (FET) triggers.  L0T and FET together form the High Level Software Trigger (HLST) of the experiment.  The event selection in the HLST triggers proceeds in steps for feature extraction ({\it e.g.}, fiducialization, energy reconstruction, \LArTPC\ and veto coincidence etc.) and trigger decisions.  At the end of each step the results are checked against the requirements defined in trigger tables.  While the L0T trigger algorithm is designed to implement very simple hit counting criteria within predefined time-windows, similar to the ones used in the hardware trigger of the \DSf\ experiment but exploiting the advantage of the higher flexibility allowed by a software based trigger, the FET trigger attempts a full offline-like event reconstruction, using a version of the offline reconstruction software customized for the online environment and refined calibrations.  The HLST is designed to provide:

\begin{compactitem}
\item Full event-building and global event selection: TPC+veto;
\item Online full event reconstruction with offline-quality;
\item Prompt pre-processing of raw data before transfer to main offline data centers;
\item Tools and computing power for high level monitoring of detector reconstruction performances in real time.
\end{compactitem}

Being fully based on software, the system will guarantee high level of flexibility, allowing to expand processing capabilities by just adding cpus/storage/network links blocks in order to scale up the system as needed, and to implement different data-reduction and selection strategies and to modify them when needed.

\begin{figure*}[t!]
\centering
\includegraphics[width=\textwidth]{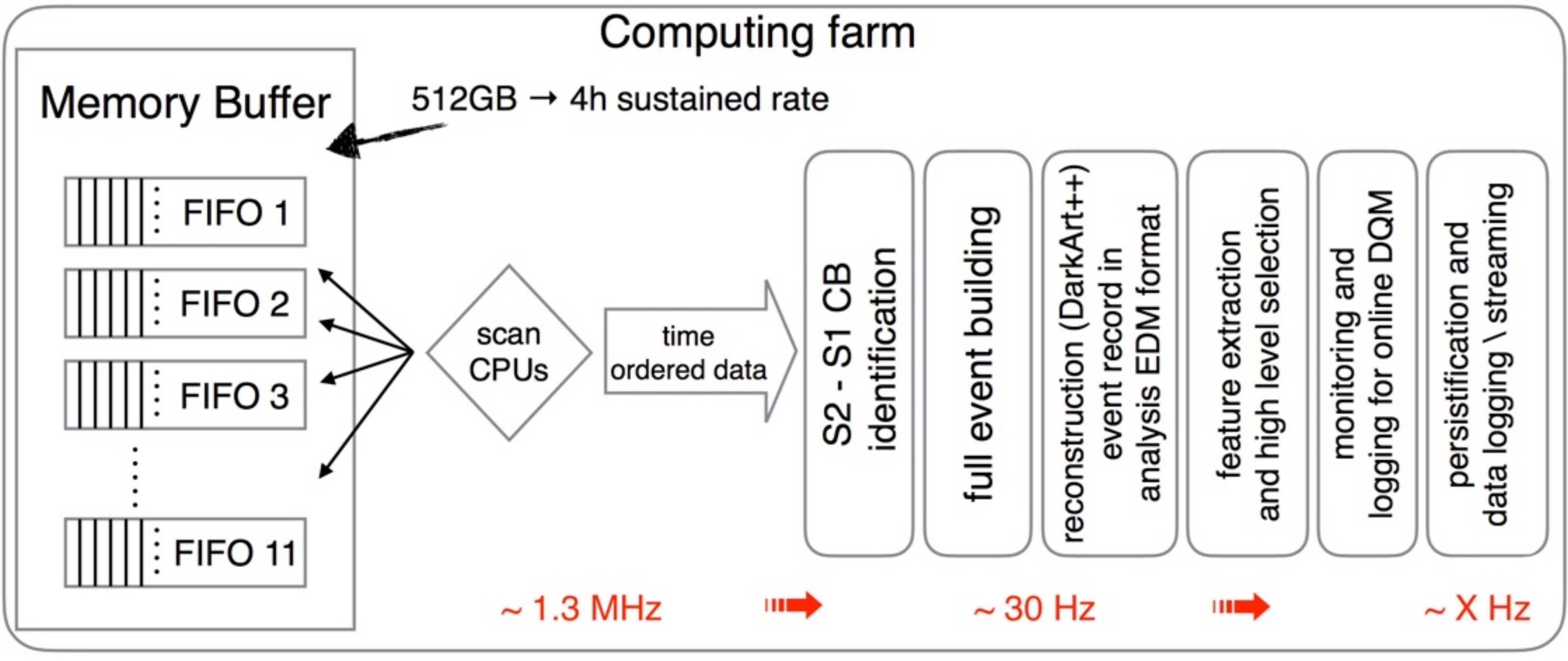}
\caption[Schematic view of the \DSk\ HLST trigger system.]{Schematic view of the \DSk\ HLST trigger system.  Only the \DSkDAQCratesNumber\ crates for \LArTPC\ are considered in this graph.}
\label{fig:DAQ-HLST}
\end{figure*}

A schematic view of the HLST trigger system is shown in Fig.~\ref{fig:DAQ-HLST}.  The L0T trigger includes a time ordering step, in which time-ordered parallel to serial conversion of data from the front-end readout system is performed.  In addition the \STwo/\SOne\ coincidence signal identification would be possible as well as the full event building step of all the veto detectors.  The FET trigger includes all the remaining logical steps (reconstruction of the events and production of the event record in analysis/offline EDM format, feature extraction and high level selection according to the trigger tables, algorithms and logging for online data quality monitoring (DQM), and finally persistification and data logging/streaming to online storage.  Raw data from the front-end readout systems, are stored in \DSkHLSTBufferNumber\ memory buffers, to match the number of \LArTPC\ and veto front-end crates, deep enough to store hours of data taking.  With the rate expected from the front-end readout design for \DSkHLSTBufferSize\ GB of memory for all buffers is enough to keep up to 4h of data at sustained rate.  L0T step is expected to reduce the trigger rate from the \DSkSingleTotRate\ input from the front-end readout below 100 Hz, while the FET step will reduce the L0T rate further and will produce several parallel output data streams (physics, monitoring, calibration, etc...)

\subsubsection{HLST Hardware and Software Options}
\label{DAQ:HLST-Options}

For the hardware architecture two equivalent options are envisioned and under study: a system based on conventional commodity CPUs computing farm, or a mixed system exploit parallel processing power of modern high-end multicore GPU based systems.  In case the CPU farm solution is chosen it is foreseen to install the HLST event selection farm in one of two racks hosting the L0T and FET processors, the local storage and the network switch apparatus.  A dedicated machine with \num{4} Xeon-E5-2650V3 class processors with \num{10} physical cores each, \SI{512}{\giga\byte} RAM used as I/O buffer and \num{2} \SI{10}{\giga\bit\per\second} ETH fiber interfaces is dedicated to run the L0T and Event building, while \num{10} 1U rack mountable machines with \num{2} Xeon-E5-2650V3 class processors, \SI{128}{\giga\byte} RAM and \num{2} \SI{10}{\giga\bit\per\second} ETH fiber interface and \SI[product-units=power]{2 x 1}{\tera\byte} NLSAS HD storage system will be able to sustain the reconstruction of events at offline level at rates up to \DSkHLSTRateMax.  In case the GPU solution is chosen, the entire CPU part of the system can be replaced by a NVIDIA high-end GPU rack-mountable system, like for example the NVIDIA DGX-1 system, with \num{2} Intel Xeon CPUs with 512 GB RAM and \num{8} Tesla P100 class boards (30720 physical cores) with \SI{16}{\giga\byte} HBM2 memory each, connected with NVLink GPU to GPU bidirectional link at \SI{10}{\giga\byte\per\second}, \SI{7}{\tera\byte} SSD storage ad dual \SI{10}{\giga\bit\per\second} Ethernet interface.  Both systems will be able to fulfill the requirements needed for the HLST trigger system. The GPU solution allows in addition to implement highly parallel techniques for \STwo\ signal reconstruction from the stored waveforms, for 3D reconstruction of the position of the interaction via multivariate pattern-recognition techniques, and, if needed, for offline optimal filtering of the SPE  waveform to extract a precise timing information for \SOne\ and \PSD.

L0T and FET reconstruction and event selection algorithms run inside the processing units (CPU or GPU physical cores).  Each processing unit can process events in parallel in concurrent worker-threads.  Multi-threading minimizes overheads from context-switching and avoids stalling the CPU when waiting for requested data from the L0T/Event Builder system, or when publishing monitoring information.  While this ansatz allows for an efficient use of multi-CPU and multi-core processor resources, it requires that all software running in the HLST is thread-safe.  This is handled by the data flow software itself, including creation and deletion of threads and synchronization of resources. As software framework for the data flow software two possible solution are envisaged: one based on the \artdaq\ online software framework form Fermilab, already used in the \DSf\ experiment, or, a system based on the FELIX~\cite{Anderson:2015er} framework and the ATLAS experiment software trigger steering software, exploiting the developments for the integration of GPU and parallel computing systems in software trigger systems ongoing for the next phase of operations of the LHC experiments.

\subsection{Trigger schemes for global and local running}
\label{DAQ:Trigger-Schemes}

In normal running condition a trigger would be issued by a signal above threshold in the \TPC. The \DAQ\ would then record data corresponding to the reconstructed event in the \TPC\ along with \PMTs\ signal following (and proceeding) the \TPC\ trigger, in order to be able to perform the prompt and delayed neutron tagging offline.  The veto should also be able to trigger on its own activity for calibration and monitor purposes. Muons crossing the \WCV\ should also be used as trigger input for the \TPC\ and \LSV\ during regular data taking, to check the cosmogenic backgrounds.

A data acquisition window which at least includes the veto delayed window must be opened in the veto at each \TPC\ trigger. The data acquisition window should also record data for few tens of microseconds before the prompt window, and for fifty microsecond after the delayed window, in order to record and monitor the accidental backgrounds. Random triggers could also be generated with the same goal. The flexibility of the HLST and the common \DAQ\ design for all the subdetectors, should also allow easily to generate  triggers from signal in the veto without asking explicitly for a   signal in the \TPC\ in order to understand inefficiencies in an unbiased sample.

Stand-alone runs for the various sub-detectors would be possible by partitioning the HLST and configuring its processes appropriately for test or calibration purposes.

\subsection{Auxiliary Systems and Infrastructure}
\label{DAQ:Infrastructure}

An electronics room located in the vicinity of the water tank will host most of the \DAQ\ electronics for all the \DSk\ subsytems, with a possible exception for \LArTPC\ signal receivers and digitizer for which it is crucial to stay as close as possible to the \TPC\ external flanges in order to minimize cable length and hence noise pickup. 

While the exact number of crates and racks will be fixed at a later stage depending on exact technological choice, it is anticipated that the room will need \DSkDAQCratesNumber\ racks to host \LArTPC\, veto electronics, a clock distribution system and the HLST.  Power supply and HV system for the \LArTPC\ Detector Modules and \SiPMs\ will be also located in the same infrastructure. Given the power distribution system scheme described in~\ref{sec:PhotoElectronics-PowerDistribution} it is envisaged that around \DSkSMsNumber\ low and high voltage power supply modules would be hosted in addition in the electronics room.  Commercial, off the shelf, solutions will be adequate for such systems and possible candidates are the CAEN A1542(H) and CAEN A1518B models, respectively. 

An efficient cooling system, analogous in size to the current Borexino one, will be installed in order to control the temperature inside the electronics room given the amount of power dissipation expected from the \DAQ\ and power supply modules.

\subsection{Slow-Control System}
\label{sec:DAQ-SlowControl}

\begin{figure}[t!]
\centering
\includegraphics[width=0.5\columnwidth]{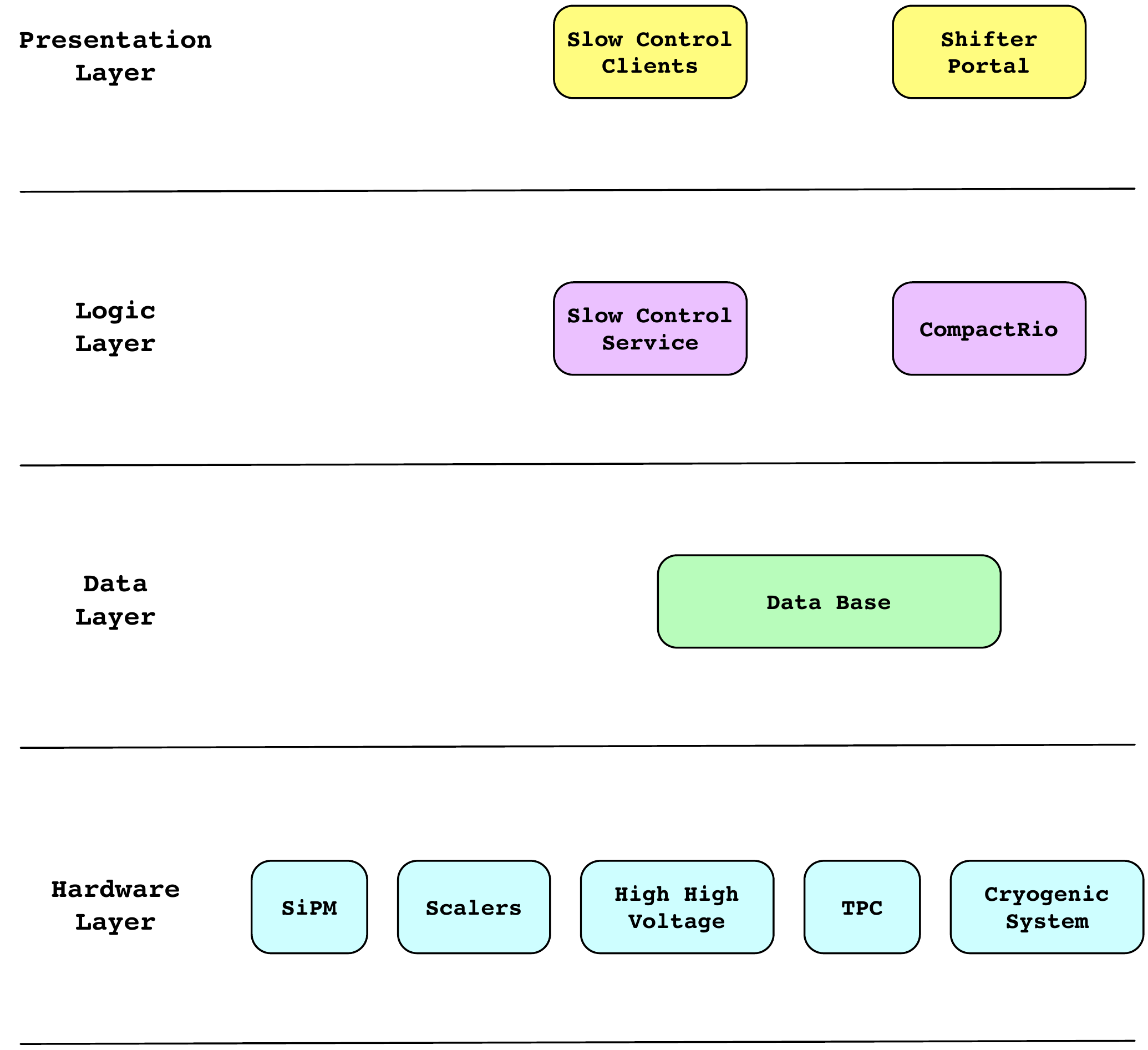}
\caption{\DSk\ software multilayered architecture.}
\label{fig:DAQ-SlowControl}
\end{figure}

The slow control system (SCS) is a \DAQ-independent system that will provide control and a real time monitoring of different hardware systems including: \SiPM, TPC electric field HHV, scalers, \LArTPC\, cryogenic system and auxiliary facilities.  The SCS is being designed in order to make all the subsystems and services interoperable to achieve a reliable and robust system.  The concepts of scalability and usability through a multilayered architecture and abstraction concepts are taken into account throughout the development of the slow control to allow for easy further customization, configuration and reuse.  The whole software will be developed using LabView (Laboratory Virtual Instrument Engineering Workbench), a system-design platform and development environment for a visual programming from National Instruments.  The SCS architecture as illustrated in Fig.~\ref{fig:DAQ-SlowControl} is based on a distributed layered architecture that allows integrating many services that communicate with different devices and different hardware through various protocols.

The architecture consists of four separate layers:

\begin{compactenum}
\item Presentation layer;
\item Logic layer (services and CompactRIO);
\item Data layer;
\item Hardware and devices layer.
\end{compactenum}

The presentation layer is a set of graphical user interfaces (GUIs) which allow controlling and monitoring all the subsystems, it is the layer which users can access directly.  The logic layer is a set of components, the majority of them are deployed as services and they communicate either to servers application or directly to devices.  Any service may run on any node of the \DSk\ network, a reliable private network where all the SCS services will be deployed.  The SCS monitors all the critical parameters and sends alarm messages to shifters and experts according to a set of thresholds. Notifications are sent via short message service (SMS) and via e-mail.  In this layer there is also a dedicated service to control the cryogenic system and it is deployed inside the CompactRIO.  The CompactRIO is a combination of a real-time controller, reconfigurable IO Modules (RIO), FPGA module and an Ethernet expansion chassis.  The real-time controller is a powerful processor with a wide range of clock frequencies for implementing the control algorithm.  The Data Layer implements the SCS data persistency.  The DataBase used for storage and persistence is PostgreSQL, an open-source object-relational database management system with an emphasis on extensibility and standards compliance.

The Veto system in \DSf\ was equipped  with a slow control implemented in \LabVIEW, the same system could be deployed for \DSk.  The Veto slow control will set parameters like the high voltage of the \PMTs\ and the thresholds of the discriminators of the front end electronics, and it will read the \PMT\ currents and dark noise rates, among other parameters.  The set and read parameters will be stored into a database and alarm conditions will be set in case of failure or values outside specified ranges.  A remotely accessible display of the relevant parameters will be created.  The gains of the \PMTs\ will be monitored by building the charge spectra of single photoelectrons using a pulsed laser and optical fibers.  A software code will adjust the high voltage of each \PMT\ in order to equalize their gain.

\section{Computing}
\label{sec:Computing}

\subsection{Introduction}
The data storage and offline processing system must support transfer, storage, and analysis of the data recorded by the \DAQ\ system, for the entire lifespan of the experiment.  It must also provide for production and distribution of simulated data, and access to conditions and calibration information and other non-event data and the physics analysis activities of the collaboration. Additional, necessary components of the data storage and offline system are provision of the software framework and services, the data management system, user-support services, and the world-wide data access and analysis job-submission system.  The design of the system is built upon the knowledge acquired in the construction and operations of the \DSf\ detector.   

\subsubsection{Requirements}
The large number of channels in the \LArTPC\ makes it impractical to digitize and save the full waveform of each channel, as done in \DSf.  Nevertheless, the charge and hit time information that will be saved will preserve all the necessary details about the amplitude and time evolution of the signals generated in the target. With appropriate filtering and compression, in addition to the expected background reduction described elsewhere in this document, the amount of data selected for recording in \DSk\ is expected to be only a few times that of \DSf.  The \DSf\ experiment typically collects about \DSfDataTotalRate, of which \DSfDataLaserRate\ come from laser calibrations, with the remaining \DSfDataDMSRate\ from the dark matter search.  The size of a \DSf\ laser calibration event is about \DSfDataLaserSize, while the size of a dark matter search data event is about \DSfDataDMSSize.  This is expected to decrease by a factor of \DSkDSfDataReductionFactor\ in \DSk, despite the much larger number of channels.  Taking into account the event size and data rate, the improved background rejection and data filtering, and using the experience from \DSf\, the short-term storage required at the experimental site is expected to be less than \DSkDataStorageLNGSShortDisk.  Long-term data storage will be at the \INFN\ Tier2 computing center in Rome and at Fermilab for offline processing and analysis.  The total storage inventory required for the experiment  at both locations is expected to be on the order of \DSkDataStorageTotalDisk.  This includes the storage needed for simulated and reconstructed events, which take much less space than raw data.

\subsection{Computing Model}

\subsubsection{Computing systems and data workflow}
The primary event processing occurs at the experimental site in the software trigger farm described elsewhere in this document. Pre-processed data is archived on the temporary storage at the experimental site and copied to central computing centers (Tier-1/Tier-2) developed and used for LHC computing activities located in Italy and USA. These facilities archive the pre-processed data, provide the reprocessing capacity, provide access to the various processed versions, and allow analysis of the processed data. Derived datasets produced in the physics analyses are also copied to the Tier-1/Tier-2 facilities for further analysis and long-term storage. The Tier-1/Tier-2 facilities also provide the simulation capacity for the \DSk\ experiment. 

Bulk data processing is expected to be performed using low cost commodity cluster computing based on commercial \CPUs\ available both at the Italian Tier2 facility and at Fermilab.  Final data analysis will be performed either directly at the Italian Tier-1/Tier-2 centers and Fermilab or on commercial \CPUs\ hosted at institutes participating in the collaboration.  PNNL also has resources available and is already established as a Tier-1 data handling site for the Belle II experiment.  Contributions from PNNL could include distributed data management, data workflow and data processing and storage, and expertise of PNNL collaborators with substantial experience in grid (distributed) computing and advanced architectures.  There is also the option of using the considerable free resources available on the Open Source Grid.  Many experiments are currently using the Open Source Grid to run their reconstruction and analysis jobs, and have access to more than one million CPU hours per week.  

The amount of short term storage currently available at \LNGS\ for \DSf\ consist of \DSfDataStorageLNGSShortDisk\ of front-end storage used as temporary buffer and located in the underground laboratory, plus \DSfDataStorageLNGSLongDisk\ of disk space in the above ground computing center for short- and long-term storage of \DSf\ data.  From there, raw data are copied to \CNAF\ and Fermilab for reprocessing and analysis.  \CNAF\ is making available \DSfDataStorageCNAFDisk\ of disk storage and \DSfDataStorageCNAFTape\ of tape storage.  At Fermilab, there is \DSfDataStorageFNALDisk\ of fault-tolerant disk storage and about \DSfDataStorageFNALTape\ on the dCache-based tape system for long-term storage.  It is expected that much of this inventory will be recycled for the \DSk\ experiment, aside from what is necessary for ongoing storage of \DSf\ raw data.  Any necessary additional tape storage will be purchased and installed. The total amount of storage for the five years of data-taking, including calibration and simulated data, is \SI{2} PB of disk storage and \SI{1} PB of tape storage. The processing power currently used for reprocessing and analysis of \DSf\ data includes a farm of \DSfCPUsLNGS\ at \LNGS\ for production and validation, plus \DSfCPUsCNAF\ at \CNAF\ and \DSfCPUsFNAL\ (soon to increase to \DSfCPUsFNALUpgrade) on the Fermilab grid system.  The latter is extensible up to several thousand slots when the Fermigrid cluster load is light. As for the mass storage, a large fraction of this is expected to continue to be available for \DSk.  Finally, the previously-mentioned option of using the Open Source Grid would dramatically increase the amount of available processing power. 

Currently, a single \DSf\ event takes about a half of a second to reconstruct on a typical \SI{2.8}{\giga\hertz} processor, meaning that \DSkOffLineCoresRealtimeReconNumber\ dedicated cores can maintain reconstruction in realtime for an event rate up to \DSkComputingRealtimeMaxEventRate.  Assuming a factor two increase in the cpu time needed to reconstruct an event,  a factor ten to simulate a full event, a real data plus calibration event rate of  \DSkRealtimeCalibrationEventRate, and a sample of simulated events of the same dimension of the real-data one, \DSk\ needs \DSkOffLineCoresRealtimeReconNumber\ dedicated cores to maintain reconstruction in realtime of the collected real data + calibration events, and \DSkOffLineCoresSimulationNumber\ dedicated cores to produce the simulated samples. Moreover \DSkOffLineCoresReprocessOneYearNumber\ physical cores are sufficient to reprocess in three months all physics events collected in one year. This amount of processing power can be achieved using either the Italian Tier-1/Tier-2 centers, or the Fermilab resources alone and exploiting the Open Source Grid/Cloud available resources. 

\subsection{Software Environment}
\DSk\ will adopt an object-oriented approach to software, based primarily on the C++ programming language,  with some components implemented using other high level languages (Python etc.). A software framework has been built up during the \DSf\ experiment which provides flexibility in meeting the basic processing needs of the experiment, as well as in  responding to changing requirements. In order to support code reuse, common user access to low-level algorithms used for I/O, and data persistency, 
 and to build a system optimized for both the offline and software trigger environments, the C++ code will make heavy use of object oriented abstract interfaces techniques.

\subsubsection{Simulation}
\GFDS\ is a \Geant-based simulation toolkit specifically developed for \DS . The modular architecture of the code was developed in order to describe the energy and time responses of all the detectors belonging to the \DS\ program, namely \DSt\, \DSf\, and \DSk . For each of them, \GFDS\ provides a rich set of particle generators, detailed geometries, opportunely tuned physical processes, and the full optical propagation of the photons produced by scintillation in liquid argon and by electroluminescence in gaseous argon.
 
The main goals of \GFDS\ are: the accurate description of the light response, to calibrate the energy responses in \SOne\ and \STwo\ and the time response expressed by the \FNine\ variable; the tuning of the analysis cuts and their efficiency estimation; the prediction of the electron and nuclear recoil backgrounds; and the definition of the signal acceptance band.  

\GFDS\ tracks photons up to their conversion in photoelectrons, which occurs when a photon reaches the active region of a photosensor and survive to the quantum efficiency. The conversion of the photoelectron into a charge signal is handled by the electronic simulation, an independent, custom-made C++ code. The electronic simulation embeds all the effects induced by the photosensors (e.g. after-pulse and cross talk) and by the electronics itself (e.g. saturation). The electronic simulation also has the option to overlap simulated events and real baselines in order to provide more realistic simulations.  As output, it produces waveforms for each channel with the same data format of the real data in order to be processed by the reconstruction code, the same used by the real data. 
 
\GFDS\ can also track events generated by \FLUKA\ and \TALYS\ simulation codes, opportunely developed for the \DS\ program.  The \FLUKA\ simulation is mostly used to study cosmogenic isotope productions, while \TALYS\ for the (alpha,n) reactions, and hence the prediction of the nuclear recoil background. 

\subsubsection{Reconstruction}
The role of reconstruction is to derive from the stored raw data a reduced set of physics parameters and auxiliary information necessary for physics analysis. The reconstruction combines information from the \TPC\ and the veto detectors. A typical reconstruction algorithm takes one or more collections of information from the event data model (EDM) raw data stream as input, calls a set of modular tools, and outputs one or more collections of reconstructed objects. Common tools are shared between reconstruction algorithms, exploiting abstract interfaces to reduce dependencies.

\subsubsection{Calibration}
Analysis of calibration data will be performed within the reconstruction and simulation software environment. Details on requirements and expected performances are 
described in Sec.~\ref{sec:Calibrations}.

\subsubsection{Databases}
\DSk\ will produce roughly \DSkAnnualDataProduction\ of data annually, combining the data processing, simulation, calibration and distributed analysis activities.  A data storage and management infrastructure is necessary to allow efficient storage and access to all this data.  Two types of data storage are foreseen: file-based data and relational-database-resident data.  File storage is used for bulky items such as event data (physics data, calibration data and simulation data). Database storage is used for other types of information, including: technical data like detector production, installation, survey and geometry data; online/TDAQ databases; conditions databases (online and offline); offline processing configuration and bookkeeping information; and to support distributed data and database management services. File-based storage of C++ objects will be implemented through the use of ROOT I/O, which provides high performance and highly scalable object serialization to random-access files. Database storage will be based on SQL-based relational databases (MySQL and SQLite). All these technologies are widely used and well tested in HEP experiments. 

\subsubsection{GRID tools and services}
Computing resources will be accessed by through Grid middleware components and services.  These provide services for software installation and publication, production operations and data access for analysis through a uniform security and authorization infrastructure, as well as interfaces for remote job submission and data retrieval, and job scheduling tools designed to optimize utilization of computing resources. The Grid infrastructure will be based on the infrastructure and the software tools and services developed for the LHC Computing Grid (LCG) project.

\subsubsection{Use of the offline software for online HLST and monitoring}
The High-Level Software Trigger (HLST) provides the online event selection. The  trigger is based on an online version of the \DSf\ reconstruction software, which has been optimized and tailored for the \DSk\ online environment running on farms of Linux PCs and/or GPUs farms.  Overall, the HLST has to provide the required rate reduction and pre-processing of raw data, including the production of global event records in analysis EDM format. The HLST will use the offline computing environment, allowing \DSk\ considerable commonality in the design and management of the selection software itself.  This also allows the HLST to use various offline software components, like detector description, calibrations, EDM, and reconstruction algorithms.  The same infrastructure can be used for data monitoring by simply replacing or augmenting the selection algorithms with those for monitoring.

\subsubsection{Code management and distributions, quality assurance and documentation}
The computing group will provide and maintain the software development environment, including: code management tools (Git, Github, etc.); managing the use of external software (for example CERN Root, \Geant, theory interpretation codes, etc.); providing scripting for building software releases; producing code distribution kits; and providing documentation such as web, wiki pages, bug reporting etc. The computing group plans to make use of the typical quality assurance and quality control tools used in HEP large experiments (i.e. LHC experiments). Code documentation will be provided, using standard web pages as central starting points for finding documentation. Doxygen and Twiki web pages will be used to for code documentation and to document specific sub-projects and provide instructions and tutorials.

\subsection{Computing Operations}
\DSk\ software builds will be organized into several categories: nightly builds, developer releases, production releases, and bug-fixes to releases. The production releases hold the stable code, used for production and end-user analysis. Production release builds will be managed and supervised by a Release Coordinator and a team of librarians. The responsibility of the Release Coordinator will be to ensure that all software components work together in a coherent way, and to patch the releases and the software distribution with bug-fixes when needed. Production releases will be installed on the Grid using the methods developed for the LCG. 

\subsubsection{Commissioning of the system}
Data processing in the initial phase of data taking is likely differ from the steady state case.  Initial development of primary calibrations and reconstruction algorithms will lead to an intensive need for raw data access.   While the distribution and access to the data should be well prepared and debugged by various data challenge tests based on \DSf\ data and on the \DSp\ data,  it is envisaged that in the early stages much of the data will be taken in raw-data format (i.e. without HLST pre-processing). The storage and processing power has also been dimensioned and staged to accommodate this phase with, for example, repeated re-processing of the initial data.

\subsection{Development Testing}
Risks associated with the deployment of the computing system are related to possible delays in the various computing sub-projects which might prevent \DSk\ from 
preparing a complete and coherent software suite and computing environment in time for the beginning of data taking. To mitigate these risks, the plan is to organize a series of ``data challenges'' to test the development of the software framework and the computing infrastructure at progressively larger and larger scales. These data challenges will start very early in the development of the computing system, testing the following, in order of increasing complexity: the ability to simulate a sizable number of events with the detector geometry; the ability to reconstruct and to produce physics-analysis level objects; the ability to perform simulation and reconstruction using the Grid infrastructure; and finally a general test of all the components of the computing model including all the software components (simulation, reconstruction, databases, distributed computing etc..).

\section{Materials and Assay}
\label{sec:MatAssay}

Trace radioactivity in detector materials is a major (and often the dominant) source of background in direct dark matter searches.  Betas and alphas from decays of these contaminants can be backgrounds from materials in contact with the active \LAr, while gammas and neutrons (from spontaneous fission and $(\alpha,n)$ interactions) can produce background from more distant sources.

A strategy for selecting and employing materials that maximizes the physics reach of the detector by controlling backgrounds from such radioactivity has been developed.  This will be accomplished by developing a ``background budget'' to identify materials purity and assay requirements, developing cleaning and handling processes to meet radiopurity requirements, and performing assays to identify clean materials and to validate processes and sources during construction and commissioning of detector components.

\subsection{Background Budget \& Assay Priorities}
\label{sec:MatAssay-RadiopurityBudget}

\begin{table*}
\rowcolors{3}{gray!35}{}
\centering
\caption[Primary \DSk\ detector materials list with masses and assumed activities.]{Primary detector materials list with masses and assumed activities.  For \ce{^238U}, when two activities are given, they refer to the upper (top) and lower (bottom) parts of the chain, and secular equilibrium is not assumed.  The titanium cryostat will have less than half the mass of the stainless steel one, while the copper cryostat will have approximately double the mass.  \DSf\ measurements (``GDMS'' =  glow discharge mass spectroscopy, ``Ge'' = gamma counting, ``NAA'' = neutron activation analysis) and \DSk\ measurements (``\ICPMS'' = mass spectrometry) are given as indicated.}
\begin{tabular}{lrrrrrcc}
Material
			&\minitab{c}{Mass}{[kg]}
						&\minitab{c}{\ce{^238U}}{[mBq/kg]}
									&\minitab{c}{\ce{^232Th}}{[mBq/kg]}
													&\minitab{c}{\ce{^60Co}}{[mBq/kg]}
																&\minitab{c}{\ce{^40K}}{[mBq/kg]}
																			&\minitab{c}{\DS}{Measurement}
																								&\minitab{c}{Other}{Reference}\\
\hline
Steel Cryostat	&\num{8479}	&\minitab{r}{2.4}{\num{0.4(2)}}
									&\num{0.8(3)}		&\num{13(1)}	&\num{<0.03}	&GDMS, Ge			&\textendash\\
\ce{Ti} Cryostat
			&\num{1400}	&\minitab{r}{\num{4.9(12)}}{\num{<0.37}}		
									&\num{<0.8} 		&\textendash 	&\num{<1.6}	&\textendash			&\cite{Akerib:2015is}\\
\ce{Cu} Cryostat
			&\num{11960}	&\num{<0.06}	&\num{<0.02} 		&\textendash 	&\num{0.12}	&GDMS				&\textendash\\
\DAr\ (active)	&\num{23000}	&\textendash 	&\textendash 		&\textendash 	&\textendash 	&\cite{Agnes:2016fz}		& Sec.~\ref{sec:Argon}\\
Copper Parts	&\num{2795}	&\num{<0.06}	&\num{<0.02} 		&\textendash 	&\num{0.12}	&GDMS				&\textendash\\
\PTFE\ Parts	&\num{1445}	&\num{<0.07}	&\num{<0.004}		&\textendash 	&\num{0.10(4)}	&NAA				&\cite{Leonard:2008bk}\\
Fused Silica Parts
			&\num{189}	&\num{0.008} 	&\num{0.01} 		&\textendash 	&\textendash	&\ICPMS, NAA			&\textendash\\
\SiPMs		&\num{11}		&\num{<0.025} 	&\num{<0.003}		&\textendash 	&\textendash 	&\ICPMS				&\cite{AguilarArevalo:2015hf}\\
Sapphire Substrate
			&\num{69}		&\num{<0.30} 	&\num{0.12(3)}		&\textendash 	&\num{<0.21} 	&\textendash			&\cite{Leonard:2008bk}\\
\end{tabular}
\label{tab:Materials-Activity}
\end{table*}

The Materials and Assay working group (\MAWG) will maintain the background budget of \DSk, giving the radiopurity requirements on each material or component of critical systems, chiefly the TPC, cryostat, photodetectors, and neutron veto.  A crucial input to the background budget comes from Monte Carlo simulations that, when normalized by measured or assumed activities, give estimates of expected background from particular sources.  Another input is the expected background rejection.  Rejection of electron-recoil backgrounds relies primarily on Pulse Shape Discrimination (see Sec.~\ref{sec:Physics}), while neutrons are rejected using the \LSV\ (see Sec.~\ref{sec:Vetoes}).  Additional rejection is provided by the fiducial-volume and single-scatter cuts.  The background budget will clearly evolve as assays are performed and more components' activities are determined, the detector design is updated, the Monte Carlo detector geometry and physics models improve, and background rejection estimates improve.  Also, the background budget is iterative; lower achieved activity in one component affects the targets for other components.

Initially, the background budget is made with activities known to be achievable based on the experience in \DSf\ and published measurements from other experiments.  An example is given in Table~\ref{tab:Background}, which serves as an indication of the feasibility of the experiment with existing materials, as well as providing initial targets for both radiopurity and background rejection.  Moving forward, focus will be put on the major sources of potential background: typically those that are present in large masses ({\it e.g.}, cryostat titanium or steel, PTFE  reflectors in the TPC, copper field cage rings, the \LAr\ itself), those that are in contact with or very close to the active \LAr\ ({\it e.g.}, \PTFE, fused silica windows, \SiPM\ arrays, wavelength-shifter), and those that have high $(\alpha,n)$ neutron yields (such as \PTFE).  For the \LSV, the steel of the tank, the PMTs, and the components of the scintillator itself must be assayed, primarily to ensure that the rate in the \LSV\ will be low enough to allow the low energy thresholds needed for efficient neutron vetoing (see Sec.~\ref{sec:Vetoes-BaselineDesign}).

The background studies leading to Table~\ref{tab:Background} used the activities in Table~\ref{tab:Materials-Activity}, measured by \DS\ and other experiments, for normalization. They indicate that the biggest sources of neutrons are the cryostat (titanium or steel) and the \PTFE\ reflector panels.  The resulting neutron background rates from these components are comparable, but with {\it caveats} like the fact that the PTFE U/Th activities in Table~\ref{tab:Materials-Activity} are upper limits.  This gives a high priority to searches for suppliers of lower-activity stocks of these three materials.  Early indications are that such searches may be successful -- for example, \ICPMS\ results from the Skobeltsyn Institute for Nuclear Physics at Moscow State University (MSU) demonstrate the potential for having Ti with an order of magnitude lower \ce{^238U} and a factor of a few lower \ce{^232Th} than those assumed in the preparation of Table~\ref{tab:Materials-Activity} and  achieved by other groups (e.g., Ref.~\cite{Akerib:2015is}).  The ability to source such material in quantity will obviously inform the titanium versus stainless-steel decision to be made for the cryostat.  In some cases, such as PTFE, assays with higher sensitivity than those used up to now (in particular, for \ce{^210Pb}) may be needed to allow characterization of especially clean samples.  Other particularly high-priority materials are the sapphire  substrates for the \SiPMs\ and the PMTs for the outer detector systems.  All of these high-priority assay campaigns have begun; a few specific cases are discussed below.

The plan is to assay {\it all} materials or items selected to reside within the TPC cryostat.  This becomes more straightforward as the major materials discussed above are finalized; the criterion for other materials becomes ``negligible impact on the overall background.''  As samples are acquired, they will be queued for assay as described below.  It is also part of the \MAWG's responsibility to prioritize assays {\it before} the other working groups acquire samples; potentially problematic materials, say with high $(\alpha,n)$ yields, must have samples acquired early for screening.

\subsection{Managing Assay Capabilities}
\label{sec:MatAssay-RadiopurityManagement}

The \DSk\ collaboration possesses diverse assay capabilities with significant throughput capacity.  Critical to taking advantage of this large capacity is a thorough organization of the assay effort.  Often in low-background experiments the responsibility for obtaining radiopurity assay of a candidate material has fallen on the material user, which is a model fraught with inefficiencies.  In \DSk\ the responsibility is delegated to the \MAWG, a single working group within the collaboration, comprised of experts for each of the assay methods and representatives from each institution that hosts assay capability.  This has given the collaboration a single point of contact with the capability and capacity for \DSk's entire assay needs.

\begin{figure}[t!]
\centering
\includegraphics[width=0.75\columnwidth]{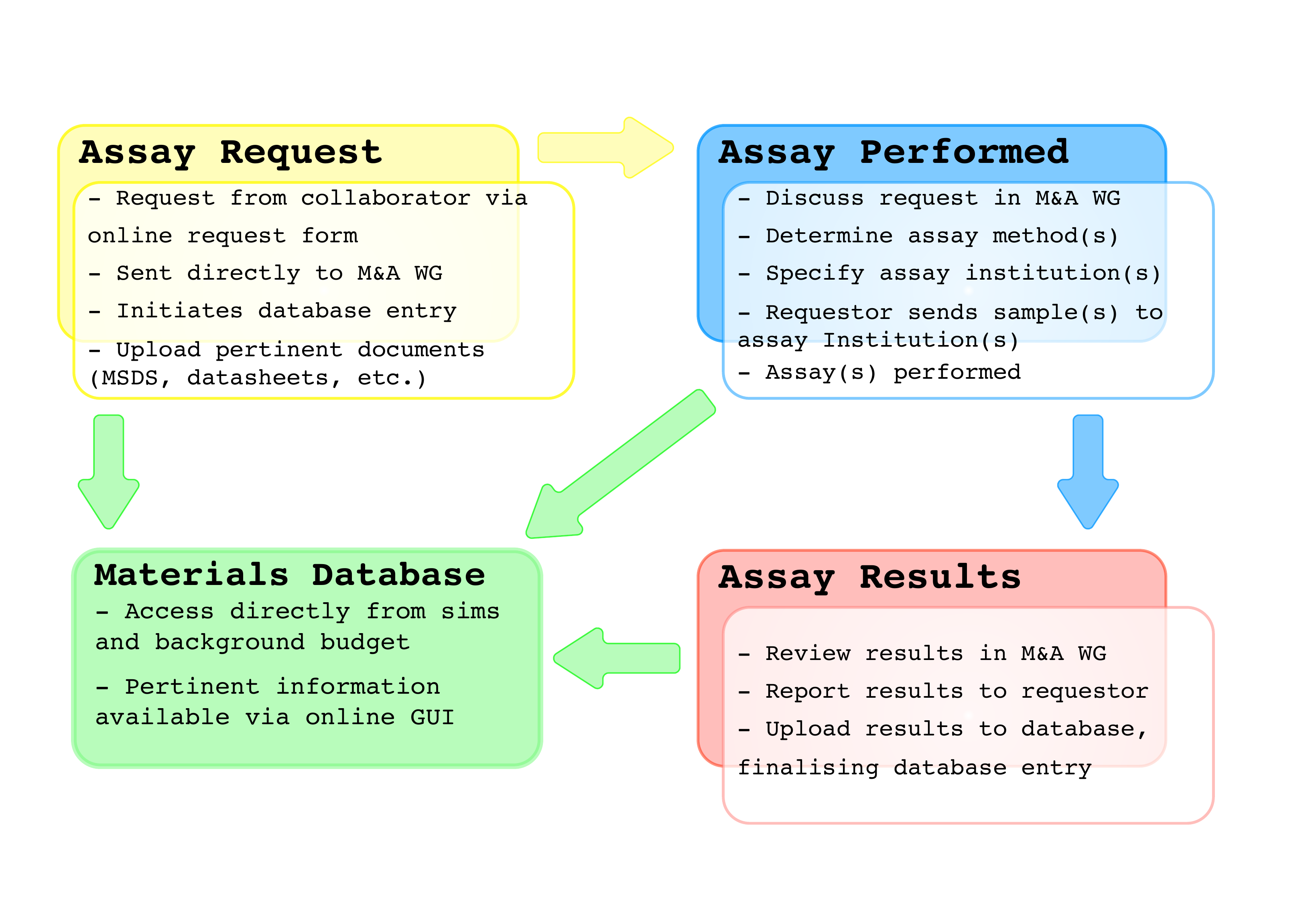}
\caption{Diagram illustrating \DSk\ workflow for radiopurity assay.}
\label{fig:MatAssay-OrganizationalFlowDiagram}
\end{figure}

The assay of materials is a well-defined  process that begins with a formal request and ends with an official assay report and recording of results in the materials database.  The process is sketched in Fig.~\ref{fig:MatAssay-OrganizationalFlowDiagram}.  The initial request is initiated with an online request form that asks for:
\begin{enumerate}
\item requester information;
\item sample details (supplier, lot number, etc.);
\item assay requests ({\it i.e.}, analytes, activity levels, number of samples, etc.).
\end{enumerate}
Submission of this form also initiates an entry in a database of materials, which is updated as the assay process progresses.  It is also possible to include other pertinent information about the material ({\it e.g.}, material properties, etc.), as well as relevant documentation ({\it e.g.}, datasheets).

The request is reviewed within the \MAWG\ to determine both the limit of activity that the material must meet and the best method for assay based on such things as material properties, type of analyte, and whether destructive assay is permitted.  The institution within the \MAWG\ that has the necessary assay capability, the requisite sensitivity, and the availability to perform the assay is then assigned the assay task.  The material ``owner'' is instructed where to send the material and the assay is performed.  The \MAWG\ then reviews the results and an official report is sent to the material owner.

Throughout the process, the database entry for the material is updated with assay information as it becomes available, including such things as assay method and institution.  When the final results are available and have been reviewed by the \MAWG, they are input into the database and the entry for that material is finalized.

An example of the success of this organizational model comes from the assay of a \KMThreePMTSize\ \PMT\ that is a candidate for use in the water tank.  This was an unusual assay because of the \PMT's large size and the need to assay it nondestructively.  The owner of the PMT contacted the \MAWG\ about assaying the \PMT\ and submitted the online request form.  After some discussions with the owner and review of the assay request, it was determined that both PNNL and Temple could reconfigure existing counting chambers to assay the device via \gr\ spectroscopy.  Within six business days of receiving the online assay request, an institution was chosen to perform the assay and the PMT owner was instructed to ship the PMT to Temple.  In less than six weeks after submitting the request, the PMT was shipped to the United States from Italy, assayed with a  \HPGe\ counter in two different configurations, and the results were reported and recorded.

The streamlined organization of the materials and assay effort ensures that materials for \DSk\ will be screened in the most efficient way possible.  Additionally, when there are materials or parts of the experiment that require unique assays, the group surveys its institutional points of contact to identify solutions quickly.  The organizational structure described here removes the burden of assaying materials from individual collaborators and other working groups and effectively places it on the \MAWG, the most capable group of collaborators for evaluating and performing radiopurity assay.

\subsection{\DSk\ Assay Resources}
\label{sec:MatAssay-AssayResources}

To ensure the radiopurity of all detector materials to the levels defined by the background model, the \MAWG\ has developed a radiopurity assay program that takes advantage of facilities throughout the collaboration.  Overall, it is anticipated that this program will span approximately three years and will involve a few thousand assays, including searches for radiopure materials, development and validation of cleaning and handling procedures, and screening of all detector components.  The collaboration has extensive and diverse assay capabilities that are more than sufficient to complete this program on schedule, with extra capacity to handle additional unforeseen assays.  In the following, estimates of the  capacity available to \DSk\ for each assay type are provided.  The assay capabilities are being organized throughout the collaboration and a detailed assay schedule will be developed as detector design matures.  The assay challenges are organized into five focus areas: mass spectrometry,  radon emanation, direct gamma assay, surface assay, and materials handling and process development.

\subsubsection{Mass Spectrometry}
\label{sec:MatAssay-AssayResources-ICPMS}

\begin{table}
\rowcolors{3}{gray!35}{}
\normalsize
\centering
\caption{ICP-MS \ce{U} and \ce{Th} detection limits attained at PNNL for \DSk\ candidate materials.}
\begin{tabular}{lcc}
Material						&\minitab{c}{\ce{^238U}}{[\SI{}{\mu\becquerel\per\kilo\gram}]}
														&\minitab{c}{\ce{^232Th}}{[\SI{}{\mu\becquerel\per\kilo\gram}]}\\
\hline
Titanium						&\num{6}				&\num{2}\\
Copper~\cite{LaFerriere:2015hy}	&\num{0.131}			&\num{0.034}\\
PTFE							&\num{1}				&\num{1}\\
Fused Silica					&\num{1.3}				&\num{0.38}\\
\end{tabular}
\label{tab:MatAssay-ICPMS}
\end{table}

For some radioactive contaminants, {\it e.g.} actinides like uranium and thorium, radiochemistry and mass spectrometry will provide superior sensitivity and will ultimately allow the assay requirements suggested by the background model to be met.  Inductively coupled plasma mass spectrometry (\ICPMS) is a highly sensitive technique that has detection limits in the parts-per-trillion (\si{ppt}) range and lower.  For reference, \SI{1}{ppt} is equivalent to \MatAssayUActivityPptEquiv\ (\MatAssayThActivityPptEquiv) for \ce{^238U} (\ce{^232Th}).  Samples are typically introduced to the plasma in the form of a nebulized aerosol of liquid droplets.  The \MatAssayPlasmaTemperature\ plasma effectively atomizes and ionizes most elements of the periodic table (including \ce{K}, \ce{Th}, and \ce{U}) to the \MatAssayICPMSChargeState\ charge state.  The ions are then funneled into a higher vacuum region and focused into the mass spectrometer using ion optics where they are separated by mass and detected.  \ICPMS\ has been around for over three decades.  Its sensitivity is particularly good for high mass elements and has improved significantly over the last decade.  High sensitivity and relatively quick sample throughput (compared to counting methods) make \ICPMS\ a mainstay in the ultra-low-background community.  To reach the low detection limits required, however, sample preparation requires development of  assiduously clean wet-chemistry methods with validated efficiency.

Within the collaboration, many \ICPMS\ instruments with different capabilities are available for coordinated assay campaigns.  Radiochemistry and mass-spectrometry capabilities across the collaboration will be coordinated and scheduled to allow sufficient capacity to address assay needs for the most routine as well as the most demanding of samples.

At PNNL, substantial expertise and resources are available for \DSk\ assay.  There are two  \ICPMS\ instruments that are used to analyze only ultra-low background materials -- this in order to keep the backgrounds at extremely low levels.  They have the capability to routinely measure sub-ppt \ce{Th} and \ce{U}, as well as sub-\si{ppb} \ce{K}, in a variety of materials, including polymers, \ce{Ti}, \ce{Cu}, and fused silica; current detection limits for candidate \DSk\ materials are shown in Table~\ref{tab:MatAssay-ICPMS}.  With new instrumentation, PNNL plans to automate much of the technical processes involved in the most sensitive materials assays, primarily to improve throughput.  Additionally, their newest instrument incorporates a one-of-a-kind triple quadrupole \ICPMS, which greatly improves analysis capabilities.  The triple quadrupole function allows the removal of problematic interferences so that increased throughput and improved detection limits can be attained.

BHSU, CIEMAT, LNGS, MSU, and PoliMi also have \ICPMS\ instrumentation that will aid in the screening of candidate materials.  MSU has an \ICPMS\ setup dedicated to \ce{U}/\ce{Th} assay in \ce{Ti} and has performed hundreds of assays with \si{ppt}-level sensitivity.  All institutions are capable of processing samples in cleanrooms using a variety of chemical methods ({\it e.g.}, separations, electrochemical purification, dry ashing, etc.).  As previously mentioned, PNNL can attain sub-\si{ppt} detection limits for most samples.  BHSU, CIEMAT, LNGS, MSU, and PoliMi can attain ppt to ppb detection limits.  Each of the laboratories can process several samples per week.  These sensitivities and throughput give the collaboration the ability to effectively and efficiently screen materials for the \DSk\ assay program.

Other \ICPMS\ capabilities throughout the collaboration include \ICPMS\ with sample introduction by laser ablation (\LA), available at BHSU and PNNL.  \LA\ is an \ICPMS\ introduction method that uses a high powered laser to ablate a sample's surface.  An aerosol of sample particulates is formed during the ablation process that is then introduced into the \ICPMS\ via a carrier gas for atomization, ionization, and mass determination.  \LAICPMS\ has several advantages.  Little or no sample preparation is required, which can substantially increase throughput while also reducing the potential for contamination associated with sample processing.  Also, \LAICPMS\ allows for investigation of the contaminant distribution along the surface of the sample.  One drawback is that quantitative analysis with \LAICPMS\ requires a reference material of the same composition as the sample and with a certified concentration of the analyte of interest in order to directly compare responses, and the number of certified reference materials is quite limited for elements like \ce{Th} and \ce{U} in polymers and metals.  Nevertheless, useful semi-quantitative assays can be performed as part of the overall \DSk\ assay program.

\subsubsection{Radon Emanation}
\label{sec:MatAssay-AssayResources-RadonEmanation}

Rn-222 is a ubiquitous noble gas that is sufficiently long-lived (\RnTwoTwoTwoHalfLife\ half-life) to be a background concern.  This is due to its high mobility ({\it i.e.}, chemically inert gas) and the wide energy range and variety of radiations emitted by its decay products.  In addition, radioactive equilibrium of the \ce{^222Rn} decay chain is often broken in materials because of the long half-life (\PbTwoOneZeroHalfLife) of the \ce{^210Pb} daughter.  This makes predictions of background caused by this isotope challenging.  In many cases, the most direct and sensitive approach for controlling and minimizing an experiment's radon-related backgrounds is via radon emanation screening of the materials of construction.

All parts of the detector in contact with the argon target may be considered as radon emanation sources.  All such argon-wetted components, including the cryostat, all TPC parts -- photodetectors, insulators, cables and electronics -- and the gas circulation loop ({\it i.e.}, tubing, getter, liquefier, charcoal trap, etc.) will be screened.  It is also planned to perform a final emanation test of the complete detector, as was done for \DSf.  Using a portable cryogenic radon detector, collaborators will investigate on-site emanation rates for all vacuum subsystems: pipes, sensors, vacuum gauges and the cryostat.  Studies of \ce{^220Rn} emanation have never been performed in this context; the plan is to systematically investigate relevant materials like stainless steel, copper and lead.  Further, because welds are known to be emanation sources, different welding procedures will be tested by measuring materials before and after welding.
   
A maximum tolerable radon concentration in the \LAr\ fiducial volume of the TPC will be established through detailed Monte Carlo simulations, leading to determination of emanation limits for specific components.  One should include here effects of reduced emanation at \LAr\ temperature and removal of \ce{^222Rn} by the cooled charcoal trap in the circulation loop.  These effects are (generally) expected to relax the emanation requirements, especially for non-metal components for which emanation is driven mostly by diffusion and is therefore strongly suppressed at lower temperatures.

The measured \DSf\ radon concentration upper limit of \DSfRnTwoTwoTwoSpecificActivityLimit\ can be used as a conservative limit for \DSk\ and corresponds to an equilibrium radon activity of \DSfRnTwoTwoTwoActivityLimit\ in the \LAr\ fiducial volume.  As discussed in Sec.~\ref{sec:BackgroundFree}, the \bg\ rejection power demonstrated with \DSf\ is more than sufficient to exclude background associated with radon at this level.  In addition to \DSf, several xenon-based experiments have demonstrated the feasibility of a dual-phase \TPC\ with such a low radon concentration ({\it e.g.}, LUX~\cite{Akerib:2015is} and EXO-200~\cite{Albert:2015IM}), and it is worth noting that the Xe detectors operate at a higher temperature and without charcoal traps in their circulation loops.  Finally, relative to \DSf\ and the Xe-based experiments, \DSk\ has a smaller surface-area to volume ratio that is expected to result in a relatively smaller radon concentration in the \LAr.  Nevertheless, radon emanation is a critical component of the \DSk\ assay program because all argon-wetted materials must still be screened to ensure that the target radon level is achieved, and because radon emanation measurements provide radiopurity information that is complementary to (and sometimes more sensitive than) the other assay methods.

Emanation assays will be conducted with existing radon emanation facilities at the Smoluchowski Institute of Physics, Jagiellonian University in Krakow and a system currently under development at PNNL.  First samples of high-purity titanium sponge have already been screened for \ce{^220Rn} and \ce{^222Rn}.  The PNNL system includes a \MatAssayPNNLChamberTwoVolume\ emanation chamber and a detection system that uses ultra-low background proportional counters~\cite{Seifert:2012ip}.  The expected detection limit is in the range of a few tens of \si{\mu\becquerel} per sample.  The Krakow system consists of three ultra-low background emanation chambers (\MatAssayKrakowRnChamberOneVolume, \MatAssayKrakowRnChamberTwoVolume\ and \MatAssayKrakowRnChamberThreeVolume) coupled to a custom radon detector in which radon isotopes are collected via cryo-adsorption onto a small cold surface and their subsequent decays are detected with an alpha counter that faces the collector.  This cryogenic radon detector is portable and can be attached to any vacuum vessel.  In order to reduce background from self-emanation some special solutions have been implemented.  The vessels were made according to ultra-high vacuum (\UHV) standards, have electropolished inner surfaces, and are equipped with only metal-sealed valves, flanges and connectors.  All chambers are equipped with \UHV\ pumping and heating systems to allow investigation of radon emanation as a function of temperature (controlled to within \MatAssayKrakowRnChamberTemperaturePrecision) for positive temperatures up to \MatAssayKrakowRnChamberTemperatureMax.  Additionally, for radon emanation studies at cryogenic temperatures, Krakow has a dedicated system in which the temperature of the sample can be controlled from \LINNormalTemperature\ to \MatAssayKrakowRnChamberTemperatureHigh.  All chambers have large openings for insertion of voluminous samples.  The Krakow system is unique world-wide with respect to the design and achieved parameters; for the first time it is possible to investigate emanation of not only \ce{^222Rn} but also the short-lived \ce{^220Rn} ($\HalfLife\ = \RnTwoTwoZeroHalfLife$) from the thorium chain.  Extremely low background rates allow for registration of single atoms emanated from samples.  As demonstrated in~\cite{Zuzel:2003im}, a combination of highly sensitive radon emanation and diffusion measurements can also be used to determine \ce{^226Ra} content in low-density materials with significantly better sensitivity than available with the best \gr\ spectrometers.

To assay one sample close to the detection limit ($\sim$\SI{1}{\mu\becquerel} per sample) about one month is needed.  The Krakow group currently operates three chambers with a possibility to extend the system with a fourth, with at least two vessels able to be exclusively dedicated to \DS\ screening.  The resulting emanation assay capacity is at least two samples per month at the ultimate sensitivity, which would be further augmented by including the PNNL system.  This will be sufficient to perform all the necessary tests for the \DSk\ experiment in a timely manner, including on-site tests of the gas system components.

\subsubsection{\HPGe\ \gr\ Counting}
\label{sec:MatAssay-AssayResources-HPGe}

\begin{table*}
\rowcolors{3}{gray!35}{}
\normalsize
\centering
\caption[\HPGe\ detectors at \DSk\ collaborating institutions or otherwise available.]{\HPGe\ detectors at \DSk\ collaborating institutions or otherwise available.  The total number of detectors is given in the second column, and columns 3--5 indicate the subset that have the indicated background-reduction feature.  The final column provides estimates (in detector-months per year) of the assay capacity available to \DSk.}
\begin{tabular}{lcccccc}
Institution	 					&\minitab{c}{Number of}{Detectors} 
						 					&\minitab{c}{Low-BG}{Cryostat}
						 								&\minitab{c}{Cosmic}{Veto}
						 											&\minitab{c}{Radon}{Purge}
						 														&\minitab{c}{Depth}{[m.w.e.]}
						 																	&\minitab{c}{DS-20k Access}{det-months/yr}\\
\hline
Canfranc~\cite{Alvarez:2013bn}	&\num{8}		&\num{8}	&\num{0}	&\num{8}	&\num{2450}	&\num{12}\\
Black Hills State University		&\num{1}		&\num{1}	&\num{0}	&\num{1}	&\num{4200}	&\num{3}\\
Jagiellonian University			&\num{1}		&\num{1}	&\num{1}	&\num{1}	&\num{0}		&\num{8.5}\\
\LNGS						&\num{8}		&\num{8}	&\num{0}	&\num{8}	&\num{3800}	&\num{6}\\
Petersburg NPI					&\num{1}		&\num{0}	&\num{1}	&\num{0}	&\num{0}		&\num{6}\\
PNNL surface					&\num{>20}	&\num{2}	&\num{0}	&\num{0}	&\num{0}		&\num{12}\\
PNNL underground				&\num{3}		&\num{3}	&\num{3}	&\num{3}	&\num{30}		&\num{3}\\
Temple University				&\num{2}		&\num{1}	&\num{1}	&\num{1}	&\num{0}		&\num{12}\\
\end{tabular}
\label{tab:Screening-HPGe}
\end{table*}

\HPGe\ spectroscopy is one of the most powerful tools for identifying \gr-emitters in the \ce{U}/\ce{Th} chains and isotopes such as \ce{^60Co}, \ce{^40K} and \ce{^210Pb}.  Sensitivity for relatively short-lived isotopes extends to concentrations far below detections limits obtainable by other methods.  The strength of this technique comes from its excellent energy resolution (few keV at MeV scales) and the high purity of germanium diode detectors, which results in very low intrinsic backgrounds.  External backgrounds can be effectively reduced by surrounding with layers of radiopure shielding ({\it e.g.}, lead and copper) and locating the spectrometers in relatively deep underground laboratories to attenuate the cosmic-ray flux.  High detection sensitivities can be achieved with large Ge crystals and large sample chambers.  Detection efficiencies are usually determined by Monte Carlo simulations supported by test measurements for selected sample geometries.  Rare-event searches such as \DSk\ require the lowest possible radioactivity for their detector components, and low-level \gr\ spectroscopy with \HPGe\ detectors has been one of the main tools for low-background materials screening over the last few decades.

\HPGe\ spectroscopy has the advantage of being a nondestructive method that does not require complex sample treatment.  Although sometimes less sensitive than \ICPMS\ or neutron activation analysis, it provides more complete information about the screened sample because the measured energy spectrum registers contributions from a variety of radioisotopes.  The most important feature, however, is the ability to directly detect isotopes relevant to understanding the \DSk\ background budget.  Specifically, measurement of \gr-emitting progeny near the top ({\it e.g.}, \ce{^228Ac} and \ce{^{234m}Pa}) and near the bottom ({\it e.g.}, \ce{^208Tl} and \ce{^214Bi}) of the \ce{^232Th} and \ce{^238U} decay chains makes it possible to investigate secular equilibrium conditions, thereby providing important complementary information to the other assay techniques described in this section.

Using MAX-R\&D funds starting in \MatAssayTempleYear, a \DS-dedicated \HPGe\ facility has been under development at Temple University.  This facility is now nearing completion.  The detector is an ORTEC ``\MatAssayTempleEff'' GMX detector of  n-type Ge, allowing for a very thin entrance dead layer to accept low-energy radiation ({\it e.g.}, the \PbXRayEnergy\ x-ray of \ce{^210Pb}).  The cryostat is ORTEC's ``J-type'' in which the cylindrical detector is mounted with its axis vertical at the end of a \MatAssayTempleAxisLength\ long horizontal cold finger.  This allows the best shielding geometry for very low-background work.   A carbon-fiber endcap is fitted, again enhancing the efficiency at low energy.  ORTEC's ultra-low background cryostat design locates the potentially radioactive preamplifier and HV filter at the origin of the cold finger, far from the detector.  The detector resolution in ORTEC's tests was \MatAssayTempleResolution\ FWHM at \MatAssayTempleEnergy.

Before assembly, every internal part of the detector assembly was sent to Temple from ORTEC under an NDA and radio-assayed at LNGS.  Several internal parts were found to be active and were replaced with known-good equivalents.  These included the temperature sensor (\MatAssayTempleTempSensorTh\ \ce{Th} and \MatAssayTempleTempSensorU\ \ce{U}, replaced with much cleaner \DSf\ type), alumina insulating ring (\MatAssayTempleAluminaU\ U, replaced with specially fabricated sapphire ring), and the stainless-steel end-cap base (\MatAssayTempleEndCap\ \ce{^60Co}, replaced with one fabricated from pre-war ship steel provided by LNGS).  In addition, a shield of ultra-low activity \ce{Pb} was added between the FET (\MatAssayTempleFET\ \ce{Th}) and the detector.  An initial three-day background spectrum taken at Temple in a provisional shield was free of visible background peaks except for \PositronAnnihilationGammaEnergy\ annihilation radiation, possibly from cosmogenic \ce{^68Ge} in the detector itself.

A sophisticated shielding assembly was constructed to house the detector.  Interlocking lead bricks of virgin Doe Run lead formed a \MatAssayTempleLeadThickness\ thick outer shield, surrounding a \MatAssayTempleCavityVolume\ sample cavity lined with \MatAssayTempleCuThickness\ of OFHC copper.  The copper was cleaned and passivated at FNAL to remove and avoid radon daughter deposition, using a PNNL recipe~\cite{Hoppe:2007gd}.  Radon is excluded from the sample cavity by enclosing the entire shield in an airtight envelope flushed with boil-off nitrogen.  Samples are inserted into the outer chamber of a load lock, in which radon and other short-lived activities can decay.  A built-in glove box then allows samples to be transferred from the load lock, through the interlocking, motor driven doors of the shield, and into the sample cavity.  Yet to be completed is the 4$\pi$ outer shield of borated polyethylene and a set of plastic scintillator muon-veto counters.

\subsubsection{Surface Screening}
\label{sec:MatAssay-AssayResources-Surface}

Plans are being made to carry out large-area assay at the Canfranc Underground Laboratory (LSC) in Spain with the \BiPo\ detector, which allows dedicated measurements of \ce{^208Tl} (\ce{^232Th} chain) and \ce{^214Bi} (\ce{^238U} chain) contamination in thin foils with better sensitivity than a \HPGe\ detector~\cite{Argyriades:2010do}.  \BiPo\ is a unique facility with two separate modules of thin ultra-radiopure organic plastic polystyrene-based scintillator plates coupled with a PMMA optical guide to \MatAssayBiPoPMTsize\ low-radioactivity photomultipliers.  It has a total sample surface area of \MatAssayBiPoSurfaceArea\ and has been running underground since \MatAssayBiPoStartYear\ to measure the radiopurity of selenium source foils for the SuperNEMO double-$\beta$ decay experiment.  The sample foils are installed between the two scintillators and \ce{^212Bi} (\ce{^208Tl}) and \ce{^214Bi} decays are measured.  The system exploits the so-called ``back-to-back events'' in which there is an energy deposition from a $\beta$ decay in one scintillator without a coincidence from the opposite side, followed by a delayed $\alpha$ decay in the opposite side without a coincidence in the first side.  The energy of the delayed $\alpha$ provides information on whether the contamination is on the surface or in the bulk of the foil.  The demonstrated intrinsic background is \MatAssayBiPoBackground,  corresponding to respective \ce{^208Tl} and \ce{^214Bi} sensitivities of less than \MatAssayBiPoTlSensitivity\ and \MatAssayBiPoBiSensitivity\ (at \MatAssayBiPoSensitivityCL\ confidence) for a six month measurement that utilizes the full \BiPo\ detector.  The screening campaign for SuperNEMO will be completed in \MatAssayBiPoSDSkcreeningStartYear.  There is an agreement with the BiPo collaboration to use this large-area screening facility to perform surface screening of samples for \DSk, including measurements of Al, Mylar, \PTFE, Ti and stainless-steel sheets.  The first measurements are planned to be carried out in \MatAssayBiPoScreeningYearTwo, with as much as \MatAssayBiPoScreeningPercentage\ of the counting time for both modules available to \DSk.

Another large-area surface detector, the XIA UltraLo-1800 $\alpha$ spectrometer, is operated by the Krakow group.  It uses argon as a counting gas and can accommodate samples up to \MatAssayKrakowXIASampleSurface\ in surface and up to \MatAssayKrakowXIASampleThickness\ thick.  It is especially suitable for investigations of surface and bulk specific activities of \ce{^210Po}, which is frequently a very serious background source in low-background experiments.  Surface \ce{^210Po} can be analyzed by observation of the \PoTwoOneZeroAlphaEnergy\ peak while bulk \ce{^210Po} generates a continuous spectrum from the detector threshold (\MatAssayKrakowXIAThreshold) up to the nominal peak energy.  Because the background of the detector is very low (\MatAssayKrakowXIABackground\ under the \ce{^210Po} peak) it provides a unique possibility to study natural surface contamination of various samples -- copper, stainless steel, titanium -- down to \MatAssayKrakowXIASurfaceSensitivity\ and in the bulk down to \MatAssayKrakowXIABulkSensitivity\ (or less, depending on the material).  The instrument will also be used to investigate surface cleaning techniques ({\it e.g.}, etching and electropolishing; see also Sec.~\ref{sec:MatAssay-ProcessDevelopment} below).  This capability will allow the \DS\ collaboration to establish specific procedures that are most effective for removal of $\alpha$ surface activity (mainly \ce{^210Po}) from particular materials.

Due to the low counting rates observed or expected for most samples of relevance, it is reasonable to assume that no more than three effective  measurements per month can be performed with the UltraLo-1800.  \MatAssayKrakowXIACapacityReservedForDSk\ of the detector capacity can be devoted to \DSk, therefore the collaboration can analyze at least \MatAssayKrakowXIASampleRate\ at the detector's ultimate sensitivity.  In combination with the \BiPo\ detector, this should be more than sufficient to perform all necessary studies in a timely fashion.

\subsection{Targeted Assays}
\label{sec:MatAssay-AssayResources-Targeted}

A targeted assay program is currently underway focused on three items of high priority for \DSk: \SiPMs\ produced by \FBK, titanium for the \LArTPC\ cryostat, and sapphire substrates for the detector module (see Sec.~\ref{sec:Cryogenics} and Sec.~\ref{sec:LArTPC} for details on the cryostat and TPC).  Each of these three components present unique assay challenges and together illustrate the variety of methods the \MAWG\ is using to meet the \DSk\ radiopurity requirements.

The initial assay of \SiPMs\ produced by \FBK\ was performed by three of the \DSk\ institutions that host \ICPMS\ capabilities: BHSU, LNGS and PNNL.  This exercise served as an initial screening of the \SiPMs\ and provided a cross-calibration of the assay capabilities at each site.  Each lab received a set of \SiPM\ samples originating from the same production wafer and followed a common pre-treatment procedure defined by the \MAWG, and then they assayed the devices following their own best practices.  A set of process blanks accompanied all samples from sample preparation through analysis, and each group performed multiple trials.  This enabled estimation of detection limits, testing of sample heterogeneity, and determination of the measurement precision achievable at each lab.  There were no positive detections of \ce{U} or \ce{Th} among all samples analyzed by all three labs.  \LNGS\ obtained few \MatAssaySiPMLNGSUTh\ upper limits, while BHSU measured \MatAssaySiPMBHSUU\ and \MatAssaySiPMBHSUTh\ for \ce{U} and \ce{Th}, respectively.  PNNL achieved the best sensitivity, with U and Th detection limits of \MatAssaySiPMPNNLU\ and \MatAssaySiPMPNNLTh.  These results help to inform how to distribute future assay samples to best capitalize on the strengths and capabilities of each institution.  Additionally, the Krakow group will perform a surface assay of an entire \SiPM\ wafer using their XIA detector (see Sec.~\ref{sec:MatAssay-AssayResources-Surface}) and thus provide complementary radiopurity information.

\DSk\ is evaluating titanium as the construction material for the TPC cryostat.  The fabrication of radiopure \ce{Ti} parts requires careful control throughout a multi-step production process.  The \MAWG\ has assayed samples from two of the manufacturing steps: \ce{Ti} sponge and \ce{Ti} plate.  A sponge sample was assayed using the GeMPI2 detector at LNGS, \ICPMS\ at MSU, and the Krakow radon emanation system, and all labs found the activity to be below the \DSk\ requirement.  The \ce{Ti} plate has been assayed for \ce{U}, \ce{Th}, and \ce{K} via \ICPMS\ at MSU, LNGS, and PNNL.  The results show that MSU has succeeded in preserving the low level of \ce{U}/\ce{Th} measured in the \ce{Ti} sponge through the electron beam melting and subsequent rolling process used to create the \ce{Ti} plate, with respective \ce{U} and \ce{Th} upper limits for both sponge and plate of \MatAssayTiMSUU\ and \MatAssayTiMSUTh.  Assay of the \ce{Ti} plate with PNNL's new triple quadrupole \ICPMS\ indicates \MatAssayTiPNNLK\ of \ce{^40K} and a \MatAssayTiPNNLTh\ detection of \ce{Th} (consistent with the MSU upper limit).  Deploying multiple assay tools gives the \MAWG\ the flexibility to evaluate samples from multiple steps along the titanium production process and to cross-check these results.

\subsection{Process \& Materials Development}
\label{sec:MatAssay-ProcessDevelopment}

The \DS\ collaboration will develop or modify fabrication, cleaning, handling, storage, shipping, and assembly processes that allow materials to be incorporated into the finished detector while still meeting all radiopurity goals.  This includes considering the background impact of chemical processes such as target purification and working across the collaboration to identify methods to limit background associated with exposure to dust, radon and cosmic rays.

Preliminary studies to search for metals free of bulk and surface \ce{^210Po}, including tests of various surface cleaning techniques, have already been performed.  Assay of the natural $\alpha$ surface activity is necessary for several reasons; alphas with degraded energy are a potential background source, and in cases of broken equilibrium in the \ce{^226Ra} decay chain (at \ce{^210Pb}), only direct determination of \ce{^210Po} bulk and surface activities is relevant.  Ultra-sensitive screening required for verification of \ce{^210Po} (\ce{^210Pb}) presence (removal) was performed with the Krakow XIA UltraLo-1800 $\alpha$ spectrometer (described above in Sec.~\ref{sec:MatAssay-AssayResources-Surface}).  Advanced pulse-shape discrimination and an additional veto guard electrode result in a background level in the region of interest more than two orders of magnitude lower than achievable with ultra-low background semiconductor $\alpha$ detectors operated underground~\cite{Misiaszek:2013hq}.

Two cleaning techniques for metal surfaces have been studied: etching applied to copper, stainless steel and titanium; and electropolishing of copper and stainless steel.  For each test, a fresh solution of chemicals was mixed according to material-specific recipes.  Measurement of Electrolytic-Tough-Pitch (ETP) copper sheet with the UltraLo-1800 showed some bulk \ce{^210Po} but no indication of surface \ce{^210Po} activity (with an estimated upper limit of \MatAssayCopperSurfaceActivitiyLimit).  The etching procedure had no effect on the observed bulk and surface \ce{^210Po} activities.  The same result was obtained for a grade~304 stainless steel plate.  In the case of an industrial grade~2 titanium plate, significant bulk and surface \ce{^210Po} activities were measured, orders of magnitude higher than the copper and steel samples.  After etching, a reduction factor of \MatAssayKrakowPoReductionFromTiEtching\ was observed in the surface and sub-surface \ce{^210Po} and \ce{U}/\ce{Th} activities, where the latter were gauged using alphas with energy above \MatAssayKrakowEtchingAlphaEenrgyLowerLimit.  Electropolishing of a copper surface loaded with \ce{^210Po} proved to be very effective in removal of this isotope, with a demonstrated reduction factor of \MatAssayKrakowPoReductionfromCuSSElectroPolishing; a similar result was obtained for a stainless-steel surface.

Summarizing, removal of \ce{^210Po} via etching is not effective for copper and stainless steel, confirming earlier studies with surfaces artificially loaded with \ce{^210Po}~\cite{Zuzel:2012ht}, whereas etching of titanium significantly reduces its surface and sub-surface $\alpha$ activities.  Electropolishing of copper and stainless steel reduces the \ce{^210Po} surface activity by a large factor of up to \MatAssayKrakowPoReductionfromCuSSElectroPolishing.  The results obtained so far confirm abilities to assess the natural surface and bulk $\alpha$ activities of \ce{^210Po} and constitute a solid basis for the development of the surface cleaning procedures required.  Multi-step electropolishing of copper and stainless steel are the most promising surface cleaning procedures and will also be studied in detail.  High-purity Ti will be investigated in terms of its bulk \ce{^210Po} content as well as with respect to removal of surface activity.

The impact of cosmogenic activation of detector materials will be carefully considered and cosmogenic sources that could impact otherwise radiopure materials will be identified and mitigated.  As an example, limits for \ce{^60Co} in pure copper parts ({\it i.e.}, the electric field shaping rings) will be determined.

If commercially available materials do not exist that meet materials selection criteria, or are impractical for reasons of cost or availability, it may become necessary to develop alternatives.  This might involve new materials processing from known radiopure precursor material to avoid problematic commercial production processes, or it may involve developing additional processing methods to clean commercial input materials to a level that can meet materials selection criteria,  as it was shown for the case of copper, stainless steel and titanium.  Similar work is ongoing at PNNL~\cite{Grate:2015vu}.

\subsection{Risks and Their Mitigation}
\label{sec:MatAssay-RisksMitigation}

The risks to the experiment associated with the Materials and Assay program have been assessed.  Below is a list of the major risks, including the potential effects of mitigation activities upon the Materials and Assay program.

\begin{asparaenum}
\item[\bf Schedule delays:] Project delays may arise if screening capacity is inadequate.  This risk is kept low by utilizing and organizing all of the assay capabilities within the collaboration, as discussed in Sec.~\ref{sec:MatAssay-RadiopurityManagement}.  However, the level of risk will become critical if assay capacity is substantially decreased, which could happen if major instrumentation availability is reduced for technical reasons or because a collaborating institution is unable to participate in the experiment.  Loss of the most sensitive and/or highest throughput methods would be particularly problematic.

\item[\bf Insufficient sensitivity:] Not being able to assay a material to the required level would be a significant risk to the experiment.  The most likely consequence is an increase in the experiment's expected background and a corresponding decrease in its sensitivity.  The \MAWG, however, has access to world-leading capabilities in low-level materials assay.  Currently the background model has not predicted a material activity level that is outside the reach of the assay methods available.  This risk will elevate (and could become critical) if the most sensitive assay methods become unavailable.  Additionally, this risk might increase if the background model suggests an activity requirement for a material that is below levels achievable with current technology.  The current background model, however, uses existing and already proven data for material radiopurity, and so this is not very likely.

\item[\bf No assay method available:] It is conceivable that a material cannot be assayed.  Because all materials can be gamma counted, this is related to the previous risk; if the mass is very small or if the chemistry needed for \ICPMS\ is incompatible with the material, it is possible that an insufficiently sensitive assay via gamma counting with a \HPGe\ detector is the only available option, making the assay effectively not possible.  As noted above, however, this collaboration has access to world-leading low-level materials assay with a variety of complementary techniques; the collaboration has yet to encounter a material that cannot be assayed.

\item[\bf Decay-chain equilibrium is broken:] Broken equilibrium in the \ce{^238U} and \ce{^232Th} chains is the principal concern because assay of early-chain isotopes may not be representative of late-chain activities, resulting in the potential for elevated background and a corresponding decrease in experimental sensitivity.  The lowest  activity levels can only be reached by \ICPMS, which measures the uranium and thorium content directly and does not measure the lower parts of the decay chains.  If equilibrium is broken significantly, and the lower chain contains more activity than is predicted by the uranium and thorium concentrations, then the $(\alpha,n)$ rate would be underestimated.  This risk will be minimized by performing \ICPMS, radon emanation, and \HPGe\ gamma counting for the materials expected to dominate the $(\alpha,n)$ budget ({\it e.g.}, \PTFE\, and titanium).
\end{asparaenum}

\section{Physics Reach}
\label{sec:Physics}

\begin{figure}
\includegraphics[width=\columnwidth]{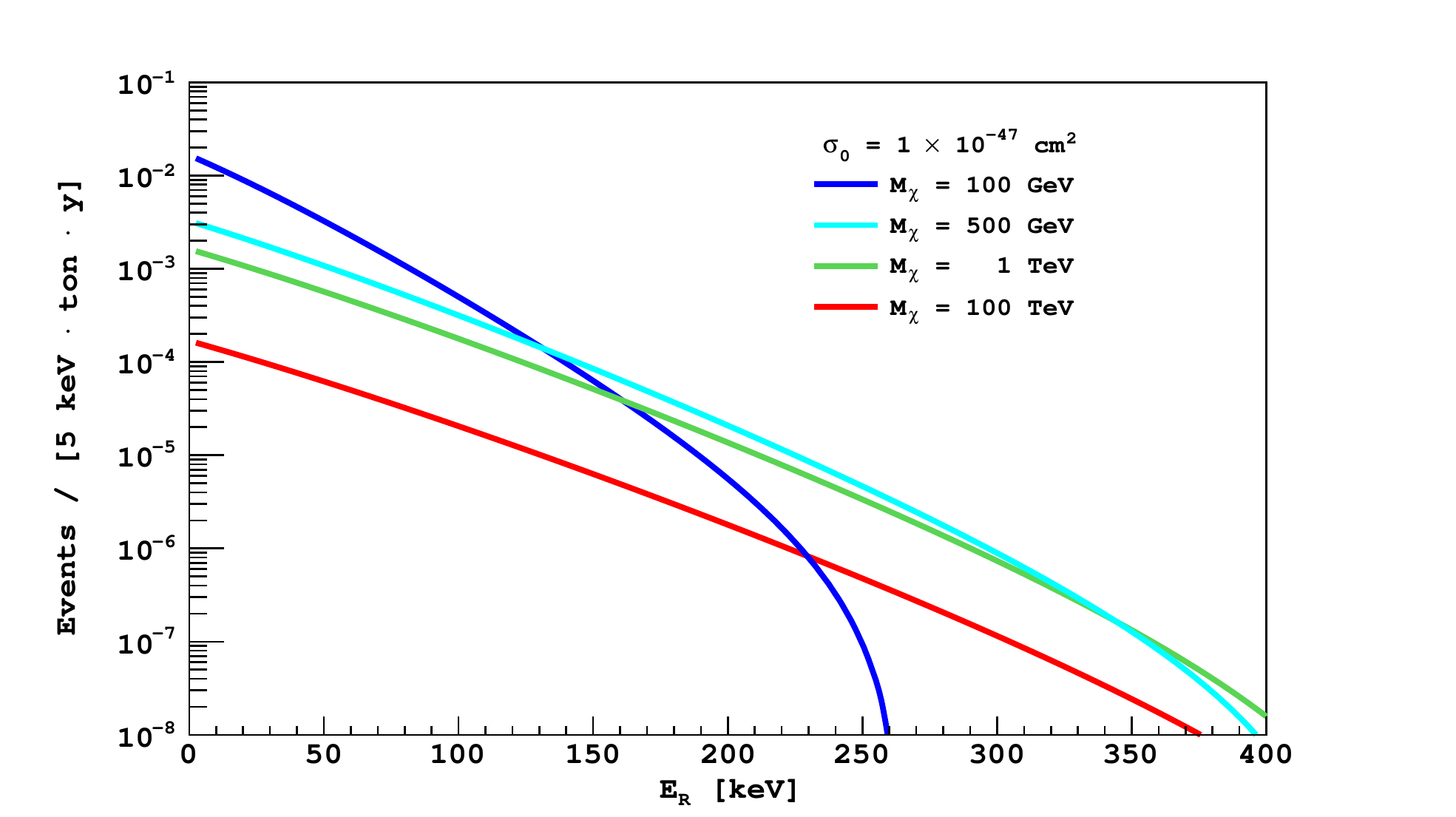}
\caption[\WIMP\ nuclear recoil spectra in \LAr.]{\WIMP\ nuclear recoil spectra in \LAr.}
\label{fig:G4DS-Rate}
\end{figure}

The ability to identify, measure, and reject background will ultimately define the sensitivity of direct dark matter searches.  \DSk\ is designed to perform a search for high mass (tens of \SI{}{\GeV\per\square\c} to hundreds of \SI{}{\TeV\per\square\c}) dark matter particles with an exposure of \DSkExposure.  \DSk\ will search for \WIMPs\ in a broad energy range from \DSkROIEnergyRange.  Fig.~\ref{fig:G4DS-Rate} shows the nuclear recoil energy spectra for \WIMPs\ interacting with argon nuclei, for various masses between \SI{100}{\GeV} and \SI{100}{\TeV}.  The steeply falling rate with increasing recoil energy means that a lower energy threshold can greatly increases sensitivity for \WIMP\ detection, as long as the detector still operates with a very low and well understood background.  As will be detailed in this section, the operation of \DSk\ in an instrumental background-free mode is possible thanks to its outstanding rejection power for \bg\ backgrounds using pulse shape discrimination (\PSD), and rejection of neutron-induced nuclear recoils by the identification of multiple scatters thanks to the intrinsic spacial resolution of the \LArTPC\ and the system of two active vetoes, similar to the ones used in \DSf.

Based on the background rejection techniques developed and demonstrated with \DSf, the proposed \DSk\ experiment will be in a position to lead the exploration of the range of \WIMP\ masses greater than \WIMPMassHundredGev, especially when approaching at masses of \WIMPMassOneTev\ and beyond (see Fig.~\ref{fig:ArgoDSkDSf-ProjectedExclusion}), with the highest sensitivity among the experiments that will start data taking within the next half-decade, as shown in Table~\ref{tab:Introduction-Sensitivity}.

\DSk's system of nested detectors is designed to efficiently reject the backgrounds described in Sec.~\ref{sec:BackgroundFree}: nuclear recoils induced by radiogenic and cosmogenic neutrons, electron recoils from gamma rays and radioactive contaminants in the \LAr, and backgrounds from radon daughters deposited on detector surfaces.  Detailed simulations projecting that \DSk\ will be operated in an instrumental background-free condition in the \WIMP\ search region at the required exposure will be discussed in Sec.~\ref{sec:Physics-Background}.  These studies have been performed using \GFDS, the \Geant-based Monte Carlo simulation tool developed by the \DS\ collaboration and described in Sec.~\ref{sec:Physics-MonteCarlo}. 

\DSk\ is an experiment designed to reach the highest sensitivity for \WIMP\ dark matter, in the high mass region of the \WIMP-nucleon cross-section vs. \WIMP\ mass parameter space.  To provide unbiased results for the dark matter search at the limits of sensitivity offered by the detector, a blind analysis procedure will be implemented, following an initial phase of data-taking devoted to the commissioning of the detector.  The blind analysis employed by \DSk\ will build upon the procedure currently being developed by the \DS\ collaboration for \DSf.  In \DSf, following the first round of publications with limited exposures collected with \AAr\ and \UAr~\cite{Agnes:2016fz,Agnes:2015gu}, the collaboration has switched to a blind analysis.  The specific procedure that has been adopted for the \DSf\ data involves the blinding of a box, defined in the \FNine-\SOne\ plane (\FNine, described in Sec.~\ref{sec:Sensitivity}, is the primary \PSD\ parameter for discrimination of electron recoils from nuclear recoils in \DSf).  The box extends well beyond the ninetieth percentile of acceptance for \WIMP\ candidates in the \WIMP\ energy \ROI.  All other data remain available for studies of background, design of cuts, and computation of the expected survival rate of backgrounds in the blinded region.  Specific background predictions developed by the analysis will be tested by gradually opening ``test strips'' within the initially blinded region but outside the \WIMP\ search region, to better understand the backgrounds that may be present in the box.  Variants of this strategy are foreseen for use in the analysis of the \DSk\ data.

\subsection{Simulation Tools Description}
\label{sec:Physics-MonteCarlo}

The Monte Carlo simulation for \DSk\ is based on the package developed for simulation of previous detectors of the \DS\ program: \DSt\ and \DSf.  \GFDS\ was designed with a modular architecture and describes the energy and time responses of each detector.  It provides a rich set of particle generators, detailed geometry descriptions, finely-tuned physical processes, and a full optical propagation model for the photons produced by scintillation in liquid argon and liquid scintillator and by electroluminescence in gaseous argon.  All the components of \GFDS\ have been subjected to very detailed validation tests by comparison with the results of the \DSf\ experiment.

\begin{figure*}
\includegraphics[width=0.45\columnwidth]{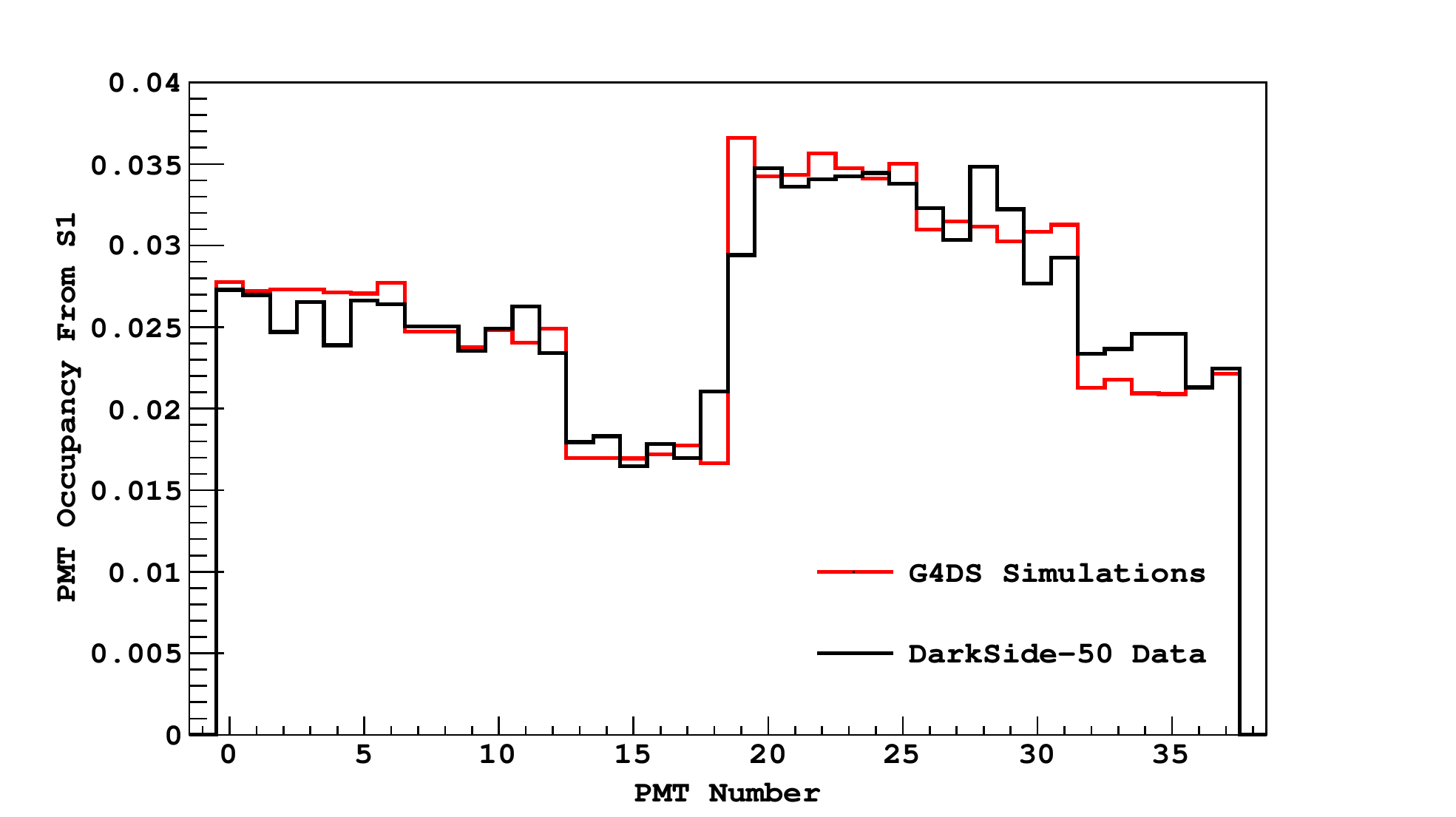}
\includegraphics[width=0.45\columnwidth]{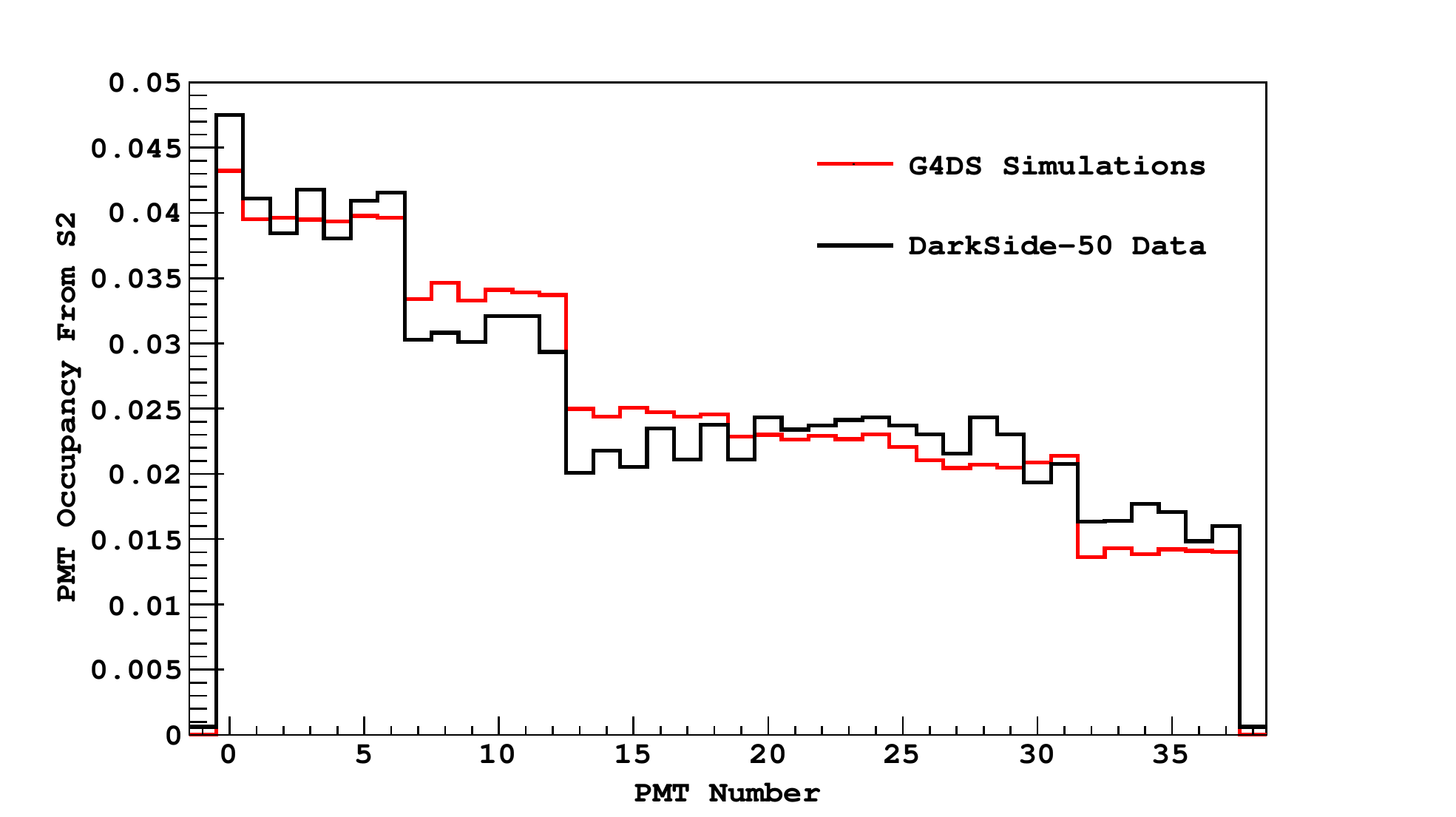}
\caption[\DSf\ \SOne\ and \STwo\ channel occupancy]{\SOne\ (left) and \STwo\ (right) channel occupancy in \DSf\ compared to the \GFDS\ prediction.}
\label{fig:G4DS-DSfOccupancy} 
\end{figure*}

\begin{figure*}
\includegraphics[width=0.45\columnwidth]{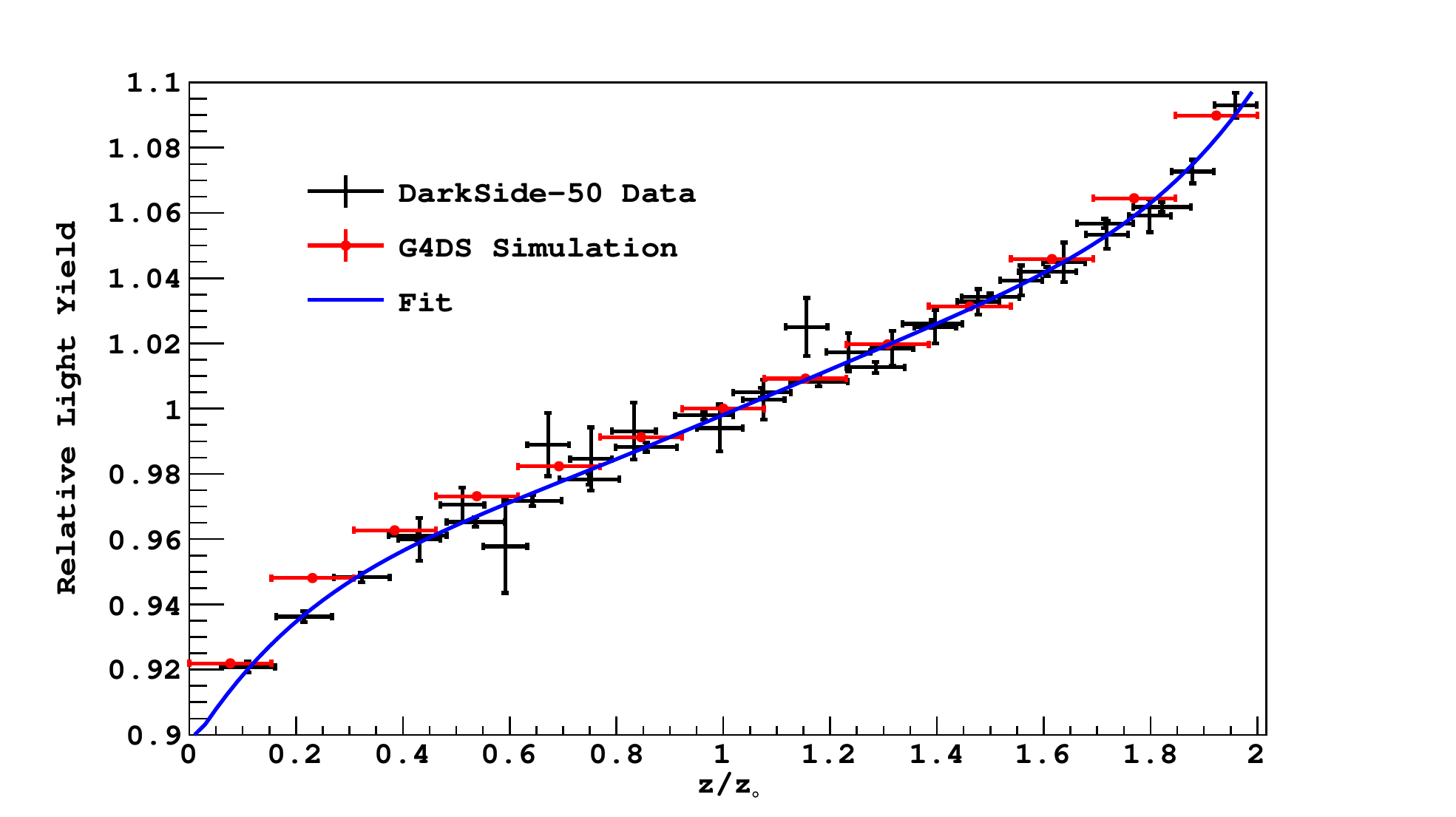}
\includegraphics[width=0.45\columnwidth]{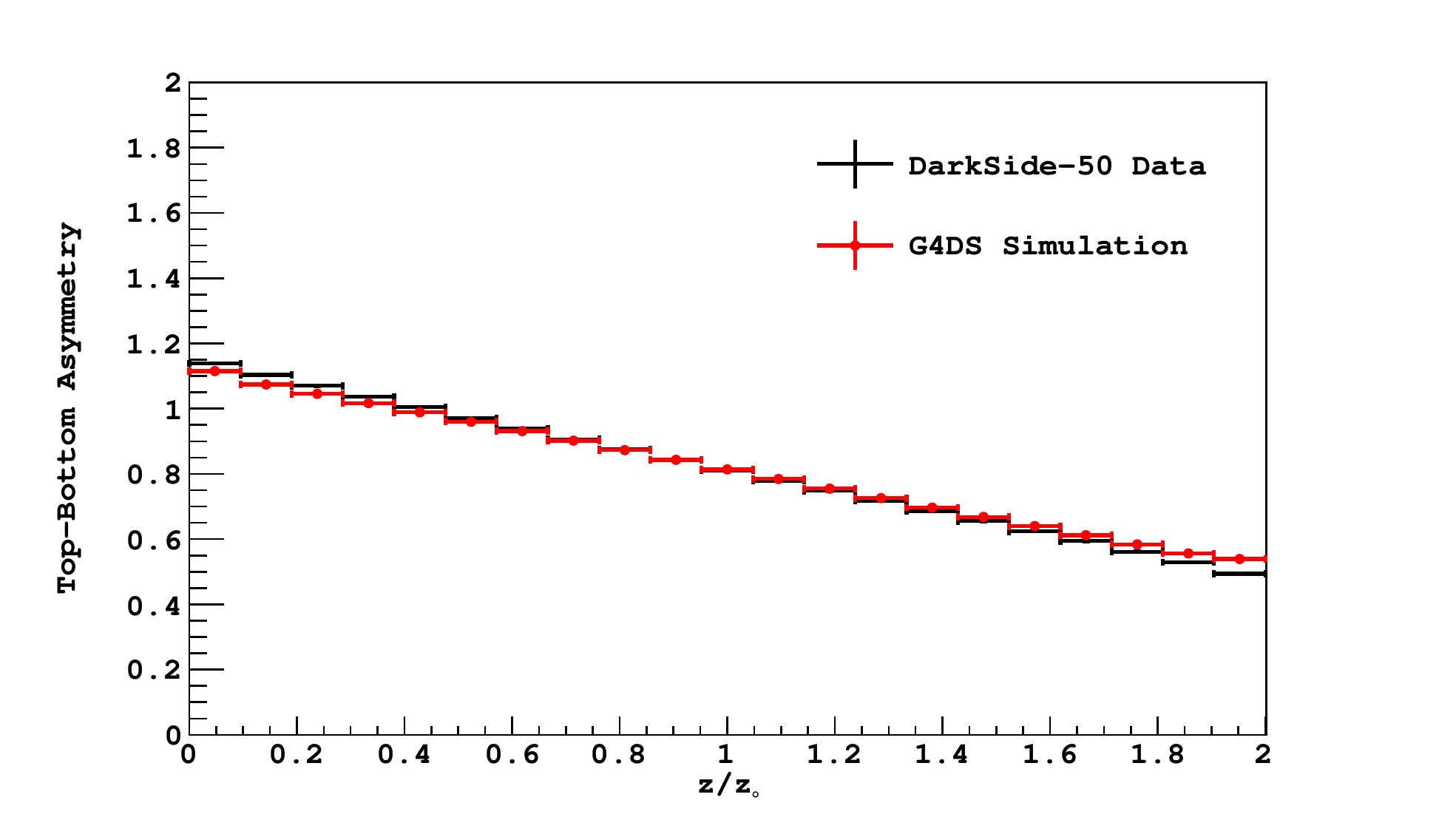}
\caption[\DSf\ light yield and top-bottom asymmetry.]{Light yield (left) and top-bottom asymmetry in light collection (right) as a function of event vertical position in \DSf\ data, compared with the \GFDS\ prediction. The vertical position in the detector shown on the horizontal scale is normalized by half the \TPC\ maximum drift distance ($z_o$).}
\label{fig:G4DS-DSfLyTBA}
\end{figure*}

\begin{figure}
\includegraphics[width=\columnwidth]{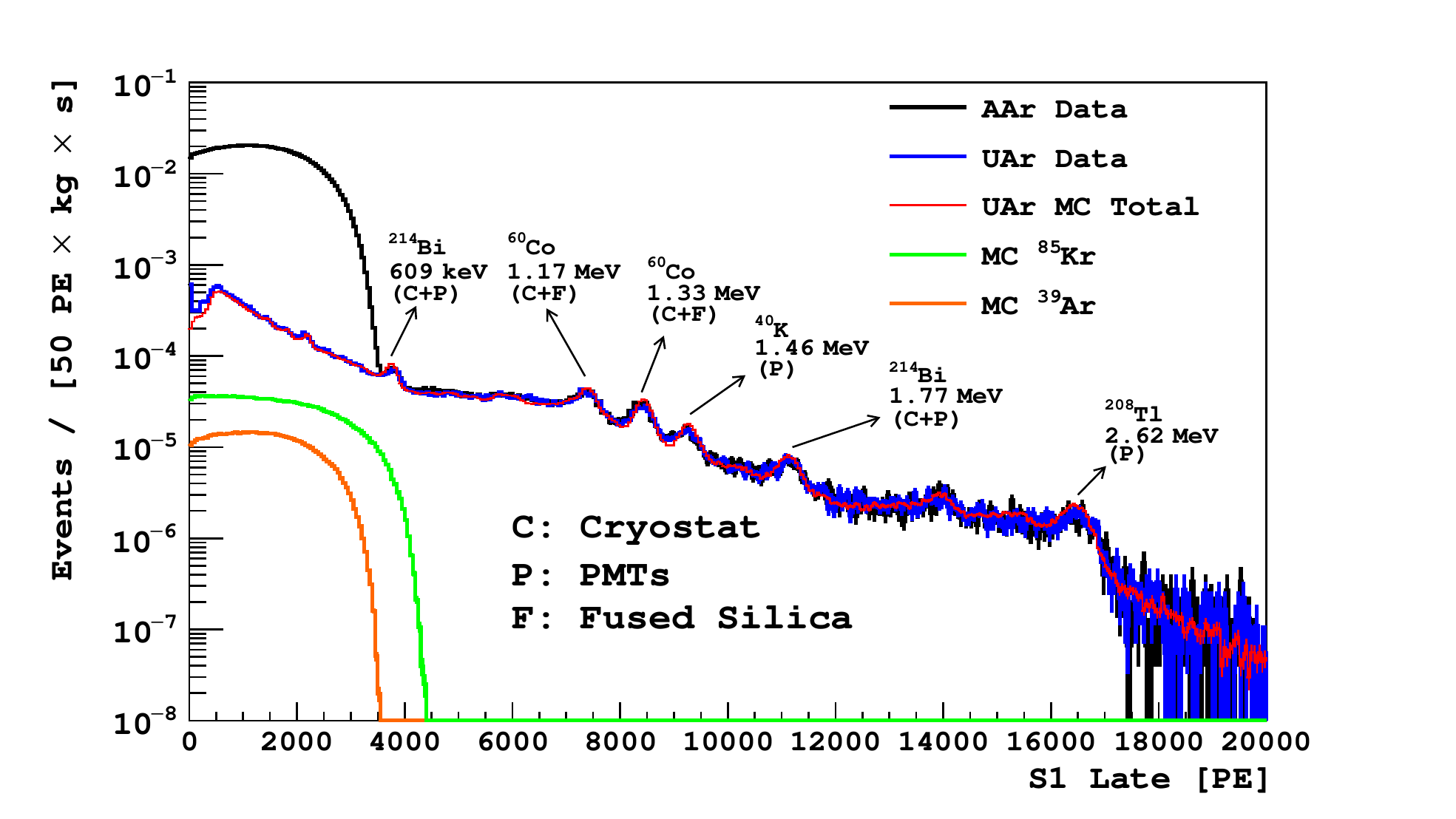}
\caption[\DSf\ \UAr\ and \AAr\ field-off spectra.]{Comparison of measured field-off spectra for the \UAr\ (blue) and \AAr\ (black) targets, normalized to exposure.  Also shown are the \GFDS\ fit to the \UAr\ data (red) and individual components from \ce{^85Kr} (green) and \ce{^39Ar} (orange) extracted from the fit. \SOne\ Late is a modified S1 variable which begins waveform integration after the first \SI{90}{\nano\second}, to avoid channel saturation.~\cite{Agnes:2016fz}}
\label{fig:DSf-AArUArSpectraNullField}
\end{figure}

For the \LArTPC, the main goal of G4DS is the accurate simulation of light production, propagation, and detection for background and signal events, in order to fully reproduce the  responses of the detector in \SOne, \STwo, and time, the three primary variables on which the discrimination of \bg\ background is based.  This allows tuning of the analysis cuts and estimation of their efficiencies, as well as prediction of the \bg\ and nuclear recoil backgrounds. 

Particular attention was given to the description of the physical processes.  For electromagnetic processes, one of the physics lists available in \Geant\ ({\em G4EmLivermorePhysics}) perfectly fits the energy range and the accuracy required by the detectors of the \DS\ family.  For the hadronic physics list, the requirement of very high accuracy neutron propagation in the energy range from a fraction of an \si{\eV} to a few \si{\MeV} required the definition of a custom list based on high-precision models for neutron interactions. The standard \Geant\ hadronic physics list is only activated to study processes such as cosmic muon propagation and production of cosmogenic isotopes.

The light generation in \LAr\ and \GAr\ is handled using a custom \Geant\ physical process, since details of atomic excitation, ionization, nuclear quenching, and electron-ion recombination effects are poorly known in argon, especially in presence of strong electric fields.  A theoretical model based on an effective description of recombination was developed, which was tuned on calibration data and is able to accurately describe the light response of \DSf\ in both \SOne\ and \STwo. 

Another critical ingredient of \GFDS\ is the tuning of optical properties of the materials through which light is propagated, and of the surfaces where light can be absorbed, reflected, or diffused.  These parameters were again tuned by comparisons with selected data samples from the \DSf\ \LArTPC.

The \DSf\ electronics response was simulated in detail by generating arrival time and hit channel information for each photoelectron, and then convolving it with the typical single photo-electron time response and adding the simulated signals to actual \PMT\ baseline responses to include noise.  The effects of the noise and resolution of \SiPMs\ and of the associated electronics are simulated (assuming that all experimental requirements are met).  The readout planes at the top and bottom of the \TPC\ are segmented according to the proposed \tile\ configuration, the arrival times of photons on each tile are registered taking into account the photon detection efficiency (\PDE) and, if the total amplitude in a given \tile\ is above a set threshold, the photon is registered as a hit on that channel, assessing a dead time for the channel of \DSkTileDeadTime.  The \SiPMs\  primary dark count rate (\DCR), direct cross talk (\DiCT), and delayed cross talk (\DeCT) are fully simulated by appropriately adding hits to the signal.  The baseline noise of the electronics is also included and its probability density function (PDF) is convolved with that of the signal.  After signal discrimination, the digitization is directly simulated in \GFDS.  The output signals can then be passed on to the full analysis chain.  

The success of \GFDS\ was validated by its ability to reproduce the background spectra in the latest analysis of the \DSf\ \UAr\ data, as well as other specific distributions measured in the the \DSf\ \LArTPC.  As one example, Fig.~\ref{fig:G4DS-DSfOccupancy} shows the occupancy of each PMT  for both \SOne\ and \STwo\ signals as measured in \DSf, and compares the distributions to the predictions of \GFDS.  Fig.~\ref{fig:G4DS-DSfLyTBA} compares measured and simulated \DSf\ \SOne\ light yield and the top-bottom asymmetry of \SOne\ (defined as  (S1$_{\text{Top}}-$S1$_{\text{Bottom}}$)/(S1$_{\text{Top}}+$S1$_{\text{Bottom}}$), where S1$_{\text{Top}}$ and S1$_{\text{Bottom}}$ are the sum of the top and bottom PMTs, respectively) as a function of the vertical position in the detector. The top-bottom asymmetry of the light yield comes from geometrical differences in the detector at the top and bottom, {\it e.g.}, the gas pocket, the grid, {\it etc.}  Once again the data and \GFDS\ simulations agree very accurately.  

Another strong validation of \GFDS\ was made when it first detected the presence of \ce{^85Kr} in the spectrum from the \UAr\ target in \DSf.  The \GFDS\ fit to the spectrum, shown in Fig.~\ref{fig:DSf-AArUArSpectraNullField}, would not converge in the region of the \ce{^39Ar} end point under the initial assumption of no other $\beta$ component with a similar end point.  The \GFDS\ working group found that only by adding a significant \ce{^85Kr} component would the fit converge to the data.  This hypothesis was confirmed by measuring the rate of coincidences due to the \KrEightFiveExcitedDecayBR\ branching ratio of \ce{^85Kr} decaying to \ce{^{85m}Rb}.  Not only did the \GFDS\ analysis identify the missing spectral component, it also measured the absolute \ce{^85Kr} activity with great accuracy, as confirmed by the coincidence measurement.  As Fig. \ref{fig:DSf-AArUArSpectraNullField} shows, \GFDS\ is designed to reproduce with great accuracy the experimental features of the \DSf\ spectrum with a \UAr\ target.

A detailed description of the \DSk\ detector geometry was developed, based on the design detailed in this document.  During the experimental design optimization process, \GFDS\ has often been used to extract predictions of the absolute light yield corresponding to trial configurations, guiding the design.

\begin{figure}
\includegraphics[width=\columnwidth]{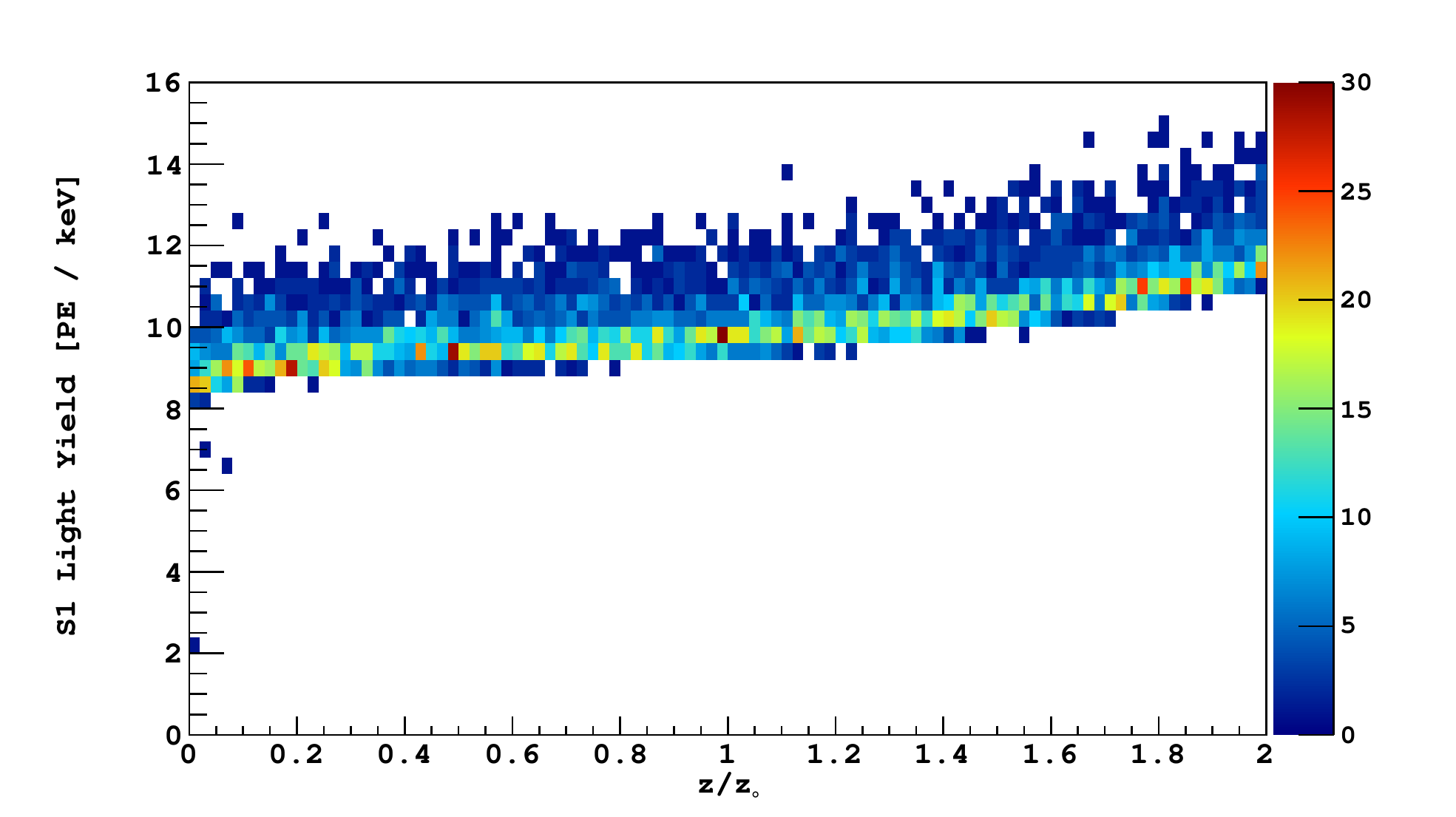}
\caption[\DSk\ light yield as a function of the vertical coordinate.]{Light yield as a function of the vertical coordinate in \DSk\ for \ce{^39Ar} with an electric field of \DSkDriftField, as obtained with the \GFDS\ simulation of the detector described in this document.
}
\label{fig:G4DS-DSkLyVsZ} 
\end{figure}

\GFDS\ has been used to calculate the light yield of \DSk\ in its final configuration, obtaining the distribution shown in Fig.~\ref{fig:G4DS-DSkLyVsZ}. This calculation used the optical parameters tuned on the \DSf\ data (plus an assumed self-absorption length of \SI{300}{\meter} for the \LAr\ scintillation light), the geometry of the detector as described in this document, and assumed the \DSkPdmPDESpecification\ overall photon detection efficiency quoted in Sec.~\ref{sec:PhotoElectronics-Introduction}.  \GFDS\ estimates a light yield of \DSkSimulationLY, larger than that obtained in \DSf, the result of many technical improvements of the \DSk\ design, larger geometrical coverage and increased photo detection efficiency for \SiPMs, compared to \PMTs.

\begin{figure*}
\centering
\includegraphics[width=0.45\columnwidth]{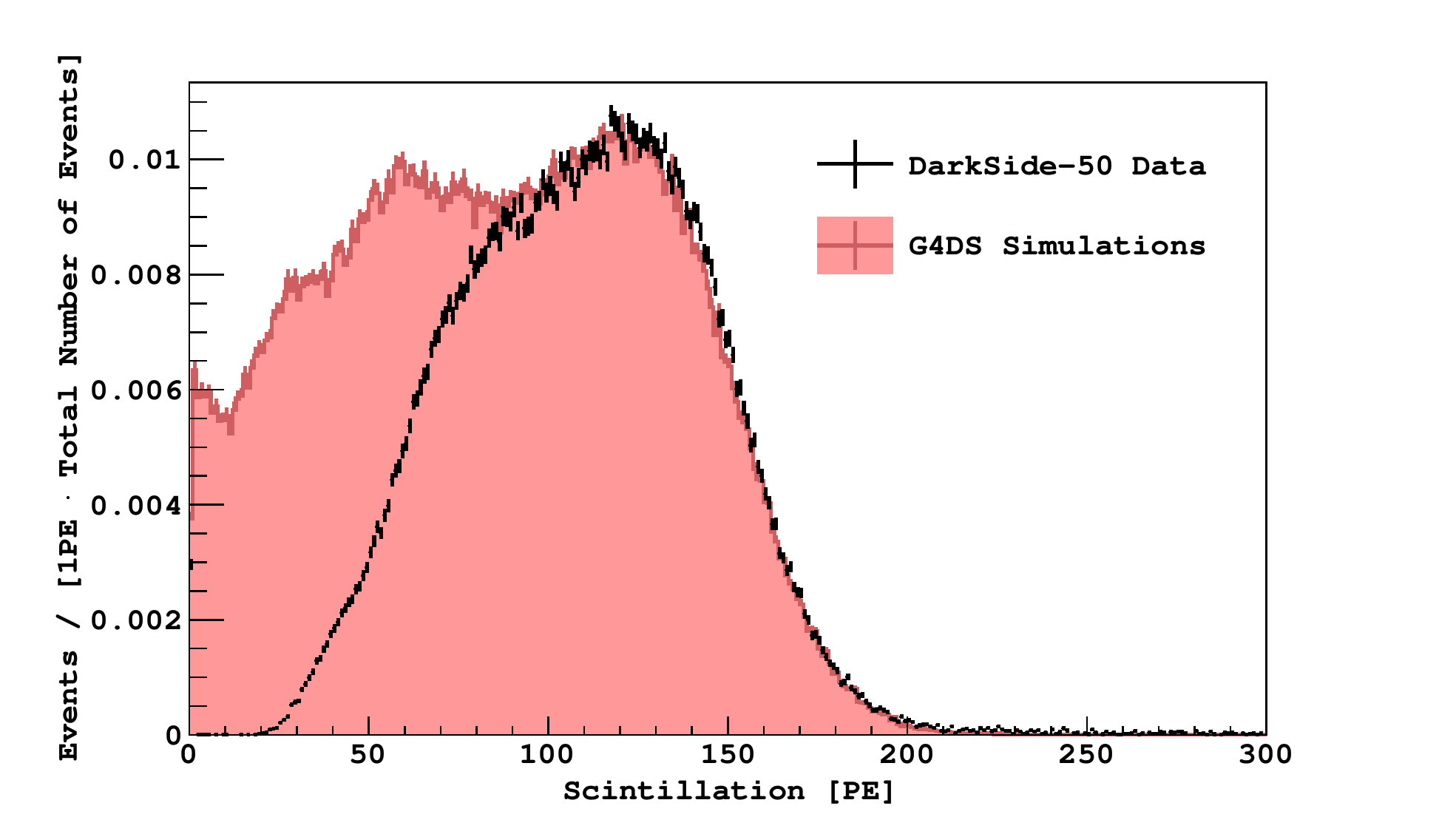}
\includegraphics[width=0.45\columnwidth]{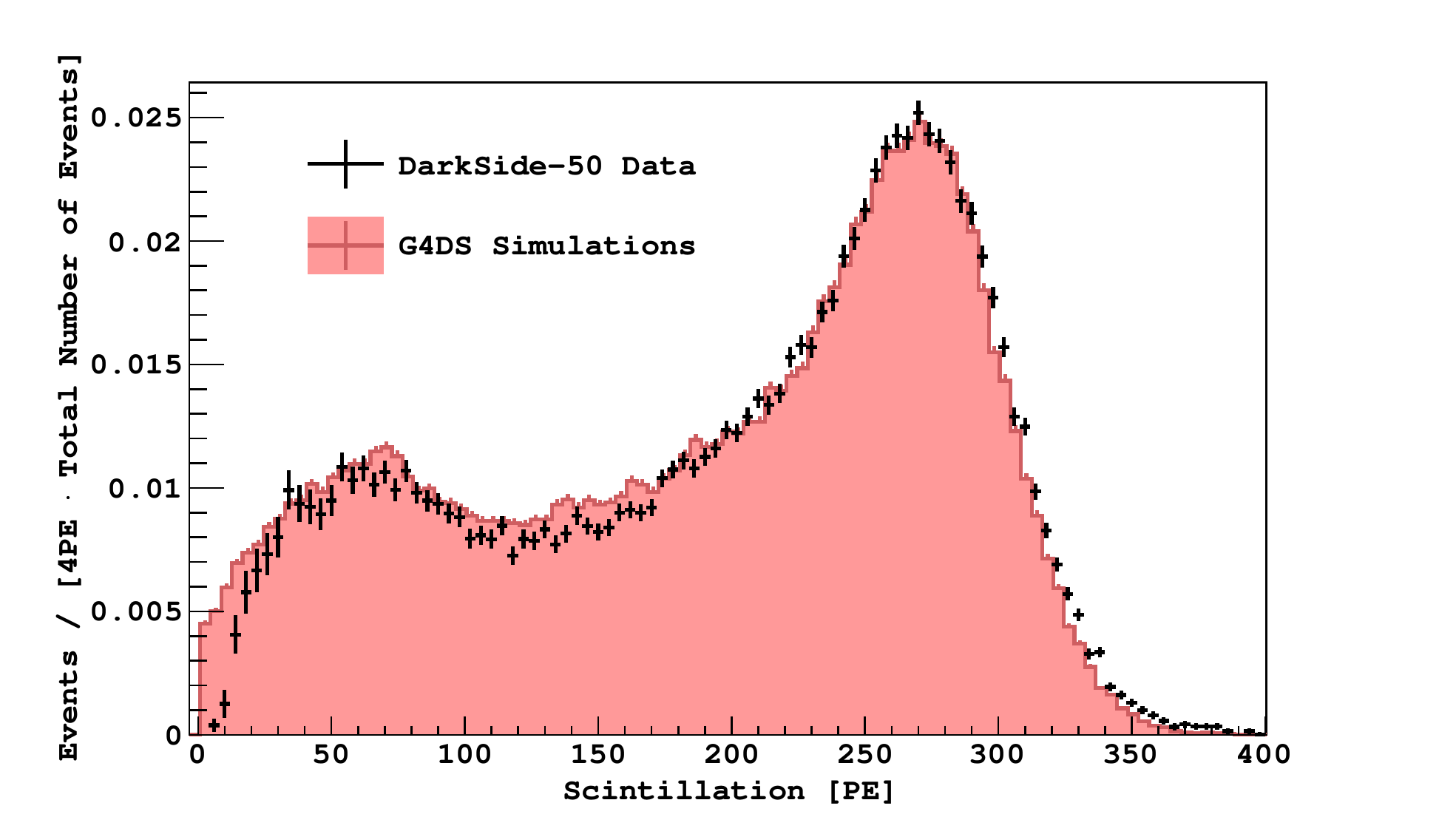}
\caption[\DSf\ \LSV\ calibration source scintillation spectrum.]{{\bf Left:} Scintillation spectrum from the \DSf\ \LSV\ in the presence of a \ce{^133Ba} calibration source, compared with the \GFDS\ simulation.  The disagreement at low energy is due to the fact that \GFDS\ does not fully simulate the majority trigger condition, requiring instead hits in a certain number of \LSV\ \PMTs.  {\bf Right:} The same for a \ce{^137Cs} calibration source, compared with the \GFDS\ simulation.}
\label{fig:G4DS-LSVLy} 
\end{figure*}

For the \LSV\ simulation, the default model for scintillation in \Geant\ is sufficient for the electromagnetic processes occurring in the scintillator.  Simulation of the \DSf\ \LSV\ photoelectron yield was tuned to accurately represent its response to external \grs\  from calibration sources, with results as shown in Fig.~\ref{fig:G4DS-LSVLy}. For full details about the \DSf\ \LSV\ tuning, see~\cite{Agnes:2016fw}.

\subsection{Studies of Backgrounds}
\label{sec:Physics-Background}

The  background sources can be divided into three broad categories: 1) ``Internal background"--ionizing events due to radioactive contamination and neutrino interactions in the active argon itself; 2) ``External background"--ionizing events originating from radioactivity in the TPC and cryostat materials (radiogenic backgrounds from outside the cryostat are negligible due to the shielding provided by the veto); 3) ``Cosmogenic background"--ionizing events originating from the residual cosmic ray flux in Hall C.

As has already been discussed, the design of \DSk\ strongly reduces many of these backgrounds by features such as the use of \UAr\ as a target material, the dual-phase \LArTPC\ design, and the nested detector system.  To ensure that the science goals are achieved, the \DS\ Collaboration will implement a thorough system of controls based on background modeling, process and materials development, and materials radiopurity assay.  The scope and prioritization of these activities will be informed by experience with \DSf~\cite{Agnes:2015gu,Agnes:2016fz}, collaborators' experience with other experiments (see Refs.~\cite{Benetti:2008kd,Simgen:2014gs,Agostini:2014dp,Abgrall:2014dk,Bellini:2014dl} and references cited therein), and the experience of the low-background community in general~\cite{Leonard:2008bk,Araujo:2012bt,Aprile:2011ds,Aprile:2013hw,Akerib:2015is}. A full background model of the detector has  been built from the \GFDS\ Monte Carlo simulation and is under continuous refinement.  The model will lead to radiological and environmental limits for all aspects of the project, and will allow identification of materials, systems, and processes that pose the greatest risks.  Much is already understood about the radiopurity of materials~\cite{Loach:2013cw}, techniques for removing low-level radioactive contamination~\cite{Zuzel:2012ht,Hoppe:2007gd}, and methods for avoiding contamination ({\it e.g.}, from exposure to radon~\cite{Street:2015fh,Wojcik:2000di} or dust~\cite{Boger:2000fd}).  

The current state of the background estimates for the proposed \DSkExposure\ exposure of \DSk\ is summarized in Table~\ref{tab:Background} and discussed in detail in this section. 

\begin{table}
\rowcolors{3}{gray!35}{}
\normalsize
\centering
\caption[Expected \bg\ and \NR\ backgrounds expected during the full \DSk\ exposure.]{Expected \bg\ and nuclear recoil (\NR) backgrounds expected during the full \DSk\ exposure, based on current data and Monte Carlo simulations.  The center column gives the total number of single-scatter events within the energy region of interest (ROI) before the application of the fiducial and veto cuts and the \PSD.  The right-most column is the total number of events surviving the veto cut, fiducial volume cut, and \PSD.  Internal \bg\ background does {\it not\/} include the \ce{^39Ar} depletion expected from Aria.  External backgrounds are shown for a stainless steel cryostat.}
\begin{tabular}{lcc}
Background					&\minitab{c}{Events in ROI}{[\DSkExposure]$^{-1}$}
															&\minitab{c}{Background}{[\DSkExposure]$^{-1}$}\\
\hline
Internal \bgs					&\DSkUArArThreeNineROIBare		&\num{0.06} \\
Internal \NRs					&negligible						&negligible \\
$e^-$-$\nu_{pp}$ scatters	&\num{\DSkROIPPBare}						&negligible \\
External \bgs					&\num{E7}						&\num{<0.05} \\
External \NRs					&\num{<81}  						&\num{<0.15} \\
Cosmogenic \bgs				&\num{3E5}						&\num{\ll0.01} \\
Cosmogenic \NRs				&\textendash						&\num{<.1} \\
$\nu$-Induced \NR				&\DSkNuInducedBackgroundBare		&\textendash \\
\end{tabular}
\label{tab:Background}
\end{table}

\subsubsection{Internal background}

\begin{asparaenum}
\item[\bf \ce{^39Ar} and \ce{^85Kr}:]
Recent data from \DSf\ running with low-radioactivity argon  shows  that the \UAr\ contains \ce{^39Ar} and \ce{^85Kr}  activity totaling slightly less than \DSfUArActivityLessThan.  As stated previously, the \DSf\ run with \AAr\ had an exposure of \DSfAArExposure\ and accumulated \DSfAArROIEventsNumberleft\ \ce{^39Ar} events in the WIMP energy ROI after all standard analysis cuts except \PSD.  Every one of these events was eliminated when the \PSD\ was applied, leaving the experiment free from background and producing a lower limit for the reduction factor of \ce{^39Ar} of $>$\DSfAArROIEventsNumberleft.

A detector with similar \PSD\ performance and filled with \UAr\ would have to run for  \DSfUArBackgroundFreeExtrapolatedExposure\ to accumulate the same number of \ce{^39Ar} events, and would also be expected to be free of \ce{^39Ar} background after \PSD.  Because the planned exposure is a factor \DSkDArThreeNineAbatementMissing\ larger than this, assuming the \DSf-measured lower limit for the reduction factor of the \ce{^39Ar} leads to a case in which the experiment would not remain background-free.  Therefore the real rejection factor that can be expected must be estimated using alternative methods, such as Monte Carlo simulations and analytic modeling.  This has been done for the case of \DSk, and is detailed later along with the explanation of the sensitivity calculation.  The details of the sensitivity estimate are presented in Sec.~\ref{sec:Sensitivity}.  In that section, the conservative assumption of the use of \UAr\ with no further depletion of the \ce{^39Ar} is made, and those results are also quoted in Table~\ref{tab:Background}.

Additionally, assuming that two passes of the \UAr\ through \SeruciTwo\ will reduce the \ce{^39Ar} by a factor of \AriaDepletionPerTwoPass, there will be only about \DSkDArArThreeNineROIBare\ residual decays of these isotopes in the \DAr\ itself in the \WIMP\ ROI during the full exposure of \DSkExposure.  This would further extend the sensitivity of \DSk\ beyond that which has been estimated when using just the \UAr, and would mark the beginning of the procurement of the target for a much larger detector, such as \Argo.

Production of \ce{^39Ar} in the \UAr\ by cosmic rays during collection, transportation, and storage of the \UAr\ before its deployment into the \DSk\ detector have also been considered.  Two codes that calculate the production and decay yields of isotopes from nuclear reactions induced by cosmic rays were used to calculate the production of \ce{^39Ar} in the \UAr\ during a realistic altitude and exposure history covering its extraction and purification, shipment to the Aria site, processing there, and transportation to the underground laboratory.  COSMO~\cite{Martoff:1992bg} and ACTIVIA~\cite{Back:2008gr} predicted \ce{^39Ar} specific activities of \AriaCOSMOActivationRate\ and \AriaACTIVIAActivationRate, respectively, much too small to be of concern.

\item[\bf Solar Neutrino-induced Electron Scatters:]
The dominant solar neutrino interaction rate would be from neutrino-electron scattering by \PP\ neutrinos.  These occur at a rate of \DSkROIPPRate\ within the \DSk\ energy ROI, giving \DSkROIPPUnit\ in the \DSkExposure\ exposure.  The \bg\ rejection power demonstrated with \DSf\ in its \AAr\ run~\cite{Agnes:2015gu} was better than one part in \DSfAArROIEventsNumber, more than sufficient to reject this \PP\ neutrino background in the \DSk\ exposure.  This would also hold true for the \ArgoExposure\ exposure planned for \Argo.

\item[\bf Neutrino-induced Nuclear Scatters:]
As discussed in Sec.~\ref{sec:BackgroundFree}, the coherent scattering of atmospheric neutrinos from argon nuclei is an as-yet-unobserved physics process as well as an irreducible physics background to \WIMP\ searches.  \DSkNuInducedBackgroundUnit\ are expected from this source in the \DSkExposure\ exposure.

\item[\bf \ce{^238U}, \ce{^232Th}, and Daughters:]
The radioactive noble element radon may enter and diffuse in the \LAr\ after being produced in the  \ce{^238U} and \ce{^232Th} decay chains.  The most important background contribution comes from $\beta$ decays of the radon daughters whose spectra fall partly within the energy ROI.  The \ce{^222Rn} specific activity in \DSf\ was measured and found to be below \DSfRnTwoTwoTwoSpecificActivityLimit. With its much smaller surface-to-volume ratio, the \ce{^222Rn} concentration in \DSk\ is expected to be much lower than in \DSf, since radon will be coming from the detector surfaces and not the bulk.  However, even if the \ce{^222Rn} contamination were at the \DSf\ level,  only \DSkROIBiTwoOneFourRate\ from \ce{^214Pb} $\beta$ decays would be expected in the energy ROI, giving \DSkROIBiTwoOneFourUnit\ in the total \DSkExposure\ exposure.  Once again the \bg\ rejection power demonstrated with \DSf\ in its \AAr\ run~\cite{Agnes:2015gu}, better than one part in \DSfAArROIEventsNumber, is more than sufficient to eliminate this source of background from the \WIMP\ search region at the level required for a \DSk\ result free from instrumental background.
\end{asparaenum}

\subsubsection{External background}

To ensure that external backgrounds are low enough to achieve the experiment goals, the \DS\ Collaboration will implement a careful program of background modeling, process and materials development, and materials radiopurity assay.  The starting point for the external background budget from radioactivity is the table of activities given in Table~\ref{tab:Materials-Activity}.

\begin{asparaenum}
\item[\bf Radiogenic \bgs:]
The energy spectrum resulting from the radioactive \bg\ decays of the detector materials was measured quite precisely in \DSf, for both the \AAr\ and \UAr\ data sets.  The measured spectra were found to match very well, indicating that the target itself did not contribute to this background in a significant way, and also showed a sufficient amount of self shielding by the \LAr.  With this results in mind, and assuming the same level of activities as in \DSf, \DSk\ will expect to see on the order of \num{E7} electron recoils events within the \WIMP\ ROI, coming from \bg\ decays of the detector materials.  Based on the discrimination power of the PSD technique to distinguish between ER and NR events that has been estimated, it is expected that this background will not cause any leakage events to appear in the \WIMP\ search region, over the entire exposure of \DSk.

\item[\bf Radiogenic neutrons:]
\label{subsec:VetoPerformance}
Radiogenic neutrons are a potentially dangerous background since they cannot be removed by \PSD.  They are produced by decays of radioisotopes in the materials of the detector through ($\alpha$,n) reactions and spontaneous fission, particularly from radioisotopes in the \ce{^238U} and \ce{^232Th} decay chains.  The very low levels of  spontaneous fission occurring in the materials selected for \DSk\ also produce neutrons, but \GFDS\ simulations with activity levels measured in \DSf\ or expected for \DSk\ show that these will be efficiently vetoed by the \LSV, given the fission neutron multiplicity and the coincident  fission \grs, leaving ($\alpha$,n) as the chief background.  

Early \GFDS\ studies indicated that the ($\alpha$,n) reactions in the Teflon (PTFE) reflector and the cryostat were likely to be the largest sources of neutrons, so further studies focused on these, testing several different configurations with \DSkSimulationGeantNeutronsPerChainDecayUTh\ neutrons generated for each configuration.  
The resulting probability for mis-identifying a radiogenic NR as a \WIMP\ candidate is then normalized to the \ce{^238U} and \ce{^232Th} activities reported in Table~\ref{tab:Materials-Activity} and the ($\alpha$,n) yields in these materials, calculated with \TALYS\ and \SRIM\ and reported in Table~\ref{tab:VetoNeutronSimulation}, to give the number of expected \WIMP-like NRs appearing during the full \DSkExposure\ of \DSk. The neutrons generated in the PTFE and steel have been simulated using ($\alpha$,n) spectra from TALYS. The TMB concentration has been varied from \SI{0}{\percent} to \LSVOldTMBConcentration\ to find the most efficient \LSV\ composition.  Fig.~\ref{fig:Vetoes-NeutronsSpectrum} shows the simulated quenched neutron energy deposited in the \LSV, including both proton recoil and neutron capture and assuming the same alpha quenching observed in \DSf.

\begin{table}
\rowcolors{3}{gray!35}{}
\normalsize
\centering
\caption[Uranium and thorium decay chain ($\alpha$,n) yields, per equilibrium decay of the entire chain.]{Uranium and thorium decay chain ($\alpha$,n) yields, per equilibrium decay of the entire chain, used in neutron background estimates.  The yields were calculated with \TALYS\ and \SRIM.  Equilibrium in the \ce{^238U} chain is often broken, and so it is divided in the simulations.}
\begin{tabular}{lccc}
			&\ce{^238U} Upper	&\ce{^238U} Lower	&\ce{^232Th}\\
\hline
Steel			&\num{1.3E-9}		&\num{5.4E-7}		&\num{1.95E-6}\\
Teflon (\PTFE)	&\num{1.2E-5}		&\num{8.9E-5}		&\num{1.3E-4}\\
\end{tabular}
\label{tab:VetoNeutronSimulation}
\end{table}

\begin{figure}[t!]
\includegraphics[width=\columnwidth]{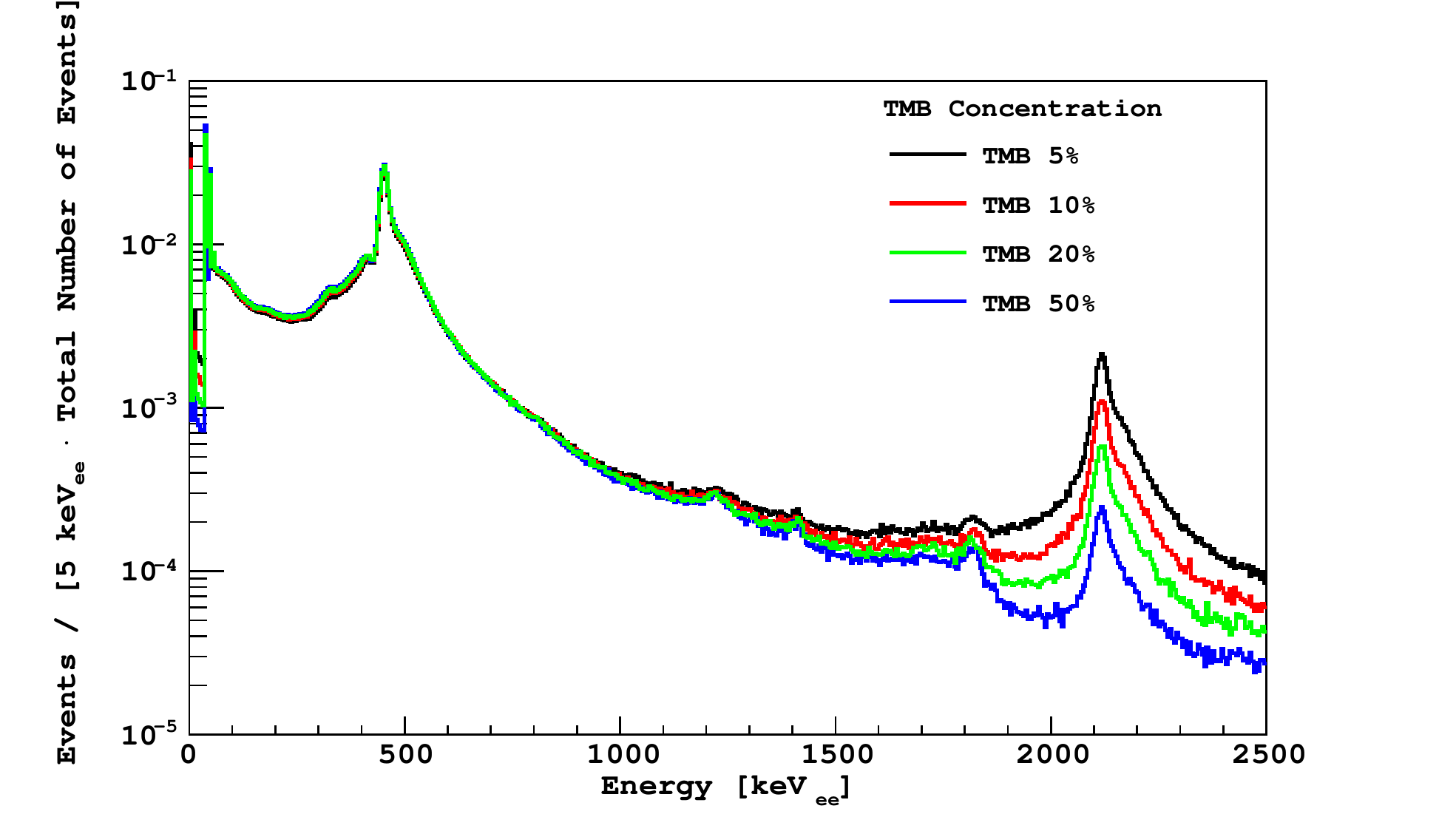}
\caption[Simulated neutron energy spectrum in the \LSV.]{Simulated neutron energy spectrum in the \LSV.  The first two peaks are the capture on \ce{^10B}, and the third, smaller, peak is the capture on hydrogen. The spectrum takes into account quenching, and it has been simulated for different TMB concentrations. }
\label{fig:Vetoes-NeutronsSpectrum}
\end{figure} 

To use the simulation to evaluate \LSV\ performance, a simulation of a typical \TPC\ analysis, which would require a single, nuclear-recoil-like scatter in a chosen fiducial volume, is performed.  These cuts are themselves quite effective in reducing neutron-induced background.  The following is required:

\begin{compactitem}
\item The event has a single resolved interaction in the \TPC;
\item The interaction is nuclear-recoil-like;
\item The event is contained in the WIMP energy region;
\item The event is inside the fiducial volume chosen, varied in these studies.
\end{compactitem}

The radial profile of the neutron captures in simulated events passing the \TPC\ selection shows little dependence on the TMB concentration and indicates that all neutrons capture well before reaching the \LSV\ \PMTs. While the capture location is independent of TMB concentration, the capture time depends on the TMB concentration, as can be seen in Fig.~\ref{fig:Vetoes-NeutronsTimeProfile}.

\begin{figure}[t!]
\includegraphics[width=\columnwidth]{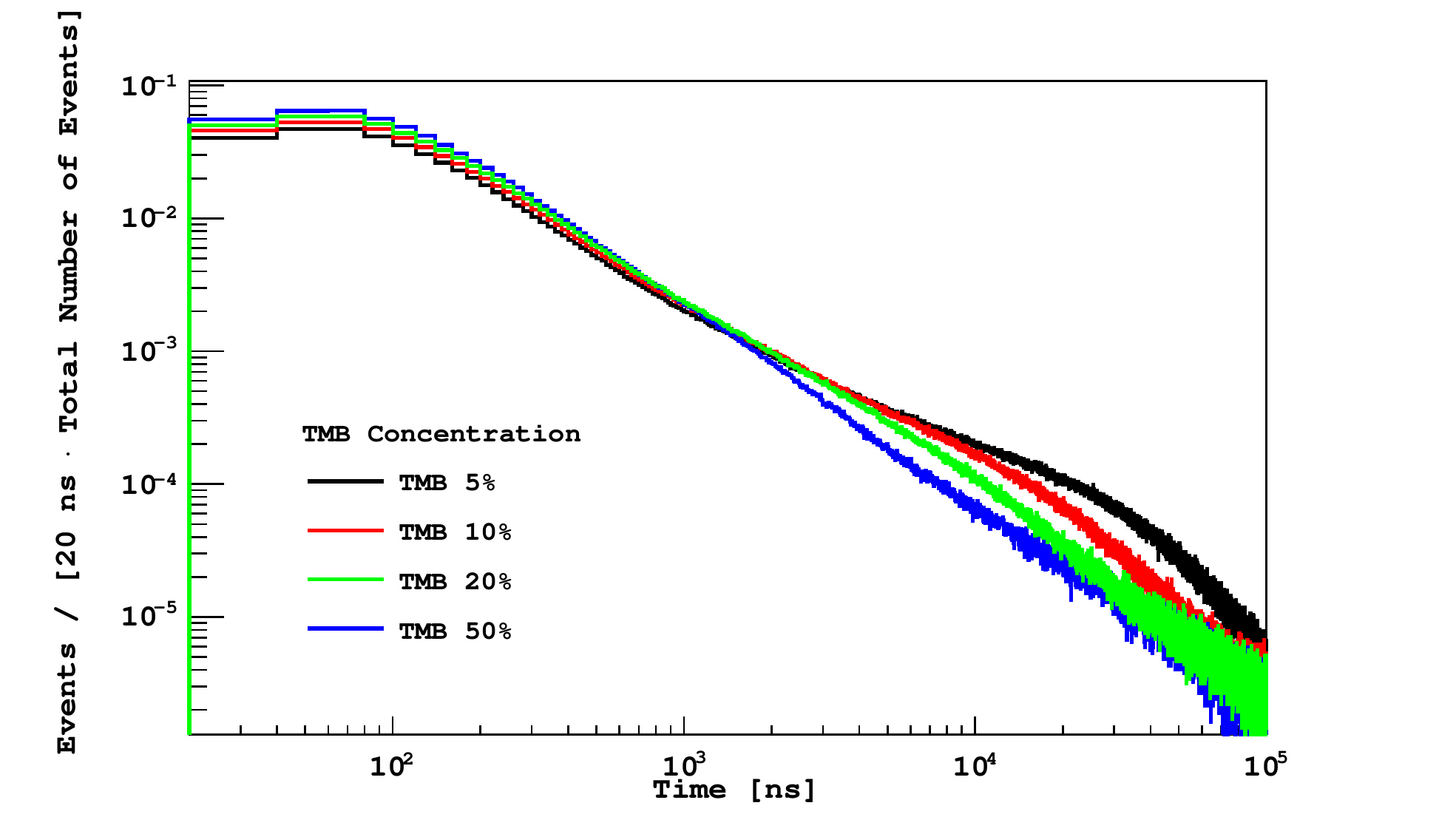}
\caption[Capture time of neutrons giving ``\WIMP-like" \TPC\ events.]{Capture time of neutrons giving ``\WIMP-like" \TPC\ events, for neutrons produced in \ce{^238U} decay chain ($\alpha$,n) interactions in the \TPC\ PTFE reflector panels.}
\label{fig:Vetoes-NeutronsTimeProfile}
\end{figure}  

If the total \LSV\  rate is low enough, the cuts to tag the prompt and delayed neutron signals can have low energy thresholds and an acceptable accidental loss. Adding more TMB shortens the capture time and thus the \LSV\ delayed window, lowering the thresholds at a fixed accidental rate or alternatively reducing the accidental rate with a fixed threshold. In this evaluation of neutron background it is assumed that \LSV\ thresholds can be low enough to not affect the neutron tagging efficiency.  In the simulation, it is assumed that events with quenched energy in the \LSV\ \DSkSimulationGeantLSVQuenchedEnergyCut\ pass the veto cuts.

The simulations for neutrons originating in the Teflon reflector panels and stainless steel cryostat give the results shown in Table~\ref{tab:neutronBackground} for a TMB concentration of \DSkLSVTMBFraction\ and a fiducial cut of \DSkFiducialCut.  Fig.~\ref{fig:Vetoes-NeutronsNumber} shows the surviving neutron background as a function of both the TMB concentration and the fiducial cut.

\begin{table*}
\rowcolors{3}{gray!35}{}
\normalsize
\centering
\caption[Expected neutron background.]{Expected neutrons produced before the application of any cuts, fraction passing \TPC\ cuts, fraction of events passing \TPC\ cuts that also pass \LSV\ cuts, and surviving neutron background events inside the \WIMP\ search region for a \DSk\ exposure of \DSkExposure.  Errors are statistical only.} 
\begin{tabular}{lcccc}
Source					&\minitab{c}{Neutrons produced}{in \DSkRunTimePlanned}
									&\minitab{c}{Fraction passing}{\TPC\ cuts} 
												&\minitab{c}{Fraction passing}{\LSV\ cuts}
															&\minitab{c}{Surviving n}{background in \DSkExposure}\\
\hline
Teflon reflector panels		&\num{<1717}	&\num{0.0072}	&\num{0.0075}	&\num{<0.093}\\
Stainless steel cryostat		&\num{2384}	&\num{0.0019}	&\num{0.0133}	&\num{0.060}\\
\hline
Total						&			&			&			&\num{<0.153}\\
\end{tabular}
\label{tab:neutronBackground}
\end{table*}

\begin{figure}[t!]
\includegraphics[width=\columnwidth]{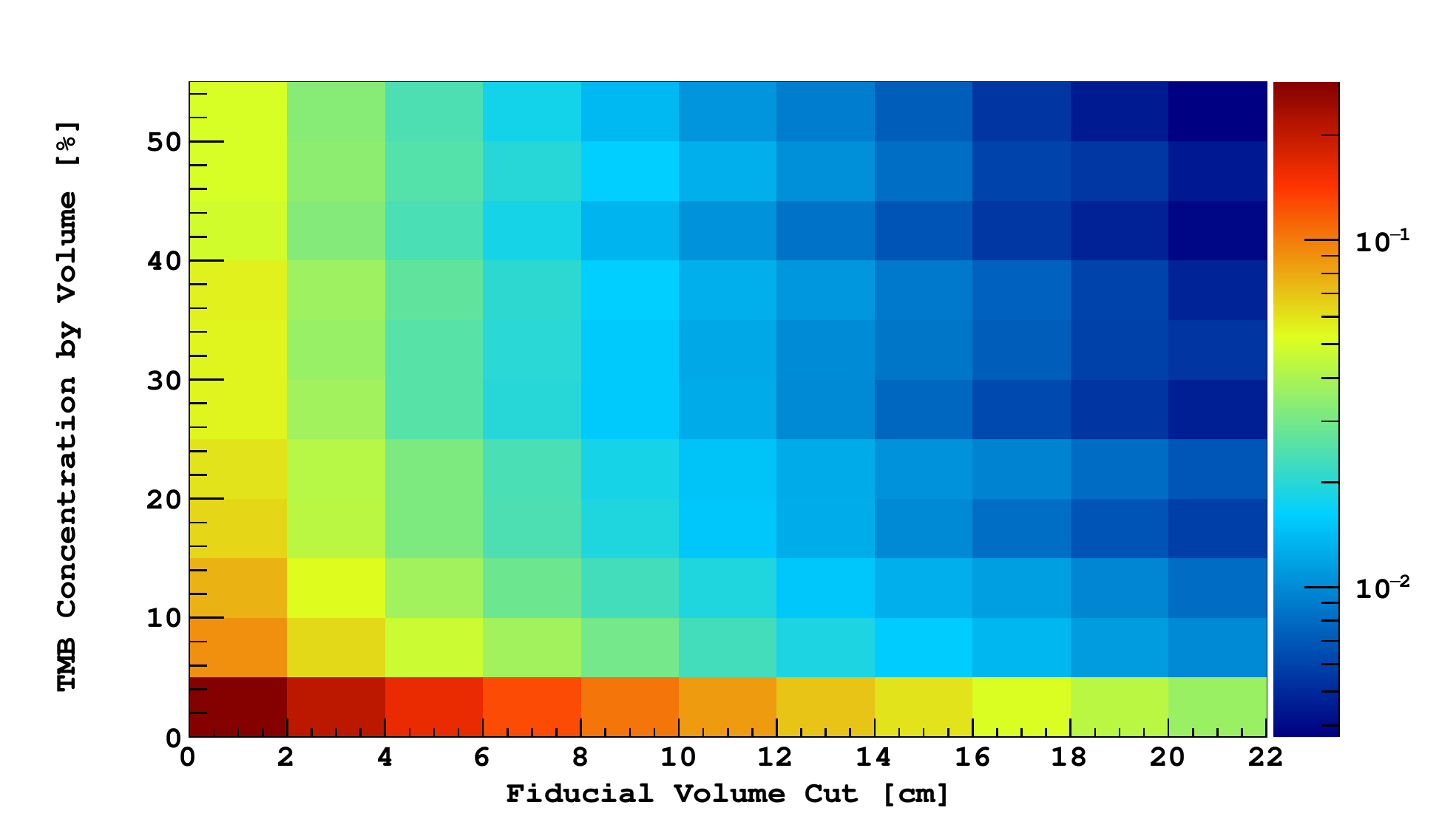}
\caption[WIMP-like neutron recoils induced by ($\alpha$,n) reactions.]{\WIMP-like neutron recoils per year from both \ce{U} and \ce{Th} induced by ($\alpha$,n) reactions in the stainless steel cryostat and in the PTFE panels.}
\label{fig:Vetoes-NeutronsNumber}
\end{figure}

\item[\bf Surface backgrounds:]
$\alpha$-decays on the surfaces of the active volume can also be a difficult background. About half the time, the $\alpha$ goes deeper into the surface, and the daughter nucleus recoils into the active volume, mimicking a \WIMP-induced nuclear recoil.  Surface contamination will be minimized by carrying out precision cleaning and assembly of the \TPC\ in the Rn-suppressed clean rooms already constructed and used for \DSf. The inner surfaces of the \TPC\ are coated with a wavelength shifter, which will be applied by vacuum evaporation carried out within an extension of the radon-suppressed clean rooms. Thus the coated surfaces will only be exposed to air that has been scrubbed of radon. \DSf\ studied the light yield of alpha interaction in \LAr~\cite{Agnes:2017cl}, but in the end these events were all removed from the final data sample using simple cuts, not including a radial fiducial volume cut.  As specified in Sec.~\ref{sec:BackgroundFree}, recoils from any remaining surface activity will be sufficiently suppressed by position reconstruction and the \DSkFiducialCut\ fiducial volume cut, along with advanced cuts looking at \SOne\ and \STwo\ light distributions and/or a long-lived component of the signal coming from nuclear recoil interactions with the TPB coated PTFE panels of the \TPC.

\item[\bf Cherenkov background:]
Cherenkov light  is produced by \bg\ interactions in materials surrounding the \LAr\ such as the PTFE reflector or the acrylic windows of the \TPC.  Overlaps of this prompt light in time with an electron recoil event in the argon, for example from a gamma with multiple Compton scatters, can appear to have a larger fast \SOne\ component, resembling that of a nuclear recoil.  Such events are rare given the use of ultra-pure materials, and are further suppressed by analysis cuts on the spatial distribution of the detected \SOne\ light, which the Monte Carlo indicates to be very effective, and additionally by the fiducial cut.
\end{asparaenum}

\subsubsection{Cosmogenic background}

Nuclear interactions of cosmic ray muons with the rock surrounding the experimental halls produce neutrons and other particles that potentially give backgrounds in \DS-type detectors.  Detailed simulation studies of these backgrounds were performed by members of the \DS\ Collaboration for Borexino, for \DSf\ with its veto system, and for a \DSkSimulationFLUKAMass\ \LArTPC\  with the \DSf\ veto system, as described in Refs.~\cite{Bellini:2013kr,Empl:2014ih,Empl:2014wa}.  In these simulations, \FLUKA\ was used to predict the particle fluxes produced in the rock and incident on the experiments,  starting from measured characteristics of the incident muon flux at LNGS.  

The most dangerous class of background events for the \WIMP\ search are neutron-induced nuclear recoils in the sensitive \LAr\ volume resulting from events which deposit little or no energy in the vetoes. To study this category, Ref.~\cite{Empl:2014ih} singled out all events for which at least one neutron (but $<50$ coincident particles) reached the active \LAr\ region in their \DSkSimulationFLUKAMass\ \LArTPC.  A very conservative criterion (dE$_{LSV}>1$~MeV) was defined to select events which should be considered detectable by the \LSV.  

A portion of the Ref.~\cite{Empl:2014ih} simulation was repeated for the proposed geometry using the \Geant-based simulation package.  The most dangerous category of cosmogenic events was again singled out, {\it i.e.}, cosmogenic neutrons entering the water tank without accompanying charged particles (all other events are easily rejected by the \WCV).  \DSk\ is more efficient at rejecting such external cosmogenic neutrons because of the increased sizes of the \WCV\ and \LSV.  The results for an exposure greater than \DSkSimulationGeantExposure\ are compatible with no cosmogenic neutron reaching the \LArTPC\ and surviving application of the veto cuts.

The background estimate described above was based on cosmic ray muon-induced secondaries which are prompt and thus can be vetoed by the outer detector systems.  The cosmic ray showers can also produce radioisotopes with lifetimes of seconds to minutes ({\it i.e.}, sources of non-prompt background) in the detector and target materials.  Production and decay of these isotopes were also simulated, and their decay products handed off to a \Geant-based Monte Carlo for simulation.  Events that deposited a significant amount of energy in the veto detectors, interacted at more than one site inside the \LArTPC, or fell outside the energy ROI and the fiducial volume were removed.  On the order of \num{3E5} events survive during a full exposure of \DSkExposure, all of which were \bg\ events.  This number is easily managed by the \LAr\ \PSD. 

\vspace{6mm}

Based on current data and Monte Carlo simulations, the expected numbers of electron and nuclear recoil background events surviving analysis cuts, for the full \DSkExposure\ exposure, is given in Table~\ref{tab:Background}.  Note that the Table gives an estimated External NR background greater than 0.1~event.  This rate is dominated by ($\alpha$,n) production in the PTFE reflector panels assuming the activities in Table~\ref{tab:Materials-Activity}.  Actual measurements of the activity of the PTFE in \DSf\ produced only upper limits.  The background estimate can thus be taken as an indication that more sensitive measurements of the PTFE must be made, and that the activity will be required to be at least a factor of $\sim 2$ below the achieved upper limits.  A campaign to improve the sensitivity  is underway (see Sec.~\ref{sec:MatAssay-RadiopurityBudget}).

\subsection{\DSk\ sensitivity to WIMPs}
\label{sec:Sensitivity}

\begin{figure*}
\centering
\includegraphics[width=0.49\columnwidth]{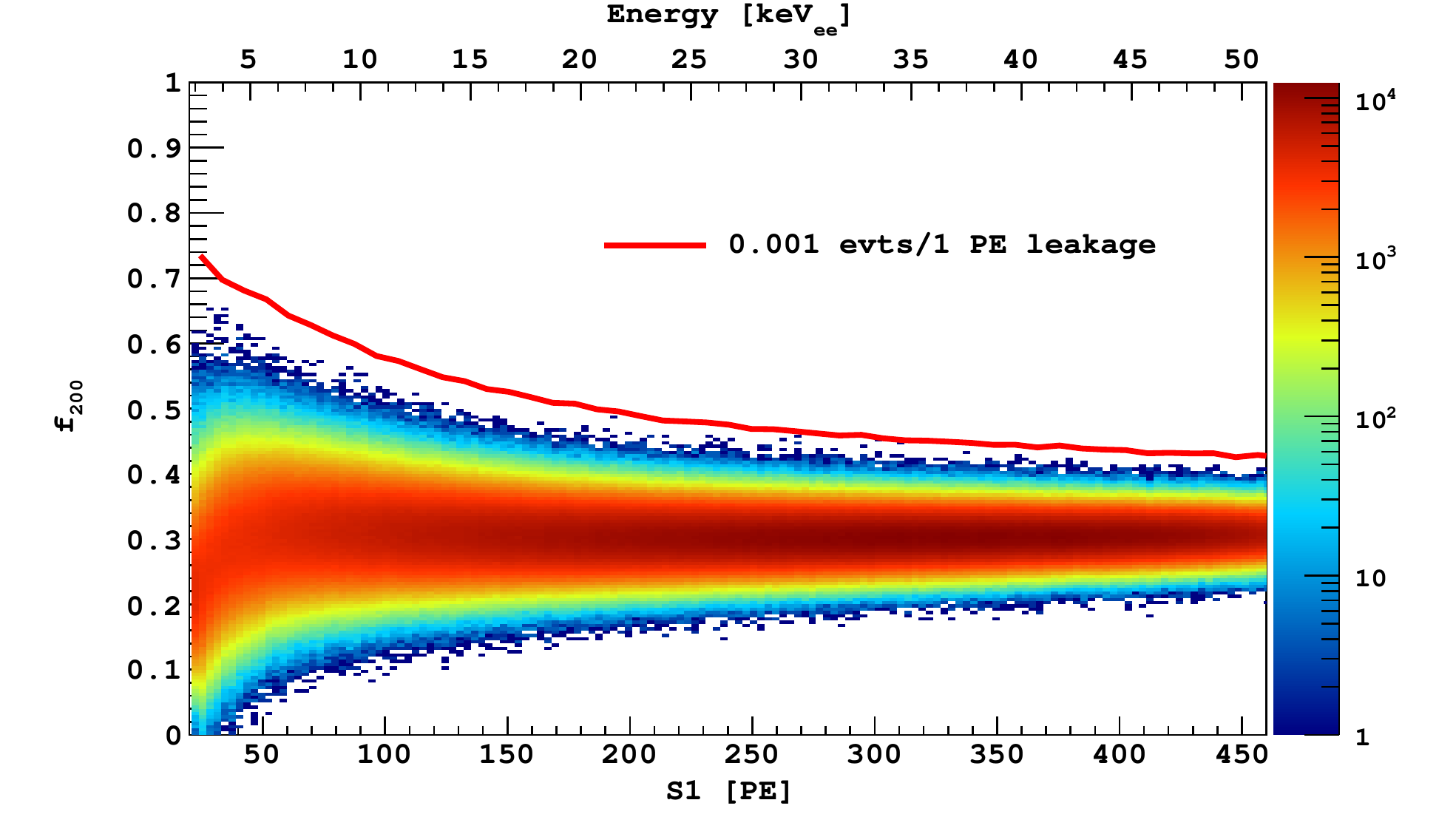}
\includegraphics[width=0.49\columnwidth]{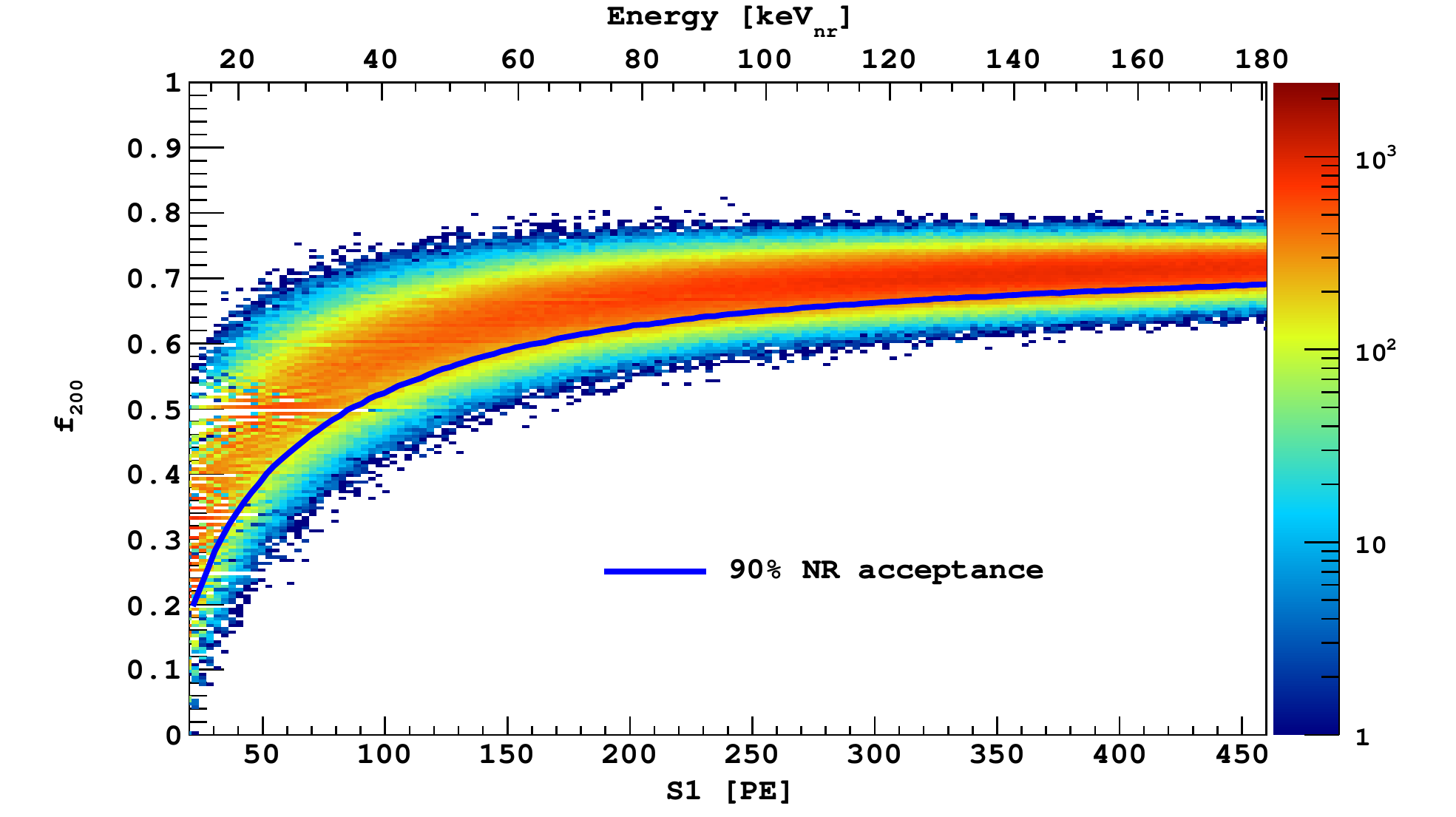}
\caption[Distribution of \FTwoZero\ as a function of \SOne\ for \ce{^39Ar} betas and \NRs.]{{\bf Left:} Distribution of \FTwoZero\ as a function of \SOne\ for \ce{^39Ar} \btas.  The red line on the left plot shows the leakage curve for \btas\ generated from the requirement of \DSkDMSArThreeNineBackgroundCondition.  {\bf Right:} Distribution of \FTwoZero\ as a function of \SOne\ for \NRs.  The blue curve delimits the \DSkDMSArThreeNinePSDAcceptanceMax\ \NR\ acceptance region.  The \WIMP\ search region is the region above both curves, red and blue.  Note: for both plots, the energy scale atop the plots is only approximate.}
\label{fig:G4DS-DSkFTwoZero}
\end{figure*}

The success of the experiment depends not only on background suppression, but also on a high detection efficiency for nuclear recoil (\NR) events from \WIMPs.  The \WIMP\ acceptance region is designed to have a leakage of electron recoils smaller than \DSkDMSArThreeNineBackgroundCondition.

To determine the expected acceptance region for \DSk, \DSkSimulationPrimaries\ \NRs\ in the range \DSkSimulationEnergyRange\ were simulated.  The simulation included the effect of the measured quenching factor for nuclear recoils in argon reported in Ref.~\cite{Cao:2015ks}, which matches the theoretical prediction of Ref.~\cite{Mei:2008ca}.  With this quenching factor, the above \NR\ \ROI\ corresponds to the energy interval \DSkROIElectronEquivalentEnergyRangeSim\ (in \si{\pe}) for \btas.  Additionally, \DSkROISimulatedBetaBare\ \ce{^39Ar} \bta-decays were simulated in the energy range of \DSkROIElectronEquivalentEnergyRange.  The number of simulated \ER\ corresponds to \DSkROISimulatedFractionBeta\ the expected number of \ce{^39Ar} decays in the \WIMP\ \ROI\ for the \DSkExposure\ exposure using \UAr\ (assuming the \ce{^39Ar} reduction factor measured in \DSf).  The \GFDS\ simulation was performed using the optical parameters, such as the index of refraction and the reflectivity of the different surfaces, that had been tuned with the \DSf\ data.  The light yields of \btas\ and \NRs\ were corrected for the position within the \LArTPC, based on a light yield map obtained with \ce{^{83m}Kr} \GFDS\ simulations.  The average light yield predicted for \DSk\ in this study was \DSkNullFieldLightYieldProjected\ for \bgs\ at null field, corresponding to \DSkWithFieldLightYieldProjected\ at a field of \DSkDriftField.

\begin{figure}
\includegraphics[width=\columnwidth]{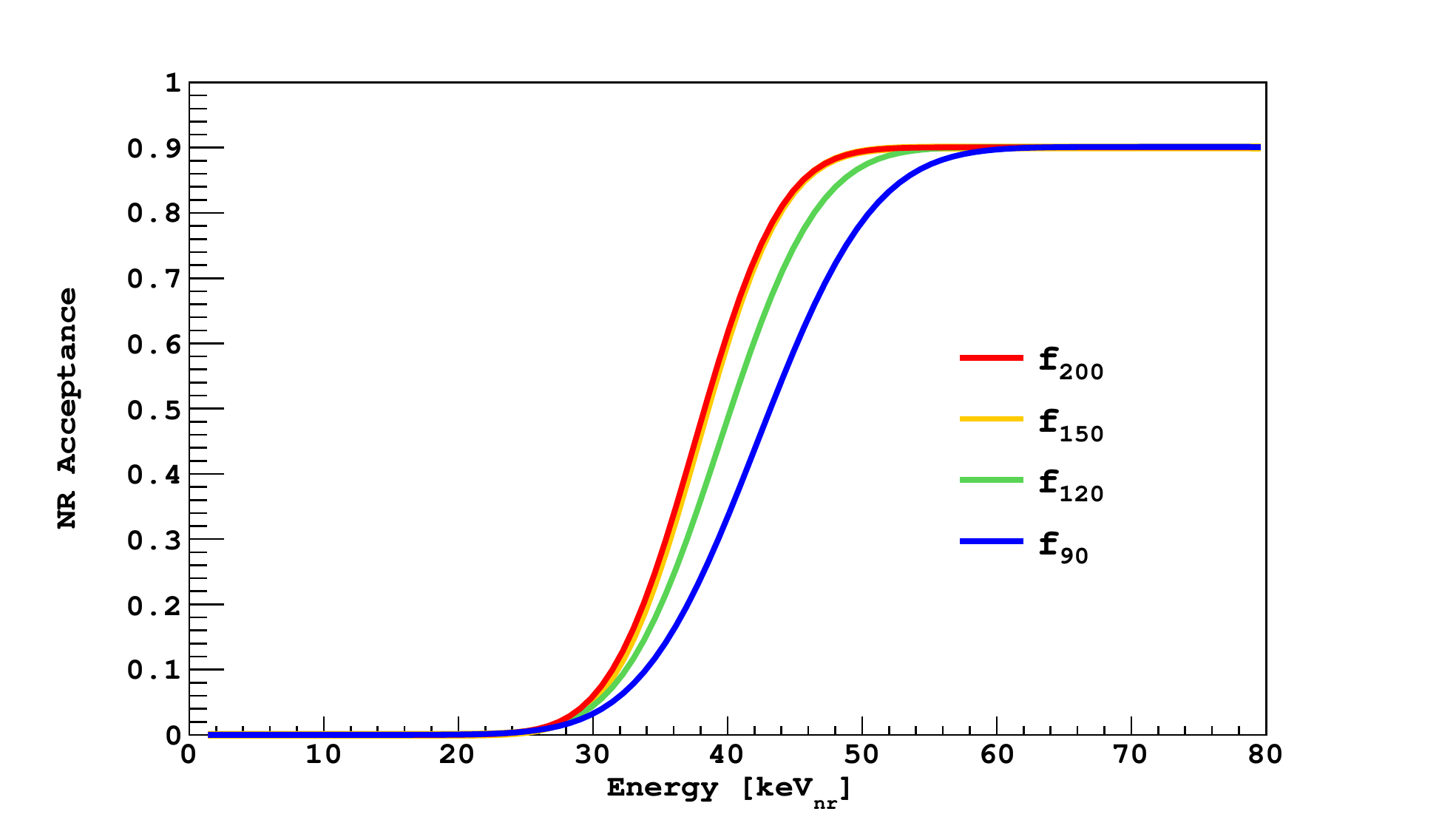}
\caption[\NR\ \FNine, \FOneFive, and \FTwoZero comparisons for \DSk.]{Comparison of nuclear recoil acceptance bands for \FNine, \FOneFive, and \FTwoZero.}
\label{fig:G4DS-NRAcceptance}
\end{figure}

The discrimination between \bg\ and \NR\ was then determined using the \PSD\ of \LAr.  In \DSf, the discrimination parameter \FNine\ was used, but due to the \DSk\ larger size and longer average distance the light has to travel before reaching the photosensors, \FNine\ might not be the optimal distribution, thus the \PSD\ performance of \DSk\ was simulated using four different parameters: \FNine, \FOneTwo, \FOneFive, and \FTwoZero, defined as the fraction of scintillation light collected in the first \WindowFNine, \WindowFOneTwo, \WindowFOneFive, and \WindowFTwoZero\ respectively.

The large simulated sample of \ER\ described above is still not sufficient to locate a \PSD\ cut contour for the full exposure.  To design the contour, an analytical model (to be described in a future paper) of the three \PSD\ parameters was fit to the simulated sample and used to extrapolate to the required background rejection.  For all three parameters, an \NR\ acceptance region was defined by requiring that it contain a leakage of less than \DSkDMSArThreeNineBackgroundCondition\ \ER\ events (total ER background then sums to \BackgroundFreeRequirement\ in the \WIMP\ search region), extrapolating to the full statistics of \ce{^39Ar} events expected in \DSk.  Note that in this study, the use of \UAr\ was assumed, with the depletion factor measured in \DSf\ and {\it not} the additional depletion anticipated from Aria.  The acceptance region was further bounded from below by the \DSkDMSArThreeNinePSDAcceptanceMax\ \NR\ detection-efficiency contour.  As an example, the distribution of \FTwoZero\ as a function of \SOne\ for \btas\ and \NRs\ is shown in Fig.~\ref{fig:G4DS-DSkFTwoZero}.  The lines on those plots correspond to the leakage curve for \bg\ and to the 90\% acceptance contour for \NR.

For each of the four discrimination parameters considered, the \NR\ acceptance band, as determined with the procedure detailed above, is converted into a function of energy and shown in Fig.~\ref{fig:G4DS-NRAcceptance}.  The comparison clearly favors \FTwoZero.  There is no gain in further lengthening the window dedicated to the counting of the fast photoelectrons beyond the \WindowFTwoZero\ of \FTwoZero.  The resulting equivalent reduction factor for \ce{^39Ar} decays inside \DSk\ is found to be \DSkUArPSDRejectionFromGFourDS, more than sufficient to maintain background-free operation for more than \DSkExtendedExposure.

The acceptance band of Fig.~\ref{fig:G4DS-NRAcceptance} is used to evaluate the sensitivity of \DSk\ to \WIMP\ recoils.  With the best performing parameter, \FTwoZero, a projected sensitivity of \DSkSensitivityOneGeVUnit\ for a WIMP mass of \WIMPMassOneTev\ (\DSkSensitivityTenGeVUnit\ for \WIMPMassTenTev) is obtained, as shown in Fig.~\ref{fig:ArgoDSkDSf-ProjectedExclusion}.  \DSk\ could also extend its operation to \DSkExtendedRunTimePlanned, increasing the exposure to \DSkExtendedExposure, reaching a sensitivity of \DSkExtendedSensitivityOneGeVUnit\ (\DSkExtendedSensitivityTenGeVUnit) for \WIMPs\ of \WIMPMassOneTev\ (\WIMPMassTenTev) mass.  As shown in Table~\ref{tab:Introduction-Sensitivity}, \DSk\ will be capable of achieving the highest sensitivity for high mass dark matter particles of any experiment currently ``on the books'', extending just beyond the reach of the two-phase liquid xenon \TPC\ experiments.

\section{Facilities and Detector Assembly}
\label{sec:Facilities}

\subsection{Introduction}
\label{sec:Facilities-Introduction}

The \DSk\ facility incorporates all plants, systems and subsystems involved in preparation, construction, assembly, commissioning and operations of the experiment.  The major components are represented by the three detectors: 1) the \WCV, 2) the \LSV, and 3) the \LArTPC\ contained inside the cryostat, togerther with their readout electronics and calibration systems. The clean rooms, radon abatement system and the fluid handling plants and storage areas are considered essential facilities of the experiment.  The fluid handling and storage plants will allow for the storage, recovery, purificartion and handling of liquids such as \LAr, \LIN, scintillator and water.  

The concentric design of the detector was successfully implemented in \DSf.  For \DSf, stainless steel-lined clean rooms were built and equipped with an active radon abatement system to reduce the radon content in the air to the record level of \RadonCRHLimit, a factor \RadonCRHSuppression\  below the activity in the air of the Hall~C.  For \DSk, the plan is to use the floor of the water tank as a clean room assembly space.  The flow of radon-suppressed air into the water tank will be provided by the radon abatement system of the \DSf\ clean rooms.


\subsection{The \LNGS\ Underground Laboratory: Hall~C}
\label{sec:Facilities-LNGSUndergroundLab}

The Gran Sasso Underground Lab consists of three large experimental halls: Hall~A, Hall~B, and Hall~C, and a series of small interconnecting tunnels and service galleries.  The Laboratory is equipped with central services for distribution of fresh air, power, cooling water, and compressed air.  Hall~C currently hosts experiments such as \BX, \DSf\, and the MiBeta test facility and has been identified by \LNGS\ management as the candidate site to host the \DSk\ experiment.  Hall~C is \LNGSHallCLength\ long, \LNGSHallCWidthHeight\ wide, and has a vaulted ceiling with \LNGSHallCWidthHeight\ height.  

Hall~C is equipped with three cranes: the double-hook (vaulted type) \LNGSHallCDoubleHookCraneLoadBare+\LNGSHallCDoubleHookCraneLoad\ crane; the \LNGSHallCLargeCraneLoad\ crane and the \LNGSHallCSmallCraneLoad\ crane.  The perimeter of the hall is supplied with fluids spill and smoke detection systems.  A volatile organic carbon detector with multiplexed sample points and a fixed, foam-based fire extinguisher system are also installed in the hall.

The floor of the Hall~C is waterproofed with a curb on its perimeter for mitigating any possible flooding due to the release of fluids from the experiments.  A metal grid placed on the floor at the south side of Hall~C allows a collection up to \LNGSHallCLiquidCollectionRate\ of liquid into a containment pit placed beneath the pavement.  The ventilation system provides at least \LNGSHallCFreshAirRate\ of fresh air and keeps Hall~C overpressure at a few \si{\milli\bar}.  Fresh air is delivered on the south side and removed from the north side.  All entrances to Hall~C (on north, south and east sides) are equipped with fire-proof stainless steel doors.

\begin{figure*}[t!]
\centering
\includegraphics[width=0.75\textwidth]{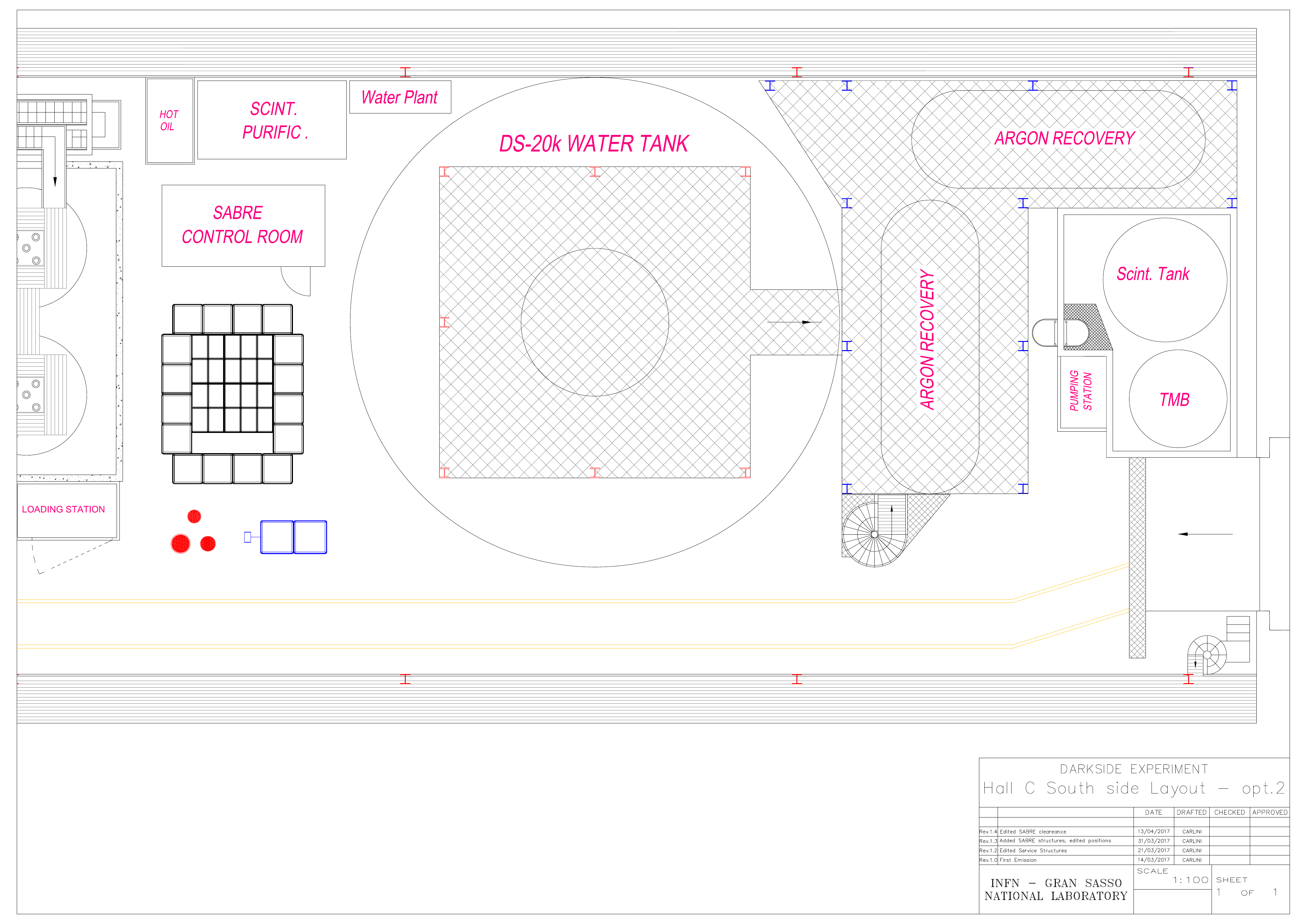}
\caption{Possible layout of the south part of Hall~C for the \DSk\ installation.}
\label{fig:Facilities-DSk2DLayout}
\end{figure*}

\begin{figure}[t!]
\centering
\includegraphics[width=0.75\columnwidth]{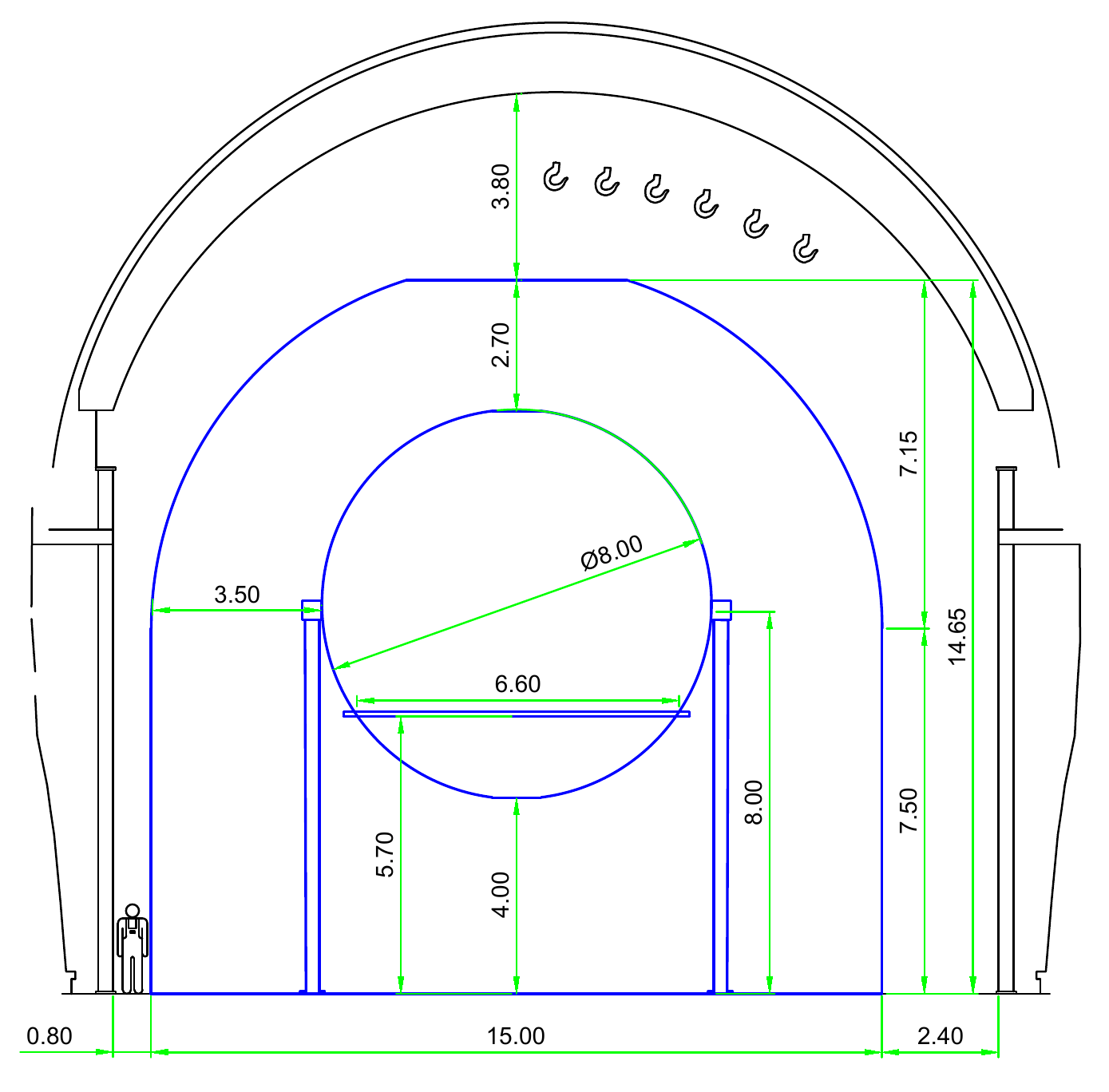}
\caption{Cross section of \DSk\ with dimensions of \WCV\ and \LSV.  All dimensions in \si{\meter}.}
\label{fig:Facilities-DSk2DVetoesDimensions}
\end{figure}

The likely location of \DSk\ will be in the south part of the Hall~C, where the OPERA experiment was placed before its decommissioning.  This area will also host the SABRE experiment.  This location would be optimal from a logistics standpoint, as it grants direct access to the TIR tunnel through the south entrance of Hall~C, equipped with a double-containment fire-proof door that allows for truck access.

\begin{figure*}[t!]
\centering
\includegraphics[width=0.75\textwidth]{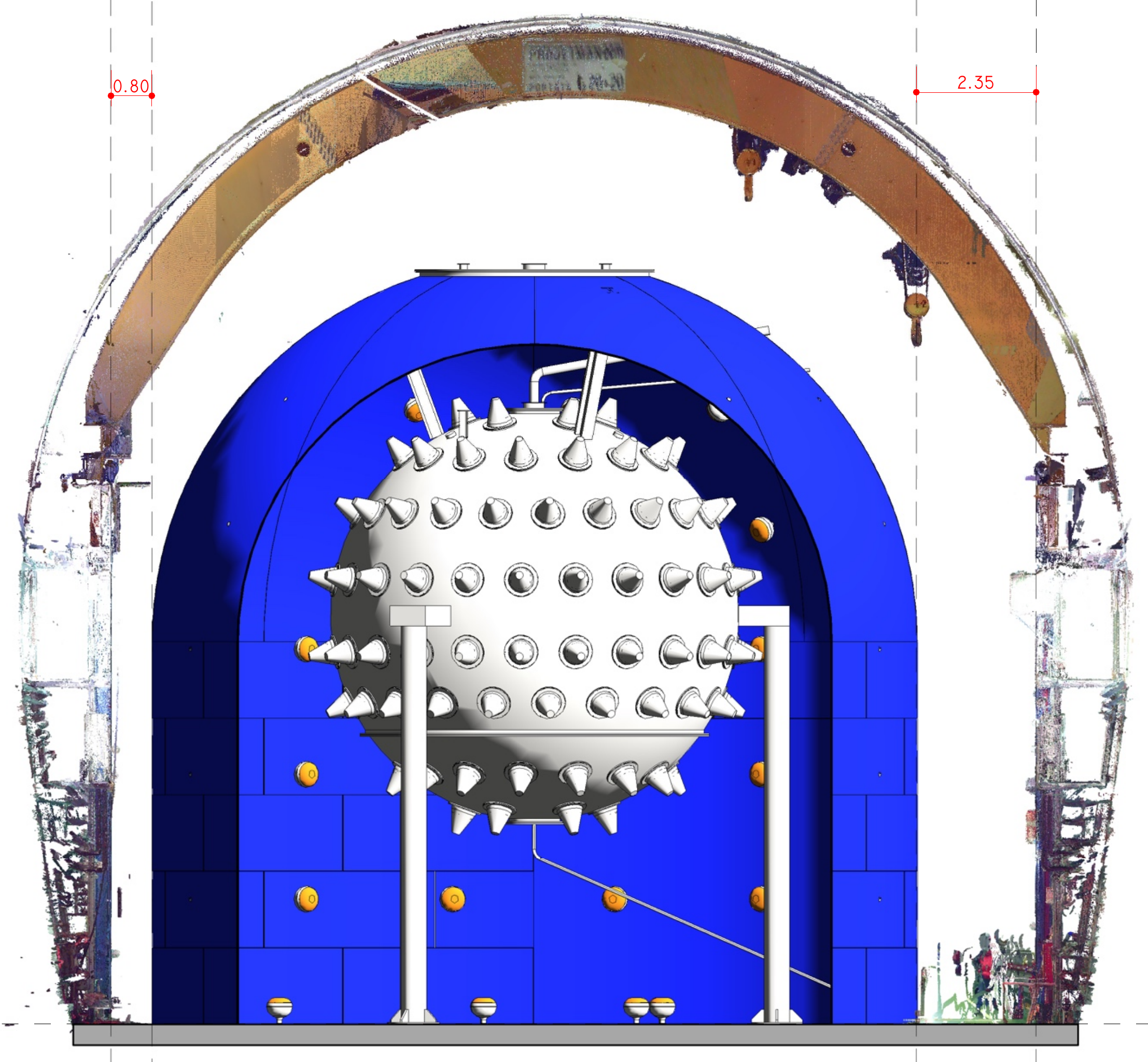}
\caption[3D view of the \DSk\ veto detectors.]{3D view of the \DSk\ veto detectors, showing the water tank and the \WCV\ detector, along with the stainless steel sphere and \LSV\ \PMTs.}
\label{fig:Facilities-DSkVetoSection}
\end{figure*}

One preliminary layout of the south part of the Hall~C is shown in Fig.~\ref{fig:Facilities-DSk2DLayout}.  The exact layout is still under discussion with the LNGS management, but as can be seen from the figure, the site will host the water tank (\DSkWCVDiameter\ diameter and \DSkWCVTotalHeight\ tall), two argon recovery tanks (\SI{9}{\meter} long and \SI{3.2}{\meter} in diameter) and the building for the cryogenic system (cooling tower) and all \DAQ\ electronics (\LArTPC\ plus two veto detectors).  A cross section of the \DSk\ detector with dimensions is shown in Fig.~\ref{fig:Facilities-DSk2DVetoesDimensions}.  Fig.~\ref{fig:Facilities-DSkVetoSection} shows a 3D rendering of the two veto detectors (\WCV\ and \LSV). 

The water tank will have a \DSkWCVManHoleSize\ manhole as a main entrance, to provide maximal access.  The bottom of the tank will be equipped with a \SI{8}{\centi\meter} thick stainless steel slab in order to provide additional shiedling against gamme-rays from the Hall~C floor.  The stainless steel sphere will have a \DSkLSVTopFlangeDiameter\ diameter flange at the top for the process tubing, and a \DSkLSVBottomFlangeDiameter\ diameter flange to provide access for the insertion of the \LArTPC\ (see Fig.~\ref{fig:Facilities-DSk2DVetoesDimensions} for dimensions).  At the bottom of this \DSkLSVBottomFlangeDiameter\ flange, a \DSkLSVRapidFlangeDiameter\ diameter flange will be added for the feedthroughs of the liquid cryogenic and scintillator lines.  

The design of the cryostat follows closely that of the successful \DSf\ cryostat: a $4\pi$ vacuum-insulated vessel made of three separate parts, the top assembly (two concentric caps), the inner cryostat vessel, and the vacuum insulation vessel (external sheel).  The baseline material for the cryostat is stainless steel, however, the collaboration is actively pursuing building the cryostat from ULR titanium, more details on the cryostat design and the ULR titanium can be found in Sec.~\ref{sec:Cryogenics-Cryostat}.

The cryostat supporting and leveling system will mirror the successful experience of \DSf: the cryostat will be suspended from a system of three rods, going all the way up from the cryostat to the top of the water tank.  Leveling of the cryostat will be adjusted by turning the rods, with the precision of a fraction of a \si{\milli\meter} over the diameter of the \LArTPC.  A study of the response of the system to seismic events is ongoing together with careful analysis of the mechanical design optimization in order to guarantee full compliance with the anti-seismic regulations.

The \DS\ detectors require extraordinarily low background levels.  

\begin{figure*}[t!]
\centering
\includegraphics[width=\textwidth]{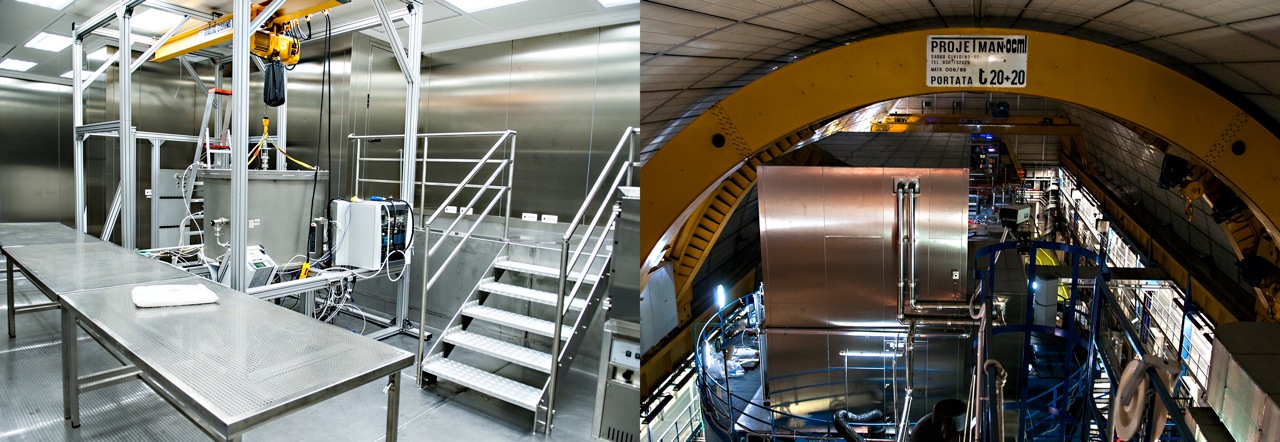}
\caption[The \DSf\ radon suppressed clean-rooms in Hall~C.]{The \DSf\ radon suppressed clean-rooms in Hall~C .  {\bf Left:}\label{fig:g2-facilities-cr} The interior of the CR1 clean room, with the UHV evaporator for \TPB\ coating of detectors' internals visible at left of center.  {\bf Right:} The exterior of the assembly clean room CRH on the top of the \CTF\ water tank, housing \DSf.}
\label{fig:Facilities-RadonSuppressedCleanRooms}
\end{figure*}

Thanks to essential support from the NSF, DOE and INFN, the \BX\ and \DS\ Collaborations were able to build a set of facilities that allowed the development and operation of such low background detectors underground.  These facilities playes a crucial role in the scientific success of \DSf.  Below is the list of the facilities which the \DSk\ experiment intends to use:
\begin{asparaenum}

\item[\bf High Purity Water Plant:] The \BX\ water purification plant is intended to be used to fill the \SI{2000}{\tonne} water tank. The plant consists of a special low pressure degassing system, a cascade of reverse osmosis columns, a deionizer, a set of ultra-Q filters, and a radon stripping column.  It is capable of producing ultra-high purity water at the rate of \SI{1}{\cubic\meter\per\hour}.  

\item[\bf Cleaning Module:] The cleaning module is a dedicated plant used to perform pickling, passivation, and precision cleaning, by recirculating heated acid and chelating solutions through the components to be cleaned.

\item[\bf Scintillator Storage:] The \BX\ scintillator storage consists of four \SI{100}{\cubic\meter\per\hour} vessels formerly used for storage of the \DS\ scintillator.  Two empty tanks will be enough to store the scintillator for the \LSV.

\item[\bf Purification Plants:] Over the span of 20 years, the \BX\ collaboration has developed a very sophisticated set of process-plants for scintillator purification.  In order to prevent possible contamination of these delicate \BX\ systems, the \DS\ collaboration procured an independent set of small purification units (distillation, \ce{N_2} stripping) capable of treating  the scintillator (or its individual components) at a rate of \SI{0.1}{\cubic\meter\per\hour}.  The plan is to keep using the same unit for scintillator purification, with the possibility for upgrading the present system.  

\item[\bf Buildings:] The \BX\ Big Building West (BBW) houses the \DS\ control room and the counting room for the muon and neutron veto.  The \BX\ Big Building East (BBE) houses the \DS\ scintillator purification units, the \DS\ scintillator buffer tank, and the interconnection system between the \DS\ scintillator loop and the \BX\ scintillator purification plants.

\item[\bf Radon Abatement System:] The \DS\ collaboration has installed and commissioned an air treatment unit which processes Hall~C air at a rate of \SI{230}{\cubic\meter\per\hour}, reducing radon by a factor greater than \RadonFilterSuppression.  This radon-suppressed air is then delivered to the \DS\ clean rooms with residual \ce{^222Rn} activity below the level of \RadonFilterLimit.

\item[\bf Radon-Suppressed Clean Rooms:] Nuclear recoils from $\alpha$ decays on the innermost surfaces of dark matter detectors are a pernicious source of background for direct dark matter searches, and the daughters of radon on these surfaces are the biggest source.  The \DS\ collaboration has built two radon-suppressed clean rooms in Hall~C, see Fig.~\ref{fig:Facilities-RadonSuppressedCleanRooms}.  These rooms receive air from the radon abatement system and are lined with stainless steel panels to limit radon emanation from the walls.  The concentration of radon in the clean rooms is below \RadonCRHLimit\ (a factor \RadonCRHSuppression\ suppression relative to the Hall~C air).  

The first \ce{^222Rn} suppressed clean-room (\RadonPrincetonCRLimit\ of air) in the world was built at Princeton University in 1998--99 for the construction of the \BX\ nylon vessels, achieving surface activities of $<$10\,$\alpha$'s/(m$^2$$\cdot$d)~\cite{Benziger:2007fm}, also achieved later for the SNO~NCD detectors~\cite{Aharmim:2008jj}) and the NEMO experiment~\cite{Nachab:2007br}.  The \DS\ clean rooms have reached a \ce{^222Rn} level lower than the Princeton clean room by more than two orders of magnitude.  The clean room CR1 contains the equipment used for the cleaning and preparation of the \DSf\ \LArTPC\ parts: the already mentioned HDPE tank and ultrasonic cleaning bath; an Ultra-High-Vacuum (UHV) evaporator (vacuum chamber: \SI{127}{\cm} diameter, \SI{84}{\cm} height, baseline vacuum: \SI{E-8}{\milli\bar}) to coat the innermost surface of the \LArTPC\ with TPB wavelength shifter; and a vacuum oven (\SI[product-units=power]{66 x 66 x 66}{\cubic\cm}).  The clean room CRH is located on top of the water tank and gives direct access into the vetoes through their top flanges. It has a \SI{6}{\meter} vertical clearance, and is served by a \SI{5}{\tonne} crane to permit assembly and installation of the \DSf\ \LArTPC\ in a radon-free environment.

\item[\bf Radon Detector] The \DS\ collaboration, has installed and commissioned an electrostatic \ce{^222Rn} detector (see Refs.~\cite{Takeuchi:1999gr,Kiko:2001ks}).  The detector consists of a stainless steel chamber in which a silicon diode $\alpha$ detector is electrically biased to act as an ion collection electrode, attracting mobile $\alpha$-emitting radon daughters. Samples are introduced via intake and exhaust plumbing, a sampling pump (which can also be run in recirculation mode) and a drier for the input air stream.  The care in design and construction allowed the sensitivity goal of about \SI{100}{\micro\becquerel\per\cubic\meter} needed to monitor the Radon-suppressed air to be achieved.  The chamber is connected through a manifold to the output of the radon abatement system and to the CR1 and CRH clean rooms, so that it can continuously monitor \ce{^222Rn} levels at the output of the radon abatement system or directly in the two clean rooms.

\item[\bf Gas Exhaust Plant:] All gas exhausts coming from within the lab are connected to a negative pressure which directs all the exhaust gases through a carbon filter and purified before being released to the environment (motor way tunnel).
	
\item[\bf Blow Down:] All liquid scintillator vapors coming from bursting discs of the liquid scintillator storage tanks and other containers are collected and recovery into a tank filled with \LNGSHallCPCRecoveryTankSize\ of water and a no-structural packing metal ring.

\item[\bf Nitrogen Supply System:] Hall~C is equipped with three different \LIN\ supplies, shared by the \BX\ and \DSf\ experiments: regular nitrogen (RN2), high purity nitrogen (HPN), and low \ce{Ar}/\ce{Kr} Nitrogen (LAKN).  All the \LIN\ systems are housed just outside of Hall~C in the underground lab.  The RN2 storage is a \SI{30}{\cubic\meter} conventional liquid nitrogen storage tank with a heater to provide boil off nitrogen for pneumatic instrumentation, routine purging of gas lines.  The HPN is mainly used for the gas blankets in the \BX/\DSf\ \PC\ storage area, in the gas stripping operations of the water, as well as in other operations. Radon is being removed from the regular nitrogen by adsorption onto a low temperature adsorber (LTA), {\it i.e.} high purity activated carbon adsorbent at \LINNormalTemperature, to reduce the \ce{^222Rn} concentration in the nitrogen by approximately a factor of \num{100}. The LTA facility is installed in the entryway to Hall~C.  The LAKN is special nitrogen tested to have much lower levels of \ce{Ar} and \ce{Kr} than found in standard commercial liquid nitrogen. The nitrogen does not need any further purification and is supplied to the Hall~C gasified by a heater. This nitrogen is used for blanketing of the scintillator, for gas stripping of pseudocumene during purification and as a service nitrogen for the \DSf\ cryogenic system.

\end{asparaenum}

\subsection{Detector Assembly and Installation Sequence}
\label{sec:Facilities-AssemblySequence}

\begin{figure*}[t!]
\includegraphics[width=\textwidth]{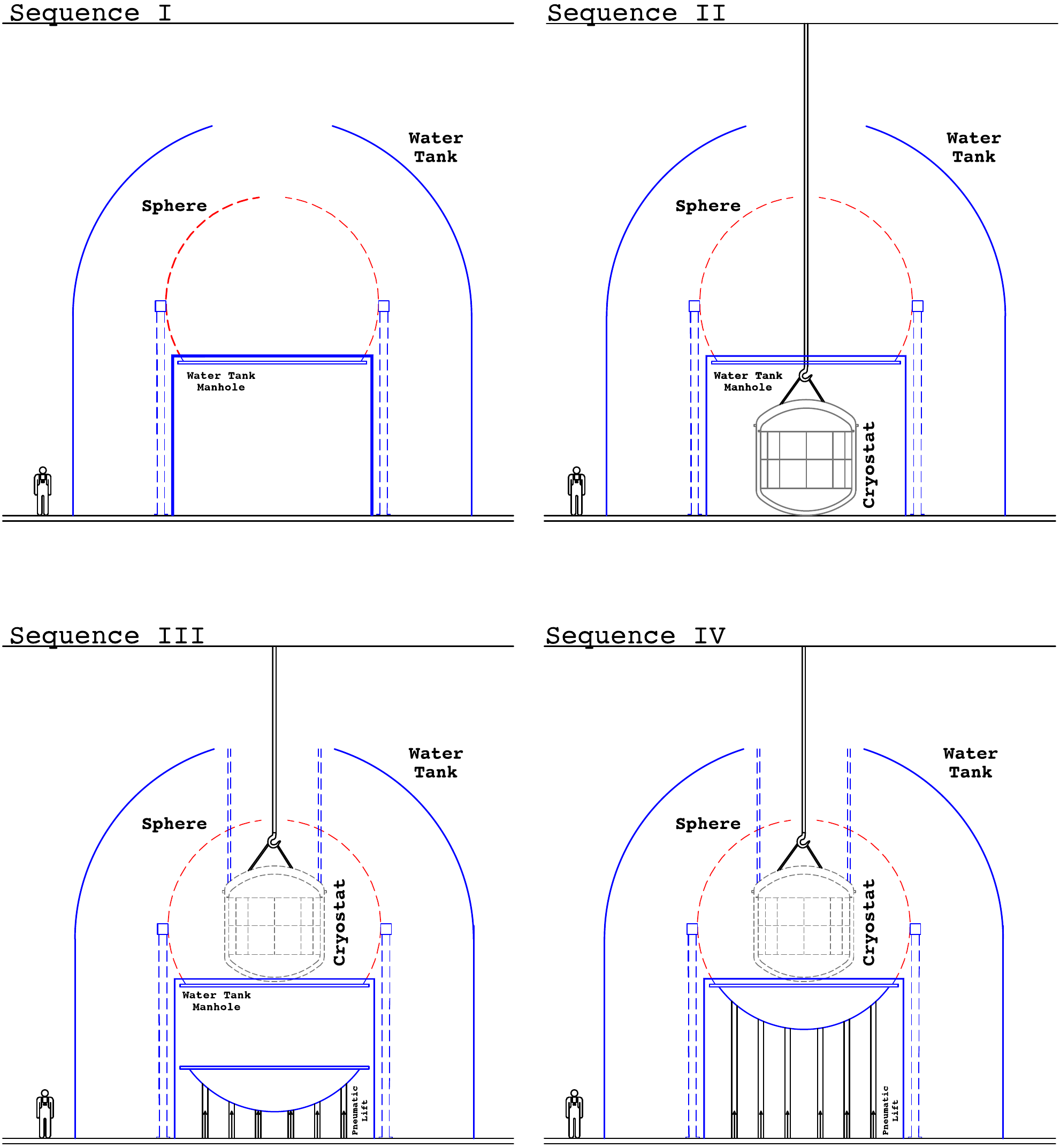}
\caption{Schematic drawing of the cryostat installation sequence.}
\label{fig:Facilities-DSk2DInstallationSequence}
\end{figure*}

The complete assembly of the detectors (water tank, stainless steel sphere, and cryostat) will start with the realization of all the foundation anchor points for the water tank, followed by the construction of the water tank itself.  The \LSV\ stainless steel sphere will then be built inside the water tank, by welding in place pre-formed stainless steel sheets brought into the water tank through the manhole, like was done during the \DSf\ construction.  The material for the stainless steel sphere, legs and reinforced beams will be carefully screened, as done for \DSf.

As was mentioned before, considering the larger size of the \LArTPC, it is foreseen to turn the water tank into a clean room in order to house the final assembly.  Therefore, once the water tank and the stainless steel sphere will be ready for the \LArTPC\ assembly and cryostat insertion, they will be connected with the radon abatement system already installed in Hall~C.  Special removable tents and HEPA filters will be utilized to create a clean environment before entering the two main vessels.  

Just before the assembly of the TPC and its insertion into the cryostat, the interior walls of the stainless steel sphere will be covered with highly-reflective Lumirror, the same material placed on the interior of the \DSf\ \LSV.  The water tank interior walls and the exterior walls of the stainless steel sphere will be covered with the reflective material Tyvek.  The cabling installation, the water tank \DSkVetoPMTDiameter\ \PMTs, the Tyvek installation to cover the internal surface and all the remaining components will be installed during the final phase of installation and before starting the liquid fill tests.

The \DSkLSVBottomFlangeDiameter\ \LSV\ flange will be located on the bottom of the sphere, allowing for the \TPC\ assembly to be done at the ground level of the water tank.  The installation sequence for the cryostat is pictured in Fig.~\ref{fig:Facilities-DSk2DInstallationSequence}.  In order to move the cryostat, it is foreseen to use a temporary rail system to transport the cryostat first into the water tank, then under the stainless steel sphere, where it would then be anchored to the main \LNGSHallCDoubleHookCraneLoadBare+\LNGSHallCDoubleHookCraneLoad\ crane to be lifted in its position at the center of the stainless steel sphere.  At that point the connections of the \LArTPC\ with the cryogenic,\DAQ\ systems will be performed.  The \PMTs\ installation in the water tank and in the stainless steel sphere and the cabling installation will be performed during the final phase of detector installation and before global leak teass and liquid fill tests are performed.

\section{Broader Impacts of the DarkSide Collaboration}
\label{sec:BIOR}
\label{sec:BIOR-BI}

The technologies developed by the \DS\ Collaboration for the \DSf, \DSk, and \Argo\ programs benefit society at large in a variety of ways. Listed below are major broader impact items from the \DS\ program and related activities:

\begin{asparaenum}
\item[\bf Technologies for Medical Diagnostics:] INFN and Princeton University filed patent P137IT00 for an innovative high-definition 3D Positron annihilation vertex Imager: \ThreeDPi.  The \ThreeDPi\ Project has been conceived to develop the technology described in this patent, aimed at overcoming the limitations of Positron Emission Tomography (\PET) and Time-Of-Flight PET (\TOFPET), nuclear imaging techniques used in the fight against cancer and imaging of other metabolic processes, such as neurological-imaging and cardio-imaging.  The market for \PET\ and \TOFPET\ machines is valued in the hundreds of million of dollars per year and is expanding rapidly.  With traditional \PET\ and \TOFPET\ machines, poor resolution in the time of flight measurement of the two \grs\ from positron annihilation inhibits the direct reconstruction of the 3D position of the individual ``vertex'' - the point where the positron annihilates and the gamma-rays are generated.  Therefore \PET\ and \TOFPET\ machines reconstruct clinical images by 2D tomography: first determining the 2D projections in surfaces perpendicular to the \grs\ line of flight, then combining and fitting the 2D projections obtained for different angles to reconstruct the 3D image.  As a consequence, resolution, contrast, and brightness of clinical images obtained with \PET\ and \TOFPET\ machines are sub-optimal.  A \ThreeDPi\ machine would, event-by-event, directly reconstruct the 3D position of each individual vertex, i.e. the location where positron annihilation occurs.  If developed and commercialized, a \ThreeDPi\ machine would turn into an extremely powerful weapon in the fight against cancer.  The key technological advances of \ThreeDPi\ are:
\begin{compactenum}[i)]
\item the replacement of crystal scintillators with doped liquid argon, which couples fast decay time with excellent scintillation light yield;
\item the replacement of traditional \PMTs\ with \SiPMs;
\item and the operation of \SiPMs\ at cryogenic temperature (\LArNormalTemperature), optimal to minimize the noise of \SiPMs.
\end{compactenum}

A resolution of \ThreeDPiTimeResolution\ in the Time-Of-Flight measurement with \ThreeDPi\ is anticipated.  The development of \ThreeDPi\ is synergistic with the \DS\ program with which is shares the \LAr\ target and cryogenic \SiPMs.  A \ThreeDPiTimeResolution\ timing resolution is  the key element enabling the direct, 3D reconstruction of each individual positron annihilation vertex.  Reaching this long-sought milestone would be nothing short of a quantum jump for nuclear imaging.  

The unique capabilities unlocked by \ThreeDPi\ include:
\begin{compactenum}
\item Ability to obtain clinical images with \si{\sim\mm} resolution, allowing oncologists to detect smaller  tumors and to  more precisely define their position, size and shape;
\item Ability to improve the \SNR\ of commercial \TOFPET\ units, producing  brighter and higher contrast clinical images.  This would allow the radiation doses administered to patients to be reduced, typically to \SI{0.1}{\milli\sievert} instead of \SI{10}{\milli\sievert} per clinical exam.  This is of special significance for pediatric patients, who may be subjected to multiple checkup imaging exams in their post-surgery life span;
\item Ability to precisely identify position of misfolded proteins (Amyloid-beta, Tau, Alpha-Synuclein, etc.) with long-latency (decades) accumulation time, possibly responsible for the onset of neurodegenerative disorders, such as Alzheimer and Parkinson diseases.  Very low-dose tracers for these proteins could be used in healthy subjects for preventive and therapeutic strategies;
\item Ability to precisely determine the differential dose delivered to cancerous vs. healthy cells during hadrotherapy sessions.  This revolutionary capability, described in patent P137IT00, would be achieved by injecting patients with special, cancer cell-targeting tracers prior to the therapy session.  These tracers would be  loaded with stable (non-radioactive) isotopes selected for the propensity to be activated into positron emitters by the hadron beam delivered during therapy.
\end{compactenum}

\item[\bf \SiPMs\ for Science and Society:] The focus on development of \SiPMs\ for \DSk\ will drive substantial improvement in this recent and very promising technology.  \SiPMs\ have the potential of positively affecting many applications in a wide variety of fields:
\begin{inparadesc}
\item \SiPMs\ could replace \PMTs\ in many particle physics experiments, especially those with the option of operating \SiPMs\ at cryogenic temperatures, in magnetic environments and in the presence of strong electric fields;
\item \SiPMs\ could outfit future generation of detectors in use for national security purposes;
\item The functional unit of \SiPMs, the \SPAD, finds application in fast light detectors such as those in use for distance sensors in cars.
\end{inparadesc}
Support for \DSk\ will directly and indirectly support this entire range of technologies.

\item[\bf \ce{^4He}:] Helium is a non-renewable resource, which is essential for many high-tech industries and for scientific research.  It plays a strategic role, as it is an indispensable resource for national defense projects and the space exploration industry.  Helium is abundantly used at university laboratories and by high-tech industry, with no viable substitute.  The US Bureau of Land Management (BLM) operates the National Helium Reservoir at Amarillo, the sole government helium storage reservoir, and its webpage lists industrial processes depending on helium and identifies government users that are potentially affected by helium shortages~\cite{USDepartmentofInteriorBureauofLandManagement:2015ui}.  With helium in high and rising demand since the end of the financial crisis, many U.S. universities and laboratories are experiencing rising prices and cutbacks in their helium deliveries.  The \DS\ Collaboration has extracted \UAr\ from underground \ce{CO2} wells at the Kinder Morgan Doe Canyon facility in southwestern Colorado since 2008, measured its He and \ce{^39Ar} content, and enabled the realization of the first commercial, large-scale extraction of underground helium.  Air Products built a helium production plant treating the entire \ce{CO2} stream produced by Kinder Morgan at Cortez~\cite{AirProductsandChemicalsInc:2015tv}.  Production started in \UraniaHeStartDate\ and will replace a \UraniaHeNationalReserveFractionEquivalentRate\ fraction of the declining helium production from the National Helium Reservoir.

\item[\bf \ce{^3He}:] \DS\ researchers have started research on methods to separate \ce{^3He} from massive streams of helium, e.g. the one at DOE Canyon inaugurated in 2015.  If successful, this effort could help solve possible future shortages of \ce{^3He}~\cite{Shea:2010vz}.  \ce{^3He} is very rare and finds applications in nuclear fusion and neutron detection.  It is also a hard to replace component of dilution refrigerators with base temperature in the a few~\si{\milli\kelvin} range.  In addition, Happer and colleagues developed the technique of lung imaging with hyperpolarized \ce{^3He}~\cite{Happer:1984il}, whose practical deployment, which crucially depends on developing large scale \ce{^3He} production processes, could significantly enhance early detection of different types of lung diseases.

\item[\bf \ce{^37Ar}:] The noble gas radioisotope \ce{^37Ar} is of great interest in the detection of underground nuclear tests.  The production of \ce{^37Ar} via the reaction \ce{^40Ca}($n$,$\alpha$)\ce{^37Ar} has a relatively high cross section and is expected to provide a signature of large numbers of neutrons interacting with the soil~\cite{Riedmann:2011ht}.  As a noble gas, \ce{^37Ar} is expected to migrate to the surface following an underground nuclear test, without chemical effects.  With a \ArThreeSevenHalfLife\ half-life, it is long-lived enough to allow time for soil-gas sampling to occur, but is expected to be present at relatively low levels in normal soil gas.  It is present in the atmosphere only in trace amounts.  \ce{^37Ar} detection is quite challenging, as it decays via electron capture, emitting low-energy Auger electrons and X-rays.  Perhaps the most well-known high-sensitivity measurement of \ce{^37Ar} was performed as a means of measuring the solar neutrino flux incident on the Earth~\cite{Cleveland:1998er}.  The major emissions of \ce{^37Ar} decay are summarized in that work.  Nominally, only two K-capture decay channels are important, both with total energy summing to \ArThreeSevenKCaptureXRaysEnergy.  In one, \ArThreeSevenKOneBR\ of decays yield only Auger electrons summing to \ArThreeSevenKCaptureXRaysEnergy.  This is the easiest decay channel to detect, as the electrons will not range far in gases used in typical proportional counters.  Thus, full-energy detection efficiency for the Auger electron emissions will be high.  In the second decay channel of interest, an additional \ArThreeSevenKTwoToFourBR\ of decays yield \ArThreeSevenKCaptureXRaysEnergy\ mostly or completely as X-rays.  The efficiency for detecting the full energy of these X-ray emissions will vary with the proportional counter operating pressure and geometry; at \SIrange{7}{10}{\bar} argon pressure of the PNNL measurements, the detection efficiency is high.  The ideal laboratory capability would be for an \ce{^37Ar} measurement to extend into the expected soil gas background range of about \SIrange{1}{200}{\milli\becquerel\per\cubic\meter} of air~\cite{Riedmann:2011ht}, and would provide sufficient capacity so that worldwide background characterizations could be supported.  Several laboratories worldwide have this capability.  In the last decade, a need has been identified for a U.S.-based effort, focused on \ce{^37Ar} and having both the desired sensitivity and capacity for many parallel measurements.  Filling this gap was one of the motivations for the recent development of a shallow underground laboratory~\cite{Aalseth:2012jl} and new proportional counter design~\cite{Aalseth:2013cg} at PNNL.  In this treaty verification application, the chemistry of argon recovery and purification is important for preparing soil-gas samples for measurement.  This chemistry is synergistic with the challenge of recovering and purifying geologic argon for use in a dark matter detector.  PNNL's focus on \ce{^37Ar} for detection of underground nuclear tests~\cite{Aalseth:2013cg}, as well as their focus on \ce{^39Ar} as an age-dating tracer of aquifer residence time, are natural technical complements to a scientific interest in detecting and elucidating the nature of dark matter with a \LAr\ detector.

\item[\bf \ce{^39Ar}:] Internal-source argon gas-proportional counters are used to detect isotopes useful as environmental radiotracers.  This is one of the most sensitive methods for routine assay of challenging radionuclides like tritium~\cite{Theodorsson:1999dn} and \ce{^39Ar} for the age-dating of water~\cite{Martoff:1992bg}.  As with dark matter detection, the \ce{^39Ar} background in atmospheric-sourced argon becomes an important limit to sensitivity.  The availability of geologic argon from methods developed for \DS\ will extend the reach of these low-level measurements and thus their application for tracers of environmental processes such as groundwater residence time.  A U.S.-based effort has been established at PNNL for ultra-low-level proportional counter measurements of argon for water reserve characterization ({\it e.g.}, groundwater mean residence time) using argon radioisotopes.  \ce{^39Ar} is a cosmogenic isotope found at a constant, but low~\cite{Benetti:2007fg}, concentration in atmospheric argon, which makes up about \SI{1}{\percent} of the atmosphere.  It is thus present as a dissolved gas in rainwater and snow which recharges aquifers.  Once in an aquifer, and thus isolated from exchange with the atmosphere, \ce{^39Ar} radioactive decay decreases its abundance.  The 269-year half-life of \ce{^39Ar}, compared to the 12.4-year half-life of tritium, can allow determination of the age of groundwater sources over an important range spanning 50 to 1000\,\,years.  This allows study of effects on the time-scale of impacts expected due to modern climate change patterns.  Geologic argon samples from the \DS\ collaboration R\&D effort have been recently used at PNNL to characterize ultra-low-background proportional counter~\cite{Aalseth:2009je,Seifert:2012ip} backgrounds.  The minimal level of \ce{^39Ar} activity in the low-radioactivity \UAr\ allows detector background signals to be elucidated.  This established the magnitude of the effect of the \ce{^39Ar} signal versus limiting backgrounds, and allowed for age-dating to be estimated for modest-sized samples, see Fig.~\ref{fig:BroaderImpact-ArThreeNineDating}~\cite{Hall:2016dw}.  The geologic argon development work central to the physics reach of DarkSide will significantly enhance the ability of researchers worldwide to employ \ce{^39Ar} as an environmental radiotracer for hydrologic transport.  An enduring U.S. source of geologic argon to support environmental science applications does not currently exist, but is an opportunity that can be developed from the DarkSide argon recovery and purification progress.  Researchers at PNNL are currently supported by the U.S. National Nuclear Security Agency to develop plans to address this need for access to geologic argon, in collaboration with \DS.

\begin{figure}[h!]
\centering
\includegraphics[width=0.5\columnwidth]{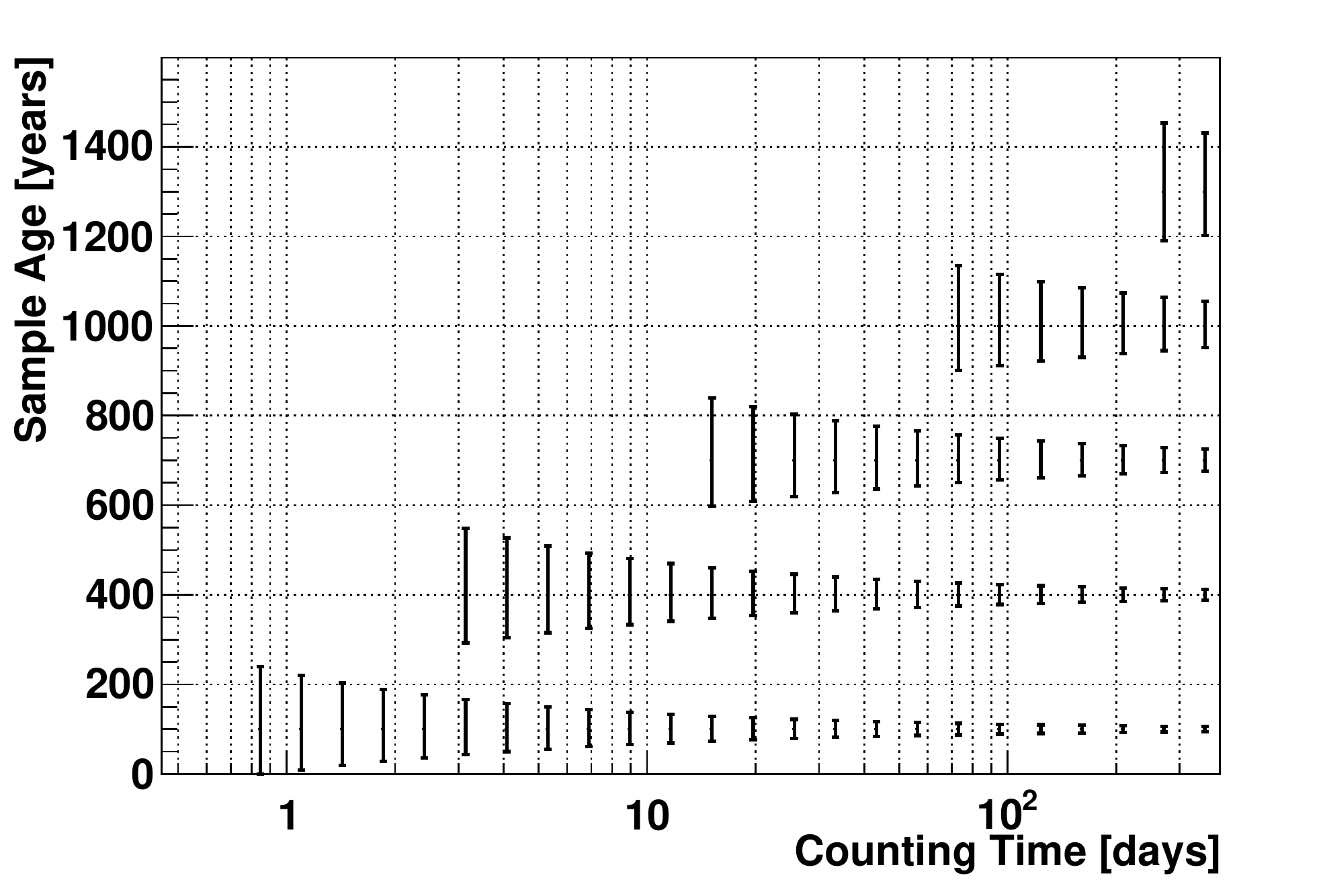}
\caption[Age-dating reach for small argon samples to be recovered from groundwater.]{Age-dating reach for small (\SI{630}{\standard\cubic\cm}) argon samples to be recovered from groundwater.  These measurements depend on using geologic argon as an age-dating endpoint, {\it i.e.}, a detector background reference, and as a ``make-up'' gas for smaller samples.  Bars indicate 1$\sigma$ uncertainty in sample age for five different groundwater age targets.}
\label{fig:BroaderImpact-ArThreeNineDating} 
\end{figure}

\item[\bf Ultra-Pure Gases:] Technologies for gas purification are extremely important in some market segments, such as the pharmaceutical and semiconductor industries.  The Aria distillation columns have a unique capability to produce high-purity gases thanks to their thousands of equilibrium stages.  Perfection of this technology could help special applications of interest to the pharmaceutical and semiconductor industries.

\item[\bf Electronics and Microelectronics:] With the ever-shrinking size of transistors in memory elements and processors, multi-bit flip due to natural and cosmogenic radioactivity is a growing concern of the semiconductor industry.  The radon-abated \CROne\ and \CRH\ clean rooms, with their world-leading \RadonCRHLimit\ limit on \ce{Rn} activity, could benefit studies to characterize and minimize surface $\alpha$-emitter contamination on silicon chips.  A viable measurement technique for surface $\alpha$-emitter contamination that makes use of silicon CCDs was recently introduced in Ref.~\cite{AguilarArevalo:2015hf} and could be useful for this purpose.

\item[\bf IV Generation Nuclear Plants:] The \Aria\ cryogenic distillation columns for isotopic separation will have the ability to separate \ce{^15N} from atmospheric nitrogen gas, nitric oxide or ammonia.  Uranium nitride (\ce{U_n}\ce{^15N_m}) is among the best candidates for  fueling of IV Generation nuclear reactors, due to its superior thermal and mechanical properties~\cite{Zakova:2012dy,Youinou:2014dv,Jaques:2015cw}.  The main drawback is that uranium nitride must be synthesized from \ce{^15N} with purity greater than \SI{99}{\percent}, to avoid neutron absorption on \ce{^14N} and obtain an overall optimal neutron economy of operation. The  advantage of  uranium nitride fuels is that they allow  a decreased frequency of refueling shutdowns and thus higher up-time and greater economy.  The  higher density, higher melting temperature, better thermal conductivity, and lower heat capacity~\cite{Hayes:1990hz,Hayes:1990go} compared with e.g. oxides, help  improve the safety margin in reactor design~\cite{Zhao:2014ia}.   The adoption of uranium nitride as the fuel of choice for IV Generation nuclear reactors would create a new market for \ce{^15N}, valued in the hundreds of millions of dollars per year.  

\item[\bf Precision cleaning:] The custom precision cleaning facility developed for cleaning \DS\ mechanical parts in the clean rooms, achieved benchmarks for cleanliness and cleanliness certification. The European market for precision cleaning is comparatively underdeveloped. This represents a unique opportunity for the creation of a high-technology startup in Italy, in particular in the L'Aquila district, based on technology already proven and demonstrated. This technology is not only of interest in the aerospace industry, but in the pharmaceutical one as well.

\item[\bf ULR \ce{Ti}] The techniques for producing ultra-low radioactivity (ULR) \ce{Ti} being explored for use in \DS\ may yield results of interest for the aerospace, electronics, and optics industries. It is intended to exploit the potential of these technologies in cooperation with private partners in Italy, specifically in the Abruzzo Region.

\item[\bf Nuclear Medicine:] There has been great interest in augmented means of production for \ce{^18O} and \ce{^13C}, used  as precursors of tracer isotopes for tumor therapy, clinical studies, and development of new drugs.  The development of the Aria cryogenic distillation columns for isotopic separation will lead the way to new methods of producing these important isotopes.

\end{asparaenum}

\begin{acknowledgement}
The \DS\ Collaboration would like to thank LNGS and its staff for invaluable technical and logistical support. This report is based upon work supported by the U.~S.~National Science Foundation (NSF) (Grants No.~PHY-0919363, No.~PHY-1004054, No.~PHY-1004072, No.~PHY-1242585, No.~PHY-1314483, No.~PHY- 1314507, associated collaborative grants, No.~PHY-1211308, No.~PHY-1314501, No.~PHY-1455351 and No.~PHY-1606912, as well as Major Research Instrumentation Grant No.~MRI-1429544), the Italian Istituto Nazionale di Fisica Nucleare (Grants from Italian Ministero dell`Istruzione, Universit\`a, e Ricerca Progetto Premiale 2013 and Commissione Scientific Nazionale II). We acknowledge the financial support from the UnivEarthS Labex program of Sorbonne Paris Cit\'e (Grants ANR-10-LABX-0023 and ANR-11-IDEX-0005-02), the S\~ao Paulo Research Foundation (Grant FAPESP-2016/09084-0), and the Russian ScienceÊFoundationÊ(Grant No.~16-12-10369 and Grant No.~16-19-10535).  The authors were also supported by the ``Unidad de Excelencia Mar\'ia de Maeztu: CIEMAT - F\'isica de part\'iculas'' (Grant MDM-2015-0509) and the Polish National Science Centre (Grant No.~UMO-2014/15/B/ST2/02561) and the Foundation for Polish Science (Grant No.~TEAM/2016-2/17).  We also wish to acknowledge the support from Pacific Northwest National Laboratory, which is operated by Battelle for the U.S. Department of Energy under Contract No.~DE-AC05-76RL01830.
\end{acknowledgement}
\clearpage

\clearpage
\bibliographystyle{ds}
\bibliography{ds}
\end{document}